\documentclass{phdthesis}

\title{Field space parametrization\\ in quantum gravity and the identification of a unitary conformal field theory\\
 at the heart of 2D Asymptotic Safety}
\author{Andreas Nink}
\birthplace{Dernbach}
\date{12.\ September 2016}
\dateofexam{19.\ Januar 2017}%{\underline{\hspace{8em}}}

\newcommand{\e}{\mathrm{e}}
\newcommand{\Tr}{\operatorname{Tr}}
\newcommand{\STr}{\operatorname{STr}}
\newcommand{\Trp}{\operatorname{Tr^\prime}}
\newcommand{\detp}{\operatorname{det^\prime}}
\newcommand{\tr}{\operatorname{tr}}
\newcommand{\Rk}{\mathcal{R}_k}
\newcommand{\Rz}{R^{(0)}}
\newcommand{\p}{\partial}

\newcommand{\sg}{\sqrt{g}}
\newcommand{\sbg}{\sqrt{\bar{g}}}
\newcommand{\sbgx}{\sqrt{\bar{g}(x)}}
\newcommand{\sbgy}{\sqrt{\bar{g}(y)}}
\newcommand{\sbgz}{\sqrt{\bar{g}(z)}}
\newcommand{\bg}{\bar{g}}
\newcommand{\hg}{\hat{g}}
\newcommand{\shg}{\sqrt{\hg}}
\newcommand{\shgx}{\sqrt{\hg(x)}}
\newcommand{\shgy}{\sqrt{\hg(y)}}
\newcommand{\shgz}{\sqrt{\hg(z)}}
\newcommand{\tPhi}{\widetilde{\Phi}}
\newcommand{\ZL}{Z_\UV}
\newcommand{\hS}{\hat{S}}
\newcommand{\mn}{{\mu\nu}}
\newcommand{\rs}{{\rho\sigma}}
\newcommand{\bR}{\bar{R}}
\newcommand{\bD}{\bar{D}}
\newcommand{\bB}{\bar{\Box}}
\newcommand{\bz}{\bar{z}}
\newcommand{\bh}{\bar{h}}
\newcommand{\mO}{\mathcal{O}}
\newcommand{\mD}{\mathcal{D}}
\newcommand{\mF}{\mathcal{F}}
\newcommand{\mG}{\mathcal{G}}
\newcommand{\mA}{\mathcal{A}}
\newcommand{\mC}{\mathcal{C}}
\newcommand{\mS}{\mathscr{S}}
\newcommand{\mZ}{\mathcal{Z}}
\newcommand{\mL}{\mathcal{L}}
\newcommand{\mFpq}{\mF_{(p,q)}}
\newcommand{\tFpq}{\widetilde{\mF}_{(p,q)}}
\newcommand{\SymT}{\Gamma\big(S^2 T^* M\big)}
\newcommand{\td}{\text{d}}
\newcommand{\dd}{\td^d}
\newcommand{\dex}{\td^{2+\varepsilon}x}
\newcommand{\mat}{\mathcal{M}}
\newcommand{\Mpq}{\mat_{(p,q)}}
\newcommand{\matrices}{\mathbb{R}^{d\times d}}
\newcommand{\GLd}{\operatorname{GL}(d)}
\newcommand{\Opq}{\operatorname{O}}
\newcommand{\On}{\operatorname{O}(n)}
\newcommand{\Od}{\operatorname{O}(d)}
\newcommand{\diag}{\text{diag}}
\newcommand{\cmat}{c_\text{m}}
\newcommand{\ccr}{c_\text{m}^\text{crit}}
\newcommand{\bo}{{\bar{o}}}
\newcommand{\fg}{\mathfrak{g}}
\newcommand{\fh}{\mathfrak{h}}
\newcommand{\fm}{\mathfrak{m}}
\newcommand{\fD}{\mathfrak{D}}
\newcommand{\Ad}{\mathrm{Ad}}
\newcommand{\hR}{\hat{R}}
\newcommand{\hD}{\hat{D}}
\newcommand{\hB}{\hat{\Box}}
\newcommand{\hV}{\hat{V}}
\newcommand{\hRL}{\widehat{\mathcal{R}}_\UV}
\newcommand{\hRk}{\widehat{\mathcal{R}}_k}
\newcommand{\Ye}{Y_\varepsilon}

\newcommand{\vp}{\varphi}
\newcommand{\cg}{\check{g}}
\newcommand{\cl}{\check{\lambda}}

\newcommand{\ns}{N}
\newcommand{\UV}{\Lambda}
\newcommand{\Ldim}{\mathlarger{\curlywedge}}
\newcommand{\SB}{S_\UV}
\newcommand{\Sgf}{S^\text{gf}}
\newcommand{\Sgh}{S^\text{gh}}

\newcommand{\bP}{\bar{\Phi}}
\newcommand{\hP}{\hat{\Phi}}
\newcommand{\hvp}{\hat{\varphi}}
\newcommand{\hh}{\hat{h}}
\newcommand{\bp}{\bar{\varphi}}
\newcommand{\xb}{\bar{\xi}}
\newcommand{\bx}{\bar{\xi}}
\newcommand{\dmu}{\text{d}\mu\big[\hat{\varphi},C,\bar{C}\mkern1mu\big]}
\newcommand{\bGamma}{\bar{\Gamma}}
\newcommand{\Gk}{\Gamma_k}
\newcommand{\Gg}{\Gamma_k^\text{grav}}
\newcommand{\Gm}{\Gamma_k^\text{m}}
\newcommand{\gsc}{\bar{g}_k^\text{sc}}
\newcommand{\Asc}{\bar{A}_k^\text{sc}}
\newcommand{\ve}{\varepsilon}
\newcommand{\rG}{\mathring{G}}
\newcommand{\rL}{\mathring{\Lambda}}
\newcommand{\rg}{\mathring{g}}
\newcommand{\rl}{\mathring{\lambda}}
\newcommand{\rGk}{\mathring{G}_k}
\newcommand{\rgk}{\mathring{g}_k}
\newcommand{\rLk}{\mathring{\Lambda}_k}
\newcommand{\rlk}{\mathring{\lambda}_k}
\newcommand{\rgs}{\mathring{g}_*}
\newcommand{\rls}{\mathring{\lambda}_*}
\newcommand{\rgDyn}{\mathring{g}^\text{Dyn}}
\newcommand{\rlDyn}{\mathring{\lambda}^\text{Dyn}}
\newcommand{\rrlDyn}{\mathring{\mathring{\lambda}}^\text{Dyn}}
\newcommand{\rgB}{\mathring{g}^\text{B}}
\newcommand{\rlB}{\mathring{\lambda}^\text{B}}
\newcommand{\Ggd}{\Gamma_k^\text{grav,2D}}
\newcommand{\GgdN}{\Gamma_k^\text{grav,2D,NGFP}}
\newcommand{\GL}{\Gamma_k^\text{L}}
\newcommand{\tg}{\tilde{g}}
\newcommand{\stg}{\sqrt{\tg}}
\newcommand{\tR}{\tilde{R}}

\newcommand{\tB}{\tilde{\Box}}
\newcommand{\cgr}{c_\text{grav}^\text{NGFP}}
\newcommand{\SEH}{S^\text{EH}}
\newcommand{\mku}{\mkern1mu}
\newcommand{\mkuu}{\mkern2mu}
\newcommand{\NEH}{N_\text{EH}}
\newcommand{\gfull}{\e^{2\phi}\mku\hg}
\newcommand{\Zm}{Z_\text{matter}}
\newcommand{\Yg}{Y_\text{grav}^\text{NGFP}}
\newcommand{\Gi}{\Gamma^\text{ind}}
\newcommand{\ob}{\mathscr{O}}
\newcommand{\bob}{\bar{\mathscr{O}}}
\newcommand{\Zmc}{Z_\text{m}^{(c)}}
\newcommand{\mP}{\mathcal{P}}
\newcommand{\GkL}{\Gamma_{k,\UV}}
\newcommand{\St}{S_\text{tot}}
\newcommand{\Id}{\mathds{1}}
\newcommand{\PrFull}{\text{Pr}_{\perp(\sg,\sg\mkern1mu R)}}
\newcommand{\PrDiv}{\text{Pr}_{\perp(\text{div})}}
\newcommand{\RL}{\mathcal{R}_{\UV}}
\newcommand{\GLL}{\Gamma_{\UV,\UV}}
\newcommand{\SW}{S_{\UV}^\text{W}}
\newcommand{\DSk}{\Delta S_k}
\newcommand{\cZ}{\check{Z}}
\newcommand{\cx}{\check{\xi}}
\newcommand{\cgamma}{\check{\gamma}}
\newcommand{\cb}{\check{b}}
\newcommand{\cm}{\check{\mu}}
\newcommand{\Nmax}{N_\text{max}}
\newcommand{\ca}{\check{\alpha}}
\newcommand{\cu}{\check{u}}
\newcommand{\cV}{\check{V}}

\newcommand{\GkLiou}{\Gamma_k^\text{L}}
\newcommand{\Mma}{\textit{\textsf{Mathematica}}}
\newcommand{\ntr}{n_\text{tr}}
\def\undertilde#1{\mathord{\vtop{\ialign{##\crcr
$\hfil\displaystyle{#1}\hfil$\crcr\noalign{\kern1.5pt\nointerlineskip}
$\hfil\tilde{}\hfil$\crcr\noalign{\kern1.5pt}}}}}

\newcommand{\gDyn}{g^\text{Dyn}}
\newcommand{\lDyn}{\lambda^\text{Dyn}}
\newcommand{\gB}{g^\text{B}}
\newcommand{\lB}{\lambda^\text{B}}
\newcommand{\dQ}{\delta^\text{Q}}
\newtheorem{theorem}{Theorem}[chapter]
\newtheorem{lemma}[theorem]{Lemma}

\begin{document}

%----------------------------------------------------------------------------------------------------------------------

\frontmatter
\maketitlepage
%----------------------------------------------------------------------------------------------------------------------
\begin{abstract}

\noindent
Although only little is known about the precise quantum nature of the gravitational interaction, we can impose several
essential requirements a consistent theory of quantum gravity must meet by all means: It must be renormalizable in
order to remain well defined in the high energy limit, it must be unitary in order to admit a probabilistic
interpretation, and it must be background independent as the spacetime geometry should be an outcome of the theory
rather than a prescribed input.
Being nonrenormalizable from the traditional, perturbative point of view, for a usual quantum version of general
relativity already the first of these conditions seems to be ruled out.
In the Asymptotic Safety program, however, a more general, nonperturbative notion of renormalizability is proposed, on
the basis of which quantum gravity could be defined within the framework of conventional quantum field theory. The
key ingredient to this approach is given by a nontrivial renormalization group fixed point governing the high energy
behavior in such a way that the infinite cutoff limit is well defined.
While there is mounting evidence for the existence of a suitable fixed point by now, investigations of background
independence are still in their infancy, and the issue of unitarity is even more obscure.

\noindent
In this thesis we extend the existing Asymptotic Safety studies by examining all three of the above conditions and
their compatibility. We demonstrate that the renormalization group flow and its fixed points are sensitive to changes
in the metric parametrization, where different qualified parametrizations, in turn, are seen to correspond to different
field space connections. A novel connection is proposed, and the renormalization group flow resulting from the
associated parametrization and a particular ansatz for the effective average action is shown to possess the decisive
nontrivial fixed point required for nonperturbative renormalizability. For two special parametrizations we argue that
background independence can be achieved in the infrared limit where all quantum fluctuations are completely integrated
out.
In order to study the question of unitarity in an asymptotically safe theory we resort to a setting in two spacetime
dimensions. We provide a detailed analysis of an intriguing connection between the Einstein--Hilbert action in $d>2$
dimensions and Polyakov's induced gravity action in two dimensions. By establishing the 2D limit of an
Einstein--Hilbert-type effective average action at the nontrivial fixed point we reveal that the resulting fixed point
theory is a conformal field theory, where the associated central charge, shown to be $c=25$, guarantees unitarity.
Further properties of this theory and its implications for the Asymptotic Safety program are discussed.
In the last part of this work we present a strategy for conveniently reconstructing the bare theory pertaining to a
given effective average action. For the Einstein--Hilbert case we prove the existence of a nontrivial fixed point in
the bare sector and exploit the dependence of the bare action on the underlying functional measure to simplify the maps
between bare and effective couplings. Applying this approach to 2D asymptotically safe gravity coupled to conformal
matter we uncover a number of surprising consequences, for instance for the gravitational dressing of matter field
operators and the KPZ scaling relations.

\end{abstract}
%----------------------------------------------------------------------------------------------------------------------

\cleardoublepage

%----------------------------------------------------------------------------------------------------------------------
\begin{abstractgerman}

\noindent
Auch wenn über den genauen Quantencharakter der gravitativen Wechselwirkung bislang nur wenig bekannt ist, können wir
einige Forderungen aufstellen, die eine konsistente Theorie der Quantengravitation zwingend erfüllen muss:
Sie muss renormierbar sein, um auch im Hochenergielimes wohldefiniert zu bleiben, sie muss unitär sein, um eine
Wahrscheinlichkeitsinterpretation zuzulassen, und sie muss hintergrundunabhängig sein, da die Raumzeitgeometrie keine
Vorgabe, sondern ein Ergebnis der Theorie sein sollte. Da eine gewöhnliche Quantenversion der allgemeinen
Relativitätstheorie aus störungstheoretischer Sicht nicht-renormierbar ist, scheint bereits die erste dieser
Bedingungen ausgeschlossen.
Das Asymptotic-Safety-Programm schlägt jedoch einen allgemeineren, nicht-störungstheoretischen Begriff von
Renormierbarkeit vor, anhand dessen Quantengravitation im Rahmen konventioneller Quantenfeldtheorie definiert werden
könnte. Die Grundidee basiert auf einem nicht-trivialen Renormierungsgruppenfixpunkt, an dem der Limes des unendlichen
Cutoffs gebildet werden kann, sodass das Hochenergieverhalten in diesem Zugang wohldefiniert bleibt.
Während es inzwischen vermehrt Hinweise für die Existenz eines geeigneten Fixpunktes gibt, haben die Untersuchungen
zur Hintergrundabhängigkeit gerade erst begonnen, und das Unitaritätsproblem ist derzeit noch unklarer.

\noindent
In der vorliegenden Arbeit werden die bisherigen Studien zu Asymptotic Safety erweitert, indem alle drei der obigen
Bedingungen sowie deren Kompatibilität untersucht werden. Wir zeigen, dass der Renormierungsgruppenfluss und dessen
Fixpunkte von der Parametrisierung der Metrik abhängen, wobei unterschiedliche Parametrisierungen wiederum auf
unterschiedliche Zusammenhänge im Feldraum zurückgeführt werden können. Im Hinblick darauf schlagen wir einen neuen,
eigens konstruierten Zusammenhang vor und weisen nach, dass der Renormierungsgruppenfluss, der sich aus der zugehörigen
Parametrisierung und einem speziellen Ansatz für die effektive Mittelwertwirkung ergibt, einen für die
nicht-störungstheoretische Renormierbarkeit erforderlichen Fixpunkt aufweist.
Für zwei bestimmte Parametrisierungen legen wir dar, dass im Infrarotlimes, in dem alle Quantenfluktuationen
vollständig ausintegriert sind, Hin\-ter\-grund\-un\-ab\-häng\-ig\-keit tatsächlich erreicht werden kann.
Um die Frage nach Unitarität in einer asymptotisch sicheren Theorie zu erörtern, bedienen wir uns eines Szenarios in
einer $2$-dimensionalen Raumzeit. Hierbei decken wir einen verblüffenden Zusammenhang zwischen der
Einstein--Hilbert-Wirkung in $d>2$ Dimensionen und Polyakovs induzierter Gravitationswirkung in zwei Dimensionen auf.
Indem wir den 2D-Limes einer effektiven Mittelwertwirkung des Einstein--Hilbert-Typs am nicht-trivialen Fixpunkt
bilden, können wir zeigen, dass die resultierende Fixpunkttheorie eine konforme Feldtheorie ist, und dass die
entsprechende zentrale Ladung, die wir zu $c=25$ berechnen, Unitarität gewährleistet. Darüber hinaus diskutieren wir
weitere Eigenschaften dieser Theorie sowie die Implikationen für das Asymptotic-Safety-Programm.
Im letzten Teil der Arbeit stellen wir eine Strategie vor, mittels derer die nackte (mikroskopische) Theorie zu einer
gegebenen effektiven Mittelwertwirkung zweckmäßig rekonstruiert werden kann. Für den Einstein--Hilbert-Fall beweisen
wir die Existenz eines nicht-trivialen Fixpunktes auf nackter Ebene und nutzen die Abhängigkeit der nackten
Wirkung von dem zugrundeliegenden Funktionalmaß aus, um die Abbildungen zwischen den nackten und den effektiven
Kopplungen zu vereinfachen. Durch Anwenden dieser Methode auf 2D asymptotisch sichere Gravitation, die an konforme
Materie gekoppelt ist, enthüllen wir eine Reihe überraschender Konsequenzen, die sich beispielsweise für den
gravitativen Effekt auf Materiefeldoperatoren und für die KPZ-Relationen ergeben.

\end{abstractgerman}
%----------------------------------------------------------------------------------------------------------------------

\cleardoublepage

\begin{Spacing}{1.15}

\tableofcontents

%----------------------------------------------------------------------------------------------------------------------
\chapter*{List of abbreviations}
\addcontentsline{toc}{chapter}{List of abbreviations}
%----------------------------------------------------------------------------------------------------------------------

\vspace{1em}
\begin{center}
\begin{tabular}{ll}
\hline
Abbreviation \hspace{1em} & Meaning $\vphantom{\Big|}$ \\
\hline
2D & $2$-dimensional \\
BRST & Becchi--Rouet--Stora--Tyutin \\
CDT & causal dynamical triangulation \\
CFT & conformal field theory \\
EAA & effective average action \\
EH & Einstein--Hilbert \\
FRG & functional renormalization group \\
FRGE & functional renormalization group equation \\
GR & general relativity \\
IR & infrared \\
KPZ & Knizhnik--Polyakov--Zamolodchikov \\
LC & Levi-Civita \\
LHS & left hand side \\
NGFP & non-Gaussian fixed point \\
QED & quantum electrodynamics \\
QEG & Quantum Einstein Gravity \\
QFT & quantum field theory \\
RG & renormalization group \\
RHS & right hand side \\
UV & ultraviolet \\
VDW & Vilkovisky--DeWitt \\
WI & Ward identity \\
\hline
\end{tabular}
\end{center}

%----------------------------------------------------------------------------------------------------------------------

\mainmatter

%----------------------------------------------------------------------------------------------------------------------
\chapter{Introduction}
\label{chap:Intro}
%----------------------------------------------------------------------------------------------------------------------

%Acquiring a deeper understanding of the quantum nature of gravitation is one of the most fascinating and challenging
%open problems in theoretical physics.
It is one of the most fascinating and challenging open problems in theoretical physics to acquire a deeper
understanding of the quantum nature of gravitation. Remarkably enough, the two apparent pillars of quantum gravity,
quantum field theory on the one hand and Einstein's classical theory of gravity on the other hand, are among the most
accurately verified theories in physics and lead to strikingly precise predictions such as, for instance, the anomalous
magnetic moment in quantum electrodynamics, and the perihelion precession of Mercury in general relativity.
However, the perturbative nonrenormalizability of Einstein gravity prevents a straightforward unification of the two
concepts and seems to curtain the fundamental theory at the heart of quantum gravity \cite{Kiefer2012,Hamber2009}.

These difficulties do not imply a defect of quantum field theory or gravity per se, but rather hint at the limitations
of perturbation theory. A particularly interesting approach following this possibility is based on a more general,
nonperturbative notion of renormalizability, referred to as \emph{Asymptotic Safety} \cite{Weinberg1976,Weinberg1979}.
The key idea of this program consists in that the underlying coupling constants governing the strength of interactions
are not plagued by unphysical singularities at high energies but converge to finite, not necessarily small fixed point
values instead.

During the past two decades, Asymptotic Safety matured from a hypothetical scenario to a theory with a realistic chance
to describe the structure of spacetime and the gravitational interaction consistently and predictively, even on the
shortest length scales possible. In particular, there is mounting evidence supporting the existence of the decisive
nontrivial renormalization group (RG) fixed point in the space of coupling constants \cite{NR06,Percacci2009,RS07,%
RS12,Nagy2014,NRS13,CPR09}.

Apart from these promising results concerning nonperturbative renormalizability there are several further properties a
fundamental quantum theory of gravity \emph{must} possess. The two most important ones are \emph{background
independence} and \emph{unitarity}.
A background independent theory is characterized by the absence of any prescribed geometrical background structure: The
structure of spacetime, usually encoded in a dynamical metric, must be an outcome of the theory rather than an input.
Unitarity refers to the absence of unphysical states with negative norm; only under this condition the probabilistic
interpretation of quantum mechanics and quantum field theory can be maintained.

In the light of these considerations a virtually inevitable question suggests itself: Is there a theory of the
gravitational field with the correct classical limit that combines all three crucial properties at the same time,
i.e.\ is there a theory that is nonperturbatively renormalizable \emph{and} background independent \emph{and} unitary?

Although giving a final answer to this question seems to be out of reach with the methods presently at hand, we may
shed some light on the issue by decomposing it into smaller subsets which are more easily accessible.
First, we can study the compatibility of Asymptotic Safety and the requirement for background independence. Second, we
can investigate whether Asymptotic Safety can be reconciled with unitarity in principle.
Finding positive answers in both cases would mark another important step for the Asymptotic Safety program.

It turns out that, for both technical and conceptual reasons, a quantum field theoretical description of Einstein
gravity actually requires the introduction of a background field \cite{DeWitt2003}. This does not necessarily imply a
violation of
the principle of background independence, though. It is perfectly possible that the background field serves merely as
an auxiliary tool during the intermediate steps of calculation, and in the end all physical predictions are independent
of it. This is precisely the approach we pursue in this thesis. We introduce a background metric $\bg_\mn\mku$, use it
to define a scale dependent version of the effective action, the effective average action $\Gamma_k\mku$, and aim at
demonstrating, at least for a special case, that the essential part of $\Gamma_k$ is $\bg_\mn$-independent in the limit
of vanishing RG scale $k$, that is, when all quantum fluctuations have been integrated out completely.

Before proceeding along these lines, however, we shall discuss another as yet unsolved structural problem. It
originates from the fact that, despite its name, the RG is rather a semigroup since the number of degrees of freedom
decreases during each RG step. In general, the flow direction (from ultraviolet to infrared scales) cannot be reversed.
Hence, without further assumptions (such as fixing the types of variables during the RG evolution) we have no direct
access to the physics at short distances, and the \emph{fundamental variables are unknown} in principle. In the case of
gravity they may or may not be given by a metric field. Furthermore, there may be several different ways to parametrize
them in terms of the background field and some sort of fluctuations.

In this work we study in detail two particular parametrizations of the dynamical metric $g_\mn\mku$, the
\emph{linear split}
\begin{equation}
 g_\mn = \bg_\mn + h_\mn \, ,
\label{eq:LinFirst}
\end{equation}
and the \emph{exponential parametrization}
\begin{equation}
 g_\mn=\bg_{\mu\rho}\mku\big(\e^h\big)^\rho{}_\nu \, ,
\label{eq:ExpFirst}
\end{equation}
where in both cases the fluctuations are given by a symmetric tensor field, $h_\mn=h_{\nu\mu}\mku$, and indices are
raised and lowered by means of the background metric. Although these two parametrizations have already been employed
in the literature on Asymptotic Safety, they have merely been considered as convenient choices for performing
calculations so far. We will argue, however, that they have a much more fundamental meaning which we discuss on
the basis of connections and geodesics on field space. Interestingly enough, \eqref{eq:LinFirst} and
\eqref{eq:ExpFirst} do \emph{not even parametrize the same object}:
The set of tensor fields that can be represented by the linear parametrization is larger than the set of tensor fields
that can be written in the form \eqref{eq:ExpFirst}. This will lead to differences of the respective RG
flows, whereas the discussion and the main results concerning background independence are essentially the same for both
parametrizations. It is remarkable that even universal (i.e.\ cutoff scheme independent) quantities like the fixed
point value of the running Newton constant near two dimensions can depend on the way the metric is parametrized.

From the Asymptotic Safety perspective the two-dimensional setting is particularly interesting: The mass dimension of
the running Newton constant, $[G_k]=2-d$, vanishes in exactly $d=2$ spacetime dimensions, and a perturbative treatment
becomes feasible. This approach involves computing the $\beta$-functions (i.e.\ the vector field which drives the RG
flow) in $d=2+\ve > 2$ dimensions and expanding them in terms of $\ve$. A general consideration \cite{Weinberg1979}
shows that the $\beta$-function of the dimensionless Newton constant, $g_k\equiv k^{d-2} G_k\mku$, must be of the form
\begin{equation}
 \beta_g = \ve\mku g_k - b\mku g_k^2 \,,
\label{eq:betagIntro}
\end{equation}
with a positive constant $b$. Notably, this $\beta$-function possesses a nontrivial RG fixed point, defined by the
zero, $\beta_g(g_*)=0$, resulting in the fixed point value %$g_* = \ve/b$.
\begin{equation}
 g_* = \ve/b \,.
\label{eq:NGFPproptoVe}
\end{equation}
Hence, already the perturbative analysis
demonstrates the applicability of the As\-ymp\-to\-tic Safety program in principle.
In fact, eq.\ \eqref{eq:betagIntro} can be reproduced also non\-per\-tur\-ba\-tive\-ly. This is what makes the
($2+\ve$)-dimensional case so special; it allows us to \emph{test nonperturbative results perturbatively}.

Note that the structure of the gravitational $\beta$-function in $2+\ve$ dimensions agrees with the one of an
$\operatorname{SU}(N)$ Yang--Mills theory in $4+\ve$ dimensions, where the running of $\alpha_s(k)\equiv
\frac{g_s^2(k)}{4\pi}$, with $g_s(k)$ the dimensionless version of the strong coupling constant, is given by $k\p_k
\alpha_s(k) = \beta_\alpha = \ve\mku\alpha_s(k) - b_s\mku \alpha_s^2(k)$ \cite{RW94a}. The positive coefficient
$b_s = \frac{11\mku N}{6\pi}$ entails asymptotic freedom in exactly $d=4$ dimensions, while there is a nontrivial fixed
point for $d>4$.

We show in this thesis that the crucial coefficient $b$ in \eqref{eq:betagIntro} depends on the choice of the
underlying metric parametrization. Although it remains positive, its numerical value changes when switching between
\eqref{eq:LinFirst} and \eqref{eq:ExpFirst}. In spite of this parametrization-dependence, $g_*$ at lowest order is
always proportional to $\ve$.

The significance of a suitable RG fixed point for the Asymptotic Safety scenario justifies a closer look to its
properties. After having chosen a metric parametrization we may ask the question about the precise nature of the action
functional which describes this fixed point. In which way exactly does it depend on the metric, the background metric,
and the Faddeev–Popov ghosts? Is it local? What are the structural properties of the fixed point theory, i.e.\ the one
defined directly at the fixed point itself rather than being defined by an RG trajectory running away from it?  Is this
theory a conformal field theory?

Since conformal invariance implies scale invariance, any conformal field theory in a theory space governed by the RG
must be located at a fixed point as, by definition, only fixed points are unaffected by changes of the RG scale. The
reverse, on the other hand, seems to hold only in two spacetime dimensions: Under a few technical assumptions,
\emph{scale invariant 2D quantum field theories are necessarily conformally invariant} \cite{Nakayama2015}. In four
dimensions, however, it is still unclear whether (and under what conditions) scale invariant fixed point theories
possess the full conformal symmetry.
For this reason we shall focus on the 2D case when discussing the conformal character of a fixed point theory. If,
indeed, we identified a conformal field theory, the issue of unitarity could then be studied in a straightforward way
by making use of well-known arguments which are established for generic conformally invariant theories in two
dimensions \cite{Ginsparg1988}.

It may be somewhat unexpected that taking the 2D limit of an action defined in $d>2$ dimensions can be a formidable
task, in fact, depending on the behavior of the coupling constants and the geometrical properties of the invariants
appearing in that action. As for gravity, we are mainly interested in (effective average) actions of the
\emph{Einstein--Hilbert} type:
\begin{equation}
 \Gamma_k^\text{EH}[g] = \frac{1}{16\pi G_k} \int \dd x \sg \,\big( -R + 2\mku\Lambda_k\big)\,,\qquad d>2\,,
\label{eq:EHIntro}
\end{equation}
where $R$ is the scalar curvature, and $G_k$ and $\Lambda_k$ denote the dimensionful running Newton and cosmological
constant, respectively. The key point is that, according to eq.\ \eqref{eq:NGFPproptoVe}, $G_k$ is proportional to
$\ve=d-2$ in the vicinity of the fixed point, and we will see later on that $\Lambda_k\propto\ve$, too. Hence, the
cosmological term in \eqref{eq:EHIntro} remains finite in the limit $\ve\to 0$, while the curvature term seems to
diverge as it contains the factor $G_k^{-1}\propto\ve^{-1}$. On the other hand, in exactly $d=2$ dimensions, the
integral $\int\td^2 x\sg\,R$ becomes trivial in the sense that it is purely topological and fully independent of the
metric. Loosely speaking, the combination of the integral and the prefactor $\propto G_k^{-1}\mku$ thus leads to the
problematic limit $\frac{1}{16\pi G_k}\int\td^{2+\ve} x\sg\, R \to \text{``$0/0\mkuu$''}$ for $\ve\to 0$.
We will demonstrate that it is actually possible to make sense of this limit. Remarkably enough, its essential part
amounts to a nontrivial, finite, \emph{nonlocal} functional which is proportional to the \emph{induced gravity action}
\begin{equation}
 I[g]\equiv \int\td^2 x\sg\,R\,\Box^{-1} R\,,
\end{equation}
where $\Box^{-1}$ is the inverse of the Laplacian. It is this limit action that is used to investigate the conformal
properties of the fixed point theory. In this manifestly two-dimensional setting, the question concerning unitarity
has a precise answer.

By writing the metric $g_\mn$ in terms of a conformal factor and a reference metric, $g_\mn = \e^{2\phi}\hg_\mn\mku$,
the fixed point functional can be expressed as a \emph{Liouville action},
\begin{equation}
 \GL[\phi;\hg] = (-2a_1)\int\td^2x\shg \left({\frac{1}{2}}\hD_\mu\phi\hD^\mu\phi
  + {\frac{1}{2}}\hR\mku\phi - {\frac{a_2}{4}}\e^{2\phi} \right) ,
\label{eq:LiouvilleIntro}
\end{equation}
plus a term that is independent of the conformal mode $\phi$. Actions of the type \eqref{eq:LiouvilleIntro} play an
important role in 2D quantum gravity and noncritical string theory \cite{Nakayama2004}. Here, the coupling constants
$a_1$ and $a_2$ depend on the properties of the fixed point. The requirement for unitarity of the microscopic theory
will be seen to impose the constraint $a_1>0$.
However, if this is indeed satisfied, the kinetic term of $\phi$ has the ``wrong'' sign, apparently leading to an
instability of the conformal mode. Thus, unitarity on the one hand and stability of $\phi$ on the other hand are
\emph{mutually exclusive}. We will discuss in detail whether or not this circumstance is problematic from the physics
point of view.

Finally, we address ourselves to an analysis of the microscopic (``classical'') system corresponding to a given RG
trajectory and a fixed point. Most nonperturbative studies on Asymptotic Safety are based upon the effective average
action rather than the bare action. In this context, RG trajectories are fully determined by the respective initial
conditions and an RG evolution equation alone, dispensing with the need for a bare action and a functional integral.
While all physically relevant quantities like $n$-point functions are already contained in the effective average
action, gaining insight into the bare theory might nonetheless be of interest in certain cases, for instance when a
connection between Asymptotic Safety and other approaches to quantum gravity is to be established. After choosing an
appropriately regularized functional measure we show that \emph{the bare action can be ``reconstructed''} from the
effective theory in such a way that the corresponding functional integral reproduces the prescribed effective average
action.

We reconstruct the bare action for two different underlying systems: for an effective average action of the
Einstein--Hilbert type, eq.\ \eqref{eq:EHIntro}, and one of the Liouville type given by eq.\ \eqref{eq:LiouvilleIntro}.
In this manner we obtain mappings from RG trajectories on the effective side to trajectories in the space of bare
couplings, parametrized by some ultraviolet cutoff scale. For the Einstein--Hilbert case we discuss whether the RG
fixed point always has a counterpart on the bare side. As a direct application of this consideration, the path integral
for a gravity+matter theory in $d=2$ dimensions is constructed explicitly. It can be used to investigate the
gravitational dressing of matter field operators when asymptotically safe gravity is coupled to conformal matter. In
this regard, it would be particularly interesting to see if the well-known Knizhnik--Polyakov--Zamolodchikov (KPZ)
scaling can be observed in this system, too.
\medskip

This work is organized as follows. Apart from Chapter \ref{chap:Framework}, a preparatory chapter introducing the
fundamentals of the functional renormalization group, Asymptotic Safety, and conformal field theory, the body of the
thesis consists of three major parts: the study of \textbf{(1)} parametrization dependence in quantum gravity,
\textbf{(2)} the 2D limit of asymptotically safe gravity, and \textbf{(3)} the reconstruction of bare theories.
\smallskip

\noindent
\textbf{(1)}
Chapter \ref{chap:SpaceOfMetrics} contains a thorough analysis of the space of metrics. Making use of
methods from differential geometry and group theory we define several connections on this space. In that context,
different metric parametrizations correspond to geodesics based on different connections. We advocate one specific
connection which is adapted to the structure of the space of metrics. In a discussion on global geodesics we carefully
distinguish between Euclidean and Lorentzian metrics. This chapter is the most mathematical one.

While Chapter \ref{chap:SpaceOfMetrics} illuminates different metric parametrizations from the mathematics point of
view, Chapter \ref{chap:ParamDep} focuses on their physical implications. Choosing an effective average action as in
eq.\ \eqref{eq:EHIntro}, supplemented by suitable gauge fixing and ghost terms, we determine the running of the
dimensionless Newton constant $g_k$ and the dimensionless cosmological constant $\lambda_k$ by means of functional
RG methods, while paying particular attention to the existence and parametrization dependence of nontrivial fixed
points suitable for the Asymptotic Safety program. The question about background independence is addressed in a
so-called bimetric computation.
\smallskip

\noindent
\textbf{(2)}
In Chapter \ref{chap:EHLimit} we consider the local Einstein--Hilbert action \eqref{eq:EHIntro} which describes
quantum gravity in $d>2$ dimensions and construct its limit of exactly two dimensions. Exploiting the fact that the
Newton constant is of the order $\ve=d-2$ we find that this limit action is a nonlocal functional of the metric. We
discuss the influence of zero modes of the Laplacian and comment on a potential generalization to four dimensions.

Chapter \ref{chap:NGFPCFT} concerns the nature of the 2D limit of the fixed point theory following from the results
obtained in Chapter \ref{chap:EHLimit}. We examine if it represents a conformal field theory and if it is unitary.
Furthermore, the conformal factor problem is put in perspective by making a point on physical state conditions and the
compatibility with unitarity.
\smallskip

\noindent
\textbf{(3)}
In Chapter \ref{chap:Bare} we demonstrate that there is a one-loop relation between the effective average action
and the bare action provided that the measure of the associated functional integral is fixed. As an example, we map the
RG flow pertaining to eq.\ \eqref{eq:EHIntro} onto its counterpart in the space of bare coupling constants. We explain
how this mapping can be simplified by choosing the functional measure appropriately. Under the assumption that there is
a fixed point on the effective side we show that there exists also a bare fixed point.

Chapter \ref{chap:FullReconstruction} is devoted to the bare side of the 2D fixed point theory and a to comparison of
Asymptotic Safety to other approaches to 2D gravity. For that purpose we reconstruct the functional integral describing
asymptotically safe gravity coupled to conformal matter and investigate whether or not KPZ scaling occurs. We discuss
similarities and differences compared with noncritical string theory and Monte Carlo simulations in the causal
dynamical triangulation approach.

Chapter \ref{chap:BareLiouville} is a first attempt to reconstruct the bare action for a Liouville-type effective
average action. Several ans\"{a}tze for the bare action are made to determine the corresponding bare couplings, and
various criteria such as Ward identities for testing their consistency are suggested.
\medskip

Each chapter begins with an executive summary stating its motivation and most important results. If its content is
based on already published, own material, we provide the corresponding reference. Finally, a concluding
discussion and an outlook is presented in Chapter \ref{chap:Conclusion}.

The main chapters are supplemented with a number of appendices. While Appendices \ref{app:Variations} --
\ref{app:Cutoffs} cover numerous general relations that are used throughout this thesis, Appendices \ref{app:ExpParam}
-- \ref{app:WeylMeasureCutoff} are assigned to specific chapters. They consist of additional material like detailed
calculations and proofs.

%----------------------------------------------------------------------------------------------------------------------
\chapter{Theoretical foundations}
\label{chap:Framework}
%----------------------------------------------------------------------------------------------------------------------

\begin{summary}
This chapter introduces three essential pieces of equipment that are needed for our subsequent discussions: the
functional renormalization group, Asymptotic Safety and conformal field theory. (i) After reviewing the general concept
of the renormalization group, we show how the ideas can be formulated in a functional language by defining a scale
dependent effective action and stating the corresponding evolution equation. In order to apply this machinery to
gravity we employ the background field method. (ii) The Asymptotic Safety program suggests that the unphysical
ultraviolet divergences occurring in conventional perturbative quantum gravity can be circumvented by means of a
nontrivial renormalization group fixed point. (iii) Anticipating that there is a connection between the 2D limit of
asymptotically safe gravity and 2D conformal field theory, we present a brief introduction to the latter theory, with
a special focus laid on the issue of unitarity.

\noindent
\textbf{Based on:} Partially Ref.\ \cite{NRS13}.
\end{summary}

%----------------------------------------------------------------------------------------------------------------------
\section{The functional renormalization group}
\label{sec:FRG}
%----------------------------------------------------------------------------------------------------------------------

%----------------------------------------------------------------------------------------------------------------------
\subsection{General concept}
\label{sec:FRGgen}
%----------------------------------------------------------------------------------------------------------------------

In the early stages of its development, ``renormalization'' was regarded merely as a tool to tame infinities in Feynman
diagrams. This understanding changed with the advent of the renormalization group (RG), though. Following the idea that
\emph{scale} determines the perception of the world, it has been realized that coupling constants can vary rather than
being strictly constant, and that their change is described by renormalization group equations which relate couplings
at different (momentum/cutoff) scales \cite{SP53,GL54}.

Inspired by Kadanoff's block spin transformations \cite{Kadanoff1966}, Wilson formalized the concept of scale
transformations in the language of functional integrals \cite{Wilson1971a,Wilson1971b,WK74}, paving the way for the
functional renormalization group (FRG). It governs the change of a physical system due to smoothing or averaging
out microscopic details when going to a lower resolution. Wilson's version of the FRG is implemented by means
of a \emph{scale dependent bare action}, the Wilson action $\SW$, which is defined in such a way that lowering the
scale from $\UV$ to $\UV'<\UV$ amounts to integrating out those modes in the functional integral whose momenta are
restricted by $\UV'^{\mku 2}\le p^2\le \UV^2$, giving rise to a new action $S^\text{W}_{\UV'}$ defined at the scale
$\UV'$. The variation of $\SW$ with respect to $\UV$ is then dictated by RG equations. While there is no simple
representation of these RG equations in Wilson's original formulation which relies on a sharp cutoff, the
generalization to smooth cutoffs allows for deriving them in a compact form, the Polchinski equation
\cite{Polchinski1984}.

From a practical point of view, using the Wilson action as the fundamental object has the disadvantage that extracting
physical information requires performing the remaining functional integration (over modes with momenta between $\UV'$
and $0$ in the above example) in order to obtain the corresponding effective action, see Refs.\
\cite{BB01,BTW02,Delamotte2012} for reviews.
Working with a \emph{scale dependent effective action}, on the other hand, would be more intuitive and more appropriate
for calculations, in particular in the context of gauge theories. It is this latter type of action, the \emph{effective
average action}, that we employ throughout this thesis.

%----------------------------------------------------------------------------------------------------------------------
\subsection{The effective average action and its FRGE}
\label{sec:EAAFRGE}
%----------------------------------------------------------------------------------------------------------------------

In order to clarify the concept, we start by formally defining the effective average action (EAA) by means of
functional integrals. Here, ``formally'' refers to the fact that this approach depends on the precise definition of the
functional measure. Later on we will obtain the EAA as a solution of its RG equation rather than employing a functional
integral-based construction, so we dispense with the need for specifying a measure and an ultraviolet (UV)
regularization prescription.\footnote{A precise knowledge of the functional measure becomes necessary only if the bare
action is of interest. This situation is discussed in more detail in Chapter \ref{chap:Bare} and Appendix
\ref{app:Measure}.}
\smallskip

\noindent
\textbf{(1) Effective average action.}
The basic method is demonstrated for scalar fields in the following, while the generalization to the gravitational
field is discussed in Subsection \ref{sec:FRGgrav}. Let $\chi$ denote a scalar field, $J$ its corresponding source, and
$S[\chi]$ the bare action. We employ the condensed notation $J\cdot\chi$ for a spacetime integration: $J\cdot\chi\equiv
\int\dd x\sg \, J(x)\mku\chi(x)\mku$. The key idea behind the EAA is to modify the standard partition function such
that \emph{high momentum modes are integrated out} while \emph{low momentum modes are suppressed}, see Figure
\ref{fig:ModeSupression}.
(It is implied that fields are expanded in terms of eigenmodes of the covariant Laplacian, $-D^2$, and squared
``momenta'' refer to the corresponding eigenvalues.) To this end, we add a ``cutoff action'' $\Delta S_k[\chi]$ in the
exponent of the integrand, leading to the definition
\begin{equation}
 Z_k[J] \equiv \int\mD\chi\; \e^{-S[\chi] - \Delta S_k[\chi] + J\cdot\chi\mku}\,,
\label{eq:DefZk}
\end{equation}
where the cutoff action can be written as $\Delta S_k[\chi]\equiv \frac{1}{2}\mkuu\chi\cdot\Rk\mku\chi$ with the cutoff
operator $\Rk\equiv\Rk\big(-D^2\big)$.
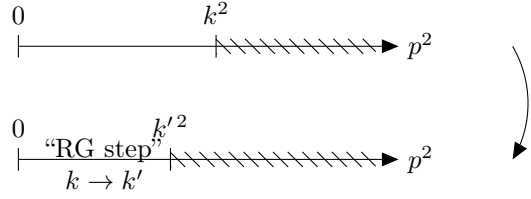
\begin{figure}[tp]
\centering
\small
\tikzset{hstyle/.style={solid}}
\begin{tikzpicture}
 \draw[-] (0,1.45) -- (0,1.75);
 \draw[-] (0,1.6) node[above=0.5em]{$0$} -- (2.6,1.6);
 \draw[-] (2.6,1.45) -- (2.6,1.75);
 \draw[-triangle 45] (2.6,1.6) node[above=0.5em]{$k^2$} -- (5,1.6) node[right]{$p^2$};
 \tkzhatchrect[0.1935](2.6,1.5)(4.7,1.7);
 %\fill[GridSize=5pt, pattern=wide north east lines, pattern color=darkgray] (2.6,1.5) rectangle (4.7,1.7);
 \draw[-] (0,-0.05) -- (0,0.25);
 \draw[-] (0,0.1) node[above=0.5em]{$0$} -- (2,0.1);
 \draw[-] (2,-0.05) -- (2,0.25);
 \draw[-triangle 45] (2,0.1) node[above=0.5em]{$k'^{\mkuu\mku 2}$} -- (5,0.1) node[right]{$p^2$};
 \tkzhatchrect[0.1935](2,0)(4.7,0.2);
 %\fill[GridSize=5pt, pattern=wide north east lines, pattern color=darkgray] (2,0) rectangle (4.7,0.2);
 \draw[-triangle 45] (6.5,1.6) arc[radius=1.5, start angle=30, end angle=-30] node[midway, right=0.1em]{%
  \begin{tabular}{c} ``RG step'' \\ $k\to k'$ \end{tabular}};
\end{tikzpicture}
\caption{In the modified functional integral \eqref{eq:DefZk} modes with momenta satisfying $p^2 \gtrsim k^2$ are
 integrated out, as indicated by the hatched area, while those with $p^2 \lesssim k^2$ are suppressed, where squared
 momenta refer to eigenvalues of $-D^2\mku$ (upper ray).
 Lowering the scale from $k$ to $k'$ amounts to integrating out additional modes correspondingly (second ray).}
\label{fig:ModeSupression}
\end{figure}
We require $\Rk$ to act effectively as an \emph{infrared cutoff}. This is achieved by choosing a cutoff profile similar
to the one sketched in Figure \ref{fig:CutoffProfile}, which leaves the high momentum modes unaffected, i.e.\ they are
integrated out in \eqref{eq:DefZk}, while it plays the role of a mass-like cutoff for infrared modes.
For convenience we write $\Rk$ in terms of a dimensionless function $R^{(0)}$: $\Rk\equiv\mZ_k\mkuu k^2\mku R^{(0)}
\big(-D^2/k^2\big)$, where $\mZ_k$ is a constant (that may carry internal indices in the case of general fields).
Several possible choices for the shape function $R^{(0)}$ are specified in Appendix \ref{app:Cutoffs}.

Defining $W_k[J]\equiv \ln Z_k[J]$ we can express the (scale dependent) field expectation value as $\phi\equiv \langle
\chi\rangle = \delta W_k/\delta J$. This relation is now formally solved for the source, $J\equiv J_k[\phi]$, viewing
$\phi$ as an independent argument henceforth. Finally, the effective average action $\Gamma_k$ is defined as the
Legendre transform of $W_k[J]$ with the cutoff action subtracted \cite{RW93a,RW93b,Wetterich1993,RW94a,RW94b}:
\begin{equation}
 \Gamma_k[\phi] \equiv J\cdot\phi - W_k[J] - \frac{1}{2}\,\phi\cdot\Rk\mku\phi\,.
\label{eq:DefGammak}
\end{equation}
The EAA describes a family of effective field theories labeled by the scale $k$. By construction, it approaches
the standard quantum effective action in the limit $k\to 0$: $\Gamma_{k=0}=\Gamma$.
In the UV limit, on the other hand, it is closely related to the bare action \cite{MR09,VZ11,MS15,NR16a}. We will
investigate this latter property in more detail in Chapter \ref{chap:Bare}.
\begin{figure}[tp]
 \centering
 \vspace{1.2em}
 \includegraphics[width=0.32\columnwidth]{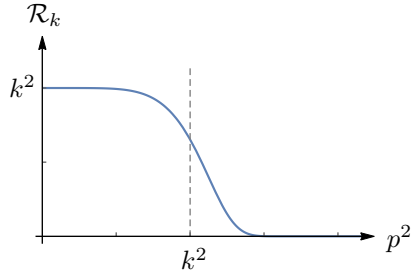}
 \begin{textblock}{5}(1.22,-1.88)
   \small
   $\Rk$
 \end{textblock}
 \begin{textblock}{5}(1,-1.37)
   \small
   $k^2$
 \end{textblock}
 \begin{textblock}{5}(2.72,-0.14)
   \small
   $k^2$
 \end{textblock}
 \begin{textblock}{5}(4.78,-0.3)
   \small
   $p^2$
 \end{textblock}
 \vspace{0.2em}
\caption{Illustration of a suitable cutoff profile $\Rk(p^2)$. It should be chosen such that high momentum modes with
 $p^2\gtrsim k^2$ are almost unaffected, while low momentum modes with $p^2\lesssim k^2$ are suppressed in
 \eqref{eq:DefZk}. This leads to the following two requirements a generic cutoff operator should satisfy: $\Rk\to 0$
 for UV modes and $\Rk\to k^2$ for IR modes.}
\label{fig:CutoffProfile}
\end{figure}
\smallskip

\noindent
\textbf{(2) Functional RG equation.}
A particularly important feature of the EAA is its transformation behavior under the RG action. Differentiating
\eqref{eq:DefGammak} with respect to the scale $k$ shows that the RG flow of $\Gamma_k$ is governed by the
\emph{functional renormalization group equation} (FRGE) \cite{Wetterich1993,Morris1994,RW94a,Reuter1998}
\begin{equation}[b]
 k\p_k\Gamma_k = \frac{1}{2}\STr\left[\big(\Gamma_k^{(2)}+\Rk\big)^{-1}\,k\p_k\Rk\right] .
\label{eq:FRGE}
\end{equation}
Here, $\Gamma_k^{(2)}$ denotes the Hessian of $\Gamma_k$ with respect to the fluctuating field. The
supertrace `$\STr$' comprises an operator trace that takes into account all field types involved, weighting standard
fields with a plus sign and Grassmann-valued fields with a minus sign. For the scalar field example `$\STr$' thus boils
down to the usual operator trace `$\Tr$'.

The FRGE \eqref{eq:FRGE} has a couple of remarkable properties: It is
\emph{fully nonperturbative} and does not rely on the smallness of any coupling, it is \emph{exact} (as it involves no
approximation), it is \emph{UV finite} (due to the presence of $k\p_k\Rk$ in the numerator on the RHS), and it is
\emph{IR finite} (due to the appearance of $\Rk$ in the denominator), to mention but a few. Moreover, it does no longer
involve any functional integral. Therefore, it may even serve as a starting point for an RG analysis: Possible
candidates for the EAA are now given by \emph{solutions to the FRGE} rather than being based on a functional integral
construction.
\smallskip

\noindent
\textbf{(3) Theory space.}
In the aforementioned approach, the only input data to be fixed at the beginning are, first, the kinds of quantum
fields carrying the theory's degrees of freedom, and second, the underlying symmetries. This information determines the
stage the RG dynamics takes place on, the so-called \emph{theory space}, consisting of all possible action functionals
that respect the prescribed symmetry. A prime example is given by the theory space of \emph{Quantum Einstein Gravity}
(QEG). QEG is the generic name for a quantum field theory that takes the metric as the dynamical field variable and
whose symmetry is given by diffeomorphism invariance.

Henceforth, we assume that any point in a given theory space, i.e.\ any admissible action functional,
can be expanded as a linear combination of field monomials, $\Gamma_k[\phi] = \sum_{\alpha=1}^\infty C_\alpha(k)\mku
P_\alpha[\phi]$, where $\{P_\alpha\}$ denotes a set of $k$-independent basis invariants. The corresponding (possibly
dimensionful) coupling constants $C_\alpha(k)$ can always be made dimensionless by multiplying them with a suitable
power of the RG scale: $c_\alpha(k) \equiv k^{d_\alpha} C_\alpha(k)\mku$, with $d_\alpha$ the canonical mass dimension
of $P_\alpha[\phi]$. Then the scale dependence of $\Gamma_k$ is completely determined by (infinitely many)
$\beta$-functions describing the RG ``running'' of the dimensionless couplings: 
\begin{equation}
 k\p_k\mku c_\alpha(k) = \beta_\alpha(c_1,c_2,\dotsc).
\end{equation}

\noindent
\textbf{(4) Truncations.}
In order to find approximate solutions to the FRGE \eqref{eq:FRGE} one usually resorts to truncations, implying a
reduction of the infinite-dimensional theory space. To this end, we may --- for instance --- set all but a finite
number of couplings to zero and consider the projection onto the subspace spanned by the reduced basis
$\{P_\alpha\}$ with $\alpha =1,\dotsc,n$. This amounts to the \emph{truncation ansatz}
$\Gamma_k[\phi]= \sum_{\alpha=1}^n c_\alpha(k)k^{-d_\alpha}\mku P_\alpha[\phi]$.
Inserting such an ansatz into \eqref{eq:FRGE} and projecting also the trace on the RHS onto the
truncated theory space yields a system of $n$ ordinary differential equations, $k\p_k\mku c_\alpha(k)=
\beta_\alpha(c_1,\dotsc,c_n)$, for each $\alpha\in\{1,\dotsc,n\}$.\footnote{Note that even the case $n=\infty$ may be
considered, e.g.\ an $f(R)$-type truncation \cite{%LR02c,CPR08,CPR09,FLNR13,OP14,FLNR16a,DM15,OP16,
MS08,DSZ12,BC12,Benedetti2013,DM13a,DM13b,DSZ15a,DSZ15b,Eichhorn2015,OPV15,OPV16,FLNR16b,FO16}.}
Although giving rise to an approximation of the
exact RG flow, these $\beta$-functions inherit the full nonperturbative character of the FRGE.
In the next subsection we present a concise step-by-step instruction how to systematically compute them.

%----------------------------------------------------------------------------------------------------------------------
\subsection[How to extract \texorpdfstring{$\beta$}{beta}-functions]%
{How to extract \texorpdfstring{\bm{$\beta$}}{beta}-functions}
\label{sec:Recipe}
%----------------------------------------------------------------------------------------------------------------------

The following is a recipe for calculating $\beta$-functions on the basis of the FRGE, assuming that the
theory space is fixed, i.e.\ field types and symmetries are known.
\smallskip

\noindent
\textbf{(1)}
We start by choosing an appropriate truncation ansatz. The number and the kind of invariants that are included in the
ansatz should be such that the resulting approximation of the exact flow is as good as possible in order to capture
the essential physics but also such that the calculation is still technically manageable. For gravity the prime example
is the Einstein--Hilbert truncation, $\frac{1}{16\pi G_k} \int \! \dd x \sg \,\big( -R + 2\Lambda_k \big)$, which
consists of the classical Einstein--Hilbert action with the couplings replaced by running ones, enabling us to describe
both the classical and the UV regime.

When considering gauge theories, this first step also involves choosing a suitable gauge fixing action and constructing
the corresponding ghost action.
\smallskip

\noindent
\textbf{(2)}
We insert the truncation ansatz for $\Gamma_k$ into the LHS of the FRGE \eqref{eq:FRGE} and differentiate it with
respect to the RG scale $k$. This derivative acts on the (dimensionful) coupling constants, the only $k$-dependent
pieces in $\Gamma_k\mku$.
\smallskip

\noindent
\textbf{(3)}
In order to process the RHS of \eqref{eq:FRGE}, we first compute the Hessian $\Gamma_k^{(2)}$, i.e.\ the second
functional derivative of $\Gamma_k$ with respect to the fluctuating field. Typically, it is of the form $\Gamma_k^{(2)}
=-\Box+U$ (dropping all prefactors, $k$-dependences and internal indices), with the Laplacian $\Box\equiv D_\mu
D^\mu$ and a potential $U$. In the case of gravity with an EAA composed of the metric, it can be obtained by making use
of the list of variations of geometric quantities given in Appendix \ref{app:Variations}.
\smallskip

\noindent
\textbf{(4)}
We write the argument of `$\STr$' in \eqref{eq:FRGE} as function of $-\Box$. In all cases considered in this thesis the
FRGE can then be expressed as $k\p_k\mku\Gamma_k = \frac{1}{2}\sum_i\Tr\big[W_i(-\Box)\big]$, where the sum is
over different field types.
(If there are uncontracted derivatives this step might require choosing an appropriate gauge \cite{Reuter1998} or more
general techniques \cite{GSZ11} in order to evaluate the trace.)
\smallskip

\noindent
\textbf{(5)}
Writing $W_i$ formally as a Laplace transform, $W_i(-\Box)=\int_0^\infty \td s\;\e^{\mku s\mku\Box}\,
\widetilde{W}_i(s)$, allows us to apply the trace to $\e^{\mku s\mku\Box}$ and expand it by means of heat kernel
techniques, see Appendix \ref{app:Heat}. In this way, we can project the trace onto those invariants which are
contained in the truncation. By eqs.\ \eqref{eq:Heat1} and \eqref{eq:Heat3} such an expansion reads
$\Tr\big[W_i(-\Box)\big] = (4\pi)^{-d/2}\,\tr(\Id) \left\{ Q_{d/2}[W_i]\int\!\sg + \frac{1}{6}\,
Q_{d/2-1}[W_i]\int\!\sg\,R + \cdots \right\}$, where $Q_n[W_i]$ denotes the generalized Mellin transform of $W_i\mku$,
cf.\ Appendix \ref{app:Cutoffs}.
\smallskip

\noindent
\textbf{(6)}
After having expanded the trace on the RHS of the FRGE \eqref{eq:FRGE}, we can compare the coefficient of each
invariant with the corresponding one on the LHS, yielding the $\beta$-functions for the dimensionful couplings.
\smallskip

\noindent
\textbf{(7)}
Finally, we rewrite the result in terms of dimensionless couplings, leading to a system of ordinary differential
equations, $k\p_k\mku c_\alpha(k) = \beta_\alpha(c_1,\dotsc,c_n)$, $\mkuu\alpha=1,\dotsc,n$.
\medskip

We follow the above instructions for all EAA-based RG investigations performed in this thesis, in particular for
the RG flow studies in Chapter \ref{chap:ParamDep}.

%----------------------------------------------------------------------------------------------------------------------
\subsection{The background field formalism}
\label{sec:BFF}
%----------------------------------------------------------------------------------------------------------------------

Any theory of quantum gravity must comply with the principle of \emph{background independence} \cite{DeWitt1967b,MR10}:
When setting up the theory, no special background geometry should play a distinguished role or be put in by hand. The
actual spacetime metric, $g_\mn\mku$, should rather arise as the expectation value of a quantum field, say $\gamma_\mn
\mku$, with respect to some state: $g_\mn=\langle\gamma_\mn\rangle$. By way of contrast, conventional quantum field
theories require a nondynamical (rigid) metric as an indispensable background structure, i.e.\ the metric has the
status of an external input. In this latter approach the metric is crucial for introducing a notion of time and
causality (necessary for defining equal time commutation relations, for instance), for constructing actions that
consist of covariant and ``nontopological'' terms, and for defining a length scale which is required for the
application of the aforementioned RG techniques (as they are based upon the eigenvalues of the Laplacian).

There are two different strategies for resolving these conceptual difficulties and implementing background independence
in quantum gravity.
(i) One could abandon the traditional route of quantum field theory and try to set up the theory without ever defining
a background metric at all, an example being loop quantum gravity \cite{Ashtekar1991,Rovelli2004}.
(ii) One introduces a nondynamical, arbitrarily chosen background metric, $\bg_\mn\mku$, during the intermediate steps
of the calculation, but shows in the end that no physical prediction depends on the choice of $\bg_\mn\mku$. Using this
bootstrap method one can apply the concepts of conventional quantum field theory, where the background metric defines
the ``arena'' all invariants of the theory can be constructed in.

In this thesis we would like to consider a field theoretical description of quantum gravity, that is, we have to take
the second path. As a consequence, the introduction of a background field is unavoidable. The approach presented in the
following, the \emph{background field method}, has first been established for gravity, but it can more generally be
applied to other field theories as well \cite{DeWitt1967b,Hooft1975,DeWitt1982,Boulware1981,Abbott1981,Abbott1982}.

In the standard formulation of this method, the dynamical quantum metric $\gamma_\mn$ is decomposed into the background
field $\bg_\mn$ and a fluctuating field $\hat{h}_\mn$ in a linear way:
\begin{equation}
 \gamma_\mn = \bg_\mn + \hat{h}_\mn \,.
\label{eq:QuantumLinearSplit}
\end{equation}
Note that the fluctuations $\hat{h}_\mn$ are not assumed to be small compared to $\bg_\mn\mku$ but can become
arbitrarily large. If $h_\mn\equiv\big\langle \hat{h}_\mn\big\rangle$ denotes the associated expectation value, the
full spacetime metric reads $g_\mn \equiv \langle \gamma_\mn \rangle = \bg_\mn + h_\mn\mku$. These definitions allow us
to employ the FRG techniques of Section \ref{sec:EAAFRGE}, where $\gamma_\mn$ corresponds to the quantum field $\chi$,
and length scales and the Laplacian are based on the background metric $\bg_\mn\mku$.

Motivated by general relativity, the microscopic (bare) action $S[\gamma]$ is assumed to be invariant under general
coordinate transformations,
\begin{equation}
 \delta\gamma_\mn=\mL_X \gamma_\mn \,,
\end{equation}
where the vector fields $X$ generate diffeomorphisms on the manifold considered, the Lie derivative $\mathcal{L}_X$
appearing in their infinitesimal representation. Due to the fact that the description depends on two fields now, there
is some freedom in splitting the gauge transformation: both $\bar{g}_{\mu\nu}$ and $\hat{h}_{\mu\nu}$ can be
transformed independently as long as the sum $\delta\bar{g}_{\mu\nu}+\delta \hat{h}_{\mu\nu}$ equals $\delta
\gamma_{\mu\nu}$. Two possible choices are the \emph{true or quantum gauge transformations},
\begin{equation}
 \delta\bar{g}_{\mu\nu}=0\,, \qquad \delta \hat{h}_{\mu\nu}=\mL_X(\bar{g}_{\mu\nu} + \hat{h}_{\mu\nu}) \,,
\label{eq:TrueGauge}
\end{equation}
and the \emph{background gauge transformations}:
\begin{equation}
 \delta \bar{g}_{\mu\nu}=\mL_X \bar{g}_{\mu\nu}\,, \qquad \delta \hat{h}_{\mu\nu} = \mL_X \hat{h}_{\mu\nu} \,.
\label{eq:BackGauge}
\end{equation}
The former are gauge fixed in the functional integral defining the effective average action, so the invariance under
\eqref{eq:TrueGauge} is explicitly broken. The latter transformations, however, leave the EAA invariant. More
precisely, $\Gamma_k[g,\bg,\xi,\bx\mku]$ (which is in fact a functional of both $g_\mn$ and $\bg_\mn\mku$, and of the
ghost fields $\xi^\mu$ and $\bx_\mu$) remains unchanged under $\{\delta \bg_\mn=\mL_X \bg_\mn\mku , \,\delta g_\mn =
\mL_X g_\mn\mku , \,\delta \xi^\mu = \mL_X \xi^\mu , \,\delta \bx_\mu = \mL_X \bx_\mu\mku\}$. Hence, \emph{at the level
of $\Gamma_k$ diffeomorphism invariance is fully intact}. Note that the true gauge transformations are accounted for by
generalized BRST Ward identities. They reduce to the usual ones at vanishing RG scale, $k=0$, but get modified for
higher scales due to the mode suppression term \cite{Reuter1998}.

We would like to point out that the relation between quantum, background and fluctuating field can be more general than
the linear split \eqref{eq:QuantumLinearSplit}. One could as well choose a nonlinear parametrization, which may be
written as $\gamma_\mn \equiv \gamma_\mn\big[\hat{h};\bg\big]$. The fact that such a generalization is indeed useful
will be motivated and explained in detail in Chapter \ref{chap:SpaceOfMetrics}. Note that it may be quite involved to
find the transformation behavior of $\hat{h}_\mn$ in the general case. Therefore, we write the rules
\eqref{eq:TrueGauge} and \eqref{eq:BackGauge} in terms of $\gamma_\mn$ and $\bg_\mn$ rather than $\hat{h}_\mn$ and
$\bg_\mn\mku$. Then the quantum gauge transformations read $\{\delta \bg_\mn=0, \,\delta \gamma_\mn = \mL_X
\gamma_\mn\}$, while the background gauge transformations are expressed as $\{\delta \bg_\mn=\mL_X \bg_\mn\mku ,
\,\delta \gamma_\mn = \mL_X \gamma_\mn\}$. This will be used in Section \ref{sec:ParamDepFramework}.

%----------------------------------------------------------------------------------------------------------------------
\subsection{The FRGE for quantum gravity}
\label{sec:FRGgrav}
%----------------------------------------------------------------------------------------------------------------------

Combining the methods of Section \ref{sec:EAAFRGE} with the background field formalism (including a suitable gauge
fixing) and applying it to metric gravity yields the effective average action $\Gamma_k[g,\bg,\xi,\bx\mku]$, the
primary tool for investigating the gravitational RG flow at the nonperturbative level \cite{Reuter1998}. It is a
functional of the dynamical metric $g_\mn$ and the ghost fields $\xi^\mu$ and $\bx_\mu\mku$, but it also has an
extra $\bg_\mn$-dependence. This extra background dependence is a consequence of gauge fixing and ghost terms on
the one hand, and of regulator terms on the other hand. The latter contributions to $\Gamma_k$ vanish in the limit
$k\to 0$, while the former ones remain nonzero even in the IR limit. Consequently, since for $k\to 0$ the background
enters only the gauge parts, physical predictions derived from $\Gamma_{k=0}$ should not depend on $\bg_\mn\mku$, in
agreement with the principle of background independence. Whether this is actually confirmed by RG computations can be
investigated only by means of \emph{bimetric truncations} (whose corresponding theory subspaces contain invariants
constructed out of both metrics, requiring a careful distinction between $g_\mn$ and $\bg_\mn$ at any step of the
calculation), as discussed in Ref.\ \cite{BR14} and Section \ref{sec:bi}.

The dependence of $\Gamma_k$ on $g_\mn$ may be reexpressed as a dependence on the metric fluctuations $h_\mn\mku$,
where $h_\mn\equiv g_\mn-\bg_\mn$ in the case of the linear parametrization.
For the rewritten functional $\Gamma_k$  we employ the ``semicolon notation''
\begin{equation}
 \Gamma_k\big[h,\xi,\bx;\bg\mku\big] \equiv \Gamma_k\big[g,\bg,\xi,\bx\mkuu\mku\big]
 \equiv \Gamma_k\big[\bg+h,\bg,\xi,\bx\mkuu\mku\big].
\label{eq:SemicolonNotation}
\end{equation}
If a general metric parametrization is used, the last equivalence in \eqref{eq:SemicolonNotation} has to be stated as
$\Gamma_k \big[g,\bg,\xi,\bx\mkuu\mku\big] \equiv \Gamma_k\big[g[h;\bg],\bg,\xi,\bx\mkuu\mku\big]$, as clarified in
Section \ref{sec:Applications}.

In this thesis we use a common approximation that consists in neglecting the running of the ghost part. For
consistency, this requires setting the ghost fields $\xi^\mu$ and $\bx_\mu$ to zero \emph{after} having determined the
Hessian of $\Gamma_k$ on the RHS of the FRGE. (In a sense, the assumption of scale independent ghosts may thus be
considered part of the truncation ansatz.) In this case the supertrace in the FRGE \eqref{eq:FRGE} decomposes into a
purely gravitational part and a ghost contribution \cite{Reuter1998}:
\begin{equation}
\begin{split}
 k\p_k\Gamma_k = &\frac{1}{2}\Tr\left[\Big(\big(\Gamma_k^{(2)}\big)_{\mkern-1mu hh}+\Rk^\text{grav}\Big)^{\mkern-1mu-1}
 \, k\p_k\Rk^\text{grav}\right]\\ & - \Tr\left[\Big(\big(\Gamma_k^{(2)}\big)_{\bx\xi}+\Rk^\text{gh}\Big)^{\mkern-1mu-1}
 \, k\p_k\Rk^\text{gh}\right]\,.
\end{split}
\label{eq:FRGEgrav}
\end{equation}
Here, $\big(\Gamma_k^{(2)}\big)_{\mkern-1mu hh}\equiv \frac{\delta^2\Gamma_k}{\delta h^2}[h,0,0;\bg]$ is the
second functional derivative of $\Gamma_k$ with respect to the metric fluctuations, and
$\big(\Gamma_k^{(2)}\big)_{\bx\xi} \equiv \frac{\delta}{\delta\xi} \frac{\delta\Gamma_k}{\delta\bx}[h,0,0;\bg]$ agrees
(up to a factor minus one) with the Faddeev--Popov operator. The cutoff operators of the gravitational and the ghost
sector are denoted by $\Rk^\text{grav}$ and $\Rk^\text{gh}$, respectively.

Most standard FRG analyses rely on \emph{single-metric truncations}, obtained by projection onto such invariants
that depend on $g_\mn$ alone. During the computation of $\beta$-functions this approximation amounts to identifying
background and dynamical metric, $\bg_\mn=g_\mn\mku$, or equivalently, $h_\mn=0$, but only after the second
functional derivative appearing in the FRGE has been taken. A particularly important example is the Einstein--Hilbert
truncation whose gravitational part is given by $\Gamma_k^\text{grav}[g]\equiv\frac{1}{16\pi G_k} \int \! \dd x \sg \,
\big( -R + 2\Lambda_k \big)$. The RG behavior of the scale dependent Newton constant and cosmological constant,
$G_k$ and $\Lambda_k$, respectively, will be studied in Section \ref{sec:single}. Note that the above version of the
FRGE, eq.\ \eqref{eq:FRGEgrav}, applies to both single-metric and bimetric truncations, the only assumption that
entered its derivation being a $k$-independent ghost action. (The case of running ghosts has been considered in Refs.\
\cite{GS10,EG10,Eichhorn2013a,CDP14a}.)

%----------------------------------------------------------------------------------------------------------------------
\section{Asymptotic Safety}
\label{sec:AS}
%----------------------------------------------------------------------------------------------------------------------

According to the notion introduced in Subsection \ref{sec:EAAFRGE}, the scale dependence of an action is encoded in a
running of the coupling constants that parametrize this action, $\{c_\alpha\} \equiv \{c_\alpha(k)\}$. This gives rise
to a trajectory in the underlying theory space (RG trajectory), describing the evolution of an action functional with
respect to the scale $k$. Which of all possible trajectories is realized in Nature has to be determined by
measurements.
\smallskip

\noindent
\textbf{(1) Taking the UV limit.}
In the present context, the construction of a consistent quantum field theory amounts to finding an RG trajectory which
is infinitely extended in the sense that the action functional described by $\{c_\alpha(k)\}$ is well-behaved for all
values of the ``momentum'' scale parameter $k$, including the infrared limit $k \rightarrow 0$ and the UV limit $k
\to\infty$. The Asymptotic Safety program \cite{Weinberg1976,Weinberg1979} is a way of dealing with the latter limit.
Its fundamental requirement is the existence of a \emph{fixed point of the RG flow}. By definition this is a point
$\{c_\alpha^*\}$ in theory space where the running of all dimensionless couplings stops, or, in other words, a zero of
all $\beta$-functions: $\beta_\gamma(\{c_\alpha^*\})=0$ for all $\gamma$.\footnote{More precisely, it is only the
\emph{essential} couplings whose running is required to stop, i.e.\ only all those couplings which cannot be eliminated
by a field redefinition. Inessential, unphysical couplings may still diverge. Here we assume for the sake of the
argument that all couplings are essential.} In addition, that fixed point must have at least one \emph{UV-attractive}
direction. This ensures that there are one or more RG trajectories which run into the fixed point for increasing scale.
\smallskip

\noindent
\textbf{(2) The UV critical surface.}
The set of all points in the theory space that are ``pulled'' into the fixed point by going to larger scales is
referred to as \emph{UV critical surface}. Thus, the UV critical surface consists of all those trajectories which are
safe from UV divergences since all couplings approach finite fixed point values as $k \rightarrow \infty$,
see Figure \ref{fig:UVSurface}. The key hypothesis underlying Asymptotic Safety is that only trajectories lying
entirely within the UV critical surface of an appropriate fixed point can be \emph{infinitely extended} and thus define
a fundamental quantum field theory. (See Refs.\ \cite{NR06,Percacci2009,RS07,RS12,Nagy2014} for reviews.) This may be
thought of as a systematic search strategy which identifies physically acceptable theories as compared with the
unacceptable ones plagued by short distance singularities. Note that the existence of a fixed point allows the
asymptotically safe trajectories to stay in its vicinity for an infinitely long RG time. Since the method does not rely
on any kind of smallness of the couplings, asymptotically safe theories can be considered \emph{nonperturbatively
renormalizable}.

\begin{figure}[tp]
 \centering
 \includegraphics[width=0.99\columnwidth]{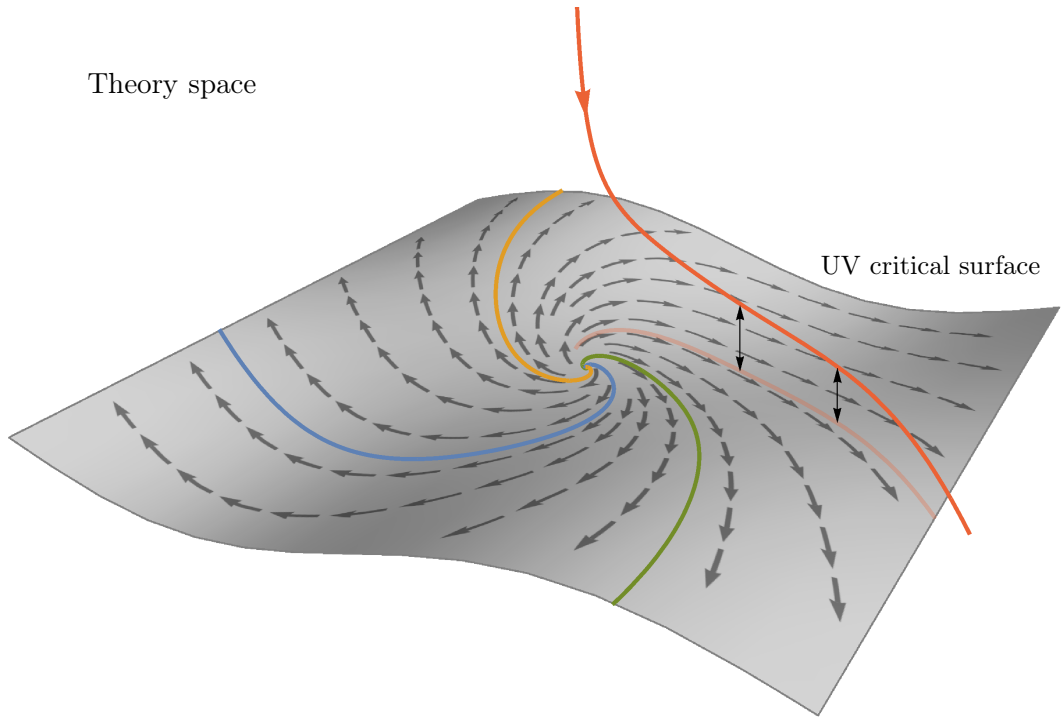}
 \begin{textblock}{5}(-0.8,-4.6)
   Theory space
 \end{textblock}
 \begin{textblock}{5}(6.8,-3.3)
   \small
   UV critical surface
 \end{textblock}
 \vspace{0.5em}
\caption{Vector field of the RG flow and some sample trajectories in theory space, parametrized by the coupling
 constants. By convention, the arrows of the vector field (and the one on the red trajectory) point from UV to IR
 scales. The set of actions which lie inside the theory space and are pulled into the fixed point under the inverse RG
 flow (i.e., going in the direction opposite to the arrows) is referred to as UV critical surface. The Asymptotic
 Safety hypothesis states that a trajectory can be realized in Nature only if it is contained in the UV critical
 surface of a suitable fixed point since only then it has a well-behaved high energy limit (green, blue, and dark
 yellow trajectories, by way of example). Unless there is another fixed point, trajectories outside this surface escape
 the theory space for $k\to\infty$ as they develop unacceptable divergences in the UV, while they approach the UV
 critical surface when going to lower scales. This situation is represented by the red trajectory which lies above the
 surface and runs away from it for increasing RG scale (opposite to the red arrow).}
\label{fig:UVSurface}
\end{figure}

\smallskip
\noindent
\textbf{(3) Predictivity of asymptotically safe theories.}
With regard to the fixed point, UV-attractive directions are called \emph{relevant}, UV-repulsive ones
\emph{irrelevant}, since the corresponding scaling fields increase and decrease, respectively, when the scale is
lowered. Therefore, the dimensionality of the UV critical surface equals the number of relevant couplings. An
asymptotically safe theory is thus \emph{the more predictive the smaller the dimensionality of the corresponding UV
critical surface is}.

For instance, if the UV critical surface has the finite dimension $n$, it is sufficient to perform only $n$
measurements in order to uniquely identify Nature's RG trajectory. Once the $n$ relevant couplings are measured, the
requirement for Asymptotic Safety \emph{fixes all other couplings} since the latter have to be adjusted in such a way
that the RG trajectory lies within the UV critical surface. In this spirit, the theory is highly predictive as
infinitely many parameters are fixed by a finite number of measurements.

Figure \ref{fig:UVSurface} illustrates the example of a three-dimensional theory space and a two-dimensional UV
critical surface. The couplings pertaining to the two relevant directions can be determined by two measurements, while
the ``vertical'' direction is fixed by requiring that the trajectory be located within the UV critical surface. On the
other hand, RG trajectories lying below or above (like the red one) are excluded in the Asymptotic Safety program.
\smallskip

\noindent
\textbf{(4) Gaussian and non-Gaussian fixed points.}
A fixed point is called ``Gaussian'' if it corresponds to a free theory. Its critical exponents agree with the
canonical mass dimensions of the corresponding operators. Usually this amounts to the trivial fixed point values
$c_\alpha^* = 0$ for all essential couplings $c_\alpha$. Thus, standard perturbation theory is applicable only in the
vicinity of a Gaussian fixed point. In this regard, Asymptotic Safety at the Gaussian fixed point is equivalent to
perturbative renormalizability plus asymptotic freedom. Clearly, this possibility is ruled out for gravity which can
not be renormalized in the perturbative way.

In contrast, a nontrivial fixed point, that is, a fixed point whose critical exponents differ from the canonical ones,
is referred to as ``non-Gaussian''. Usually this requires $c_\alpha^* \neq 0$ for at least one essential $c_\alpha$. It
is such a non-Gaussian fixed point (NGFP) that provides a possible scenario for quantum gravity. Most studies on
Asymptotic Safety thus mainly focus on establishing the existence of a suitable NGFP.
\smallskip

\noindent
\textbf{(5) The bare action.}
As opposed to other approaches, a bare action which should be promoted to a quantum theory is not needed as an input
here. It is the theory space and the RG flow equations that determine possible fixed points with the desired UV
behavior. Since such a fixed point, in turn, acquires the status of the corresponding bare action, one can consider the
\emph{bare action} a \emph{prediction} in the Asymptotic Safety program \cite{MR09}, the precise connection being
discussed in Chapter \ref{chap:Bare}.
\medskip

To sum up, the concept of Asymptotic Safety is based upon two essential ingredients: (i) a suitable fixed point for
taming the UV behavior and (ii) a UV critical surface of reduced dimensionality for reasons of predictivity.

%----------------------------------------------------------------------------------------------------------------------
\section{Conformal field theory}
\label{sec:CFT}
%----------------------------------------------------------------------------------------------------------------------

This section contains a brief introduction to conformal field theory. We explain conformal transformations, their
generators, and the Virasoro algebra with its corresponding representations, paying particular attention to the
question about unitarity. More detailed reviews and primers are given in Refs.\
\cite{Ginsparg1988,Schellekens1996,FMS97,Cardy1987,Wipf1990,Schottenloher2008}, for instance.
\smallskip

\noindent
\textbf{(1) Weyl transformations.}
A Weyl transformation is a local rescaling of the metric (and of other fields, if present), leaving the coordinates
unchanged. Since we have to exclude sign changes and disappearances of the metric during this operation, the scaling
factor must be a strictly positive function, and we write Weyl transformations in the form
\begin{equation}
 g_\mn(x)\to\e^{2\mku\sigma(x)}\mku g_\mn(x) \,,
\end{equation}
where $\sigma$ is an arbitrary smooth function.

If $S$ is an action that is invariant under Weyl transformations, the corresponding stress-energy (energy-momentum)
tensor, defined by $T^\mn(x)\equiv \frac{2}{\sqrt{g(x)}} \frac{\delta S}{\delta g_\mn(x)}\mku$, is traceless:
$T^\mu{}_\mu(x)=0$. On the other hand, if an action has a traceless stress-energy tensor, then it is Weyl invariant.
(Note that the invariance of an action under general coordinate transformations, $x\to x'$, leads to a conserved
stress-energy tensor: $D_\mu T^\mn=0$. This explains the important role of $T^\mn$ for studying symmetries.)
\smallskip

\noindent
\textbf{(2) Conformal transformations.} Let us consider two (semi-) Riemannian manifolds $(M,g)$ and $(\tilde{M},
\tilde{g})$ of the same dimension as well as two open subsets $U\subset M$, $V\subset \tilde{M}$. Then a smooth
mapping $f:U\to V$ of maximal rank is called conformal transformation, if there is a smooth function $\sigma:U\to
\mathbb{R}$ such that $f^*\tilde{g}=\e^{2\sigma}g$, where $f^*\tilde{g}(X,Y)\equiv\tilde{g}\big(\td f(X),\td f(Y)\big)$
denotes the pullback of $\tilde{g}$ by $f$. If the two manifolds agree, $(M,g)=(\tilde{M},\tilde{g})$, the defining
relation reads $f^*g=\e^{2\sigma}g$.

Now, a general coordinate transformation $x\to x'$ within a given manifold induces a transformation of the metric
according to $g\to g'\equiv f^*g$, where $f$ is the inverse of the coordinate change, $x'=f^{-1}(x)$. In local
coordinates this amounts to the usual tensorial transformation behavior, $g_\mn(x)\to g'_\mn(x')=
\frac{\p x^\alpha}{\p x'^\mu} \frac{\p x^\beta}{\p x'^\nu}g_{\alpha\beta}(x)$. Thus, a conformal transformation is
defined by the property
\begin{equation}
 g_\mn(x)\to g'_\mn(x')=\frac{\p x^\alpha}{\p x'^\mu}\frac{\p x^\beta}{\p x'^\nu}g_{\alpha\beta}(x)
 =\e^{2\sigma(x)}g_\mn(x).
\label{eq:DefConfTrans}
\end{equation}
In other words, \emph{a conformal transformation is a coordinate transformation which acts on the metric as a Weyl
transformation}. Since the angle between two vectors $X$ and $Y$ is determined by the normalized scalar product
$\frac{g(X,Y)}{||X||\, ||Y||}$, such transformations are \emph{angle-preserving}.

In the remainder of this section we will work in flat Euclidean space (unless otherwise stated), with $g_\mn=
\delta_\mn\mku$. Note that a theory in flat spacetime with a conserved and traceless stress-energy tensor is
\emph{invariant under general coordinate transformations and Weyl transformations}, respectively, and thus it is
\emph{conformally invariant} in flat space: Consider a coordinate transformation with the property
\eqref{eq:DefConfTrans}. Due to coordinate invariance it does not change the value of the underlying action,
but only the fields inside, including the metric. Then Weyl invariance can be used to transform the metric back to its
original form. Such combined transformations leave the metric unchanged, i.e.\ we stay in flat space, and the action is
invariant. From this point of view a conformal transformation is a transformation acting only on the remaining fields.
We will come back to this interpretation in a moment.

Since an infinitesimal coordinate transformation $x'^\mu=x^\mu+\epsilon^\mu$ is conformal if and only if eq.\
\eqref{eq:DefConfTrans} is satisfied, we can use this equation to
infer conditions for the function $\epsilon^\mu(x)$. This way we obtain
two differential equations, $\p_\mu\epsilon_\nu+\p_\nu\epsilon_\mu = \frac{2}{d} g_\mn\mku\p_\alpha\epsilon^\alpha$
and $(d-2)\p_\mu\p_\nu\p_\alpha\epsilon^\alpha=0$, fixing the general form of a conformal transformation. In two
dimensions the latter constraint is absent, though, and the group of conformal transformations, or more precisely, the
number of its generators, is much larger then.

In $d>2$ dimensions one finds that $\epsilon^\mu(x)$ is at most quadratic in $x$, leading to four different kinds of
conformal transformations whose infinitesimal versions are given by: $x^\mu\to x^\mu+\alpha^\mu$ (translations), $x^\mu
\to x^\mu+\omega^\mu{}_\nu x^\nu$ with $\omega_\nu{}^\mu=-\omega^\mu{}_\nu$ (Lorentz transformations/rotations),
$x^\mu\to x^\mu+\lambda x^\mu$ (scale transformations), and $x^\mu\to x^\mu+b^\mu x^2-2 x^\mu\mku b\cdot x$
(special conformal transformations). The global version of the special conformal transformations reads
\begin{equation}
 x^\mu\to x'^\mu=\frac{x^\mu+b^\mu\mku x^2}{1+2\mku b\cdot x+b^2 x^2}\,,
\end{equation}
an example being illustrated in Figure \ref{fig:ConfTrans}.
\begin{figure}[tp]
 \begin{minipage}{0.34\columnwidth}
  \includegraphics[width=\columnwidth]{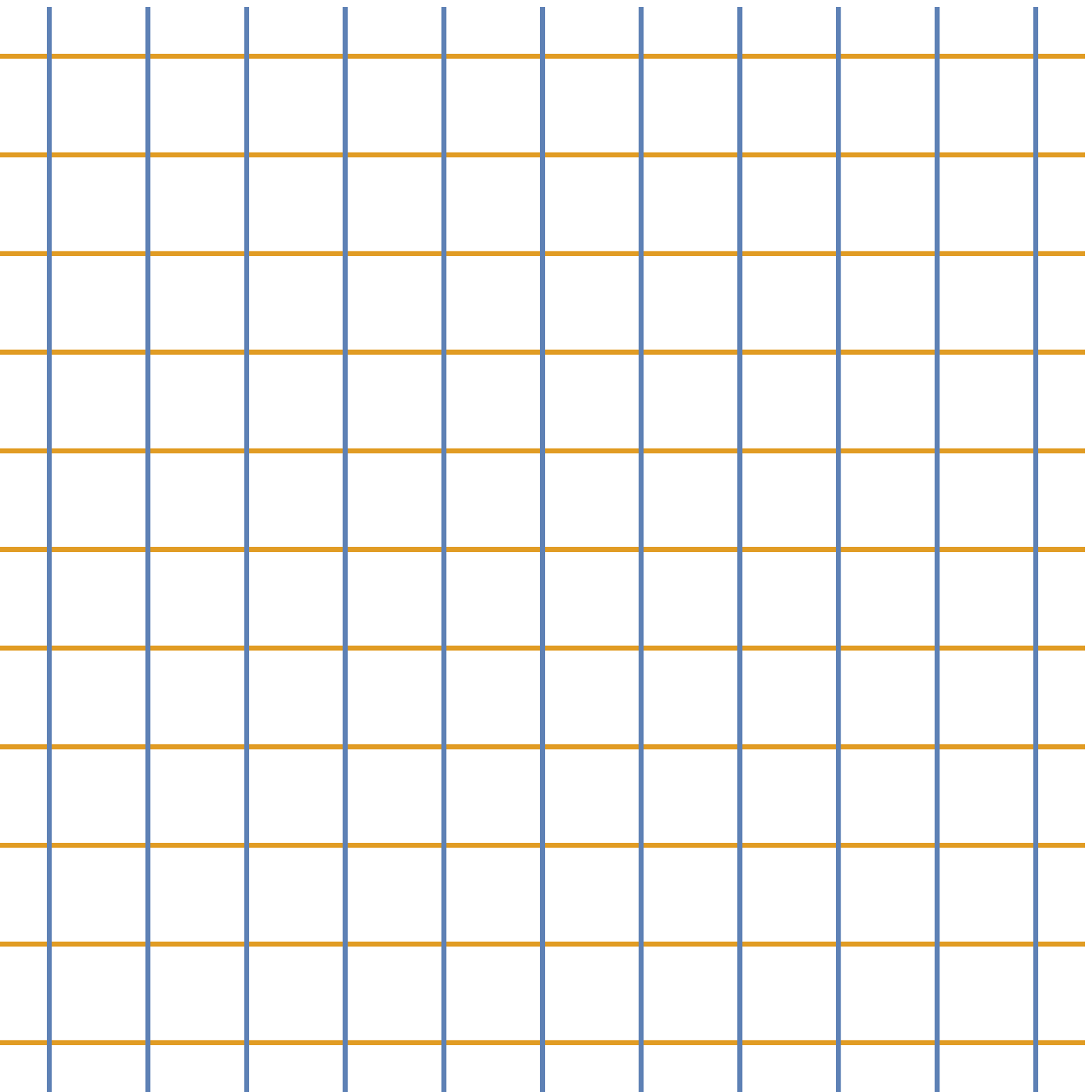}
 \end{minipage}
 \hfill
 \begin{minipage}{0.3\columnwidth}
 \centering
 \small
 \begin{tikzpicture}
  \draw[-triangle 45] (0,0) -- (3.5,0) node[midway,above=0.8em]{$x'^{\mku\mu}=\frac{x^\mu+b^\mu\mku x^2}{1+2\mku b\cdot x
   +b^2 x^2}$} node[midway,below=1.23em]{\footnotesize $(b^0,b^1)=(2,0)$};
 \end{tikzpicture}
 \end{minipage}
 \hfill
 \begin{minipage}{0.34\columnwidth}
  \includegraphics[width=\columnwidth]{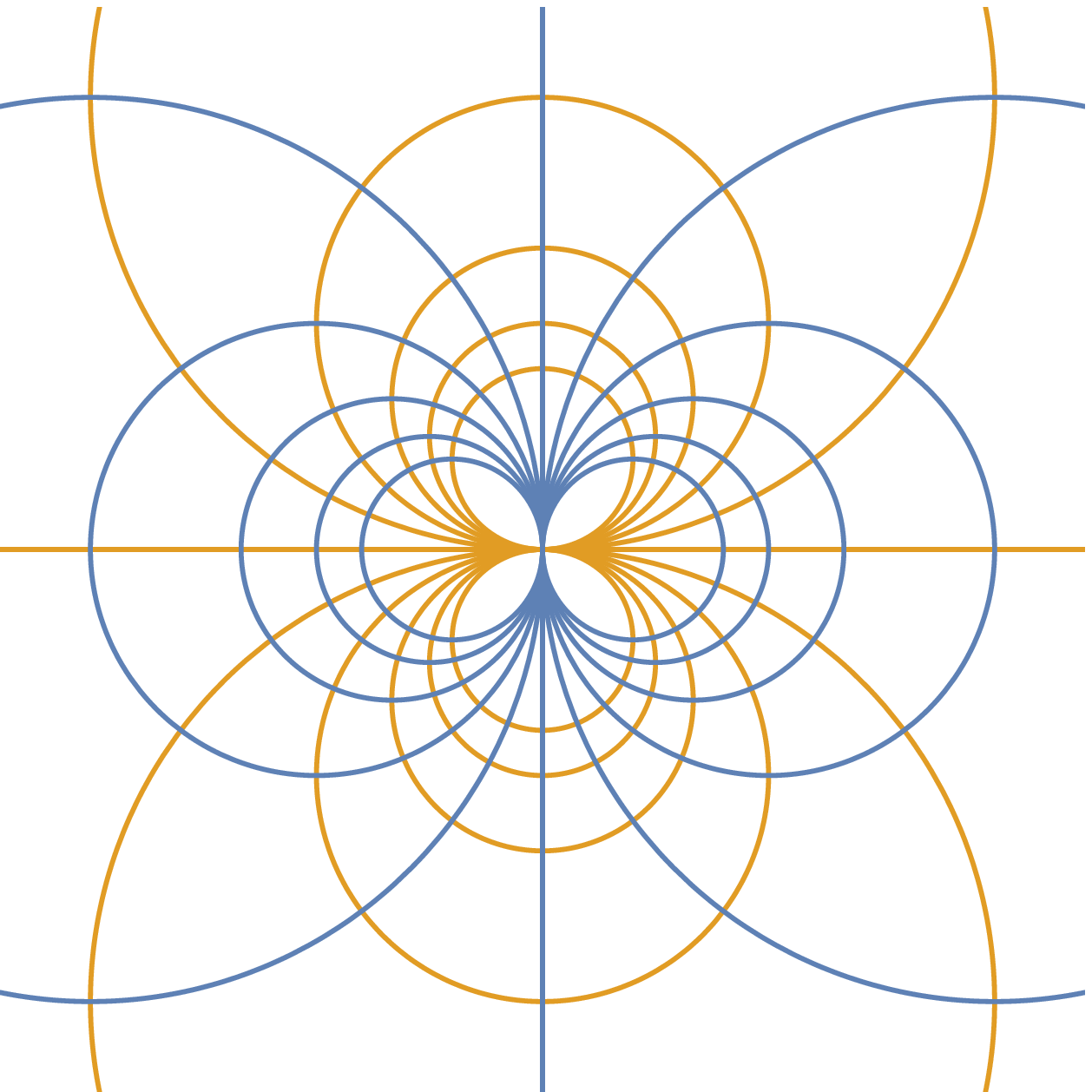}
 \end{minipage}
 \vspace{0.8em}
\caption{Effect of a special conformal transformation on a couple of sample grid lines. Like any other conformal
 transformation, this map is angle-preserving.}
\label{fig:ConfTrans}
\end{figure}
The number of corresponding generators for all four kinds of transformations, $\frac{1}{2}(d+1)(d+2)$, agrees with the
dimension of the conformal group, which is isomorphic to $\operatorname{SO}(d+1,1)$.
\smallskip

\noindent
\textbf{(3) Conformal transformations in $\bm{d=2}$ dimensions.} It is convenient to parametrize the points $(x^1,x^2)
\in\mathbb{R}^2$ by a complex number $z\in\mathbb{C}$ (and its complex conjugate $\bz$), using the identification
$z,\bz=x^1\pm ix^2$. We have seen previously that in $d>2$ dimensions there are two differential equations constraining
the function $\epsilon^\mu(x)$ such that the map $x^\mu\to x^\mu+\epsilon^\mu(x)$ is conformal. In $d=2$ dimensions, on
the other hand, there is only one constraint left which, in terms of $\epsilon,\bar{\epsilon}=\epsilon^1\pm i
\epsilon^2$, boils down to $\p_{\bz}\epsilon=0$ and $\p_z\bar{\epsilon}=0$. That is, $z\to z+\epsilon$ and $\bz\to\bz+
\bar{\epsilon}$ represent a conformal transformation if and only if $\epsilon\equiv\epsilon(z)$ is an arbitrary
infinitesimal meromorphic (i.e.\ holomorphic up to isolated points, here $0$ and $\infty$) function that depends only
on $z$, and analogously for $\bar{\epsilon}\equiv\bar{\epsilon}(\bz)$.
(Note that $\epsilon$ and $\bar{\epsilon}$ are usually viewed as being independent rather than complex
conjugates of each other. By imposing a reality condition at the end of calculations one obtains the correct result.)
The corresponding global versions of this coordinate change, i.e.\ the conformal transformations on the Riemann sphere
$\mathbb{C} \cup\{\infty\}$, are given by
\begin{equation}
 z\to f(z),\qquad \bz\to\bar{f}(\bz),
\label{eq:ConfTrans2D}
\end{equation}
where $f$ and $\bar{f}$ are arbitrary meromorphic functions.

Such meromorphic functions, and hence the conformal transformations, are generated by the operators $\ell_n\equiv
-z^{n+1}\p_z$ and $\bar{\ell}_n\equiv -\bz^{n+1}\p_{\bz}$ with $n\in\mathbb{Z}$. They span the \emph{Witt algebra} and
satisfy the commutation relations $[\ell_n,\ell_m]=(n-m)\ell_{n+m}\mku$, $[\bar{\ell}_n,\bar{\ell}_m]=(n-m)
\bar{\ell}_{n+m}$ and $[\ell_n,\bar{\ell}_m]=0$.

The only conformal transformations which are defined globally without singularities on the entire Riemann sphere are
generated by the subalgebra $\{\ell_{-1},\ell_0,\ell_1\}$ and the corresponding barred operators.
This gives rise to the group of M\"{o}bius transformations which is isomorphic to $\operatorname{SL}(2,\mathbb{C})/
\mathbb{Z}_2$ and to $\operatorname{SO}(3,1)$. The latter group is precisely the one encountered in point \textbf{(2)}.
Therefore, the conformal transformations in 2D include translations, Lorentz transformations, scale transformations and
special conformal transformations. The full algebra, however, is infinite-dimensional.
\smallskip

\noindent
\textbf{(4) Conformal fields in 2D.} Tensors in complex coordinates can be obtained from their counterparts in
$\mathbb{R}^2$ by $V_z=\frac{\p x^1}{\p z}V_1 + \frac{\p x^2}{\p z}V_2 = \frac{1}{2}(V_1+i V_2)$ and $V_{\bz} =
\frac{1}{2}(V_1-i V_2)$, and analogously for tensors with more indices. Here we adopt the common notation where $z$
($\bz$) denotes both the coordinate and the corresponding index. The metric $g_\mn=\delta_\mn\mku$, for instance,
transforms to $g_{zz}=\frac{1}{4}(g_{11}+ig_{12}+ig_{21}-g_{22})=0=g_{\bz\bz}$ and $g_{z\bz}=\frac{1}{4}(g_{11}+ig_{12}
-ig_{21}+g_{22})=\frac{1}{2}=g_{\bz z}\mku$. For the stress-energy tensor, tracelessness translates into
$T_{z\bz}=0=T_{\bz z}\mku$, while its conservation reads $\p_{\bz}T_{zz}=0=\p_z T_{\bz\bz}\mku$.

A tensor field $\phi\equiv\phi_{z,\dotsc,z,\bz,\dotsc,\bz}(z,\bz)$ is called \emph{primary field} or \emph{conformal
field} of weight $(h,\bar{h})$ if it transforms as
\begin{equation}
 \phi(z,\bz) \to \bigg(\frac{\p f}{\p z}\bigg)^{\!\! h} \bigg(\frac{\p \bar{f}}{\p \bz}\,
 \bigg)^{\!\!\bar{h}}\,\phi\big(f(z),\bar{f}(\bz)\big),
\label{eq:DefConfField}
\end{equation}
under the conformal transformation $z\to f(z)$, $\bz\to\bar{f}(\bz)$. Usually, the number $\Delta\equiv h+\bar{h}$ is
referred to as \emph{scaling weight}, and $s\equiv h-\bar{h}$ is the \emph{conformal spin}. The infinitesimal version
of \eqref{eq:DefConfField} reads
\begin{equation}
 \delta_{\epsilon,\bar{\epsilon}}\mku\phi(z,\bz)=\big((h\p\epsilon+\epsilon\p)+(\mku
\bar{h}\bar{\p}\bar{\epsilon}+\bar{\epsilon}\bar{\p})\big)\phi(z,\bz),
\label{eq:DefConfFieldInf}
\end{equation}
under $z\to z+\epsilon$ and $\bz\to\bz+\bar{\epsilon}$.
\smallskip

\noindent
\textbf{(5) Conformal invariance and the conformal bootstrap.} Since the correlation functions $G^{(n)}(z_1,
\dotsc,z_n,\bz_1,\dotsc,\bz_n)\equiv\langle\phi_1(z_1,\bz_1)\cdots\phi_n(z_n,\bz_n)\rangle$ in a conformally invariant
theory are supposed to be invariant under \eqref{eq:DefConfFieldInf}, we have $\delta_{\epsilon,\bar{\epsilon}}\mkuu
G^{(n)}=0$. This equation \emph{constrains the correlation functions considerably}. For $n=2$ and $n=3$, for instance,
it determines the form of $G^{(2)}$ and $G^{(3)}$ completely \cite{Polyakov1970,BPZ84}: If $h_1\neq h_2$ or $\bh_1\neq
\bh_2\mku$, then $G^{(2)}(z_1,z_2,\bz_1,\bz_2)=0$, while for $h_1= h_2$ and $\bh_1= \bh_2\mku$:
\begin{equation}
 G^{(2)}(z_1,z_2,\bz_1,\bz_2)=C_{12}\, z_{12}^{\mku -2h}\,\bz_{12}^{\mkuu -2\bar{h}}\,,
 \qquad\quad h\equiv h_1=h_2\,,\quad \bh\equiv\bh_1=\bh_2\,,
\end{equation}
where $(h_1,\bh_1)$ and $(h_2,\bh_2)$ are the conformal weights of $\phi_1$ and $\phi_2$, respectively. Furthermore,
\begin{equation}
 G^{(3)}(z_i,\bz_i)=C_{123}\,z_{12}^{h_3-h_1-h_2}\, z_{23}^{h_1-h_2-h_3}\, z_{13}^{h_2-h_3-h_2}\,
 \bz_{12}^{\bh_3-\bh_1-\bh_2}\, \bz_{23}^{\bh_1-\bh_2-\bh_3}\, \bz_{13}^{\bh_2-\bh_3-\bh_2}\,.
\end{equation}
Here, $C_{12}$ and $C_{123}$ are constants, and $z_{ij}$ and $\bz_{ij}$ are defined by the differences $z_{ij}\equiv
z_i-z_j$ and $\bz_{ij}\equiv \bz_i-\bz_j\mku$, respectively.
This procedure of determining correlation functions (and the exploitation of further symmetry constraints) is known as
the conformal bootstrap.

Note that under some technical assumptions like Poincar\'{e} invariance and unitarity (which are satisfied by most
relevant examples of 2D quantum field theories) any scale invariant quantum field theory in $d=2$ dimensions necessarily
possesses the enhanced conformal symmetry \cite{Zamolodchikov1986,Polchinski1988,Nakayama2015}.
\smallskip

\noindent
\textbf{(6) Quantization in 2D conformal field theory.} Let $T(z)\equiv T_{zz}(z)$ and $\bar{T}(\bz)\equiv
\bar{T}_{\bz\bz}(\bz)$ denote the two nonvanishing components of the stress-energy tensor. Then the currents associated
with an infinitesimal conformal transformation are given by $J(z)=T(z)\epsilon(z)$ and $\bar{J}(\bz)=\bar{T}(\bz)
\bar{\epsilon}(\bz)$. The corresponding conserved charge becomes
\begin{equation}
 Q_{\epsilon,\bar{\epsilon}}\mkuu=\frac{1}{2\pi\mku i}\oint\Big(\td z\, T(z)\epsilon(z) + \td\bz\, \bar{T}(\bz)
 \bar{\epsilon}(\bz)\Big).
\label{eq:ConfTransCharge}
\end{equation}
As usual, conserved charges can be used to generate the transformation from which they were derived: At the quantum
level we have
\begin{equation}
 \delta_{\epsilon,\bar{\epsilon}}\mku\phi(w,\bar{w})=\big[Q_{\epsilon,\bar{\epsilon}}\mku,\phi(w,\bar{w})\big],
\label{eq:DeltaPhiQ}
\end{equation}
where radial ordering (cf.\ \cite{Ginsparg1988,Schellekens1996} for instance) is implied. By comparing eq.\
\eqref{eq:DeltaPhiQ} with \eqref{eq:DefConfFieldInf} one can infer an expansion for the (radially ordered) operator
product $T(z)\mku\phi(w,\bar{w})$, namely $T(z)\mku\phi(w,\bar{w})=\frac{h}{(z-w)^2}\phi(w,\bar{w})+\frac{1}{z-w}\mku
\p_w\phi(w,\bar{w})+\mO\big((z-w)^0\big)$, and an analogous expansion for $\bar{T}(\bz)\mku\phi(w,\bar{w})$.
In a similar manner one can show that
\begin{equation}
 T(z)T(w) = \frac{c/2}{(z-w)^4}+\frac{2}{(z-w)^2}\, T(w)+\frac{1}{z-w}\,\p_w T(w),
\label{eq:TTOpProd}
\end{equation}
and analogously for the barred counterpart. The constant $c$ is called \emph{central charge} and its value depends on
the theory under consideration.
\smallskip

\noindent
\textbf{(7) The Virasoro algebra.} The significance of the stress-energy tensor for generating the conformal
transformations justifies a closer look to $T(z)$ and $\bar{T}(\bz)$. Introducing the operators
$L_n\equiv\oint\frac{\td z}{2\pi\mku i}\,z^{n+1}\mku T(z)$ and $\bar{L}_n\equiv\oint\frac{\td \bz}{2\pi\mku i}\,
\bz^{n+1}\mku\bar{T}(\bz)$ we can express $T(z)$ and $\bar{T}(\bz)$ as a Laurent series:
\begin{equation}
 T(z) = \sum_{n\in\mathbb{Z}} z^{-n-2}\mkuu L_n\,,\qquad
 \bar{T}(\bz) = \sum_{n\in\mathbb{Z}} \bz^{-n-2}\mkuu \bar{L}_n\,.
\label{eq:TasL}
\end{equation}
The commutator algebra satisfied by the modes $L_n$ and $\bar{L}_n$ can be computed by inserting their definitions,
taking into account the correct order of contours during the integration, and finally using equation
\eqref{eq:TTOpProd}. The result reads
\begin{equation}
 \big[L_n,L_m\big] = (n-m)L_{n+m}+\frac{c}{12}(n^3-n)\mku \delta_{n+n,0}\,,
\label{eq:VirasoroAlgebra}
\end{equation}
and $\big[\bar{L}_n,\bar{L}_m\big] = (n-m)\bar{L}_{n+m}+\frac{\bar{c}}{12}(n^3-n)\mku\delta_{n+n,0}\mku$, as well as
$\big[L_n,\bar{L}_m\big]=0$. This defines two copies of an infinite-dimensional algebra which is called the
\emph{Virasoro algebra}. It is a central extension of the Witt algebra with central charge $c$. As we discuss in the
next point, $L_n$ and $\bar{L}_n$ can be used to systematically construct the field space. Note that the requirements
that $T(z)$ and $\bar{T}(z)$ be Hermitian operators dictate the relations $L_n^\dagger=L_{-n}$ and $\bar{L}_n^\dagger
=\bar{L}_{-n}\mku$.
\smallskip

\noindent
\textbf{(8) Highest weight representations of the Virasoro algebra.} A \emph{highest weight state} is an eigenstate of
$L_0$ and $\bar{L}_0$ corresponding to the \emph{smallest eigenvalues}, $h$ and $\bh$, respectively. Such a state can
be constructed according to
\begin{equation}
 \big|h,\bh\big\rangle \equiv \phi(0,0)|0\rangle\,,
\end{equation}
where $\phi(z,\bz)$ is a conformal field with weights $h$ and $\bh$. Here, the vacuum $|0\rangle$ is defined by the
condition that it respects a maximal number of symmetries, i.e.\ it must be annihilated by as many $L_n$ (and
$\bar{L}_n$) as possible. The largest possible set with this property that does not conflict with the Virasoro
commutation relations is given by $\{L_n\,|\,n\ge -1\}$, that is, $L_n|0\rangle=0$ for all $n\ge -1$. There is a barred
analogue of this result (and the subsequent results), but we restrict our discussion to the non-barred objects
henceforth.

Based on the definition of $L_n$ and the operator product expansion of $T(z)\mku\phi(w,\bar{w})$ given in point
\textbf{(6)}, one can verify the relation $[L_n,\phi(w,\bar{w})]=h(n+1)w^n\mku\phi(w,\bar{w})+w^{n+1}\p_w
\phi(w,\bar{w})$. Hence, $L_n$ commutes with $\phi(0,0)$ for all $n>0$, and we find
\begin{equation}
 L_n\big|h,\bh\big\rangle = [L_n,\phi(0,0)]\mkuu |0\rangle + \phi(0,0)\mku L_n|0\rangle=0\qquad \text{for } n>0,
\end{equation}
while the case $n=0$ leads to
\begin{equation}
 L_0\big|h,\bh\big\rangle = h \big|h,\bh\big\rangle\,.
\end{equation}
For $n<0$, on the other hand, we obtain a new nonvanishing state $L_n\big|h,\bh\big\rangle$. It is an eigenstate of
$L_0$ again, where the corresponding eigenvalue has increased:
\begin{equation}
 L_0\mku L_n\big|h,\bh\big\rangle = \big([L_0,L_n] + L_n\mku L_0\big)\big|h,\bh\big\rangle = (h-n)L_n\big|h,\bh\big
 \rangle\,.
\end{equation}
Therefore, the $L_n$ with $n<0$ act as \emph{raising operators} while the $L_n$ with $n>0$ play the role of
\emph{lowering operators}, and $\big|h,\bh\big\rangle$ is indeed an $L_0$-eigenstate with the lowest eigenvalue.

This consideration shows that ground states of Virasoro representations are generated by conformal fields. The
new states obtained by acting with one or more raising operators on $\big|h,\bh\big\rangle$ are called
\emph{descendants}. We observe that there is in general more than one way of constructing a state at the
excitation level $n>0$ (i.e.\ with the $L_0$-eigenvalue $h+n$), namely all linear combinations of states of the type
\begin{equation}
 L_{-n_1}\cdots L_{-n_k}\mku \big|h,\bh\big\rangle \,,\qquad \sum_{i=1}^k n_i=n\,,
\end{equation}
with all $n_i$ positive. The collection of all such linear combinations for all $n\ge 0$ is called the \emph{Verma
module} of $\big|h,\bh\big\rangle$. By construction, the set of states in the Verma module is closed with respect to
the action of the Virasoro generators.
\smallskip

\noindent
\textbf{(9) Unitarity.} We refer to a representation of the Virasoro algebra as \emph{unitary} if it does not contain
any negative norm states (and only one zero norm state), i.e.\ if the state space is a \emph{(positive) Hilbert space}.
For the simplest descendants we find
\begin{equation}
\begin{split}
 \big|\big| L_{-n}\big|h,\bh\big\rangle \big|\big| &= \big\langle h,\bh\big| L_n\mku L_{-n} \big|h,\bh\big\rangle
 = \big\langle h,\bh\big| [L_n, L_{-n}] \big|h,\bh\big\rangle \\
 &= \bigg[\frac{c}{12}\big(n^3-n\big)+2nh\bigg] \big\langle h,\bh\big|h,\bh\big\rangle \,.
\end{split}
\label{eq:NormLn}
\end{equation}
Thus, the unitarity requirement $\big|\big| L_{-n}\big|h,\bh\big\rangle \big|\big|\raisebox{0pt}[0pt]{${}
\stackrel{!}{\ge}{}$} 0$ demands $c\ge 0$ (due to the large-$n$ behavior) as well as $h\ge 0$ (following from the case
$n=1$). These are necessary conditions. A careful consideration of all mixed states shows, however, that there are
negative norm states even if $c\ge 0$ and $h\ge 0$. The preferred tool for studying these cases is provided by the
\emph{Kac determinant}. There is one such determinant at each excitation level, and the general definition can be best
understood by means of the second level example: At the level $n=2$ there are two basis states, $L_{-2}\big|h,\bh\big
\rangle$ and $(L_{-1})^2\big|h,\bh\big\rangle$. The corresponding Kac determinant reads
\begin{equation}
 \det \begin{pmatrix}
       \big\langle h,\bh\big| L_{-2}^\dagger\mku L_{-2} \big|h,\bh\big\rangle &
       \big\langle h,\bh\big| L_{-2}^\dagger\mku L_{-1}\mku L_{-1} \big|h,\bh\big\rangle \\
       \big\langle h,\bh\big| (L_{-1}\mku L_{-1})^\dagger\mku L_{-2} \big|h,\bh\big\rangle &
       \big\langle h,\bh\big| (L_{-1}\mku L_{-1})^\dagger\mku L_{-1}\mku L_{-1} \big|h,\bh\big\rangle
      \end{pmatrix}.
\label{eq:KacDet2}
\end{equation}
For $n>2$, there is an analogous construction involving all possible basis states of the level considered. By using the
commutation relations \eqref{eq:VirasoroAlgebra} the Kac determinants can be computed explicitly. They are functions
depending on $c$ and $h$. For instance, the determinant in \eqref{eq:KacDet2} amounts to $2\big(16h^3-10h^2+2h^2c+hc
\big)\big\langle h,\bh\big| h,\bh\big\rangle^2$.

Now, the key idea is that a negative or a zero determinant automatically means that there is a negative or a zero norm
state. For large $c$ and $h$ the Kac determinants are positive, and there are no negative norm states. Decreasing $c$
and/or $h$ one might encounter points in the $(c,h)$-space where one or more Kac determinants become zero, indicating a
transition into a region that admits negative norm states. This has been worked out in Refs.\ \cite{FQS84,FQS85,FQS86},
revealing the following results.

For $c\ge 1$, the Kac determinant analysis forms no obstacle to the existence of unitary representations of the
Virasoro algebra as long as $h\ge 0$. In particular, this space, $\big\{(c,h)\mkuu |\, c\ge 1,\, h\ge 0\big\}$, is
\emph{continuous}.

For $0\le c < 1$, on the other hand, there is only a \emph{discrete} set of points $(c,h)$ that allow unitary
representations. These points are given by
\begin{align}
 c &= 1-\frac{6}{m(m+1)}\,,\qquad m\ge 2,
\label{eq:DUMc}\\
\intertext{and}
 h &= \frac{[(m+1)p-mq]^2-1}{4m(m+1)}\,,\qquad p=1,\dotsc,m-1,\quad 1\le q \le p\,.
\label{eq:DUMh}
\end{align}
Figure \ref{fig:UnitaryReps} illustrates how the points are distributed in the $(c,h)$-space.
\begin{figure}[tp]
 \centering
 \includegraphics[width=0.6\columnwidth]{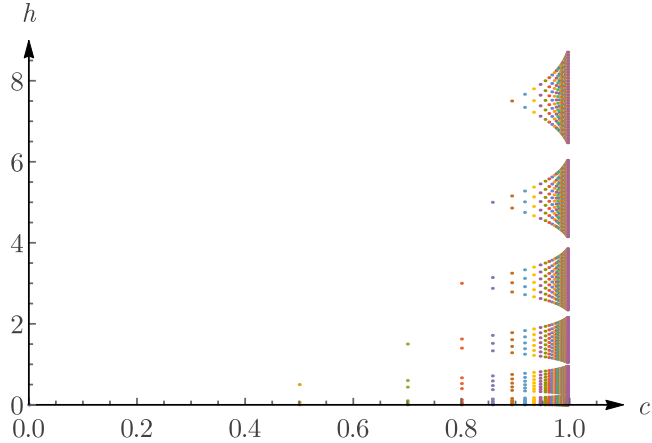}
\caption{Values of $c$ and $h$ in the region $0\le c<1$ that admit unitary Virasoro representations, according to
 eqs.\ \eqref{eq:DUMc} and \eqref{eq:DUMh} with $2\le m\le 40$.}
\label{fig:UnitaryReps}
\end{figure}

All other values of $c$ and $h$ (in the region $0\le c<1$) lead to negative norm states. It has been shown in Ref.\
\cite{GKO86} that the conditions for $c$ and $h$, eqs.\ \eqref{eq:DUMc} and \eqref{eq:DUMh}, respectively, are actually
sufficient for the existence of unitary representations. The importance of eqs.\ \eqref{eq:DUMc} and \eqref{eq:DUMh}
lies in the fact that they allow us to describe the possible scaling dimensions of fields in 2D CFTs, and thereby the
possible critical exponents of $2$-dimensional systems at their critical points. There is a complete classification
that identifies the discrete series of $c$- and $h$-values with \emph{statistical mechanical models at their second
order phase transitions}, for instance the Ising model ($m=3$) and the three-state Potts model ($m=4$)
\cite{FQS84,ABF84,Huse1984}.

For $c=0$ there is no interesting unitary Virasoro representation: By \eqref{eq:DUMc}, $c=0$ requires $m=2$ which, in
turn, dictates the trivial value $h=0$. From eq.\ \eqref{eq:NormLn} it then follows that all states $L_{-n}\big|h,\bh
\big\rangle$ would have zero norm. Hence, unitarity for $c=0$ can be achieved only if all the $L_n$ are represented by
$0$.

To sum up, \emph{a conformal field theory can be unitary} (corresponding to a nontrivial unitary Virasoro
representation) \emph{only if its central charge is positive}, $c>0$. If $c$ is even greater or equal to $1$, unitary
representations exist for any positive value of $h$.
\smallskip

\noindent
\textbf{(10) Final remarks.}
As an aside we would like to mention that the value $c=25$ plays a special role. The computation of the Kac determinant
involves the parameter $m=-\frac{1}{2}\pm\frac{1}{2}\sqrt{\frac{25-c}{1-c}}$ (which agrees with eq.\ \eqref{eq:DUMc}
solved for $m$, but now we allow general $c$ and $m$). For $1<c<25$ it becomes complex-valued, whereas for $c\ge 25$
it is strictly real, implying that all eigenvalues of the Kac determinant are positive. In Section
\ref{sec:ParamDepIntro} we present another argument justifying the name ``critical central charge'' for the value
$c=25$.

Finally, we note that, if a conformal field theory is quantized in an arbitrary external gravitational field, i.e.\ if
it is embedded in a curved background space, the length scale provided by the local scalar curvature $R$ breaks scale
invariance, and the expectation value of the stress-energy tensor is no longer traceless:
\begin{equation}
 \langle T^\mu{}_\mu\rangle = g_\mn\mkuu\frac{2}{\sg}\mkuu\frac{\delta\Gamma}{\delta g_\mn}=-\frac{c}{24\pi}\mkuu R \,,
\label{eq:ConfAnomaly}
\end{equation}
where $\Gamma$ denotes the effective action. This is referred to as \emph{trace anomaly} or \emph{conformal anomaly}.
In fact, eq.\ \eqref{eq:ConfAnomaly} can be used to determine the central charge of a theory if its effective action is
known (cf.\ Chapter \ref{chap:NGFPCFT}). By combining these ideas with FRG methods one can define a running
$c$-function \cite{CDP14b,CD15,CDP15}. At any fixed point, this $c$-function is constant and agrees with the central
charge of the corresponding conformal field theory, while at all other points it is a decreasing function w.r.t.\ the
RG scale (from the UV to the IR), demonstrating the irreversibility of the RG flow \cite{Zamolodchikov1986}.

%----------------------------------------------------------------------------------------------------------------------
\chapter[The space of metrics and the role of different parametrizations]%
{Towards quantum gravity: the space of metrics and the role of different parametrizations}
\label{chap:SpaceOfMetrics}
%----------------------------------------------------------------------------------------------------------------------

\begin{summary}
It is an open question how the fundamental microscopic field variables in quantum gravity look like. Motivated by the
classical formulation of general relativity we consider the case where the fundamental field is given by a proper
metric. Furthermore, we discuss a generalization to arbitrary symmetric rank-$2$ tensor fields. It turns out that the
most straightforward way to construct a reparametrization invariant effective (average) action is based on a geometric
formalism involving geodesics on the underlying field space. Here we propose a new connection on the space of metrics,
giving rise to a simple parametrization of geodesics. We demonstrate that this connection is adapted to the fundamental
geometric structure of the space of metrics. Special emphasis is laid upon the differences between Euclidean and
Lorentzian metric signatures. Finally, we compare the results with the closely related Vilkovisky--DeWitt method, and
we use the geometric language to set up reparametrization invariant, covariant quantities.

\noindent
\textbf{What is new?} Novel connection on the space of metrics (Secs.\ \ref{sec:DetConnections} \&
\ref{sec:GroupTheory}), its relation to the canonical connection (Secs.\ \ref{sec:GroupTheory} \& \ref{sec:CompConn}),
the role of the exponential metric parametrization as a geodesic (Sec.\ \ref{sec:GroupTheory}), a discussion on
peculiarities with Lorentzian metrics (Sec.\ \ref{sec:EuLor}).

\noindent
\textbf{Based on:} Refs.\ \cite{Nink2015} and \cite{DN15}.
\end{summary}

%----------------------------------------------------------------------------------------------------------------------
\section{Motivation and preliminaries}
\label{sec:SOMMotivation}
%----------------------------------------------------------------------------------------------------------------------

Metrics on a manifold $M$ are given by the covariant, symmetric, nondegenerate, smooth rank-$2$ tensor
fields.\footnote{In this chapter, a metric $g_\mn$ may refer to both quantum field and expectation value, cf.\ Sec.\
\ref{sec:BFF}, the actual status being either irrelevant for the respective argument or clear from the context.} In
local coordinates, a metric at some point $x\in M$ can be viewed as a symmetric matrix with prescribed signature
$(p,q)$:
\begin{alignat}{2}
 &\text{(i)} && g_\mn(x)\in\GLd\quad\forall\,x\in M,
 \label{eq:MetLocGLd}\\
 &\text{(ii)} && g_{\nu\mu}(x) = g_\mn(x)\quad\forall\,x\in M,\\
 &\text{(iii)}\qquad\quad && g_\mn(x)\text{ has $p$ positive and $q$ negative eigenvalues},\qquad\quad
 \label{eq:MetLocpq}
\end{alignat}
where $d=p+q$ is the dimension of $M$. The matrix representation $g_\mn(x)$ depends on the chosen basis of the tangent
space $T_x M$. By Sylvester's law of inertia, however, the numbers $p$ and $q$ are independent of the choice of basis,
and due to smoothness and nondegeneracy they are independent of the point $x$ as well, leading to a constant metric
signature. It is this fact that allows a global definition. 

In general, the set of all field configurations is referred to as field space, henceforth denoted by $\mF$. In the
present case, $\mF$ is the \emph{set of all metrics} on $M$ that have signature $(p,q)$. It is globally defined by
\begin{equation}[b]
 \mF \equiv \mF_{(p,q)} \equiv \Big\{g\in\SymT\,\Big |\; g \text{ has signature } (p,q)\Big\},
\label{eq:Fpq}
\end{equation}
where $\SymT$ is the space of symmetric type-$(0,2)$ tensor fields on $M$. (The notation ``$\Gamma\mku$'' indicates that
metrics are sections, $g: M\to S^2 T^* M$.) It can be shown that $\mF$ by itself exhibits the structure of an (infinite
dimensional) manifold \cite{Ebin1968,Ebin1970,FG89,GM91,Blair2000}.

In the conventional formulation of classical general relativity (GR) it is in fact the metric which is used as the
fundamental object to describe the geometry of the spacetime manifold $M$. Hence, classical GR admits only those
elements of $\SymT$ as candidates for $g$ that \emph{satisfy the fixed signature
constraint}.\footnote{For generalizations of classical GR that include signature changes, see Refs.\
\cite{DEHM97,WWV10}, for instance.} As we will see, this requirement restricts the full space
$\SymT$ considerably.

In quantum gravity the situation is different. The properties of the microscopic degrees of freedom are not known, in
particular it is unclear whether the fundamental field variables are given by symmetric rank-$2$ tensor fields at
all. A counterexample is provided by the vielbein formalism \cite{HR12,DP13} whose field variables are tetrads, and which
gives rise to (an equivalent version of) Einstein's equations at the classical level. Henceforth we will assume that
the fundamental field variable is given by an element of $\SymT$, though.

Even with this assumption we still do not know if the space $\SymT$ is to be constrained further:
It is a notoriously difficult question in virtually all functional integral based approaches to quantum gravity
whether, or to what extent, \emph{degenerate, wrong-signature} or even \emph{vanishing} tensor fields should be
included \cite{Witten1988,Percacci1991}.\footnote{It is well known that standard 1D configuration space functional
integrals are dominated by nondifferentiable paths since the set of differentiable ones has measure $0$. The basic
laws of quantum mechanics, noncommutativity of positions and momenta, force us to include these classically forbidden
nondifferentiable trajectories in the path integral \cite{GCP15}. Similarly, a consistent gravitational path integral
might require integrating over ``metrics'' which have further nonclassical features to a degree that is to be found
out.}
Since the set of pure metrics, $\mF$, forms a nonempty \emph{open} subset in $\SymT$ \cite{FG89,DN15},
there is no a priori reason to expect that $\mF$ has vanishing functional measure (nor that its complement has
vanishing functional measure), and so this question has no obvious answer.\footnote{In local coordinates the argument
can be clarified as follows. Metrics at some spacetime point correspond to symmetric matrices with signature $(p,q)$,
see eqs.\ \eqref{eq:MetLocGLd}--\eqref{eq:MetLocpq}. Embedding the space of all symmetric $d\times d$-matrices into
$\mathbb{R}^D$, with $D=\frac{1}{2}d(d+1)$, its subset of symmetric signature-$(p,q)$ matrices has nonvanishing
Lebesgue measure.}
It is known, however, that ``sufficiently different'' choices will lead to inequivalent theories \cite{AL98}.
Note that the class of actions one usually considers is constructed out of invariants of the type $\int\td^d x\sg$,
$\int\td^d x\sg R$, where for degenerate metrics the volume element $\sg$ could vanish and the inverse metric
required to raise indices could be nonexistent/divergent.

In this chapter we will demonstrate that the two options, $g\in\SymT$ vs.\ $g\in\mF$, can be
described in a simple way by using \emph{different parametrizations} for $g$.

As mentioned in Section \ref{sec:BFF}, all approaches to quantum gravity that are based on conventional quantum field
theory methods require the introduction of a non\-dynamical background metric, $\bg$, which is indispensable for the
construction of (nontopological) covariant objects. The metric fluctuations, denoted by $h$, then ``live'' on the
background geometry. There is, however, no unique way to parametrize the full, dynamical metric $g$ in terms of $\bg$
and $h$. Note that $h$ belongs to the tangent space to the space of all $g$. For the two options discussed above we
have
\begin{alignat}{2}
 &h \in T_g\mku\mF = \SymT\quad &&\text{if } g\in\mF,\\
 &h \in T_g\mkuu \SymT = \SymT\quad &&\text{if } g\in\SymT.
\end{alignat}
Hence, \emph{in both cases the fluctuating field $h$ is a symmetric type-$(0,2)$ tensor field}.\footnote{If $M$ is
noncompact, the $h$-space generalizes to $\big\{h\in\SymT\mku\big|\text{ $h$ has compact support}\big\}$ \cite{GM91}.}

We will see that there is a \emph{natural connection on $\SymT$} (namely the trivial connection),
and a \emph{natural connection on $\mF$} (which can be referred to as ``enhanced canonical connection''). Based on
these connections, the relation
\begin{equation}
 g_\mn(x) = \bg_\mn(x) + h_\mn(x),
\label{eq:StdParamX}
\end{equation}
formulated in local coordinates, parametrizes a \emph{geodesic on $\SymT$}, while
\begin{equation}
 g_\mn(x) = \bg_{\mu\rho}(x)\big(\e^{\mku\bg^{-1}(x)\mku h(x)}\big)^\rho{}_\nu
\label{eq:ExpParamX}
\end{equation}
parametrizes a \emph{geodesic on $\mF$}, respectively. Here $\e^{\mku\bg^{-1}\mku h}$ denotes the matrix exponential.
Indices are raised and lowered with the background metric. Note that since the signature requirement in the
definition of $\mF$ is a \emph{nonlinear} constraint, $\mF$ is not a vector space, whereas $\SymT$
is. The following sections focus on a closer investigation of $\mF$ in order to reveal its basic properties.

Since eqs.\ \eqref{eq:StdParamX} and \eqref{eq:ExpParamX} are pointwise relations, we drop the argument $x$ henceforth
if not explicitly needed. We refer to
\begin{equation}[b]
 g_\mn = \bg_\mn + h_\mn
\label{eq:StdParam}
\end{equation}
as the \emph{linear parametrization} (or standard parametrization), and to
\begin{equation}[b]
 g_\mn = \bg_{\mu\rho}\big(\e^h\big)^\rho{}_\nu
\label{eq:ExpParam}
\end{equation}
as the \emph{exponential parametrization}. In \eqref{eq:ExpParam} we adopted the usual notation
\cite{KKN93a,KKN93b,KKN93c,KKN96,AKKN94,NTT94,AK97} dropping the inverse background metric in the exponent,
cf.\ eq.\ \eqref{eq:ExpParamX}, as the index position $(\cdot)^\rho{}_\nu$ already indicates the involvement of $\bg$.
For later use, let us rewrite equation \eqref{eq:ExpParam} in matrix notation, too: With $h^T=h\in
\text{Sym}_{d\times d}$ it reads
\begin{equation}
 g = \bg\,\e^{\mku\bg^{-1}h}.
\label{eq:ExpParamMatrix}
\end{equation}

The remainder of this chapter is organized as follows. In Section \ref{sec:DetConnections} we derive connections on
$\SymT$ and $\mF$ whose associated geodesics are parametrized by \eqref{eq:StdParam} and \eqref{eq:ExpParam},
respectively. We investigate in Section \ref{sec:RepInv} if, or, on what conditions, \eqref{eq:ExpParam} can be
interpreted as a reparametrization of \eqref{eq:StdParam}. The main part is contained in Section \ref{sec:GroupTheory}:
We uncover the fundamental geometric structure of $\mF$, giving rise to a connection which emerges in the most natural
way and which agrees with the one derived in Section \ref{sec:DetConnections}. Notice the two opposed approaches:
In Section \ref{sec:DetConnections} we start out from the parametrizations, \emph{require} that they describe geodesics
and deduce the corresponding connections, while in Section \ref{sec:GroupTheory} the form of the geodesics is
\emph{derived} from the geometric properties inherent in the space of metrics. Furthermore, we point out significant
differences between the space of Euclidean metrics (which have signature $(p,q)=(d,0)$) and the space of Lorentzian
metrics (with mixed signature), see Section \ref{sec:EuLor}. The results are reviewed in general terms in Section
\ref{sec:CompConn} by comparing the new connection with the Levi-Civita connection and the Vilkovisky--DeWitt
connection. Finally, we discuss the exponential parametrization in the context of covariant Taylor expansions and
split-Ward (or Nielsen) identities in Section \ref{sec:Applications}.

%---------------------------------------------------------------------------------------
\section{Determining connections by reverse engineering}
\label{sec:DetConnections}
%---------------------------------------------------------------------------------------

Usually, considering geodesics requires some knowledge about the geometric details of the space, in particular
about the underlying connection. In this section, however, we take another path: For a moment we disregard the
information we have concerning the geometry of the spaces $\SymT$ and $\mF$. We rather take the view that we are
given the parametrizations \eqref{eq:StdParam} and \eqref{eq:ExpParam}, and we \emph{assume} that they parametrize
geodesics. Based on this assumption we would like to determine connections on $\SymT$ and $\mF$, respectively, such
that their corresponding geodesic equations are compatible with the parametrizations.

In the current section we follow this ``reverse logic'' for historical reasons. The parametrizations
\eqref{eq:StdParam} and \eqref{eq:ExpParam} have been used extensively in the literature (see for instance
\cite{Reuter1998,NR06,RS07,Percacci2009,CPR09,RS11,NRS13} for the linear parametrization and
\cite{KKN93a,KKN93b,KKN93c,KKN96,AKKN94,NTT94,AK97} for the exponential parametrization) without any clear declaration
if they are considered as geodesics or what spaces they are defined in. They have been applied rather due to their
advantages at the technical level in calculations. Let summarize some nongeometric arguments that motivate the use of
\eqref{eq:StdParam} and \eqref{eq:ExpParam}, the detailed geometric approach being postponed to Section
\ref{sec:GroupTheory}.
\medskip

\noindent
\textbf{(1) Motivation for the use of the linear parametrization}. It is evident that the parametrization $g_\mn=
\bg_\mn+h_\mn$ is the simplest implementation of the background field method, cf.\ Section \ref{sec:BFF}. Since the
background field is indispensable in the setting considered here, the use of eq.\ \eqref{eq:StdParam} introduces the
least amount of additional complexity in our calculations. By way of example, let $F[g]$ be a functional of the
metric. Then its functional derivatives w.r.t.\ $g_\mn$ agree with those w.r.t.\ $h_\mn\mku$:
$\frac{\delta}{\delta g_\mn} F[g] = \frac{\delta}{\delta h_\mn} F[\bg+h]$, and similarly for higher derivatives.

With regard to the above discussion concerning the space of symmetric rank-$2$ tensors, $\SymT$, as opposed to the
space of metrics, $\mF$, we find that $g=\bg+h$ in fact parametrizes elements of $\SymT$ since
$\bg\in\mF\subset\SymT$ and $h\in\SymT$, and since $\SymT$ is a vector space. Hence, using this parametrization admits
a $g$-space that is larger than $\mF$, including wrong-signature and vanishing tensor fields.

The linear parametrization has led to many important results in asymptotically safe gravity, both at the perturbative
and at the nonperturbative level, see Refs.\ \cite{Weinberg1979} and \cite{NR06}, for instance.
As this parametrization is the standard one, we refrain from going into more detail here.
\medskip

\noindent
\textbf{(2) Motivation for the use of the exponential parametrization}. Apart from its geometric meaning,
the parametrization $g_\mn = \bg_{\mu\rho}\big(\e^h\big)^\rho{}_\nu$ entails the following interesting consequences.
\begin{enumerate}
 \item We show in Appendix \ref{app:ExpParam} that eq.\ \eqref{eq:ExpParam} gives rise to proper metrics only:
  Provided that $\bg\in\mF$ and $h\in\SymT$ we find that $g=\bg\,\e^{\mku\bg^{-1} h}\in\mF$. Hence, the restriction to
  proper metrics (nowhere vanishing, correct signature) is an intrinsic feature of the exponential parametrization.
 \item The use of parametrization \eqref{eq:ExpParam} allows for an easy separation
  of the conformal mode from the fluctuations: When splitting $h_\mn$ into trace and traceless contributions,
  $h_\mn=\hat{h}_\mn+\frac{1}{d}\mku\bg_\mn\mku \phi$, with $\phi=\bg^\mn h_\mn$ and $\bg^\mn\hat{h}_\mn=0$, the trace
  part gives rise to a conformal factor in \eqref{eq:ExpParam}:
  \begin{equation}
   g_\mn = \e^{\frac{1}{d}\mku\phi}\, \bg_{\mu\rho}\big(\e^{\hat{h}}\mku\big)^\rho{}_\nu \,.
  \end{equation}
  Remarkably enough, the volume element on the spacetime manifold depends only on $\phi$, while the traceless part of
  $h_\mn$ drops out completely: 
  \begin{equation}
   \sg=\sbg\,\e^{\frac{1}{2}\phi} \,.
  \label{eq:sgExpParam}
  \end{equation}
  In the context of gravity this means that the cosmological constant occurs as a coupling only
  in the conformal mode sector. This will become explicit in the calculations performed in the next chapter.
\item Partially related to the previous point, there are certain cases where computations are simplified or become
  feasible only if parametrization \eqref{eq:ExpParam} is used. Let us briefly mention four examples.
  (a) In the search of scaling solutions in scalar-tensor gravity, infrared singularities occurring in standard
  calculations \cite{NP10,NR10} can be avoided by employing the exponential parametrization \cite{PV15,LPV16}.
  (b) The RG flow of nonlocal form factors appearing in a curvature expansion of the effective average action $\Gk$ is
  divergent in the limit $d\to 2$ for small $k$ when based on \eqref{eq:StdParam} \cite{SCM10}, but it has a meaningful
  limit when based on \eqref{eq:ExpParam} \cite{CD15}.
  (c) The exponential parametrization provides an easy access to unimodular quantum gravity
  \cite{Eichhorn2013b,Eichhorn2015}.
  (d) The use of \eqref{eq:ExpParam} ensures gauge independence at one-loop level without resorting to the
  Vilkovisky--DeWitt method \cite{Falls2015a,Falls2015b} (cf.\ also Section \ref{sec:CompConn}).
\item Our main motivation for parametrization \eqref{eq:ExpParam} arises from its apparent connection to conformal
  field theory: CFT studies show that there is a critical number of scalar fields in a theory of gravity coupled
  to conformal matter, referred to as the critical central charge, at which the conformal mode $\phi$ decouples. It
  amounts to $c_\text{crit}=25$ \cite{David1988,DK89,Polchinski1989,Watabiki1993}. Notably, this result is correctly
  reproduced in the Asymptotic Safety program when using the exponential parametrization
  \cite{KKN93a,KKN93b,KKN93c,KKN96,AKKN94,NTT94,AK97,Nink2015,CD15}, while the linear relation \eqref{eq:StdParam}
  gives rise to $c_\text{crit}=19$ \cite{Tsao1977,Brown1977,KN90,JJ91,Reuter1998,Nink2015,CD15}. This will be discussed
  in detail in Chapter \ref{chap:ParamDep}.
\end{enumerate}
\medskip

\noindent
\textbf{(3) Connections, geodesics and DeWitt's notation}. Geodesics on a differentiable manifold --- parametrized by
means of an \emph{exponential map}\footnote{Note that, a priori, the exponential parametrization is unrelated to the
exponential map.} --- are fixed by the choice of an affine connection. In this context, different connections lead to
different exponential maps. Above we have discussed the relevance of the linear and the exponential metric
parametrizations. Now we aim at finding connections on $\SymT$ and $\mF$ in such a way that the corresponding
exponential maps are given by \eqref{eq:StdParam} and \eqref{eq:ExpParam}, respectively.

In order to introduce the method in general terms, we employ \emph{DeWitt's condensed notation} \cite{DeWitt1964} where
each Latin index represents both discrete and continuous (e.g.\ spacetime) labels, $i\equiv (\mu,\nu,x)$, for instance.
Let $\vp$ denote a generic field.
Then $\vp^i$ can be regarded as the local coordinate representation of a point in field space $\big($here $\SymT$ or
$\mF\mku\big)$, so we identify\footnote{Note that $\mu,\nu$ are covariant (lower) indices referring to the dual of the
tangent space to $M$, while $i$ is a contravariant (upper) index referring to the tangent space to $\SymT$ or $\mF$.}
\begin{equation}
 \vp^i\equiv g_\mn(x)\,.
\end{equation}
Repeated condensed indices are interpreted as summation over discrete and integration over continuous indices: $a^i\mku
b_i \equiv \int_x\mku a_\mn(x)\mkuu b^\mn(x)$, with $\int_x \equiv \int\dd x$. By $\bp^i$ we will denote a fixed but
arbitrary background field.

Our starting point for the derivation of the desired connections will be an expansion of $\vp^i$ in terms of tangent
vectors, determined by a \emph{geodesic connecting $\bp^i$ to $\vp^i$}. Let $\vp^i(s)$ denote such a geodesic, i.e.\
a curve with
\begin{equation}
\vp^i(0)=\bp^i\quad \text{and}\quad \vp^i(1)=\vp^i,
\end{equation}
that satisfies the geodesic equation
\begin{equation}
 \ddot{\vp}^i(s)+\Gamma^i_{jk}\,\dot{\vp}^j(s)\dot{\vp}^k(s)=0,
\label{eq:GeodEqu}
\end{equation}
where the dots indicate derivatives w.r.t.\ the curve parameter $s$, and $\Gamma^i_{jk}$ is the Christoffel symbol
evaluated at $\vp^i(s)$, that is, $\Gamma^i_{jk}\equiv\Gamma^i_{jk}[\vp^i(s)]$. We assume for a moment that the
geodesic $\vp^i(s)$ lies entirely in one coordinate patch. As we will see, the two connections determined below give
rise to only such geodesics that automatically satisfy this assumption. In that case we can expand the local
coordinates as a series,
\begin{equation}
 \vp^i(s)=\sum\limits_{n=0}^\infty\frac{s^n}{n!}\left(\frac{\td^n}{\td s^n}\vp^i(s)\Big|_{s=0}\right).
\label{eq:PhiExpansion0}
\end{equation}
We observe that it is possible to express all higher derivatives in \eqref{eq:PhiExpansion0} in terms of
$\dot{\vp}^i$ by using equation \eqref{eq:GeodEqu} iteratively. If $h^i\equiv\dot{\vp}^i(0)$ denotes the tangent
vector at the point $\bp$ in the direction of the geodesic, we obtain the following relation for $\vp^i=\vp^i(1)$:
\begin{equation}
  \vp^i= \bp^i+h^i -{\textstyle\frac{1}{2}}\mku\bGamma^i_{jk}\,h^j h^k +{\textstyle\frac{1}{6}}\big(\bGamma^i_{mj}
  \bGamma^m_{lk}+\bGamma^i_{km}\bGamma^m_{lj}-\bGamma^i_{jk,l}\big)h^j h^k h^l +\mO(h^4),
\label{eq:PhiExpansion}
\end{equation}
where we used the abbreviations $\bGamma^i_{jk}=\Gamma^i_{jk}[\bp]$ and $\bGamma^i_{jk,l}\equiv
\frac{\delta}{\delta\bp^l}\bGamma^i_{jk}$ for the connection and its derivatives at the point $\bp$.

By construction, any geodesic from $\bp^i\equiv \vp^i(0)$ to $\vp^i\equiv \vp^i(1)$ with initial velocity
$\dot{\vp}^i(0)=h^i$ satisfies equation \eqref{eq:PhiExpansion}. On the other hand, if we start with an arbitrary
parametrization of $\vp^i$ in terms of $\bp^i$ and $h^i$, say
\begin{equation}
 \vp^i= f(\bp^i,h^i),
\end{equation}
with $f(\bp^i,0)=\bp^i$, and we require that it be a geodesic, then we can expand $f(\bp^i,h^i)$ in terms of $h^i$ and
compare it with \eqref{eq:PhiExpansion} in order to determine a suitable connection. It is this approach that we pursue
in the remainder of this section. Note that the connection  $\bGamma^i_{jk}$ can be read off already from the second
order term in \eqref{eq:PhiExpansion} and in the expansion of $f(\bp^i,h^i)$. In standard index notation equation
\eqref{eq:PhiExpansion} amounts to
\begin{equation}
 g_\mn(x)= \bg_\mn(x)+h_\mn(x) -{\textstyle\frac{1}{2}}\int_y\int_z
  \bGamma^{\alpha\beta\,\rho\sigma}_\mn(x,y,z)h_{\alpha\beta}(y) h_{\rho\sigma}(z) +\mO(h^3).
\label{eq:gExpansion}
\end{equation}
\smallskip

\noindent
\textbf{(4) Deriving a connection compatible with the linear parametrization}. We would like to determine a connection
$\bGamma^i_{jk} \equiv \bGamma^{\alpha\beta\,\rho\sigma}_\mn(x,y,z)$ on $\SymT$ in such a way that it is compatible
with the linear parametrization,
\begin{equation}
 g_\mn(x) = \bg_\mn(x) + h_\mn(x).
\label{eq:tmpStdParamX}
\end{equation}
To this end we compare \eqref{eq:tmpStdParamX} with \eqref{eq:gExpansion}. As the equality must hold for any $h_\mn$,
we conclude $\bGamma^{\alpha\beta\,\rho\sigma}_\mn(x,y,z)=0$. Moreover, since the background metric is arbitrary, the
connection must vanish everywhere. This proves that the trivial (flat) connection,
\begin{equation}[b]
 \Gamma^{\alpha\beta\,\rho\sigma}_\mn(x,y,z)=0 \quad \text{on } \SymT,
\end{equation}
leads to geodesics on $\SymT$ that are parametrized by the linear relation \eqref{eq:tmpStdParamX}. Although this
connection has been obtained from the second order term in \eqref{eq:gExpansion}, the equality
\eqref{eq:tmpStdParamX}${}={}$\eqref{eq:gExpansion} holds at all orders as all higher order terms vanish.
\medskip

\noindent
\textbf{(5) Deriving a connection for the exponential parametrization}. Analogously, for the space of metrics, $\mF$,
equation \eqref{eq:gExpansion} is to be compared with the exponential metric parametrization \eqref{eq:ExpParam}, which
can be written as the pointwise series
\begin{equation}
  g_\mn(x) = \bg_\mn(x)+h_\mn(x)+{\textstyle\frac{1}{2}}\mku\bg^{\rho\sigma}(x)h_{\mu\rho}(x)h_{\nu\sigma}(x) +\mO(h^3).
\label{eq:gExpSeries}
\end{equation}
The connection $\bGamma^{\alpha\beta\,\rho\sigma}_\mn(x,y,z)$ can again be read off from the second order terms in
\eqref{eq:gExpansion} and \eqref{eq:gExpSeries}. Here we must take into account that any affine connection
maps two vector fields to another vector field. In our current setup we have to ensure that the connection maps to
the space of symmetric tensors. Thus, we require: $\bGamma(X,Y)=Z\in\SymT$ for $X,Y\in\SymT$. In terms of local
coordinate relations, this requirement can be implemented by symmetrizing indices adequately.\footnote{By convention,
round brackets indicate symmetrization, for instance, $a_{(\mn)}\equiv\frac{1}{2}(a_{\mn}+a_{\nu\mu})$.} We obtain
$\bGamma^{\alpha\beta\,\rho\sigma}_\mn(x,y,z)=-\delta^{(\alpha}_{(\mu}\,
\bg^{\raisebox{0.2ex}{$\scriptstyle\beta)(\rho$}}(x)\,\delta^{\sigma)}_{\nu)}\; \delta(x-y)\delta(x-z)$.
Since the result is valid for arbitrary base points $\bg_\mn$, we can proceed to its unbarred version, i.e.\ to the
connection evaluated at $g_\mn$, yielding
\begin{equation}[b]
  \Gamma^{\alpha\beta\,\rho\sigma}_\mn(x,y,z)=-\delta^{(\alpha}_{(\mu}\,
  g^{\raisebox{0.2ex}{$\scriptstyle\beta)(\rho$}}(x)\,
  \delta^{\sigma)}_{\nu)}\; \delta(x-y)\delta(x-z) \quad \text{on } \mF.
\label{eq:NDConnection}
\end{equation}
This is the main result of this section.

It remains to be shown that the connection \eqref{eq:NDConnection} inserted into \eqref{eq:gExpansion} is consistent
with \eqref{eq:gExpSeries} not only at second order but also at all higher orders. It is straightforward to convince
oneself that the third order terms do in fact agree. For a complete proof at all orders, however, we proceed
differently. The idea is to find exact solutions to the geodesic equation \eqref{eq:GeodEqu} based on the connection
\eqref{eq:NDConnection}.

Before doing so, let us make an important remark. Since $\Gamma^{\alpha\beta\,\rho\sigma}_\mn(x,y,z)$ is proportional
to $\delta(x-y)\delta(x-z)$, all integrations implicit  in \eqref{eq:GeodEqu} are trivial. Therefore, the geodesic
equation is \emph{effectively pointwise} with respect to spacetime. This means that geodesics on $\mF$ starting at
$\bg_\mn(x)$ at some spacetime point $x$ can only go to metrics of the type $g_\mn(x)$ at the same point $x$; it can
never reach, say, $g_\mn(x')$ if $x'\neq x$, nor can it give rise to nonlocal expressions involving spacetime
integrations.
As already stated above, any metric in local coordinates at a given point $x$ can be considered an element of the set
of symmetric matrices with signature $(p,q)$. The latter is an \emph{open} and connected subset in the vector space of
symmetric matrices (cf.\ discussion in Section \ref{sec:GroupTheory}), and thus it can be \emph{covered with one
coordinate chart}. Therefore, geodesics corresponding to \eqref{eq:NDConnection} \emph{stay indeed in one chart}, in
agreement with the assumption that led to eq.\ \eqref{eq:PhiExpansion0}.

Due to the pointwise character of the geodesic equation, the spacetime dependence is not written explicitly in the
following. Based on the connection \eqref{eq:NDConnection}, equation \eqref{eq:GeodEqu} boils down to
\begin{equation}
 \ddot{g}_\mn -\delta^{(\alpha}_{(\mu}\, g^{\raisebox{0.2ex}{$\scriptstyle\beta)(\rho$}}\,\delta^{\sigma)}_{\nu)}
 \dot{g}_{\alpha\beta} \dot{g}_{\rho\sigma}
 = \ddot{g}_\mn -g^{\beta\rho}\dot{g}_{\mu\beta} \dot{g}_{\rho\nu}=0.
\label{eq:NewGeodEqu}
\end{equation}
Upon multiplication with $g^{\nu\lambda}$ we observe that \eqref{eq:NewGeodEqu} can be brought to the form
\begin{equation}
\frac{\td}{\td s}\left(\dot{g}_\mn g^{\nu\lambda}\right)=0,
\end{equation}
that is, $\dot{g}_\mn g^{\nu\lambda}=c^\lambda_\mu=\text{const}$. In matrix notation this reads
\begin{equation}
 \dot{g}(s)=c\mku g(s).
\label{eq:MatDiffEq}
\end{equation}
Equation \eqref{eq:MatDiffEq} is known to have the unique solution $g(s)=\e^{sc}\mkuu g(0)$. Using the initial
conditions $g(0)=\bg$ and $h=\dot{g}(0)=c\mkuu g(0)=c\mkuu\bg$ we obtain $g(s)=\e^{s\mku h\bg^{-1}}\mkuu \bg$, which
finally leads to
\begin{equation}
 g(s)=\bg\,\e^{s\mku \bg^{-1} h}.
\label{eq:GeodInM}
\end{equation}
Setting $s=1$ and switching back to index notation, this is precisely the exponential relation \eqref{eq:ExpParam} for
the metric. Hence, we have proven that geodesics corresponding to the connection \eqref{eq:NDConnection} are uniquely
parametrized by $g_\mn = \bg_{\mu\rho}\big(\e^h\big)^\rho{}_\nu\mku$. As a result, \eqref{eq:gExpansion} and
\eqref{eq:gExpSeries} agree indeed at all orders.

In conclusion, there is a connection that defines a structure on the field space $\mF$, the set of all metrics,
entailing a simple exponential parametrization of geodesics on $\mF$. Here it has been derived by starting with the
parametrization and assuming that it describes geodesics. Whether there is a more fundamental geometric motivation for
this connection, for instance a field space metric, will be discussed in Section \ref{sec:GroupTheory}.

%----------------------------------------------------------------------------------------------------------------------
\section{A note on reparametrization invariance}
\label{sec:RepInv}
%----------------------------------------------------------------------------------------------------------------------

Let us briefly discuss as to why the choice of parametrization is relevant at all. A priori, there seems to be no
reason to prefer one parametrization over another one. In fact, field redefinitions in a path integral for the
partition function do not change $S$-matrix elements, a statement known as the equivalence theorem
\cite{Borchers1960,CWZ69,KT73}. Hence, all physical quantities are invariant under field redefinitions. The point we
want to make here is that switching between the linear and the exponential relation for the metric is not a genuine
reparametrization, in the sense that it is not a one-to-one correspondence.
\medskip

\noindent
\textbf{(1)}
As discussed above and proven in Appendix \ref{app:ExpParam}, the exponential parametrization gives rise to only proper
metrics satisfying the signature constraint, while the linear parametrization admits also wrong-signature and vanishing
tensor fields: $g=\bg\,\e^{\mku\bg^{-1}h}\in\mF$ and $g=(\bg+h)\in\SymT$, respectively. Therefore, \emph{the
exponential parametrization cannot be obtained from the linear parametrization by means of a field redefinition.} There
exist infinitely many $g\in\SymT$ that can be expressed as $g=\bg+h$, but not as $g=\bg\,\e^{\mku\bg^{-1}h}$. Put
another way, the addition in $g=\bg+h$ with $\bg\in\mF$ and $h\in\SymT$ can result in ``leaving'' the space $\mF$.

However, it is possible to constrain the $h$-space when the linear parametrization is used such that $\bg+h$ becomes a
proper metric. The constrained $h$-space, henceforth denoted by $H_{\bg}$, is a subset of the space of symmetric
tensors, $H_{\bg}\subset\SymT$, and it depends on the background metric $\bg\mku$: $H_{\bg}\equiv\big\{h\in\SymT\,\big|
\,(\bg+h)\in\mF\big\}$. Note that it has similar nonlinear properties to $\mF$. Only with this restriction, the linear
relation
\begin{equation}
 g = \bg + h'\,, \quad \text{with } h'\in H_{\bg}\,,
\end{equation}
can be a reparametrization of
\begin{equation}
 g=\bg\,\e^{\mku\bg^{-1}h}\,, \quad \text{with } h\in\SymT.
\end{equation}

\noindent
\textbf{(2)}
Although the restriction to $H_{\bg}$ is possible in principle, it is usually not applied to calculations in the
pertinent quantum gravity literature since one prefers to integrate over linear spaces.\footnote{This way, it is easier
to evaluate Gaussian integrals \cite{Mottola1995}, for instance.} Hence, in all standard approaches the exponential and
the linear parametrization describe different objects after all. This justifies our discussion concerning field
parametrization dependent results, see also Chapter \ref{chap:ParamDep}. Even if we assume for a moment that
restriction to $H_{\bg}$ is applied, the question about reparametrization invariance is more involved than it seems at
first sight: While the equivalence theorem is based on the use of the equations of motion, we argue in the following
that the (off shell) effective action $\Gamma$ in the usual formulation does still depend on the choice of the
parametrization. This is a crucial observation since there are many important physical applications involving off shell
quantities, e.g.\ $\beta$-functions and the existence of fixed points in RG studies (see below), or the effective
potential part of the effective action in the context of spontaneous symmetry breaking \cite{RW94b,BTW95}. Choosing the
parametrization appropriately may be a powerful tool to simplify the underlying computations. For points
\textbf{(3)} and \textbf{(4)} we continue assuming that there is a one-to-one correspondence between the
parametrizations.
\medskip

\noindent
\textbf{(3)}
Pioneered by Vilkovisky \cite{Vilkovisky1984} and DeWitt \cite{DeWitt1987}, there is a way to construct an effective
action, $\Gamma^\text{VDW}$, which is reparametrization invariant, gauge invariant and gauge independent both off and
on shell.\footnote{``Gauge independence'' denotes the invariance of the effective action under changes of the gauge
condition, while ``gauge invariance'' refers as usual to its invariance under gauge transformations.} However,
the price one has to pay for this invariance is a nontrivial dependence of $\Gamma^\text{VDW}$ on the background
metric, encoded in modified Ward identities (sometimes also referred to as modified Nielsen identities) relating
$\delta\Gamma^\text{VDW}/\delta g_\mn$ to $\delta\Gamma^\text{VDW}/\delta \bg_\mn\mku$ \cite{BK87,Kunstatter1992},
cf.\ Section \ref{sec:Applications}. Unlike the conventional effective action, the Vilkovisky--DeWitt (VDW) effective
action does not generate the 1PI correlation functions, and since it entails new nonlocal structures, calculations are
generically much more involved. Furthermore, $\Gamma^\text{VDW}$ can have a remaining dependence on the chosen
configuration space metric \cite{Odintsov1991}. Ultimately, it depends on the desired application whether or not a
reparametrization invariant approach is useful.
\medskip

\noindent
\textbf{(4)}
RG studies (without the VDW method) show that $\beta$-functions and fixed points do indeed vary when the
parametrization is changed \cite{Wegner1974,BW75,RGN85,Golner1986,Morris1998}. A similar example of off shell
noninvariance is provided by the frame dependence in cosmology \cite{KS15}. Moreover, reparametrization invariance
is violated even on shell when truncations, e.g.\ derivative expansions, are considered \cite{Morris1998}.
In the context of asymptotically safe gravity there is, in principle, the interesting possibility that a non-Gaussian
fixed point exists in parametrization A, giving rise to a well defined UV limit, while there is no such fixed point in
parametrization B. Clearly, such a result would have to be tested for stability under extensions of the truncation.

Combining RG techniques with the ideas of Vilkovisky and DeWitt leads to the geometrical effective average action,
$\Gamma_k^\text{VDW}$, which --- by analogy with $\Gamma^\text{VDW}$ --- is reparametrization and gauge invariant
as well as gauge independent, and which is constrained by modified Ward identities
\cite{BMV03,Pawlowski2003}. Therefore, again, the benefits entailed by this construction can be obtained only at the
expense of nontrivial dependencies on the background, and, on the technical side, computations are of increased
complexity \cite{DP12}. This constitutes one of the major drawbacks of the VDW method. 

The path we will take in the following is a compromise between the VDW and the conventional approach. We avoid the
aforementioned nonlocalities by choosing a geometric formalism (taking into account the nonlinear character of $\mF$)
that leads to a reparametrization invariant and (background) gauge invariant but not gauge independent effective
(average) action. This will reduce the complexity of calculations considerably. In Sections \ref{sec:CompConn} and
\ref{sec:Applications} we clarify the idea in more detail and compare our results with those of the VDW method.
\medskip

\noindent
\textbf{(5)}
Let us come back to the usual case where the exponential parametrization is not a proper field redefinition of the
linear one. Due to the problem of finding appropriate physical observables in gravity,\footnote{If accessible,
considering physical observables is of course preferable as these should not exhibit any parametrization or gauge
dependence. In quantum gravity, however, it is not even clear what physically meaningful observable quantities are,
and so far there is no experiment for a direct measurement of quantum gravity effects \cite{Woodard2009}. Based on
effective field theory arguments it is possible to compute the leading quantum corrections to the Newtonian potential
\cite{Donoghue1994a,Donoghue1994b,DM97,Kirilin2007}, but the effect is unobservably small and the description is valid
only in the low energy regime, so it cannot be considered a fundamental theory of the gravitational field.}
the best one can do with a candidate theory of quantum gravity is to test it for self-consistency, check the classical
limit, and compare it with other approaches. In this regard, too, studying off shell quantities like $\beta$-functions
is of substantial interest. Their parametrization dependence might then be exploited to simplify the comparison between
different theories. In fact, we will see in Chapters \ref{chap:NGFPCFT} and \ref{chap:FullReconstruction} that it is
the exponential parametrization that establishes a connection of our approach to conformal field theory and bosonic
string theory.
\medskip

To sum up, we have argued that the choice of parametrization plays an important role, both from a technical and from a
fundamental perspective, even if only proper (i.e.\ one-to-one) field redefinitions are considered. In our setup, the
latter could be achieved by restricting the $h$-space for the linear parametrization to $H_{\bg}$. However, such a
restriction is inconvenient, and we will not apply it in the remainder of this thesis. Thus, by employing the
exponential parametrization as compared with the linear one we describe a different fundamental field, possibly giving
rise to a different theory at the quantum level.

%----------------------------------------------------------------------------------------------------------------------
\section[The canonical connection and its geodesics]%
{The fundamental geometric structure of the space of metrics: the canonical connection and its geodesics}
\label{sec:GroupTheory}
%----------------------------------------------------------------------------------------------------------------------

We have already discussed that the space of symmetric rank-$2$ tensors is a vector space. Its most natural
connection is the flat one, and the corresponding geodesics are straight lines described by the linear parametrization.
This section, on the other hand, addresses solely the space of metrics, $\mF\equiv\mFpq$, defined in eq.\
\eqref{eq:Fpq}.

We would like to show that, from a group theory and differential geometry perspective, $\mF$ possesses a fundamental
structure which does not rely on any further external input like the definition of a connection, but which singles
out one particular connection instead. Thus, unlike in Section \ref{sec:DetConnections} we \emph{derive} a connection
from a few principles to be stated in a moment, rather than adapt it to a specific parametrization.
While most of the arguments presented in Subsection \ref{sec:GenDesc} are well known (see for instance Refs.\
\cite{ONeill1983,KN69}, cf.\ also \cite{DeWitt1967a},\cite{FG89} and \cite{PV15}), the connection in $\mF$ that
eventually derives from them, as well as its geodesics, represent new results \cite{DN15}. By reviewing the foundations
in Subsection \ref{sec:GenDesc} we also intend to reconcile the mathematical with the physical literature. In
Subsection \ref{sec:EuLor} we distinguish carefully between Euclidean and Lorentzian metrics, pointing out some
important issues related to the exponential parametrization in the Lorentzian case.

%----------------------------------------------------------------------------------------------------------------------
\subsection{General description}
\label{sec:GenDesc}
%----------------------------------------------------------------------------------------------------------------------

As observed in Section \ref{sec:SOMMotivation}, any metric $g\in\mF$ at a given spacetime point can be considered a
symmetric matrix. More precisely, if $g$ has signature $(p,q)$, then in any chart $(U,\vp)$ for the spacetime manifold
$M$ the metric \emph{in local coordinates} is a map
\begin{equation}
\label{eq:mapping}
 g\big|_U:U\to \mat \, , \quad x\mapsto g_\mn(x),
\end{equation}
where $\mat\equiv\Mpq$ denotes the set of \emph{real invertible symmetric $d\times d$ matrices with signature
$(p,q)$},
\begin{equation}[b]
 \mat \equiv \Mpq \equiv \left\{ A \in \GLd \big| \, A^T=A,\; A \text{ has signature }(p,q)\right\}.
\label{eq:DefMatrices}
\end{equation}
Due to this local appearance there is a simple illustration of the full space $\mF$ whose rigorous definition in terms
of sections of a fiber bundle, given by eq.\ \eqref{eq:Fpq}, is rather abstract: We may think of $\mF$ as a topological
product,
\begin{equation}
 \mF \simeq \prod_{x\in M} \mat\,,
\label{eq:FtoM}
\end{equation}
supplemented by additional requirements that guarantee continuity.

In this section we focus on the properties of $\mat$. By eq.\ \eqref{eq:FtoM} most topological and differential
geometrical features carry over from $\mat$ to $\mF$.

There is one important constraint which will underly our discussion concerning geodesics on $\mF$: We restrict
ourselves to \emph{local} geodesics. Here ``local'' refers to ``local w.r.t.\ spacetime''. This means that, loosely
speaking, a geodesic on $\mF$ connecting $\bg_\mn(x)$ to $g_\mn(x)$ for $x\in M$ ``stays'' in $x$ for all points of
the geodesic, and it is independent of all other spacetime points.\footnote{Note the distinction between spacetime
points, $x\in M$, and points on geodesics, $g\in\mF$.} In particular, the construction of geodesics does not contain
any spacetime integrations involving the background metric or tangent vectors, for instance. Only then geodesics on
$\mat$ can be \emph{lifted} straightforwardly to geodesics on $\mF$. In order to guarantee this locality we have to
make a simple \emph{assumption for the class of connections we admit}: We allow only such connections that are
\emph{spacetime-diagonal} in local coordinates, i.e.
\begin{equation}
 \Gamma^{\alpha\beta\,\rho\sigma}_\mn(x,y,z) \propto \delta(x-y)\delta(x-z)\,.
\end{equation}
Based on this assumption the analysis of geodesics on $\mF$ can be done pointwise, cf.\ also \cite{FG89}. Hence, we can
reduce our discussion to the matrix space $\mat$.\footnote{Note that the Vilkovisky--DeWitt connection does not fall
into the class of considered connections as it is nondiagonal w.r.t.\ spacetime. Moreover, it is nonlocal w.r.t.\ the
field space $\mF$.} Once we have found a geodesic on $\mat$ parametrized by a tangent vector, we obtain a geodesic on
$\mF$ by using the same parametrization but promoting the tangent vector to an $x$-dependent field. Continuity of the
geodesic with respect to $x$ is then ensured by continuity of the vector field.

At this point we can specify the principles our derivation of a connection on $\mF$ will be based on: (a) The
connection is required to be spacetime-diagonal, and (b) it is to be adapted to the natural geometric structure of
$\mF$. The first requirement is needed to reduce the discussion to $\mat$, while the second one will uniquely
single out one connection. 

Let us discuss the properties of $\mat$ now. We will denote points in $\mat$ by $o$ and $\bo$ rather than $g$ and $\bg$
in order to avoid confusion with elements of $\mF$, and since the symbol $g$ will be used for group elements in
accordance with the standard literature, here $g\in G\equiv\GLd$. Unless otherwise specified, the following arguments
are \emph{valid for all $p,q\ge 0$} satisfying $p+q=d$, i.e.\ for both Euclidean and Lorentzian metrics.
\medskip

\noindent
\textbf{(1) The set $\bm{\mat}$ as a homogeneous space.}
We find that $\mat$ is a smooth manifold since it is an \emph{open subset} in the vector space of all symmetric
matrices,\footnote{Proof: Any matrix $o\in\Mpq$ has nonvanishing determinant, $\det(o)\neq 0$. Continuity of the
determinant implies that all symmetric matrices in a sufficiently small neighborhood of $o$ (with respect to some
matrix norm) must also have nonvanishing determinant: $\det(o+\epsilon X)=\det(o)\det(\Id+\epsilon\mku o^{-1}X)=\det(o)
\big[1+\epsilon\mku\Tr(o^{-1}X) + \mO(\epsilon^2)\big]\neq 0$ for $\epsilon$ small enough. As the (real) eigenvalues of
symmetric matrices change continuously, too, the matrices $o+\epsilon X$ in the neighborhood of $o$ cannot have any
zero eigenvalue and the number of positive and negative eigenvalues cannot change, so $(o+\epsilon X)\in\Mpq$.
Hence, $\Mpq$ is an open subset of $S_d\mku$.}
\begin{equation}
 S_d \equiv \left\{A\in\matrices\big|A^T=A\right\}.
\end{equation}
Hence, the tangent space at any point $o\in\mat$ is given by $T_o\mat=S_d$. In what follows we aim at describing $\mat$
as a homogeneous space. For this purpose we recognize that the group $G\equiv\GLd$ acts transitively on $\mat$ by
\begin{equation}
\begin{split}
 \phi:G\times\mat &\rightarrow\mat, \\
 (g,o) &\mapsto \phi(g,o)\equiv g*o \equiv (g^{-1})^T o\mku g^{-1}.
\end{split}
\label{eq:groupAction}
\end{equation}
The fact that $g*o$ belongs indeed to $\mat$ and that the action is transitive
(i.e.\ $\forall\ o_1,o_2\in\mat\; \exists\ g\in G: g*o_1=o_2$) is a consequence of Sylvester's law of inertia.
Note that $\phi$ is a left action, that is, $g_1*(g_2*o)=(g_1g_2)*o$.
Let us consider a fixed but arbitrary base point $\bo\in\mat$ now. It is most convenient to think of $\bo$ as
\begin{equation}
I_{(p,q)}= \begin{pmatrix}\mathds{1}_{p\times p} & \\ & -\mathds{1}_{q\times q}\end{pmatrix},
\label{eq:Ipq}
\end{equation}
although the subsequent construction is independent of that choice. The \emph{isotropy group} (stabilizer) of $\bo$
is given by\footnote{Note that $h^T\bo\, h=\bo$ is equivalent to $h*\bo \equiv (h^{-1})^T\bo\, h^{-1}=\bo$.}
\begin{equation}
H \equiv H_\bo \equiv \Opq_\bo(p,q)\equiv \left\{ h\in\matrices\big|\, h^T\bo\mku h=\bo\right\},
\end{equation}
which is conjugate to the semi-orthogonal group, and which is a closed subgroup of $G\equiv\GLd$.
This makes $\mat$ a \emph{homogeneous space}, and we can write
\begin{equation}[b]
 \mat \simeq G/H,
\label{eq:MasGandH}
\end{equation}
where $G/H$ are the \emph{left} cosets of $H$ in $G$.
Defining the \emph{canonical projection}
\begin{equation}
 \pi: G \rightarrow \mat,\; g\mapsto \pi(g)\equiv (g^{-1})^T\bo\mku g^{-1},
\label{eq:CanProj}
\end{equation}
we see that $(G,\pi,\mat,H)$ becomes a \emph{principal bundle with structure group $H$}. Figure \ref{fig:PBundle}
illustrates this relation.
\begin{figure}[tp]
 {\small
 \begin{textblock}{1}(4.59,1.26)
  $g$
 \end{textblock}
 \begin{textblock}{2}(4.73,2.3)
 \begin{tikzpicture}[node distance=1em,auto,>=latex]
  \draw[->] (0,0.65) -- (0,0);
  \node [right] at (0,0.325) {$\,\pi:\, G \to \mat$};
 \end{tikzpicture}
 \end{textblock}
 \begin{textblock}{1}(4.57,2.85)
  $o$
 \end{textblock}
 \begin{textblock}{2}(6.96,0.2)
  $G\equiv\GLd$
 \end{textblock}
 \begin{textblock}{2}(7.01,2.65)
  $\mat\simeq G/H$
 \end{textblock}
 \begin{textblock}{2}(4.82,0.4)
  $H\simeq \Opq(p,q)$
 \end{textblock}
 }
 \centering
 \includegraphics[width=0.45\columnwidth]{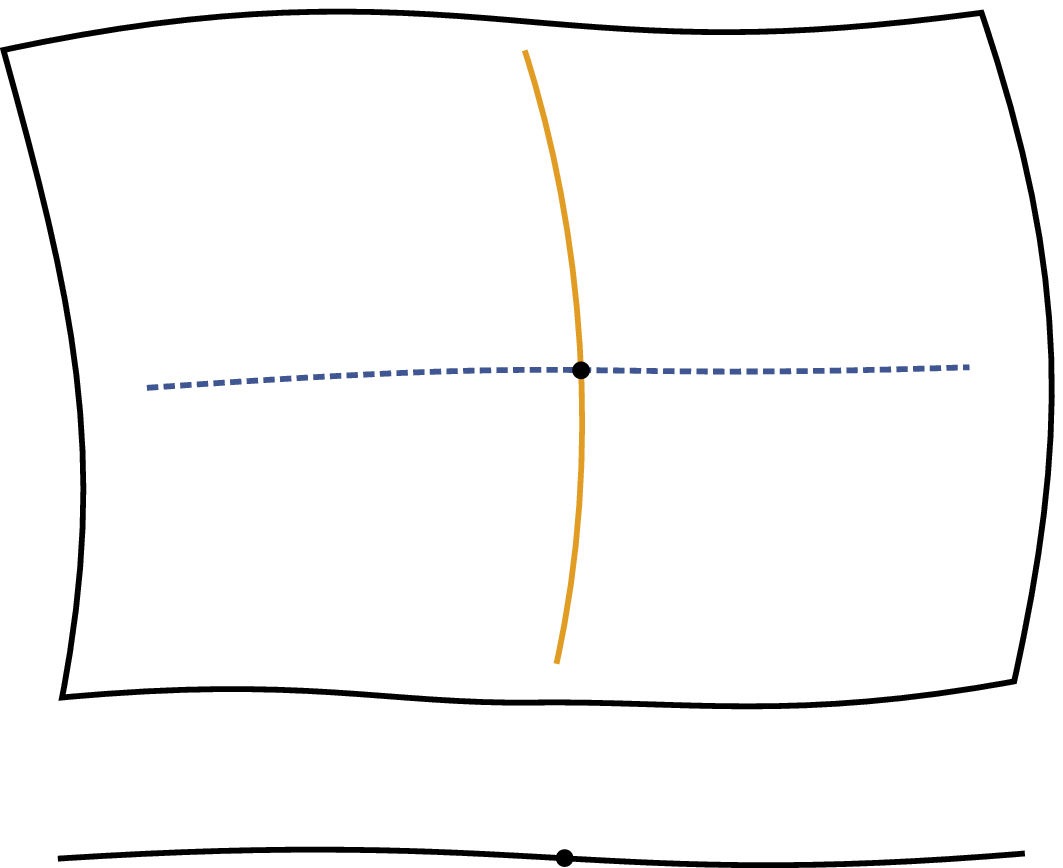}%
 \phantom{$G\equiv\GLd$}
 \vspace{1em}
 \caption{The space of real symmetric matrices with signature $(p,q)$, $\mat$, interpreted as base space of the
 principal bundle $(G,\pi,\mat,H)$. In the tangent space to this bundle, the vertical direction is determined by the
 structure group $H$, while the horizontal direction, indicated by the blue dashed line, is not fixed until a
 connection is chosen.}
 \label{fig:PBundle}
\end{figure}
\medskip

\noindent
\textbf{(2) Geometric interpretation.}
Before setting up a connection on the principal bundle let us briefly illustrate the geometric notion behind
this construction. Consider $d$ linearly independent vectors in $\mathbb{R}^d$. This frame can be represented
as a matrix $B\in\GLd$. Now we \emph{fix} a metric $\eta$ by \emph{declaring} the frame to be orthonormal:
\begin{equation}
 \eta(B_{(i)},B_{(j)}) \stackrel{!}{\equiv} \delta^{(p,q)}_{ij} \equiv (I_{(p,q)})_{ij}\, ,
\label{eq:fixEta}
\end{equation}
where $B_{(i)}$ denotes the $i$-th column of $B$, and $I_{(p,q)}$ is given by \eqref{eq:Ipq}. Writing \eqref{eq:fixEta}
in matrix notation and solving for $\eta$ yields
\begin{equation}
 \eta = (B^{-1})^T I_{(p,q)}(B^{-1}),
\label{eq:fixEtaMat}
\end{equation}
so $\eta$ is indeed determined by $B$. We see, however, that the RHS of equation \eqref{eq:fixEtaMat} is invariant
under multiplications of the type $B\rightarrow B\mku O^{-1}$, where $O\in\Opq(p,q)\equiv\{A\in\matrices|A^T I_{(p,q)}A
= I_{(p,q)}\}$. Thus, two frames that differ by a semi-orthogonal transformation define the same metric, so the set of
all metrics is given by $\GLd/\Opq(p,q)$. If a general background metric is used instead of $I_{(p,q)}$ on the RHS of
\eqref{eq:fixEta}, say, $\eta(B_{(i)},B_{(j)}) \equiv \bo_{ij}$, then $\Opq(p,q)$ is to be replaced with $H$,
reproducing \eqref{eq:MasGandH}.
\medskip

\noindent
\textbf{(3) The canonical connection on the principal bundle.}
In order to find a connection on $(G,\pi,\mat,H)$ adapted to the bundle structure we consider the corresponding Lie
algebras. In the following, Lie brackets are given by the commutator of matrices. The Lie algebra $\fg$ of $G$ is the
space of all real, square matrices,
\begin{equation}
 \fg=\matrices .
\end{equation}
The Lie algebra of $H$ is the space of ``$\bo$-antisymmetric'' matrices,
\begin{equation}
 \fh = \left\{A\in\matrices\big|\;A^T\bo=-\bo A\right\}.
\label{eq:defLieh}
\end{equation}
By $\Ad:G\rightarrow\mathrm{Aut}(\fg)$ we denote the adjoint representation of the group $G\mku$:
\begin{equation}
 \Ad(g)(X)=gXg^{-1}\,,\quad g\in G,\, X\in\fg.
\label{eq:AdRep}
\end{equation}
We find that its restriction $\Ad(H)$ keeps $\fh$ invariant, i.e.\footnote{Proof: Let $h\in H$ and $X\in\fh$, so we
have $h^T\bo\mku h=\bo$ and $X^T\bo=-\bo\mku X$. We define $Y\equiv \Ad(h)(X)\equiv hXh^{-1}$. Then: $Y^T\bo =
(h^{-1})^T X^Th^T\bo \mku h\mku h^{-1} = (h^{-1})^T X^T\bo\mku h^{-1}=-(h^{-1})^T \bo\mku X\mku h^{-1}=-(h^{-1})^T
\bo\mku h^{-1} h\mku X \mku h^{-1}=-\bo\mku Y$. Hence $Y\in\fh$, proving $\Ad(h)(\fh)\subset\fh$. Since the map
$X\mapsto Y=hXh^{-1}$ is bijective, we conclude that the reverse direction, $\fh\subset \Ad(h)(\fh)$, holds true,
too.\label{fn:ProofInv}}
\begin{equation}
 \Ad(h)(\fh) = \fh\quad \forall\, h\in H.
\end{equation}
Let us further define $\fm$ as the space of ``$\bo$-symmetric'' matrices,
\begin{equation}
 \fm \equiv \left\{A\in\matrices\big|\;A^T\bo=\bo A\right\}.
\end{equation}
This defines a vector space complement of $\fh$ in $\fg$,
\begin{equation}[b]
 \fg = \fm \oplus \fh,
\end{equation}
and $\fm$ is called \emph{Lie subspace} for $G/H$. (Note, however, that $\fm$ is not a Lie algebra since
$[m_1,m_2]\in \fh\quad \forall\, m_1,m_2 \in \fm$.) It is straightforward to show that $\fm$ is invariant under
$\Ad(H)$, too,\footnote{For the proof we proceed as in Footnote \ref{fn:ProofInv}, but taking $h\in H$ and $X\in\fm$
instead. This way we find that $\Ad(h)(X)\in\fm$. Bijectivity of the map $X\mapsto\Ad(h)(X)$ then implies $\Ad(h)(\fm)
=\fm$.}
\begin{equation}
 \Ad(h)(\fm) = \fm\quad \forall\, h\in H.
\end{equation}
Therefore, the homogeneous space $G/H$ is \emph{reductive}.

We use the differential of the canonical projection at the identity $e$ in $G$ in order to make the transition from
the Lie algebra $\fg$ to the tangent space of $\mat$ at $\bo=\pi(e)$,
\begin{equation}
 \td\pi_e: T_e G\equiv \fg \rightarrow T_\bo\mat.
\end{equation}
Since $\td\pi_e$ is surjective and has kernel $\fh$, the restriction $\td\pi_e|_\fm$ is an isomorphism on
the complement $\fm$. Thus, we can \emph{identify} $\fm$ with $T_\bo\mat$.

By means of the left translations $L_g:G\rightarrow G$ we can push forward the Lie subspace $\fm$ to any point $g$
in order to define a distribution on $G$, namely the \emph{horizontal distribution}
\begin{equation}[b]
 \mathcal{H}_g=\td L_g \fm.
\end{equation}
This defines a \emph{connection} on the principal bundle since it is invariant under the right
translations of $H$:
\begin{equation}
\begin{split}
 \td R_h(\mathcal{H}_g)&=\td R_h \td L_g \fm =\td L_g \td R_h \fm =\td L_g \td L_h \Ad(h^{-1}) \fm \\
 &= \td L_g \td L_h \fm = \td L_{gh} \fm = \mathcal{H}_{gh}.
\end{split}
\end{equation}
It is called the \emph{canonical connection} of the principal bundle $(G,\pi,\mat,H)$.
\medskip

\noindent
\textbf{(4) The induced connection on the tangent bundle of $\bm{\mat}$.}
The canonical connection, in turn, \emph{induces a connection on the tangent bundle} $T\mat$ which is associated to the
principal bundle \cite{KN69},\footnote{Eq.\ \eqref{eq:DefTM} comprises an implicit reduction of the frame
bundle: Generically, the tangent bundle is associated to the frame bundle, $\operatorname{GL}(\mat)$, according to
$T\mat\simeq\operatorname{GL}(\mat)\times_{\operatorname{GL}(D)}\mathbb{R}^D$, where $D\equiv\text{dim}(\mat)=
\frac{1}{2}d(d+1)$. Since the adjoint representation \eqref{eq:AdRep} maps $H$ to $\operatorname{GL}(D)$ (up to an
isomorphism) and since it is possible to find a principal bundle homomorphism $G\rightarrow\operatorname{GL}
(\mat)$ (with $\mat$ as common base space) compatible with the $H$-action, the structure group is reduced and we have
$\operatorname{GL}(\mat)\times_{\operatorname{GL}(D)}\mathbb{R}^D \simeq G \times_H \fm$.}
\begin{equation}
T\mat \simeq G \times_H \fm \equiv (G \times \fm)/H \,,
\label{eq:DefTM}
\end{equation}
where $h\in H$ acts on $G \times \fm$ by $(g,X)\mapsto (gh^{-1},\Ad(h)X)$. This induced connection is often referred to
as the \emph{canonical linear connection} of the homogeneous space $\mat \simeq G/H$.
As we will see below, it can be derived from a metric on $\mat$. In the following we use only the term ``canonical
connection'' since it is clear from the context whether a connection on the principal bundle or on the tangent
bundle is meant.
\medskip

\noindent
\textbf{(5) Torsion.}
In general, the torsion tensor following from the canonical connection is given by $T(X,Y)=-\mathrm{pr}_\fm ([X,Y])$
for $X,Y \in\fm$, where $\mathrm{pr}_\fm$ denotes the projection onto $\fm$ (see e.g.\ Reference \cite{KN69}). Here,
however, we have $[\fm,\fm]\subset\fh$. To see this, let us consider $m_1\in\fm$ and $m_2\in\fm$, i.e.\ by definition
$m_1^T\bo=\bo\mku m_1$ and $m_2^T\bo=\bo\mku m_2$. Then the commutator satisfies
\begin{equation}
 \begin{split}
 [m_1,m_2]^T\bo &= m_2^T m_1^T\bo - m_1^T m_2^T\bo = m_2^T \bo\mku m_1 - m_1^T \bo\mku m_2 \\
 &= \bo\mku(m_2 m_1 - m_1 m_2) = -\bo\mku [m_1,m_2],
 \end{split}
\end{equation}
so $[m_1,m_2]\in\fh$. Thus, $\mathrm{pr}_\fm ([X,Y])=0$ for all $X,Y \in\fm$, implying that the canonical connection is
\emph{torsion free}.
\medskip

\noindent
\textbf{(6) A metric on $\bm{\mat}$ and its Levi-Civita connection.}
It is possible to define a $G$-\emph{invariant metric on} $\mat$, denoted by $\gamma$. For any
$X,Y\in T_\bo \mat=S_d$ we set
\begin{equation}[b]
  \gamma_\bo(X,Y) \equiv \tr(\bo^{-1}X \mku\bo^{-1}\mku Y) + \frac{c}{2} \tr(\bo^{-1}X)\tr(\bo^{-1}\mku Y),
\label{eq:metricOnMat}
\end{equation}
with an arbitrary constant $c$. The metric \eqref{eq:metricOnMat} can be considered a generalization of the Killing
form for $\fg$. It is the most general $G$-invariant metric on $\mat$ up to a global factor. Here, $G$-invariance means
that the group action \eqref{eq:groupAction} of $G$ on $\mat$, $\phi_g(o)\equiv \phi(g,o)=(g^{-1})^T o\mku g^{-1}$, is
\emph{isometric} with respect to this metric: With $(\td\phi_g)_\bo X = (g^{-1})^T X\mku g^{-1}$, we have
\begin{equation}
  \gamma_{\phi_g(\bo)} \big(\mku (\td\phi_g)_\bo \mkuu X ,\mku (\td\phi_g)_\bo \mkuu Y \mku \big) = \gamma_\bo(X,Y) \,,
\label{eq:GInvMet}
\end{equation}
for all $X,Y\in T_\bo \mat$.

In combination with the $G$-invariance of the canonical connection (w.r.t.\ left translations), $\td L_{g_1}
\mathcal{H}_{g_2}=\mathcal{H}_{g_1 g_2}$, equation \eqref{eq:GInvMet} has the consequence that the covariant derivative
obtained from the canonical connection \emph{preserves the metric} \eqref{eq:metricOnMat} \cite{KN69}. Thus, we
conclude that \emph{the canonical connection is the Levi-Civita connection} on $T\mat$ with respect to $\gamma$.

Applying the principle of minimum energy as in Ref.\ \cite{GM91} leads to the geodesic equation corresponding to the
Levi-Civita connection for the metric \eqref{eq:metricOnMat}: We minimize the energy functional $E_\bo[o]\equiv
\frac{1}{2}\int_0^t \gamma_\bo\big(\dot{o}(s),\dot{o}(s)\big)\td s\mku$ with respect to the curves
$o:\mathbb{R}\to\mat$, $s\mapsto o(s)$, resulting in the differential equation
\begin{equation}
 \ddot{o}(s)-\dot{o}(s)\mku\bo^{-1}\mku \dot{o}(s)=0.
\label{eq:GeodEquOnMat}
\end{equation}
Comparing this expression to the generic geodesic equation $\ddot{o}(s)+\Gamma_\bo\big(\dot{o}(s),\dot{o}(s)
\big)=0$, we can conclude that $\Gamma_\bo(X,X) = -X\mku \bo^{-1}X$ for $X\in T_\bo\mat$. Finally, symmetrizing
appropriately yields, for $X,Y\in T_\bo \mat$, the Levi-Civita connection
\begin{equation}[b]
 \Gamma_\bo(X,Y) = -\frac{1}{2}\big(X\bo^{-1}Y+Y\bo^{-1}X\big).
\label{eq:LCOnMInMatrixForm}
\end{equation}

For the sake of completeness we mention that for any point $\bo\in\mat$ there is a symmetry $s_\bo$, i.e.\ a map
$s_\bo:\mat\rightarrow\mat$ which is an element of the isometry group of the metric $\gamma$ and which has the
reflection properties, $s_\bo(\bo)=\bo$ and $(\td s_\bo)_\bo= -\text{Id}$. It is given by the involution $s_\bo(o)
\equiv\bo\mku o^{-1}\bo$ and makes $\mat$ a \emph{symmetric space}.
\medskip

\noindent
\textbf{(7) Geodesics w.r.t.\ the canonical connection.}
With the above groundwork it is straightforward to construct geodesics through the point $\bo$. For that purpose we
have to find the \emph{exponential map} on the manifold $\mat$ with base point $\bo$, here denoted by $\exp_\bo$. On
the matrix Lie group $G$ the exponential map is given by the standard matrix exponential, $\exp$, where we also write
$\exp A=\e^A$. As shown in References \cite{ONeill1983,KN69}, the map $\exp_\bo \circ\, \td\pi_e:\fm\rightarrow\mat$ is
a local diffeomorphism, and it holds
\begin{equation}
 \exp_\bo \circ\, \td\pi_e = \pi \circ\, \exp \,.
\end{equation}
Hence, geodesics on $\mat$ are determined by
\begin{equation}
 \exp_\bo X = \pi\big( \e^{\td\pi_e^{-1}X}\big),
\end{equation}
for $X\in T_\bo\mat=S_d$. From equation \eqref{eq:CanProj} we obtain $\td\pi_e^{-1} X = -\frac{1}{2} \bo^{-1} X$,
resulting in
\begin{equation}
 \exp_\bo X = \pi\big( \e^{-\frac{1}{2} \bo^{-1} X}\big)
  = \big( \e^{\frac{1}{2} \bo^{-1} X}\big)^T \bo \; \e^{\frac{1}{2} \bo^{-1} X} \,.
\end{equation}
Using $\bo \; \e^{\frac{1}{2} \bo^{-1} X}\,\bo^{-1}= \e^{\frac{1}{2}X\mku \bo^{-1}}$ as well as $X^T=X$ and $\bo^T=\bo$
we finally obtain
\begin{equation}[b]
 \exp_\bo X = \bo\, \e^{\bo^{-1} X}\,.
\label{eq:GeodesicsParametrization}
\end{equation}
The same result can be derived directly from eq.\ \eqref{eq:GeodEquOnMat}.
With the identifications $\bo=\bg(x)$ and $X=h(x)$ this equals precisely the metric parametrization
\eqref{eq:ExpParamMatrix}.\footnote{This is to be contrasted with the geodesics found in Reference \cite{FG89} (see
also \cite{DeWitt1967a}) which are based on the LC connection induced by the DeWitt metric in $\mF$ (rather than
$\mat$). This is equivalent to determining geodesics on $\mat$ with respect to the LC connection of the metric $\sg\,
\gamma$, i.e.\ of our metric \eqref{eq:metricOnMat} times $\sg$. The resulting parametrization of geodesics has a more
involved form than \eqref{eq:GeodesicsParametrization}. In the referenced calculations, the authors decompose $\mat$
into a product of $\mat_\mu$ and $\mathbb{R}^+$, where $\mat_\mu$ are all elements of $\mat$ with determinant $\mu$.
Remarkably, geodesics on $\mat_\mu$ based on $\sg\,\gamma$ have the same structure as our result
\eqref{eq:GeodesicsParametrization} that describes geodesics on $\mat$ based on $\gamma$. As will be discussed in
Section \ref{sec:CompConn}, this can be traced back to the factor $\sg$ which is constant in $\mat_\mu$.}

That is the main result of this section. \emph{The exponential parametrization describes geodesics with respect to the
canonical connection}.
\medskip

\noindent
\textbf{(8) The metric and the canonical connection in local coordinates.}
At last, we would like to determine the form of $\gamma$ defined in \eqref{eq:metricOnMat} in local coordinates.
Symmetrizing adequately we obtain
\begin{equation}
\begin{split}
 \gamma_{\bo}(X,Y)&=\tr(\bo^{-1}X\mku \bo^{-1}Y) + \frac{c}{2} \tr(\bo^{-1}X) \tr(\bo^{-1}Y) \\
  &= \left( \bo^{\mu(\rho} \bo^{\sigma)\nu} + \frac{c}{2}\, \bo^\mn \bo^{\rho\sigma} \right) X_\mn Y_{\rho\sigma} 
  \stackrel{!}{=} \gamma^{\mu\nu\rho\sigma}X_\mn Y_{\rho\sigma}.
\end{split}
\end{equation}
Thus, we can read off
\begin{equation}
 \gamma^{\mu\nu\rho\sigma}=\bo^{\mu(\rho} \bo^{\sigma)\nu} + \frac{c}{2}\, \bo^\mn \bo^{\rho\sigma}\,. 
\label{eq:MatMetLoc}
\end{equation}

Moreover, the corresponding Christoffel symbols follow directly from equation \eqref{eq:LCOnMInMatrixForm}: The
canonical connection in local coordinates is given by
\begin{equation}[b]
 (\Gamma_\bo)^{\alpha\beta\,\rho\sigma}_\mn = -\delta^{(\alpha}_{(\mu}\; \bo^{\,\raisebox{0.2ex}{$\scriptstyle\beta)
 (\rho$}}\;\delta^{\sigma)}_{\nu)}
\label{eq:CanonicalConnection}
\end{equation}
It is to be emphasized that this result is independent of the parameter $c$. Remarkably enough, the tensor structure
of \eqref{eq:CanonicalConnection} agrees with the one of eq.\ \eqref{eq:NDConnection}. This crucial observation will be
discussed in more detail in the next section where we analyze how the canonical connection on $T\mat$ can be lifted to
a connection on $T\mF$.
\medskip

To sum up, we have seen that the canonical connection arises in a very straightforward way from the basic structure of
$\mat\simeq G/H$ interpreted as the base space of a principal bundle, so its associated geodesics, given by
\eqref{eq:GeodesicsParametrization}, are adapted to this structure, too. 
The extension from $\mat$ to $\mF$, worked out in Section \ref{sec:CompConn}, leads to the exponential
parametrization \eqref{eq:ExpParamX}, which can thus be considered the most natural way to parametrize pure metrics.

%----------------------------------------------------------------------------------------------------------------------
\subsection{Euclidean vs.\ Lorentzian signatures}
\label{sec:EuLor}
%----------------------------------------------------------------------------------------------------------------------

Next, we specify some topological and geometrical properties of $\mat\equiv\Mpq$, defined by equation
\eqref{eq:DefMatrices}, in combination with the canonical connection, where it turns out crucial in certain cases to
distinguish between different signatures. For the sake of brevity, not all of the following statements will be proven
in detail, but they follow from the results of the previous subsection and from the theorems of Appendix
\ref{app:ExpParam}. Let us start by giving and illustrating two important definitions, which will be needed for a
classification of $\Mpq$.
\medskip

\noindent
\textbf{Definition: Geodesic completeness.} A semi-Riemannian manifold $M$ equipped with an arbitrary connection is
geodesically complete if, for all $x\in M$, the corresponding exponential map $\exp_x$ is defined for all $v\in T_x M$,
i.e.\ if every maximal geodesic is defined on the entire real line $\mathbb{R}$.

Broadly speaking, this means that geodesics ``stay'' in $M$ rather than running into the boundary or a singularity.
\medskip

\noindent
\textbf{Definition: Geodesic connectedness.} A semi-Riemannian manifold $M$ equipped with an arbitrary connection is
geodesically connected if any two points in $M$ can be connected by a geodesic.
\medskip

The geodesics in both of these definitions depend on the underlying connection. Therefore, ``geodesic completeness''
and ``geodesic connectedness'' are not properties of the manifold alone but of the manifold and the connection.
We see by way of example that the two properties are fully independent: They are illustrated in Figure
\ref{fig:IllGeoComAndConn} where they appear in different combinations. Note that geodesic connectedness implies
connectedness (and path connectedness), while the opposite direction is not true. We would like to emphasize that even
path-connectedness plus geodesic completeness does not imply geodesic connectedness.
\begin{figure}[tp]
 \small
 \begin{minipage}[t]{0.47\columnwidth}
  \includegraphics[width=\columnwidth]{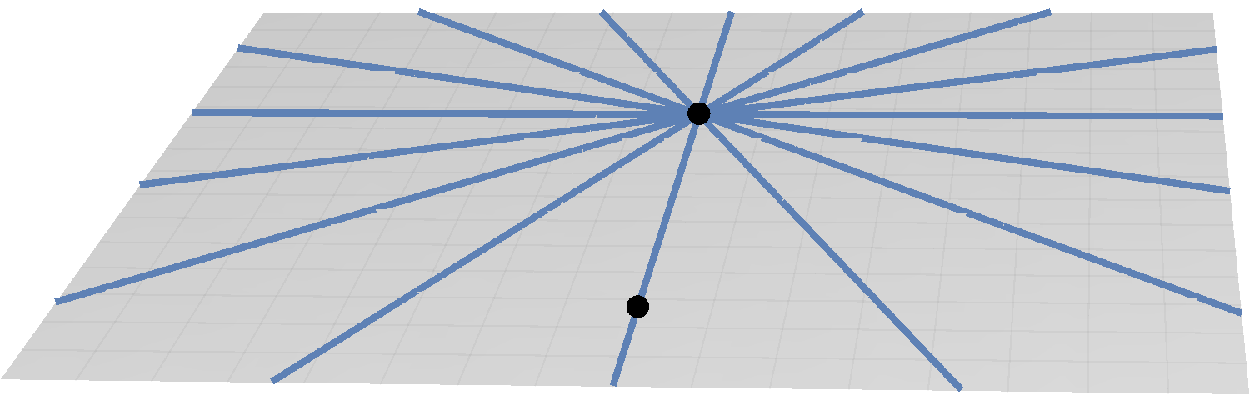}\\[0.5em]
  (a) The flat plane, $\mathbb{R}^2$, with vanishing connection: Both geodesically complete and geodesically
  connected.
 \end{minipage}
 \hfill
 \begin{minipage}[t]{0.47\columnwidth}
  \includegraphics[width=\columnwidth]{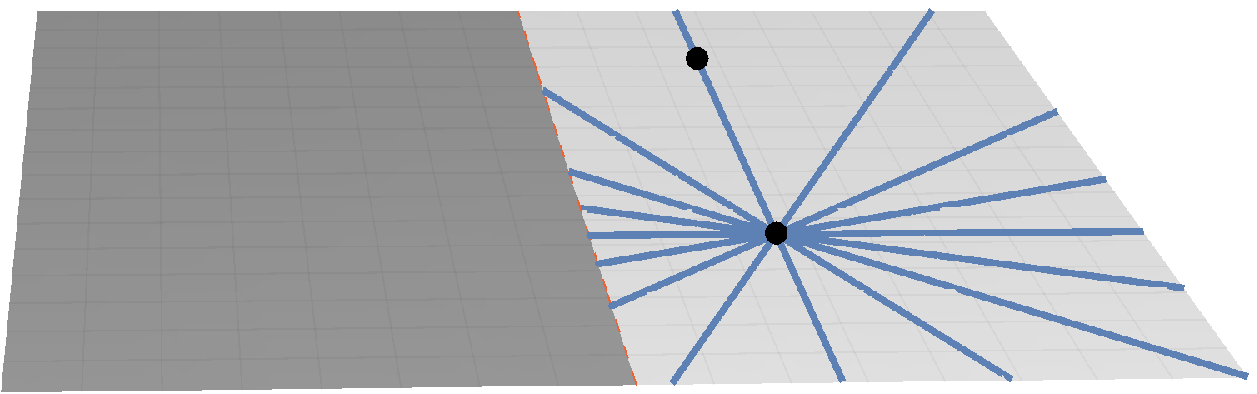}\\[0.5em]
  (b) The half plane, $\{x\in\mathbb{R}^2\, |\, x_1>0 \}$, with vanishing connection: Not geodesically complete but
  geodesically connected.
 \end{minipage}
 
 \vspace{1.8em}
 \begin{minipage}[t]{0.47\columnwidth}
  \includegraphics[width=\columnwidth]{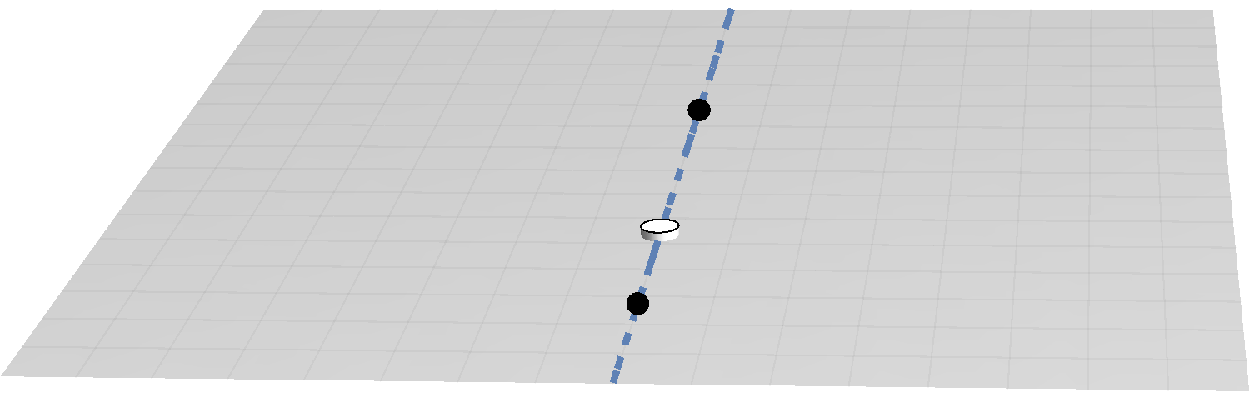}\\[0.5em]
  (c) The punctured plane, $\mathbb{R}^2\backslash\{0\}$, with vanishing connection: Neither geodesically complete nor
  geodesically connected.
 \end{minipage}
 \hfill
 \begin{minipage}[t]{0.47\columnwidth}
  \includegraphics[width=\columnwidth]{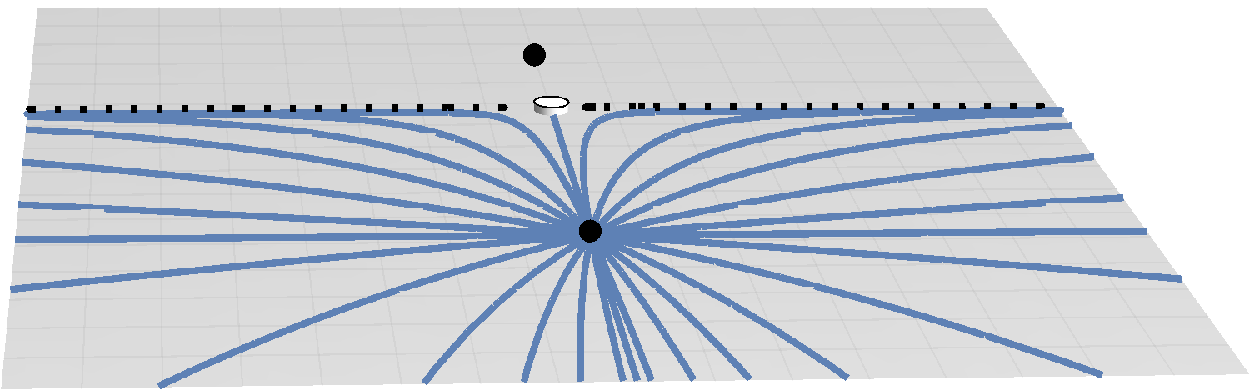}\\[0.5em]
  (d) The punctured plane, $\mathbb{R}^2\backslash\{0\}$, with a certain nontrivial connection: Geodesically complete
  (and path-connected) but not geodesically connected.
 \end{minipage}
 \vspace{1.2em}
\caption{Four examples illustrating the meaning of geodesic completeness and geodesic connectedness. The blue curves
represent geodesics starting at one point (marked as a black dot), and it is sketched whether or not they can reach the
second marked point. In (a) -- (c), geodesics are based on the trivial connection, i.e., they are straight lines. The
connection in (d), on the other hand, is (artificially designed) such that geodesics bend away from the singularity at
$x=0$ and never reach the upper half plane. The single geodesic in (d) running towards the singularity does not run
into $x=0$ at any finite $t$ but approaches it only in the limit $t\to\infty$, guaranteeing geodesic completeness.}
\label{fig:IllGeoComAndConn}
\end{figure}

Let us come to classify $\Mpq$ now. In the following, ``for all $p,q$'' refers to ``for all $p,q\in\mathbb{N}_0$ with
$p+q=d\mku$''.
\medskip

\noindent
\textbf{(1) Properties of \bm{$\Mpq$} valid for all \bm{$p,q$}.}
\begin{itemize}[]
 \item As already stated above, $\Mpq$ is an \emph{open subset} in the space of symmetric matrices. This has the
 important consequence that it can be covered with one chart only.
 \item Irrespective of the signature it is \emph{noncompact}. (If $o\in\Mpq$, then $\alpha\mku o\in\Mpq$, too, where
 $\alpha\in\mathbb{R}^{+}$. Considering the limit $\alpha\to\infty$ disproves compactness.)
 \item It is \emph{path-connected}. (Note that $G=\GLd$ is nonconnected, but the subgroup $H$ has
 elements in both of the connected components of $G$. Hence, $\Mpq\simeq G/H$ is connected. Since it is an open subset,
 it is even path-connected.)
 \item The scalar curvature $R_\mat$ of $\Mpq$ is a negative constant: Independent of $p$, $q$ and the metric parameter
 $c$, we deduce from eq.\ \eqref{eq:CanonicalConnection} that
 \begin{equation}
  R_\mat = -\frac{1}{8}d(d-1)(d+2).
 \end{equation}
 \item Remarkably enough, the space $\Mpq$ furnished with the canonical connection \eqref{eq:CanonicalConnection}
 is \emph{geodesically complete}. In Appendix \ref{app:ExpParam} it is shown algebraically that $\bo\, \e^{\bo^{-1} X}$
 stays in $\Mpq$ for all $X\in S_d$. Note, however, that an algebraic proof is not even necessary here since geodesic
 completeness is already guaranteed by construction: $\Mpq$ is a homogeneous space, and by homogeneity the exponential
 map corresponding to the canonical connection is defined on the entire tangent space.
\end{itemize}
\smallskip

\noindent
\textbf{(2) Properties of \bm{$\Mpq$} specific to both \bm{$(p,q)=(d,0)$} and \bm{$(p,q)=(0,d)$}.} These are the
positive definite matrices (i.e.\ Euclidean signatures) and the negative definite matrices, respectively, to which we
can attribute four interesting additional properties.
\begin{itemize}[parsep=2pt plus 2pt]
 \item The spaces $\mat_{(d,0)}$ and $\mat_{(0,d)}$ are \emph{simply connected}. (This can be seen by noting that they
 are \emph{convex}: If $A,B\in\mat_{(d,0)}$, then $x^TA\mku x>0$ and $x^TB\mku x>0$ for all $x\neq 0$, implying
 $x^T[sA+(1-s)B]x>0$ for all $x\neq 0$ and all $s\in[0,1]$. The case $(p,q)=(0,d)$ follows analogously.)
 \item The space $\Mpq$ exhibits a \emph{Riemannian structure} provided that $c> -\frac{2}{d}$ since the metric
 $\gamma$ given by equation \eqref{eq:metricOnMat} is positive definite: For both $(p,q)=(d,0)$ and $(p,q)=(0,d)$ one
 can show that
 \begin{equation}
  \gamma_\bo(X,X)=\tr\big((\bo^{-1}X)^2\big)+\frac{c}{2}\big(\tr (\bo^{-1}X)\big)^2 >0 ,
 \label{eq:gammaPosDef}
 \end{equation}
 for all $X\in T_\bo\mat=S_d$ with $X\neq 0$, and for $c> -\frac{2}{d}$. In the case $c= -\frac{2}{d}\,$ ($c< -\frac{2}{d}$)
 $\,\gamma$ becomes positive semidefinite (indefinite). As an aside we would like to mention that passing over from
 $\Mpq$ to $\mFpq$ leads to a surprising statement: The natural metric in the space of negative definite metrics is
 positive definite. 
 \item Our most important observation is that both $\mat_{(d,0)}$ and $\mat_{(0,d)}$ are \emph{geodesically connected}.
 There are two ways to prove this.
 
 \noindent
 (i) In Appendix \ref{app:ExpParam} it is shown that for any $\bo\in\Mpq$ and any $o\in\Mpq$, with $(p,q)=(d,0)$ or
 $(p,q)=(0,d)$, there exists an $X\in S_d$ satisfying $o=\bo\,\e^{\mku\bo^{-1}X}$. Since we know from Subsection
 \ref{sec:GenDesc} that the latter relation describes geodesics, this proves that any two points in $\Mpq$ can be
 connected by a geodesic.
 
 \noindent
 (ii) By eq.\ \eqref{eq:gammaPosDef} $\Mpq$ has a Riemannian structure for $c> -\frac{2}{d}$. Therefore, the
 Hopf--Rinow theorem is applicable, which implies in turn that $\Mpq$ is geodesically connected. Since we have shown
 that the canonical connection is independent of the parameter $c$, see \eqref{eq:CanonicalConnection}, the resulting
 geodesics do not depend on $c$ either. Thus, the statement of geodesic connectedness remains true even for
 $c \le -\frac{2}{d}$.
 \item The exponential map, $\exp_{\bo}: T_{\bo}\Mpq\equiv S_d\to\Mpq,\; X \mapsto o=\bo\,\e^{\mku\bo^{-1}X}$, is a
 \emph{global diffeomorphism}, i.e.\ there is a \emph{one-to-one correspondence} between $o\in\Mpq$ and $X\in S_d$.
\end{itemize}
\smallskip

\noindent
\textbf{(3) Properties of \bm{$\Mpq$} specific to \bm{$p\ge 1,\, q\ge 1$}.} These are the indefinite matrices
(corresponding to Lorentzian, i.e.\ mixed, signatures), which exhibit fundamentally different features.
\begin{itemize}
 \item When considering mixed signatures, $\Mpq$ is \emph{not simply connected}. (This can be proven by means of the
 long exact homotopy sequence. For the special case $d=2$ we will see it in a moment by means of an illustrative
 example.)
 \item Independent of $c$, the space $\Mpq$ has a \emph{semi-Riemannian structure}: For $p\ge 1$ and $q\ge 1$ the
 expression $\gamma_\bo(X,X)$ can become both positive and negative, depending on $X$, so $\gamma$ is indefinite. As
 an example let us consider $\bo=\diag(-1,1,\cdots)$, where the numbers abbreviated by the dots are chosen to be
 consistent with the signature. Furthermore, we set
 \begin{equation}
  X\equiv\begin{pmatrix}
  1&0\\
  0&1\\
  &&0\\
  &&&\scalebox{0.7}[1]{$\ddots$}\,\\
 \end{pmatrix}
 \quad\text{and}\quad
  Y\equiv\begin{pmatrix}
  0&1\\
  1&0\\
  &&\scalebox{0.7}[1]{$\ddots$}\\
  &&&0\mku
 \end{pmatrix}\,.
 \end{equation}
 Using \eqref{eq:metricOnMat}, this choice results in $\gamma_\bo(X,X)=2>0$ and $\gamma_\bo(Y,Y)=-2<0$ for all $c$. For
 different base points $\bo$ similar examples can be found. Hence, $\gamma$ is indefinite.\footnote{It is possible to
 define a different metric for $p\ge 1$, $q\ge 1$ that makes $\Mpq$ Riemannian. However, such a metric would not be
 $G$-invariant, its Levi-Civita connection would not be the canonical connection, and it would not extend to a
 covariant metric in field space $\mF$. In particular, corresponding geodesics would not be given by the simple
 exponential parametrization.}
 \item For $p,q\ge 1$ the space $\Mpq$ is \emph{not geodesically connected}, so the exponential map $\exp_\bo\mku$ is
 \emph{not surjective}. This is the most important difference as compared with the positive and negative definite
 matrices discussed in point \textbf{(2)}, and it establishes the main result of this subsection. Before proving the
 statement, we notice that its basic cause lies in the fact that $\Mpq$ is \emph{semi}-Riemannian. Hence, the
 Hopf--Rinow theorem is not applicable.
 
 In order to disprove geodesic connectedness it is sufficient to find appropriate counterexamples. The general case
 is treated in Appendix \ref{app:ExpParam}. Here, we sketch the idea by means of a simple counterexample for
 $2\times 2$-matrices, that is, for $p=1$ and $q=1$. We try to connect the base point
 \begin{equation}
  \bo=\begin{pmatrix}1&0\\0&-1\end{pmatrix} \text{ to another point }
  o =\begin{pmatrix}-2&0\\0&1\end{pmatrix},
 \label{eq:CounterexPts2}
 \end{equation}
 both of which belong to $\Mpq$. According to eq.\ \eqref{eq:GeodesicsParametrization} we have to find an
 $X\in T_\bo\Mpq\equiv S_d$ that solves the equation
 \begin{equation}
  \bo^{-1}o = \begin{pmatrix}-2&0\\0&-1\end{pmatrix} = \e^{\bo^{-1}X}.
 \label{eq:Counterex2}
 \end{equation}
 There is an existence theorem \cite{Culver1966}, however, which states that a real square matrix has a \emph{real}
 logarithm if and only if it is nondegenerate and each of its Jordan blocks belonging to a negative eigenvalue occurs
 an even number of times. Thus, since the matrix in the middle of equation \eqref{eq:Counterex2} has two distinct
 negative eigenvalues, it does not have a real logarithm, so there is no $X\in T_\bo\Mpq$ that solves
 \eqref{eq:Counterex2}. This proves that the exponential map is not surjective.
 \item Even the restriction of $\Mpq$ to the image of $\exp_\bo$ to guarantee surjectivity does not turn $\exp_\bo$
 into a global diffeomorphism since it is also \emph{not injective}. Again, the general case is proven in Appendix
 \ref{app:ExpParam}, while we specify a simple counterexample in $d=2$ dimensions here. Let us consider the base point
 \begin{equation}
  \bo=\begin{pmatrix}1&0\\0&-1\end{pmatrix},
 \end{equation}
 and the one-parameter family of tangent vectors, i.e.\ symmetric matrices,
 \begin{equation}
  X_\alpha=\begin{pmatrix}0&\alpha\\\alpha&0\end{pmatrix} \in T_\bo\Mpq\,.
 \end{equation}
 Inserting these matrices into the exponential map yields
 \begin{equation}
 \begin{split}
  o_\alpha &\equiv \exp_\bo(X_\alpha)= \bo\, \e^{\bo^{-1}X_\alpha}
  = \begin{pmatrix}1&0\\0&-1\end{pmatrix}\,\exp\left[\begin{pmatrix}0&\alpha\\-\alpha&0\end{pmatrix}\right] \\
  &= \begin{pmatrix}1&0\\0&-1\end{pmatrix}\,\begin{pmatrix}\cos\alpha&\sin\alpha\\-\sin\alpha&\cos\alpha\end{pmatrix}
  = \begin{pmatrix}\cos\alpha&\sin\alpha\\\sin\alpha&-\cos\alpha\end{pmatrix},
 \end{split}
 \end{equation}
 which is periodic, and thus not injective. In particular, we find $\exp_\bo(X_\alpha)= \bo$ for all $\alpha\in\{2\pi
 \mku k\, |\, k\in\mathbb{Z}\}$.
\end{itemize}

Let us briefly summarize our main insights. Whether or not the space $\Mpq$, equipped with the canonical connection, is
geodesically connected depends highly on the signature $(p,q)$. For positive definite and negative definite matrices,
i.e.\ for $(p,q)=(d,0)$ and $(p,q)=(0,d)$, respectively, any two points in $\Mpq$ can be connected by a geodesic. The
exponential map $\exp_\bo$ \emph{``reaches'' every point in $\Mpq$ once and only once}. For indefinite matrices,
$p\ge 1$, $q\ge 1$, on the other hand, there are points in $\Mpq$ that can \emph{never be reached by any of the
geodesics starting at the base point} $\bo$, while there are other points that are \emph{reached infinitely many times
by a single geodesic}.
\medskip

\noindent
\textbf{(4) Illustration of \bm{$\Mpq$}.} 
Finally, we would like to visualize our results. It is particularly interesting to find out how geodesics on the space
of indefinite matrices look like and how a geodesically complete space can be geodesically nonconnected at all.
In the case of $2\times 2$-matrices the space $\Mpq$ can be illustrated by means of three-dimensional plots.
It will turn out convenient to parametrize arbitrary symmetric matrices by
\begin{equation}
 \begin{pmatrix} z-x & y \\ y & z+x \end{pmatrix},
\label{eq:ParamSymMat}
\end{equation}
since the various subspaces assume simple geometric shapes then. Any symmetric matrix is thus mapped to a point in
$\mathbb{R}^3$. The eigenvalues of \eqref{eq:ParamSymMat} are given by
\begin{equation}
 \lambda = z \pm \sqrt{x^2+y^2}.
\end{equation}
Hence, the condition for positive definite, negative definite or indefinite matrices, i.e. both eigenvalues positives,
negative or mixed, respectively, leads to a condition for $x$, $y$ and $z$, which can be displayed graphically.
For instance, positive definiteness implies two positive eigenvalues, i.e. $z + \sqrt{x^2+y^2} > 0$ and
$z - \sqrt{x^2+y^2} > 0$, which boils down to the single condition
\begin{equation}
 z > \sqrt{x^2+y^2}\,.
\end{equation}
This representation gives rise to an open cone embedded into $\mathbb{R}^3$. Analogously, we find $z <
-\sqrt{x^2+y^2}$ for negative definite matrices, and $-\sqrt{x^2+y^2} < z < \sqrt{x^2+y^2}$ for indefinite matrices.
\begin{figure}[tp]
  \centering
  \includegraphics[width=.7\columnwidth]{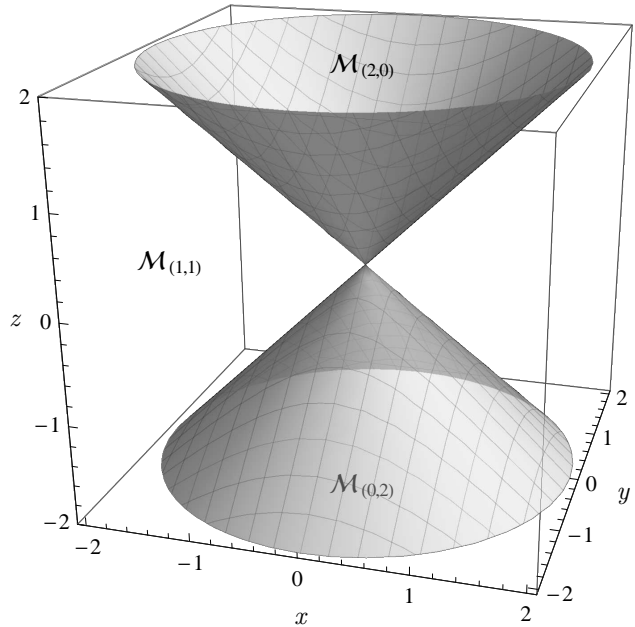}
  \caption{Using parametrization \eqref{eq:ParamSymMat} the space of symmetric $2\times 2$-matrices decomposes
    into positive definite matrices $\mat_{(2,0)}$ (interior of the cone with positive $z$), negative definite matrices
    $\mat_{(0,2)}$ (interior of the cone with negative $z$), and symmetric matrices with signature $(1,1)$
    ($\mathbb{R}^3$ with the closure of the two cones cut out). The cones extend to $z\rightarrow\pm\infty$. We observe
    that $\mat_{(1,1)}$ is not simply connected.}
  \label{fig:PosDef}
\end{figure}

The analysis shows that the set of all nondegenerate symmetric $2\times 2$-matrices decomposes into three open sets,
$\mat_{(2,0)}$, $\mat_{(1,1)}$ and $\mat_{(0,2)}$. This is depicted in Figure \ref{fig:PosDef}. The set of positive
definite matrices, $\mat_{(2,0)}$, is represented by the inner part of a cone which is upside down and has its apex
at the origin. Note that it extends to $z\rightarrow\infty$. The negative definite matrices, $\mat_{(0,2)}$, are
merely a reflection of this cone through the origin. Finally, $\mat_{(1,1)}$ is mapped to $\mathbb{R}^3$ from which
two cones are cut out. The surfaces of the cones belong to neither of the three sets but rather to degenerate
symmetric matrices.

\begin{figure}[tp]
 \centering
 \vspace{-1.8em}
 \includegraphics[width=.7\columnwidth]{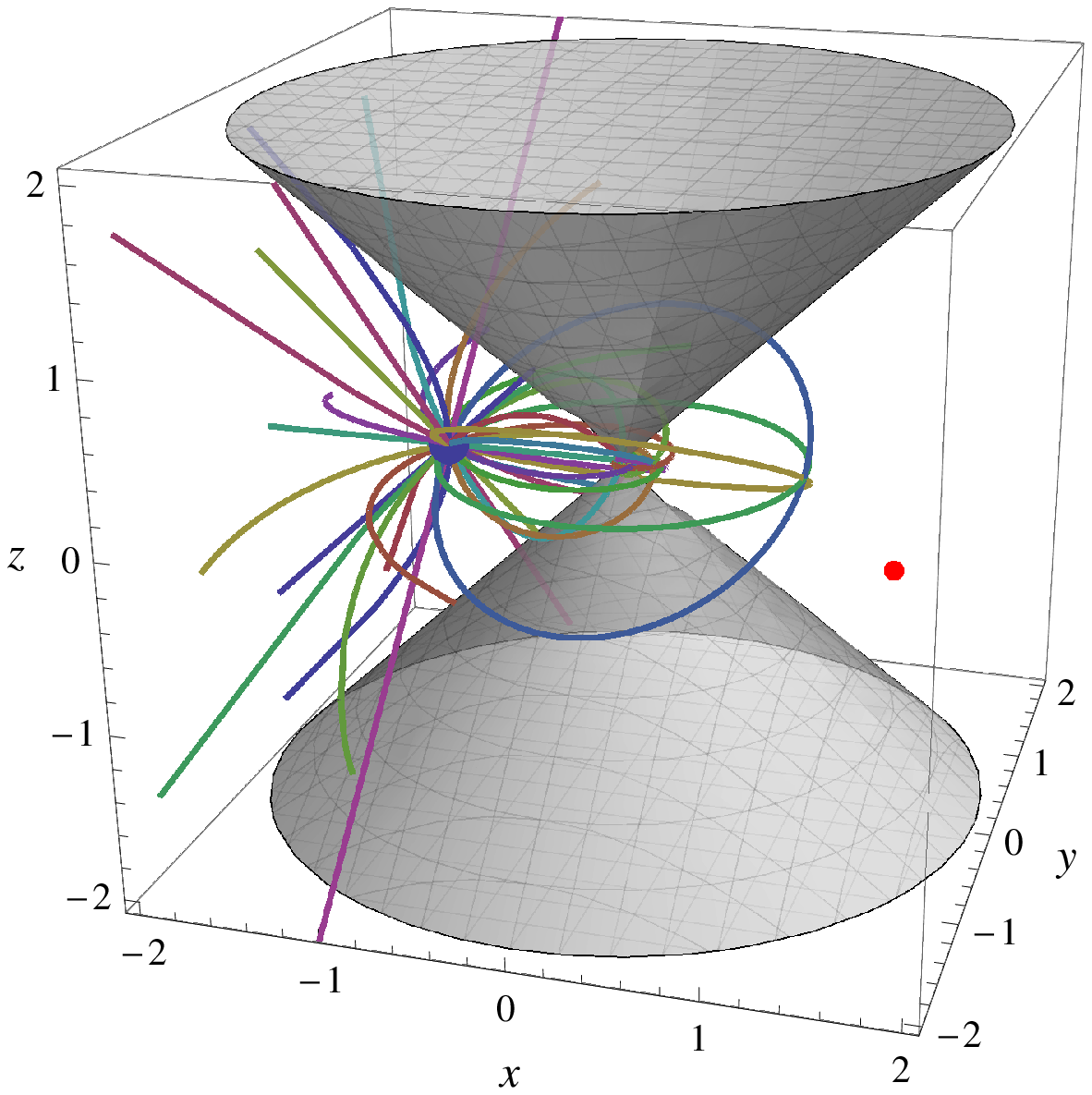}
 \vspace{-1em}
 \caption{Geodesics on $\mat_{(1,1)}$, starting at $(x,y,z)=(-1,0,0)$, where $\mat_{(1,1)}$ is given by the
  white space without the gray cones. As opposed to the case of positive definite matrices, we find periodic solutions
  here. Moreover, whenever a geodesic traverses the $yz$-plane on the positive $x$ side, it crosses the half-line
  $\{(x,0,0)\in\mathbb{R}^3|x>0\}$. There is no geodesic connecting the base point to the point marked in red at
  $(x,y,z)=\big(\frac{3}{2},0,-\frac{1}{2}\big)$, for instance.
  \label{fig:Geodesics}}
 \vspace{0.8em}
 \includegraphics[width=.7\columnwidth]{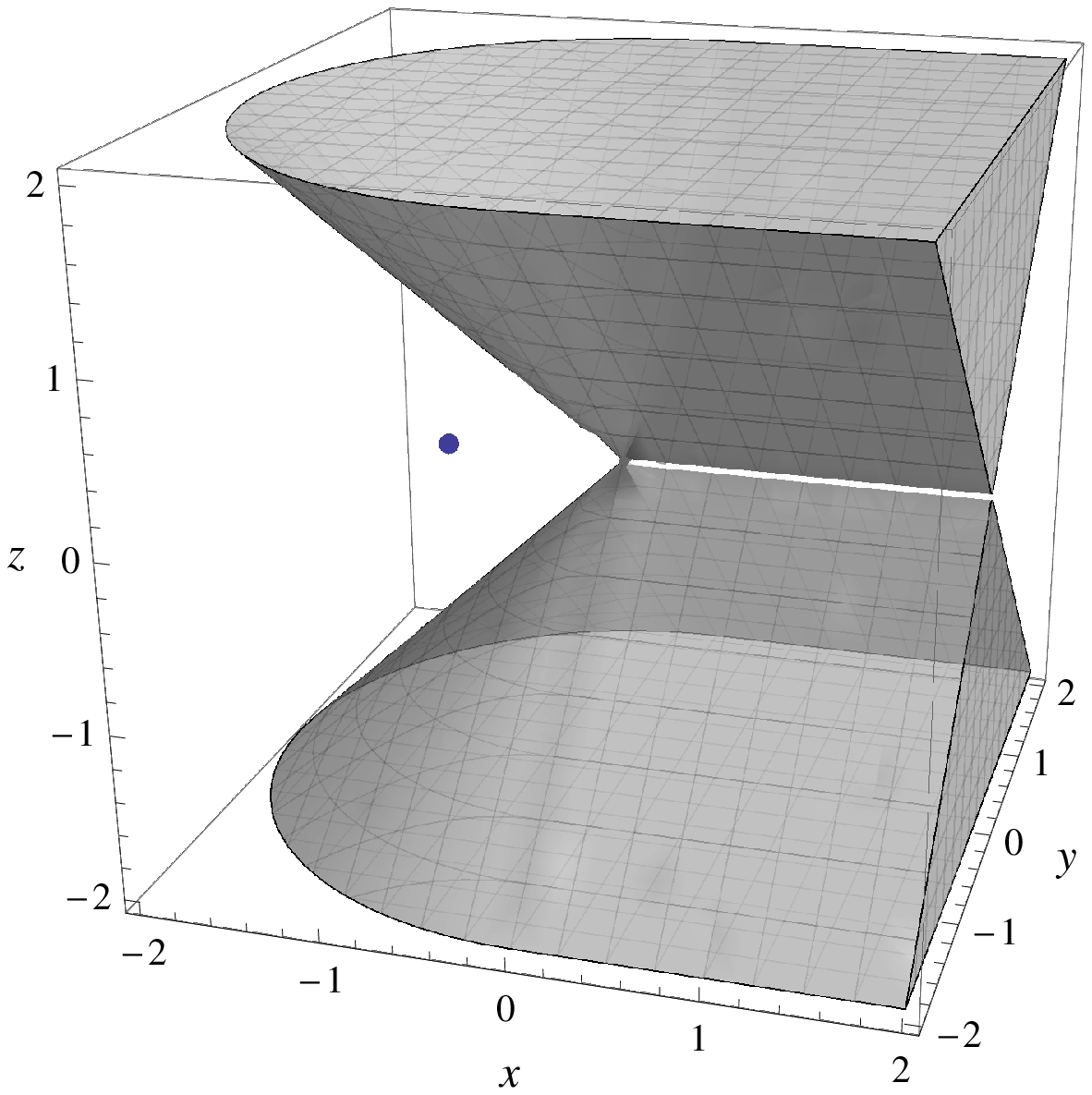}
 \vspace{-1em}
 \caption{The white region shows the space within $\mat_{(1,1)}$ that can be reached by a geodesic starting from the
  base point at $(x,y,z)=(-1,0,0)$.
  \label{fig:Reachable}}
\end{figure}

At last, we illustrate geodesics on $\mat_{(1,1)}$. This helps to understand how it can be possible that every maximal
geodesic is defined on the entire real line, while still not all points can be reached by geodesics starting from a base
point. Figure \ref{fig:Geodesics} shows what happens. By way of example, we choose a base point $\bo\in\mat_{(1,1)}$
with the parametrization $(x,y,z)=(-1,0,0)$ and some random tangent vectors that give rise to corresponding geodesics.
We observe that most of the example geodesics lie entirely in the half space with negative $x$. However, those entering
the positive $x$ half space have in common that they run through the same axis: Whenever they cross the $yz$-plane at
positive $x$ they intersect the $x$-axis. This holds for all geodesics starting at $\bo$, that is, at $x>0$ they
can never reach points in the $yz$-plane with $z>0$ or $z<0$. Furthermore, we see the aforementioned periodic solutions
in Figure \ref{fig:Geodesics} as geodesics circling around the origin.

In order to visualize that part of $\mat_{(1,1)}$ which cannot be reached by geodesics starting at $\bo$ we can make
use of the existence theorem for real logarithms \cite{Culver1966} again: Following the same logic as the one
underlying the above discussion around eqs.\ \eqref{eq:CounterexPts2} and \eqref{eq:Counterex2}, these geodesically
unconnected points are all those $o\in\mat_{(1,1)}$ for which the product $\bo^{-1}o$ has two distinct negative
eigenvalues. The result is shown in Figure \ref{fig:Reachable}. Points that can be reached from the base point $\bo$
by a geodesic are given by the white region. It can be observed that the two cones effectively shield the space behind
them.
\medskip

To sum up, for Euclidean signatures there is a one-to-one correspondence between tangent vectors and points in $\Mpq$,
while for Lorentzian signatures there is none. In order to ``cure'' the latter case, we would have to start from
several base points and restrict the corresponding tangent spaces at the same time such that all points in $\Mpq$ are
reached once and only once. As our results carry over from $\Mpq$ to $\mFpq$, this peculiarity has to be taken into
account when considering functional integrals over Lorentzian metrics.

%----------------------------------------------------------------------------------------------------------------------
\section{Comparison of connections on field space}
\label{sec:CompConn}
%----------------------------------------------------------------------------------------------------------------------

So far, we have studied the space $\mat\equiv\Mpq$, the local manifestation of the field space $\mF\equiv\mFpq$.
In this section we will show how the results derived previously for $\mat$ transition into properties of $\mF$.
To this end, we will lift the metric \eqref{eq:MatMetLoc}, the connection \eqref{eq:CanonicalConnection} and the
corresponding geodesics from their matrix form to tensor field expressions. Note that it is perfectly admissible to use
the parametrization $o=\exp_\bo(X)$ given by \eqref{eq:GeodesicsParametrization} and replace $o,\bo\in\mat$ and
$X\in T_\bo\mat$ by the $x$-dependent tensor fields $g(x)$, $\bg(x)$ and $h(x)$, respectively, where continuity of $g$
with respect to $x$ is ensured by continuity of $\bg$ and $h$. The question is rather if this parametrization still
describes geodesics on $\mF$ associated to the Levi-Civita connection. In this regard we discuss and compare
different connections on the space of metrics.
\medskip

\noindent
\textbf{(1) The underlying manifolds.}
Apart from the spacetime manifold $M$ and the space $\mat$ of symmetric matrices with signature $(p,q)$, we will see in
a moment that $\mF$ can be equipped with a metric, too. Thus, we consider three (semi-)Riemannian manifolds in total,
which we distinguish carefully:
\begin{align}
(M,g), \qquad (\mat,\gamma),\qquad (\mF,G)\,,
\end{align}
where, in local coordinates, $g_{\mu\nu}$ is the spacetime metric, $\gamma^{\mu\nu\rho\sigma}$ denotes the metric in
$\mat$, and $G_{ij}$ is the field space metric in DeWitt notation.\footnote{The DeWitt notation has been introduced in
point \textbf{(3)} of Section \ref{sec:DetConnections}. The DeWitt label $i$ represents all indices a tensor field
possesses, including the spacetime coordinate, here $i\equiv(\mu,\nu,x)$.} Note that $g_\mn$ represents also a point in
$\mF$. We would like to find the most natural form of $G_{ij}$ and discuss its relation to $\gamma^{\mu\nu\rho\sigma}$
in the following.
\medskip

\noindent
\textbf{(2) The DeWitt metric.}
The field space metric $G_{ij}$ is part of the definition of the theory under consideration. Nevertheless, it can be
fixed if a few requirements adapted to the space of metrics, $\mF$, are made.

First, we want to take into account that gravity is a gauge theory. The classical action is invariant under
diffeomorphisms, and so are all physical quantities. This leads to the reasonable requirement that the metric
$G_{ij}$ on $\mF$ be \emph{gauge invariant}, too, i.e.\ that the action of the gauge group on $\mF$ be an
isometry. In general terms, a gauge transformation can be written as
\begin{equation}
 \delta \vp^i = K_\alpha^i[\vp] \delta\epsilon^\alpha\, ,
\label{eq:GaugeTrafo}
\end{equation}
where $\delta\epsilon^\alpha$ parametrizes the transformation and the $\mathbf{K}_\alpha$ are the generators of the
gauge group, henceforth denoted by $\mG$. In the case of gravity, equation \eqref{eq:GaugeTrafo} reads $\delta g_\mn=
\mathcal{L}_{\delta\epsilon}g_\mn$, with the Lie derivative $\mathcal{L}$ along a vector field $\delta
\epsilon^\alpha$. The action of $\mG$ on $\mF$ induces a principal bundle structure \cite{Ebin1968,Ebin1970}. Points
that are connected by gauge transformations are physically equivalent while the space of orbits $\mF/\mG$ contains all
physically nonequivalent configurations. Now, if the gauge group is to generate isometric motions in $\mF$, then the
field space metric $G_{ij}[\vp]$ must satisfy Killing's equation, i.e.\ our first requirement reads
\begin{equation}
K^k_{\alpha,i} G_{jk} + K^k_{\alpha,j} G_{ik} + K^k_{\alpha} G_{ij,k} =0\,,
\end{equation}
where commas denote functional derivatives with respect to the field $\vp^i$.

Second, we require that $G_{ij}[\vp]$ be \emph{ultralocal}, i.e. that it involve only undifferentiated $\vp$'s, and
that it be \emph{diagonal} in $x$-space.

There is a unique one-parameter family of field space metrics satisfying all requirements, which is known as
\emph{DeWitt metric} \cite{DeWitt1967a}. It reads
\begin{equation}
  G^{\mu\nu\,\rho\sigma}(x,y)[g] = \sg \left(g^{\mu(\rho} g^{\sigma)\nu} + \frac{c}{2}\, g^{\mu\nu}
  g^{\rho\sigma}\right) \delta(x-y) \,,
\label{eq:DeWitt_metric}
\end{equation}
where the $x$-dependence of $g_\mn$ is implicit. This metric on $\mF$ is our starting point. On $T_g\mF\equiv\SymT$
it induces the inner product
\begin{equation}
 G_g(h,h')\equiv \int\dd x\,\dd y\; G^{\mu\nu\,\rho\sigma}(x,y)[g]\,h_\mn(x)\mku h'_\rs(y)\,.
\end{equation}
By comparing the DeWitt metric on $\mF$ with the metric $\gamma^{\mu\nu\rho\sigma}$ on $\mat$, given by
\eqref{eq:MatMetLoc}, we observe an identical tensor structure. The factor $\sg$ in \eqref{eq:DeWitt_metric} is needed
merely to make $G^{\mu\nu\,\rho\sigma}(x,y)$ a bitensor density of correct weight. Hence, the DeWitt metric can be
written as
\begin{equation}
  G^{\mu\nu\,\rho\sigma}(x,y)[g] = \sqrt{g(x)}\, \gamma^{\mu\nu\,\rho\sigma} (g(x))\, \delta(x-y) \,.
\label{eq:ultra_local_metric}
\end{equation}
\smallskip

\noindent
\textbf{(3) The Levi-Civita connection on \bm{$\mF$}.}
The Levi-Civita (LC) connection on $\mat$ w.r.t.\ the metric $\gamma$ is given by the canonical connection, and it has
already been computed in the previous section. In order to compare it with the LC connection on $\mF$ induced by the
DeWitt metric, let us introduce another convenient notation: In the following, capital Latin indices refer to pairs of
spacetime indices but not to spacetime coordinates, e.g.\ $I\equiv(\mu,\nu)$, and we write $o^I\equiv o_\mn$ for points
in $\mat$ and $g^I(x) \equiv g_{\mu\nu}(x) \equiv g^i$ for points in $\mF$.

Let $\left\{ {}^{K}_{IJ} \right\}$ denote the Christoffel symbols of the LC connection on $(\mat,\gamma)$. Then, by
definition,
\begin{equation}
 \left\{ {}^{K}_{IJ} \right\} = \frac{1}{2}\,\gamma^{KL}\left(\gamma_{IL,J}+\gamma_{JL,I}
 -\gamma_{IJ,L}\right)\,.
\label{eq:Levi-CivitaOnMDef}
\end{equation}
As computed in Section \ref{sec:GroupTheory}, they read
\begin{equation}
 \left\{ {}^{K}_{IJ} \right\} \equiv \big\{ {}^{\alpha\beta\,\rho\sigma}_{\mu\nu} \big\}
 = -\delta^{(\alpha}_{(\mu}\, g^{\raisebox{0.2ex}{$\scriptstyle\beta)(\rho$}}\,\delta^{\sigma)}_{\nu)} \,.
\label{eq:Levi-CivitaOnM}
\end{equation}

With this in mind, let us construct connections on field space $\mF$ now. For that purpose we start out from the LC
connection w.r.t.\ the DeWitt metric \eqref{eq:DeWitt_metric}. Its Christoffel symbols are denoted by $\big\{ {}^k_{ij}
\big\}$, and they follow from the usual definition:
\begin{equation}
 \big\{ {}^k_{ij} \big\} \equiv \frac{1}{2}\, G^{kl} \left(G_{il,j}+ G_{jl,i} - G_{ij,l}\right) \,.
\label{eq:Levi-Civita}
\end{equation}
Their precise form in terms of field space coordinates $g_\mn$ has been determined in Refs.\ \cite{DeWitt1967a,HKLT88}.
We will specify them in a moment.

Now, a \emph{generic} connection on $\mF$ can always be written as 
\begin{equation}
 \Gamma_{ij}^k = \big\{ {}^k_{ij} \big\} + A_{ij}^k\,.
\label{eq:Christoffels}
\end{equation}
The last term in \eqref{eq:Christoffels}, $A_{ij}^k$, is an arbitrary smooth bilinear bundle homomorphism, and
different connections on $\mF$ merely differ in that term.

We would like to emphasize that, \emph{although by equation \eqref{eq:ultra_local_metric} $G^{\mu\nu\,\rho\sigma}(x,y)$
is proportional to $\gamma^{\mu\nu\,\rho\sigma}$, the corresponding LC connections are not}. The field space LC
connection rather contains additional terms. We find that it decomposes into two pieces,
\begin{equation}[b]
  \big\{ {}^k_{ij} \big\} = \left(\left\{ {}^K_{IJ} \right\} + T^K_{IJ}\right)\!(x)\,\delta(x-y)\delta(x-z)\,,
\label{eq:LCComparison}
\end{equation}
where the first term is given by equation \eqref{eq:Levi-CivitaOnM} with $g_\mn$ replaced by $g_\mn(x)$, and
$T^K_{IJ}\equiv T_\mn^{\alpha\beta\,\rho\sigma}$ reads \cite{DeWitt1967a,HKLT88}
\begin{equation}
\begin{split}
 T_\mn^{\alpha\beta\,\rho\sigma} = \quad&\frac{1}{4}\, g^{\alpha\beta}\delta^\rho_{(\mu}\delta^\sigma_{\nu)}
  - \frac{1}{2(2+dc)}\, g_\mn g^{\alpha(\rho} g^{\sigma)\beta}\\
  + &\frac{1}{4}\, g^{\rho\sigma}\delta^\alpha_{(\mu}\delta^\beta_{\nu)}
  - \frac{c}{4(2+dc)}\, g_\mn g^{\alpha\beta} g^{\rho\sigma} \,.
\end{split}
\label{eq:LCConTTerms}
\end{equation}
Clearly, the reason for this difference between the LC connections on $\mat$ and $\mF$ can be traced to a
\emph{nonconstant proportionality factor} relating the underlying metrics, i.e.\ to the volume element $\sqrt{g}$
in \eqref{eq:ultra_local_metric}.
When taking functional derivatives of $G_{ij}$ they act both on $\sqrt{g}$ and on $\gamma^{\mu\nu\,\rho\sigma}$ in
\eqref{eq:ultra_local_metric}. Thus, the second term in \eqref{eq:LCComparison} contains \emph{only contributions due
to derivatives acting on the volume element}. This is a special characteristic of gravity. In other theories, like in
nonlinear sigma models for instance \cite{Friedan1980,Friedan1985,HPS88}, proportionality of a field space metric to a
metric in (the equivalent of) $\mat$ results in proportional LC connections. There the volume element is a prescribed
external ingredient, while it depends on the field in the case of gravity.
\medskip

\noindent
\textbf{(4) Lifting the canonical connection from \bm{$\mat$} to \bm{$\mF$}.}
The naive approach to lifting geodesics w.r.t.\ \eqref{eq:Levi-CivitaOnM} from $\mat$ to $\mF$ consists in making
the Levi-Civita connection \eqref{eq:Levi-CivitaOnM} spacetime dependent. This can be achieved by multiplying it with
appropriate $\delta$-functions, and by replacing $g_\mn$ with $g_\mn(x)$, leading to the result
$-\delta^{(\alpha}_{(\mu}\, g^{\raisebox{0.2ex}{$\scriptstyle\beta)(\rho$}}(x)\,\delta^{\sigma)}_{\nu)}\; \delta(x-y)
\delta(x-z)$, which would reproduce exponentially parametrized geodesics as desired. We have to make sure, though,
that this expression defines a proper connection on $\mF$. To this end, we want to write it as in eq.\
\eqref{eq:Christoffels} in terms of the Levi-Civita connection on $\mF$ w.r.t.\ the DeWitt metric.

As argued in the previous point, the LC connection on $(\mF,G)$ contains additional terms originating from the volume
element. Thus, we merely have to remove these terms in order to obtain a connection on $\mF$ that is proportional to
\eqref{eq:Levi-CivitaOnM}. This can easily be achieved by choosing a bundle homomorphism $A_{ij}^k$ in
\eqref{eq:Christoffels} which takes the form
\begin{equation}[b]
 A_{ij}^k = -T^K_{IJ}\,\delta(x-y)\delta(x-z) \,,
\label{eq:AForND}
\end{equation}
with $T^K_{IJ}$ as in eqs.\ \eqref{eq:LCComparison} and \eqref{eq:LCConTTerms}.
That choice is perfectly admissible: All terms in $T^K_{IJ}$ are properly symmetrized, so it maps two symmetric tensors
to a symmetric tensor again. Therefore, $A_{ij}^k$ represents a valid bundle homomorphism. As a result, we obtain
indeed
\begin{equation}
 \Gamma_{ij}^k\equiv \Gamma^{\alpha\beta\,\rho\sigma}_\mn(x,y,z)=-\delta^{(\alpha}_{(\mu}\,
 g^{\raisebox{0.2ex}{$\scriptstyle\beta)(\rho$}}(x)\,
 \delta^{\sigma)}_{\nu)}\; \delta(x-y)\delta(x-z)
\label{eq:NDConnectionAgain}
\end{equation}
as a natural connection on $\mF$. Remarkably enough, this agrees precisely with the connection
\eqref{eq:NDConnection}, determined in Section \ref{sec:DetConnections}. It is to be emphasized, however, that in
Section \ref{sec:DetConnections} the connection was designed artificially such that it leads to geodesics given by the
exponential parametrization, while here it was derived from the requirement that it be adapted to the geometric
structure of the space of metrics.
\medskip

\noindent
\textbf{(5) The Vilkovisky--DeWitt connection.}
For comparison, we would like to mention another famous choice for $A_{ij}^k$ which is due to Vilkovisky
\cite{Vilkovisky1984} and DeWitt \cite{DeWitt1987}. It is adapted to the principal bundle structure of
$\mF$ induced by the gauge group. The basic idea is to define geodesics on the physical base space
$\mF/\mG$ of the bundle and horizontally lift them to the full space $\mF$. In this manner,
coordinates in field space are decomposed into gauge and gauge-invariant coordinates. The resulting
Vilkovisky--DeWitt connection is obtained by using \eqref{eq:Christoffels} with the bundle homomorphism
\cite{PT09}
\begin{equation}
 A_{ij}^k = \eta^{\alpha\rho} \eta^{\beta\sigma} K_{\alpha\mku i} K_{\beta j} K^l_{(\rho} K^k_{\sigma);l}
 - \eta^{\alpha\beta} K_{\alpha\mku i} K^k_{\beta;j} - \eta^{\alpha\beta} K_{\alpha\mku j} K^k_{\beta;i} \,,
\label{eq:VDW-connection}
\end{equation}
Here, $K_{\alpha\mku i} \equiv G_{ij} K_\alpha^j$, involving the generators $K_\alpha^j$ of the gauge group,
$\eta^{\alpha\beta}$ is the inverse of $\eta_{\alpha\beta}\equiv K_\alpha^i\mku G_{ij} K_\beta^j$, and semicolons
denote covariant derivatives in field space corresponding to the LC connection \eqref{eq:Levi-Civita}. In contrast to
\eqref{eq:NDConnectionAgain}, the Vilkovisky--DeWitt connection is highly nonlocal, containing infinitely many
differential operators \cite{PT09}. Based on this connection, it is possible to construct a reparametrization invariant
and gauge independent effective action \cite{Vilkovisky1984,DeWitt1987}.
\medskip

To sum up, we have discussed three different connections on the space of metrics, $\mF$, all of which have the form
$\Gamma_{ij}^k = \big\{ {}^k_{ij} \big\} + A_{ij}^k\mku$, where they are characterized by different choices for the
bundle homomorphism $A_{ij}^k\mku$.
\begin{itemize}[itemsep=0.4em, topsep=0.4em,parsep=0em]%[itemsep=0pt]
 \item Setting $A_{ij}^k=0$ yields the LC connection induced by the DeWitt metric. Its associated geodesics were
 calculated in \cite{DeWitt1967a,FG89,GM91}. Although these geodesics are local and possess an explicit representation
 in terms of tangent vectors, their structure is more involved than the one of the exponential parametrization.
 \item Choosing relation \eqref{eq:VDW-connection} for $A_{ij}^k$ gives rise to the Vilkovisky--DeWitt connection,
 which takes into account the principal bundle character of the field space $\mF$ with the gauge group as structure
 group. It can be used in principle to construct reparametrization invariant and gauge independent quantities (even off
 shell). The corresponding geodesics are highly nonlocal, though, and they cannot be represented by an explicit
 formula.
 \item The choice \eqref{eq:AForND} for $A_{ij}^k$ leads to the novel connection \eqref{eq:NDConnectionAgain}. It is
 adapted to the geometric structure of the space of metrics. Furthermore, it generates geodesics which are local and
 possess a simple representation: the exponential metric parametrization.
\end{itemize}

%----------------------------------------------------------------------------------------------------------------------
\section{Covariant Taylor expansions and Ward identities}
\label{sec:Applications}
%----------------------------------------------------------------------------------------------------------------------

Taking the geometric path advocated previously, involving connections and geodesics on field space, allows for the
construction of covariant objects, in particular, of a geometric effective (average) action. Here, ``covariance'' has
a double meaning as it denotes both covariance w.r.t.\ spacetime and covariance w.r.t.\ field space. It is the latter
property, also referred to as reparametrization covariance, that we will focus on in this section. We will briefly
review the approach and discuss specifically the implications of the connection \eqref{eq:NDConnectionAgain}. A more
detailed introduction to the geometric formalism can be found, for instance, in Ref.\ \cite{PT09}.
\medskip

\noindent
\textbf{(1) Covariant Taylor expansions.}
Having some connection $\Gamma^k_{ij}$ on $\mF$ at hand, the key idea is to define coordinate charts based on
geodesics. We start by selecting an arbitrary base point $\bp$ in field space and using $\Gamma^k_{ij}$ to construct
geodesics that connect $\bp$ to neighboring points $\vp$.\footnote{We assume here that such geodesics exist. This
assumption is valid for Euclidean metrics, but metrics with Lorentzian signatures have to be handled with more care,
see Section \ref{sec:EuLor}.} As in Section \ref{sec:DetConnections}, let $\vp^i (s)$ denote such a geodesic in local
coordinates connecting $\vp^i (0) = \bp^i$ to $\vp^i(1) = \vp^i$. The vector which is tangent to the geodesic at the
starting point $\bp^i$ is given by $\frac{\td \vp^i(s)}{\td s}\big|_{s=0} = h^i[\bp,\vp]$. It depends on both base
point and end point. We have already argued that $\mF$ is geodesically complete, and that geodesics are determined by
the exponential map. Since the exponential map is a local diffeomorphism, we see that
$\exp_{\bp} : T_{\bp} \mF\rightarrow\mathcal{U}\subseteq\mF$ with $h \mapsto \vp[h;\bp]$ constitutes
a coordinate chart. We refer to this chart as geodesic coordinates. In this sense, the field $ h^i[\bp,\vp]$ plays a
twofold role, as a tangent vector located at $\bp$, and as the coordinate representation of the point $\vp$.

On the basis of geodesic coordinates it is possible to perform (field space-) covariant expansions which can eventually
be used to define a reparametrization invariant effective action. Let $A[\varphi]$ be any scalar functional of the
field $\varphi^i$, and let $\vp^i (s)$ be a geodesic as above. Then the functional $A[\varphi]$ can be expanded as a
Taylor series according to
\begin{equation}
 A[\vp] = A[\vp(1)] = \sum_{n=0}^{\infty} \frac{1}{n!} \left.\frac{\td^n }{\td s^n}\right|_{s=0} A[\vp(s)]\,.
\end{equation}
By iteratively making use of the geodesic equation as in Section \ref{sec:DetConnections}, this relation can be
rewritten as \cite{Honerkamp1972}
\begin{equation}[b]
 A[\vp] = \sum_{n=0}^{\infty} \frac{1}{n!}\, A_{i_1\dots i_n}^{(n)}[\bar{\varphi}]\, h^{i_1} \cdots h^{i_n} \,,
\label{eq:cov_exp}
\end{equation}
where $A_{i_1\dots i_n}^{(n)} [\bp] \equiv \mD_{(i_n} \dots \mD_{i_1)} A[\bar{\varphi}]$ denotes the $n$-th covariant
derivative (induced by the field space connection) with respect to $\vp$ evaluated at the base point $\bp$, and the
$h^i$'s are the coordinates of the tangent vector $h \equiv h[\bp,\vp] \in T_{\bp} \mF$. Relation \eqref{eq:cov_exp}
constitutes a \emph{covariant expansion of} $A[\vp]$ \emph{in powers of tangent vectors}.
\medskip

\noindent
\textbf{(2) Covariant derivatives expressed as partial derivatives.}
By viewing $h^i$ as the coordinate representation of the point $\vp$ (based on geodesic coordinates), $\vp \equiv
\vp[h;\bp]$, any scalar functional $A[\vp]$ depends parametrically on $h$ and on the base point $\bp$. Let us denote
functionals interpreted this way with a tilde, so in geodesic coordinates we have
\begin{equation}
A\big[\vp[h;\bp]\big] \equiv \tilde{A}[h;\bp] \,.
\label{eq:functionalNotation}
\end{equation}
Expansion \eqref{eq:cov_exp} implies a useful relation connecting partial and covariant derivatives which reads
\begin{equation}[b]
 \left.\frac{\delta^n}{\delta h^{i_1}\dots\delta h^{i_n}} \tilde{A}[h;\bp]\right|_{h=0}
 = \mD_{(i_n} \dots \mD_{i_1)} A[\bp] \,.
\label{eq:relatingDs}
\end{equation}
The significance of equation \eqref{eq:relatingDs} comes from the fact that the right hand side is manifestly covariant, so it
can be used to construct reparametrization invariant objects, while covariance is hidden on the left hand side. Hence, \emph{we
observe that} $\left(\frac{\delta}{\delta h}\right)^n A[\exp_{\bp}(h)]\big|_{h=0}$ \emph{is covariant}.
\medskip

\noindent
\textbf{(3) Covariance in \bm{$\mF$} and \bm{$\mat$}.}
Employing the connection \eqref{eq:NDConnection} with its diagonal character in $x$-space, \emph{a covariant derivative
in the field space} $\mF$ \emph{reduces to a covariant derivative in the target space} $\mat$, which we will denote by
\begin{equation}
\mD_k h^i \equiv \fD_K h^I \delta(x-y) \equiv \fD^{\alpha\beta} h_{\mu\nu}\, \delta(x-y),
\label{eq:PropertyCovDer}
\end{equation}
where capital Latin labels denote again pairs of spacetime indices, $h^I(x)\equiv h_{\mu\nu}(x)$. Assuming that the
functional $A$ can be written as $A[\varphi] = \int\td^d x\,\mathcal{L}(\varphi)$, expansion \eqref{eq:cov_exp} becomes
\begin{equation}
 A[\vp] = \int \td^d x\mkuu \sum_{n=0}^{\infty} \frac{1}{n!}\, \fD_{(I_n}\dots \fD_{I_1)} \mathcal{L}[\bar{\varphi}]\;
 h^{I_1}(x) \cdots h^{I_n}(x) \,.
\label{eq:ReductionCovDer}
\end{equation}
Thus, with the connection \eqref{eq:NDConnection}, covariant expansions in $\mat$ can be lifted to covariant expansion
in $\mF$ \emph{in a minimal way}. In fact, this applies to all spacetime-diagonal connections, while there is no such
mechanism for other connections. In particular, the Vilkovisky--DeWitt connection does not give rise to reductions of
the type \eqref{eq:ReductionCovDer}. Note that, in gravity, derivatives act also on the volume element $\sg$
which usually occurs inside $\mathcal{L}$, in contrast to the case of nonlinear sigma models.
\medskip

\noindent
\textbf{(4) The geometric effective action.}
Let us turn to the quantum theory now. Based on the conventional definition, the effective action $\Gamma$ is
determined by a functional integro-differential equation,
\begin{equation}
 \e^{-\Gamma[\vp]} = \int\mathcal{D}\hat{\vp}\;\e^{-S[\hat{\vp}] + (\hat{\vp}^i-\vp^i)
 \frac{\delta\Gamma}{\delta\vp^i}} \,,
\label{eq:IntegroDiff}
\end{equation}
where $S$ denotes the classical (bare) action, and the integration variable is given by the quantum field $\hat{\vp}$.
By construction, the argument $\vp$ of the effective action agrees with the expectation value, $\vp= \langle\hat{\vp}
\rangle$. In the case of gauge theories, the functional integral involves an additional integration over ghost fields,
and gauge fixing and ghost action terms are added in the exponent on the RHS of \eqref{eq:IntegroDiff}. This may
require the introduction of a background field $\bp$ which then appears as an additional argument of $\Gamma$.
A discussion of the functional measure $\mathcal{D}\hat{\vp}$ can be found in Appendix \ref{app:Measure}, cf.\ also
Ref.\ \cite{Mottola1995}.

The key point we want to make is that $\Gamma$ fails to be reparametrization invariant. As already
noticed by Vilkovisky \cite{Vilkovisky1984}, the reason for noncovariance in the naive definition originates from the
source term $(\hat{\vp}^i-\vp^i)J_i$ with $J_i= \delta\Gamma/\delta\vp^i\mku$: Since $\hat{\vp}^i$ and $\vp^i$ are
merely coordinates in a nonlinear space, their difference is not defined, and thus, such a source term makes no sense
from a geometrical point of view. However, by employing the powerful tools of Riemannian geometry it is possible to
define the path integral covariantly.

The idea is to \emph{couple sources to tangent vectors} which are determined by geodesics connecting $\vp$ to
$\hat{\vp}$. That means, the source term in \eqref{eq:IntegroDiff} must be of the form $S^\mathrm{source} = J_i\mku
\hat{h}^i \equiv J_i\mku \hat{h}^i[\vp,\hat{\vp}]$, where the fluctuation field $\hat{h}$ is an element of $T_{\vp}\mF$
now, and the source field $J$ is a cotangent vector, $J\in T^*_{\vp} \mF$. Moreover, the field space metric can be used
to include the volume factor $\sqrt{\det G_{ij}}$ in the functional integral such that the combination $\mathcal{D}
\hat{\vp}\sqrt{\det G_{ij}[\hat{\vp}]}$ and its analog in terms of $\mathcal{D}\hat{h}$ are manifestly covariant
\cite{DeWitt2003}. This procedure allows for the construction of a reparametrization invariant effective action
\cite{Vilkovisky1984}, referred to as the \emph{geometric effective action}.
As it is a functional of $h$ and $\bp$, we employ the notation of eq.\ \eqref{eq:functionalNotation} and label it with
a tilde: $\tilde{\Gamma}[h;\bp]$. Its full definition can be obtained from eq.\ \eqref{eq:GeoFuncInt} in Appendix
\ref{app:SplitWard} by setting $k=0$.

The corresponding functional $\Gamma[\vp,\bp]$ can then be defined by means of the tangent vector to the geodesic
connecting $\bp$ to $\vp$, say, $h\equiv h[\bp,\vp]$, which is inserted into $\tilde{\Gamma}$ thereafter:
$\Gamma[\vp,\bp]\equiv\tilde{\Gamma}\big[h[\bp,\vp];\bp\big]$. In general, in particular for gauge theories, $\Gamma$
cannot be written as a functional of $\vp$ alone, but it contains an extra $\bp$-dependence. This is discussed in more
detail in a moment. Within the geometric approach to defining the effective action, the equation $h=\big\langle\hh
\big\rangle$ is satisfied by construction (since it is $\hh$ that is coupled to the source), while we have $\vp\neq
\langle\hvp\rangle$ for a general field space connection; the relation between the dynamical field and an expectation
value is rather given in terms of a geodesic, according to $\vp\equiv\vp[h;\bp]=\vp\big[\big\langle\hh\big\rangle; \bp
\big]$.

In the remainder of this section we would like to review some properties of the geometric effective action, $\Gamma$,
and its generalization to the geometric effective average action, $\Gamma_k$, which takes into account scale dependence
according to the renormalization group. The following statements are not restricted to a particular connection, say,
the Vilkovisky--DeWitt connection, but they are valid for any field space connection, in particular for the one given
by equation \eqref{eq:NDConnection}.
\medskip

\noindent
\textbf{(5) Loop expansion.} Like in the standard (``nongeometric'') case, the geometric effective action
$\Gamma[\vp,\bp]\equiv\tilde{\Gamma}[h;\bp]$ in a Euclidean quantum field theory can be expressed in terms of an
$\hbar$-expansion:
\begin{equation}
 \tilde{\Gamma} [h;\bp] = \tilde{S}[h;\bp] + \frac{\hbar}{2}\STr \log \tilde{S}^{(2)}[h;\bp] + \mO(\hbar^2)\,,
\label{eq:GammaExpansion}
\end{equation}
where $\tilde{S}^{(2)}_{ij}[h;\bp] \equiv \tfrac{\delta^2 \tilde{S}[h;\bp]}{\delta h^j \delta h^i}$ is the Hessian
of $\tilde{S}$ with respect to $h$. We derive a similar relation for $\Gamma_k$ in Chapter \ref{chap:Bare}.
\medskip

\noindent
\textbf{(6) The geometric effective average action.} By adding a covariant infrared cutoff term of the type
$-\frac{1}{2}\hat{h}^i(\mathcal{R}_k[\bp])_{ij}\,\hat{h}^j$ with the scale $k$ to the exponent on the RHS of
\eqref{eq:IntegroDiff} and applying the same modifications to the functional integral as in point \textbf{(4)} in order
to achieve covariance, it is possible to construct a generalization of the geometric effective action, denoted by
$\Gamma_k[\vp,\bp]\equiv \tilde{\Gamma}_k[h;\bp]$, which is referred to as geometric effective average action
\cite{BMV03,Pawlowski2003,DP12}. Its running is governed by an FRGE similar to the standard one given by eq.\
\eqref{eq:FRGE} \cite{Pawlowski2003}:
\begin{equation}
\p_k\tilde{\Gamma}_k[h;\bp] = \frac{1}{2}\STr \Big[\big(\tilde{\Gamma}_k^{(2)}[h;\bp] + \Rk\big)^{-1}\p_k \Rk\Big] \,.
\label{eq:geoFRGE}
\end{equation}
Both in \eqref{eq:GammaExpansion} and in \eqref{eq:geoFRGE} the effective (average) action depends additionally on the
base point $\bp$. As mentioned previously, an extra $\bp$-dependence generally remains when switching from geodesic
coordinates based on $h$ to a $\vp$-based coordinate chart, $\Gamma_k[\vp,\bp]\equiv\tilde{\Gamma}_k[h;\bp]$. This
extra dependence stems from gauge fixing, ghost and cutoff terms. It is constrained by generalized Ward identities,
though, as we will clarify in points \textbf{(8)} and \textbf{(9)}. Note that a single-field effective (average) action
is usually obtained by taking the coincidence limit $\vp \to \bp$, or equivalently, $h \to 0$.
\medskip

\noindent
\textbf{(7) Constructing covariant expressions from \bm{$\Gamma_k$}.}\footnote{The same arguments apply to $\Gamma$,
too.} In practice, RG flow computations based on the EAA usually resort to the method of truncations, i.e.\
$\tilde{\Gamma}_k[h;\bp]$ is constructed out of a restricted set of possible invariants, as explained in Section
\ref{sec:EAAFRGE}. Most studies based on the functional RG deal with single field truncations, where the effective
average action is approximated by functionals of the form $\tilde{\Gamma}_k[h;\bp] = \Gamma_k [\vp(h;\bp)]$ without
extra $\bp$-dependence (apart from gauge fixing and ghost terms possibly). In this case, after taking the field
coincidence limit we can make use of relation \eqref{eq:relatingDs} on the right hand side of \eqref{eq:geoFRGE},
where we write
\begin{equation}
 \frac{\delta^2 \tilde{\Gamma}_k[h;\bp]}{\delta h^i \delta h^j} \bigg|_{h=0} = \mD_{(i}\mD_{j)} \Gamma_k [\bp] \,,
\end{equation}
thus yielding a fully covariant expression. In fact, the statement remains true when going back from $\tilde{\Gamma}_k$
to a general $\Gamma_k\mku$: Upon inserting $\vp=\exp_{\bp}(h)$ into $\Gamma_k[\vp,\bp]$, the \emph{partial derivatives
with respect to $h$ comprised by the Hessian are equivalent to covariant derivatives in $\mF$ with respect to $\vp$}.

In particular, this result applies to the use of connection \eqref{eq:NDConnection} and the associated exponential
parametrization. A direct calculation reveals the reason for covariance: By means of equation \eqref{eq:PhiExpansion}
we can expand $g=\bg\,\e^{\mku\bg^{-1}h}$ inside $\Gamma_k$ in terms of $h$, that is, schematically we have
$\Gamma_k\big[\bg\,\e^{\mku\bg^{-1}h},\bg\big]=\Gamma_k\big[\bg+h-\frac{1}{2}\bar{\Gamma}\mku hh+\mO(h^3),\bg\big]$.
Thanks to the appearance of the connection, \emph{a subsequent expansion of} $\Gamma_k$ \emph{in terms of} $h$ \emph{is
covariant in} $\mF$, in contrast to an expansion of $\Gamma_k[\bg+h,\bg]$ with the linear split \eqref{eq:StdParam}
which is covariant only in $\SymT$ with vanishing connection. This is a very important property of the exponential
parametrization. In uncondensed notation we have
\begin{equation}[b]
 \left.\frac{\delta^2 \Gamma_k[\mku\bg \, \e^{\mku\bg^{-1}h},\bg ]}{\delta h_{\mu\nu}(x) \delta h_{\alpha\beta}(y)}
 \right|_{h=0} = \mD^{\mu\nu}_{(x)} \mD^{\alpha\beta}_{(y)} \Gamma_k[g,\bg]\Big|_{g=\bg} \;,
\label{eq:CovariantHessian}
\end{equation}
where the covariant derivatives act on the first argument of the effective average action, and symmetrization is
ensured by the connection \eqref{eq:NDConnection}. 
\medskip

\noindent
\textbf{(8) Split-Ward identities} (also referred to as modified Nielsen identities).
Above, we have mentioned the extra $\bp$-dependence of the effective (average) action. However, $\tilde{\Gamma}[h;\bp]$
only seemingly depends on two fields. As discussed in Refs.\
\cite{BK87,Kunstatter1992,BMV03,Pawlowski2003,MR10,MRS11a,MRS11b,BDM14,DM15,BR14}, it rather depends on a
certain combination of the two fields $h$ and $\bp$ since $\tilde{\Gamma}[h;\bp]$ has to satisfy the
split-Ward identities
\begin{align}
\label{eq:NielsenID}
\frac{\delta \tilde{\Gamma}}{\delta \bar{\vp}^i} 
  + \big\langle\bar{\mD}_i \hat{h}^j\big\rangle\frac{\delta \tilde{\Gamma}}{\delta h^j} = 0 \,,
\end{align}
in the case of non-gauge theories. The tangent vector $\hat{h}^j$ appearing inside the expectation value corresponds to
the geodesic connecting the base point $\bp$ to the integration variable $\hat{\vp}$, i.e.\ we have $\hat{h}^j \equiv
\hat{h}^j[\bp,\hat{\vp}]$. The barred covariant derivative in \eqref{eq:NielsenID} is with respect to the base point,
$\bar{\mathcal{D}}_i \hat{h}^j[\bp,\hat{\vp}] = \tfrac{\delta \hat{h}^j}{\delta \bp^i} + \Gamma^j_{ik}[\bp] \hat{h}^k$.
Relation \eqref{eq:NielsenID} implies that $\bp^i$ \emph{and} $h^i$ \emph{can simultaneously be varied in such a way
that} $\tilde{\Gamma}[h;\bp]$ \emph{is left unchanged}. This is particularly important, as it guarantees that the
effective action and, consequently, all physical quantities are in fact \emph{independent of the choice of the base
point}. The statement can be phrased in terms of $\vp$ and $\bp$, too, where $\Gamma_k[\vp,\bp]$ depends only on a
combination of $\vp$ and $\bp$.

In a flat field space $\mF$ and in Cartesian coordinates we have $\hat{h}^i[\bp,\hat{\vp}] = \hat{\vp}^i - \bp^i$ and
thus $\big\langle\bar{\mathcal{D}}_i \hat{h}^j\big\rangle = - \delta^j_i$. In this special case, relation
\eqref{eq:NielsenID} reduces to the simple identity
\begin{equation}
\frac{\delta \tilde{\Gamma}}{\delta \bp^i} =\frac{\delta \tilde{\Gamma}}{\delta h^j} \,,
\end{equation}
implying a linear split, $\tilde{\Gamma}[h;\bp] = \Gamma[\bp + h] = \Gamma[\vp]$.

In the case of gauge theories there may be additional terms on the right hand side of \eqref{eq:NielsenID} due to
ghosts and gauge fixing: If a general field space connection is considered, the split-Ward identities read
\begin{equation}
 \frac{\delta \tilde{\Gamma}}{\delta \bar{\vp}^i} 
 + \big\langle\bar{\mD}_i \hat{h}^j\big\rangle\frac{\delta \tilde{\Gamma}}{\delta h^j} = \left\langle
 \frac{\delta \Sgf}{\delta \bar{\vp}^i} \right\rangle 
 + \left\langle \frac{\delta \Sgh}{\delta \bar{\vp}^i} \right\rangle \, ,
\label{eq:NielsenIDGauge}
\end{equation}
while they reduce to \eqref{eq:NielsenID} if the Vilkovisky--DeWitt connection is used \cite{BK87,Kunstatter1992}.
A derivation of \eqref{eq:NielsenIDGauge} can be found in Appendix \ref{app:SplitWard}.
\medskip

\noindent
\textbf{(9) Split-Ward identities for \bm{$\Gamma_k\mku$}.}
The corresponding relation for the effective average action receives further contributions due to the presence of the
regulator. As shown in Appendix \ref{app:SplitWard} for a general connection, the counterpart of eq.\
\eqref{eq:NielsenIDGauge} is given by
\begin{equation}
 \frac{\delta \tilde{\Gamma}_k}{\delta \bp^i}+\big\langle\bar{\mathcal{D}}_i\hat{h}^j\big\rangle
 \frac{\delta \tilde{\Gamma}_k}{\delta h^j} = \frac{1}{2} \Tr G_k\mku \bar{\mD}_i\mku\Rk
 + \Tr \Rk G_k\mku \frac{\delta \big\langle\bar{\mathcal{D}}_i \hat{h}\big\rangle }{\delta h}
 + \left\langle \frac{\delta \Sgf}{\delta \bar{\vp}^i} \right\rangle 
 + \left\langle \frac{\delta \Sgh}{\delta \bar{\vp}^i} \right\rangle ,
\label{eq:ModNielsenID}
\end{equation}
with the propagator $G_k =\big(\mku\tilde{\Gamma}_k^{(2)}[h;\bg] + \Rk\big)^{-1}$. When using the Vilkovisky--DeWitt
connection, on the other hand, the gauge fixing and ghost contributions in \eqref{eq:ModNielsenID} are absent
\cite{Pawlowski2003}.
In the limit $k\to 0$ the identity \eqref{eq:ModNielsenID} reduces to \eqref{eq:NielsenIDGauge}, as it should be.
Another instructive limit is $\big\langle\bar{\mathcal{D}}_i \hat{h}^j\big\rangle\to -\delta_i^j$ resulting from a flat
field space, where the second trace term in \eqref{eq:ModNielsenID} vanishes.

Similar to the corresponding identities for $\tilde{\Gamma}$, equation \eqref{eq:ModNielsenID} is of primary
importance for the discussion of background independence. The split-Ward identities state that any change of the
background field $\bp$ can be compensated for by a suitable change of $h$. This result guarantees that \emph{physical
predictions obtained from $\tilde{\Gamma}_k$ do not depend on the choice of the background field}.

Recently, the first steps towards a computation of RG flows satisfying split-Ward identities like
\eqref{eq:ModNielsenID} have been taken \cite{Pawlowski2003,DP12,MR10,MRS11a,MRS11b,BDM14,DM15,BR14}.
However, such considerations are possible only for special cases and approximations. As yet, a fully general treatment
seems to be out of reach. In this thesis, we will mainly be focused on single-field (single-metric) truncations where
the field is identified with the background field, so the split-Ward identities are suspended. They become accessible
only in the bimetric case. As an example, we will check $\Gamma_k$ for split-symmetry restoration in the limit $k\to 0$
in the bimetric analysis performed in Section \ref{sec:bi}.

%---------------------------------------------------------------------------------------
\section{Summarizing remarks}
\label{sec:ConcRemarks}
%---------------------------------------------------------------------------------------

\noindent
\textbf{(1)}
We have considered two possibilities for the type of the fundamental field variable in quantum gravity: pure metrics
with a fixed signature, $g\in\mF$, versus arbitrary symmetric rank-$2$ tensor fields, $g\in\SymT$.

\noindent
\textbf{(2)}
The space $\SymT$ is a vector space, i.e.\ it is linear. Hence, the most natural connection on its tangent bundle is
the flat one, and geodesics are straight lines, parametrized by $g=\exp_{\bg}(h)=\bg+h$.

\noindent
\textbf{(3)}
On the other hand, $\mF$ is a nonlinear space. Locally, at each spacetime point it is isomorphic to a homogeneous space
$\mat$, where the most natural connection, the canonical connection on $T\mat$, is adapted to the geometric structure
of $\mat$. This connection determines a connection on $T\mF$ in turn, giving rise to geodesics parametrized by
$g=\exp_{\bg}(h)=\bg\,\e^{\mku\bg^{-1}h}\mku$.

\noindent
\textbf{(4)}
Looking at it the other way round, the linear parametrization describes elements of $\SymT$, while the exponential
parametrization produces only pure metrics which strictly satisfy the signature constraint.
Hence, the exponential parametrization is not a proper (one-to-one) field redefinition of the linear parametrization.
The equivalence theorem for $S$-matrix elements does not apply.

\noindent
\textbf{(5)}
Restricting the tangent space for the linear parametrization such that the sum $\bg+h$ ``stays'' in $\mF$ is possible
but uncommon, and it would require the introduction of a nontrivial Jacobian in the functional integral \cite{PV15}.
By not considering such restrictions in this thesis, we take the point of view that $g=\bg\,\e^{\mku\bg^{-1}h}$ is not
a proper reparametrization of $g=\bg+h$.

\noindent
\textbf{(6)}
As suggested by the previous points, we expect different results for the linear and the exponential parametrization
when RG quantities like $\beta$-functions, fixed point values and critical exponents are computed. This will be
confirmed in the subsequent chapter.

\noindent
\textbf{(7)}
Using a geometric formalism based on geodesics it is possible to construct a reparametrization invariant and
gauge invariant effective average action, $\Gamma_k\mku$. For a special connection, the Vilkovisky--DeWitt connection,
$\Gamma_k$ is even gauge independent, but its associated geodesics are nonlocal and do not possess an explicit
representation. The connection derived in this chapter seems to combine the best of both worlds, though: (i)
Reparametrization and gauge invariance are guaranteed by construction. (ii) Corresponding geodesic are given by the
simple parametrization $g=\bg\,\e^{\mku\bg^{-1}h}$ which is local in spacetime. (iii) Remarkably enough, the use of the
exponential parametrization is already sufficient to ensure gauge independence at one-loop level for the
Einstein--Hilbert truncation \cite{Falls2015a,Falls2015b}.

\noindent
\textbf{(8)}
Gravity shares many properties with nonlinear sigma models, e.g.\ the homogeneous space structure of the respective
field space \cite{Friedan1980,Friedan1985,HPS88}. There is a crucial difference, though, which is due to the volume
element $\sg\mkuu$ inevitably occurring in all spacetime integrals and field space metrics $G_{ij}$. In gravity, this
introduces an extra field dependence, giving rise to additional terms in the Levi-Civita connection on the field space.

\noindent
\textbf{(9)}
For Euclidean metrics (and also for negative definite ones), the space $\mF$ equipped with the connection determined
in this chapter is geodesically complete and geodesically connected. There is a one-to-one correspondence between
metrics $g$ and tangent vectors $h$, i.e.\ the exponential map is a global diffeomorphism.

For Lorentzian signatures, $\mF$ is geodesically complete but not geodesically connected. The exponential map is
neither surjective nor injective. In a gravitational path integral this fact can be dealt with by applying two steps.
(i) One should sum over several background metrics such that any metric can be reached. (ii) The tangent spaces should
be restricted such that each metric is integrated over once and only once.

\noindent
\textbf{(10)}
In the Euclidean case, convexity of $\mF$ guarantees that the expectation value of a quantum metric is again an
element of $\mF$ with the correct signature: Let $\gamma\in\mF$ denote a quantum metric and $\hat{h}\in\SymT$ the
corresponding fluctuating tangent vector, i.e.\ $\gamma=\bg\,\e^{\mku\bg^{-1}\hat{h}}$, where $\bg\in\mF$ is given.
Then $\langle\gamma\rangle\equiv\big\langle\bg\,\e^{\mku\bg^{-1}\hat{h}}\big\rangle$ defines a proper metric again.
(This statement is independent of the above result that $g$ defined by $g\equiv\bg\,\e^{\mku\bg^{-1}h}$ with
$h=\big\langle\hh\big\rangle\in\SymT$ is a proper metric. Note here that $g\neq\langle\gamma\rangle$ for a
general field space connection.)

On the other hand, whether or not Lorentzian quantum metrics lead to expectation values $\langle\gamma\rangle$ that
can again be interpreted as Lorentzian metrics depends on the underlying action.

Nonetheless, the fact that in both the Euclidean and the Lorentzian case the field $g\equiv\bg\,\e^{\mku\bg^{-1}h}$
defines a metric with the correct signature justifies the use of the exponential metric parametrization also within the
argument of the effective average action, in addition to its possible appearance in a functional integral.

%----------------------------------------------------------------------------------------------------------------------
\chapter{Parametrization dependence in asymptotically safe gravity}
\label{chap:ParamDep}
%----------------------------------------------------------------------------------------------------------------------

\begin{summary}
After having seen in the previous chapter that the linear metric parametrization, $g_\mn=\bg_\mn+h_\mn$, and the
exponential one, $g_\mn=\bg_{\mu\rho}\mku(\e^{h})^\rho{}_\nu$, are not reparametrizations of each other, we expect this
fact to be reflected in different results for $\beta$-functions and their associated fixed points.
The current chapter is dedicated to confirming this conjecture. We perform a careful RG analysis based on a
single-metric Einstein--Hilbert truncation of the EAA for both the linear and the exponential
parametrization. Differences concerning flow diagrams and fixed point properties will be pointed out.
Motivated by conformal field theory studies the implications of our findings near two spacetime dimensions, where
the $\beta$-function of Newton's constant is closely related to a central charge, are of particular interest:
Only the exponential parametrization reproduces the well known critical central charge $c=25$.
The distinguished status of exponentials is explained by observing that they emerge in a natural way in the 2D limit.
Finally, we compute the $\beta$-functions in a bimetric setting on the basis of a twofold Einstein--Hilbert
truncation. For the linear parametrization it is known that background independence can be restored in the infrared
and reconciled with Asymptotic Safety in the UV.
Here we investigate if the exponential parametrization features this crucial property, too.

\noindent
\textbf{What is new?} Detailed RG analysis with the exponential parametrization for a single-metric truncation (Secs.\
\ref{sec:singleExpd}, \ref{sec:singleExp4} \& \ref{sec:singleExp2}) and a bimetric truncation (Sec.\ \ref{sec:biExp});
flow diagrams near 2D for the linear parametrization (Sec.\ \ref{sec:singleLin2}); argument for the special role of the
exponential parametrization (Sec.\ \ref{sec:Birth}).

\noindent
\textbf{Based on:} Ref.\ \cite{Nink2015}.
\end{summary}

%----------------------------------------------------------------------------------------------------------------------
\section{An introductory example}
\label{sec:ParamDepIntro}
%----------------------------------------------------------------------------------------------------------------------

All standard FRG analyses of metric gravity (for reviews see Refs.\
\cite{NR06,Percacci2009,RS07,RS12,CPR09,Litim2011,Nagy2014}) are based on the linear parametrization,
\begin{equation}
g_\mn = \bg_\mn + h_\mn \, .
\label{eq:stdParam}
\end{equation}
In respect of the previous chapter, however, it seems crucial to examine if the main results of these analyses remain
valid when the metric is parametrized by
\begin{equation}
 g_\mn=\bg_{\mu\rho}\mku\big(\e^h\big)^\rho{}_\nu \, ,
\label{eq:newParam}
\end{equation}
as only the latter choice guarantees that $g_\mn$ is a proper metric. Further benefits of the exponential
parametrization have already been discussed in Section \ref{sec:DetConnections}. In particular, we have mentioned the
possibility to compare our approach with conformal field theory by establishing its connection to the central charge.
Let us elaborate on this in more detail now. It will provide a first example of parametrization dependence.

We begin by recalling the results of the conformal field theory side, or, more precisely, of Polyakov's formulation
of bosonic string theory \cite{Polyakov1981,Polyakov1987,KPZ88}. To this end, we consider a path integral for
two-dimensional gravity coupled to conformal matter (i.e.\ to a matter theory that is conformally invariant when the
metric is fixed to be the flat one) with central charge $\cmat$. Here it is sufficient to regard such matter actions
that are constructed out of scalar fields. In this case, $\cmat$ is merely the number of these scalar fields. As shown
by Polyakov, integrating out the matter fields induces a nonlocal gravitational action, $\Gi\mku$, and the full path
integral decomposes into an integral over the conformal mode $\phi$ with a Liouville-type action times a
$\phi$-independent part, where the kinetic term for $\phi$ is found to be proportional to the number $\cmat$.
Performing the integration over the Faddeev-Popov ghosts corresponding to the conformal gauge, this factor gets
modified to $(\cmat-26)$, reflecting the famous critical dimension of bosonic string theory. If, finally, the implicit
$\phi$-dependence of the path integral measure is shifted into the action, the kinetic term for $\phi$ undergoes
another change and becomes proportional to $(\cmat-25)$ \cite{David1988,DK89,Polchinski1989}. For this reason we call
\begin{equation}
 \ccr \equiv 25 
\end{equation}
the \emph{critical central charge} at which the conformal mode $\phi$ decouples.

How is this related to the FRG studies of gravity and Asymptotic Safety? By definition, the running of the
dimensionless version of Newton's constant, $g_k$, is encoded in its $\beta$-function: $k\p_k g_k = \beta_g(g_k)$. Now
the essential point is that, in $d=2$ dimensions, the $\beta$-function, denoted by $\beta_g\equiv\beta_g(g)$, is of the
form
\begin{equation}
 \beta_g = -\frac{2}{3}\,c_\text{grav}\mku g^2,
\label{eq:betaG2DCGrav}
\end{equation}
up to higher orders in $g$. The coefficient $c_\text{grav}$ can be interpreted as a \emph{gravitational central
charge} since it can be read off from an action of the same type as the one occurring in the aforementioned string
theory example, the induced gravity action $\Gi$, although it is not induced by scalar fields this time but rather
represents a combined gravity+matter contribution to the gravitational fixed point action (cf.\ Chapter
\ref{chap:FullReconstruction}). Relation \eqref{eq:betaG2DCGrav} has been proven within the FRG framework by means of
scaling arguments applied to the gravitational functional integral \cite{CD15} and by means of a generalized nonlocal
ansatz for the effective average action \cite{CDP14b}.\footnote{Note that the definition of the gravitational central
charge in Refs.\ \cite{CD15,CDP14b} includes a minus sign as compared with our convention. See also the discussion in
Chapter \ref{chap:NGFPCFT}, in particular eq.\ \eqref{eq:CLCGrav}.}

Going slightly away from two dimensions, $d=2+\ve>2$, it is still possible to determine the general form of the
$\beta$-function of Newton's constant. Already a perturbative treatment \cite{Weinberg1979} shows --- and the
nonperturbative approach will be seen to confirm --- that $\beta_g$ can be written as
\begin{equation}
 \beta_g = \ve\mku g - b\mku g^2 ,
\label{eq:betaeps}
\end{equation}
up to the order $\mathcal{O}(g^3)$. For positive $b$, this implies the existence of a non-Gaussian fixed point at
\begin{equation}
 g_* = \ve/b \,,
\label{eq:bTogStar}
\end{equation}
which is crucial for the Asymptotic Safety scenario. Clearly, eq.\ \eqref{eq:betaG2DCGrav} can be obtained from
\eqref{eq:betaeps} by taking the limit $\ve\to 0$, and the gravitational central charge can be read off from the second
order term. This way we obtain the rule
\begin{equation}
 c_\text{grav} = \frac{3}{2}\,b\,.
\end{equation}
We will rederive this relation between $b$ and the central charge in Chapter \ref{chap:NGFPCFT} as a direct result
of the 2D limit, without having to insert the induced gravity action by hand as in Refs.\ \cite{CD15,CDP14b}.

It turns out that the coefficient $b$ \emph{depends on the underlying pa\-ra\-me\-tri\-za\-tion} of the metric.
Perturbative calculations based on the linear parametrization \eqref{eq:stdParam} yield $b=\frac{38}{3}$ for pure
gravity and $b=\frac{2}{3}(19-\cmat)$ for gravity coupled to $\cmat$ scalar fields
\cite{Weinberg1979,Tsao1977,Brown1977,KN90,JJ91}. This gives rise to the central charge
\begin{align}
 c_\text{grav} &= 19 - \cmat \qquad\text{(for the linear parametrization)}.
 \label{eq:ccritStdPert}
 \\
 \intertext{%
 If, on the other hand, parametrization \eqref{eq:newParam} underlies the computation of $\beta$-functions, then the
 critical central charge amounts to
 }
 c_\text{grav} &= 25 - \cmat \qquad\text{(for the exponential parametrization)},
 \label{eq:ccritNewPert}
\end{align}
as was first obtained within a perturbative framework in Refs.\ \cite{KKN93a,KKN93b,KKN93c,KKN96,AKKN94,NTT94,AK97}.
Hence, only for the exponential parametrization the pure gravity part of the central charge amounts to $25$. In this
case the critical number of scalar fields is given by $\ccr=25$ again. Here, ``critical'' refers to the fact that
the non-Gaussian fixed point in the small coupling regime does not exist any longer if $\cmat>25$. In this sense, only
the exponential parametrization reproduces the result known from conformal field theory.

We would like to emphasize that the above argument is by no means a statement about the ``correctness of a
parametrization''. The discrepancy between \eqref{eq:ccritStdPert} and \eqref{eq:ccritNewPert} is rather a
manifestation of the fact that \eqref{eq:stdParam} and \eqref{eq:newParam} parametrize different objects and may
describe different theories after all. We can merely conjecture that the exponential parametrization is more
appropriate for a comparison with conformal field theory.

After having seen this first example of parametrization dependence in perturbation theory we would like to investigate
in this chapter whether the results concerning central charges can be reproduced by the fully \emph{nonperturbative}
FRG methods introduced in Section \ref{sec:FRG}. For this purpose, we derive $\beta$-functions in arbitrary spacetime
dimensions using the exponential parametrization and an effective average action on the basis of the single-metric
Einstein--Hilbert truncation, and we expand them in terms of $\ve=d-2$. Also, we review the corresponding results for
the linear parametrization, add new insights and point out the main differences.

While the $(2+\ve)$-dimensional case serves as a playground which is particularly appropriate for a comparison with 2D
conformal field theory, it seems equally important to study the implications of a change of parametrization for a
$4$-dimensional world. In Section \ref{sec:single} we perform an RG analysis that takes into account the regulator
dependence, ultimately leading to characteristic flow diagrams in the space of $g_k$ and the cosmological constant
$\lambda_k$. Particular attention is paid to the existence and properties of non-Gaussian fixed points in the context
of Asymptotic Safety.
In Section \ref{sec:Birth} we consider a conformally reduced setting to show that there is a distinguished form of
the conformal factor whose 2D limit agrees precisely with the exponential parametrization.
 
Finally, in Section \ref{sec:bi} we conduct a bimetric analysis where we proceed along similar lines to the
single-metric case: We begin by reviewing the known results for the linear parametrization before we perform the
corresponding calculations based on the exponential parametrization. We will see that for both parametrizations the
concept of Asymptotic Safety can be reconciled with the requirement for background independence.

%----------------------------------------------------------------------------------------------------------------------
\section{Effective average action and gauge fixing}
\label{sec:ParamDepFramework}
%----------------------------------------------------------------------------------------------------------------------

\noindent
\textbf{(1) How the parametrization enters technically.}
In order to derive $\beta$-functions we choose a truncation of the effective average action $\Gamma_k$ and follow the
recipe given in Section \ref{sec:Recipe}. As outlined in Section \ref{sec:BFF}, our formalism requires the introduction
of a background metric, so $\Gamma_k$ is a functional of both $g_\mn$ and $\bg_\mn$ in
general: $\Gamma_k\equiv\Gamma_k[g,\bg]$. If we want to reexpress this as a functional of the tangent vector $h_\mn$
and the background metric $\bg_\mn$ instead of $g_\mn$ and $\bg_\mn$, the two parametrizations give rise to
\begin{equation}
\Gamma_k^\text{linear}[h;\bg] \equiv \Gamma_k[\bg+h,\bg] ,
\label{eq:GammaStandard}
\end{equation}
as opposed to
\begin{equation}
\Gamma_k^\text{exponential}[h;\bg] \equiv \Gamma_k\big[\bg \, \e^{\mku\bg^{-1}h},\bg\big] .
\label{eq:GammaNew}
\end{equation}
(As usual we adopt the comma notation for functionals of two metric fields, e.g.\ $\Gamma_k[g,\bg]$, and the semicolon
notation if the list of arguments contains the tangent vector and the background metric as in $\Gamma_k[h;\bg]$. Since
this notation is sufficient for a clear distinction, we omit the tilde on $\Gamma_k[h;\bg]$, unlike in Section \ref{sec:Applications}.)
The difference between \eqref{eq:GammaStandard} and \eqref{eq:GammaNew} is crucial; switching from one parametrization
to the other results in a modification of some terms in the FRGE \eqref{eq:FRGEgrav}.

This can most easily be seen at the level of
the corresponding Hessians, $\Gamma_k^{(2)}$. As the second derivatives are with respect to $h$, the two
parametrizations lead to different terms because, according to the chain rule,
\begin{equation}
\begin{split}
 \Gamma_k^{(2)}(x,y) &\equiv \frac{1}{\sbgx\sbgy}\;\frac{\delta^2\Gamma_k}{\delta h(x) \, \delta h(y)} \\
 &= \frac{1}{\sbgx\sbgy}\int\! \dd u\int\! \dd v\; \frac{\delta^2\Gamma_k}{\delta g(u)\,\delta g(v)} \,
 \frac{\delta g(v)}{\delta h(x)}\,\frac{\delta g(u)}{\delta h(y)} \\
 &  \qquad +\frac{1}{\sbgx\sbgy}\int\! \dd u\; \frac{\delta \Gamma_k}{\delta g(u)} \,
 \frac{\delta^2 g(u)}{\delta h(x)\,\delta h(y)} \; ,
\end{split}
\label{eq:2ndVar}
\end{equation}
where we suppressed all spacetime indices for the sake of clarity. The first term on the RHS of equation
\eqref{eq:2ndVar} is the same for both parametrizations, at least at lowest order in $h$, since
\begin{equation}
 \frac{\delta g_\mn(x)}{\delta h_{\rho\sigma}(y)}
 = \begin{cases}
 \delta^\rho_{(\mu} \, \delta^\sigma_{\nu)} \, \delta(x-y) &\text{(linear)},\\[0.5em]
 \delta^\rho_{(\mu} \, \delta^\sigma_{\nu)} \, \delta(x-y) +\mO(h) \qquad\quad &\text{(exponential)},
 \end{cases}
\end{equation}
where round brackets enclosing index pairs denote symmetrization.

The last term in \eqref{eq:2ndVar}, however, vanishes identically for parametrization \eqref{eq:stdParam} because
\begin{equation}
 \frac{\delta^2 g_\mn(u)}{\delta h_{\rho\sigma}(x) \, \delta h_{\lambda\gamma}(y)} = 0\,,
\end{equation}
whereas the exponential relation \eqref{eq:newParam} entails
\begin{equation}
 \frac{\delta^2 g_\mn(u)}{\delta h_{\rho\sigma}(x) \, \delta h_{\lambda\gamma}(y)}
 = {\textstyle \frac{1}{2}}\, \left(\bg^{\lambda(\sigma} \delta^{\rho)}_{(\mu} \, \delta^\gamma_{\nu)}
 + \bg^{\rho(\gamma} \delta^{\lambda)}_{(\mu}\, \delta^\sigma_{\nu)} \right) \delta(u-x) \delta(u-y) + \mO(h)\,.
\end{equation}
As a consequence, the latter case implies \emph{additional contributions to the FRGE} \eqref{eq:FRGEgrav}. We would
like to point out that these new contributions are proportional to the first variation of $\Gamma_k$ in
\eqref{eq:2ndVar}. Therefore, since $\delta\Gamma_k/\delta g_\mn\big|_\text{on shell}=0$, \emph{the exponential
parametrization gives the same result for the Hessian as the linear one when going on shell}. Nonetheless, due to the
inherent off shell character of the FRGE, we expect differences in $\beta$-functions and the corresponding RG flow.
\medskip

\noindent
\textbf{(2) The transformation behavior of \bm{$h_\mn\mku$}.}
As we want to comment on gauge invariance and gauge fixing, we have to know how the field $h_\mn$ transforms under
diffeomorphisms provided that both $g_\mn$ and $\bg_\mn$ transform as usual tensor fields, i.e.\ they satisfy
$\delta g_\mn = \mL_\xi g_\mn$ and $\delta \bg_\mn = \mL_\xi \bg_\mn\mku$. Here, $\xi$ is the vector field which
generates the diffeomorphism and $\mL_\xi$ denotes a Lie derivative along $\xi$.

For the linear parametrization the answer is rather obvious: The defining relation $g_\mn=\bg_\mn+h_\mn$ implies that
$h_\mn$ transforms as a tensor field, too:
\begin{equation}
 \delta h_\mn = \delta(g_\mn-\bg_\mn) = \mL_\xi(g_\mn-\bg_\mn) = \mL_\xi h_\mn\,.
\end{equation}

For the exponential parametrization such a conclusion is not as straightforward as it seems at first sight. Starting
out from relation \eqref{eq:newParam}, we observe that $(\e^h)^\rho{}_\nu$ must transform as a tensor field under
general coordinate transformations if $g_\mn$ and $\bg_{\mu\rho}$ transform as tensor fields. However, since
$\delta h_\mn$ does not commute with $h_\mn$ in general, we cannot write $\delta (\e^h)^\rho{}_\nu$ in the form
$(\e^h)^\rho{}_\sigma\mku \delta h^\sigma{}_\nu\mku$, which would directly entail the simple tensorial transformation
behavior for $h_\mn\mku$. Nevertheless, such a behavior can still be shown by a more careful analysis: We
prove in Appendix \ref{app:Trans} that $h_\mn$ transforms indeed as an ordinary tensor field, too, that is
\begin{equation}
\delta h_\mn = \mL_\xi h_\mn \,.
\end{equation}

Hence, background gauge transformations, introduced in Section \ref{sec:BFF}, are induced by the usual transformation
laws $\delta g_\mn = \mL_\xi g_\mn$, $\delta \bg_\mn = \mL_\xi \bg_\mn$ and $\delta h_\mn = \mL_\xi h_\mn$ \emph{for
both parametrizations}. It is these transformations under which the effective average action is invariant.
\medskip

\noindent
\textbf{(3) Quantum gauge transformation.}
Let us briefly recall the arguments of Section \ref{sec:BFF}. In the process of the (functional integral based)
construction of the effective average action we must ensure that we pick only one ``point'' (field configuration) per
gauge orbit during the integration, i.e.\ we have to fix the gauge, which is usually accomplished by adding a gauge
fixing action in the exponent of the integrand. The bare action $S[\gamma]$ (with $\gamma_\mn$ the quantum metric) is
invariant under the transformation $\gamma_\mn\to \gamma_\mn+\delta \gamma_\mn=\gamma_\mn+\mL_\xi\gamma_\mn\mku$.
Viewing $\gamma_\mn$ as a function of $\bg_\mn$ and the quantum tangent vector $\hat{h}_\mn$ (cf.\ discussion on
geodesics in the space of metrics in Chapter \ref{chap:SpaceOfMetrics}), we have the freedom to distribute the full
change $\delta \gamma_\mn=\mL_\xi \gamma_\mn$ among $\delta \bg_\mn$ and $\delta \hat{h}_\mn\mku$.
One particular choice is the quantum or true gauge transformation, here denoted by $\dQ\mku$, which is
characterized by $\dQ\bg_\mn=0$.
As an example, let us consider the linear parametrization, $\gamma_\mn=\bg_\mn+\hat{h}_\mn\mku$. Choosing
\begin{align}
 \dQ \bg_\mn &= 0\,,\\
 \dQ \hat{h}_\mn &= \mL_\xi\big(\bg_\mn+\hat{h}_\mn\big)=\mL_\xi \gamma_\mn\,,
\end{align}
we observe that the transformation behavior of the quantum metric $\gamma_\mn$ is unchanged:
\begin{equation}
 \dQ \gamma_\mn = \dQ \bg_\mn + \dQ \hat{h}_\mn = \mL_\xi\big(\bg_\mn+\hat{h}_\mn\big) = \mL_\xi \gamma_\mn\,.
\end{equation}

For the exponential parametrization $\gamma_\mn=\bg_{\mu\rho}\big(\e^{\hat{h}}\big)^\rho{}_\nu\mku$, on the other hand, it is
much more involved to find the quantum gauge transformation law for $\hat{h}_\mn\mku$, i.e.\ to solve the requirements
$\dQ\bg_\mn=0$ and $\dQ \gamma_\mn=\mL_\xi\gamma_\mn$ for $\dQ \hat{h}_\mn\mku$. Making use of Lemmas
\ref{lem:MatrixLog} and \ref{lem:VarMatrixExp} finally leads to the integral representation (in matrix notation)
\begin{equation}
 \dQ \hat{h} = \int_0^\infty\td s \int_0^1\td t\; \e^{-ts\mku \gamma\mku\bg^{-1}}\mku \mL_\xi \gamma\;
 \e^{-(1-t)s\mku\bg^{-1}\gamma} \,.
\label{eq:DeltaQh}
\end{equation}
Using this expression as a basis for the construction of a ghost action (after having chosen the underlying gauge
fixing action) would lead to an unusual form of the Faddeev-Popov operator. Therefore, we will proceed differently in
the following.
\medskip

\noindent
\textbf{(4) The \bm{$g_\mn$}-type gauge fixing method.}
In order to be as close to the standard calculations based on \eqref{eq:stdParam} as possible \cite{Reuter1998}, we
slightly adapt the gauge fixing procedure. The standard gauge fixing condition for the linear parametrization is of
the form $F_\alpha\equiv\mF_\alpha^\mn[\bg]\,\hat{h}_\mn=0$, and the corresponding ghost action is proportional to
\begin{equation}
 \int\dd x\,\bar{C}_\mu\mkuu\bg^\mn \frac{\p F_\nu}{\p \hat{h}_{\alpha\beta}}\,\dQ \hat{h}_{\alpha\beta}
 = \int\dd x\,\bar{C}_\mu\mkuu\bg^\mn \frac{\p F_\nu}{\p \hat{h}_{\alpha\beta}}\,\mL_C\big(\bg_{\alpha\beta}
 +\hat{h}_{\alpha\beta}\big),
\label{eq:propGhAction}
\end{equation}
with the ghost fields $\bar{C}_\mu$ and $C^\mu$. At this point we make the unsurprising but crucial observation that
\emph{$\hat{h}_\mn$ in the gauge fixing condition can be replaced by $\gamma_\mn\mku$}: We employ the most convenient
class of $\mF$'s where $\mF_\alpha^\mn[\bg]$ contains only such terms which are proportional to the covariant
derivative $\bD_\mu$ corresponding to the background metric, and therefore, since $\bD_\mu\mku \bg_{\alpha\beta}=0$,
\begin{equation}
 0 = \mF_\alpha^\mn[\bg]\,\hat{h}_\mn = \mF_\alpha^\mn[\bg]\big(\bg_\mn+\hat{h}_\mn\big)
 = \mF_\alpha^\mn[\bg]\, \gamma_\mn \,,
\end{equation}
for the linear parametrization. That is, we can always write the gauge condition as $\mF_\alpha^\mn[\bg]\,\gamma_\mn=0$
instead of $\mF_\alpha^\mn[\bg]\,\hat{h}_\mn = 0$. Henceforth, we refer to this as the ``metric version'' of the gauge
fixing condition. Similarly, the ghost action \eqref{eq:propGhAction} can be expressed as
\begin{equation}
 \int\dd x\,\bar{C}_\mu\mkuu\bg^\mn \frac{\p F_\nu}{\p \gamma_{\alpha\beta}}\,\mL_C \gamma_{\alpha\beta}\,.
\label{eq:propGhAction2}
\end{equation}
The advantage of \eqref{eq:propGhAction2} is that it does not involve $\dQ \hat{h}_\mn\mku$. By construction, for the
linear parametrization the metric versions of the gauge condition and the ghost action are completely equivalent to the
standard versions.

Passing on to the exponential parametrization, we can choose the metric version of the gauge condition, too,
\begin{equation}
\mathcal{F}_\alpha^\mn[\bg] \, \gamma_\mn = 0 \,,
\end{equation}
along with the ghost action \eqref{eq:propGhAction2}. This form is preferred to the $\hat{h}_\mn$-version because (a)
avoiding the use of $\dQ \hat{h}_\mn$ given by \eqref{eq:DeltaQh} reduces the complexity of computations, and (b) the
metric version leads to the \emph{same Faddeev-Popov operator} as in the standard case \cite{Reuter1998}.

As discussed in Section \ref{sec:Recipe}, the standard FRG approach consists in choosing a suitable truncation ansatz
for $\Gamma_k$ rather than evaluating a functional integral. Such a truncation ansatz includes gauge fixing and ghost
contributions, the usual choice being motivated by possible gauge fixing actions and ghost actions as they would appear
in the exponent of the corresponding functional integral. Therefore, at the level of $\Gamma_k\mku$, we have to specify
the gauge fixing and ghost action in terms of $h_\mn$ (or $g_\mn$) rather than $\hat{h}_\mn$ (or $\gamma_\mn$). For the
above discussion including point \textbf{(3)} and \textbf{(4)} this means that we can employ the same arguments, but
applied to $h_\mn$ and $g_\mn$ this time. In particular, we use a gauge fixing condition of the form
\begin{equation}
\mathcal{F}_\alpha^\mn[\bg] \, g_\mn = 0 \,.
\end{equation}
We will refer to this choice as ``$g_\mn$-type'' gauge fixing condition. Its use implies that the Faddeev-Popov
operator is independent of the metric parametrization. As a consequence, \emph{all contributions to the FRGE coming
from gauge fixing and ghost terms are the same for both parametrizations considered.}
By virtue of the one-to-one correspondence between $g_\mn$ and $h_\mn$ (see Appendix \ref{app:ExpParam}) this gauge
fixing method is perfectly admissible for the exponential parametrization.
\medskip

\noindent
\textbf{(5) Choice of the gauge condition.}
Both for the single-metric computation presented in Section \ref{sec:single} and for the bimetric analysis shown in
Section \ref{sec:bi} we employ the harmonic coordinate condition (de Donder gauge): $\mF_\alpha^\mn[\bg] \, g_\mn = 0$
with
\begin{equation}
 \mF_\alpha^\mn[\bg] = \delta^\nu_\alpha \,\bg^{\mu\rho}\bar{D}_\rho - \frac{1}{2} \,\bg^\mn \bar{D}_\alpha \,,
\label{eq:GaugeCond}
\end{equation}
(corresponding to $\beta=\frac{d}{2}-1$ in Ref.\ \cite{GKL15}). As for the gauge parameter $\alpha$ appearing in the
gauge fixing action, we choose a Feynman-type gauge, $\alpha=1$, in the single-metric case, while the bimetric results
are obtained by employing the ``$\Omega$ deformed $\alpha=1$ gauge'' introduced in Ref.\ \cite{BR14}. This
allows us to compare the subsequent calculations based on the exponential parametrization with the standard results
\cite{Reuter1998,BR14}.

%----------------------------------------------------------------------------------------------------------------------
\section{RG analysis for a single-metric truncation}
\label{sec:single}
%----------------------------------------------------------------------------------------------------------------------

In this section we aim at determining the RG running of the Newton constant and the cosmological constant.
As usual, we resort to a truncation of the full theory space, i.e.\ we determine the RG flow within a subspace of
reduced dimensionality. In what follows, we choose a subspace that consists only of such invariants which are
constructed out of one single metric. More precisely, our computations are based on the Einstein--Hilbert truncation
\cite{Reuter1998}:
\begin{equation}
 \Gamma_k\big[g,\bg,\xi,\bx\, \big] = \Gamma_k^\text{grav}\big[g,\bg\big] + \Gamma_k^\text{gf}\big[g,\bg\big]
 +\Gamma_k^\text{gh}\big[g,\bg,\xi,\bx\, \big].
\label{eq:EHtrunc}
\end{equation}
with
\begin{equation}
 \Gamma_k^\text{grav}\big[g,\bg\big] \equiv \frac{1}{16\pi G_k} \int \! \dd x \sg \,\big( -R + 2\Lambda_k \big).
\label{eq:EHtruncGrav}
\end{equation}
Here $G_k$ and $\Lambda_k$ are the dimensionful Newton constant and cosmological constant,
respectively, and
\begin{equation}
 \Gamma_k^\text{gf}\big[g,\bg\big] \equiv \frac{1}{2\alpha}\frac{1}{16\pi G_k}\int\dd x \sbg \,\bg^{\alpha\beta}
 \big(\mF_\alpha^\mn[\bg] g_\mn\big)\big(\mF_\beta^\rs[\bg]g_\rs\big)
\label{eq:GaugeFixingAction}
\end{equation}
is the gauge fixing action, where $\alpha=1$ and $\mathcal{F}_\alpha^\mn[\bg]$ is given by eq.\ \eqref{eq:GaugeCond}.
Furthermore, $\Gamma_k^\text{gh}$ denotes the associated ghost action with the ghost fields $\xi$ and $\bx$. After
having inserted the respective metric parametrization into the EAA \eqref{eq:EHtrunc}, the corresponding
$\beta$-functions are obtained by following the steps of Section \ref{sec:Recipe}.

In order to determine critical central charges in the upcoming Sections \ref{sec:singleLin2} and \ref{sec:singleExp2}
we add a matter action to the ansatz given by eq.\ \eqref{eq:EHtrunc}: We consider the truncation
$\Gamma_k\big[g,\bg,A,\xi,\bx\, \big] = \Gamma_k^\text{grav}\big[g,\bg\big] + \Gamma_k^\text{m}\big[g,\bg,A\big]
+ \Gamma_k^\text{gf}\big[g,\bg\big] + \Gamma_k^\text{gh}\big[g,\bg, \xi,\bx\, \big]$, where the matter contribution is
given by a multiplet of $N$ scalar fields,\footnote{Note that, in order to avoid confusion between the gravitational
and the matter central charge, we denote the number of matter fields by $N$ instead of $\cmat$ henceforth.}
$A = (A^i)$, with $i = 1,\dotsc,N$, minimally coupled to the full, dynamical metric:
\begin{equation}
 \Gamma_k^\text{m}\big[g,\bg,A\big]\equiv \frac{1}{2} \sum_{i=1}^N\int\!\dd x\sg\; g^\mn\,\p_\mu A^i\mku\p_\nu A^i\,.
\label{eq:matter}
\end{equation}
Note that the matter action contains no running parameters in the present truncation.\footnote{In fact, with the action
defined in eq.\ \eqref{eq:matter} the RHS of the FRGE \eqref{eq:FRGE} can generate terms proportional to
$\p_\mu A^i\mku\p_\nu A^i$, so $\Gamma_k^\text{m}$ is $k$-dependent in general. Here, however, we are interested only
in the running of the Newton constant and the cosmological constant, while the $k$-dependence of $\Gamma_k^\text{m}$
can be neglected. In this sense, $\Gamma_k^\text{m}$ may be considered always at its fixed point. On the technical
level, this behavior is achieved by setting $A^i$ to zero after having determined the Hessian.

For the analysis performed in this chapter, we could couple the scalar fields to the background metric as well: If
$\Gamma_k^\text{m}$ in \eqref{eq:matter} were a functional of $\bg_\mn$ instead of $g_\mn\mku$, the FRGE would not
generate any terms that could lead to a running of $\Gamma_k^\text{m}$. In this case $\Gamma_k^\text{m}$ would be
strictly $k$-independent. Within a single-metric truncation, where $\bg_\mn$ is identified with $g_\mn$ after
functional derivatives have been taken, the two points of view give rise to equivalent results.}
Thus, we can write $\Gamma_k^\text{m}\big[g,\bg,A\big] \equiv \Gamma^\text{m}\big[g,A\big]$.

In the following six subsections we would like to investigate the parametrization dependence of fixed points, critical
exponents and other qualitative features of flow diagrams. Apart from the phase portraits in $d=2+\ve$ dimensions,
shown in Section \ref{sec:singleLin2}, the results for the linear parametrization are well known, so we refrain from
repeating the underlying computation. We merely present a collection of the most important facts (Secs.\
\ref{sec:singleLin4} and \ref{sec:singleLin2}). Afterwards we derive the differences entailed by the use of the
exponential parametrization (Secs.\ \ref{sec:singleExpd}, \ref{sec:singleExp4} and \ref{sec:singleExp2}), where the
details of the calculation are specified in Appendix \ref{app:single}.

%----------------------------------------------------------------------------------------------------------------------
\subsection[The linear parametrization in \texorpdfstring{$d=4$}{d=4} dimensions]%
{The linear parametrization in \texorpdfstring{\bm{$d=4$}}{d = 4} dimensions}
\label{sec:singleLin4}
%----------------------------------------------------------------------------------------------------------------------

For comparison with the exponential parametrization, we begin with a brief summary of known results for the linear
parametrization.

The $\beta$-functions of the dimensionless couplings,
\begin{equation}
 g_k \equiv k^{d-2}G_k\,,\qquad \lambda_k \equiv k^{-2}\Lambda_k\,,
\end{equation}
have been derived in Ref.\ \cite{Reuter1998} for general dimensions $d$. In the special case $d=4$ they
give rise to the flow diagram shown in Figure \ref{fig:StdSingle}. In addition to the Gaussian fixed point at the
origin, \emph{there exists a non-Gaussian fixed point} (NGFP) with a positive Newton constant, suitable for
the Asymptotic Safety scenario. Its critical exponents have positive real parts, so it has \emph{two UV-attractive
directions}. Furthermore, we make the crucial observation that there are trajectories emanating from the NGFP and
passing the classical regime close to the Gaussian fixed point. This type of trajectories is believed to be realized
in Nature \cite{RW04}. In Figure \ref{fig:StdSingle} they lie to the right of the \emph{separatrix}, the trajectory
connecting the non-Gaussian to the Gaussian fixed point.

The red, dashed curve in Figure \ref{fig:StdSingle} indicates that the $\beta$-functions diverge at these points. Thus,
trajectories approaching this boundary/singularity line are not defined beyond or below a certain RG scale. This holds
in particular for type IIIa trajectories (based on the classification proposed in Ref.\ \cite{RS02}) which, by
definition, emanate from the NGFP and run into the singularity line at positive $\lambda$ towards IR scales. They lie
entirely in the first quadrant, mainly to the right of and below the separatrix. The aforementioned trajectory realized
in Nature falls into this class. It is believed that the singularity line is merely a truncation artifact \cite{RW04}:
In a less truncated or untruncated theory space trajectories are expected to be defined at all scales down to $k=0$.
For the present analysis the most important message is that the singularity line does not ``block'' the separatrix.

\begin{figure}[tp]
\centering
\includegraphics[width=.78\columnwidth]{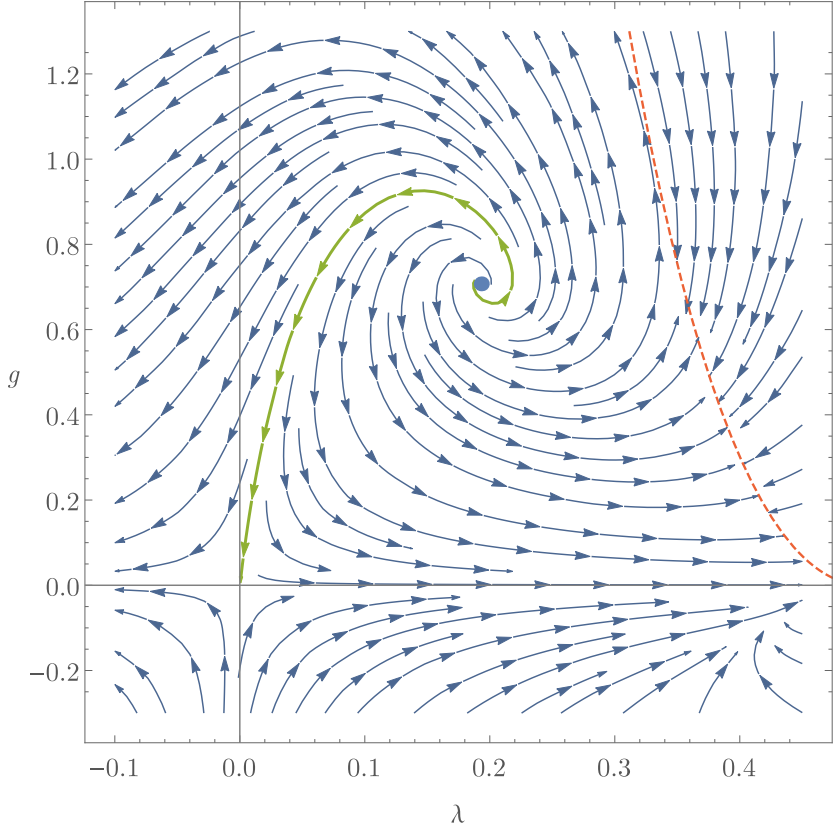}
\caption{Flow diagram for the Einstein--Hilbert truncation in $d=4$ based on the \emph{linear parametrization} (first
 obtained in \cite{RS02} for a sharp cutoff; here for the optimized cutoff \cite{Litim2001}). There is a non-Gaussian
 fixed point at positive $g$ and $\lambda$, indicated by the blue dot in the middle of the spiral. The separatrix
 connecting the non-Gaussian to the Gaussian fixed point follows the green arrows. On the red, dashed curve the
 $\beta$-functions become divergent. Note that, by convention, arrows point from the UV (``$k\to\infty\mku$'') to the
 IR (``$k\to 0\mku$'').}
\label{fig:StdSingle}
\end{figure}

It has turned out that the qualitative picture (existence of the NGFP, number of relevant directions, connection
between NGFP and classical regime) is extremely stable under many kinds of
modifications of the setup, for instance under changes of the truncation ansatz (like the inclusion of higher order
curvature terms \cite{LR02c,CPR08,MS08,CPR09,FLNR13,FLNR16a,BC12,Benedetti2013,DM13a,DM13b,DSZ12,DSZ15a,DSZ15b,DM15,%
OP14,Eichhorn2015,OPV15,OP16,OPV16,FLNR16b,FO16}, matter fields \cite{DP98,BMS10,DEP14,DEP15,DELP16} or running ghosts
\cite{EG10,GS10}), the gauge fixing action and the cutoff scheme; for reviews see
\cite{NR06,RS07,Percacci2009,CPR09,Litim2011,RS12}.
In particular, changes in the cutoff shape function do not alter the picture, except for insignificantly shifting
numerical values like fixed point coordinates. The very existence of the NGFP is found for all realistic settings
investigated so far.

%----------------------------------------------------------------------------------------------------------------------
\subsection[The linear parametrization in \texorpdfstring{$d=2+\ve$}{d = 2 + epsilon} dimensions]%
{The linear parametrization in \texorpdfstring{\bm{$d=2+\ve$}}{d = 2 + epsilon} dimensions}
\label{sec:singleLin2}
%----------------------------------------------------------------------------------------------------------------------

In $d=2+\ve$ dimensions the form of $\beta$-functions implies that the Newton constant and the cosmological constant
at the NGFP are of first order in $\ve\mku$: $\mku g_*=\mO(\ve)$ and $\lambda_*=\mO(\ve)$, respectively. Hence, unless
we consider points too far away from the NGFP, we can assume $g=\mO(\ve)$ and $\lambda=\mO(\ve)$, too. Inserting this
back into the $\beta$-functions yields the following expansion in terms of the couplings, which is also an expansion in
terms of $\ve\mku$:
\begin{align}
 \beta_g &= \ve g - b g^2 \,,
\label{eq:betag2dFRGE}\\
 \beta_\lambda &= -2\mku\lambda -2\mku\Phi_1^1(0)\mku g\,,
\label{eq:betal2dFRGE}
\end{align}
up to higher orders, where the threshold functions of the type $\Phi_n^p(w)$ are defined in Appendix \ref{app:Cutoffs}.
We observe that the $\beta$-function of the Newton constant has the same structure as in the perturbative analysis,
see equation \eqref{eq:betaeps}, $\beta_g = \ve g - b g^2$. It is possible to show \cite{Reuter1998} that the
coefficient $b$ is a universal number, i.e.\ it is independent of the cutoff shape function, and its value is given by
$b=\frac{38}{3}$ for pure gravity. Positivity of $b$ implies the existence of a non-Gaussian fixed point with
positive Newton constant, here $g_*=\frac{3}{38}\mku\ve$. The fixed point value of the cosmological constant is not
universal, though, since the threshold function $\Phi_1^1(0)$ depends on the cutoff. It can be argued, however, that
$\Phi_1^1(0)$ is positive and of order $1$ for all standard cutoff shapes. For the optimized shape function
\cite{Litim2001} we obtain $\lambda_*=-\frac{3}{38}\mku\ve$.

If, additionally, scalar fields are included in the analysis by taking into account the matter action
\eqref{eq:matter}, then the coefficient $b$ becomes $b=\frac{2}{3}(19-N)$ for all cutoff shapes. Thus, the linear
parametrization gives rise to the \emph{universal} result
\begin{equation}
 c_\text{grav} = 19 - N\,,
\end{equation}
leading to the \emph{critical central charge} $\ccr\equiv N^\text{crit}=19$, in agreement with the per\-tur\-ba\-tive
result \eqref{eq:ccritStdPert}.

Finally, we would like to visualize the RG flow corresponding to the full $\beta$-functions \cite{Reuter1998} in
$d=2+\ve$ without relying on any expansion of the type \eqref{eq:betag2dFRGE} and \eqref{eq:betal2dFRGE}. To this end
we introduce the normalized couplings
\begin{equation}
 \rl \equiv \lambda/\ve\,,\qquad \rg \equiv g/\ve\,,
\end{equation}
whose fixed point values, $\rls\mku ,\mkuu\rgs\mku$, remain finite in the limit $\ve\to 0$. In this representation,
even the flow diagram and its associated RG trajectories approach a ``finite'' form for $\ve\to 0$. The situation
is illustrated in Figure \ref{fig:Flow2D}, where we show several diagrams at different values of $\ve$. Each diagram
contains four sample trajectories, all of which run into the UV fixed point for $k\to\infty$. The initial conditions
for the respective trajectories, i.e.\ their starting points in the infrared, are the same for all diagrams.

\begin{figure}[tp]
\begin{minipage}{0.47\columnwidth}
 \centering
 \small$\ve=0.35$\\
 \includegraphics[width=\columnwidth]{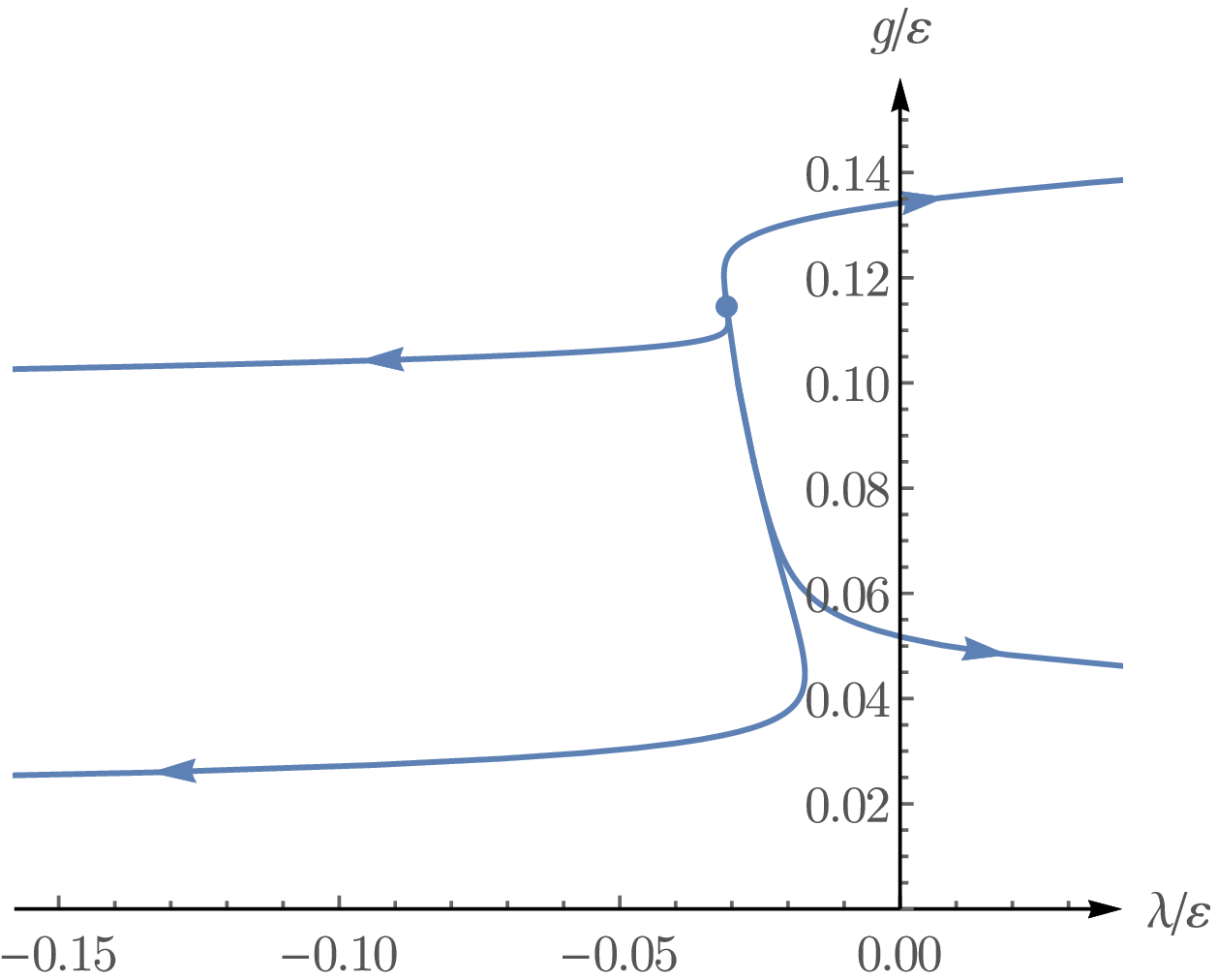}
\end{minipage}
\hfill
\begin{minipage}{0.47\columnwidth}
 \centering
 \small$\ve=0.2$\\
 \includegraphics[width=\columnwidth]{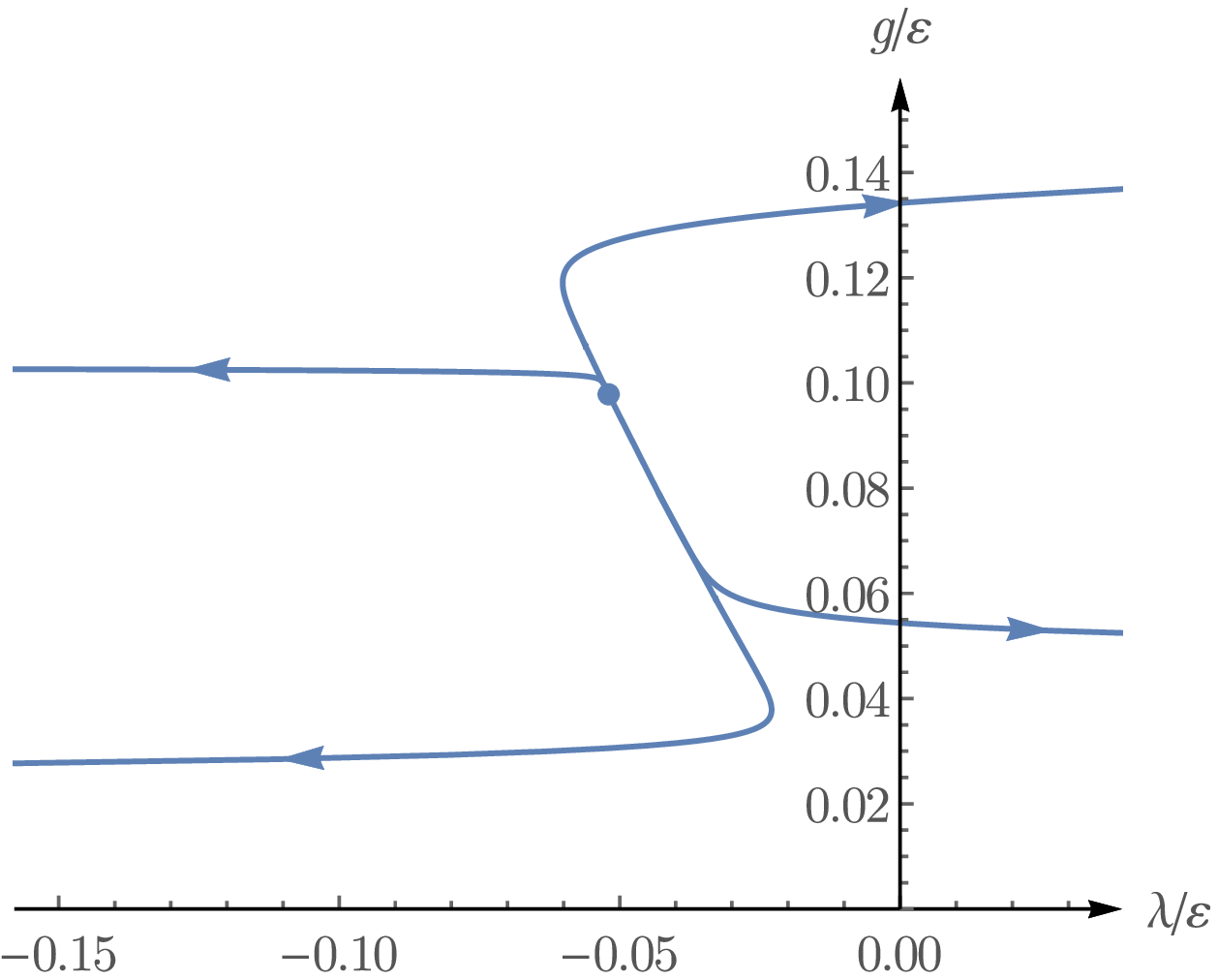}
\end{minipage}

\vspace{2.5em}
\begin{minipage}{0.47\columnwidth}
 \centering
 \small$\ve=0.05$\\
 \includegraphics[width=\columnwidth]{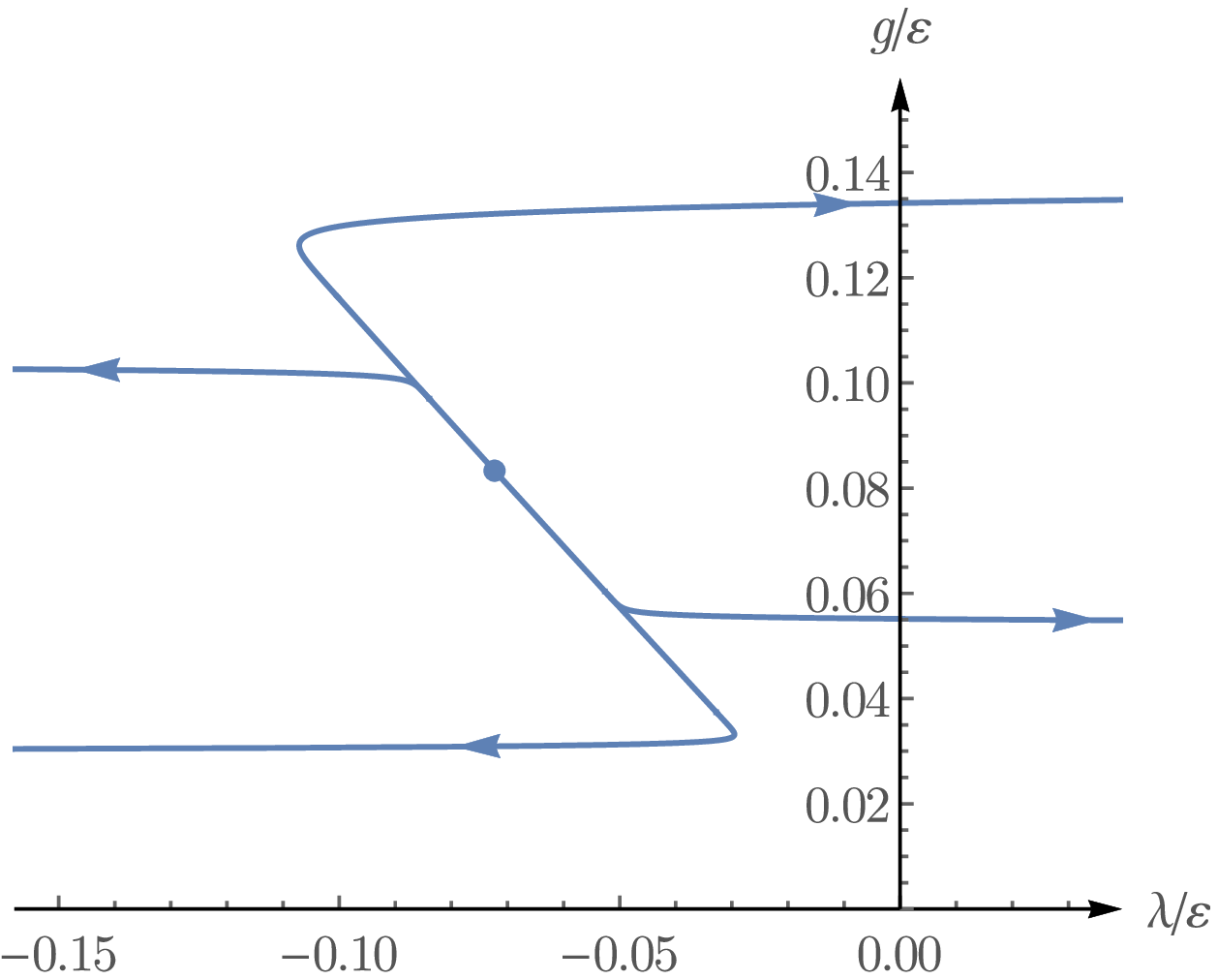}
\end{minipage}
\hfill
\begin{minipage}{0.47\columnwidth}
 \centering
 \small$\ve=0.005$\\
 \includegraphics[width=\columnwidth]{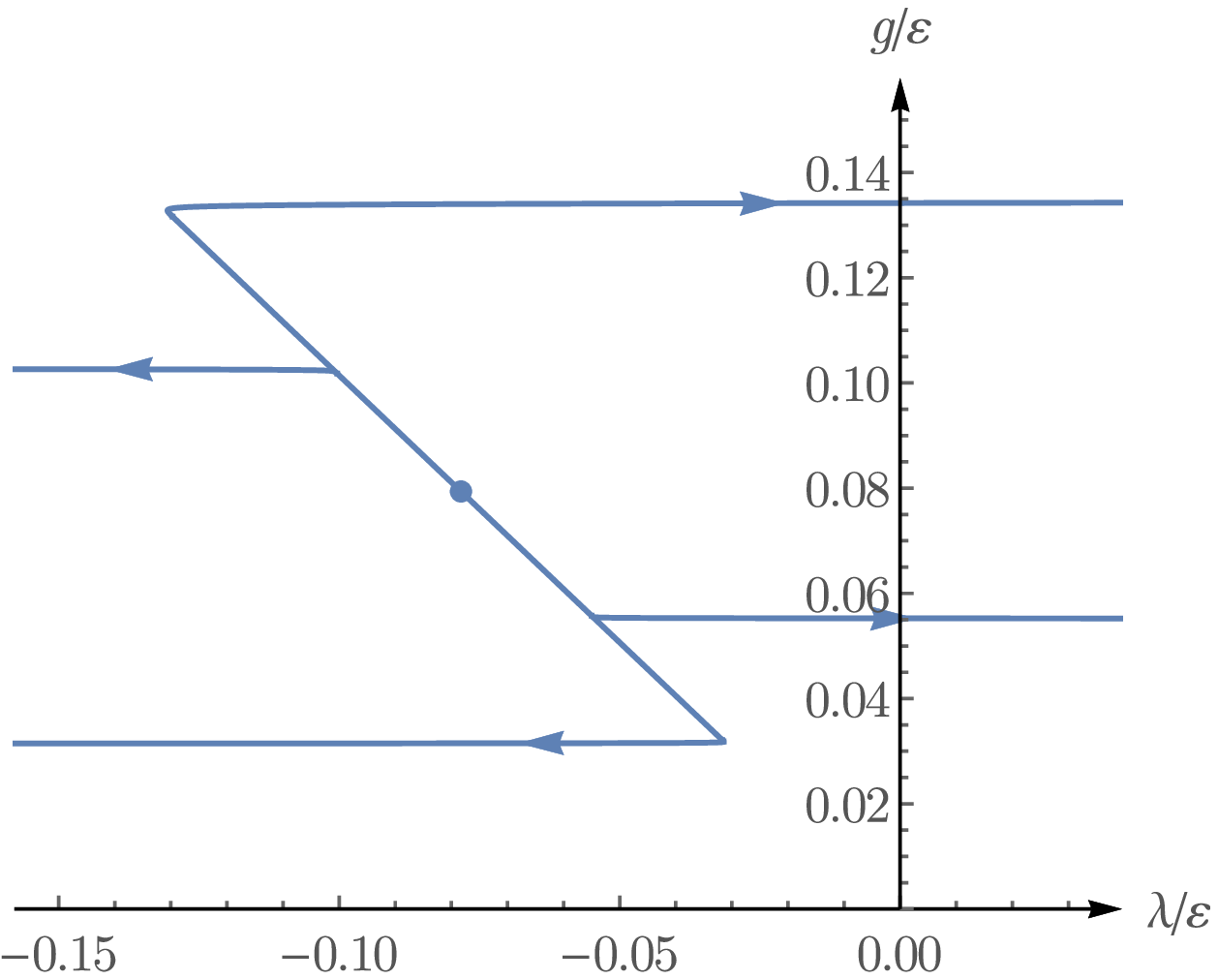}
\end{minipage}

\vspace{1em}
\caption{RG trajectories in the space of the normalized couplings $\rl \equiv \frac{\lambda}{\ve}$ and $\rg \equiv
 \frac{g}{\ve}$, based on the Einstein--Hilbert truncation in $d=2+\ve$ dimensions with the \emph{linear
 parametrization} and the optimized cutoff. Shown are the cases $\ve=0.35$, $\ve=0.2$, $\ve=0.05$ and $\ve=0.005$, with
 four sample trajectories for each diagram. Blue dots indicate UV fixed points.}
\label{fig:Flow2D}
\end{figure}

We observe that, while trajectories are still noticeably curved for $\ve$ sufficiently large, they approach straight
lines in the limit $\ve\to 0$, containing only one sharp bend: Let $ON$ denote the straight line through the origin and
the NGFP. Then, in the limit $\ve\to 0$ trajectories appear as \emph{perfect horizontal lines} at infrared and medium
scales, until they hit $ON$ as $k$ increases (i.e.\ following the inverse RG flow). There, at the crossing point, they
\emph{instantly change their direction}, from then on lying on top of $ON$ towards increasing RG scales, until they
\emph{finally run straightly into the fixed point} in the UV limit. Thus, they may be described as zigzag lines with
one sharp bend each. This result is quite remarkable, particularly with regard to the fact that in terms of the
unnormalized couplings the non-Gaussian fixed point collapses into the Gaussian one for $\ve\to 0$, and the
corresponding flow diagram loses its characteristic structure.

We would like to point out that the singularity line, present in the 4D diagram shown in Figure \ref{fig:StdSingle}, is
shifted to infinity for the normalized couplings when the limit $\ve\to 0$ is taken, so trajectories are well
defined at all scales.

In conclusion, we have seen that the RG flow diagrams in $d=2+\ve$, based on the linear parametrization and normalized
couplings, approach a rigid structure in the small $\ve$ limit, featuring a non-Gaussian fixed point at $\rgs=3/38$.

%----------------------------------------------------------------------------------------------------------------------
\subsection{The exponential parametrization in general dimensions}
\label{sec:singleExpd}
%----------------------------------------------------------------------------------------------------------------------

In this subsection and the two following ones, we investigate to what extent the above results pertaining to the linear
parametrization change when choosing the exponential parametrization instead. As argued in Section
\ref{sec:ParamDepFramework}, point \textbf{(1)}, the nonlinear character of the exponential parametrization entails
additional terms contributing to the Hessian of $\Gamma_k\mku$. The $\beta$-functions are obtained by a careful
analysis along the steps proposed in Section \ref{sec:Recipe}. While the calculation is performed in Appendix
\ref{app:single}, we focus on presenting results and consequences in the following.

For a general dimension $d$ the $\beta$-functions of the dimensionless couplings $g_k \equiv k^{d-2}G_k$ and
$\lambda_k \equiv k^{-2}\Lambda_k$ are given by equations \eqref{eq:beta_g_FRG} and \eqref{eq:beta_lambda_FRG}. Before
studying in detail their implications in $d=4$ and $d=2+\ve$ dimensions, an important remark concerning the appearance
of the cosmological constant is in order.

We have seen in Section \ref{sec:DetConnections}, in particular in eq.\ \eqref{eq:sgExpParam}, that the volume element
$\sg\mku$ is independent of the traceless part of the field $h_\mn\mku$: Upon splitting $h_\mn$ into trace and
traceless contributions, $h_\mn=\hat{h}_\mn+\frac{1}{d}\mku\bg_\mn\mku \phi$, with $\phi=\bg^\mn h_\mn$ and
$\bg^\mn\hat{h}_\mn=0$, we observe that the volume element depends only on $\phi$, while $\hat{h}_\mn$ drops out
completely: 
\begin{equation}
 \sg=\sbg\,\e^{\frac{1}{2}\phi} \,.
\label{eq:sgForExpParam}
\end{equation}
Hence, \emph{the cosmological constant can occur as a coupling only in the trace sector}. This is reflected both in the
Hessian of $\Gamma_k$, determined by eq.\ \eqref{eq:HessianTraceTraceless}, and in the $\beta$-functions: Those
contributions to $\beta_\lambda$ and $\beta_g$ that stem from the trace part involve threshold functions (cf.\ Appendix
\ref{app:Cutoffs}) of the form $\Phi_n^p(-\mu\lambda)$, while those originating from the traceless part contain only
threshold functions of the form $\Phi_n^p(0)$, see eqs.\ \eqref{eq:B1SingleExpParam} -- \eqref{eq:beta_lambda_FRG}.
This result is in distinction from the one for the linear parametrization where $\lambda$ occurred in both cases.

Another difference is given by the argument of the threshold functions: For the linear parametrization all threshold
functions that involve the cosmological constant are of the form $\Phi_n^p(-2\lambda)$ or $\tPhi_n^p(-2\lambda)$,
independent of the dimension $d$. For the exponential parametrization, on the other hand, they are replaced by
$\Phi_n^p(-\mu\lambda)$ and $\tPhi_n^p(-\mu\lambda)$, respectively, where $\mu\equiv \frac{2d}{d-2}$. This change turns
out to be particularly significant: All threshold functions become singular when their argument approaches $-1$. That
is, for the linear parametrization they have a pole at $\lambda=1/2$, while for the exponential parametrization the
pole is located at $\lambda=1/\mu$. This pole marks the starting point (at $g=0$, $\lambda=\frac{1}{2}$ or $\lambda=
\frac{1}{\mu}$) of the singularity line discussed in Section \ref{sec:singleLin4}. Since $\mu>2$, however, \emph{the
singularity line is shifted towards smaller values of $\lambda$ when the exponential parametrization is used}. We
expect to see this behavior in the corresponding flow diagrams, to be determined in the next section in the 4D case.

%----------------------------------------------------------------------------------------------------------------------
\subsection[The exponential parametrization in \texorpdfstring{$d=4$}{d=4} dimensions]%
{The exponential parametrization in \texorpdfstring{\bm{$d=4$}}{d = 4} dimensions}
\label{sec:singleExp4}
%----------------------------------------------------------------------------------------------------------------------

Let us consider the special case of four dimensions now. Inserting $d=4$ into the $\beta$-functions
\eqref{eq:beta_g_FRG} and \eqref{eq:beta_lambda_FRG} yields
\begin{align}
 \beta_g &= (2+\eta_N)\mku g \, ,
\label{eq:betagSingleExp4D}\\
 \beta_\lambda &= -(2-\eta_N)\lambda + \frac{g}{4 \pi } \Big[ 2\mku\Phi_2^1(-4 \lambda )+2\mku\Phi_2^1(0)
 -\eta_N \mku\tPhi_2^1(-4 \lambda ) - 9\mku\eta_N \mku\tPhi_2^1(0)\Big] ,
\label{eq:betalSingleExp4D}
\end{align}
where the anomalous dimension of Newton's constant, $\eta_N \equiv G_k^{-1}\mku k\mku\p_k G_k\mku$, is given by
\begin{equation}
 \eta_N = \frac{2\mku g \Big[\Phi_1^1(-4 \lambda)-3 \Phi_2^2(-4 \lambda )+\Phi_1^1(0)-21 \Phi_2^2(0)\Big]}%
 {12 \pi + g\Big[\tPhi_1^1(-4 \lambda )-3\mku\tPhi_2^2(-4 \lambda )+9\mku\tPhi_1^1(0)-9\mku\tPhi_2^2(0)\Big]}\,.
\end{equation}
The threshold functions, $\Phi_n^p(w)$, $\tPhi_n^p(w)$, are defined (and evaluated for several cutoff shapes) in
Appendix \ref{app:Cutoffs}. Due to the form of their arguments, $-4\lambda$, we find that they have a pole at $\lambda
=1/4$. Thus, the influence of the cutoff shape function on $\beta$-functions and fixed points might be increased
already at small $\lambda$ as compared with the situation for the linear parametrization where the pole lies at
$\lambda=1/2$. In the following we confirm this conjecture by considering global properties of the RG
flow for different shape functions.
\medskip

\noindent
\textbf{(1) Optimized cutoff.} An numerical evaluation of the $\beta$-functions \eqref{eq:betagSingleExp4D} and
\eqref{eq:betalSingleExp4D} gives rise to the flow diagram shown in Figure \ref{fig:NewSingleOpt}.

\begin{figure}[tp]
\centering
\includegraphics[width=.78\columnwidth]{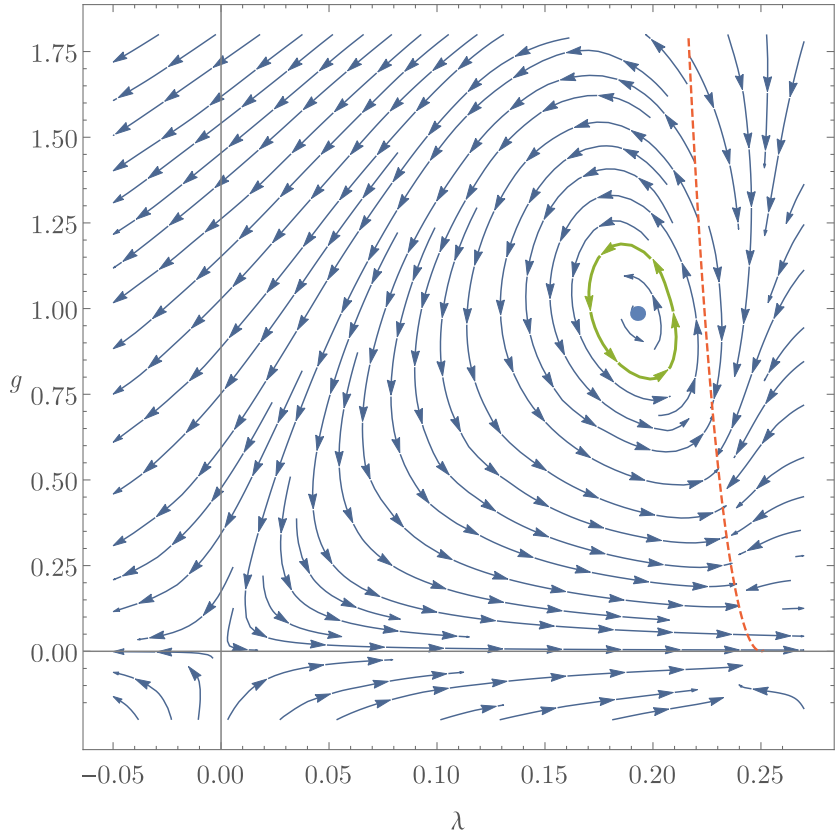}
\caption{Flow diagram for the Einstein--Hilbert truncation in $d=4$ based on the \emph{exponential parametrization} and
 the optimized cutoff. There is a limit cycle, indicated by the green arrows, whose inside contains a non-Gaussian
 fixed point (blue dot). The singularity line is shown as a red, dashed line. As usual, arrows point from the UV to the
 IR.}
\label{fig:NewSingleOpt}
\end{figure}

The result is fundamentally different from what is known for the linear pa\-ra\-me\-tri\-za\-tion (cf.\ Figure
\ref{fig:StdSingle}). Although we find again the Gaussian fixed point at the origin and a \emph{non-Gaussian fixed
point at positive $g$ and positive $\lambda$}, we encounter new properties of the latter. The NGFP is
\emph{UV-repulsive} in both directions now since its critical exponents have negative real parts. Furthermore, it is
surrounded by a \emph{closed limit cycle}. This limit cycle by itself is UV-attractive: Trajectories both inside and
outside approach the cycle for $k\to\infty$, unless they run into a singularity.

As expected, the singularity line (marked by the dashed, red curve in Figure \ref{fig:NewSingleOpt}), on which
$\beta$-functions diverge and beyond which the truncation ansatz is no longer reliable, has been shifted to smaller
values of $\lambda$. It prevents the existence of globally defined trajectories emanating from the limit cycle and
passing the classical regime, i.e.\ there is no connection between the limit cycle and the Gaussian fixed point.
Clearly, \emph{there cannot be a separatrix} either as the limit cycle ``shields'' its inside from its outside, not
allowing any crossing trajectories.

Trajectories inside the limit cycle may be considered asymptotically safe in a generalized sense since they approach
the cycle in the UV, while they hit the NGFP in the infrared. However, they can never reach a classical region, so they
cannot be realized in Nature. Note that the limit cycle is similar to those found in References \cite{HR12,DP13} which
are based on different but also nonlinear metric parametrizations.
\medskip

\noindent
\textbf{(2) Sharp cutoff.} Next, we repeat the analysis for the sharp cutoff. The corresponding flow diagram is shown
in Figure \ref{fig:NewSingleSharp}.
At first sight it seems to resemble the one of Figure \ref{fig:StdSingle} (pertaining to the linear parametrization and
the optimized cutoff) much more than the one of Figure \ref{fig:NewSingleOpt} (exponential parametrization and
optimized cutoff): Figure \ref{fig:NewSingleSharp} features the Gaussian and a \emph{non-Gaussian fixed point} as
previously, where the NGFP is \emph{UV-attractive} in both $g$- and $\lambda$-direction. In particular, there is
\emph{no limit cycle}.

\begin{figure}[tp]
\centering
\includegraphics[width=.78\columnwidth]{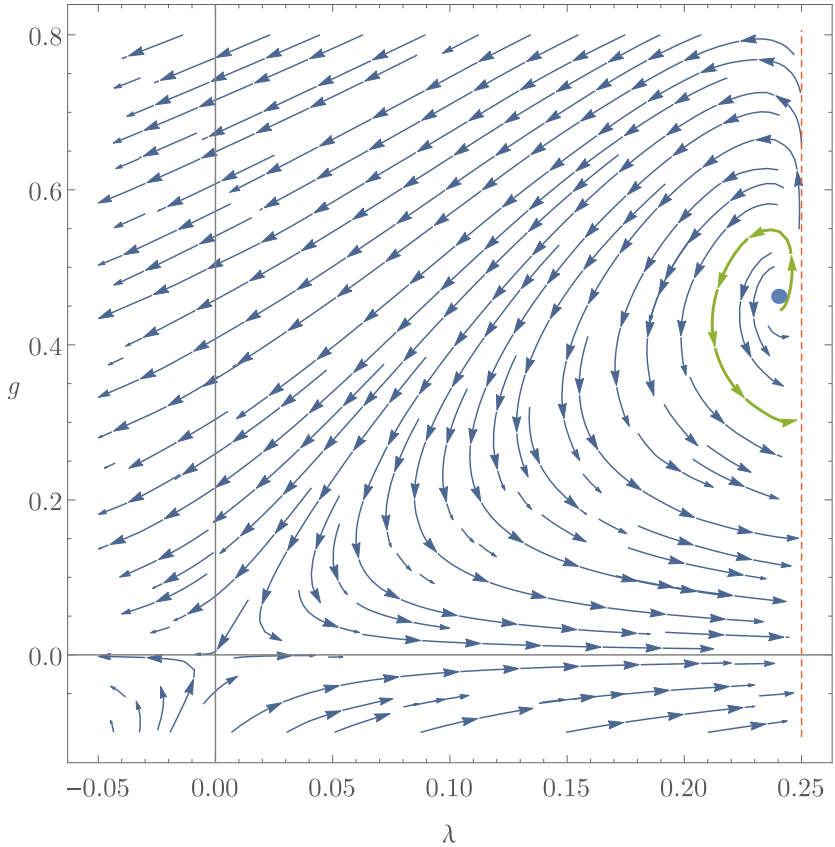}
\caption{Flow diagram for the Einstein--Hilbert truncation in $d=4$ based on the \emph{exponential parametrization} and
 the sharp cutoff. As indicated by the green arrows, all trajectories emanating from the NGFP (blue dot) run into the
 singularity line (red, dashed curve) towards the infrared so that they cannot come close to the Gaussian fixed point.}
\label{fig:NewSingleSharp}
\end{figure}

We observe an important difference between Figure \ref{fig:NewSingleSharp} and Figure \ref{fig:StdSingle}, though: Due
to the singularity line, \emph{there is no separatrix} in Figure \ref{fig:NewSingleSharp}, and hence, there is no
trajectory emanating from the NGFP that has a sufficiently extended classical regime close to the Gaussian fixed point.
This can be understood as follows.
The singularity line is too close to the NGFP such that all asymptotically safe trajectories eventually terminate at
some finite scale $k$ when going from the UV towards the IR, i.e.\ they run into the singularity line, and thus, they
have no chance to reach the vicinity of the Gaussian fixed point.
\medskip

\noindent
\textbf{(3) Exponential cutoff.} The exponential cutoff as introduced in Appendix \ref{app:Cutoffs} with generic values
of the parameter $s$ gives rise to a flow diagram that shares features with both Figure \ref{fig:NewSingleOpt} and
Figure \ref{fig:NewSingleSharp}. Here, we refrain from depicting diagrams for several $s$ since they do not provide
much further insight. We rather describe the result.

For cutoff parameters $s>0.93$ \emph{there exists an NGFP} at positive $g$ and positive $\lambda$. This fixed point is
\emph{UV-repulsive}, as it is for the optimized cutoff. However, this time there is \emph{no closed limit cycle}.
Although a relict of the cycle is still present, it does not form a closed line, but rather runs into the singularity
line. Again, there is \emph{no separatrix} connecting the fixed points. Varying $s$ amounts to shifting the coordinates
of the NGFP.

For $s\leq 0.93$ \emph{the fixed point even vanishes}, or, more precisely, it is shifted beyond the singularity,
leaving it \emph{inaccessible} by shielding it from trajectories that have a classical regime. Thus, the NGFP that
seemed to be indestructible for the linear parametrization can be made disappear with the exponential parametrization.
\medskip

In summary, some fundamental qualitative features of the RG flow like the signs of the real parts of critical
exponents, the existence of limit cycles, or the existence of suitable non-Gaussian fixed points seem to have a
stronger cutoff dependence when the exponential parametrization is used. None of the above flow diagrams corresponding
to the exponential parametrization contains a trajectory that describes a complete and consistent quantum theory, or to
put it another way, that can be realized in Nature. However, this conclusion holds true only within the scope of our
simplified setting which is based on the Einstein--Hilbert truncation (without field redefinitions, cf.\ Sec.\
\ref{sec:singleExpRemarks}) and a specific choice for the gauge. We will discuss in Section
\ref{sec:singleExpRemarks} that it is in fact the exponential parametrization that leads to the most reliable results
after all.

%----------------------------------------------------------------------------------------------------------------------
\subsection[The exponential parametrization in \texorpdfstring{$d=2+\ve$}{d = 2 + epsilon} dimensions]%
{The exponential parametrization in \texorpdfstring{\bm{$d=2+\ve$}}{d = 2 + epsilon} dimensions}
\label{sec:singleExp2}
%----------------------------------------------------------------------------------------------------------------------

Inserting $d=2+\ve$ into the $\beta$-functions \eqref{eq:beta_g_FRG} and \eqref{eq:beta_lambda_FRG} we find that there
is a non-Gaussian fixed point whose coordinates are of order $\ve$: $\mku\lambda_*=\mO(\ve)$, $g_*=\mO(\ve)$.
Thus, for all points $(\lambda,g)$ not too far away from the NGFP we have $\lambda=\mO(\ve)$ and $g=\mO(\ve)$, too.
This can be used to expand the $\beta$-functions in terms of $\ve$, yielding
\begin{align}
 \beta_g &= \ve g - b g^2 \,,
\label{eq:betag2d}\\
 \beta_\lambda &= -2\lambda + 2 g\Big[-2\mku\Phi_1^1(0)+\Phi_1^1\big(-{\textstyle\frac{4}{\ve}}\lambda\big)\Big],
\label{eq:betalambda2d}
\end{align}
up to higher orders in $\lambda$, $g$ and $\ve$. Here, the coefficient $b$ is given by
\begin{equation}
 b=\frac{2}{3}\Big[2\mku \Phi_0^1(0)+24\mku\Phi_1^2(0)-\Phi_0^1\big(-\textstyle\frac{4}{\ve}\lambda\big)\Big].
\label{eq:coeffbExp}
\end{equation}
Some of the threshold functions $\Phi_n^p$ appearing in \eqref{eq:coeffbExp} are independent of the underlying cutoff
shape function  $R^{(0)}(z)$: As specified in Appendix \ref{app:Cutoffs}, we have $\Phi_n^{n+1}(0)=1$ for any cutoff,
hence $\Phi_0^1(0)=1$ and $\Phi_1^2(0)=1$.

Furthermore, for all standard shape functions satisfying $R^{(0)}(z=0)=1$ we find $\Phi_0^1\big(
-\frac{4}{\ve}\lambda\big)=\big(1-\frac{4}{\ve}\lambda\big)^{-1}$. Due to the occurrence of $\ve^{-1}$ in the argument
of $\Phi_0^1$, the $\lambda$-dependence does not drop out of $\beta_g$ at lowest order. Rather, the combination
$\lambda/\ve$ results in a finite correction.

By contrast, the sharp cutoff \cite{RS02} does not fall into the class of standard cutoffs (cf.\ Appendix
\ref{app:Cutoffs}): It becomes infinitely large at vanishing argument, leading to the constant function $\Phi_0^1\big(
-\frac{4}{\ve}\lambda\big) = 1$ for all $\lambda$.\footnote{For the sharp cutoff, $\Phi^1_n(w)=-\frac{1}{\Gamma(n)}
\ln(1+w)+\vp_n$ is determined up to a constant $\vp_n$, which, for consistency, is chosen such that $\Phi^1_n(w=0)$
agrees with $\Phi^1_n(0)$ corresponding to some other cutoff \cite{RS02}, cf.\ Appendix \ref{app:Cutoffs}.
In the limit $n\rightarrow 0$, however, the $w$-dependence drops out completely, and $\Phi^1_0(w)^\text{sharp}=
\Phi^1_0(0)^\text{other}$. Since $\Phi^1_0(0)=1$ for any cutoff, we find $\Phi^1_0(w)^\text{sharp}=1 \;\, \forall w$.}

Collecting the above results, we find
\begin{equation}
 b = \begin{cases}
 \,\frac{2}{3}\Big[26 - \big(1-\frac{4}{\ve}\lambda\big)^{-1} \Big] \qquad\; &\text{for all standard cutoffs},\\[0.8em]
 \,\frac{50}{3} &\text{for the sharp cutoff}.
\end{cases}
\label{eq:bcoeff}
\end{equation}
Note that even if $b$ has the same form for all standard cutoffs, it does not give rise to a universal fixed point
coordinate. This can be seen as follows: The threshold functions of the type $\Phi_1^1(w)$ occurring in eq.\
\eqref{eq:betalambda2d} are cutoff dependent everywhere, even at $w=0$. Hence, $\beta_\lambda$ inevitably depends on
the cutoff shape, and so does $\lambda_*\mku$. Since $b$ depends on $\lambda_*$ in turn, its value at the fixed point
is not universal. As a consequence, both $\lambda_*$ and $g_*$ depend on the cutoff shape function.

In order to calculate critical central charges as in Section \ref{sec:singleLin2}, we include the matter action
\eqref{eq:matter} in the ansatz for the EAA, amounting to $N$ minimally coupled scalar fields in addition.
In this case, the $\beta$-functions are given by eqs.\ \eqref{eq:beta_g_FRG_N} and \eqref{eq:beta_lambda_FRG_N}.
Again, an expansion in terms of $\ve$ yields $\beta_g = \ve g - b g^2$ up to higher orders, where the coefficient
$b$ is changed into
\begin{equation}[b]
\; b = \begin{cases}
 \,\frac{2}{3}\Big[26 - \big(1-\frac{4}{\ve}\lambda\big)^{-1} -N\Big] \qquad\; &\text{for all standard cutoffs},\\[0.8em]
 \,\frac{2}{3}\big[25-N\big] &\text{for the sharp cutoff}.
\end{cases}
\label{eq:bcoeffN}
\end{equation}
As discussed in Section \ref{sec:ParamDepIntro}, the gravitational central charge is given by $c_\text{grav} =
\frac{3}{2}\,b$. The critical value of $N$, determined by the zero of $c_\text{grav}$ at the NGFP, can be computed for
different cutoff shape functions now.

Before considering the general case, we would like to compare our result to the perturbative one, specified in eq.\
\eqref{eq:ccritNewPert}. To this end, we have to set $\lambda=0$ by hand in \eqref{eq:bcoeffN} since the perturbative
studies that led to \eqref{eq:ccritNewPert} did not take into account the impact of the cosmological constant on the
$\beta$-function of the Newton constant \cite{KKN93a,KKN93b,KKN93c,KKN96,AKKN94,NTT94,AK97}.
As a result, eq.\ \eqref{eq:bcoeffN} boils down to
\begin{equation}[b]
 c_\text{grav} = 25-N \qquad \text{for all cutoffs if }\lambda = 0\,.
\end{equation}
Hence, we obtain the critical value $\ccr=N^\text{crit}=25$, \emph{reproducing the critical central charge of the
matter sector that was found perturbatively}.

If, however, the cosmological constant is not set to zero by hand, the cutoff dependent fixed point value $\lambda_*$
enters the coefficient $b$ for all standard cutoffs, according to eq.\ \eqref{eq:bcoeffN}. Thus, the critical central
charge depends on the cutoff shape in this case. We confirm these general arguments by evaluating the threshold
functions numerically for various cutoff shape functions (cf.\ Appendix \ref{app:Cutoffs}) and computing the
corresponding fixed point coordinates. Specifically, we obtain $\lambda_*\approx -0.0729$ for the optimized cutoff,
$\lambda_*\approx -0.1226$ for the sharp cutoff, $\lambda_*\approx -0.1426$ for the exponential cutoff with $s=0.5$,
$\lambda_*\approx -0.1187$ for the exponential cutoff with $s=1$, $\lambda_*\approx -0.0892$ for the exponential cutoff
with $s=5$, and $\lambda_*\approx -0.0806$ for the exponential cutoff with $s=20$. These numbers lead to the critical
central charges listed in Table \ref{tab:ccrit}, the main result of this subsection. We observe that although the value
of $\ccr$ is not universal, it is close to $25$ for all cutoffs considered. As seen above, the number $25$ becomes an
exact and universal result when the cosmological constant is left aside, making contact to the CFT result.

{\renewcommand{\arraystretch}{1.2}
\begin{table}[tp]
\centering
\begin{tabular}{cc}
 \hline
 Cutoff shape & $\ccr$ \\
 \hline
 $\quad$Any cutoff, but setting $\lambda=0 \quad$ & $25$ \\
 Optimized cutoff & $\quad 25.226 \quad$ \\
 Sharp cutoff & $25$ \\
 Exponential cutoff ($s=0.5$) & $25.363$ \\
 Exponential cutoff ($s=1$) & $25.322$ \\
 Exponential cutoff ($s=5$) & $25.263$ \\
 Exponential cutoff ($s=20$) & $25.244$ \\
 \hline
\end{tabular}
\caption{Cutoff dependence of the critical central charge for the exponential parametrization. (In case of the linear
 parametrization we had $\ccr=19$ for all cutoff shapes.)}
\label{tab:ccrit}
\end{table}%
}%

At last, we want to visualize the RG flow corresponding to the full (nonexpanded) $\beta$-functions
\eqref{eq:beta_g_FRG} and \eqref{eq:beta_lambda_FRG} in $d=2+\ve$ dimensions for several values of $\ve$. As in Section
\ref{sec:singleLin2}, we employ the normalized couplings
\begin{equation}
 \rl \equiv \lambda/\ve\,,\qquad \rg \equiv g/\ve\,,
\end{equation}
which lead to finite fixed point values, $\rls$ and $\rgs$, respectively, when the limit $\ve\to 0$ is taken. The 
associated RG trajectories are illustrated in Figure \ref{fig:FlowExp2D}, showing four diagrams at different values of
$\ve$ with four sample trajectories each.

\begin{figure}[tp]
\begin{minipage}{0.47\columnwidth}
 \centering
 \small$\ve=0.35$\\
 \includegraphics[width=\columnwidth]{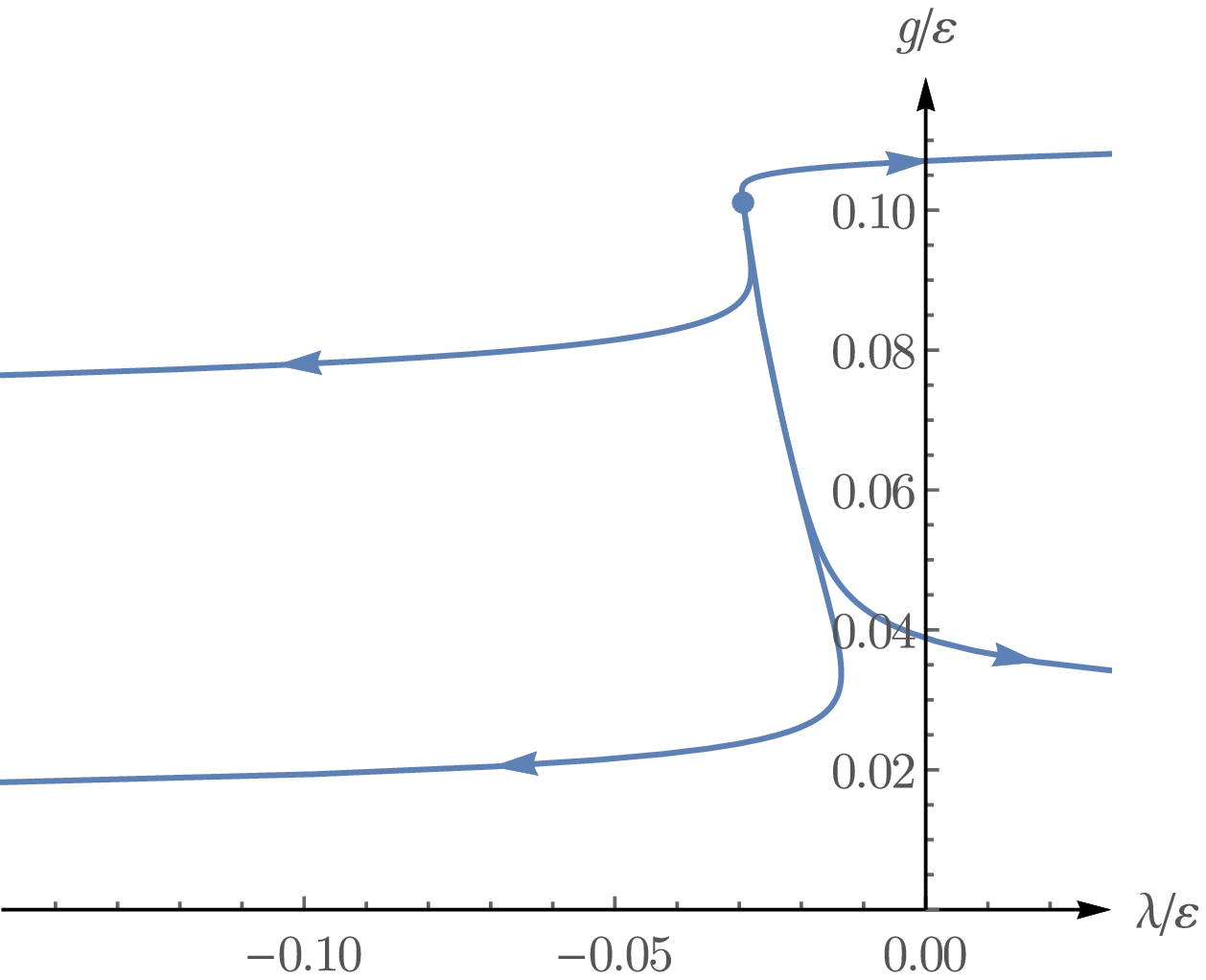}
\end{minipage}
\hfill
\begin{minipage}{0.47\columnwidth}
 \centering
 \small$\ve=0.2$\\
 \includegraphics[width=\columnwidth]{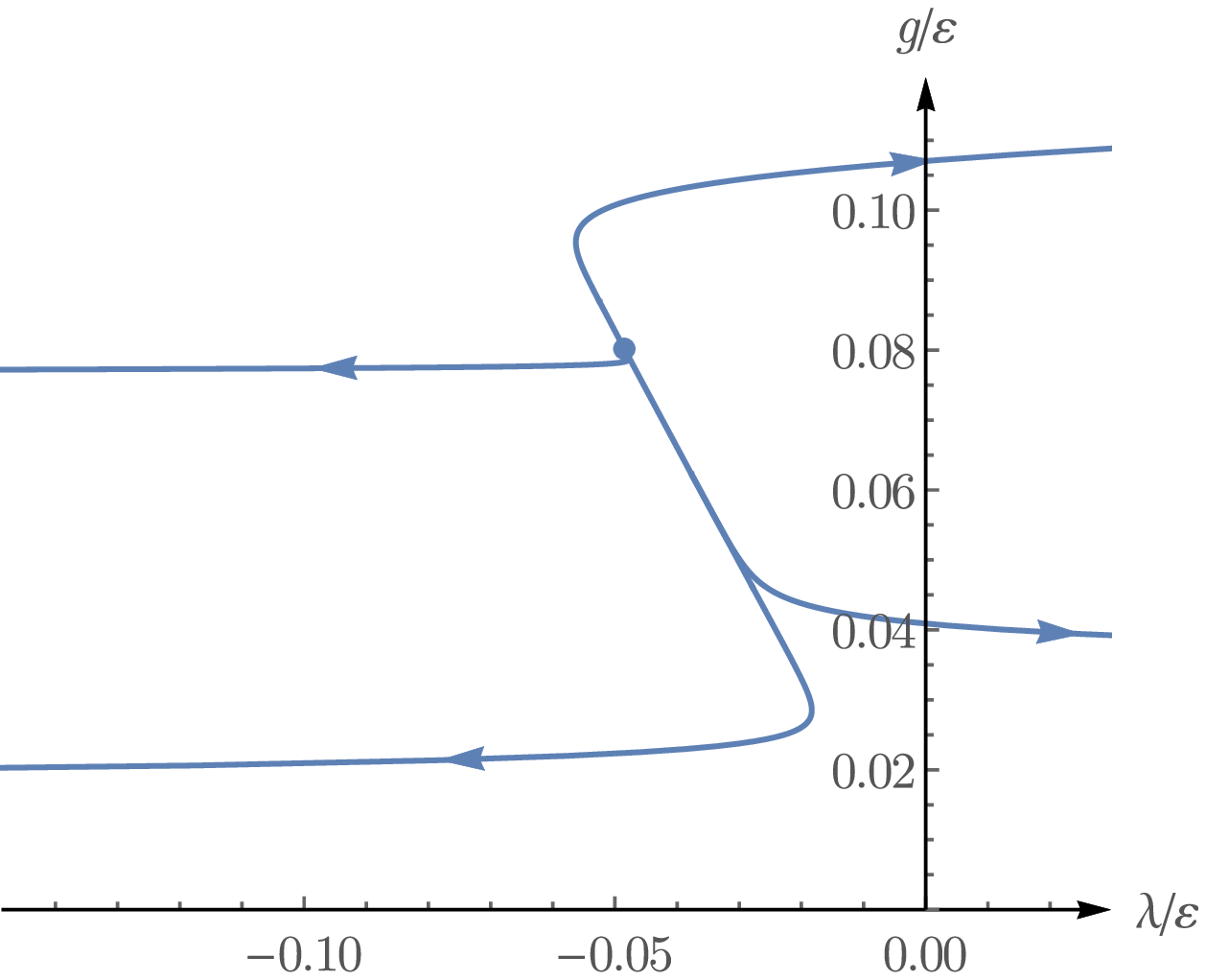}
\end{minipage}

\vspace{2.5em}
\begin{minipage}{0.47\columnwidth}
 \centering
 \small$\ve=0.05$\\
 \includegraphics[width=\columnwidth]{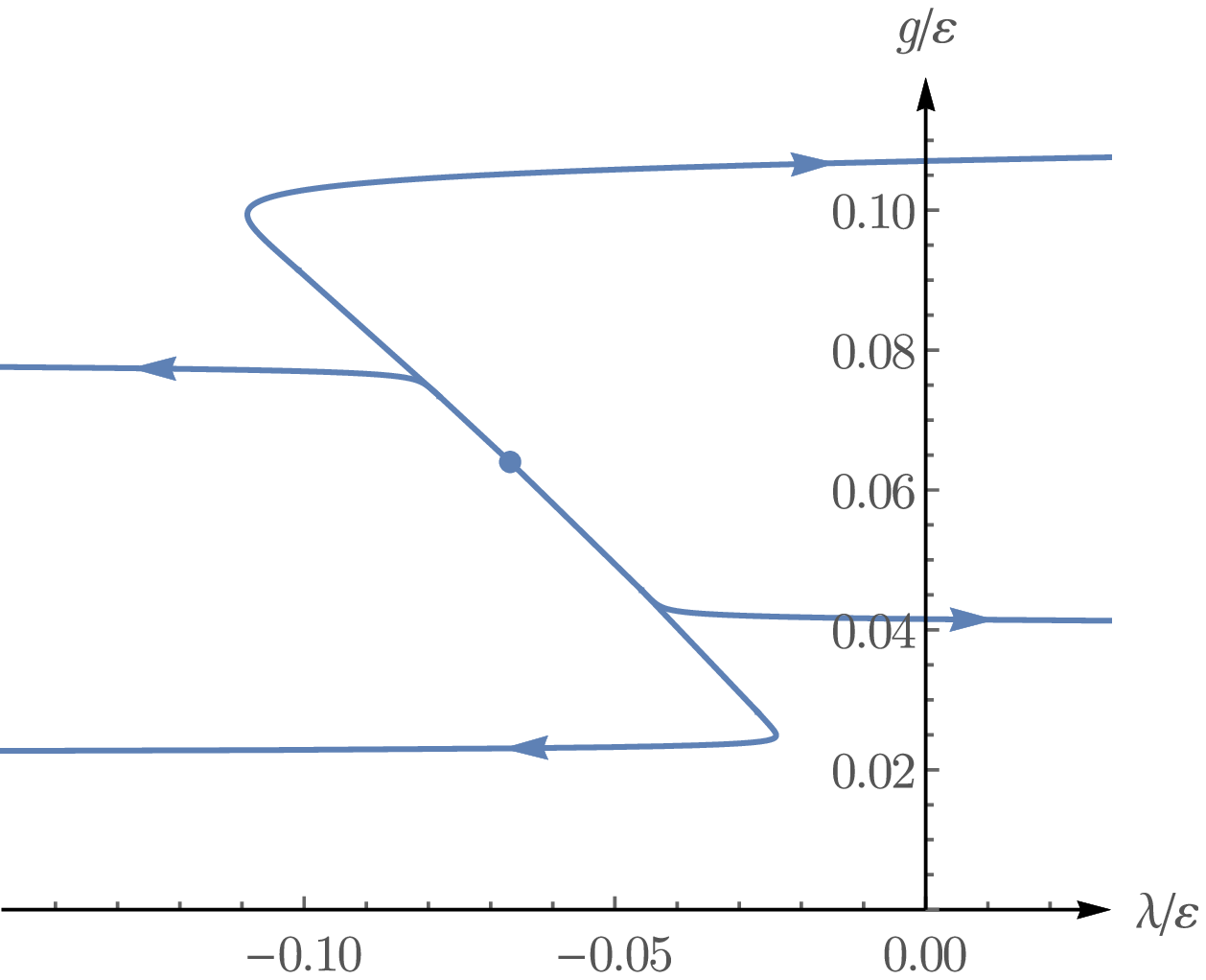}
\end{minipage}
\hfill
\begin{minipage}{0.47\columnwidth}
 \centering
 \small$\ve=0.005$\\
 \includegraphics[width=\columnwidth]{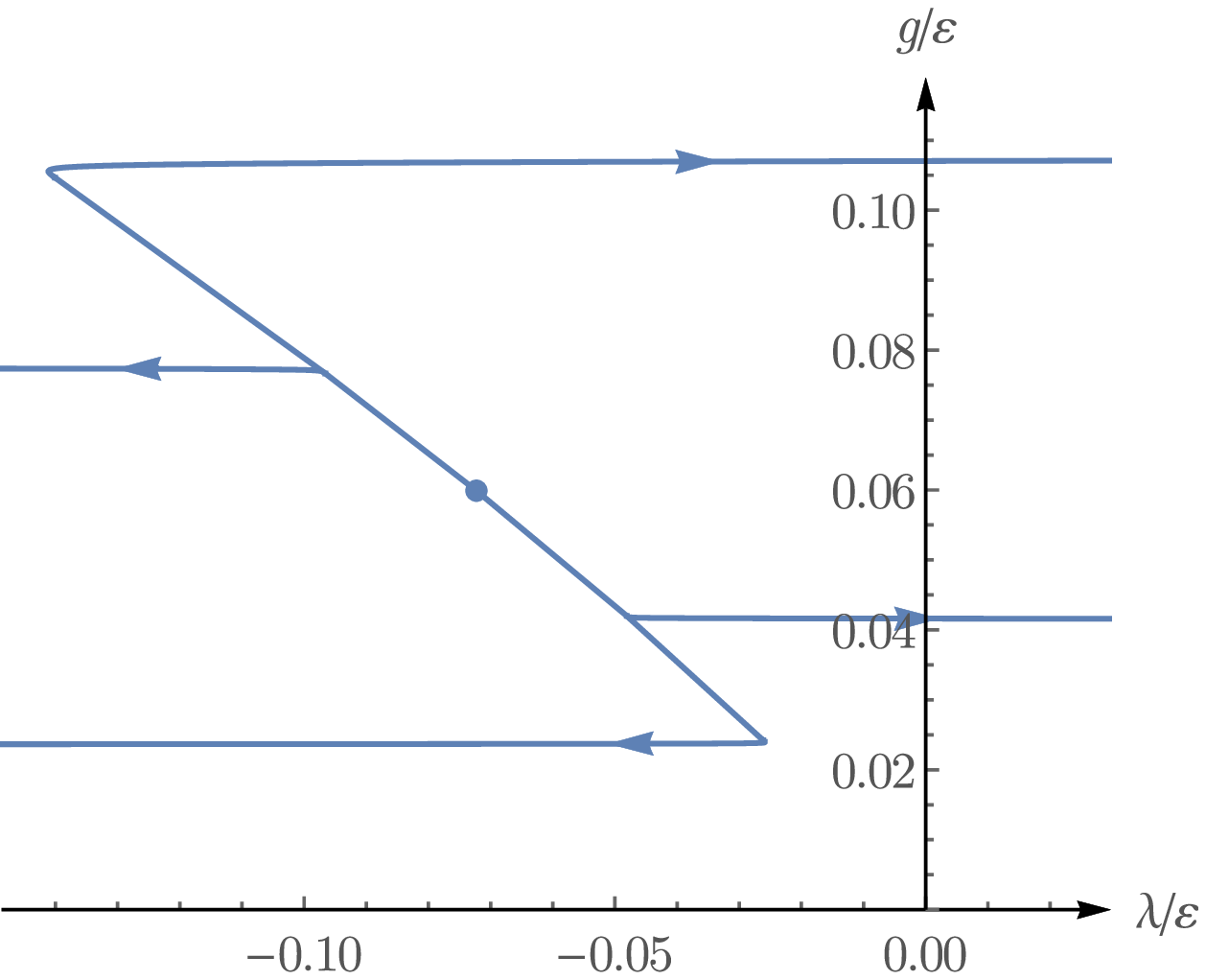}
\end{minipage}

\vspace{1em}
\caption{RG trajectories in the space of the normalized couplings $\rl \equiv \frac{\lambda}{\ve}$ and $\rg \equiv
 \frac{g}{\ve}$, based on the Einstein--Hilbert truncation in $d=2+\ve$ dimensions with the \emph{exponential
 parametrization} and the optimized cutoff. As in Figure \ref{fig:Flow2D}, we show the cases $\ve=0.35$, $\ve=0.2$,
 $\ve=0.05$ and $\ve=0.005$. In the limit $\ve\to 0$ a rigid zigzag structure is approached.}
\label{fig:FlowExp2D}
\end{figure}

It is remarkable how much Figure \ref{fig:Flow2D} (linear parametrization) and Figure \ref{fig:FlowExp2D} (exponential
parametrization) resemble each other. They both feature a \emph{UV-attractive non-Gaussian fixed point}
(at slightly different positions as the numerical values of the coordinates have changed). Furthermore, the structure
the diagrams approach in the limit $\ve\to 0$  is very similar for the two parametrizations: In the infrared,
trajectories appear as horizontal lines which become perfectly straight for $\ve\to 0$. Once these lines hit the
connecting line through the origin and the NGFP, they instantly change their direction, now heading straightly towards
the NGFP for increasing RG scale. In the UV limit they finally approach the NGFP. Thus, following the RG flow direction
(from high to low scales) each trajectory becomes a zigzag ray starting at the NGFP in the UV, having one sharp bend at
intermediate scales, and proceeding indefinitely in the IR. Like for the linear parametrization, the singularity line
present in Figures \ref{fig:NewSingleOpt} and \ref{fig:NewSingleSharp} is shifted to infinity in Figure
\ref{fig:FlowExp2D} in the limit $\ve\to 0$,\footnote{The mechanism of removing the singularity line is different for
the exponential parametrization, though. In the case of the linear parametrization, the singularity line has a zero at
$\lambda=1/2$ because of the involvement of $\Phi_n^p(-2\lambda)$. In terms of normalized couplings this is shifted to
$\rl=1/(2\mku\ve)\to\infty$ for $\ve\to 0$. Since $g$ is rescaled, too, $\rg\equiv g/\ve$, the line itself is scaled
upwards to $\rg=\infty$. For the exponential parametrization, on the other hand, there are threshold functions of
the form $\Phi_n^p(-4\lambda/\ve)$ leading to a pole (which is a zero of the singularity line at the same time) at
$\lambda=\ve/4$. In terms of normalized couplings this pole is located at $\rl=1/4$ for all $\ve$, i.e.\ it is not
shifted to infinity for $\ve\to 0$. However, the $\beta$-functions are such that all divergent contributions of the
threshold functions in combination actually converge to a finite limit. Thus, effectively there is no singularity when
$\rl$ passes the point $\rl=1/4$. For $\rl\neq 1/4$, the coordinates of all points with potentially divergent
$\beta$-functions are again scaled to $\rg=\infty$ due to the rescaling $\rg\equiv g/\ve$.
} and trajectories in the $(\rl,\rg)$-space are well defined at all scales.
\medskip

To sum up Subsections \ref{sec:singleExp4} and \ref{sec:singleExp2}, we recovered many results known for the linear
parametrization, like the existence of a non-Gaussian fixed point. The stronger cutoff dependence observed for the
exponential parametrization seems to indicate that the corresponding results are less reliable. However, there are two
points in favor of the exponential parametrization: (i) It reproduces the correct value of the critical central charge,
$\ccr=25$, known from conformal field theory. (ii) The high cutoff dependence is mainly due to the closer singularity
line which is believed to be merely a truncation artifact \cite{RW04}. Hence, using extended truncations, different
gauge choices and/or field redefinitions will most probably lead to more stable results. We will argue in the next
subsection that it is actually the exponential parametrization that features a higher reliability after all.

%----------------------------------------------------------------------------------------------------------------------
\subsection{Remark about recent results}
\label{sec:singleExpRemarks}
%----------------------------------------------------------------------------------------------------------------------

The results presented in this chapter (and published in Ref.\ \cite{Nink2015}) have triggered a couple of follow-up
investigations concerning the exponential metric parametrization
\cite{CD15,PV15,Falls2015a,Falls2015b,DN15,OP16,OPV15,GKL15,OPV16,DELP16,OPP16}. Here, we would like to briefly review
two recent contributions, Refs.\ \cite{GKL15} and \cite{OPP16}.
\medskip

\noindent
\textbf{(1)}
The idea behind Ref.\ \cite{GKL15} is based on the principle of minimum sensitivity, which is applied as follows. The
critical exponents $\theta_i$ should be universal quantities. Also, it is believed that the product $g_*\lambda_*$ is
physically observable and thus universal \cite{LR02a}. Therefore, testing the cutoff and gauge dependence of
$\theta_i$ and $g_*\lambda_*$ constitutes a quantitative criterion for the reliability of approximate results.
This test can be applied to any parametrization now. To this end, the authors of Ref.\ \cite{GKL15} exploit that the
difference between the linear and the exponential parametrization originates entirely from the second order term in an
expansion of $g_\mn\mku$: Recalling that $g_\mn^\text{exp} \equiv \bg_{\mu\rho}\big(\e^h\big)^\rho{}_\nu=\bg_\mn+h_\mn
+\frac{1}{2}\mku h_{\mu\rho}h^\rho{}_\nu+\mO(h^3)$, we can introduce the general parametrization
\begin{equation}
 g_\mn = \bg_\mn+h_\mn+\frac{\tau}{2}\mkuu h_{\mu\rho}h^\rho{}_\nu \,.
\end{equation}
Up to quadratic order, this expression interpolates smoothly between the linear pa\-ra\-me\-tri\-za\-tion ($\tau=0$)
and the exponential parametrization ($\tau=1$). Furthermore, a two-parameter family of gauge fixing actions is chosen:
The gauge condition \eqref{eq:GaugeCond} is generalized to $\mF_\alpha^\mn[\bg] = \delta^\nu_\alpha \,\bg^{\mu\rho}
\bar{D}_\rho - \frac{1+\beta}{d} \,\bg^\mn \bar{D}_\alpha$, and the parameter $\alpha$ appearing in eq.\
\eqref{eq:GaugeFixingAction} is not set to one this time but left arbitrary. Based on this approach, it can now be
tested for which value of $\tau$ the results for $\theta_i$ and $g_*\lambda_*$ exhibit the least dependence on $\alpha$
and $\beta$.

In addition to that, it is possible to study the influence of particular field redefinitions: The metric fluctuations
$h_\mn$ can be split according to the York decomposition into transverse traceless tensor modes, a transverse vector
mode and two scalar modes. This change of variables usually introduces Jacobians in the underlying functional integral.
Choosing a certain nonlocal field redefinition \cite{DP98,LR02a}, however, its associated Jacobians cancel against
those from the York decomposition, provided that a maximally symmetric background is considered. Since rigorous
arguments about the form of the fundamental variables of quantum gravity are still lacking, it is unclear whether or
not such a field redefinition should be used. Thus, the minimum sensitivity analysis described above is performed for
both original and redefined fields in Ref.\ \cite{GKL15}.

Without field redefinition, the characteristic variables $\theta_i$ and $g_*\lambda_*$ depend on the gauge parameters
to a much larger extent for the exponential parametrization ($\tau=1$) than for the linear one ($\tau=0$). Hence, the
exponential parametrization leads to less reliable results, confirming our observations of the previous subsections.

Employing a field redefinition, on the other hand, both parametrizations feature an extended range for the gauge
parameters that leads to very stable results. This indicates an even level of reliability.

Moreover, Ref.\ \cite{GKL15} contains an analysis with fixed gauge parameters but varying parameter $\tau$. The outcome
is quite remarkable: The most stable results are found for $\tau\approx 1.22$, which is clearly closer to $\tau=1$
corresponding to the exponential parametrization. The values of $\theta_i$ and $g_*\lambda_*$ for $\tau\approx 1.22$
are close to the ones found for $\tau=1$, while those for $\tau=0$ deviate considerably.

Finally, we would like to emphasize that there is one particularly suitable choice of the gauge parameter $\beta$. We
already know that the traceless sector of the metric fluctuations is independent of the cosmological constant if the
exponential parametrization is used. If we choose $|\beta|\to\infty$ now, the cosmological constant drops out of the
flow equations completely. In this case the $\beta$-function of the Newton coupling is independent of $\lambda$.
With regard to eq.\ \eqref{eq:bcoeffN} we obtain $b=\frac{2}{3}\big[25-N\big]$ \emph{for all cutoffs}, leading to the
universal gravitational central charge $c_\text{grav}=25-N$. Besides, in the limit $|\beta|\to\infty$ all results
become independent of $\alpha$.
\medskip

\noindent
\textbf{(2)}
In Ref.\ \cite{OPP16} the parametrization is generalized even further: The fundamental variable is not given by the
metric $g_\mn\mku$, but rather by a tensor density $\gamma_\mn$ of a certain weight, or even by some densitized inverse
metric $\gamma^\mn$. The relation between $g_\mn$ ($g^\mn$) and $\gamma_\mn$ ($\gamma^\mn$) is given by
\begin{equation}
 g_\mn = (\det(\gamma_\mn))^m\mkuu\gamma_\mn\,,\qquad g^\mn = (\det(\gamma_\mn))^{-m}\mkuu\gamma^\mn\,.
\end{equation}
Then $\gamma_\mn$ (and also $\gamma^\mn$) can be parametrized in different ways, the linear and the exponential
parametrization being special cases. Putting everything together and expanding the metric $g_\mn$ up to quadratic order
yields
\begin{equation}
 g_\mn = \bg_\mn + h_\mn + m\mku \bg_\mn h + \omega\mku h_{\mu\rho}h^\rho{}_\nu + m\mku h\mku h_\mn
 + m\left(\textstyle\omega-\frac{1}{2}\right)\bg_\mn h^{\alpha\beta}h_{\alpha\beta}
 + \textstyle\frac{1}{2}\mku m^2\bg_{\mu\nu}h^2 ,
\end{equation}
with $h\equiv \bg^\mn h_\mn\mku$. Here, the choice $\omega=0$ corresponds to the linear expansion of the metric,
$\omega=1/2$ corresponds to the exponential expansion, and $\omega=1$ corresponds to the linear expansion of the
inverse metric.

Based on these definitions, the dependence of the RG flow on $m$ and $\omega$ as well as on the gauge parameters
$\alpha$ and $\beta$ is investigated in \cite{OPP16}. It turns out that the exponential parametrization ($\omega=1/2$)
leads to the most stable results, which is reflected in an independence of $m$ in particular. The choice $\omega=1/2$
and $|\beta|\to\infty$ automatically eliminates all dependence on $m$, $\alpha$, and on the cosmological constant.
This is a very favorable situation since it reduces the amount of uncertainty of results considerably.
\medskip

In conclusion, we have seen that a simple modification of the gauge condition (by implementing the parameter
$\beta$ and considering the limit $\beta\to\pm\infty$) and/or a field redefinition can substantially increase the
degree of reliability of the results obtained with the exponential parametrization.

%----------------------------------------------------------------------------------------------------------------------
\section{The birth of exponentials in 2D}
\label{sec:Birth}
%----------------------------------------------------------------------------------------------------------------------

We emphasize that the above results do not imply any statements about the ``correctness'' of certain parametrizations.
For the time being, it is not clear whether the exponential and the linear parametrization, respectively, describe the
same physics at the exact level. As argued in Chapter \ref{chap:SpaceOfMetrics}, the former gives rise to pure metrics
only, while the latter includes degenerate, wrong-signature and vanishing tensor fields.\footnote{The latter would be
in the spirit of Ref.\ \cite{Witten1988}, and one might expect to find a phase of unbroken diffeomorphism invariance,
among others.} We cannot fully exclude the possibility that both of them are equally correct, but probe instead two
\emph{different universality classes}. If so, we conjecture that these classes would then be represented by
$c_\text{grav}=25$ for the exponential parametrization (in the pure gravity case) and by $c_\text{grav}=19$ for the
linear one.

But why is it the former choice that reproduces the results of standard conformal field theory, while the latter one
fails to do so? In the following we will argue that the exponential parametrization is a particularly appropriate
choice in the 2D limit. More precisely, we will see that there is a \emph{distinguished parametrization} in any
dimension $d$ which approaches an exponential form as $d\to 2$. Although this does not mean that the exponential
parametrization should be preferred over the linear one in general, we can at least understand its compatibility with
2D conformal field theory. In any case, the issue of parametrization dependence should always be reconsidered when a
better truncation becomes technically manageable.\footnote{A first indication pointing towards the possibility of
different universality classes might be contained in recent results from the $f(R)$-truncation in 4D where an
apparently parametrization dependent number of relevant directions was observed \cite{OPV15,OPV16}.}

The argument presented in this Section (cf.\ Ref.\ \cite{NR16a}) considers only such dynamical metrics $g_\mn$ that are
conformally related to a fixed reference metric $\hg_\mn$, and only their relative conformal factor is quantized. The
resulting ``conformally reduced'' setting \cite{RW09a,RW09b} amounts to the exact theory in 2D, but it is an
approximation in higher dimensions. Accordingly, ``exponential parametrization'' refers to the form of the conformal
factor in the following. Now, among all possible ways of parametrizing the conformal factor there exists one
distinguished choice in each dimension $d$.

\medskip
\noindent
\textbf{(1) Distinguished parametrizations.} Let us consider the conformal reduction of the Einstein--Hilbert action
$\SEH[g]\equiv -\frac{1}{16\pi G}\int\dd x\sg\,(R-2\Lambda)$ in any number of dimensions $d>2$. That is, we evaluate
$\SEH$ only on metrics which are conformal to a given $\hg$ consistent with the desired topology. But how should we
write the factor relating $g$ and $\hg$ now? Assume, for instance, the reduced $\SEH$ plays the role of a bare action
under a functional integral over a certain field $\Omega$ representing the conformal factor, how then should the latter
be written in terms of $\Omega\,$? Clearly, infinitely many parametrizations of the type $g_\mn = f(\Omega)\mku\hg_\mn$
are possible here, and depending on our choice the reduced $\SEH$ will look differently.

There exists a distinguished parametrization, however, which is specific to the dimensionality $d$, having the property
that $\int\!\sg\,R$ \emph{becomes quadratic in} $\Omega$. Starting out from a power ansatz, $g_\mn=\Omega^{2\nu}\,
\hg_\mn$, the integral $\int\!\sg\,R$ will in general produce a potential term $\propto \hR$ times a particular power
of $\Omega$, and a kinetic term $\propto \big( \hD\Omega \big)^2$ times another power of $\Omega$. The exponent of the
latter turns out to be zero, yielding a kinetic term quadratic in $\Omega$, precisely if \cite{JNP05}
\begin{equation}
 \nu = 2/(d-2)\,,\qquad g_\mn = \Omega^{4/(d-2)}\,\hg_\mn \,.
\label{eq:ExponentChoice}
\end{equation}
In this case, the potential term $\propto \hR$ is found to be quadratic as well, and one obtains \cite{JNP05,RW09a}
\begin{equation}[b]
\begin{aligned}
 &\SEH\big[g = \Omega^{4/(d-2)}\,\hg\big]\\[0.1em]
 &= -\frac{1}{8\pi G}\int\dd x\shg\left[ \frac{1}{2\,\xi(d)}\,\hD_\mu\Omega\,
 \hD^\mu \Omega + \frac{1}{2}\hR\,\Omega^2-\Lambda\,\Omega^{2d/(d-2)}\right].
\end{aligned}
\label{eq:SCREH}
\end{equation}
Here, we introduced the constant
\begin{equation}
 \xi(d) \equiv \frac{(d-2)}{4(d-1)}\,.
\end{equation}
Usually, one employs $\Omega(x)-1\equiv\omega(x)$ rather than $\Omega$ itself as the dynamical field that is
quantized, i.e.\ integrated over if $\SEH$ appears in a functional integral. Then there will be no positivity issues
as long as $\omega(x)$ stays small. We emphasize, however, that the derivation of neither \eqref{eq:SCREH} nor the
related action for $\omega$,
\begin{equation}
 \SEH[\omega;\hg] = -\frac{1}{8\pi G}\int\!\dd x\shg\left[ \frac{1}{2\,\xi(d)}\,\hD_\mu\omega\,\hD^\mu \omega
  + \frac{1}{2}\hR\,(1+\omega)^2-\Lambda\,(1+\omega)^{2d/(d-2)}\right],
\label{eq:omegaAction}
\end{equation}
involves any (small field, or other) expansion. (It involves an integration by parts, though, hence there could be
additional surface contributions if spacetime has a boundary.)

\medskip
\noindent
\textbf{(2) Metric operators.} The exponent appearing in the conformal factor $\Omega^{2\nu}$ is
noninteger in general, exceptions being $d=3,4$, and $6$, see Table \ref{tab:Dimensions}.
{\renewcommand{\arraystretch}{1.2}
\begin{table}[tp]
\centering
\begin{tabular}{cccc}
 \hline
 $d$  &  $\quad 3\quad$  &  $\quad 4\quad$  &  $\quad 6\quad$\\
 \hline
 $\;$ Conformal factor $\;$  &  $\Omega^4$  &  $\Omega^2$  &  $\Omega$\\
 Volume operator &  $\Omega^6$  &  $\Omega^4$  &  $\Omega^3$\\
 \hline
\end{tabular}
\caption{Conformal factor and volume operator for the distinguished parametrization.}
\label{tab:Dimensions}
\end{table}%
}%
The virtue of a quadratic action needs no mentioning, of course. As long as the cosmological constant plays no
role --- $\Lambda$ will always give rise to an interaction term --- the computation of the RG flow will be easiest and
\emph{most reliable} if we employ the distinguished parametrization.\footnote{The RG flow of
the conformally reduced Einstein--Hilbert truncation (``CREH'') with the distinguished parametrizations has been
computed in \cite{RW09a}, an LPA-type extension was considered in \cite{RW09b}, see also \cite{MP09}.}

One should be aware that there is a conservation of difficulties also here. Generically the conformal factor depends on
the quantum field \emph{nonlinearly}. Hence, canonically speaking, even if the action is trivial (Gaussian), the
construction of a \emph{metric operator} amounts to defining $\Omega^{2\nu}$ or $(1+\omega)^{2\nu}$ as a composite
operator. And in fact, the experience with models such as Liouville theory \cite{OW86,DO94,KN94} shows how extremely
difficult this can be.

At present, we are just interested in comparing the relative degree of reliability of two truncated RG flows, based upon
different field parametrizations. For this purpose it is sufficient to learn from the above argument that the ``most
correct'' results should be those from the distinguished parametrization \eqref{eq:ExponentChoice} since then the
theory is free (for $\Lambda=0$). But what is the distinguished parametrization in $2$ dimensions?

\medskip
\noindent
\textbf{(3) The limit \bm{$d\rightarrow 2$}}. As we lower $d=2+\ve$ towards two dimensions, the distinguished form of
the conformal factor, $(1+\omega)^{4/(d-2)}$, develops into a function which increases with $\omega$ faster than any
power. At the same time the constant $\xi(d)$ goes to zero, and \eqref{eq:omegaAction} becomes
\begin{equation}
\begin{split}
 \SEH[\omega;\hg] = -\frac{1}{16\pi \rG}\int\td^{2+\ve}x\shg\,\bigg[\frac{4}{\ve^2}\,\hD_\mu\omega\,\hD^\mu \omega\,
 \big\{1+\mO(\ve)\big\} & \\
  + \frac{1}{\ve}\,\hR\,(1+\omega)^2 - 2\rL\,(1+\omega)^{2(2+\ve)/\ve}& \bigg].
\end{split}
\label{eq:omegaAction2}
\end{equation}
Here we introduced normalized couplings again, $G\equiv \rG\,\ve$ and $\Lambda\equiv \rL\,\ve$, assuming that
$\rG,\rL=\mO(\ve^0)$. We see that in order to obtain a meaningful kinetic term we must rescale $\omega$ by a factor of
$\ve$ prior to taking the limit $\ve\searrow 0$.

Introducing the new field $\phi(x) \equiv 2\omega(x)/\ve$, its kinetic term $\hD_\mu\phi\,\hD^\mu \phi\,\big\{1+
\mO(\ve)\big\}$ will have a finite and nontrivial limit. The concomitant conformal factor
$\Omega^{2\nu}$ has the limit
\begin{equation}[b]
 \lim\limits_{\ve\rightarrow 0} \, (1+\omega)^{4/\ve}
 = \lim\limits_{\ve\rightarrow 0}  \Big(1+{\textstyle\frac{1}{2}}\mku\ve\mku\phi\Big)^{4/\ve}
 = \lim\limits_{n\rightarrow\infty}  \Big(1+{\textstyle\frac{2\phi}{n}}\Big)^n = \e^{2\phi} \,.
\end{equation}
This demonstrates that \emph{the exponential parametrization} $g_\mn = \e^{2\phi}\hg_\mn$ \emph{is precisely the 2D
limit of the distinguished (power-like) parametrizations in} $d>2$.

The cosmological term in \eqref{eq:omegaAction2} involves the same exponential for $d\rightarrow 2$, and the originally
quadratic potential $\hR(1+\omega)^2$ turns into a linear one for $\phi$. Taking everything together the Laurent series
of $\SEH$ in $\ve$ looks as follows:
\begin{equation}
 \SEH[\phi;\hg] = -\frac{1}{16\pi\rG}\bigg\{ \frac{1}{\ve}\!\int\!\td^{2+\ve} x\shg\,\hR + \int\!\td^2 x\shg\,
 \Big(\hD_\mu\phi\,\hD^\mu \phi + \hR\,\phi - 2\rL\,\e^{2\phi}\Big)\bigg\} + \mO(\ve).
\label{eq:SEHphiAction}
\end{equation}
The first term on the RHS is $\phi$-independent and involves a purely topological contribution proportional to the
Euler characteristic, $\chi\equiv\frac{1}{4\pi}\int\td^2 x\sg\,R$, which will be discussed in more detail in Section
\ref{sec:IndGravityFromEH}. Obviously, from eq.\ \eqref{eq:SEHphiAction} we obtain \emph{Liouville theory as the
intrinsically 2D part of the Einstein--Hilbert action}, but this is perhaps not too much of a surprise (as will also be
seen in Chapter \ref{chap:EHLimit}).

What is important, though, is that in this derivation, contrary to the standard argument, the exponential field
dependence of the conformal factor was not put in by hand, we rather \emph{derived} it.

Here, our input were the following two requirements: First, the scaling limit of $\SEH$ should be both nonsingular and
nontrivial, and second, it should go through a sequence of actions which, apart from the cosmological term, are at most
quadratic in the dynamical field. Being quadratic implies that when $\SEH[\omega;\hg]$ is used as the (conformal
reduction of the) Einstein--Hilbert truncation, this truncation is ``perfect'' at any $\ve$.

Therefore, we believe that using the exponential parametrization already in slightly higher dimensions $d>2$ yields
more reliable results for the $\beta$-functions and their 2D limits than using the linear parametrization in $d>2$ and
taking the 2D limit of the corresponding $\beta$-functions afterwards.
(There is still a minor source of uncertainty due to the ghost sector. In either parametrization there are
ghost-antighost-graviton interactions which are not treated exactly by the truncations considered here.)

The basic difference between the two parametrizations can also be seen quite directly. If we insert $g=\e^{2\phi}\hg$
into $\SEH$, the resulting derivative term reads exactly, i.e.\ without any expansion in $\ve$ and/or $\phi$ and
rescaling of $\phi$:
\begin{equation}
 -\frac{(d-1)}{16\pi\rG}\int\dd x\shg\;\e^{(d-2)\phi}\big(\hD\phi\big)^2 \,.
\end{equation}
For $d\rightarrow 2$ this term has a smooth limit (we did use $G=\rG\,\ve$ after all) and this limit is quadratic in
$\phi$.

On the other hand, inserting the linear parametrization $g=(1+\omega)\mku\hg$ into $\SEH$ we obtain again exactly,
i.e.\ without expanding in $\ve$ and/or $\omega$ and rescaling $\omega$:
\begin{equation}
 -\frac{(d-1)}{64\pi\rG}\int\dd x\shg\;(1+\omega)^{(d-2)/2}\,\frac{\big(\hD\omega\big)^2}{(1+\omega)^2} \;.
\label{eq:KinTermOmega}
\end{equation}
The term \eqref{eq:KinTermOmega}, too, has a smooth limit $d\rightarrow 2$, but it is not quadratic in the dynamical
field. This renders the $\omega$-theory interacting and makes it a nontrivial challenge for the truncation.

\medskip
\noindent
\textbf{(4) The dimension \bm{$d=6$}}. As an aside we mention that according to Table \ref{tab:Dimensions} the case
$d=6$ seems to be easiest to deal with since in the preferred field parametrization the conformal factor is linear in
the quantum field, and so there is no need to construct a composite operator. The kinetic term \eqref{eq:KinTermOmega}
becomes quadratic exactly at $d=6$.

It is intriguing to speculate that this observation is related to the following rather surprising property enjoyed by
the $\beta$-functions derived from the bimetric Einstein Hilbert truncation (see Appendix A.1 of Ref.\ \cite{BR14}):
If $d=6$, and if in addition the dimensionful dynamical cosmological constant $\Lambda^\text{Dyn}$ is zero, then
\emph{the gravity contributions to the $\beta$-functions of both $\Lambda^\text{Dyn}$ and the dimensionful dynamical
Newton constant $G^\text{Dyn}$ vanish exactly}. (There are nonzero ghost contributions, though.)

\medskip
\noindent
\textbf{(5) Summary}. On the basis of the above arguments we conclude that \emph{most probably the exponential
pa\-ra\-me\-tri\-za\-tion is more reliable in 2D than the linear one}. We believe in particular that $c_\text{grav}=25$
is more likely to be a correct value of the central charge at the pure gravity fixed point than its competitor `19'.
Depending on the reliability of the linear parametrization, the `19' could be a poor approximation to `25', or a hint
at another universality class.

%----------------------------------------------------------------------------------------------------------------------
\section{RG analysis for a bimetric truncation}
\label{sec:bi}
%----------------------------------------------------------------------------------------------------------------------

As argued above, the full effective average action $\Gamma_k$ is inherently a functional of \emph{two} metrics, $g_\mn$
and $\bg_\mn$. Hence, unless further conditions (e.g.\ a single-metric truncation) are imposed on an ansatz for
$\Gamma_k$, it can contain all kinds of invariants: those constructed out of $g_\mn$ alone, out of $\bg_\mn$ alone, or
out of mixed terms like $\int\dd x\sbg\,R$, $\int\dd x\sg\,\bR$, etc. Truncations which do not involve the
identification $g_\mn=\bg_\mn$ but keep both metrics separately are referred to as bimetric \cite{MR10,MRS11a,MRS11b}.
Being more general, it can be expected that a bimetric truncation of a given order (of derivatives, for instance) is a
better approximation to the exact EAA than a single-metric truncation of the same order.

At the technical level, calculations become more complex in the bimetric case, and the standard approach for deriving
$\beta$-functions, introduced in Section \ref{sec:Recipe}, is no longer applicable: The Hessian $\Gamma_k^{(2)}$
w.r.t.\ the dynamical field can contain all kinds of second order derivative operators like $\Box$,
$\bB$, $D_\mu \bD^\mu$, and even uncontracted ones like $\bD_\mu D_\nu$, and so forth. Thus, employing the standard
recipe, which is based on a heat kernel expansion and relies on the occurrence of only one type of covariant derivative
(either $D_\mu$ or $\bD_\mu$), is not an option here. As yet, there are only a few approximate techniques at our
disposal that cope with this difficulty. Here, we employ the \emph{conformal projection technique} \cite{MRS11b}. It
consists in conformally relating the two metrics $g_\mn$ and $\bg_\mn$ as follows:
\begin{equation}
 g_\mn(x) = \e^{2\Omega}\mku\bg_\mn(x),
\label{eq:ConfPrRelation}
\end{equation}
where $\Omega$ is an $x$-independent number which can be used as a bookkeeping parameter. Since any metric
parametrization (including the linear and the exponential one) can be expanded as $g_\mn=\bg_\mn+h_\mn+\mO(h^2)$, and
since eq.\ \eqref{eq:ConfPrRelation} implies $g_\mn=\bg_\mn+2\mku\Omega\mku\bg_\mn+\mO(\Omega^2)$, we find that the
terms of an expansion of $\Gamma_k[h;\bg] \equiv \Gamma_k[g,\bg]$ linear in $h_\mn$ can be filtered out by inserting
\eqref{eq:ConfPrRelation} into $\Gamma_k[g,\bg]$ and projecting onto the terms linear in $\Omega$. Although the choice
\eqref{eq:ConfPrRelation} amounts to a restriction of the full theory space, it is still possible to differentiate
between invariants that stem from different metrics, at least within the truncation ansatz considered in this section.
The advantage of this method resides in the fact that there is only one kind of covariant derivative left, $\bD_\mu$,
such that a heat kernel expansion is applicable. Then the accessible ``bimetric information'' can be reconstructed by
disentangling terms of the order $\Omega^0$ and terms of the order $\Omega^1$. (See Refs.\ \cite{MRS11b,BR14} for
further details).

For the subsequent RG analysis we consider the bimetric truncation ansatz
\begin{equation}
\begin{split}
 \Gamma_k\big[g,\bg,\xi,\bx\, \big] = {} &\frac{1}{16\pi G_k^\text{Dyn}} \int\! \dd x \sg
 \big(\! -R + 2\mku\Lambda_k^\text{Dyn} \big) \\
 & +\frac{1}{16\pi G_k^\text{B}} \int\! \dd x \sbg \big(\! -\bar{R} + 2\mku\Lambda_k^\text{B}
 \big)\\[0.1em]
 & + \Gamma_k^\text{gf}\big[g,\bg \big] + \Gamma_k^\text{gh}\big[g,\bg,\xi,\bx\, \big].
\end{split}
\label{eq:doubleEHtrunc}
\end{equation}
It consists of two separate Einstein--Hilbert terms belonging to the dynamical ('Dyn') and the background ('B') metric
and their corresponding couplings. In order to extract $\beta$-functions from the FRGE \eqref{eq:FRGEgrav}, we proceed
along the lines of Ref.\ \cite{BR14}: We choose the gauge parameter $\alpha$ in the most convenient way, referred to as
the ``$\Omega$ deformed $\alpha=1$ gauge'', and we employ the conformal projection technique. Both of these choices
simplify the Hessian $\Gamma_k^{(2)}$ considerably. For the linear parametrization the calculation has been done in
Ref.\ \cite{BR14}. As for the exponential parametrization, a detailed derivation of $\beta$-functions is contained in
Appendix \ref{app:bi}.

In Chapter \ref{chap:Intro} as well as in Section \ref{sec:BFF} we have discussed the requirement for background
independence: Physical observables must not depend on an externally prescribed background field. The most
straightforward possibility to implement this condition is to make sure that $\Gamma_k$ has no extra
$\bg$-dependence once all fluctuations are integrated out, i.e.\ the partial functional derivative
$\frac{\delta\Gamma_k[g,\bg]}{\delta\bg_\mn(x)}$ must vanish identically at the scale $k=0$. In this case, $\bg_\mn$
can enter $\Gamma_{k=0}$ only via $g_\mn\mku$, provided that $g_\mn$ is parametrized by $\bg_\mn$ and $h_\mn$, the
linear and the exponential parametrization being typical examples. Then it is always possible to vary $\bg_\mn$ and
$h_\mn$ simultaneously in such a way that $g_\mn$ remains constant. Thus, $\Gamma_{k=0}$ is invariant under such
\emph{split-symmetry transformations}, too. In other words, background independence is achieved if split-symmetry
is restored in the IR limit.

With regard to our truncation ansatz, the second line in \eqref{eq:doubleEHtrunc} containing the extra
$\bg$-dependent terms has to vanish in the limit $k\to 0$ in order to ensure background independence.\footnote{Note
that gauge fixing and ghost terms violate background independence, too, even at the scale $k=0$, This is a very mild
violation, though, since it concerns the gauge modes only, and it should disappear upon going on-shell \cite{BR14}.
Thus, for the present discussion we consider only the non-gauge parts of $\Gamma_k\mku$.} This leads to the
requirements
\begin{equation}
 \frac{1}{G_k^\text{B}} \xrightarrow{\;k\to 0\;} 0\,, \qquad\text{and}\qquad
 \frac{\Lambda_k^\text{B}}{G_k^\text{B}} \xrightarrow{\;k\to 0\;} 0\,.
\label{eq:BIRequirement}
\end{equation}

As usual, the RG analysis is mainly performed in terms of dimensionless couplings, in particular, when fixed points and
RG trajectories are concerned. They are defined as
\begin{alignat}{2}
 \gDyn_k &\equiv k^{d-2} G^\text{Dyn}_k\,,\quad &\lDyn_k &\equiv k^{-2} \Lambda^\text{Dyn}_k\,, \\
 \gB_k &\equiv k^{d-2} G^\text{B}_k\,, &\lB_k &\equiv k^{-2} \Lambda^\text{B}_k\,.
\end{alignat}
We will confirm later on that almost all trajectories are characterized in the IR by the canonical running of the
couplings. In the background sector this means $\gB_k \propto k^{d-2}$ and $\lB_k\propto k^{-2}$, implying
$1/G_k^\text{B}=\text{const}$ and $\Lambda_k^\text{B}/G_k^\text{B}=\text{const}$ for small $k$. In this case,
\eqref{eq:BIRequirement} is not satisfied.

However, if there was a fixed point $(\lB_*,\gB_*)$ in the background sector, a trajectory starting at
$(\lB_*,\gB_*)$ at some finite scale $k$ would ``stay'' in this point for $k\to 0$. For this special case, we would
have $\lB_k=\lB_*=\text{const}$ and $\gB_k=\gB_*=\text{const}$ in the IR, finally leading to
\begin{equation}
 \frac{1}{G_k^\text{B}} = \frac{1}{\gB_*}\, k^{d-2} \xrightarrow{\;k\to 0\;} 0\,, \qquad\text{and}\qquad
 \frac{\Lambda_k^\text{B}}{G_k^\text{B}} = \frac{\lB_*}{\gB_*}\, k^d \xrightarrow{\;k\to 0\;} 0\,,
\label{eq:BIRequirementFP}
\end{equation}
as it should be. We thus conclude that \emph{background independence by means of split-symmetry restoration can be
established on the basis of a suitable fixed point in the background sector}.

It is this possibility that we investigate in the following for both the linear and the exponential parametrization.
In particular, we aim at proving the existence of such RG trajectories that are asymptotically safe in the UV and
restore background independence in the IR.

Before performing explicit computations, a general remark is in order: Since the background couplings $G_k^\text{B}$
and $\Lambda_k^\text{B}$ in the truncation ansatz \eqref{eq:doubleEHtrunc} occur in terms that contain only the
background metric, they drop out when calculating the second derivative of $\Gamma_k$ with respect to $h_\mn$, and
hence, they cannot enter the RHS of the FRGE \eqref{eq:FRGEgrav}. As a consequence, there is a typical hierarchy of
coupling constants. This becomes explicit on the level of the $\beta$-functions: Independent of the parametrization,
they have the general form
\begin{equation}
\begin{split}
 \beta_g^\text{Dyn} &\equiv \beta_g^\text{Dyn}\big( \gDyn, \lDyn \big)\,, \\
 \beta_\lambda^\text{Dyn} &\equiv \beta_\lambda^\text{Dyn}\big( \gDyn, \lDyn \big)\,, \\
 \beta_g^\text{B} &\equiv \beta_g^\text{B}\big( \gDyn, \lDyn, \gB \big)\,, \\
 \beta_\lambda^\text{B} &\equiv \beta_\lambda^\text{B}\big( \gDyn, \lDyn, \gB, \lB \big)\,. \\
\end{split}
\label{eq:BiHierarchy}
\end{equation}
In particular, we observe that \emph{the RG flow of the dynamical coupling sector is decoupled} as the
$\beta$-functions of $\lambda^\text{Dyn}$ and $g^\text{Dyn}$ constitute a closed system. Thus, one can solve the RG
equations of the 'Dyn' couplings independently at first.

On the other hand, the background $\beta$-functions depend on both dynamical and background couplings. Therefore, the
RG running of $g_k^\text{B}$ and $\lambda_k^\text{B}$ can be determined only if a solution of the 'Dyn' sector is
picked. With regard to the Asymptotic Safety program we would like to choose a 'Dyn' trajectory which emanates from a
NGFP and passes the classical regime near the Gaussian fixed point. This trajectory is then inserted into the
$\beta$-functions of the background sector, \emph{making them explicitly $k$-dependent}. Therefore, the vector field
these $\beta$-functions give rise to depends on $k$, too, and possible ``fixed points'', i.e.\ simultaneous zeros of
$\beta_\lambda^\text{B}$ and $\beta_g^\text{B}$, become \emph{moving points}. We will refer to a UV-attractive ``moving
NGFP'' as \emph{running attractor} \cite{BR14}. One might think of such a running attractor as a moving magnet:
Starting at a given point in the background coupling sector, its RG evolution is such that it is trailed behind the
running attractor. If the running attractor approaches a finite limit for $k\to\infty$, it finally becomes an ordinary
(i.e.\ nonmoving) UV fixed point.

%----------------------------------------------------------------------------------------------------------------------
\subsection{Results for the linear parametrization}
\label{sec:biLin}
%----------------------------------------------------------------------------------------------------------------------

In this subsection we quote a couple of known results for the linear parametrization, first obtained in
Ref.\ \cite{BR14}. The hierarchy \eqref{eq:BiHierarchy} of the coupling constants, which was derived from very general
arguments, is indeed found by an explicit calculation. Consequently, it is possible to solve the `Dyn' system first,
select a suitable trajectory, and insert it into the `B' system.

%----------------------------------------------------------------------------------------------------------------------
\subsubsection[The linear parametrization in \texorpdfstring{$d=4$}{d = 4} dimensions]%
{The linear parametrization in \texorpdfstring{\bm{$d=4$}}{d = 4} dimensions}
%----------------------------------------------------------------------------------------------------------------------

We pick a `Dyn' trajectory which is asymptotically safe in the UV, passes the vicinity of the Gaussian fixed point at
classical scales, and then runs towards large positive values of the cosmological constant in the IR. By the
classification of Ref.\ \cite{RS02}, such a trajectory belongs to the type IIIa trajectories. The $k$-dependent
solution,
\begin{equation}
 k\mapsto(\lDyn_k,\gDyn_k),
\end{equation}
is inserted into the $\beta$-functions of the background couplings now, yielding an effectively \emph{nonautonomous
system}:
\begin{equation}
\begin{split}
 \beta_g^\text{B} &\equiv \beta_g^\text{B}(\gB,k)\,,\\
 \beta_\lambda^\text{B} &\equiv \beta_\lambda^\text{B}(\lB,\gB,k)\,.
\end{split}
\end{equation}
The corresponding $k$-dependent vector field with its ``fixed points'' is depicted in Figure
\ref{fig:LinParamMovingFP}. (All diagrams that belong to the background sector will be drawn in dark yellow.) We show
the vector field at six different values of $t\equiv\ln(k/k_0)$ with some reference scale $k_0$.
\begin{figure}[tp]
 \centering
 \vspace{-0.53em}
 \begin{minipage}{0.3\columnwidth}
  \includegraphics[width=\columnwidth]{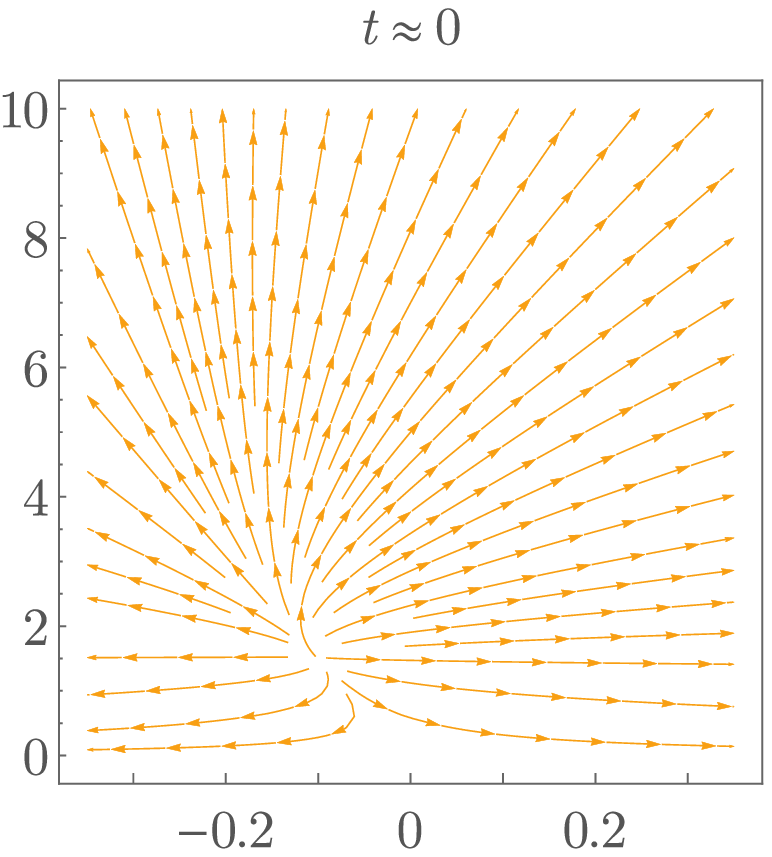}
 \end{minipage}
 \hfill
 \begin{minipage}{0.3\columnwidth}
  \includegraphics[width=\columnwidth]{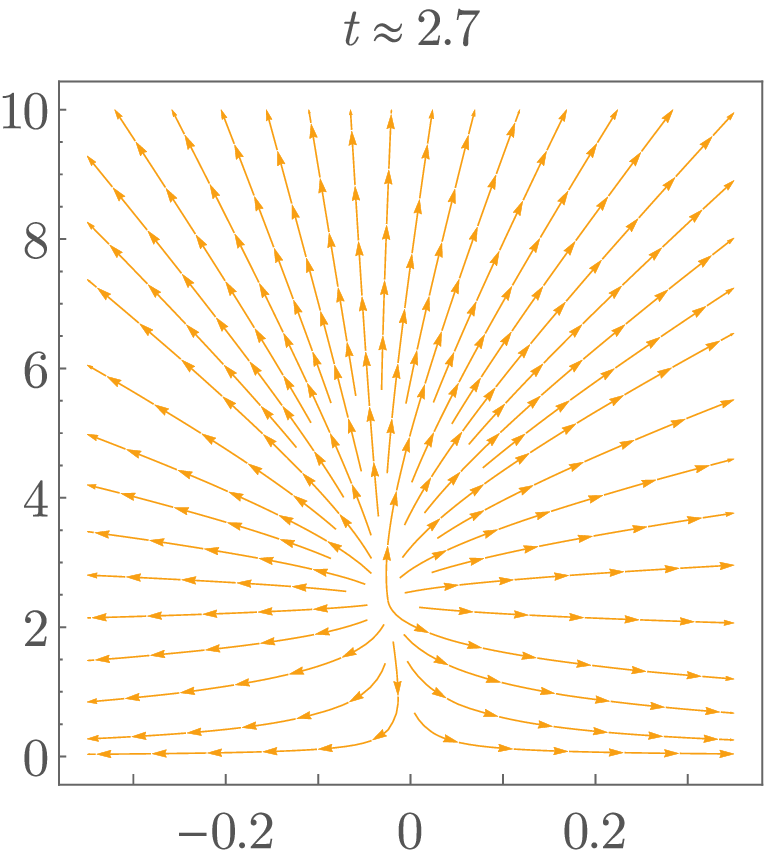}
 \end{minipage}
 \hfill
 \begin{minipage}{0.3\columnwidth}
  \includegraphics[width=\columnwidth]{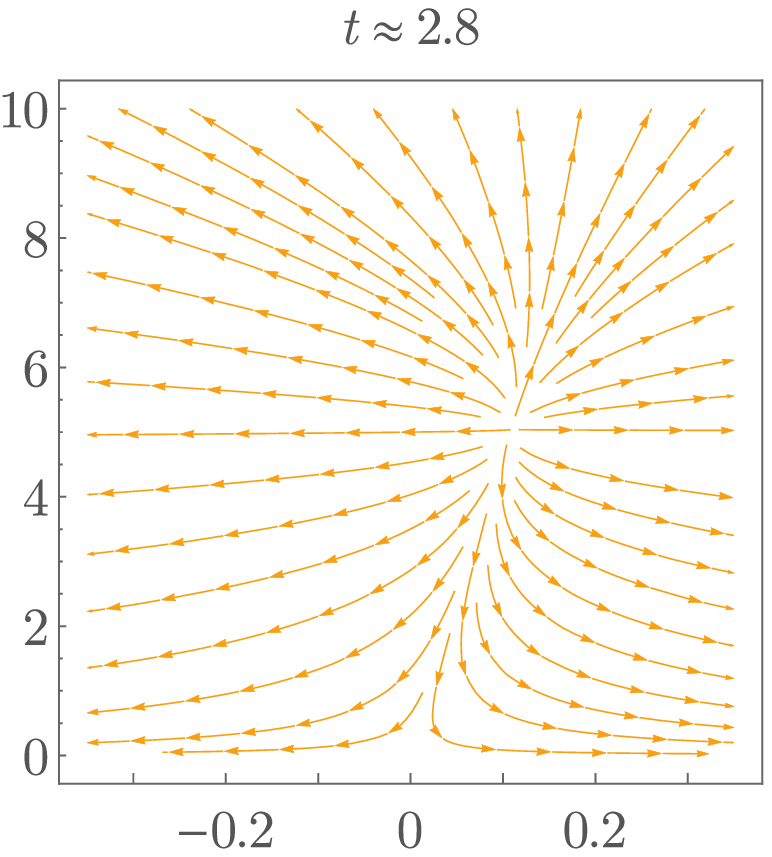}
 \end{minipage}
 
 \vspace{0.7em}
 \begin{minipage}{0.3\columnwidth}
  \includegraphics[width=\columnwidth]{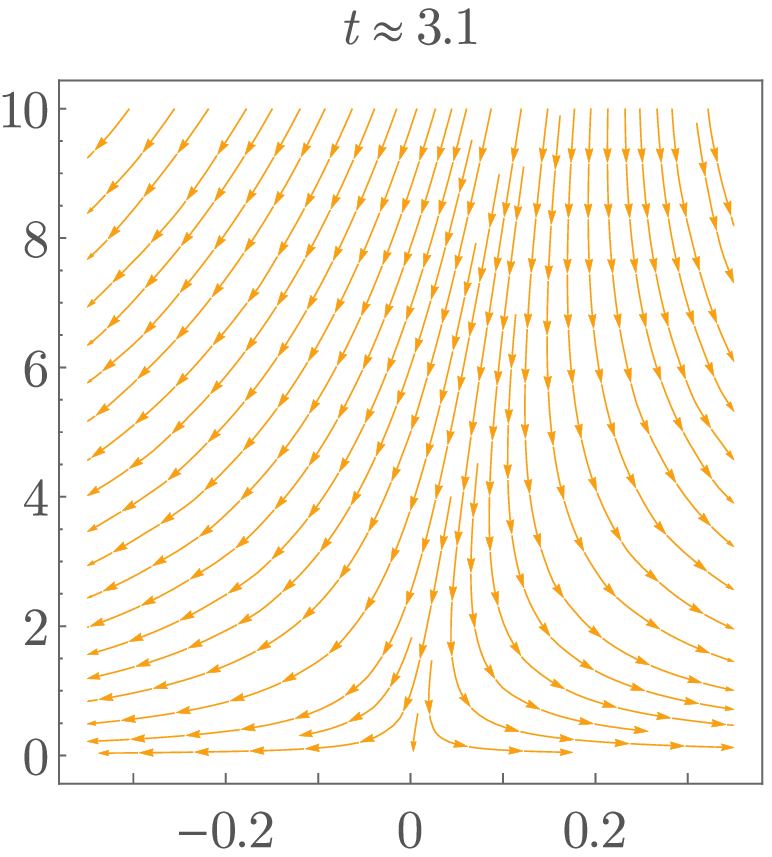}
 \end{minipage}
 \hfill
 \begin{minipage}{0.3\columnwidth}
  \includegraphics[width=\columnwidth]{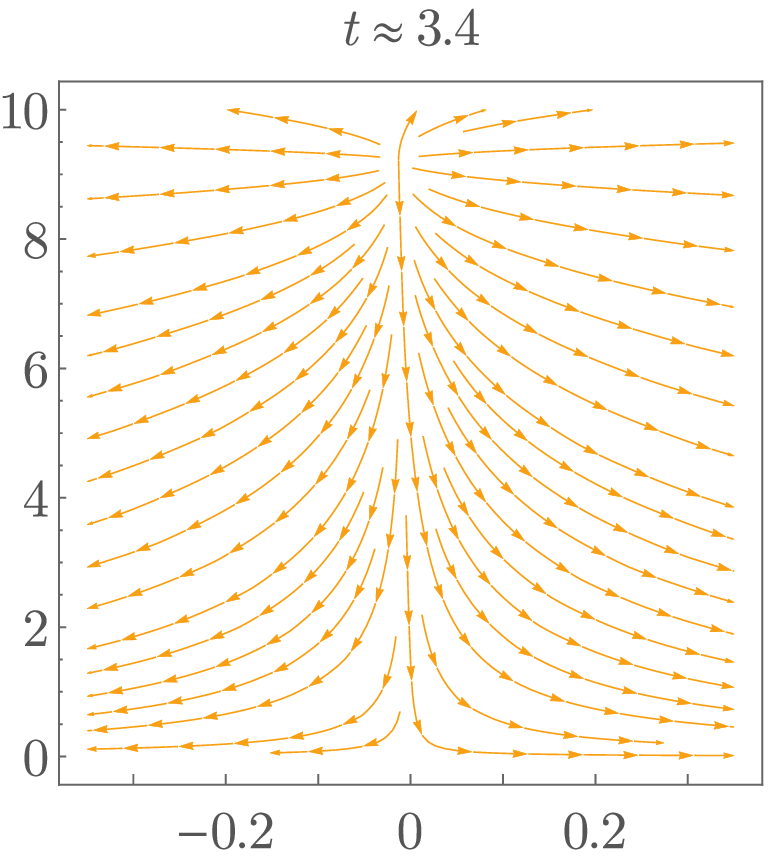}
 \end{minipage}
 \hfill
 \begin{minipage}{0.3\columnwidth}
  \includegraphics[width=\columnwidth]{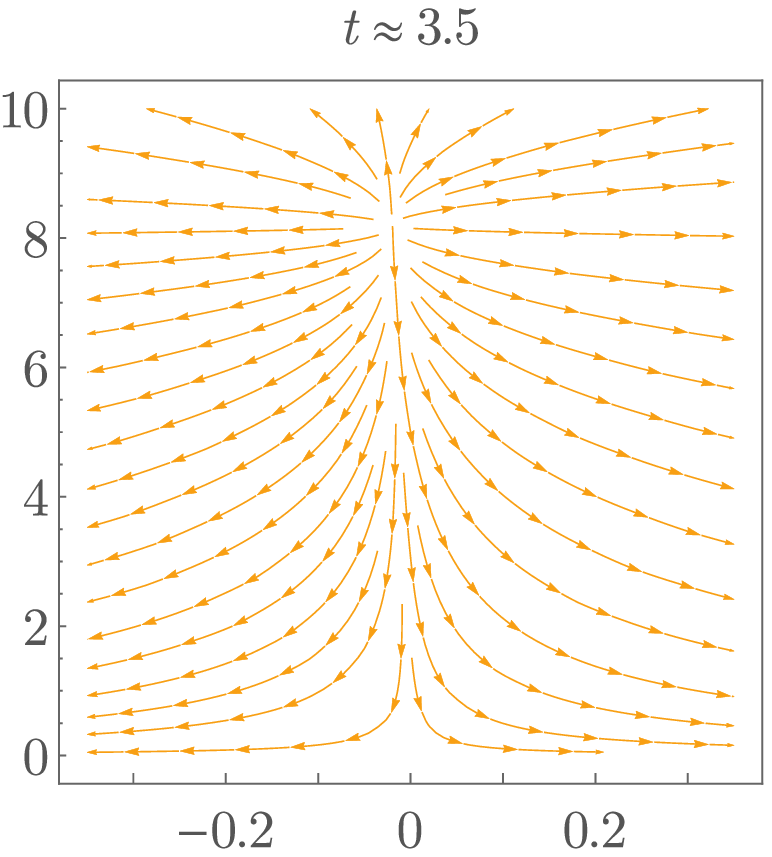}
 \end{minipage}
\vspace{0.4em}
\caption{Flow diagrams of the background sector for the linear parametrization at several finite RG times
$t\equiv\ln(k/k_0)$. Horizontal axes show the background cosmological constant, $\lB$, while vertical axes show the
background Newton constant, $\gB$. There is a moving non-Gaussian ``fixed point'' whose existence and position depends
on the RG parameter $t$. This ``fixed point'' is found to exist in the infrared, for small values of $t$. At
intermediate scales it disappears for a moment of time, see figure with $t\approx 3.1$ (or, more precisely, it
diverges, jumps to negative $\gB$, and jumps back to positive $\gB$). For large $t$ it is present
again, and it approaches a stable value in the limit $t\to\infty$. The diagram in the last figure ($t\approx 3.5$)
already agrees almost entirely with its final form at $t\to\infty$.}
\label{fig:LinParamMovingFP}
\end{figure}
We observe that the running attractor, i.e.\ the moving fixed point, exists at low scales, vanishes at an intermediate
scale, and exists again at high scales, in particular for $k\to\infty$. Note that the temporarily divergent running
attractor does not lead to divergent RG trajectories: Even though trajectories are attracted by a point at infinity
at those potentially problematic RG times, the trajectories themselves do not diverge since this happens only during
a finite RG time interval. Thus, all relevant trajectories stay in theory space and approach a finite point in the
limit $k\to\infty$. We emphasize that the curve given by the position of the running attractor is not an RG trajectory.

A similar picture is obtained if we choose a type Ia trajectory (characterized by negative cosmological constants in
the IR, according to the classification of Ref.\ \cite{RS02}) in the `Dyn' sector and adapt the $\beta$-functions in
the `B' sector correspondingly.

We have argued in \eqref{eq:BIRequirementFP} that background independence can be achieved at the scale $k=0$ only if
there is a suitable fixed point. It turns out that the moving fixed point observed in Figure \ref{fig:LinParamMovingFP}
has indeed the right properties.\footnote{Note that the moving fixed point depends on the choice of a suitable `Dyn'
trajectory, here selected to be of type IIIa. In fact, type IIIa trajectories might run into the singularity line (if
present) at some positive value of $\lDyn$ such that they would not possess a well defined infrared limit. However,
since the singularity line is believed to be merely a truncation artifact (cf.\ discussion in the single-metric case),
it is \emph{assumed} here as well as in Ref.\ \cite{BR14} that trajectories extend to $(\lDyn,\gDyn)\to (\infty,0)$ for
$k\to\infty$, i.e.\ the singularity at $\lDyn=1/2$ is ignored for a moment. In this limit of the `Dyn' couplings, the
corresponding moving fixed point in the `B' sector has indeed a finite limit that serves as a fixed point at $k=0$.
To increase the numerical reliability we stop the RG evolution towards the IR at some small, finite scale before
getting too close to the singularity, though. Nonetheless, this is sufficient for showing the applicability of the
mechanism in principle.}
Now, let us consider the background trajectory that starts precisely at the position of this running attractor in the
IR. What happens if the RG scale increases now? From Figure \ref{fig:LinParamMovingFP} we know that the running
attractor moves away. Being UV-attractive it trails the starting point under consideration, where the resulting RG
trajectory is given by curve of this trailed point. At all finite scales, the point lags behind the running attractor.
Finally, they both approach a common fixed point in the limit $k\to\infty$. In this manner, we obtain a trajectory that
satisfies the requirement for background independence in the IR and is asymptotically safe in the UV.

This situation is illustrated in Figure \ref{fig:StdBiOpt}. It shows the vector field in the background sector at
$k\to\infty$ and the RG trajectory (gray) that starts at the IR position of the running attractor and ends at its
$k\to\infty$ position (w.r.t.\ the inverse RG flow).
\begin{figure}[tp]
\begin{minipage}{0.78\columnwidth}
 \includegraphics[width=\columnwidth]{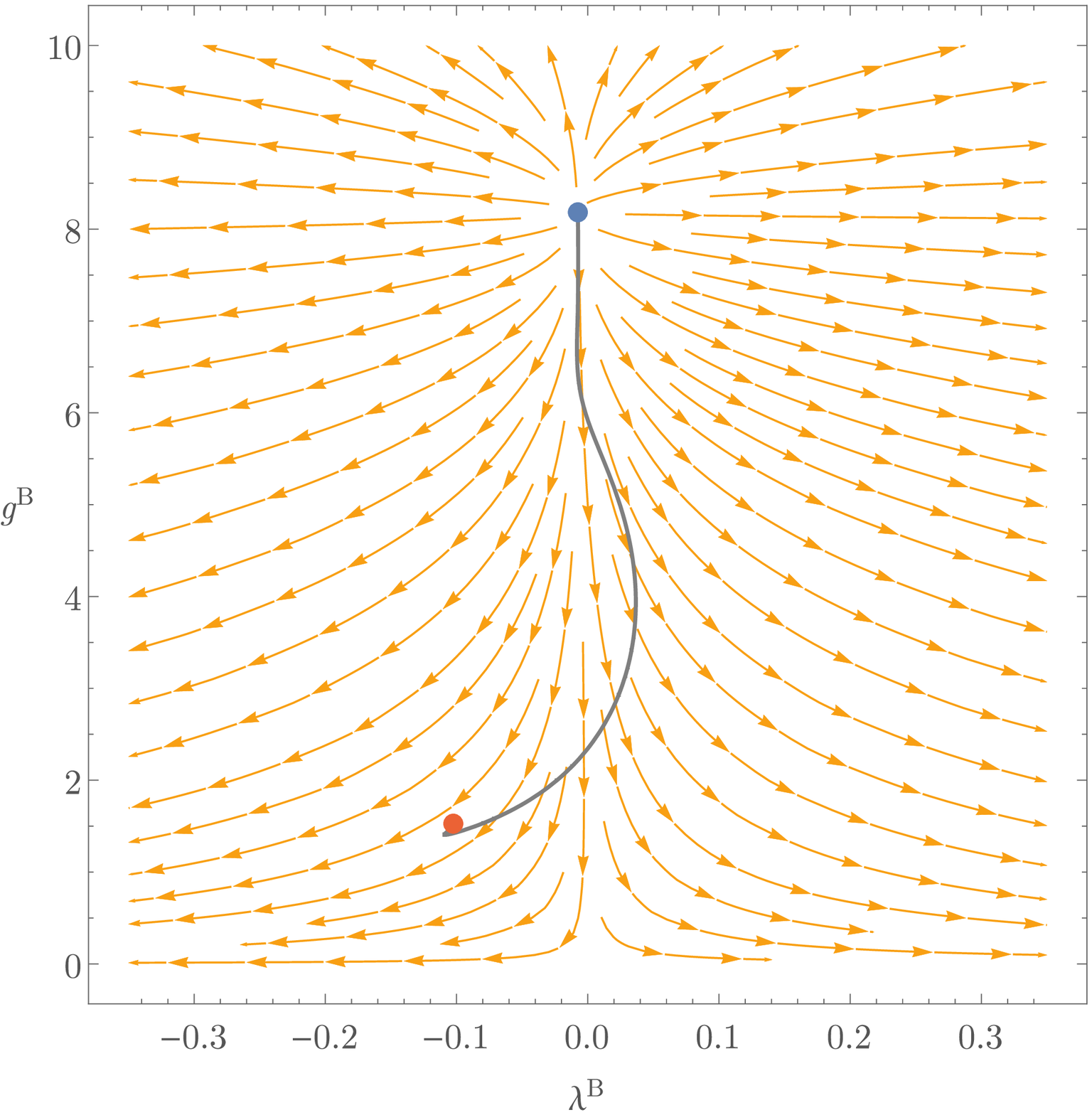}
\end{minipage}
\hfill
\hspace{-0.11\columnwidth}
\begin{minipage}{0.29\columnwidth}
 \vspace{2.5em}
 \includegraphics[width=\columnwidth]{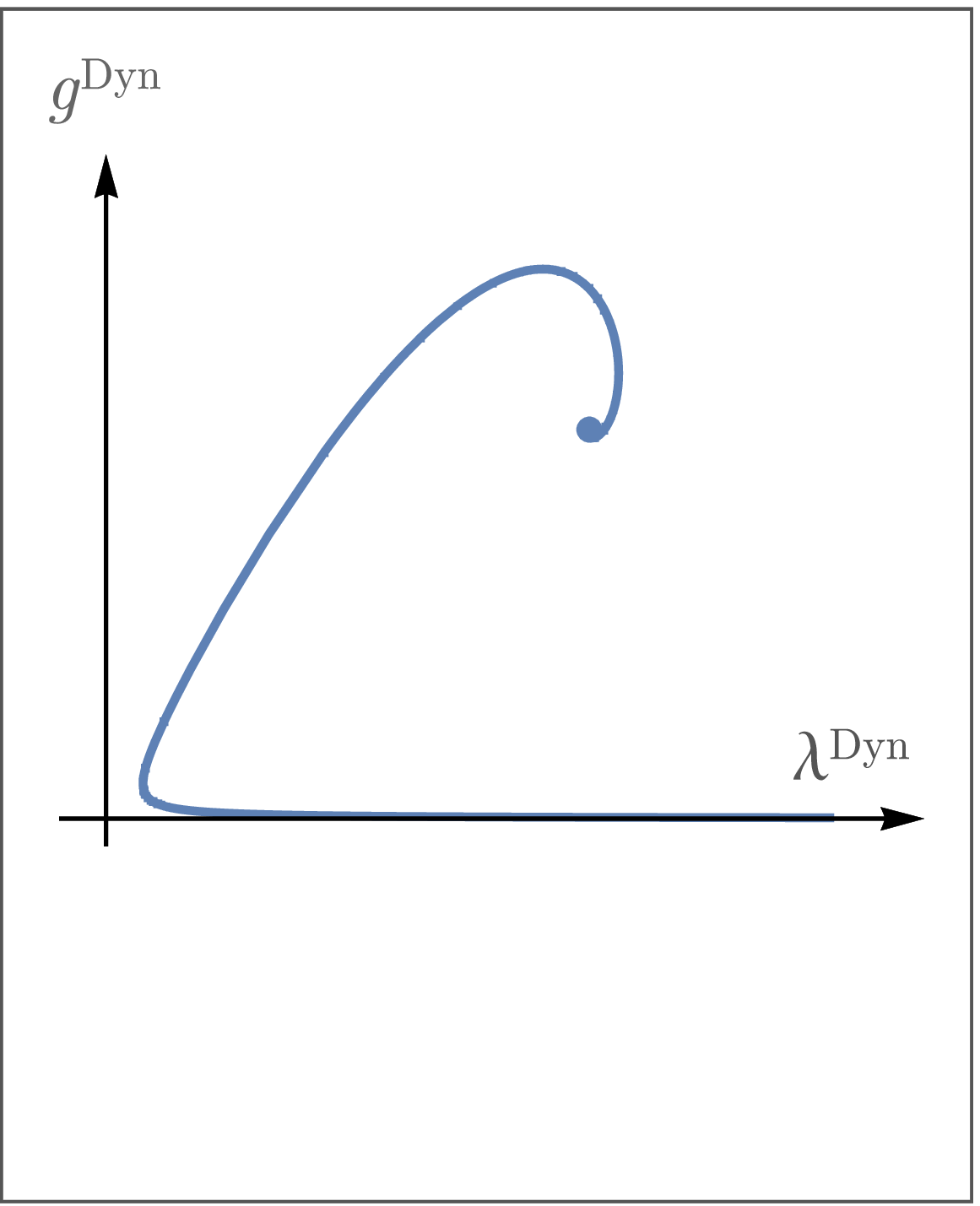}
 \begin{textblock}{2.5}(0.254,-0.8)
 \begin{center}
  \small
  Underlying 'Dyn'\\ trajectory
 \end{center}
 \end{textblock}
\end{minipage}
\caption{Vector field for the background couplings at $k\rightarrow\infty$ and RG trajectory (gray curve) that is
 asymptotically safe in the UV and restores split symmetry in the IR (left figure), and the underlying trajectory in
 the 'Dyn' sector (right figure), based on the \emph{linear parametrization} and the optimized cutoff in $d=4$.
 Note that the marked RG trajectory in the `B' diagram comprises all RG scales from the IR (red point) to the
 UV (blue point), while the vector field is in its final state in the UV limit.}
\label{fig:StdBiOpt}
\end{figure}
The main result of Ref.\ \cite{BR14} can be summarized as follows: For any appropriate choice of initial conditions in
the 'Dyn' sector \emph{there exists a unique trajectory in the 'B' sector that complies with the requirements for both
background independence and Asymptotic Safety}. This statement is independent of the chosen cutoff function.

%----------------------------------------------------------------------------------------------------------------------
\subsubsection[The linear parametrization in \texorpdfstring{$d=2+\ve$}{d = 2 + epsilon} dimensions]%
{The linear parametrization in \texorpdfstring{\bm{$d=2+\ve$}}{d = 2 + epsilon} dimensions}
%----------------------------------------------------------------------------------------------------------------------

As an interesting supplement to the single-metric results in $2+\ve$ dimensions we would like to discuss the bimetric
case now. Note that the following results deviate from those of Ref.\ \cite{BR14} which did not take into account that
$\lDyn_*$ is of the order $\ve$. Although we employ the same set of equations for the $\beta$-functions in $d$
dimensions as in Ref.\ \cite{BR14}, we carefully keep track of all potential appearances of $\ve$.

A numerical analysis based on the optimized cutoff shows that there exists an NGFP in $d=2+\ve$ whose coordinates are
of the order $\ve\,$:
\begin{align}
 \gDyn_* &=\mO(\ve)\,, & \lDyn_* &=\mO(\ve)\,, \\
 \gB_* &=\mO(\ve)\,, & \lB_* &=\mO(\ve)\,.
\end{align}
Thus, in the vicinity of the NGFP all couplings satisfy $\gDyn, \lDyn, \gB, \lB=\mO(\ve)$. For an analytical
calculation it is convenient to introduce the normalized couplings
\begin{align}
 \gDyn_k &\equiv\rgDyn_k\,\ve\,, & \lDyn_k &\equiv\rlDyn_k\,\ve\,, \\
 \gB_k &\equiv\rgB_k\,\ve\,, & \lB_k &\equiv\rlB_k\,\ve\,,
\end{align}
where $\rgDyn_k$, $\rlDyn_k$, $\rgB_k$ and $\rlB_k$ are of the order $\mO(\ve^0)$. Inserting these relations into the
$\beta$-functions and expanding in terms of $\ve$, the relevant order in the `Dyn' sector reads
\begin{align}
 \beta_g^\text{Dyn} &= \rgDyn \left(\frac{4\,\rgDyn \Big[5+6\mku \rlDyn
 \big(12\, \Phi^3_1(0)-24\, \Phi^4_2(0)-1\big)\Big]}{24\, \rgDyn\big( \tilde{\Phi}^2_1(0)-2\,
 \tilde{\phi}^3_2(0)\big)+3}+1\right) \ve^2 +\mO(\ve^3)\,,\\
 \beta_\lambda^\text{Dyn} &= \big(20\, \rgDyn\, \Phi^2_2(0)-2 \rlDyn\big)\ve + \mO(\ve^2)\,.
\end{align}
The $\beta$-functions in the background sector are not stated here in general, but in a moment we specify the result
for the optimized shape function instead. We would like to point out that the $\beta$-functions of the two Newton
couplings are of the same form as in the single-metric case: $\beta_g^\text{Dyn} = \ve\mku \gDyn - b^\text{Dyn}\mku
(\gDyn)^2$ and $\beta_g^\text{B} = \ve\mku \gB - b^\text{B}\mku (\gB)^2$, respectively, up to higher orders. Since they
contain cutoff dependent threshold functions, all $\beta$-functions are \emph{nonuniversal}.

Solving the system $\big\{\beta_\lambda^\text{Dyn}=0, \beta_g^\text{Dyn}=0\big\}$ yields the fixed point values
$\rgDyn_*$ and $\rlDyn_*\,$. For the coefficient $b^\text{Dyn}$ this leads to
\begin{equation}
 b^\text{Dyn} = -\frac{4 \Big[5+6\, \rlDyn_* \big(12\, \Phi^3_1(0)-24 \,\Phi^4_2(0)-1\big)\Big]}{3
  +24\, \rgDyn_* \big(\tilde{\Phi}^2_1(0)-2 \,\tilde{\Phi}^3_2(0)\big)} \,,
\end{equation}
together with $\rlDyn_* = 10\, \rgDyn_*\, \Phi^2_2(0)$ and $\rgDyn_*=1/b^\text{Dyn}$. By eliminating both couplings
we obtain a quadratic equation with two possible solutions for $b^\text{Dyn}$. For the optimized cutoff the first
solution is given by
\begin{equation}
 b^\text{Dyn} \approx -\frac{34.45}{3}\,,\qquad b^\text{B} \approx \frac{72.45}{3}\,,
\label{eq:bBiSol1}
\end{equation}
while the second solution reads
\begin{equation}
 b^\text{Dyn} \approx \frac{10.45}{3}\,,\qquad b^\text{B} \approx \frac{27.55}{3}\,.
\label{eq:bBiSol2}
\end{equation}

A general consideration shows that the sum of $b^\text{Dyn}$ and $b^\text{B}$ must agree with the
coefficient $b\equiv b^\text{sm}$ from the corresponding single-metric computation: Setting $g_\mn=\bg_\mn$ in
\eqref{eq:doubleEHtrunc} to project onto the single-metric truncation we see that the only remaining
Einstein--Hilbert term --- the term from which $b^\text{sm}$ can be read off --- is now proportional to
$\Big(\frac{1}{G_k^\text{Dyn}}+\frac{1}{G_k^\text{B}}\Big)$. Since the $b$-coefficients are proportional to
$\frac{1}{G_k^\text{Dyn}}$, $\frac{1}{G_k^\text{B}}$  and $\frac{1}{G_k^\text{sm}}$, respectively, in $2+\ve$
dimensions, we conclude that
\begin{equation}
 b^\text{Dyn} + b^\text{B} = b^\text{sm}\,.
\label{eq:genRuleSumB}
\end{equation}
Using \eqref{eq:bBiSol1} and \eqref{eq:bBiSol2} we find indeed
\begin{equation}
 b^\text{Dyn}+b^\text{B} = \frac{38}{3}\,,
\end{equation}
for both solutions, in perfect agreement with the single-metric result of Section \ref{sec:singleLin2}.

%----------------------------------------------------------------------------------------------------------------------
\subsection{Results for the exponential parametrization}
\label{sec:biExp}
%----------------------------------------------------------------------------------------------------------------------

In this subsection we investigate the same bimetric truncation as above, eq.\ \eqref{eq:doubleEHtrunc}, but now we
employ the exponential parametrization. The corresponding $\beta$-functions are derived in detail in Appendix
\ref{app:bi}. We find the same hierarchical structure of couplings in the $\beta$-functions as for the linear
parametrization. Again, this enables us to solve the `Dyn' system first and insert a `Dyn' solution into the
$\beta$-functions of the background couplings. This way, we obtain a nonautonomous system of evolution equations for
the `B' sector, which is analyzed similarly to the previous subsection.
As the threshold functions appearing in the $\beta$-functions \eqref{eq:betagDynBi} -- \eqref{eq:betalBBi} are of the
form $\Phi_n^p(-\mu\lDyn)$ with $\mu\equiv\frac{2d}{d-2}>2$ (rather than $\Phi_n^p(-2\lDyn)$ as for the linear
parametrization), we expect that the singularity line in the `Dyn' sector is shifted to smaller values of $\lDyn$ this
time.

%----------------------------------------------------------------------------------------------------------------------
\subsubsection[The exponential parametrization in \texorpdfstring{$d=4$}{d = 4} dimensions]%
{The exponential parametrization in \texorpdfstring{\bm{$d=4$}}{d = 4} dimensions}
%----------------------------------------------------------------------------------------------------------------------

We aim at proving the existence of asymptotically safe trajectories that respect the principle of background
independence by restoring split-symmetry in the infrared. To this end we try again to pick a type IIIa `Dyn' trajectory
(i.e.\ a trajectory that emanates from a UV fixed point and runs towards either large positive values of $\lDyn$ or a
singularity at positive $\lDyn$ in the IR) which has a sufficiently extended classical regime, that is, which passes
the vicinity of the Gaussian fixed point. It turns out that the existence of such trajectories depends on the chosen
cutoff shape, like in the single-metric case discussed in Section \ref{sec:singleExp4}. Consequently, the resulting RG
flow in the background sector is discussed only if we succeed in finding a suitable `Dyn' trajectory.
\medskip

\noindent
\textbf{(1) Optimized cutoff.} An evaluation of the $\beta$-functions in the 'Dyn' sector gives rise to the flow
diagram displayed in Figure \ref{fig:NewBiOptDyn}. We discover a non-Gaussian fixed point, but it is rather close to
the singularity line. As a consequence, all trajectories emanating from this fixed point will hit the singularity
after a short period of RG time. It is \emph{impossible to find suitably extended trajectories}: they do not pass
the classical regime, and they never come close to an acceptable infrared limit. For this reason, it is pointless to
investigate the possibility of split-symmetry restoration here. Although the background sector exhibits a
UV-attractive NGFP, too, owing to the lack of an appropriate infrared regime we refrain from showing vector fields for
the background couplings.
\begin{figure}[tp]
\centering
\includegraphics[width=.78\columnwidth]{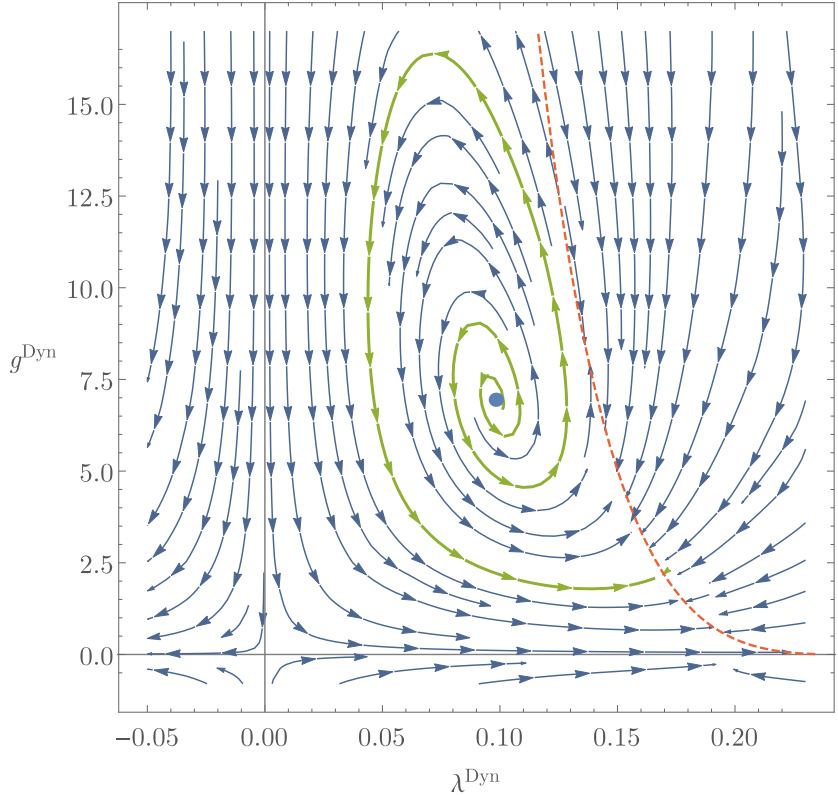}
\caption{Flow diagram of the 'Dyn' couplings in $d=4$ based on the \emph{exponential parametrization} and the optimized
 cutoff. The green arrows indicate that each trajectory that emanates from the NGFP (blue dot) finally runs into the
 (red, dashed) singularity line before it could ever pass the vicinity of the Gaussian fixed point. Note also that the
 NGFP is UV-attractive, so there is no such limit cycle as in the single-metric case.}
\label{fig:NewBiOptDyn}
\end{figure}

We emphasize, however, that the inability to establish background independence in the IR is not a flaw of the
exponential parametrization or the very mechanism, but it is merely due to the closer singularity line. Since the
singularity line is believed to disappear once the truncation is sufficiently enlarged, we expect that the above method
of restoring split-symmetry becomes applicable after all.  
\medskip

\noindent
\textbf{(2) Exponential cutoff.} We find the same qualitative picture as in Figure \ref{fig:NewBiOptDyn} which was
based upon the optimized cutoff. The exponential cutoff brings about a UV-attractive non-Gaussian fixed point for both
`Dyn' and `B' couplings. However, there are no trajectories that extend to a suitable infrared region since they run
into the singularity line. Thus, we do not discuss the possibility of restoration of background independence either.
\medskip

\noindent
\textbf{(3) Sharp cutoff.} The $\beta$-functions of the 'Dyn' couplings lead to a Gaussian and a non-Gaussian fixed
point, the latter being UV-attractive. We observe that $\beta_\lambda^\text{Dyn}$ is proportional to $\lDyn$, so `Dyn'
trajectories cannot cross the line at $\lambda^\text{Dyn}=0$. Still, there \emph{are} trajectories that connect the
NGFP to the classical regime, comparable with the ones found for the linear parametrization. Once such a `Dyn'
trajectory is chosen, the $k$-dependent solution $k\mapsto(\lDyn_k,\gDyn_k)$ is inserted into the $\beta$-functions of
the background sector, serving as a basis for further analyses of the corresponding RG flow. Similar to Subsection
\ref{sec:biLin}, we obtain a vector field in the $(\lB,\gB)$-space which varies with the RG scale. The result is shown
in Figure \ref{fig:ExpParamMovingFP} at several values of $t\equiv\ln(k/k_0)$.
\begin{figure}[tp]
 \centering
 \begin{minipage}{0.3\columnwidth}
  \includegraphics[width=\columnwidth]{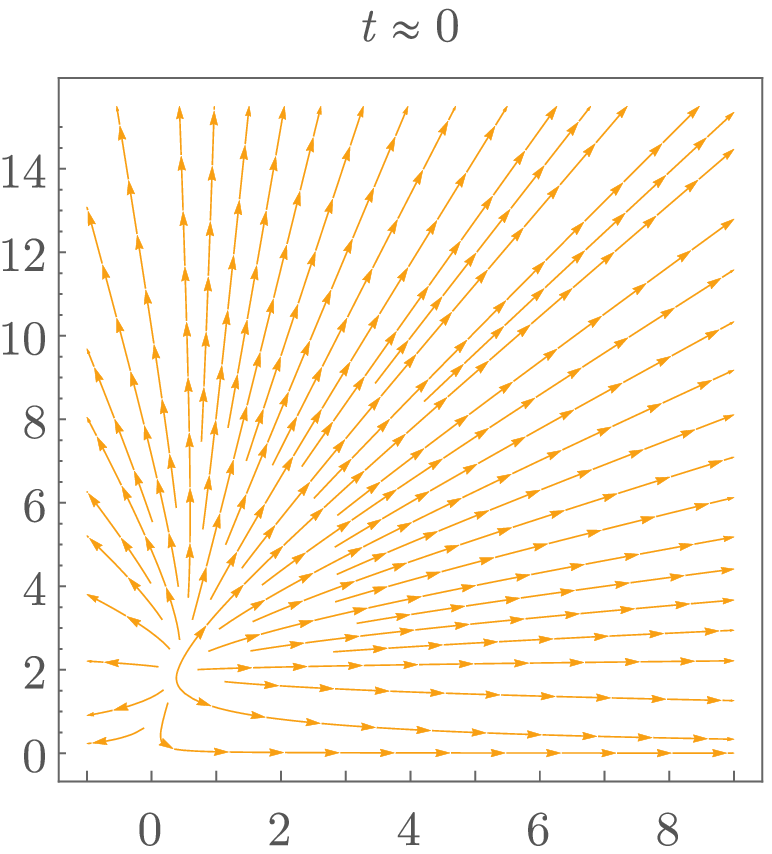}
 \end{minipage}
 \hfill
 \begin{minipage}{0.3\columnwidth}
  \includegraphics[width=\columnwidth]{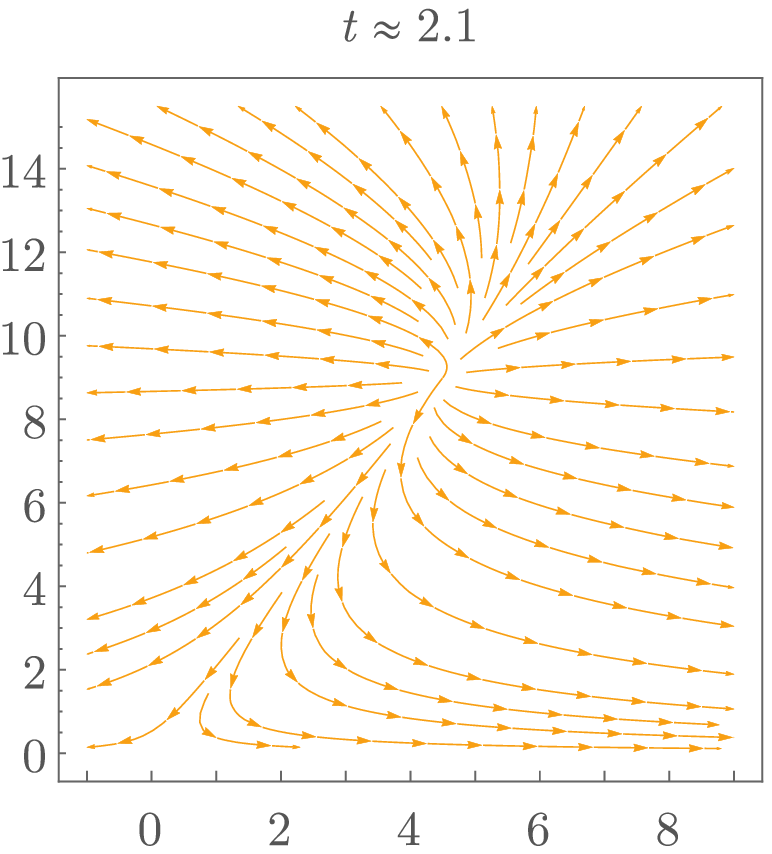}
 \end{minipage}
 \hfill
 \begin{minipage}{0.3\columnwidth}
  \includegraphics[width=\columnwidth]{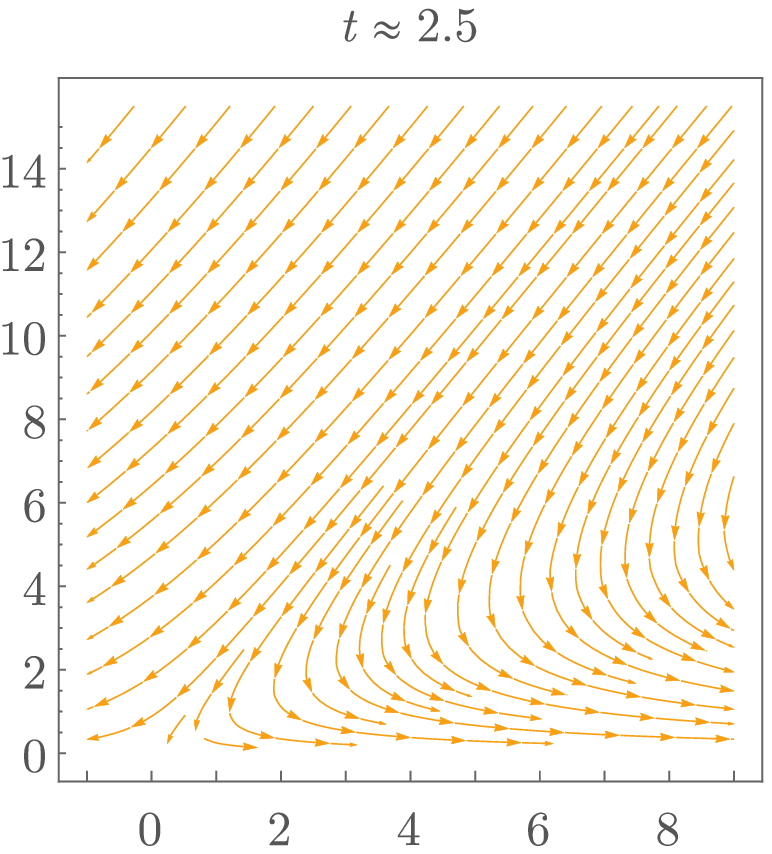}
 \end{minipage}
 
 \vspace{0.8em}
 \begin{minipage}{0.3\columnwidth}
  \includegraphics[width=\columnwidth]{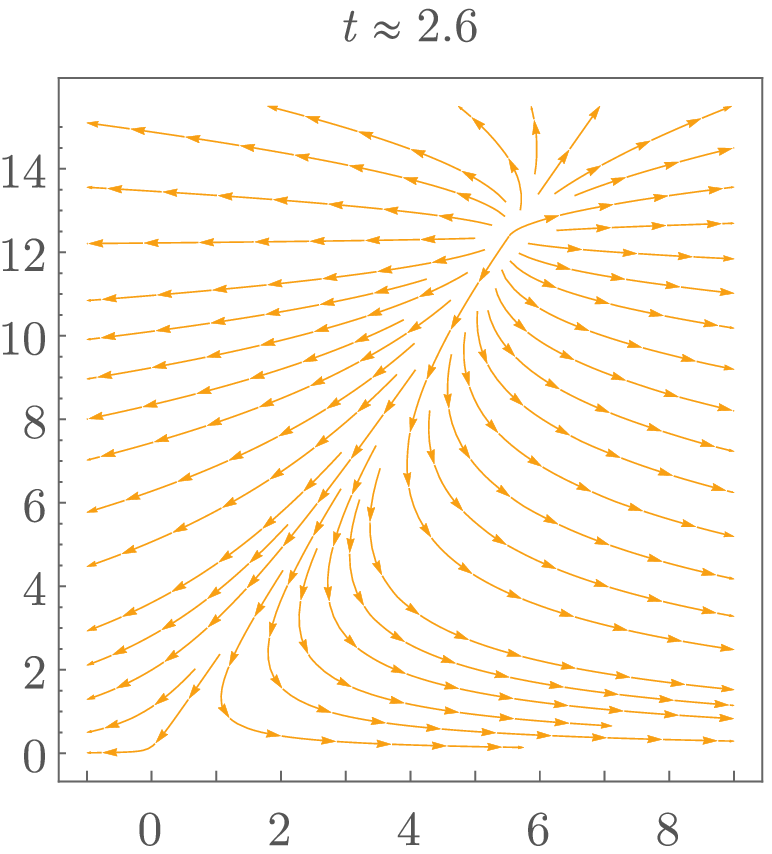}
 \end{minipage}
 \hfill
 \begin{minipage}{0.3\columnwidth}
  \includegraphics[width=\columnwidth]{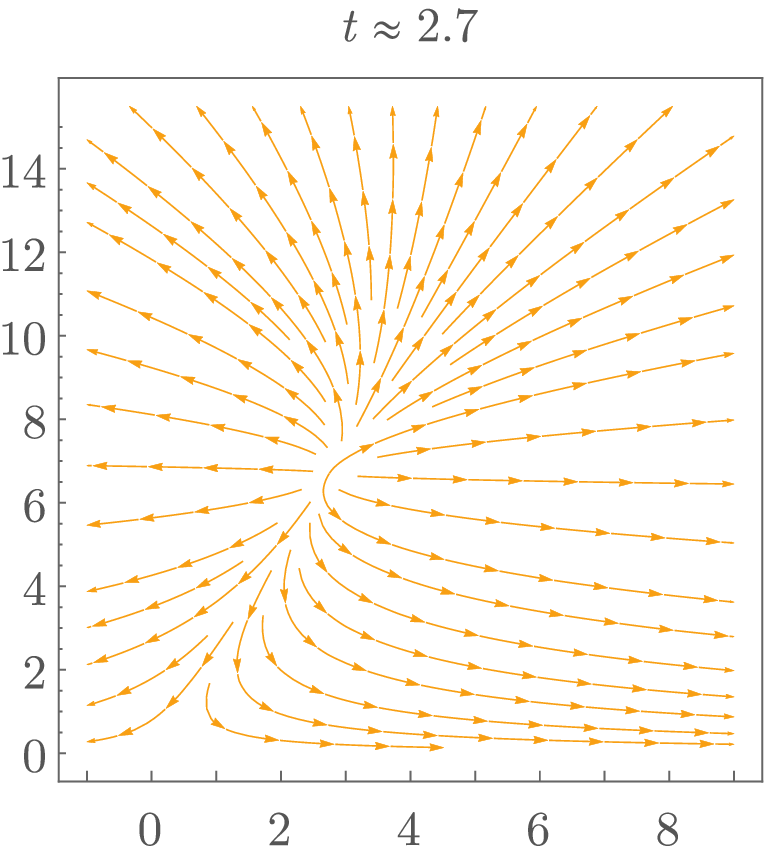}
 \end{minipage}
 \hfill
 \begin{minipage}{0.3\columnwidth}
  \includegraphics[width=\columnwidth]{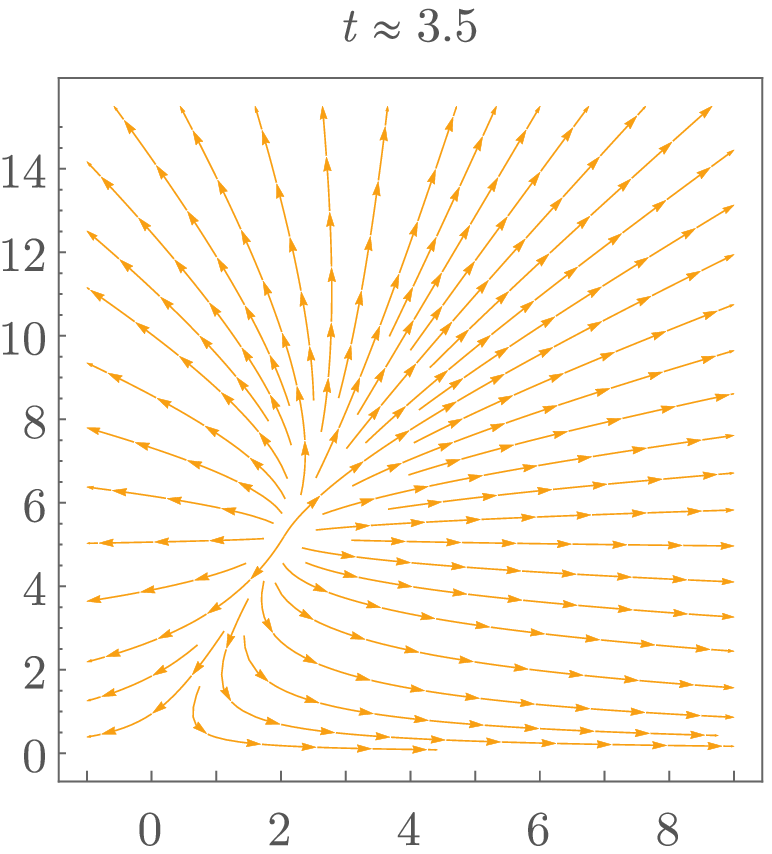}
 \end{minipage}
\vspace{0.5em}
\caption{Flow diagrams of the background sector for the exponential parametrization at several finite RG times
 $t\equiv\ln(k/k_0)$. Again, horizontal (vertical) axes show $\lB$ ($\gB$). As in Figure \ref{fig:LinParamMovingFP}
 we observe a moving, UV-attractive non-Gaussian fixed point whose existence and position depends on the RG parameter
 $t$. In the last figure ($t\approx 3.5$) the flow diagram has almost converged to its final form at $t\to\infty$.}
\label{fig:ExpParamMovingFP}
\end{figure}

In this way, we uncover the same running attractor mechanism as for the linear parametrization, based on a moving, UV
attractive non-Gaussian fixed point. In order to achieve background independence in the IR we choose the unique
trajectory in the background sector which ``starts'' (w.r.t.\ the inverse RG flow) at the IR position of the moving
fixed point.\footnote{As in Ref.\ \cite{BR14} we assume that the limit $k\to 0$ exists in order to demonstrate the
principle of the mechanism. Due to the singularity line in the `Dyn' sector, we do not ``start'' at $k=0$, though, but
rather at some finite IR scale.} This trajectory remains finite for all scales $k$, and in the limit $k\to\infty$ it
approaches the ``end position'' of the running attractor.
In Figure \ref{fig:NewBiSharp} we show the graph of this trajectory (pertaining to all scales from the IR to the UV)
as well as the final state of the `B' vector field at the scale $k\to\infty$.
\begin{figure}[tp]
\begin{minipage}{0.78\columnwidth}
 \includegraphics[width=\columnwidth]{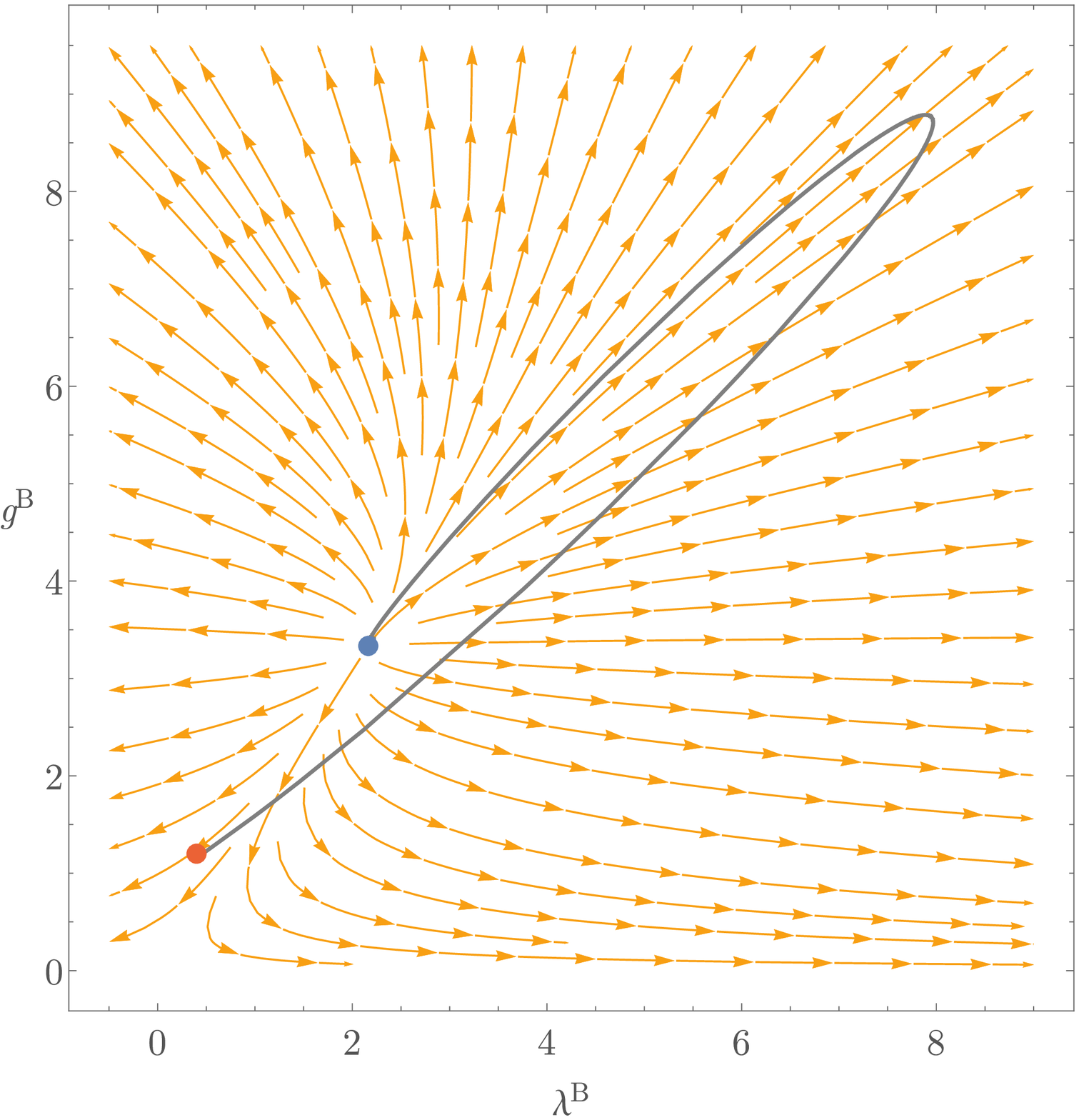}
\end{minipage}
\hfill
\hspace{-0.11\columnwidth}
\begin{minipage}{0.29\columnwidth}
 \vspace{2.5em}
 \includegraphics[width=\columnwidth]{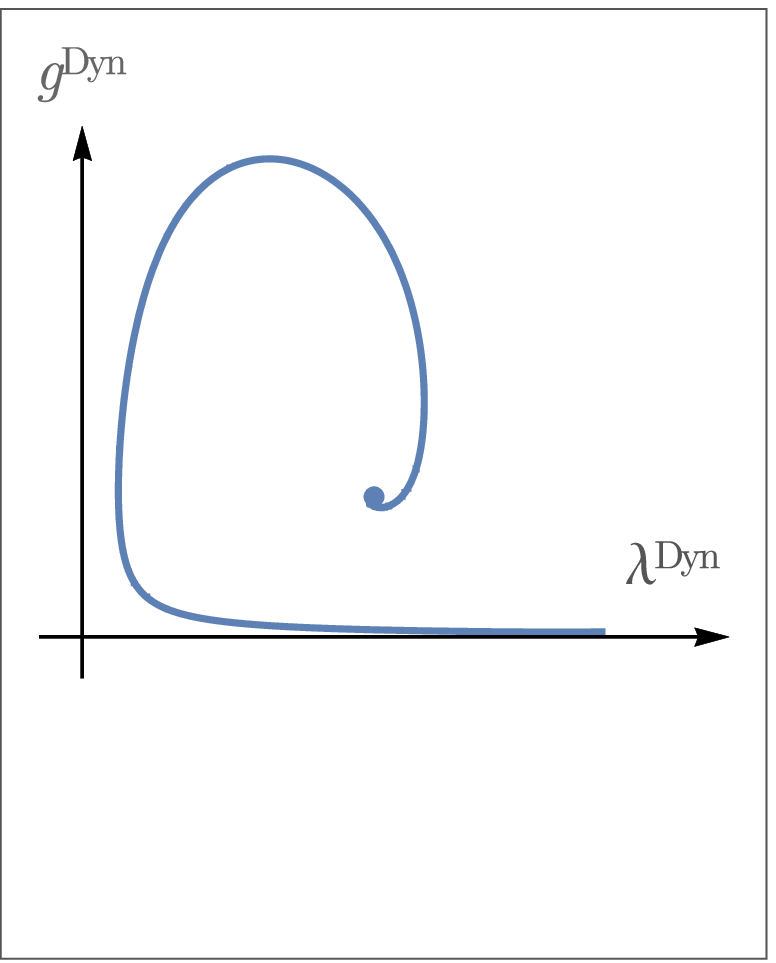}
 \begin{textblock}{2.5}(0.254,-0.8)
 \begin{center}
  \small
  Underlying 'Dyn'\\ trajectory
 \end{center}
 \end{textblock}
\end{minipage}
\caption{Vector field for the background couplings at $k\rightarrow\infty$ and RG trajectory
 that is asymptotically safe in the UV and restores split-symmetry in the IR (left
 figure), and underlying trajectory in the 'Dyn' sector (right figure), based on the
 \emph{exponential parametrization} and the sharp cutoff in $d=4$.}
\label{fig:NewBiSharp}
\end{figure}

Even though the curve of the marked trajectory in Figure \ref{fig:NewBiSharp} has a different form as compared with the
one in Figure \ref{fig:StdBiOpt}, it has the same essential properties. In particular, it \emph{restores split-symmetry
in the infrared and is asymptotically safe at the same time}, making it an eligible candidate for defining
a fundamental theory of gravity.

To summarize, the possibility to achieve background independence seems to depend in a crucial way on the underlying
cutoff shape function if the exponential parametrization is used. This cutoff dependence, however, is merely due to the
unphysical singularity line in the dynamical coupling sector, cf.\ also Section \ref{sec:singleExpRemarks}. We have
demonstrated by means of a sharp cutoff that the split-symmetry restoration mechanism works in principle for the
exponential parametrization, as it did for the linear parametrization.

%----------------------------------------------------------------------------------------------------------------------
\subsubsection[The exponential parametrization in \texorpdfstring{$d=2+\ve$}{d = 2 + epsilon} dimensions]%
{The exponential parametrization in \texorpdfstring{\bm{$d=2+\ve$}}{d = 2 + epsilon} dimensions}
%----------------------------------------------------------------------------------------------------------------------

Finally, let us discuss $\beta$-functions and fixed points in $2+\ve$ dimensions. For the exponential parametrization a
numerical analysis based on eqs.\ \eqref{eq:betagDynBi} -- \eqref{eq:betalBBi} reveals the somewhat unusual situation
that $\lDyn_*$ is of the order $\ve^2$. The remaining couplings, on the other hand, are again of the order $\ve$ at the
NGFP, so we have
\begin{align}
 \gDyn_* &=\mO(\ve)\,, & \lDyn_* &=\mO(\ve^2)\,, \\
 \gB_* &=\mO(\ve)\,, & \lB_* &=\mO(\ve)\,.
\end{align}
Consequently, for an analytical calculation in the vicinity of the NGFP we must set
\begin{align}
 \gDyn_k &\equiv\rgDyn_k\,\ve\,, & \lDyn_k &\equiv\rrlDyn_k\,\ve^2\,, \\
 \gB_k &\equiv\rgB_k\,\ve\,, & \lB_k &\equiv\rlB_k\,\ve\,,
\end{align}
where $\rgDyn_k$, $\rrlDyn_k$, $\rgB_k$ and $\rlB_k$ are of the order $\mO(\ve^0)$. When inserting this into the
$\beta$-functions and expanding in terms of $\ve$ as in Section \ref{sec:biLin} we obtain
\begin{align}
 \beta_g^\text{Dyn} &= \rgDyn\left(\frac{16\, \rgDyn\, \rrlDyn}{3}+1\right)\ve^2 + \mO(\ve^3),
 \label{eq:betagD}\\
 \beta_\lambda^\text{Dyn} &=  \rrlDyn\Big(8\, \rgDyn -2 \Big)\ve^2 + \mO(\ve^3),
 \label{eq:betalD}
\end{align}
in the `Dyn' sector, and
\begin{align}
 \beta_g^\text{B} &= \rgB\left(1-\frac{38}{3}\,\rgB\right) \ve^2 + \mO(\ve^3),
 \label{eq:betagB}\\ 
 \beta_\lambda^\text{B} &=  -2 \Big(\rgB\, \Phi^1_1(0)+\rlB\Big) \ve + \mO(\ve^2),
 \label{eq:betalB}
\end{align}
in the `B' sector, where we have already evaluated those threshold functions that are independent of the cutoff (cf.\
App.\ \ref{app:Cutoffs}). Note
that eqs.\ \eqref{eq:betagD} -- \eqref{eq:betagB} are completely cutoff independent, giving rise to
\emph{universal} fixed point values and coefficients $b^\text{Dyn}$ and $b^\text{B}$, defined by $\beta_g^\text{Dyn}
= \ve\mku \gDyn - b^\text{Dyn}\mku (\gDyn)^2$ and $\beta_g^\text{B} = \ve\mku \gB - b^\text{B}\mku (\gB)^2$,
respectively, up to higher orders. By the relations $b^\text{Dyn} = 1/\rgDyn_*$ and $b^\text{B} = 1/\rgB_*$
we obtain the \emph{universal result}
\begin{equation}[b]
 b^\text{Dyn} = \frac{12}{3} \qquad \text{and} \qquad b^\text{B} = \frac{38}{3}\,.
\end{equation}

As a test, we convince ourselves that the sum of these coefficients equals the result of the single-metric computation,
according to the general rule \eqref{eq:genRuleSumB}. We find
\begin{equation}
 b^\text{Dyn}+b^\text{B} = \frac{50}{3}\,,
\end{equation}
in agreement with the single-metric number based on the exponential parametrization, derived in Section
\ref{sec:singleExp2}.

It is highly remarkable that the \emph{background} coefficient $b^\text{B}$ of the \emph{bimetric} truncation with
the \emph{exponential} parametrization equals precisely the coefficient $b^\text{sm}$ of the \emph{single-metric}
computation based on the \emph{linear} parametrization:\footnote{The reason for this result is rather technical and can
be traced back to a surprising interplay of the conformal projection and the exponential parametrization. Like the fact
that the exponential parametrization in a single-metric truncation gives rise to additional terms as compared with the
linear parametrization, the higher levels of a conformally projected bimetric truncation represent additional terms,
too. In $d=2+\ve$ dimensions, the additional terms have the same effect in both cases (due to the similarity of the
relations $g_\mn=\bg_{\mu\rho}(\e^h)^\rho{}_\nu$ and $g_\mn=\bg_\mn\,\e^{2\mku\Omega}$). Concerning the bimetric case,
it is only the coefficient $b^\text{Dyn}$ that contains the additional terms since it is derived from the level
$\Omega^1$ in the conformal projection process. By eq.\ \eqref{eq:genRuleSumB} we have $b^\text{B}=b^\text{sm}
-b^\text{Dyn}$, so we subtract the additional terms from the full single-metric result (based on the exponential
parametrization). Hence, this difference equals precisely the single-metric coefficient for the linear
parametrization.}
$b^\text{B} = b^\text{sm} = 38/3$.

%----------------------------------------------------------------------------------------------------------------------
\section{Summarizing remarks}
%----------------------------------------------------------------------------------------------------------------------

In this chapter we have investigated the properties of the nonstandard exponential metric parametrization, in
particular with regard to the RG flow, and compared the results with the standard linear parametrization. We conclude
with a couple of general comments.
\medskip

\noindent
\textbf{(1)}
When inserting the exponential relation $g_{\mu\nu}=\bg_{\mu\rho}(\e^h)^\rho{}_\nu$ into the classical
Einstein--Hilbert action and expanding in orders of $h_\mn$ we obtain
\begin{equation}
\begin{split}
S^\text{EH}[g] &= S^\text{EH}\big[\bg \e^{\mku\bg^{-1}h}\big]=S^\text{EH}\big[\bg+h+\mO(h^2)\big] \\
&= S^\text{EH}[\bg] + \int\dd x\frac{\delta S^\text{EH}}{\delta g_\mn(x)}h_\mn(x) + \mO(h^2).
\end{split}
\end{equation}
Thus, the equations of motion are given by those of the linear parametrization,
\begin{equation}
 \frac{\delta S^\text{EH}}{\delta h_\mn}\bigg|_{g=\bg}
= \frac{\delta S^\text{EH}}{\delta g_\mn}\bigg|_{g=\bg}
= \frac{1}{16\pi G}\big(\bar{G}^\mn+\bg^\mn\Lambda\big) = 0\,,
\end{equation}
i.e.\ \emph{the two parametrizations give rise to equivalent theories at the classical level}. It is only the quantum
theory that might reveal the differences.
\medskip

\noindent
\textbf{(2)}
Since $g_\mn=\bg_\mn+h_\mn$ and $g_{\mu\nu}=\bg_{\mu\rho}(\e^h)^\rho{}_\nu$ parametrize different objects (arbitrary
signature tensor fields and pure metrics, respectively), we expect that they give rise to \emph{different quantum
theories} or that they describe \emph{different universality classes}. First evidence for this expectation is provided
by our studies of $\beta$-functions and fixed points in Sections \ref{sec:single} and \ref{sec:bi}. Most notably, we
have calculated the gravitational central charge in $d=2+\ve$ dimensions: For pure gravity, the linear parametrization
gives rise to $c_\text{grav}=19$, while the exponential parametrization reproduces the result known from conformal
field theory, $c_\text{grav}=25$.
\medskip

\noindent
\textbf{(3)}
We have explained in Section \ref{sec:Birth} why the exponential parametrization is particularly appropriate in
$d=2+\ve$ dimensions: In a conformally reduced setting there is a way of parametrizing the conformal
factor which is distinguished in that gives rise to the most natural quadratic form of the kinetic term in the action,
and whose 2D limit \emph{generates} the desired exponential. Since the conformal reduction agrees with the exact theory
in $2$ dimensions, the special status of exponentials in/near 2D is conjectured to hold more general, including the
``nonreduced'' case.
\medskip

\noindent
\textbf{(4)}
\emph{The role of Newton's constant is changed} for the exponential parametrization. This can be understood as follows.
In order to identify the Newton coupling $G_k$ with the strength of the gravitational interaction in the linear
parametrization, one usually rescales the fluctuations $h_\mn$ such that
\begin{equation}
 g_\mn = \bg_\mn + \sqrt{32\pi G_k}\, h_\mn \,.
\end{equation}
In this way, the kinetic term for $h_\mn$ does not contain any contribution from $G_k$, while each gravitational vertex
which has $n$ legs is associated with the factor $(\sqrt{32\pi G_k}\,)^{n-2}$.
For the exponential parametrization we can consider a similar rescaling of $h_\mn$, leading to the same factor
appearing in the $n$-point functions. The difference resides in the fact that there are new terms and structures in
$\Gamma_k^{(n)}$ when using the exponential parametrization. As already indicated in equation \eqref{eq:2ndVar}, these
additional contributions to each vertex are due to the chain rule. Hence, the Newton constant is associated to
different terms in the $n$-point functions.
\medskip

\noindent
\textbf{(5)}
For the exponential parametrization results depend to a larger extent on the cutoff shape function. It is somewhat
unexpected that the sharp cutoff leads to the most convincing results. We have argued, however, that this cutoff
dependence is mainly due to the closer distance between the singularity line and the NGFP. Slight modifications of the
setting may solve the issue. (a) The nonlinear relation for the metric might attach more importance to the truncated
higher order terms. More general truncations might shift or even remove the singularity such that we obtain a clearer
picture. (b) In the terminology of Ref.\ \cite{CPR09}, our calculations are based on a type I cutoff. As has been
argued in Ref.\ \cite{DP13}, in a few situations it is only the type II cutoff that leads to correct physical results,
whereas the type I cutoff does not, an example being the presence of a limit cycle (cf.\ Sec.\ \ref{sec:singleExp4}).
(c) In Section \ref{sec:singleExpRemarks} we reviewed a couple of arguments that already minor modifications in the
gauge, or (d) in the choice of basic field variables (field redefinition), lead to considerably more reliable results.
\medskip

\noindent
\textbf{(6)}
After all, the answer to the question which parametrization should be used depends on the desired application
and on which other approach the calculation is to be compared with.

%----------------------------------------------------------------------------------------------------------------------
\chapter{The 2D limit of the Einstein--Hilbert action}
\label{chap:EHLimit}
%----------------------------------------------------------------------------------------------------------------------

\begin{summary}
Classical gravity is most conveniently described by the Einstein--Hilbert action, and we have previously discussed the
significance of the Einstein--Hilbert truncation, $\frac{1}{16\pi G_k} \int \dd x \sg \,( -R + 2\mku\Lambda_k)$, for
the quantum theory. In $d=2$ dimensions, however, the term $\int \dd x \sg \, R\mku$ becomes a topological invariant.
Being independent of the metric and thus not giving rise to any equations of motion, it does no longer seem to define
an appropriate action. On the other hand, we showed in Chapter \ref{chap:ParamDep} that the Newton coupling in
$d=2+\ve$ dimensions is of the order $\ve$. Hence, the prefactor $\frac{1}{G_k}$ attaches an increasing weight to
$\int \td^{2+\ve} x \sg \, R\mku$. Loosely speaking, the action becomes more and more trivial, while its prefactor
makes it more and more important. In this chapter we show that $\frac{1}{\ve}\int \td^{2+\ve} x \sg \, R\mku$ actually
approaches a nontrivial, finite limit as $\ve\to 0$. It consists of Polyakov's induced gravity action, $\int \td^2 x
\sg\, R\, \Box^{-1}R$, as well as purely topology dependent contributions. Hence, the local Einstein--Hilbert action
has turned into a nonlocal action in the limit. Our discussion includes a consideration of zero modes of the Laplacian
which become crucial for terms involving $\Box^{-1}$.

\noindent
\textbf{What is new?} The method of establishing the 2D limit of the Einstein--Hilbert action (Secs.\
\ref{sec:IndGravityFromEH} \& \ref{sec:LimitFullEH}); taking into account zero modes (Sec.\ \ref{sec:EpsilonLimit} \&
App.\ \ref{app:Zero}).

\noindent
\textbf{Based on:} Ref.\ \cite{NR16a}.
\end{summary}
\vspace{1em}
%----------------------------------------------------------------------------------------------------------------------

In the previous chapter we studied the properties of the coupling constants, their RG evolution and, in
particular, their behavior near two dimensions. Up to this point, however, we have not discussed what happens in the 2D
limit to the underlying action itself. Does it change? If so, does it remain finite? Is it still an appropriate action?
In order to approach these questions, we again start out from the Einstein--Hilbert truncation of the EAA in
$d=2+\ve>2$ dimensions, 
\begin{equation}
 \Gamma_k^\text{grav}[g] = \frac{1}{16\pi G_k} \int \dd x \sg \,\big( -R + 2\mku\Lambda_k \big).
\label{eq:EHfunctional}
\end{equation}
As shown in the preceding chapter, the dimensionless couplings, $g_k\equiv G_k k^{d-2}=G_k k^\ve$ and $\lambda_k
\equiv k^{-2}\Lambda_k$, are of the order $\ve$ in the vicinity of the non-Gaussian fixed point, leading to $G_k
\propto\ve$ and $\Lambda_k\propto\ve$, respectively. (It can be argued that a similar relation should hold for the
classical Newton constant, too \cite{Mann1992}: $G\propto\ve$.) Hence, the pure volume part of the action,
$\frac{\Lambda_k}{8\pi G_k}\int\td^{2+\ve}x\sg\mku$, remains finite and well defined in the limit $\ve\to 0$.
It is the curvature part of $\Gamma_k^\text{grav}$, though, that requires a closer inspection. In what follows, we
investigate the nature of its $\ve\rightarrow 0$ limit, and finally construct a \emph{manifestly $2$-dimensional}
action which describes 2D Asymptotic Safety without reverting to ``higher'' dimensions in any way.

In \emph{exactly} $2$ dimensions the Gauss--Bonnet theorem states that the integral of the scalar curvature,
$\int\td^2 x\sg\,R$, is a purely
topological term,
\begin{equation}
 \int_M\td^2 x\mku \sg\,R = 4\pi\,\chi(M)\,,
\end{equation}
where $\chi$ denotes the Euler characteristic, a topological invariant that measures the number of handles of the
manifold $M$. In particular, it is independent of the metric and does not imply any local dynamics. Thus, one might
expect that the curvature part of \eqref{eq:EHfunctional} becomes trivial when $d$ approaches $2$. However, the $1/\ve$
pole entailed by the prefactor $1/G_k$ gives so much weight to $\int\td^{2+\ve} x\sg\,R$ that the limit $\ve\to 0$
in fact remains nontrivial. Making sense of this limit requires some kind of generalized L'H\^{o}pital's rule.

We will present a new argument in this chapter showing that the (local) Einstein--Hilbert action turns into a
\emph{nonlocal} action in the limit $d\rightarrow 2$ whose most essential part is given by Polyakov's induced gravity
action.

Our proof will confirm recurring speculation \cite{CD15} that the induced gravity action is the natural
$2$-dimensional analogue of the Einstein--Hilbert action in $d>2$ as both actions determine field equations for the
metric in their respective spacetime dimension.
Here we go one step further, though: We do not require that one action has to be replaced
by the other one when switching between $d=2$ and $d>2$. The idea is rather to say that there is \emph{only one} common
origin, the Einstein--Hilbert action in a general dimension $d$, and that \emph{the induced gravity action emerges
automatically when $d$ approaches $2$}.

It is this latter 2D action, analyzed at the NGFP, that establishes the contact between the Asymptotic Safety studies
within the Einstein--Hilbert truncation and $2$-dimensional conformal field theory. In Chapter \ref{chap:NGFPCFT} it
will form the basis of our investigations concerning central charges and unitarity.

We start by reviewing the special role of self-consistent backgrounds in Section \ref{sec:StressEnergy}. In particular,
we re-interpret the effective Einstein equation as a tadpole condition and the trace of the stress-energy tensor due to
metric fluctuations as a kind of classical ``trace anomaly''. Here, all calculations are performed in $2+\ve$
dimensions, and the 2D limit is taken at the very end only.
This leads us to the question if the same trace anomaly could be obtained when starting out from a strictly 2D action.
The answer to this question will be given in Section \ref{sec:IndGravityFromEH} where we compute the 2D limit of
the Einstein--Hilbert action at the NGFP and argue that it results indeed in an action with the sought-for properties.
Details of the computation, including various useful identities for Weyl transformations and a thorough discussion
of the induced gravity action in the presence of zero modes, are given in Appendix \ref{app:Weyl}.

%----------------------------------------------------------------------------------------------------------------------
\section{The 2D limit at the level of the gravitational stress-energy tensor}
\label{sec:StressEnergy}
%----------------------------------------------------------------------------------------------------------------------

In this preparatory section we collect a number of results concerning the implementation of background independence in
the EAA framework which actually does employ (unspecified) background fields, cf.\ Sec.\ \ref{sec:BFF}. In particular,
we introduce the energy-momentum tensor of metric fluctuations in a background, as well as an associated ``trace
anomaly''. The latter will be used in Chapter \ref{chap:NGFPCFT} in order to identify the conformal field theory at the
heart of Asymptotic Safety in $2$ dimensions.

%----------------------------------------------------------------------------------------------------------------------
\subsection{The effective Einstein equation re-interpreted}
\label{sec:EEE}
%----------------------------------------------------------------------------------------------------------------------

Let us consider a generic effective average action $\Gamma_k[\Phi,\bP]\equiv \Gamma_k[\vp;\bP]$ involving a multiplet of
dynamical fields $\big\langle\hP^i\big\rangle\equiv \Phi^i$, associated background fields $\bP^i$, and fluctuations $\vp^i
\equiv \langle \hvp^i\rangle = \Phi^i - \bP^i$.\footnote{For the sake of argument we consider a linear field
parametrization here. A generalization to arbitrary parametrizations, $\Phi^i=\Phi^i[\vp;\bP]$, i.e.\ $\vp^i \equiv
\langle \hvp^i\rangle = \vp^i[\Phi,\bP]$, is straightforward, cf.\ Sec.\ \ref{sec:Applications}.} The effective average
action implies a source $\leftrightarrow$ field relationship which contains an explicit cutoff term linear in the
fluctuation fields:
\begin{equation}
 \frac{1}{\sbg}\frac{\delta\Gamma_k[\vp;\bP]}{\delta\vp^i(x)}+\Rk[\bP]_{ij}\,\vp^j(x)=J_i(x) \,.
\label{eq:so-fi}
\end{equation}
By definition, self-consistent backgrounds are field configurations $\bP(x)\equiv \bP_k^\text{sc}(x)$ which allow
$\vp^i=0$ to be a solution of \eqref{eq:so-fi} with $J_i=0$. A self-consistent background is particularly ``liked'' by
the fluctuations, in the sense that they leave it unaltered on average: $\langle\hP\rangle = \bP+\langle\hvp\rangle =
\bP_k^\text{sc}$. These special backgrounds are determined by the tadpole condition $\langle\hvp^i\rangle=0$, which
reads explicitly
\begin{equation}
 \frac{\delta}{\delta\vp^i(x)}\Gamma_k[\vp;\bP]\Big|_{\vp=0,\,\bP=\bP_k^\text{sc}} = 0 \,.
\label{eq:tapo}
\end{equation}
Equivalently, in terms of the full dynamical field,
\begin{equation}
 \frac{\delta}{\delta\Phi^i(x)}\Gamma_k[\Phi,\bP]\Big|_{\Phi=\bP=\bP_k^\text{sc}} = 0 \,.
\label{eq:tapo-full}
\end{equation}

Here, we consider actions of the special type
\begin{equation}
 \Gamma_k[g,\xi,\xb,A,\bg] = \Gg[g,\bg] +\Gm[g,A,\bg] +\Gamma_k^\text{gf}[g,\bg] +\Gamma_k^\text{gh}[g,\xi,\xb,\bg] .
\label{eq:GammaAnsatzAgain}
\end{equation}
These functionals include a purely gravitational piece, $\Gg$, furthermore a (for the time being) generic matter action
$\Gm$, as well as gauge fixing and ghost terms, $\Gamma_k^\text{gf}$ and $\Gamma_k^\text{gh}$, respectively. Concerning the
latter, only the following two properties are needed at this point:
(i) The $h_\mn$-derivative of the gauge fixing functional $\Gamma_k^\text{gf}[h;\bg] \equiv 
\Gamma_k^\text{gf}[\bg+h,\bg]$ vanishes at $h_\mn=0$. This is the case, for example, for classical gauge fixing terms
$S^\text{gf} \propto \int(\mF h)^2$ which are quadratic in $h_\mn$.
(ii) The functional $\Gamma_k^\text{gh}$ is ghost number conserving, i.e.\ all terms contributing to it have an equal
number of ghosts $\xi^\mu$ and antighosts $\xb_\mu$. Again, classical ghost kinetic terms
$\propto \int\xb\mathcal{M}\xi$ are of this sort.

Thanks to these properties, $\Gamma_k^\text{gf}$ drops out of the tadpole equation \eqref{eq:tapo-full}, and it
follows that $\xi=0=\xb$ is always a consistent background for the Faddeev--Popov ghosts. Adopting this background
for the ghosts, \eqref{eq:tapo-full} boils down to the following conditions for self-consistent metric and matter
field configurations $\gsc$ and $\Asc$, respectively:
\begin{align}
 0 &= \frac{\delta}{\delta g_\mn(x)}\Big\{\Gg[g,\bg]+\Gm[g,\Asc,\bg]\Big\}\Big|_{g=\bg=\gsc} \;,
 \label{eq:selfcon} \\
 0 &= \frac{\delta}{\delta A(x)}\Gm[g,A,\bg]\Big|_{g=\bg=\gsc,\, A=\Asc} \;.
 \label{eq:selfconmatter}
\end{align}
Introducing the stress-energy (energy-momentum) tensor of the matter field,
\begin{equation}
 T^\text{m}[\bg,A]^\mn(x) \equiv \frac{2}{\sbgx} \frac{\delta}{\delta g_\mn(x)} \Gm[g,A,\bg]\Big|_{g=\bg} \;,
\end{equation}
the first condition, equation \eqref{eq:selfcon}, becomes
\begin{equation}
 0 = \frac{2}{\sbgx} \frac{\delta}{\delta g_\mn(x)} \Gg[g,\bg]\Big|_{g=\bg=\gsc} + T^\text{m}[\gsc,\Asc]^\mn(x) .
\label{eq:EFE}
\end{equation}
This relation plays the role of an effective gravitational field equation which, together with the matter
equation \eqref{eq:selfconmatter}, determines $\gsc$ and $\Asc$. Structurally, eq.\ \eqref{eq:EFE} is a generalization
of the classical Einstein equation to which it reduces if $\Gg[g,\bg]\equiv \Gg[g]$ happens to have no ``extra
$\bg$-dependence'' \cite{MR10} and to coincide with the Einstein--Hilbert action; then the
$\delta/\delta g_\mn$-term in \eqref{eq:EFE} is essentially the Einstein tensor $G_\mn$.

In this very special background-free case we recover the familiar setting of classical General Relativity where there is
a clear logical distinction between matter fields and the metric, meaning the full one, $g_\mn$, while none other appears
in the fundamental equations then. It is customary to express this distinction by putting $G_\mn$ on the LHS of Einstein's
equation, the side of gravity, and $T_\mn^\text{m}$ on the RHS, the side of matter.

In the effective average action approach where, for both deep conceptual and technical reasons \cite{MR10,BR14}, the
introduction of a background is unavoidable during the intermediate calculational steps, this categorical distinction of
matter and gravity, more precisely, matter fields and metric fluctuations, appears unmotivated. It is much more natural to
think of $h_\mn$ as a \emph{matter} field which propagates on a background spacetime furnished with the metric $\bg_\mn$.

Adopting this point of view, we interpret the $\delta/\delta g_\mn$-term in \eqref{eq:EFE} as the energy-momentum tensor of
the $h_\mn$-field, and we define
\begin{equation}[b]
 T^\text{grav}[\bg]^\mn(x) \equiv \frac{2}{\sbgx} \frac{\delta}{\delta g_\mn(x)} \Gg[g,\bg]\Big|_{g=\bg}
 = \frac{2}{\sbgx} \frac{\delta}{\delta h_\mn(x)} \Gg[h;\bg]\Big|_{h=0} \;.
\end{equation}
The tadpole equation \eqref{eq:EFE} turns into an Einstein equation with zero LHS then:
\begin{equation}[b]
 0 = T_\mn^\text{grav}\big[\gsc\big] + T_\mn^\text{m}\big[\gsc,\Asc\big] .
\label{eq:Tadpole}
\end{equation}
It states that for a background to be self-consistent, the total energy-momentum tensor of matter and metric
fluctuations, in this background, must vanish. (In the general case there could also be a contribution from the
ghosts.)

%----------------------------------------------------------------------------------------------------------------------
\subsection[The stress-energy tensor of the \texorpdfstring{$h_\mn$}{h}-fluctuations]
 {The stress-energy tensor of the \texorpdfstring{$\bm{h_\mn}$}{h}-fluctuations}
\label{sec:Stresshmunu}
%----------------------------------------------------------------------------------------------------------------------

Note that in general, $T_\mn^\text{grav}$ is not conserved, $\bar{D}_\mu T^\text{grav}[\bg]^\mn\neq 0$, since due to the
presence of two fields in $\Gg$ the standard argument does not apply. Of course, it is conserved in the special
case $\Gg[g,\bg]\equiv\Gg[g]$ when there is no extra $\bg$-dependence.

For example, choosing $\Gg[g]$ to be the single-metric Einstein--Hilbert functional \eqref{eq:EHfunctional},
the corresponding energy-momentum tensor of the $h_\mn$-fluctuations is given by the divergence-free expression
\begin{equation}
 T_\mn^\text{grav}[\bg] = \frac{1}{8\pi G_k} \Big(\bar{G}_\mn + \Lambda_k\, \bg_\mn \Big) ,
\label{eq:EHEMTensor}
\end{equation}
with $\bar{G}_\mn$ the Einstein tensor built from $\bg_\mn$. The trace of the energy-momentum tensor
\eqref{eq:EHEMTensor} reads
\begin{equation}
 \Theta_k[\bg] \equiv \bg^\mn\, T_\mn^\text{grav}[\bg] = \frac{1}{16\pi G_k}\Big[ -(d-2)\bar{R}+2d\,\Lambda_k \Big],
\end{equation}
where $\bar{R}\equiv R(\bg)$. A remarkable feature of this trace is that it possesses a completely well defined,
unambiguous limit $d\rightarrow 2$ if $G_k$ and $\Lambda_k$ are of first order in $\ve=d-2$. In terms of the finite
quantities $\rGk \equiv G_k/\ve$ and $\rLk \equiv \Lambda_k/\ve$ which are of the order $\ve^0$, we have
\begin{equation}
 \begin{split}
  \Theta_k[\bg] &= \frac{1}{16\pi\rGk} \Big[-\bar{R}+4\mku\rLk\Big] +\mO(\ve) \\
    &= \frac{1}{16\pi\rgk} \Big[-\bar{R}+4k^2\mku\rlk\Big] +\mO(\ve).
 \end{split}
\label{eq:Theta2}
\end{equation}
In the second line of \eqref{eq:Theta2} we exploited that in exactly two dimensions the dimensionful and
dimensionless Newton constant are equal, so $g_k = G_k$ and $\rgk = \rGk$, while, as always,
$\lambda_k \equiv \Lambda_k/k^2$, hence $\rlk=\rLk/k^2$.

When the underlying RG trajectory is in the NGFP scaling regime, the dimensionless couplings are scale
independent, and
\begin{equation}
 \Theta_k^\text{NGFP}[\bg] = \frac{1}{16\pi\rgs} \Big[-\bar{R}+4\mku\rls k^2 \Big].
\end{equation}
Using the representation $g_*\equiv \ve/b$ as in Chapter \ref{chap:ParamDep} and Refs.\
\cite{Weinberg1979,NR13,NR13b,NR15a,Nink2015,BR14} we obtain
\begin{equation}[b]
 \Theta_k^\text{NGFP}[\bg] = \Big({\textstyle\frac{3}{2}}b\Big) \, \frac{1}{24\pi}
  \Big[-\bar{R}+4\mku\rls k^2 \Big].
\label{eq:Theta3}
\end{equation}
Here and in the following, we consider $\Theta_k$ and $\Theta_k^\text{NGFP}$ as referring to \emph{exactly $2$
dimensions}, in the sense that the limit has already been taken, and we omit the ``$\mO(\ve)$'' symbol.

%----------------------------------------------------------------------------------------------------------------------
\subsection{The intrinsic description in exactly 2 dimensions}
\label{sec:descTwoD}
%----------------------------------------------------------------------------------------------------------------------

In this chapter we would like to describe the limit $d\rightarrow 2$ of Quantum Einstein Gravity (QEG) in an
intrinsically $2$-dimensional fashion, that is, in terms of a new functional $\Gamma_k^\text{grav,2D}$ whose arguments
are fields in strictly $2$ dimensions, and which no longer makes reference to its ``higher'' dimensional origin. Since
the Einstein--Hilbert term is purely topological in exactly $d=2$, it is clear that the sought-for action must have a
different structure.

\medskip
\noindent
\textbf{(1)}
One of the conditions which we impose on $\Gamma_k^\text{grav,2D}$ is that it must reproduce the trace $\Theta_k$
computed in $d>2$, since we saw that this quantity has a smooth limit with an immediate interpretation in $d=2$ exactly:
\begin{equation}
 2 g_\mn \frac{\delta}{\delta g_\mn} \Gamma_k^\text{grav,2D}[g,\bg]\Big|_{g=\bg}=\sbg\, \Theta_k[\bg].
\label{eq:RelEMTensors}
\end{equation}
Furthermore, if $\Gg$ is a single-metric action, we assume that $\Gamma_k^\text{grav,2D} \equiv \Gamma_k^\text{grav,2D}[g]$
has no extra $\bg$-dependence either. The condition \eqref{eq:RelEMTensors} fixes its response to an infinitesimal Weyl
transformation then:
\begin{equation}
 2 g_\mn(x) \frac{\delta}{\delta g_\mn(x)} \Gamma_k^\text{grav,2D}[\mku g] 
 \equiv \frac{\delta}{\delta \sigma(x)} \Gamma_k^\text{grav,2D}\big[\e^{2\sigma} g\big]\Big|_{\sigma=0}
 = \sqrt{g(x)}\,\Theta_k[g](x).
\label{eq:InfWeyl}
\end{equation}
For the example of the Einstein--Hilbert truncation, $\Theta_k$ is of the form
\begin{equation}
 \Theta_k[g]=a_1(-R+a_2),
\label{eq:ThetaEH} 
\end{equation}
with constants $a_1,a_2$ which can be read off from \eqref{eq:Theta2} -- \eqref{eq:Theta3} for the various cases.

\medskip
\noindent
\textbf{(2)}
It is well known how to integrate equation \eqref{eq:InfWeyl} in the conformal gauge \cite{Polyakov1981}. By setting
\begin{equation}
 g_\mn(x) = \e^{2\phi(x)}\mku \hg_\mn(x),
\end{equation}
with a fixed reference metric $\hg_\mn$ (conceptually unrelated to $\bg_\mn$), one for each topological sector, and
taking advantage of the identities listed in Appendix \ref{app:Weyl}, eq.\ \eqref{eq:InfWeyl} with \eqref{eq:ThetaEH}
is turned into
\begin{equation}
 \frac{\delta}{\delta \phi(x)} \Ggd\big[\e^{2\phi}\mku \hg\big] = a_1\shgx \Big[2\mku\hD_\mu\hD^\mu\phi(x)-\hR(x)
  +a_2 \,\e^{2\phi(x)}\Big].
\end{equation}
The general solution to this equation is easy to find:
\begin{equation}[b]
 \Ggd\big[\e^{2\phi}\mku\hg\big] = \GL[\phi;\hg]+U_k[\hg].
\label{eq:GravToLiou}
\end{equation}
Here $U_k$ is a completely arbitrary functional of $\hg$, independent of $\phi$, and $\GL$ denotes the \emph{Liouville
action} \cite{RW97}:
\begin{equation}
 \begin{split}
 \GL[\phi;\hg] &= (-2a_1)\int\td^2x\shg \left({\frac{1}{2}}\hD_\mu\phi\hD^\mu\phi
  + {\frac{1}{2}}\hR\mku\phi - {\frac{a_2}{4}}\e^{2\phi} \right) \\
 &= (-2a_1)\,\Delta I[\phi;\hg] + \frac{1}{2}a_1 a_2\int\td^2x\shg\,\e^{2\phi} \,.
 \end{split}
\label{eq:LiouvilleAction}
\end{equation}
In the last line we employed the normalized functional
\begin{equation}
 \Delta I[\phi;g] \equiv \frac{1}{2}\int\td^2x\sg\,\big(D_\mu\phi D^\mu\phi+R\mku \phi\big).
\label{eq:DeltaIDef}
\end{equation}

While this method of integrating the trace ``anomaly'' applies in all topological sectors, it is \emph{unable to find
the functional} $U_k[\hg]$. Usually, in conformal field theory or string theory this is not much of a disadvantage,
but in quantum gravity where background independence is a pivotal issue it is desirable to have a more complete
understanding of $\Ggd$. For this reason, we next discuss the possibility to take the limit
$\ve\rightarrow 0$ directly at the level of the action.

%----------------------------------------------------------------------------------------------------------------------
\section[The emergence of the induced gravity action]%
{How the induced gravity action emerges from the Einstein--Hilbert action}
\label{sec:IndGravityFromEH}
%----------------------------------------------------------------------------------------------------------------------

In this section we reveal a mechanism which allows us to regard Polyakov's induced gravity action in $2$ dimensions as
the $\ve\rightarrow 0$ limit of the Einstein--Hilbert action in $2+\ve$ dimensions. (Here and in the following
we always consider the case $\ve>0$, i.e.\ the limit $\ve\searrow 0$.) This will confirm the point of view that the
induced gravity action is fundamental in describing $2$-dimensional gravity, while it is less essential for
$d>2$ where gravity is governed mainly by an (effective average) action of the Einstein--Hilbert type. The
dimensional limit exhibits a discontinuity at $d=2$, producing a nonlocal action out of a local one.

\medskip
\noindent
\textbf{(1)}
The crucial ingredient for a nontrivial limit $\ve\rightarrow 0$ is a prefactor of the Einstein--Hilbert action
proportional to $1/\ve$. This occurs whenever the Newton constant is proportional to $\ve$. As mentioned previously,
such a behavior was found in the Asymptotic Safety related RG studies, which showed the existence of a
non-Gaussian fixed point with a Newton constant of the order $\ve$; a result that is independent of the underlying
regularization scheme and parametrization, and that is found in both perturbative and nonperturbative investigations.

In Chapter \ref{chap:Bare} we will see that this property holds not only for the effective, but also for the \emph{bare
action}: Using an appropriate regularization prescription the bare Newton constant is of first order in $\ve$, too.

This is our motivation for considering a generic Einstein--Hilbert action with a Newton constant proportional to $\ve$.
For the discussion in this section it is not necessary to specify the physical role of the action under
consideration -- the arguments apply to both bare and effective (average) actions. In both cases our aim is eventually
to make sense of, and to calculate
\begin{equation}
 \frac{1}{\ve}\int\td^{2+\ve}x\sg\,R
\label{eq:LimitInt}
\end{equation}
in the limit $\ve\rightarrow 0$.

\medskip
\noindent
\textbf{(2)}
It turns out helpful to study the transformation behavior of the Einstein--Hilbert action under \emph{Weyl rescalings}.
Under these transformations an expansion in powers of $\ve$ is more straightforward. Loosely speaking, the reason why
Weyl variations are useful in the 2D limit resides in the fact that the conformal factor is the only dynamical part of
the metric that ``survives'' when the limit $d\to 2$ is taken, i.e.\ the conformal sector captures the most essential
information also in a dimension slightly larger than two, $d=2+\ve$. This circumstance is detailed in Subsection
\ref{sec:ConfGauge}.

Weyl transformations are defined by the pointwise rescaling
\begin{equation}
 g_\mn(x) = \e^{2\sigma(x)} \hg_\mn(x) \,,
\label{eq:WeylTrans}
\end{equation}
with $\sigma$ a scalar function on the spacetime manifold. In Appendix \ref{app:Weyl} we list the transformation
behavior of all geometric quantities relevant to this section.

From \eqref{eq:WeylTrans} it follows that $g_\mn$ is invariant under the Weyl split-symmetry transformations
\begin{equation}
\hg_\mn\rightarrow \e^{2\chi}\hg_\mn \, ,\qquad \sigma\rightarrow \sigma - \chi \, .
\label{eq:SplitSymmetry}
\end{equation}
Thus, any functional of the full metric $g_\mn$ rewritten in terms of $\hg_\mn$ and $\sigma$ is invariant under
\eqref{eq:SplitSymmetry}. On the other hand, a functional of $\hg_\mn$ and $\sigma$ which is not Weyl split-symmetry
invariant cannot be expressed as a functional involving only $g_\mn$, but it contains an ``extra $\hg_\mn$-dependence''
\cite{MR10}.

Before actually calculating the 2D limit of \eqref{eq:LimitInt} in Sections \ref{sec:EpsilonLimit} and
\ref{sec:LimitFullEH} in a gauge invariant manner, we illustrate the situation in Section \ref{sec:ConfGauge} by
employing the conformal gauge, and we give some general arguments in Section \ref{sec:genRem} why and in what sense
the limit is well defined.

%----------------------------------------------------------------------------------------------------------------------
\subsection{Lessons from the conformal gauge}
\label{sec:ConfGauge}
%----------------------------------------------------------------------------------------------------------------------

In exactly $2$ spacetime dimensions any metric $g$ can be parametrized by a diffeomorphism $f$ and a Weyl scaling
$\sigma$:
\begin{equation}
 f^* g = \e^{2\sigma}\,\hg_{\{\tau\}} \,,
\label{eq:MetricDiffWeyl}
\end{equation}
where $f^*g$ denotes the pullback of $g$ by $f$, and $\hg_{\{\tau\}}$ is a fixed reference metric that depends only on
the Teichm\"uller parameters $\{\tau\}$ or ``moduli'' characterizing the underlying topology \cite{IT92}. Stated
differently, \emph{a combined}
Diff$\times$Weyl \emph{transformation can bring any metric to a reference form}. Thus, the moduli space is the
remaining space of inequivalent metrics, $\mathcal{M}_h= \mathcal{G}_h/ (\text{Diff}\times\text{Weyl})_h$, where
$\mathcal{G}_h$ is the space of all metrics on a genus-$h$ manifold.\footnote{For the topology of a sphere
$\mathcal{M}_h=\mathcal{M}_0$ is trivial, while for a torus there is one complex parameter, $\tau$, assuming values in
the fundamental region, $F_0$. Apart from such simple examples it is notoriously involved to find moduli spaces
\cite{IT92}.} Its precise form is irrelevant for the present discussion. Accordingly, if not needed we do not write
down the dependence on $\{\tau\}$ explicitly in the following. Here we consider $\hg$ a reference metric for a fixed
topological sector.

In order to cope with the redundancies stemming from diffeomorphism invariance we can fix a gauge by picking one
representative among the possible choices for $f$ in eq.\ \eqref{eq:MetricDiffWeyl}, the most natural choice being
the conformal gauge:
\begin{equation}
 g_\mn = \e^{2\sigma}\,\hg_\mn \,.
\label{eq:ConfGauge}
\end{equation}
Equation \eqref{eq:ConfGauge} displays very clearly the special role of $2$ dimensions: The metric depends only on the
conformal factor and possibly on some topological moduli parameters. Since the latter are global parameters, we see
that \emph{locally} the metric is determined only by the conformal factor.

\medskip
\noindent
\textbf{(1) Conformal flatness}.
At this point a comment is in order. By choosing an appropriate coordinate system it is always possible to bring a
2D metric to the form
\begin{equation}
 g_\mn = \e^{2\sigma} \delta_\mn\,,
\label{eq:NotConfGauge}
\end{equation}
in the neighborhood of an arbitrary spacetime point, where $\delta_\mn$ is the flat Euclidean metric (see Ref.\
\cite{DFN92} for instance). However, this is only a local property. \emph{For a general metric on a general 2D manifold
there exists no scalar function $\sigma$ satisfying \eqref{eq:NotConfGauge} globally}.\footnote{This can be understood
by means of the following counterexample. Consider the standard sphere $S^2\subset \mathbb{R}^3$ with the induced
metric. Upon stereographic projection the sphere is parametrized by isothermal coordinates, say $(u,v)$, where the
metric assumes the form $g=\frac{4}{(1+u^2+v^2)^2} (\td u^2+\td v^2)$. Setting $\sigma\equiv\ln\left(\frac{2}{1+
u^2+v^2}\right)$ we have $g=\e^{2\sigma}\hg$ with $\hg=\delta$. If we assumed that $g=\e^{2\sigma}\hg$ holds globally
for a valid scalar function $\sigma$, we could make use of identity \eqref{eq:WeylgR} to arrive at a contradiction for
the Euler characteristic $\chi=2\mku$, namely: $8\pi=4\pi\chi\equiv\int\!\sg\,R = \int\!\shg\,(\hR-2\,
\hB\mku\sigma)=-2\int\!\shg\;\hB\mku\sigma=0$, since $\hR=0$ for the flat metric, and since the sphere has vanishing
boundary. A resolution to this contradiction is to take into account that we need (at least) two coordinate patches all
of which have a boundary contributing to $\int\!\sg R$. Decomposing $S^2$ into two half spheres, $H_+$ and $H_-$, for
instance, and using $\hB\mku\sigma=-4/(1+u^2+v^2)^2$, we obtain $\int\!\sg\,R
=-2\int_{H_+}\shg\;\hB\mku\sigma-2\int_{H_-}\shg\;\hB\mku\sigma =8\pi=4\pi\chi$, as it should be.}
Rather must the reference metric in eq.\ \eqref{eq:ConfGauge} be compatible with all topological constraints, like, for
instance, the value of the integral $\int\shg\,\hR$ which is fixed by the Euler characteristic. As a consequence, we
cannot restrict our discussion to a globally conformally flat metric in general.

\medskip
\noindent
\textbf{(2) \bm{$\text{Diff}\times\text{Weyl}$} invariant functionals}.
This has a direct impact on diffeomorphism and Weyl invariant functionals $F:g\mapsto F[g]$. The naive argument
claiming that diffeomorphism invariance can be exploited to make $g_\mn$ conformally flat, and then Weyl invariance to
bring it to the form $\delta_\mn$ such that $F[g]=F[\delta]$ would be independent of the metric, i.e.\ constant, is
wrong actually. The global properties of the manifold destroy this argument.

When choosing appropriate local coordinates
to render $g$ flat up to a Weyl rescaling, there is some information of the metric implicitly encoded in the coordinate
system, e.g.\ in the boundary of each patch, giving rise to a remaining metric dependence in $F$. A combined
Diff$\times$Weyl transformation can bring the metric to unit form, but it changes boundary conditions (like periodicity
constraints for a torus) as well (see e.g.\ Ref.\ \cite{Polchinski1998}). Therefore, $F$ is in fact constant with
respect to local properties of the metric, while it can still depend on global parameters. According to eq.\
\eqref{eq:MetricDiffWeyl} these are precisely the moduli parameters. Hence, \emph{the metric dependence of any 2D
functional which is both diffeomorphism and Weyl invariant is reduced to a dependence on} $\{\tau\}$, and we can write
$F[g]=f\big(\{\tau\} \big)$ where $f$ is a function (not a functional).

\medskip
\noindent
\textbf{(3) Calculating 2D limits}.
Let us come back to the purpose of this subsection, simplifying calculations by employing the conformal gauge
\eqref{eq:ConfGauge}. Following the previous discussion we should not rely on the choice \eqref{eq:NotConfGauge}.
Nevertheless, as an example we may assume for a moment that the manifold's topology is consistent with a metric $\hg$
that corresponds to a flat space, where -- for the above reasons -- conformal flatness is not expressed in local
coordinates as in \eqref{eq:NotConfGauge} but by the coordinate free condition $\hR=0$, which is possible iff the Euler
characteristic vanishes. The general case with arbitrary topologies will be covered in Section \ref{sec:EpsilonLimit}.
We now aim at finding a scalar function $\sigma$ which is compatible with eq.\ \eqref{eq:ConfGauge} with $g_\mn$ given.
Exploiting the identities \eqref{eq:WeylR} and \eqref{eq:TransLaplace2D} given in the appendix with $\hR=0$ we obtain
\begin{equation}
 R=-2\,\Box\mku\sigma \,.
\label{eq:CondSigma}
\end{equation}
Once we have found a solution $\sigma$ to eq.\ \eqref{eq:CondSigma}, it is clear that $\sigma'=\sigma+
\text{(\emph{zero modes of}}$ $\Box\text{)}$ defines a solution, too. In particular, we can subtract from $\sigma$ its
projection onto the zero modes. This way, we can always obtain a solution to \eqref{eq:CondSigma} which is free of zero
modes. Thus, we may assume that $\sigma$ does not contain any zero modes before actually having computed it. In doing
so, relation \eqref{eq:CondSigma} can safely be inverted (cf.\ Appendix \ref{app:Weyl} for a more detailed discussion
of zero modes):
\begin{equation}
 \sigma = -\frac{1}{2}\,\Box^{-1}R
\label{eq:SigmaSolved}
\end{equation}
Note that the possibility of performing such a direct inversion is due to the simple structure of eq.\
\eqref{eq:CondSigma} which, in turn, is a consequence of $\hR=0$.

Now we leave the strictly $2$-dimensional case and try to ``lift'' the discussion to $d=2+\ve$. For this purpose we
make the assumption that we can still parametrize the metric by \eqref{eq:ConfGauge} with a reference metric $\hg$
whose associated scalar curvature vanishes, $\hR=0$. (Once again, the general case will be discussed in Section
\ref{sec:EpsilonLimit}.) In this case, by employing equation \eqref{eq:EHexpanded} we obtain the following relation
for the integral \eqref{eq:LimitInt}:
\begin{equation}
 \frac{1}{\ve}\int\td^{2+\ve}x\sg\,R=\frac{1}{\ve}\int\td^2 x\shg\,\Big[\ve\,\sigma\big(-\hB\big)\sigma\Big]+\mO(\ve).
\end{equation}
This expression can be rewritten by means of the $(2+\ve)$-dimensional analogues of eqs.\ \eqref{eq:CondSigma} and
\eqref{eq:SigmaSolved} which read $R=-2\,\Box\mku\sigma+\mO(\ve)$ and $\sigma = -\frac{1}{2}\,\Box^{-1}R+\mO(\ve)$,
respectively, and we arrive at the result
\begin{equation}
\phantom{(\hR=0)}\quad
\fbox{$\displaystyle
 \frac{1}{\ve}\int\td^{2+\ve}x\sg\,R = -\frac{1}{4}\int\td^2 x\sg\, R\,\Box^{-1} R +\mO(\ve).
$}\quad (\hR=0)
\end{equation}

Clearly, the assumption $\hR=0$ is quite restrictive. But already in this simple setting we make a crucial
observation: the emergence of a nonlocal action from a purely local one in the limit $d\rightarrow 2$. More
precisely, \emph{in the 2D limit the Einstein--Hilbert type action} $\frac{1}{\ve}\int\td^{2+\ve}x\sg\,R$ \emph{becomes
proportional to the induced gravity action}. As we will see below, a similar result is obtained for general topologies
without any assumption on $\hR$.

%----------------------------------------------------------------------------------------------------------------------
\subsection{General properties of the limit}
\label{sec:genRem}
%----------------------------------------------------------------------------------------------------------------------

\textbf{(1) Existence of the limit}. In the following we argue that $\lim_{\ve\rightarrow 0}\left(\frac{1}{\ve}
\int\td^{2+\ve}x\,\sg\,R\right)$ is indeed a meaningful quantity without restricting ourselves to a particular
topology or gauge. For convenience let us set
\begin{equation}
 \mS_\ve[g] \equiv \int\td^{2+\ve}x\,\sg\, R .
\label{eq:SEpsilon}
\end{equation}
We would like to establish that $\mS_\ve[g]$ has a Taylor series in $\ve$ whose first nonzero term which is sensitive
to the \emph{local} properties of $g_\mn$ is of the order $\ve$.

For the proof we make use of the relation $R_\mn=\frac{1}{2}g_\mn R$, valid in $d=2$ for any metric, so that the
Einstein tensor vanishes identically in $d=2$,
\begin{equation}
 G_\mn\big|_{d=2}=0 \, .
\end{equation}
Going slightly away from $2$ dimensions, $d=2+\ve$, we assume continuity and thus conclude that $G_\mn\big|_{d=2+\ve}
=\mO(\ve)$. Furthermore, the order $\ve^1$ is really the first nonvanishing term of the Taylor series with respect to
$\ve$ in general, i.e.\ $G_\mn\big|_{d=2+\ve}$ is not of the order $\mO(\ve^2)$ or higher. This can be seen by taking
the trace of $G_\mn$,
\begin{equation}
 g^\mn G_\mn=g^\mn\left(R_\mn-\frac{1}{2}g_\mn R\right)=R-\frac{d}{2}R = -\frac{1}{2}R\,\ve.
\end{equation}
Therefore, we have $g^\mn G_\mn = g_\mn\mku G^\mn \propto \ve$. (Of course, we assume $R\neq 0$ since $\mS_\ve$ would
vanish identically otherwise). But even the non-trace (tensor) parts of $G_\mn$ can be expected to be of
the order $\ve$ in general, as the following argument suggests. Let us consider a Weyl transformation of the metric,
$g_\mn=\e^{2\sigma}\hg_\mn$. The corresponding transformation of the Einstein tensor is given by equation
\eqref{eq:WeylEinstein} in the appendix. Now, let us assume that $\hg_\mn$ belongs to an Einstein manifold, i.e.\ the
corresponding Ricci tensor is proportional to the metric and the scalar curvature, $\hR_\mn =
\frac{1}{d}\mku\hg_\mn\hR\mkuu$.\footnote{In $d>2$ it is always possible to find a $\sigma$ for a given metric $g_\mn$
such that $\hg_\mn=\e^{-2\sigma}g_\mn$ leads to a space with constant scalar curvature provided that the manifold is
compact. This is known as the Yamabe problem \cite{Yamabe1960,Trudinger1968,Aubin1970,Aubin1976,Schoen1984} (while the
case $d=2$ is covered by Poincar\'{e}'s uniformization theorem). However, this statement does not imply that the
manifold is Einstein (whereas a constant \emph{sectional} curvature would imply that the manifold is Einstein). In
fact, there are known examples of metrics which are not conformal to any Einstein metric \cite{NP01}. On the other
hand, in $d=2$ any Riemannian manifold is of Einstein type.} In this case the Einstein tensor reads
\begin{equation}
 G_\mn = (d-2)\left[-\frac{1}{2d}\mkuu\hg_\mn\hR -\hD_\mu \hD_\nu\sigma + \hg_\mn\hB\sigma + \hD_\mu\sigma\hD_\nu\sigma
  + \frac{d-3}{2}\mku \hg_\mn \hD_\alpha\sigma \hD^\alpha\sigma \right],
\end{equation}
so we find $G_\mn\propto\ve$ again.

This $\ve$-proportionality is exploited now to make a statement about the Taylor series of $\mS_\ve$. For that purpose
we consider the variation of $\mS_\ve$ with respect to $g_\mn$ (assuming vanishing surface terms):
\begin{equation}
\begin{split}
 \frac{\delta \mS_\ve[g]}{\delta g_\mn(x)}= \int\td^{2+\ve}y\,\sg\left[\frac{1}{2}\mku g^\mn R -R^\mn\right]\delta(x-y)
 = -\sg \, G^\mn = \mO(\ve).
\end{split}
\label{eqn:SVariation}
\end{equation}
As a result we obtain $\mS_\ve[g]=C+\mO(\ve)$, where the constant $C$ is independent of $g_\mn$. Clearly, $C$
is obtained by computing $\mS_\ve$ in $d=2$, which is known to lead to the Euler characteristic $\chi\,$:
%\footnote{This constant vanishes when considering the topology of a torus  since $\chi=0$.}
\begin{equation}
 C=\mS_\ve\big|_{\ve=0}=4\pi\chi.
\end{equation}
That is, we have $\mS_\ve=4\pi\chi+\mO(\ve)$. (This result differs from Ref.\ \cite{CK80}, but it is in agreement with
Refs.\ \cite{MR93,Jackiw2006,GJ10}). As a consequence, the integral \eqref{eq:LimitInt} amounts to
\begin{equation}[b]
 \frac{1}{\ve}\int\td^{2+\ve}x\sg\, R = \frac{4\pi\chi}{\ve} + \text{finite}
 = \text{top.} + \text{finite},
\label{eq:expansion}
\end{equation}
where `top.' is a field independent (up to topological information) and thus irrelevant contribution to the action.
The terms in \eqref{eq:expansion} that contain the interesting information about the dynamics of the field are of order
$\mO(\ve^0)$, so the ``relevant'' part of $\frac{1}{\ve}\int\td^{2+\ve}x\,\sg\, R$ has indeed a
meaningful limit $\ve\rightarrow 0$.

\medskip
\noindent
\textbf{(2) The role of the volume form}. Next we argue that the important part of the $\ve$-dependence of
$\mS_\ve$ originates from the scalar density $\sg\,R\,$ in the integrand of \eqref{eq:SEpsilon} alone, i.e.\ loosely
speaking, it is sufficient to employ the a priori undefined fractional integration element $\td^{2+\ve}x$ at $\ve=0$.
Stated differently, all consistent definitions of ``$\td^{2+\ve}x$'' away from $\ve=0$ that one might come up with are
equivalent. The reason for that is the following.

Any integration over a scalar function on a manifold involves a volume form, i.e.\ a nowhere vanishing $d$-form (or a
density in the nonorientable case), in order to define a measure. This volume form is given by $\dd x\sg$, where $\sg$
is the square root of the corresponding Gramian determinant. If an integral is to be evaluated, the unit vectors of the
underlying coordinate system are inserted into the volume form. Since, for any $d$, these unit vectors produce a factor
of $1$ when inserted into $\dd x$, we see that it is the remaining part of the volume element that contains its
complete $d$-dependence, namely $\sg$. In particular, $\sg$ carries the canonical dimension of the volume
element.\footnote{Our conventions for the canonical mass dimensions are such that all coordinates are dimensionless,
$[x^\mu]=0$, while the metric components have $[g_\mn]=-2$, giving $\td s^2=g_\mn \td x^\mu\td x^\nu$ the canonical
dimension of an area, $[\td s^2]=-2$, regardless of the value of $d$. Hence $[\td x^\mu]=0$ and $[\sg]=-d$.

As a consequence,
the symbolic integration over the remaining ``fraction of a dimension'', $\td^\ve x$, is irrelevant even for the
dimension of $\mS_\ve[g]$.}

To summarize, for the evaluation of $\lim_{\ve\rightarrow 0} \frac{1}{\ve}\mS_\ve$ it is sufficient to consider the
$\ve$-depen\-dence of $\sg R$, while the integration can be seen as an integration over $\td^2x$. This
prescription can be considered our \emph{definition} for taking the $\ve$-limit in a well behaved way. Clearly,
the details of the domain of integration contribute some $\ve$-dependence, too. However, as we have seen in point
\textbf{(1)} in equation \eqref{eqn:SVariation}, the first relevant nonconstant, i.e.\ metric dependent, part of the
action comes from $\sg R$ alone, and any further $\ve$-dependent contributions would be of the order $\ve^2$. This
makes clear that our argument is valid in the special case of an integral over $\sg R$, but not for arbitrary
integrands.

\medskip
\noindent
\textbf{(3) Comment and comparison with related work}. As an aside we note that in Ref.\ \cite{MR93} it is argued that
the irrelevant divergent term in \eqref{eq:expansion} can be made vanish by subtracting the term $\frac{1}{\ve}
\int\dd x\sqrt{\tilde{g}}\,\tilde{R}$ from $\frac{1}{\ve}\int\dd x\sg \,R$ where the metric $\tilde{g}_\mn$ is assumed
to be $g_\mn$-dependent but chosen in such a way that the resulting field equations for $g_\mn$ do not change when $d$
approaches $2$. That means, the $g_\mn$-variation of the subtracted term (and, in turn its variation w.r.t.\
$\tilde{g}$) must vanish for $d\rightarrow 2$, leading to the requirement $\lim_{\ve\rightarrow 0} \big(\frac{1}{\ve}
\tilde{G}_\mn\big) = 0$ for the corresponding Einstein tensor.
This subtraction term would cancel the $\ve$-pole in \eqref{eq:expansion}. In \cite{MR93} it is assumed that such a
term exists for some metric $\tilde{g}_\mn$ which is conformally related to $g_\mn$. However, it remains unclear if
this is possible at all. According to the above argument in \textbf{(1)}, we would rather expect $\frac{1}{\ve}
\tilde{G}_\mn$ to remain finite in the limit $\ve\rightarrow 0$.

Unlike Ref.\ \cite{MR93}, we do not need to subtract further $g_\mn$-dependent terms from the action here, and our
discussion is valid for all metrics.

%----------------------------------------------------------------------------------------------------------------------
\subsection{Establishing the 2D limit}
\label{sec:EpsilonLimit}
%----------------------------------------------------------------------------------------------------------------------

Next we determine the first relevant order of the Taylor series of \eqref{eq:LimitInt}, providing the basis for our
main statements. Let us define the $\ve$-dependent action functional
\begin{equation}
 \Ye[g] \equiv \frac{1}{\ve}\int\dex\sg\, R\, - \frac{4\pi\chi}{\ve}\,.
\label{eq:Ye}
\end{equation}
Here, $\chi$ again denotes the metric independent Euler characteristic defined in strictly $2$ dimensions.
Corresponding to the arguments of Section \ref{sec:genRem}, $\Ye$ is well defined in the limit $\ve\rightarrow 0$
because it is of the order $\ve^0$. Therefore, $Y[g]$ defined by
\begin{equation}
 Y[g] \equiv \lim_{\ve\rightarrow 0} \Ye[g]
\label{eq:YDef}
\end{equation}
is a finite functional.

To expand the integral in \eqref{eq:Ye} in powers of $\ve$ we make use of the general transformation law of
$\int\dd x\sg R$ under Weyl rescalings, $g_\mn=\e^{2\sigma}\hg_\mn$, given by equation \eqref{eq:EHexpanded} in the
appendix. This yields
\begin{equation}
 \begin{split}
  \Ye[g] &= \frac{1}{\ve}\int\dex\shg\,\e^{\ve\sigma}\left[\hR+(1+\ve)\ve
    \big(\hD_\mu\sigma\big) \big(\hD^\mu\sigma\big)\right] - \frac{4\pi\chi}{\ve}\\
  &= \frac{1}{\ve}\int\dex\shg\, \hR - \frac{4\pi\chi}{\ve} 
    + \int\td^2 x\shg\big(\hR\sigma+\hD_\mu\sigma\hD^\mu\sigma \big) + \mO(\ve).
 \end{split}
\label{eq:YeExp1}
\end{equation}
We observe that the first two terms of the second line of \eqref{eq:YeExp1} can be combined into $\Ye[\hg]$.
Furthermore, the terms involving the parameter of the Weyl transformation, $\sigma$, are seen to agree with the
definition in \eqref{eq:DeltaIDef} and can be written as $\int\td^2x\shg\big[\hD_\mu\sigma\hD^\mu\sigma+\hR\sigma\big]
\equiv 2\, \Delta I[\sigma;\hg]$.
This, in turn, can be expressed by means of the (normalized) induced gravity functional \cite{Polyakov1981}, defined
by\footnote{If the scalar Laplacian $\Box$ has zero modes, then $\Box^{-1}$ is defined as the inverse of $\Box$ on the
orthogonal complement to its kernel, that is, before $\Box^{-1}$ acts on a function it implicitly projects onto nonzero
modes. For the arguments presented in this chapter we may assume that $\Box$ does not have any zero modes, although a
careful analysis shows that the inclusion of zero modes does not change our main results (see detailed discussion in
Appendix \ref{app:Weyl}, in particular Section \ref{app:Zero}).}
\begin{equation}
 I[g] \equiv \int \td^2 x \sg\, R\, \Box^{-1}R \,.
\label{eq:IndGrav}
\end{equation}
%which is proportional to Polyakov's induced action up to a cosmological constant term \cite{Polyakov1981}.
As shown in Appendix \ref{app:Weyl}, the change of $I$ under a finite Weyl transformation of the metric in its argument
equals precisely $-8\,\Delta I$ which therefore has the interpretation of a Wess--Zumino term, a $1$-cocycle related to
the Abelian group of Weyl transformations \cite{MM01}:\footnote{As a consequence of identity \eqref{eq:decomp1}, the
Liouville action \eqref{eq:LiouvilleAction} can be rewritten as $\GL[\phi;\hg]=\frac{a_1}{4}I[\e^{2\phi}\hg]+
\frac{1}{2}a_1 a_2\int\td^2x\sqrt{\det(\e^{2\phi}\hg)} - \frac{a_1}{4} I[\hg]$. Note that the first two terms on the
RHS of this equation depend on $\phi$ and $\hg_\mn$ only in the combination $\e^{2\phi}\hg_\mn = g_\mn$.}
\begin{equation}
 I[\e^{2\sigma}\hg] - I[\hg] = -8\,\Delta I[\sigma;\hg] \,.
\label{eq:decomp1}
\end{equation}
Inserting \eqref{eq:decomp1} into \eqref{eq:YeExp1} leads to
\begin{equation}
 \begin{split}
\Ye[g] = \Ye[\hg]+2\,\Delta I[\sigma;\hg] +\mO(\ve) 
  = \Ye[\hg] +\frac{1}{4}I[\hg]-\frac{1}{4}I[g] +\mO(\ve).
 \end{split}
\end{equation}
Rearranging terms and taking the limit $\ve\rightarrow 0$ results in the important identity
\begin{equation}
 Y[g]+\frac{1}{4}I[g]=Y[\hg]+\frac{1}{4}I[\hg].
\label{eq:YplusI}
\end{equation}
Note that the LHS of eq.\ \eqref{eq:YplusI} depends on the full metric $g=\e^{2\sigma}\hg$ while the RHS depends only
on $\hg$.

For the further analysis it is convenient to introduce the functional
\begin{equation}
 F[g] \equiv Y[g] + \frac{1}{4} I[g].
\label{eq:FDef}
\end{equation}
By construction $F$ has the following properties:
{%
\renewcommand{\theenumi}{(\roman{enumi})}%
\setlist{nolistsep}
\begin{enumerate}
 \item It is diffeomorphism invariant since it has been constructed from diffeomorphism invariant objects only.
 \item It is a functional in $d=2$ precisely since the $\ve$-limit has already been taken.
 \item It is insensitive to the conformal factor of its argument since from eq.\ \eqref{eq:YplusI} follows
 Weyl invariance:
 \begin{equation}
  F[\e^{2\sigma}\hg] = F[\hg].
 \end{equation}
\end{enumerate}%
}%
\noindent
Thanks to our preparations in Section \ref{sec:ConfGauge} we can conclude immediately that $F$ is constant apart from
a remaining dependence on some moduli $\{\tau\}$ possibly. Here it is crucial that the moduli are \emph{global}
parameters of purely \emph{topological} origin. They are insensitive to the local properties of the metric, in
particular they do not depend on a spacetime point. These arguments show that the \emph{functional} $F[g]$ becomes a
\emph{function} of the moduli, say $C\big(\{\tau\}\big)$. The precise dependence of $F$ on these moduli is irrelevant
for the present discussion since they encode only topological information. We thus have
\begin{equation}
 F[g] = C\big(\{\tau\}\big) ,
\end{equation}
i.e.\ $F$ is a metric independent constant functional, up to topological terms.

\bigskip
For the functional $Y[g]$ defined in eq.\ \eqref{eq:YDef} we obtain, using eq.\ \eqref{eq:FDef},
\begin{equation}
 Y[g]=-\frac{1}{4}I[g]+C\big(\{\tau\}\big) \, ,
\end{equation}
which leads to our final result:
\begin{equation}[b]
 \frac{1}{\ve}\int\td^{2+\ve}x\sg\, R= -\frac{1}{4}\int\td^2x\sg\,R\,\Box^{-1}R +\frac{4\pi\chi}{\ve}
  +C\big(\{\tau\}\big)+\mO(\ve).
\label{eq:LimitResult}
\end{equation}
The terms $4\pi\chi/\ve$ and $C\big(\{\tau\}\big)$ are topology dependent but independent of the local properties of
the metric, and thus they may be considered irrelevant for most purposes.

Thereby we have established that \emph{the limit $d\to 2$ of the Einstein--Hilbert action equals precisely the induced
gravity action up to topological terms}. Clearly, the most remarkable aspect of this limiting procedure is that it
leads from a local to a nonlocal action.

A similar mechanism has been discussed earlier in the framework of dimensional regularization \cite{MM01}. The result
\eqref{eq:LimitResult} is in agreement with the one of Reference \cite{Jackiw2006} where it has been obtained by means
of a different reasoning based on the introduction of a Weyl gauge potential.

We would like to emphasize that the emergence of the induced gravity action is also found for such Laplacian operators
that admit zero modes. In this case, the RHS of \eqref{eq:LimitResult} receives an additional contribution, but the
crucial term $-\frac{1}{4} I[g]$ is still present. This situation is discussed in detail in Appendix \ref{app:Zero}.

%----------------------------------------------------------------------------------------------------------------------
\section{The full Einstein--Hilbert action in the 2D limit}
\label{sec:LimitFullEH}
%----------------------------------------------------------------------------------------------------------------------

Including also the cosmological constant term, the Einstein--Hilbert truncation of the (gravitational part of
the) effective average action in $d$ dimensions reads
\begin{equation}
 \Gamma_k^\text{grav}[g] = \frac{1}{16\pi G_k} \int \dd x \sg \,\big( -R + 2\Lambda_k \big),
\label{eq:EHTruncation}
\end{equation}
with the dimensionful Newton and cosmological constant, $G_k$ and $\Lambda_k$, respectively.

\medskip
\noindent
\textbf{(1)}
As we have mentioned already, the dimensionless versions of these couplings, $g_k\equiv k^{d-2}G_k$ and $\lambda_k
\equiv k^{-2}\Lambda_k$, possess a nontrivial fixed point in $d=2+\ve$ dimensions whose coordinates are proportional
to $\ve$ (cf.\ Chapter \ref{chap:ParamDep} and Refs.\ \cite{Weinberg1979,Tsao1977,Brown1977,GKT78,CD78,KN90,JJ91,%
KKN93a,KKN93b,KKN93c,KKN96,AKKN94,NTT94,AK97,Reuter1998,NR13,NR13b,NR15a,Nink2015,CD15,Falls2015a,Falls2015b}).
Thus, at least in the vicinity of this non-Gaussian fixed point the dimensionful couplings are of the form
\begin{equation}
 G_k \equiv \ve\,\rGk \, ,\quad \Lambda_k \equiv \ve\,\rLk \,,
\end{equation}
where $\rGk$ and $\rLk$ are of the order $\mO(\ve^0)$. Making use of eq.\ \eqref{eq:LimitResult} in the
limit $\ve\rightarrow 0$ we arrive at the $2$-dimensional effective average action
\begin{equation}[b]
 \Gamma_k^\text{grav,2D}[g] = \frac{1}{64\pi \rGk} \int\td^2x\sg\,R\,\Box^{-1}R
  + \frac{\rLk}{8\pi\rGk}\int\td^2x\sg
  + \text{top}.
\label{eq:Gamma}
\end{equation}
Here 'top' refers again to topology dependent terms which are insensitive to the local properties of the metric. The
result \eqref{eq:Gamma} is quite general; it holds for any RG trajectory provided that the couplings $G_k$ and
$\Lambda_k$ in $d=2+\ve$ are of first order in $\ve$.

As an aside we note that the topological terms in \eqref{eq:Gamma} include a contribution proportional to
$\int\td^2 x\sg\,R=4\pi\mku\chi$. Thus, eq.\ \eqref{eq:Gamma} contains the induced gravity action, a cosmological
constant term, and the $\chi$-term. These are precisely the terms that were included in the truncation ansatz in Ref.\
\cite{CD15}. By contrast, in our approach they are not put in by hand through an ansatz, but they rather emerge as a
result from the Einstein--Hilbert action in the 2D limit.

\medskip
\noindent
\textbf{(2)}
If we want to consider $\Gamma_k$ exactly at the NGFP, we can insert the known fixed point values, where the one of
Newton's constant is given by $g_*=\ve/b$ according to eq.\ \eqref{eq:bTogStar}. As shown in Chapter
\ref{chap:ParamDep}, the coefficient $b$ depends on the parametrization of the metric. For the linear parametrization
it is given by \cite{Reuter1998,Weinberg1979,Tsao1977,Brown1977,KN90,JJ91,NR13,NR13b,NR15a,Nink2015,CD15}\footnote{When
the running of the Gibbons--Hawking surface term instead of the pure Einstein--Hilbert action is computed, the result
reads $b=\frac{2}{3}(1-\ns)$ \cite{GKT78,CD78}. See Refs.\ \cite{NR13,NR13b,NR15a} for a discussion.}
\begin{equation}
 b = \textstyle\frac{2}{3}\big(19-\ns\big),
\end{equation}
while the exponential parametrization leads to
\cite{KKN93a,KKN93b,KKN93c,KKN96,AKKN94,NTT94,AK97,Nink2015,CD15,DN15,Falls2015a,Falls2015b}
\begin{equation}
 b = \textstyle\frac{2}{3}\big(25-\ns\big),
\label{eq:bExp}
\end{equation}
where $N$ denotes the number of scalar fields, provided that we consider the ansatz \eqref{eq:GammaAnsatzAgain} with a
matter action of the type \eqref{eq:matter}. As the exponential parametrization was argued to be more appropriate in
the 2D limit, we will mostly state the results based on eq.\ \eqref{eq:bExp}
in the following, although the analogues for the linear parametrization can simply be obtained by replacing $25\to 19$.
Using the definition \eqref{eq:IndGrav} and combining \eqref{eq:Gamma} with \eqref{eq:bExp}, we obtain the NGFP action
\begin{equation}[b]
 \Gamma_k^\text{grav,2D,NGFP}[g] = \frac{(25-\ns)}{96\pi} \,I[g]
  + \frac{(25-\ns)}{12\pi}\,k^2 \rls\int\td^2x\sg
  + \text{top}\,,
\label{eq:GammaNGFP}
\end{equation}
where $\rls\equiv\lambda_*/\ve$ is cutoff dependent and thus left unspecified here. The actions
\eqref{eq:Gamma} and \eqref{eq:GammaNGFP} will be the subject of our discussion in Chapter \ref{chap:NGFPCFT}.

\medskip
\noindent
\textbf{(3)}
Finally, let us briefly establish the connection with Liouville theory. For this purpose we separate the conformal
factor from the rest of the metric. Inserting
\begin{equation}
 g_\mn = \e^{2\phi}\hg_\mn
\label{eq:DefConfFac}
\end{equation}
into eq.\ \eqref{eq:Gamma} for $\Gamma_k^\text{grav,2D}[g]$ and using \eqref{eq:ItoDeltaI} and \eqref{eq:DeltaI} from
the appendix yields
\begin{equation}[b]
\begin{aligned}
 \Gamma_k^\text{grav,2D}[\phi;\hg] =\; &\frac{1}{64\pi\rGk}\int\td^2x\shg\,\hR\,\hB^{-1}\hR \\
  & -\frac{1}{16\pi\rGk}\int\td^2x\shg\,\Big[\hD_\mu\phi\,\hD^\mu\phi+\hR\phi-2\mku\rLk\, \e^{2\phi}\Big]
  + \text{top} \,,
\end{aligned}
\end{equation}
where $\hg_\mn$ is a fixed reference metric for the topological sector (i.e.\ a point in moduli space) under
consideration. Hence, the effective average action for the conformal factor in precisely $2$ dimensions is nothing but
the Liouville action.

Of course, this is well known to happen if one starts from the induced gravity action, an object that lives already in
2D. It is quite remarkable and nontrivial, however, that \emph{Liouville theory can be regarded as the limit of the
higher dimensional Einstein--Hilbert theory}. Note that this result is consistent with the discussions in
Refs.\ \cite{MR93,GJ10} (cf.\ also \cite{LS94}).

\medskip
\noindent
\textbf{(4)}
To sum up, we have used the the Einstein--Hilbert action in $d>2$ to construct a manifestly $2$-dimensional action
which describes 2D Asymptotic Safety. As opposed to earlier work on the
$\ve$-expansion of $\beta$-functions the dimensional limit was taken directly at the level of the action functional.

%----------------------------------------------------------------------------------------------------------------------
\section{Aside: Is there a generalization to 4D?}
\label{sec:GenTo4D}
%----------------------------------------------------------------------------------------------------------------------

For the sake of completeness we would like to comment on a generalization of our results to $4$ dimensions. At first
sight, there seems to be a remarkable similarity. Dimensional analysis suggests that the role of the $R$-term in the
Einstein--Hilbert action near $2$ dimensions is now played by curvature-square terms in $d=4+\ve$. The gravitational
part of the action assumes the general form
\begin{equation}
 \Gamma_k^\text{grav}[g] = \Gamma_k^\text{EH}[g] + \int\td^{4+\ve}x\sg \left\{\frac{1}{a_k}E + \frac{1}{b_k}F
 + \frac{1}{c_k}R^2\right\},
\end{equation}
where $F\equiv C_{\mn\rs}C^{\mn\rs}$ is the square of the Weyl tensor. Furthermore, the term
$E\equiv R_{\mn\rs}R^{\mn\rs}-4R_\mn R^\mn+R^2+\frac{d-4}{18}R^2$ gives rise to the Gauss--Bonnet--Euler topological
invariant when integrated over in exactly $d=4$. Considerations of nontrivial cocycles of the Weyl group show that the
corresponding Wess--Zumino action in $d=4$ is generated by the $E$- and the $F$-term \cite{MM01}, analogous to the
generation of $\Delta I$ in Sec.\ \ref{sec:EpsilonLimit} due to the $R$-term. It may thus be expected that there would
be a mechanism to take the 4D limit, similar to the one of Sec.\ \ref{sec:EpsilonLimit} but now for $E$ and $F$ instead
of $R$, if the couplings $a_k$ and $b_k$ were of first order in $\ve$.

At one-loop level the $\beta$-functions in $d=4+\ve$ feature indeed a fixed point with $a_*=\mO(\ve)$, $b_*=\mO(\ve)$
and $c_*$ finite \cite{OP14}. There are, however, two crucial differences in comparison with the $2$-dimensional case:
(i) The term $\int\td^4 x\sg \,F$ is not a topological invariant, i.e.\ there is no appropriate subtraction analogous
to definition \eqref{eq:Ye}, and the limit $\ve\rightarrow 0$ remains problematic. (ii) Even if we managed to define
some 4D-functional similar to \eqref{eq:FDef} which is both diffeomorphism and Weyl invariant, this would not be
sufficient to conclude that the functional is constant since in $d=4$ the space of metrics modulo
$\text{Diff}\times\text{Weyl}$-transformations is too large and cannot be classified in terms of topological
parameters. Roughly speaking, if we found a way to circumvent problem (i), the 4D limit of the above action computed
with our methods might lead to the same nonlocal action as found in \cite{MM01}, but this would not represent the
general 4D limit since the latter must certainly contain further terms that do not originate from a variation of the
conformal factor alone.
In summary, in spite of many similarities to the 2D case there seems to be no direct generalization of our
approach of computing a nonlocal limit action to $4$ spacetime dimensions. Nevertheless, we expect that the 4D fixed
point action contains nonlocal terms, too.

%----------------------------------------------------------------------------------------------------------------------
\chapter[The non-Gaussian fixed point as a conformal field theory]%
[The non-Gaussian fixed point as a unitary conformal field theory]%
{The non-Gaussian fixed point as a unitary conformal field theory}
\label{chap:NGFPCFT}
%----------------------------------------------------------------------------------------------------------------------

\begin{summary}
We study further properties of the 2D limit of the gravitational EAA which was constructed in the
previous chapter. Directly at the fixed point, it can be written in terms of dimensionless variables as a scale
independent functional, giving rise to a conformal field theory. By means of this 2D fixed point action we discuss the
compatibility of Asymptotic Safety with Hilbert space positivity (unitarity). The corresponding central charge is
related to the fixed point value of the Newton coupling in the limit $d\to 2$. We find that the pure gravity
part is governed by a unitary conformal field theory with positive central charge $c=25$. Particular attention is paid
to the relation between the crucial sign of the central charge, the occurrence of a conformal factor instability, and
unitarity: A positive central charge implies Hilbert space positivity and an unstable conformal factor. The latter
can be seen by representing the fixed point CFT by a Liouville theory in the conformal gauge and investigating its
properties. We argue that the conformal factor instability is not only acceptable but also desired.

\noindent
\textbf{What is new?} Reconciling Asymptotic Safety with unitarity.

\noindent
\textbf{Based on:} Ref.\ \cite{NR16a}.
\end{summary}
\vspace{-2pt}

%----------------------------------------------------------------------------------------------------------------------
\section{Motivation}
\label{sec:MotNGFPCFT}
%----------------------------------------------------------------------------------------------------------------------

All studies on Asymptotic Safety carried out in the literature so far provided evidence in favor of the existence of a
suitable nontrivial RG fixed point.
In this chapter, we would like to gain further insight into the nature of the \emph{fixed point theory}, i.e.\ the
theory defined directly at the fixed point rather than by an RG trajectory running away from it. For instance, it is an
open question whether or not this is a conformal field theory.

In $2$ dimensions we are indeed used to the picture that the conformal field theories correspond to points in theory
space that are fixed points of the RG flow \cite{Nakayama2015}. In $4$ dimensions, however, Quantum Einstein Gravity
(QEG) has a scale invariant fixed point theory but it is unclear whether it is conformal.

While conformality is not known to be indispensable, we argued in the introduction that a consistent asymptotically
safe theory must possess several other properties in addition to its mere nonperturbative renormalizability (that is,
the existence of a suitable non-Gaussian fixed point), the two most important ones being background independence and
unitarity. According to Ref.\ \cite{BR14} and Section \ref{sec:bi} there are by now first promising results which
indicate that the requirements for background independence and Asymptotic Safety can be met simultaneously in
sufficiently general truncations of the RG flow. On the other hand, little is known about the status of unitarity.

In this connection the somewhat colloquial term ``unitarity'' is equivalent to ``Hil\-bert space positivity'' (cf.\
Section \ref{sec:CFT}) and is meant to express that the state space of the system under consideration contains no
vector having a negative scalar product with itself (``negative norm state''). If it does so, it is not a Hilbert
space in the mathematical sense of the word and cannot describe a \emph{quantum} system as the probability
interpretation of quantum mechanics would break down then.

At least on (nondynamical) flat spacetimes the criterion of Hilbert space positivity, alongside with the spectral
condition can be translated from the Lorentzian to the Euclidean setting where it reappears as the requirement of
reflection-, or Osterwalder--Schrader, positivity \cite{OS73,OS75,Strocchi1993,GJ87}.

Unitarity is in fact a property that is not automatic and needs to be checked in order to demonstrate the viability of
the Asymptotic Safety program based upon the effective average action. The operator formulation corresponding to the
gravitational EAA amounts to an indefinite metric (Krein space) quantization, and so the negative norm states it
contains should ultimately be ``factored out'' in order to obtain a positive (``physical'') state space, a true
Hilbert space. While this procedure is standard and familiar from perturbative quantum gravity and Yang--Mills theory,
for instance, the situation is much more involved in Asymptotic Safety. The reason is that, implicitly, this indefinite
metric quantization is applied to a bare action which is essentially given by the fixed point functional (see Refs.\
\cite{MR09,VZ11,Morris1994,MS15}, and Chapters \ref{chap:Bare} and \ref{chap:FullReconstruction}). As such it is
already in itself the result of a technically challenging nonperturbative computation which in practice can be done
only approximately, for the time being.

In the following, we explore the question of Hilbert space positivity together with a number of related issues
such as locality by analyzing the situation in $2$ dimensions where --- as we have seen --- a number of technical
simplifications occur. To this end, we employ the manifestly $2$-dimensional limit action constructed in the previous
chapter. We shall see that the non-Gaussian fixed point underlying Asymptotic Safety is governed by a conformal field
theory (CFT) which is interesting in its own right, and whose properties we shall discuss. Remarkably enough, it
turns out to possess a positive central charge, thus giving rise to a unitary representation of the Virasoro algebra
and a ``positive'' Hilbert space in the above sense.

%----------------------------------------------------------------------------------------------------------------------
\section{The unitary fixed point theory}
\label{sec:UnitaryCFT}
%----------------------------------------------------------------------------------------------------------------------

We can summarize the main message of Chapter \ref{chap:EHLimit} by saying that every trajectory $k \mapsto (g_k,
\lambda_k)\equiv (\rgk,\rlk)\ve$, i.e.\ every solution to the RG equations of the Einstein--Hilbert truncation in
$2+\ve$ dimensions, induces the following intrinsically two-di\-men\-sional running action:
\begin{equation}
 \Ggd[g] = \frac{1}{96\pi}\left(\frac{3}{2}\,\frac{1}{\rgk}\right)\left[ I[g]+8\mku\rlk\,k^2\int\td^2 x\sg \,\right],
\label{eq:GammaGravSummarized}
\end{equation}
where topological terms are left aside henceforth.
In this chapter we discuss the main properties of this RG trajectory, in particular its fixed point.

\medskip
\noindent
\textbf{(1) The fixed point functional}. Strictly speaking, the theory space under consideration comprises functionals
which depend on the \emph{dimensionless} metric $\tg_\mn\equiv k^2 g_\mn$. For any average action $\Gamma_k[g]$ we
define its analog in the dimensionless setting by $\mA_k[\tg]\equiv\Gamma_k[\tg\mku k^{-2}]$. Thus, equation
\eqref{eq:GammaGravSummarized} translates into
\begin{equation}
 \mA_k[\tg] = \frac{1}{96\pi}\left(\frac{3}{2}\,\frac{1}{\rgk}\right)\left[ I[\tg]+8\mku\rlk \int\td^2 x\stg \,\right].
\end{equation}
It is this functional that becomes strictly constant at the NGFP: $\mA_k\rightarrow \mA_*\mku$, with
\begin{equation}
 \mA_*[\tg] = \frac{1}{96\pi}\left(\frac{3}{2}\,\frac{1}{\rgs}\right)\left[ I[\tg]+8\mku\rls \int\td^2 x\stg \,\right].
\label{eq:AStar}
\end{equation}
For the exponential field parametrization we find the fixed point functional
\begin{equation}[b]
 \mA_*[\tg] = \frac{(25-\ns)}{96\pi} \int\td^2 x\stg\,\Big(\tR\,\tB^{-1}\tR+8\mku\rls\Big) .
\end{equation}
Here and in the following we usually present the results for the exponential pa\-ra\-me\-tri\-za\-tion. The
corresponding formulae for the linear parametrization can be obtained by replacing $(25-\ns)\rightarrow(19-\ns)$. (See
Chapter \ref{chap:ParamDep} for a discussion of different metric parametrizations).

While the NGFP is really a point in the space of $\mA$-functionals, it is an entire \emph{line}, parametrized by $k$,
in the more familiar dimensionful language of the $\Gamma_k$'s. Let us refer to the constant map $k\mapsto(g_*,
\lambda_*)$ $\forall\, k\in[0,\infty)$ as the ``\emph{FP trajectory}''. Moving on this trajectory, the system is never
driven away from the fixed point. According to eq.\ \eqref{eq:GammaNGFP}, it is described by the following EAA:
\begin{equation}[b]
 \Gamma_k^\text{grav,2D,NGFP}[g] = \frac{(25-\ns)}{96\pi} \left[ I[g]+8\mku\rls\, k^2\int\td^2 x\sg \,\right].
\label{eq:GgAtNGFP}
\end{equation}
As always in the EAA framework, the EAA at $k=0$ equals the standard effective action, $\Gamma =
\lim_{k\rightarrow 0}\Gamma_k$. So, letting $k=0$ in \eqref{eq:GgAtNGFP}, we conclude that the ordinary effective
action related to the FP trajectory has vanishing ``renormalized'' cosmological constant and reads
\begin{equation}
 \Gamma^\text{grav,2D,NGFP}[g] = \frac{(25-\ns)}{96\pi} \int\td^2 x\sg\,R\,\Box^{-1} R\,.
\label{eq:GgZeroAtNGFP}
\end{equation}

\medskip
\noindent
\textbf{(2) The 2D stress-energy tensor}. Differentiating $\Ggd$ of equation \eqref{eq:GammaGravSummarized} with
respect to the metric leads to the following energy-momentum tensor in the gravitational sector \cite{CR89}:
\begin{equation}
\begin{split}
 T_\mn^\text{grav}[g] = \frac{1}{96\pi}\left(\frac{3}{2}\,\frac{1}{\rgk}\right)\bigg[g_\mn\, D_\rho\big(\Box^{-1}R\big)
 D^\rho\big(\Box^{-1}R\big) + 4\, D_\mu D_\nu\big(\Box^{-1}R\big)\, &\\
 - 2\, D_\mu\big(\Box^{-1}R\big) D_\nu\big(\Box^{-1}R\big) - 4\, g_\mn R + 8\, \rlk\,k^2 g_\mn & \bigg].
\end{split}
\label{eq:Tgrav}
\end{equation}
It is easy to see that taking the trace of this tensor yields
\begin{equation}
 \Theta_k[g]=\left(\frac{3}{2}\,\frac{1}{\rgk}\right)\frac{1}{24\pi}\Big[-R+4\mkuu\rlk\,k^2 \Big],
\end{equation}
which, as it should be, agrees with the result from the Einstein--Hilbert
action in $d>2$, see equations \eqref{eq:Theta2} and \eqref{eq:Theta3}.\footnote{Note that in string theory or
conformal field theory one would usually redefine the stress-energy tensor and employ $T_\mn'\equiv T_\mn-\frac{1}{2}
g_\mn \Theta$ which is traceless at the expense of not being conserved. It is the modes of $T_\mn'$ that satisfy a
Virasoro algebra whose central extension keeps track of the anomaly coefficient then.} As for the non-trace parts of
$T_\mn^\text{grav}$, the comparatively complicated nonlocal structures in \eqref{eq:Tgrav} can be seen as the 2D
replacement of the Einstein tensor in \eqref{eq:EHEMTensor}.

In absence of matter (that is, $\Gm=0$) the tadpole equation \eqref{eq:Tadpole} boils down to $T_\mn^\text{grav}[\gsc]
=0$ with the above stress-energy tensor. Hence, self-consistent backgrounds have a constant (but $k$-dependent) Ricci
scalar:
\begin{equation}
 \Theta_k[\gsc]=0 \quad\Leftrightarrow\quad R\big(\gsc\big) = 4\mku \rlk\,k^2 \,.
\label{eq:Rsc}
\end{equation}
In terms of the dimensionless metric, $R\big(\tilde{\bar{g}}_k^\text{sc}\big)=4\mku\rlk$, in this case.

\medskip
\noindent
\textbf{(3) Intermezzo on induced gravity}. As a preparation for the subsequent discussion, we consider an arbitrary
conformal field theory on flat Euclidean space, having central charge $c_\mS$, and couple this theory to a
gravitational background field $g_\mn$, comprised in an action functional $\mS[g]$. Then the resulting (symmetric,
conserved) stress-energy tensor,
\begin{equation}
 T^{(\mS)}[g]^\mn \equiv \frac{2}{\sg}\frac{\delta\mS[g]}{\delta g_\mn} \,,
\label{eq:StressTensorGeneral}
\end{equation}
will acquire a nonzero trace in curved spacetimes, of the form
\begin{equation}
 g_\mn\, T^{(\mS)}[g]^\mn = -c_\mS\, \frac{1}{24\pi}R+\text{const} \,,
\label{eq:TCentralCharge}
\end{equation}
where ``const'' is due to a cosmological constant possibly.

\medskip
\noindent
\textbf{(3a)} Above, $\mS[g]$ can stand for either a classical or an effective action.
In the first case, $\mS[g]$ might result from a CFT of fields $\chi^I$ governed by an action $S[\chi,g]$ upon solving
the equations of motion for $\chi$, and substituting the solution $\chi_\text{sol}(g)$ back into the action:
$\mS[g] = S[\chi_\text{sol}(g),g]$. If $c_\mS \neq 0$ then the system displays a ``classical anomaly'', and Liouville
theory is the prime example \cite{DHoker1991,GM93,Nakayama2004,AADZ94}.

In the ``effective'' case, $\mS[g]$ could be the induced gravity action $S^\text{ind}[g]$ which we obtain from
$S[\chi,g]$ by integrating out the fields $\chi^I$ quantum mechanically:
\begin{equation}
 \e^{-S^\text{ind}[g]} = \int\mD\chi^I\,\e^{-S[\chi,g]}\,.
\end{equation}
Then $S^\text{ind}[g]$ is proportional to the central charge $c_\mS$,
\begin{equation}
 S^\text{ind}[g] = +\frac{c_\mS}{96\pi} I[g] + \cdots \,,
\label{eq:Sind}
\end{equation}
and by \eqref{eq:StressTensorGeneral} the action $S^\text{ind}[g]$ gives rise to a stress-energy tensor whose trace is
precisely of the form \eqref{eq:TCentralCharge}. (The dots represent a cosmological constant term.)

\medskip
\noindent
\textbf{(3b)} It is important to observe that the functional $I[g]$ is \emph{negative}, i.e.\ for any metric $g$ we
have $\int\td^2 x\sg\, R\,\Box^{-1}R <0$ . (Recall that $\Box^{-1}$ acts only on nonzero modes while it ``projects
away'' the zero modes. Since $-\Box$ is nonnegative, we conclude that $-\Box^{-1}$ has a strictly positive spectrum.)
Leaving the cosmological constant term in \eqref{eq:Sind} aside, this entails that for a positive central charge
$c_\mS>0$ the (noncosmological part of the) induced gravity action is negative, $S^\text{ind}[g]<0$.

The implications are particularly obvious in the conformal parametrization $g=\e^{2\phi} \hg$, yielding
\begin{equation}
 S^\text{ind}[\phi;\hg] = -\frac{c_\mS}{24\pi} \int\td^2 x\shg\Big(\hD_\mu\phi \hD^\mu\phi + \hR\phi\Big)
 + \frac{c_\mS}{96\pi} I[\hg] + \cdots \,.
\label{eq:instab}
\end{equation}
When $c_\mS$ is positive, the field $\phi$ is unstable, it has a ``wrong sign'' kinetic term. Stated differently,
\emph{integrating out unitary conformal matter induces an unstable conformal factor of the emergent spacetime metric.}

The 4D Einstein--Hilbert action is well known to suffer from the same conformal
factor instability, that is, a negative kinetic term for $\phi$ if the overall prefactor of $\int\!\sg R$ is adjusted
in such a way the concomitant kinetic term for the transverse-traceless (TT) metric fluctuations comes out positive,
as this befits propagating physical modes. Irrespective of all questions about the conventions in which the equations
are written down, the crucial signs are always such that
\begin{equation}
 c_\mS>0 \quad \stackrel{\mathclap{d=2}}{\Longleftrightarrow} \quad \phi \text{ unstable} \quad
 \stackrel{\mathclap{d>3}}{\Longleftrightarrow} \quad h_\mn^\text{TT} \text{ stable}.
\end{equation}
We shall come back to this point in a moment.

\medskip
\noindent
\textbf{(4) Central charge of the NGFP}. The fixed point action $\mA_*$ given by \eqref{eq:AStar} describes a conformal
field theory with central charge
\begin{equation}
 \cgr = \frac{3}{2}\,b\,,
\label{eq:RelBetweenCAndB}
\end{equation}
where $b=1/\rgs$. Depending on the parametrization it amounts to
\begin{equation}
 \cgr = \begin{cases} \;25 - \ns , \qquad\text{exponential parametrization,}\\
 \;19-\ns , \qquad\text{linear parametrization.} \end{cases}
\label{eq:CentChParams}
\end{equation}
This follows by observing that for the two field parametrizations, directly at the NGFP, the trace of the stress-energy
tensor is given by
\begin{equation}[b]
 \Theta_k[g] = \frac{1}{24\pi}\Big(-R+4\mku\rls k^2\Big)\times\begin{cases}\;25-\ns\qquad\text{(exp.),}\\
 \;19-\ns\qquad\text{(lin.)\,.}\end{cases}
\label{eq:ThetaParams}
\end{equation}
Applying the rule \eqref{eq:TCentralCharge} to eq.\ \eqref{eq:ThetaParams}, we see indeed that, first, the fixed point
theory is a CFT, and second, its central charge is given by \eqref{eq:CentChParams}.\footnote{Reading off the central
charge according to \eqref{eq:TCentralCharge} and \eqref{eq:Sind} is consistent with Refs.\ \cite{CDP14b,CD15} where
the relation between the central charge and the $\beta$-function of Newton's constant is discussed in the FRG
framework, implying a relation between $\cgr$ and $g_*\mku$. (Cf.\ also Sec.\ \ref{sec:ParamDepIntro}.)}

According to eq.\ \eqref{eq:GgAtNGFP}, the EAA related to the FP trajectory, $\Gamma_k^\text{grav,2D,NGFP}$, happens
to have exactly the structure of the induced gravity action \eqref{eq:Sind} with the corresponding central charge, for
all values of the scale parameter.

At the $k=0$ endpoint of this trajectory, the dimensionful cosmological constant $\rLk=\rls k^2$ runs to zero without
any further ado, and $\Gamma_{k\rightarrow 0}^\text{grav,2D,NGFP}$ becomes the standard effective action
\eqref{eq:GgZeroAtNGFP}. At this endpoint, by eq.\ \eqref{eq:Rsc}, self-consistent backgrounds have vanishing curvature
in the absence of matter: $R(\bg_{k=0}^\text{sc})=0$. Therefore, we have indeed inferred a central charge pertaining to
flat space by comparing \eqref{eq:ThetaParams} to \eqref{eq:TCentralCharge}.

\medskip
\noindent
\textbf{(5) Auxiliary ``matter'' CFTs}. Since the 2D gravitational fixed point action is of the induced gravity type,
we can, if we wish to, introduce a conformal matter field theory which induces it when the fluctuations of those
auxiliary matter degrees of freedom are integrated out (although such auxiliary fields are not required by our
formalism). Denoting the corresponding fields by $\chi^I$ again, and their ($k$ independent) action by
$S^\text{aux}[\chi;g]$, we then have
\begin{equation}[b]
 \e^{-\Gamma_k^\text{grav,2D,NGFP}[g]} \equiv \int\mD\chi\;\e^{-S^\text{aux}[\chi;g]}\cdot\e^{-N[g]} \; .
\label{eq:GammaSAux}
\end{equation}
Here, $N[g]\propto\int\td^2 x\sg$ is an inessential correction term to make sure that also the nonuniversal
cosmological constant terms agree on both sides of \eqref{eq:GammaSAux}; it depends on the precise definition of the
functional integral.

Clearly, the auxiliary matter CFT can be chosen in many different ways, the only constraint is that it must have the
correct central charge, $c_\text{aux}=\cgr$, that is, $c_\text{aux}=25-\ns$ or
$c_\text{aux}=19-\ns$, respectively. Let us present two examples of auxiliary CFTs:

\medskip
\noindent
\textbf{(5a) Minimally coupled scalars}.
For $c_\text{aux}>0$ the simplest choice is a multiplet of minimally coupled scalars $\chi^I(x)$,
$I=1,\cdots,c_\text{aux}$. These auxiliary fields may not be confused with the physical matter fields $A^i(x)$,
$i=1,\cdots,\ns$. The $\chi$'s and $A$'s have nothing to do with each other except that their respective numbers must
add up to $25$ (or to $19$).

\medskip
\noindent
\textbf{(5b) Feigin--Fuks theory}. The induced gravity action $I[g]$ being a nonlocal functional, it is natural to
introduce one, or several fields in addition to the metric that render the action local. The minimal way to achieve
this is by means of a single scalar field, $B(x)$, as in Feigin--Fuks theory \cite{CT74,DF84}, which has a nonminimal
coupling to the metric. Consider the following local action, invariant under general coordinate transformations applied
to $g_\mn$ and $B$:
\begin{equation}
 I^\text{loc}[g,B] \equiv \int\td^2 x\sg\,\big(D_\mu B\,D^\mu B + 2\mku R\mku B\big)\,.
\label{eq:FFAction}
\end{equation}
The equation of motion $\delta I^\text{loc}/\delta B = -2\sg\,(\Box B-R) = 0$ is solved by $B=B(g)\equiv\Box^{-1}R$
which, when substituted into $I^\text{loc}$, reproduces precisely the nonlocal form of the induced gravity action:
$I^\text{loc}[g,B(g)] = \int\sg\,R\,\Box^{-1}R\equiv I[g]$.

As $I^\text{loc}$ is quadratic in $B$, the same trick works
also quantum mechanically when we perform the Gaussian integration over $B$ rather than solve its field equation.
Hence, the exponentiated $\Gamma_k^\text{grav,2D,NGFP}$ has the representation
\begin{equation}
 \e^{-\frac{(25-\ns)}{96\pi}\, I[g] + \cdots} = \int\mD B\; \e^{-\frac{(24-\ns)}{96\pi}\int\td^2 x\sg\,(
  D_\mu B\,D^\mu B + 2\mku R\mku B + \cdots ) } \,.
\end{equation}
Here again, the dots stand for a cosmological constant which depends on the precise definition of the functional
measure $\mD B$. It is well known that thanks to the $R\mku B$-term the CFT of the $B$-field (in the limit $g_\mn\to
\delta_\mn$) has a shifted central charge \cite{FMS97,Mussardo2010}; in the present case this reproduces the values
\eqref{eq:CentChParams}.

So the conclusion is that while the fixed point action is a nonlocal functional $\propto\int\sg\,R\,\Box^{-1}R$ in
terms of the metric alone, one may introduce additional fields such that the same physics is described by a local
(concretely, second-derivative) action. In particular, $\Gamma_k^\text{grav,2D,NGFP}$ and the local functional
\begin{equation}
 \Gamma_k^\text{loc}[g,B]\equiv \frac{(24-\ns)}{96\pi}\int\td^2 x\sg\,\big(D_\mu B\,D^\mu B + 2\mku R\mku B+\cdots\big)
\end{equation}
are fully equivalent, even quantum mechanically.

\medskip
\noindent
\textbf{(6) Positivity in the gravitational sector}. Pure quantum gravity ($\ns=0$) and quantum gravity coupled to
less than $25$ (or $19$) scalars are governed by a \emph{fixed point CFT with a positive central charge}.

Clearly, this is good news concerning the pressing issue of unitarity (Hilbert space positivity) in asymptotically safe
gravity. The theories with $\cgr\ge 1$, continued to Lorentzian signature, do indeed admit a quantum
mechanical interpretation and have a state space which is a Hilbert space in the mathematical sense (no negative norm
states), supporting a unitary representation of the Virasoro algebra. In the interval $0<\cgr<1$, this can be achieved
only for discrete values of $\cgr$. In any case, we need $\cgr>0$ as a necessary condition for unitarity (cf.\ Section
\ref{sec:CFT}).

\medskip
\noindent
\textbf{(6a) Schwinger term}. Leaving the analytic continuation to the Lorentzian world aside, it is interesting to
note that already in Euclidean space the simple-looking induced gravity action ``knows'' about the fact that
$c_\text{grav}^\text{NGFP}<0$ would create a problem for the probability interpretation. By taking two functional
derivatives of the standard effective action \eqref{eq:GgZeroAtNGFP} we can compute the $2$-point function
$\langle\, T_\mn^\text{grav}(x) T_{\rho\sigma}^\text{grav}(y)\,\rangle$ and, in particular, its contracted form
$\langle\,\Theta_0(x)\Theta_0(y)\, \rangle$. Setting thereafter $g_\mn=\delta_\mn$, which, as we saw, is a
self-consistent background (assuming that we can choose a suitable, globally defined coordinate chart), we obtain the
following Schwinger term:
\begin{equation}[b]
 \langle\,\Theta_0(x)\Theta_0(y)\,\rangle = -\frac{\cgr}{12\pi}\, \p^\mu\p_\mu \delta(x-y)\,.
\label{eq:SchwingerTerm}
\end{equation}
Let us smear $\Theta_0$ with a real valued test function $f$ that vanishes at the boundary and outside of the chart
region, or, in the case where the chart is the entire Euclidean plane, falls off rapidly at infinity:\footnote{Note
that in the latter case the function $f$ has support on the entire Euclidean plane, hence we are not testing
Osterwalder--Schrader \cite{OS73,OS75} reflection positivity here \cite{Strocchi1993,GJ87}.}
$\Theta_0[f]\equiv\int\td^2 x\, f(x)\Theta_0(x)$. Then, applying $\int\td^2 x\,\td^2 y\,f(x)f(y)\cdots$ to both sides
of \eqref{eq:SchwingerTerm}, we find after an integration by parts:
\begin{equation}
 \langle\,\Theta_0[f]^2\,\rangle = 
  + \, \cgr\,\frac{1}{12\pi}\int\td^2 x\,(\p_\mu f)\delta^\mn(\p_\nu f)\,.
\label{eq:ThetaSquared}
\end{equation}
Since the integral on the RHS of \eqref{eq:ThetaSquared} is manifestly positive, we conclude that if
$\cgr<0$ the expectation value of the square $\Theta_0[f]^2$ is negative. Obviously, this would be
problematic already in the context of statistical mechanics (at least with real field variables).

\medskip
\noindent
\textbf{(6b) Induced gravity approach in 4D: a comparison}. Note that one can extract the central charge from the
Schwinger term by performing an integral $\int\td^2 x\, x^2(\cdots)$ over both sides of eq.\ \eqref{eq:SchwingerTerm}.
Since Newton's constant is dimensionless in 2D, and $\mathring{G}^{-1}=\rgs^{-1}=b=\frac{2}{3}\,\cgr$, this leads to
the following integral representation for the Newton constant belonging to the 2D world governed by the FP trajectory
\cite{Adler1982}:
\begin{equation}[b]
 \mathring{G}^{-1} = -2\pi\int\td^2 x\; x^2\langle\,\Theta_0(x)\Theta_0(0)\,\rangle \,.
\end{equation}
It is interesting to note that this representation is of precisely the same form as the relations that had been derived
long ago within the induced gravity approach in 4D, the hope being that ultimately one should be able to compute its
RHS from a matter field theory, assumed to be known (the Standard Model, say), and would then predict the value of
Newton's constant in terms of matter-related constants of Nature.

For a review and a discussion of the inherent difficulties we refer to \cite{Adler1982}. We see that, in a sense,
Asymptotic Safety was successful in making this scenario work, producing a positive Newton constant in particular, but
with one key difference: The underlying matter field theory, here the `aux' system, is no longer an arbitrary external
input, but is chosen so as to reproduce the NGFP action, an object computed from first principles.

\medskip
\noindent
\textbf{(7) Complete vs.\ gauge invariant fixed point functional}. So far we mainly focused on the gravitational part
of the NGFP functional. The complete EAA, namely $\Gamma_k = \Gg +\Gm +\Gamma_k^\text{gf} +\Gamma_k^\text{gh}$ contains
matter, gauge fixing and ghost terms in addition. But since the present truncation neglects the running of the latter
three parts, they may be considered always at their respective fixed point. Also, they have an obvious interpretation
in 2D exactly. Furthermore, our truncation assumes that neither $\Gg$ nor $\Gm$ as given in \eqref{eq:matter} has an
``extra'' $\bg$-dependence.

As a result, the sum of gravity and matter (`GM') contributions,
\begin{equation}
 \Gamma_k^\text{GM,2D}[g,A] \equiv
  \Ggd[g]+\frac{1}{2}\sum\limits_{i=1}^{\ns}\int\td^2 x\sg\;g^\mn\p_\mu A^i\p_\nu A^i \,,
\end{equation}
enjoys both background independence, here meaning literally independence of the background metric, and gauge
invariance, i.e.\ it does not change under diffeomorphisms applied to $g_\mn$ and $A^i$.

Thanks to the second property\footnote{Which might not be realized in more complicated truncations!},
we may adopt the point of view that it is actually the gauge invariant functional $\Gamma_k^\text{GM,2D}$ only which
contains all information of interest and was thus ``handed over'' alone from the higher dimensional Einstein--Hilbert
world to the intrinsically $2$-dimensional induced gravity setting.
Therefore, if in 2D the necessity of gauge fixing arises, we can in principle pick a new gauge, different from
the one employed in $d>2$ for the computation of the $\beta$-functions.\footnote{This could not be
done if one wants to combine loop or RG calculations from $d>2$ with others done in $d=2$ exactly. However, in this and
the previous chapter all dynamical calculations are done in $d>2$, i.e.\ before the 2D limit is taken.}

\medskip
\noindent
\textbf{(8) Unitarity vs.\ stability: the conformal factor ``problem''}.
Next we take advantage of the particularly convenient conformal gauge available in strictly $2$ dimensions (cf.\
Section \ref{sec:ConfGauge}), and evaluate $\GgdN[g]$ as given explicitly by eq.\ \eqref{eq:GgAtNGFP} for metrics of
the special form $g_\mn = \e^{2\phi}\mku \hg_\mn$. The result is a Liouville action as before in eqs.\
\eqref{eq:GravToLiou}, \eqref{eq:LiouvilleAction}, this time without any undetermined piece such as $U_k[\hg]$,
however:
\begin{equation}
 \GgdN\big[\e^{2\phi}\mku\hg\big] \equiv \frac{\cgr}{96\pi}\,I[\hg] + \GL[\phi;\hg]\,,
\label{eq:GammaGravILiou}
\end{equation}
with
\begin{equation}
 \GL[\phi;\hg] = \frac{\cgr}{12\pi}\int\td^2 x\shg\,\left\{-\frac{1}{2}\,\hD_\mu\phi\,\hD^\mu\phi-\frac{1}{2}\hR\phi
 +\rls\,k^2\mku\e^{2\phi} \right\} \,.
\label{eq:LiouvilleNGFP}
\end{equation}
Since $\cgr=25-\ns$ (or $\cgr=19-\ns$ with the linear parametrization), we observe that for pure gravity, and gravity
interacting with not too many matter fields, the conformal factor has a ``wrong sign'' kinetic term that might seem
to indicate an instability at first sight. If we think of the fixed point action as induced by some auxiliary CFT
with central charge $c_\text{aux}=\cgr=25-\ns>0$, we see that this is exactly the correlation mentioned in paragraph
\textbf{(3b)} above: bona fide unitary CFTs generate ``wrong sign'' kinetic terms for the conformal factor.

We emphasize that the unstable $\phi$-action is neither unexpected, nor ``wrong'' from the physics point of view,
nor in contradiction with the positive central charge of the fixed point CFT. Let us discuss these issues in turn
now.

\medskip
\noindent
\textbf{(8a) The importance of Gauss' law}. Recall the standard count of gravitational degrees of freedom in
Einstein--Hilbert gravity: In $d$ dimensions, the symmetric tensor $g_\mn$ contains $\frac{1}{2}d(d+1)$ unknown
functions which we try to determine from the $\frac{1}{2}d(d+1)$ field equations $G_\mn=\cdots\,$. Those are not
independent, but subject to $d$ Bianchi identities. Moreover, we need to impose $d$ coordinate conditions due to
diffeomorphism invariance. This leaves us with $\NEH(d)\equiv\frac{1}{2}d(d+1)-d-d=\frac{1}{2}d(d-3)$ gravitational
degrees of freedom, meaning that by solving the Cauchy problem for $g_\mn$ we can predict the time evolution of
$\NEH(d)$ functions that, (i), are related to ``physical '' (i.e.\ gauge invariant) properties of space, (ii), are
algebraically independent among themselves, and (iii), are \emph{independent of the functions describing the evolution
of matter}.

With $\NEH(4)=2$ we thus recover the gravitational waves of 4D General Relativity, having precisely $2$
polarization states. Similarly, $\NEH(3)=0$ tells us that there can be no gravitational waves in $3$
dimensions since all independent, gauge invariant properties described by the metric can be inferred already from
the matter evolution. No extra initial conditions can, or must, be imposed.

Finally $\NEH(2)=-1$ seems to suggest that ``gravity has $-1$ degree of freedom in $2$ dimensions''. Strange as it
might sound, the meaning of this result is quite clear: The quantum metric with its ghosts removes one degree of
freedom from the matter system.
If, in absence of gravity, the Cauchy problem of the matter system has a unique solution after specifying $N_\text{m}$
initial conditions, then this number gets reduced to $N_\text{m}-1$ by coupling the system to gravity.

Quantum mechanically, on a state space with an indefinite metric, the removal of degrees of freedom happens upon
imposing ``Gauss' law constraints'', or ``physical state conditions'' on the states. As a result, the potentially
dangerous negative-norm states due to the wrong sign of the kinetic term of $\phi$ are not part of the actual
(physical) Hilbert space. The latter can be built using matter operators alone, and it is in fact smaller than without
gravity.\footnote{See Polchinski \cite{Polchinski1989} for a related discussion.}

The situation is analogous to Quantum Electrodynamics (QED) in the Coulomb gauge, for example. The overall sign of the
Maxwell action $\propto F_\mn F^\mn$ is chosen such that the spatial components of $A_\mu$ have a positive kinetic
term, and so it is unavoidable that the time component $A_0$ has a negative one, like the conformal factor in
\eqref{eq:LiouvilleNGFP}. However, it is well known \cite{BD65} that the states  with negative $(\text{norm})^2$
generated by $A_0$ do not survive imposing Gauss' law $\bm{\nabla}\cdot\bm{E}=e\,\psi^\dagger\psi$ on the states. This
step indeed removes one degree of freedom since $A_0$ and $\rho_\text{em}\equiv e\,\psi^\dagger\psi$ get coupled
by an \emph{instantaneous} equation, $\bm{\nabla}^2 A_0(t,\bm{x})=-\rho_\text{em}(t,\bm{x})$.

\medskip
\noindent
\textbf{(8b) Instability and attractivity of classical gravity}. To avoid any misunderstanding we recall that in
constructing realistic 4D theories of gravity it would be quite absurd, at least in the Newtonian limit, to ``solve''
the problem of the conformal factor by manufacturing a positive kinetic term for it in some way. In taking the
classical limit of General Relativity, this kinetic term essentially descends to the $\bm{\nabla}\vp_\text{N}\cdot
\bm{\nabla}\vp_\text{N}$-part of the classical Lagrangian governing Newton's potential $\vp_\text{N}$ and therefore
fixes the positive sign on the RHS of Poisson's equation, $\bm{\nabla}^2\vp_\text{N}=+4\pi\mku G\rho$. However, this
latter plus sign expresses nothing less than the universal attractivity of classical gravity, something we certainly
want to keep.

This simple example shows that the conformal factor instability is by no means an unmistakable sign for a
physical deficiency of the theory under consideration. The theory can be perfectly unitary if there are appropriate
Gauss' law-type constraints to cut out the negative norm states of the indefinite metric state space.

\medskip
\noindent
\textbf{(8c) Central charge in Liouville theory}. Finally, we must discuss a potential source of confusion concerning
the correct identification of the fixed point's central charge. Let us pretend that the Liouville action
$\GL[\phi;\hg]$ describes a matter field $\phi$ in a ``background'' metric $\hg_\mn\mku$.\footnote{Recall, however,
that the reference metric $\hg_\mn$ that enters only the conformal parametrization of 2D metrics is to be distinguished
carefully from the true background metric $\bg_\mn$ which is at the heart of the entire gravitational EAA setting. In
this conformal parametrization, a generic bimetric action $F[g,\bg]$ translates into a functional of \emph{two}
conformal factors, $F\big[\phi,\bar{\phi};\hg\big] \equiv F\big[\e^{2\phi}\mku\hg,\mkuu\e^{2\bar{\phi}}\mku\hg\mku
\big]$.} It would then be natural to ascribe to this field the stress-energy tensor
\begin{equation}
 T_k^\text{L}[\phi;\hg]^\mn \equiv \frac{2}{\shg}\frac{\delta\GL[\phi;\hg]}{\delta\hg_\mn}\,.
\label{eq:TLiouvilleDef}
\end{equation}
Without using the equation of motion (i.e.\ ``off shell'') its trace is given by
\begin{equation}
 \Theta_k^\text{L}[\phi;\hg]\equiv \hg_\mn\,T_k^\text{L}[\phi;\hg]^\mn = \frac{\cgr}{12\pi}
  \left(\hB\phi + 2\mku\rls\,k^2\mku\e^{2\phi} \right).
\label{eq:TLiouville}
\end{equation}
Concerning \eqref{eq:TLiouville}, several points are to be noted.

\begin{enumerate}
\item Varying $\GL$ with respect to $\phi$ yields Liouville's equation $\hB\phi+2\mku\rls\, k^2\mku\e^{2\phi}
= \frac{1}{2}\hR$. With $\phi_\text{sol}$ denoting any solution to it, we obtain ``on shell'' the following
$k$-independent trace:
\begin{equation}
 \Theta^\text{L}[\phi_\text{sol};\hg] = +\cgr \frac{1}{24\pi}\hR \,.
\label{eq:ThetaLiouville}
\end{equation}
If we now compare \eqref{eq:ThetaLiouville} to the general rule \eqref{eq:TCentralCharge}, we conclude that the
Liouville field represents a CFT with the central charge
\begin{equation}
 c^\text{L} = -\cgr\,,
\label{eq:CLCGrav}
\end{equation}
which is \emph{negative} for pure asymptotically safe gravity, namely $c^\text{L} = -25$, or $-19$, respectively.
\end{enumerate}
Does this result indicate that the fixed point CFT is nonunitary, after all? The answer is a clear `no', and the
reason is as follows.
\begin{enumerate}[resume]
\item The Liouville theory governed by $\GL$ of \eqref{eq:LiouvilleNGFP} is not a faithful description of the
NGFP. According to eq.\ \eqref{eq:GammaGravILiou}, the full action contains the ``pure gravity'' term
$\frac{\cgr}{96\pi}I[\hg]$ in addition. In order to correctly identify the central charge of the NGFP, it is essential
to add the $\hg_\mn$-derivative of this term to the Liouville stress-energy tensor. Hence, the trace
\eqref{eq:TLiouville} gets augmented to
\begin{align}
 \frac{2\mku\hg_\mn}{\shg}\frac{\delta}{\delta\hg_\mn}\left(\frac{\cgr}{96\pi}I[\hg]\right)
 +{}& \Theta_k^\text{L}[\phi;\hg]
 = -\frac{\cgr}{24\pi}R(\hg)+\Theta_k^\text{L}[\phi;\hg] 
\label{eq:ThetaFull}\\
 &= \frac{\cgr}{24\pi}\left[-R(\hg)+2\mku\hB\phi+4\mku\rls\, k^2\mku\e^{2\phi}\right]
\nonumber\\
 &= \frac{\cgr}{24\pi}\left[-\e^{-2\phi}\big(R(\hg)-2\mku\hB\phi\big)+4\mku\rls\, k^2\right] \e^{2\phi}
\nonumber\\
 &= \frac{\cgr}{24\pi}\left[-R\big(\e^{2\phi}\mku\hg\big)+4\mku\rls\, k^2\right] \e^{2\phi}
\nonumber\\
 &= \e^{2\phi}\,\Theta_k\big[\mku\e^{2\phi}\mku\hg\mku\big].
\nonumber
\end{align}
In the 2${}^\text{nd}$ line of \eqref{eq:ThetaFull} we inserted \eqref{eq:TLiouville}, in going from the
3${}^\text{rd}$ to the 4${}^\text{th}$ line we exploited the identity \eqref{eq:WeylR} from the appendix, and in the
last line we used \eqref{eq:ThetaParams}. So with this little calculation we have checked that the Liouville
stress-energy tensor makes physical sense only when combined with the pure gravity piece.\footnote{In isolation,
$\Theta^\text{L}[\phi;\hg]$ is not invariant under the Weyl split-symmetry transformations \eqref{eq:SplitSymmetry},
i.e.\ not a function of the combination $\e^{2\phi}\mku\hg$ only.} If this is done, the total gravitational trace from
which the correct central charge is inferred, eq.\ \eqref{eq:ThetaParams}, is indeed recovered, as it should be. It
satisfies the relation\footnote{The explicit factor $\e^{-2\phi}$ in \eqref{eq:ThetaAdded} is simply due to the
different volume elements $\shg$ and $\sg=\shg\,\e^{2\phi}$ appearing in the definitions of the stress-energy tensors
\eqref{eq:TLiouvilleDef} and \eqref{eq:StressTensorGeneral}, respectively.}
\begin{equation}
 \Theta_k[\mku g] \equiv \Theta_k\big[\mku\e^{2\phi}\mku\hg\mku\big] 
  = \e^{-2\phi}\bigg(-\frac{\cgr}{24\pi}\,\hR+\Theta_k^\text{L}[\phi;\hg]\bigg),
\label{eq:ThetaAdded}
\end{equation}
which holds true even off shell.

\item If we take $\phi$ on shell, eq.\ \eqref{eq:ThetaLiouville} applies, and so the two terms in the brackets
of eq.\ \eqref{eq:ThetaAdded} cancel precisely. This, too, is as it should be since from eq.\ \eqref{eq:Rsc} we know
already that $\Theta_k[g]$ vanishes identically when $g\equiv\bg$ is a self-consistent background, and this is exactly
what we insert into \eqref{eq:ThetaAdded} when $\phi$ is a solution of Liouville's equation.

\end{enumerate}

Thus, taking the above points together we now understand that nothing is wrong with $c^\text{L} = -\cgr$. In fact,
$c^\text{L}<0$ for pure gravity is again a reflection of the Liouville field's ``wrong-sign'' kinetic
term\footnote{Hence, at the technical level, the wrong-sign kinetic term requires special attention (regularization,
analytic continuation, or similar) at intermediate steps of the calculation at most.} and its perfectly correct
property of reducing the total number of degrees of freedom.

%----------------------------------------------------------------------------------------------------------------------
\section{Summarizing remarks}
\label{sec:SumUpNGFPCFT}
%----------------------------------------------------------------------------------------------------------------------

In Chapter \ref{chap:EHLimit} we started from the Einstein--Hilbert truncation for the effective average action of
metric quantum gravity in $d>2$ dimensions and constructed its intrinsically $2$-dimensional limit. This limit was
taken directly at the level of the action, rather than being a mere $\ve$-expansion of $\beta$-functions. We saw that
it turns the (local, second-derivative) Einstein--Hilbert term into the nonlocal Polyakov action.

Using this result in the present chapter, we were able to conclude that in 2D the non-Gaussian fixed point underlying
Asymptotic Safety gives rise to a \emph{unitary} conformal field theory whose gravitational sector possesses the
central charge $+25$. We analyzed the properties of the fixed point CFT using both a gauge invariant description and a
calculation based on the conformal gauge where it is represented by a Liouville theory.

We close with a number of further comments.

\medskip
\noindent
\textbf{(1)}
An important step in proving the viability of the Asymptotic Safety program consists in demonstrating that Hilbert
space positivity can be achieved together with background independence and nonperturbative renormalizability. While
we consider our present result on the unitarity of the pertinent CFT as an encouraging first insight, it is clear,
however, that the 2D case is not yet a crucial test since the gravitational field has no independent propagating
degrees of freedom, and so there is no pure-gravity subspace of physical states whose positivity would be at stake. To
tackle the higher dimensional case additional techniques will have to be developed. Nevertheless, it is interesting
that at least at the purely geometric level the remarkable link between the Einstein--Hilbert and the Polyakov action
which we exploited has an analogue in all even dimensions $d=2n$. Each nontrivial cocycle of the Weyl cohomology
yields, in an appropriate limit $d\rightarrow 2n$, a well defined nonlocal action that is conjectured to be part of the
standard effective action in $2n$ dimensions \cite{MM01}.

\medskip
\noindent
\textbf{(2)}
A number of general lessons we learned here will be relevant in higher dimensions, too. We mention in particular that
the issue of unitarity cannot be settled by superficially checking for the stability of some bare action and ruling out
``wrong sign'' kinetic terms as this is sometimes implied. We saw that the CFT which is at the heart of the NGFP is
unitary \emph{even though} in conformal gauge it entails a negative kinetic energy of the Liouville field.
As we explained in Section \ref{sec:UnitaryCFT}, the background field, indispensable in our approach to quantum
gravity, plays an important role in reconciling these properties.

\medskip
\noindent
\textbf{(3)}
We showed that the crucial central charge $\cgr$ can be read off from the leading term in the $\beta$-function of
Newton's constant, and we saw that the pure gravity result is either $25$ or $19$, depending on whether the exponential
or the linear parametrization of the metric is chosen, respectively. The arguments of Section \ref{sec:Birth} suggest
accepting the result of the former, $+25$, as the correct one in the present context. Nevertheless, the
issue of parametrization dependence is not fully settled yet, and one should still be open towards the possibility that
the two sets of results, obtained from the same truncation ansatz but different choices of the fluctuating field,
might actually refer to different universality classes.

\medskip
\noindent
\textbf{(4)}
Regarding different universality classes, it is perhaps not a pure coincidence that
the ``$19$'' is also among the ``critical dimensions for noncritical strings'' which were found by Gervais
\cite{Gervais1990,Gervais1991a,Gervais1991b,Gervais1991c,Gervais1995,Gervais1996}:
\begin{equation}
 D_\text{crit}=7,13,19.
\end{equation}
They correspond to gravitational central charges $c_\text{grav}=19,13,7$,
respectively. For these special values the Virasoro algebra admits a unitary truncation, that is, there exists a
subspace of the usual state space on which a corresponding chiral algebra closes, and which is positive (in the sense
that it contains no vectors $|\psi\rangle$ with $\langle\psi|\psi\rangle<0$). The associated string theories were
advocated as consistent extensions of standard Liouville theory, which is valid only for $c<1$ and $c>25$ when gravity
is weakly coupled, into the strongly coupled regime, $1<c<25$, in which the KPZ formulae \cite{KPZ88,David1988,DK89}
would lead to meaningless complex answers.

Thus, for the time being, we cannot exclude the possibility that a better understanding of the RG flow computed with
the linear parametrization (but with more general truncations than those analyzed in this thesis) will lead to
the picture that there exists a second pure gravity fixed point compatible with Hilbert space positivity, namely at
$c_\text{grav}=19$, and that this fixed point represents another, inequivalent universality class.

We know already that this picture displays the following correlation between pa\-ra\-me\-tri\-za\-tion and universality
class, which we would then indeed consider the natural one: The exponential parametrization, i.e.\ the ``conservative''
one in the sense that it covers only nonzero, nondegenerate, hence ``more classical'' metrics having a fixed signature,
leads to $c_\text{grav}=25$ which is located just at the boundary of the strong coupling interval. In the way it is
employed, the linear parametrization, instead, gives rise to an integration also over degenerate, even vanishing tensor
field configurations not corresponding to any classical metric; typically enough, it
is this parametrization that would be linked to the hypothetical, certainly quite nonclassical theory with
$c_\text{grav}=19$ deep in the strong coupling domain.

Whatever the final answer will be, it seems premature, also in more than $2$ dimensions, to regard the exponential
parametrization merely as a tool to do calculations in a more precise or more convenient way than this would be
possible with the linear one. It might rather be that in this manner we are actually computing something else.

%----------------------------------------------------------------------------------------------------------------------
\chapter{The reconstructed bare action}
\label{chap:Bare}
%----------------------------------------------------------------------------------------------------------------------

\begin{summary}
Although it is possible to derive the FRGE from a functional integral formulation, its final manifestation given by
eq.\ \eqref{eq:FRGE} has no reminiscence of such a derivation and does not depend on any path integral. Solving the
theory amounts to solving the FRGE, and thus we dispense with the need to define a functional measure and a bare
action. However, if we want to access the microscopic degrees of freedom in more detail, a precise knowledge of the
bare action may become indispensable. In this chapter we prove a one-loop relation between the effective average
action and the bare action, the ``reconstruction formula'', and we argue that the relation becomes exact for certain
terms when the large cutoff limit is considered. We apply these results to gravity within the Einstein--Hilbert
truncation in order to determine the bare cosmological constant and the bare Newton constant. It will be shown
that the bare sector features a non-Gaussian fixed point in this framework. Finally, we reveal a mechanism how the
freedom in setting up a functional measure can be exploited to adjust bare couplings in a convenient way.

\noindent
\textbf{What is new?} Exactness beyond one-loop (Sec.\ \ref{sec:BareOneLoopExact}); existence and properties of
the bare NGFP (Secs.\ \ref{sec:BareExNGFP} \& \ref{sec:BareCritExp}); a strategy to adjust bare couplings (Sec.\
\ref{sec:BareMechanism}), used to achieve a vanishing bare cosmological constant and a bare Newton constant that
agrees with the effective one (Sec.\ref{sec:Bare2D}).
\end{summary}

%----------------------------------------------------------------------------------------------------------------------
\section{Motivation}
\label{sec:BareMotivation}
%----------------------------------------------------------------------------------------------------------------------

From a Wilson--Kadanoff point of view, the renormalization process amounts to starting from a bare action in a path
integral at some UV scale $\UV$, the Wilsonian action $\SW$, decomposing the integration field variable into high and
low momentum modes, integrating out the high momentum modes and reexpressing the remaining pieces in terms of an
``effective'' bare action, $S_{\UV'}^\text{W}$, valid at some scale $\UV'<\UV$. This procedure can be continued down to
the scale zero until all modes are integrated out, giving rise to the ordinary effective action $\Gamma$. We can think
of $\SW$ at different values of $\UV$ as a set of actions for the \emph{same system}. It is crucial that $\SW$ plays
the role of a \emph{bare action} at the scale $\UV$ as long as $\UV>0$.\footnote{When using a running bare action in
the Wilsonian sense we denote it by $\SW$. If, on the other hand, we consider a bare action at some fixed UV
scale $\UV$, we denote it by $\SB$.}

By contrast, in the effective average action (EAA), $\Gk$, there are no unintegrated fluctuations,
so inherently $\Gk$ is a standard \emph{effective action} for each $k$. In this sense, $\Gk$ describes a
family of \emph{different systems}: For each $k$ it is the ordinary effective action for a system whose
full bare action is of the form $\SB+\DSk$, where $\DSk$ denotes the mode suppression term. The corresponding
correlation functions provide an effective field theory description of the physics at scale $k$.

Having emphasized the conceptual differences between the bare/Wilsonian action and the effective average action, one
might raise the question whether the two types of actions can be transformed into each other. One ``direction'' of
such a relation is rather straightforward since the EAA can in principle be obtained by functional
integration provided that a bare action, an appropriately regularized functional measure and a mode suppression term
are given. It is the other direction that we will focus on in this chapter: Let us assume that we are given an
effective average action $\Gk$ which, upon setting $k=0$, yields the standard effective action, $\Gamma=\Gamma_{k=0}$.
This brings us to the question how a bare action $\SB$ (together with a suitably defined functional measure) has to be
chosen in order that the corresponding path integral reproduces precisely the same effective action $\Gamma$.

It is important to keep in mind that the ``derivation'' of the FRGE from a functional integral is
only formal as it ignores all difficulties specific to the UV limit of quantum field theories. In fact, rather than the
integral, the starting point of the EAA based route to a fundamental theory is the mathematically perfectly well
defined, UV cutoff-free flow equation \eqref{eq:FRGE}. In this setting, the problem of the UV limit is shifted from the
properties of the equation itself to those of its \emph{solutions}, converting renormalizability into a condition on
the existence of fully extended RG trajectories on theory space. The Asymptotic Safety paradigm is a way of achieving
full extendability in the UV and, barring other types of (infrared, etc.) difficulties, it leads to a well-behaved
action functional $\Gamma_k$ at each $k\in[0,\infty)$. Every such complete RG trajectory defines a quantum field
theory (with the cutoff(s) removed). The ``reconstruction problem'' \cite{MR09,VZ11,MS15,NR16a} consists in finding
a functional integral that reproduces a given complete $\Gamma_k$-trajectory.

The benefits of reconstructing the bare action from the effective average action are
diverse: First, the bare action provides direct access to the microscopic degrees of freedom and their fundamental
interactions. This allows reconstructing the Hamiltonian phase space formulation describing the classical system.
Second, the implementation of symmetries or constraints, the derivation of Ward identities and further general
properties can be studied more easily in a path integral setting. Third, the bare action is needed to make contact to
perturbation theory and similar approximation schemes. And finally, establishing the connection to different approaches
might require a bare action, too. In gravity, for instance, it would be interesting to know the relation between the
EAA formulation on the one hand and canonical quantum gravity, loop quantum gravity or Monte Carlo simulations of
causal dynamical triangulations (CDT) on the other hand, where the bare action plays a central role in the latter three
approaches.

There is a rule of thumb often mentioned in the literature on the EAA (see Ref.\ \cite{Wetterich1993}, for instance):
``In the large cutoff limit $\Gk$ approaches the bare action, $\Gamma_{k\to\infty} = \SB$.'' However, even if we ignore
for a moment the problems related to UV regularization, this heuristic rule cannot be complete; there are additional
correction terms.
This can be seen by critically revising the standard argument underlying the rule of thumb, which says that the mode
suppression term
\begin{equation}
 \e^{-\DSk}\equiv \e^{-\frac{1}{2}\int\sg\,(\chi-\phi)\Rk(\chi-\phi)}
\label{eq:ModeSuppressionTerm}
\end{equation}
acts effectively as a $\delta$-functional for $k\to\infty$ in a path integral over the field $\chi$.
%Then the argument of $\SB[\chi]$ would be forced to equal $\phi$ while all further terms would be made vanish.
The idea behind this argument  is based on the relation $\Rk\propto k^2$. In the limit $k\to\infty$ the term
\eqref{eq:ModeSuppressionTerm} thus fully suppresses all field contributions to the integral except for $\chi=\phi$.
The premature conclusion from this would be that \eqref{eq:ModeSuppressionTerm} is equivalent to the functional
$\delta[\chi-\phi]$ in the large $k$ limit. In fact, this is not true.

\begin{wrapfigure}{R}{0.48\textwidth}
\centering
\normalcaption
\captionwidth{0.43\textwidth}
\includegraphics[width=0.42\textwidth]{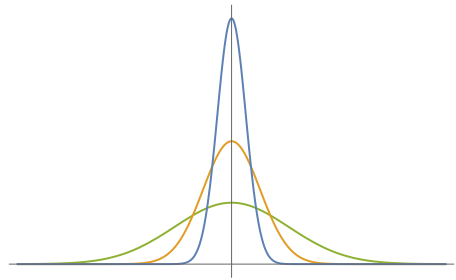}
\caption{Approximation of a delta function by a family of Gaussian curves by increasing their height and decreasing
 their width.}
\label{fig:Gaussians}
\end{wrapfigure}
Let us demonstrate the crucial issue in terms of a simple $\delta$-function which can be approximated by a family of
Gaussian curves,
\begin{equation}
 \delta_k(x)\equiv \frac{k}{\sqrt{2\pi}}\,\e^{-\frac{1}{2}k^2 x^2}\,,
\label{eq:deltaApprox}
\end{equation}
with the standard deviation $\sigma=1/k$, see Figure \ref{fig:Gaussians}. Thanks to the chosen normalization we
have $\int_{-\infty}^\infty\td x\,\delta_k(x) = 1$ for all $k$, and $\delta_k(x)$ will indeed approach a
$\delta$-function in the limit $k\to\infty$.
The key point is that $k$ enters the RHS of \eqref{eq:deltaApprox} twice: Increasing $k$ means increasing the
\emph{height} (due to the prefactor) and simultaneously \emph{squeezing} the curve (due to the exponential). Only an
appropriate combination of amplifying and squeezing will ultimately lead to a $\delta$-function.

Having said this, it is clear what prevents eq.\ \eqref{eq:ModeSuppressionTerm} from approaching $\delta[\chi-\phi]$:
The exponential leads to a squeezing of the functional for increasing $k$ which gives rise to the mode suppression, but
there is no suitable prefactor which is required to increase the height. As a consequence, we do not obtain an exact
$\delta$-functional in the large $k$ limit. Stated differently, the rule of thumb, $\Gamma_{k\to\infty}=\SB$, whose
``derivation'' relies on the validity of the $\delta$-functional argument, is incomplete.

There are two possibilities
how this problem can be cured. (1.) We could multiply \eqref{eq:ModeSuppressionTerm} by a suitable
$k$-dependent prefactor. In this way, it can be achieved that the relation $\Gamma_{k\to\infty}=\SB$ becomes exact. This
would, however, lead to a $k$-dependent path integral measure and modify the flow equation for $\Gk$. Such an approach
has been pursued in Ref.\ \cite{VZ11}, cf.\ also Ref.\ \cite{MS15}. (2.) We could stick to
\eqref{eq:ModeSuppressionTerm} without modifying the measure. This leaves the flow equation unaltered, but requires a
modification such as $\Gamma_{k\to\infty}=\SB+\text{correction}$ \cite{MR09}. In this chapter we focus on the second
possibility.

%----------------------------------------------------------------------------------------------------------------------
\section{The one-loop reconstruction formula}
\label{sec:BareOneLoop}
%----------------------------------------------------------------------------------------------------------------------

The association of a functional integral, i.e.\ a bare theory, to a $\Gamma_k$-trajectory is highly nonunique. The
first decision to be taken concerns the variables of integration: They may or may not be fields of the same sort as
those serving as arguments of $\Gamma_k$. From the practical point of view the most important situation is when the
integration variables are no (discretized) fields at all, but rather belong to a certain statistical mechanics model
whose partition function at criticality is supposed to reproduce the predictions of the EAA trajectory. Besides the
nature of the integration variables, a UV regularization scheme, a correspondingly regularized functional integration
measure, and an associated bare action $\SB$ are to be chosen. Then the information encapsulated in
$\Gamma_{k\rightarrow\infty}$ can be used to find out how the bare parameters contained in $\SB$ must depend
on the UV cutoff $\UV$ in order to give rise to a well-defined path integral reproducing the EAA-trajectory in the
limit $\UV\rightarrow\infty$.

Guided by the setting of Ref.\ \cite{MR09} we consider a reconstruction based on the following two choices:
(i) The integration variable is taken to be of the same sort as in the argument of $\Gamma_k$.
(ii) The UV regularization is implemented by means of a sharp mode cutoff.

In order to derive a reconstruction formula we have to specify in detail how the functional measure is defined.
Otherwise, it would be impossible to determine the bare action: Any shift in the bare action of the form $\SB\to\SB+X$
can be absorbed by multiplying the measure by $\e^X$, and vice versa. Thus, only the combination of measure and bare
action is a meaningful object. Appendix \ref{app:Measure} contains a thorough discussion about how the functional
measure can be defined consistently. It is shown that the definition is not unique but rather involves a parameter $M$
which labels a certain \emph{1-parameter family of measures}. The $M$-dependence of the measure translates into an
$M$-dependent bare action. This nonuniqueness signals the ``unphysicalness'' of the bare action. As we will see later
on, this fact can be exploited to adjust the bare coupling constants conveniently.

In the following subsection we review and extend the arguments of Ref.\ \cite{MR09}.

%----------------------------------------------------------------------------------------------------------------------
\subsection{Derivation}
\label{sec:BareOneLoopDer}
%----------------------------------------------------------------------------------------------------------------------

Let $\phi$ denote a (collection of) generic field(s) of unspecified type, i.e.\ $\phi$ represents scalar fields, metric
fluctuations or gauge fields, for instance. Since the line of reasoning in the subsequent computation is the same for
any kind of field, we adopt --- for the sake of readability --- the simple notation for scalar fields, bearing in mind
that an appropriate extension to other field types will in general require the use of internal indices, background
fields, as well as additional gauge fixing and ghost terms supplementing the bare action.

Starting out from the definition of the effective average action $\Gamma_{k,\UV}\mku$, given in Sec.~\ref{sec:EAAFRGE},
we can reexpress the defining equation as\footnote{Note that we state the dependence on the UV cutoff scale $\UV$
explicitly here since it enters both the bare action and the functional measure (cf.\ Appendix \ref{app:Measure}) in a
crucial way. It was dropped in Sec.\ \ref{sec:EAAFRGE} where we implicitly considered the limit $\UV\to\infty$ in the
end, in particular in the FRGE.}
\begin{equation}
 \e^{-\GkL[\phi]} \equiv \e^{-J\cdot\phi+\frac{1}{2}\phi\cdot\Rk\mku\phi}\int\!\mD_\UV\chi\;
 \e^{-\SB[\chi]+J\cdot\chi-\frac{1}{2}\chi\cdot\Rk\mku\chi}\;,
\label{eq:IntDefGammak}
\end{equation}
with the shortcuts $J\cdot\phi \equiv \int\!\dd x\sg\,J(x)\phi(x)$ and $\phi\cdot\Rk\phi \equiv \int\!\dd x\sg\,
\phi(x)\Rk(-\Box)\phi(x)$. While being irrelevant for the form of the FRGE, the explicit dependence of the functional
measure $\mD_\UV\chi$ on the UV cutoff scale $\UV$ (and on the parameter $M$) will turn out to be crucial for the
reconstruction step (cf.\ Appendix \ref{app:Measure}).
The source $J(x)\equiv J_{k,\UV}[\phi](x)$ is determined by the equation
\begin{equation}
 \GkL^{(1)}[\phi](x) \equiv \frac{1}{\sqrt{g(x)}}\frac{\delta\Gamma_k[\phi]}{\delta\phi(x)} = J(x) - \Rk\mku\phi(x).
\label{eq:EqForJ}
\end{equation}
Replacing $J$ in \eqref{eq:IntDefGammak} according to \eqref{eq:EqForJ} yields
\begin{equation}
 \e^{-\GkL[\phi]} = \int\!\mD_\UV\chi\;\e^{-\SB[\chi]
  +\GkL^{(1)}[\phi] \cdot(\chi-\phi)-\frac{1}{2}(\chi-\phi)\cdot\Rk(\chi-\phi)}\;.
\end{equation}
We can now exploit the translation invariance of the measure to make the change of variables $\chi \rightarrow f
= \chi-\phi$ and obtain
\begin{equation}
 \e^{-\GkL[\phi]} = \int\mD_\UV f\;\e^{-\St[f;\phi]}\;,
\label{eq:GkLvsStot}
\end{equation}
where we introduced the total action
\begin{equation}
 \St[f;\phi]\equiv \SB[\phi+f]- \GkL^{(1)}[\phi]\cdot f +\frac{1}{2}\mku f\cdot\Rk\mku f\,.
\end{equation}
It is convenient to reinstate $\hbar$ as a bookkeeping parameter for a moment, allowing us to systematically
count loop orders. Equation \eqref{eq:GkLvsStot} then becomes
\begin{equation}
 \e^{-\frac{1}{\hbar}\GkL[\phi]} = \int\mD_\UV f\;\e^{-\frac{1}{\hbar}\St[f;\phi]}\;.
\label{eq:GkLvsStothbar}
\end{equation}

At this point we make the assumption that $\SB$ behaves like a generic action in that it is bounded from below.
(Clearly, when the bare action has been reconstructed, one should test a posteriori if the solution $\SB$ is
consistent with this assumption.) In that case, since $\Rk$ is positive by construction, we find that $\St$, too, is
bounded from below. As a consequence, $\St[f;\phi]$ must have a minimum w.r.t.\ $f$ for fixed $\phi$, so the equation
\begin{equation}
 \frac{\delta\St}{\delta f}[f_0;\phi]=0,
\end{equation}
defining a stationary ``point'' $f_0$, is guaranteed to have a solution. This stationary point can be used in turn
to perform a \emph{saddle point expansion} in the integrand of \eqref{eq:GkLvsStothbar}: We decompose the integration
variable $f$ according to
\begin{equation}
 f = f_0 + \sqrt{\hbar}\,\frac{M}{\UV}\,\vp\,,
\end{equation}
and eq.\ \eqref{eq:GkLvsStothbar} becomes
\begin{equation}
 \e^{-\frac{1}{\hbar}\GkL[\phi]} = \int\mD_\UV \vp\,J_\UV\,\e^{-\frac{1}{\hbar}\St[f_0;\phi]
 -\frac{1}{2}\frac{M^2}{\UV^2}\int\!\sg\,\vp\left(\SB^{(2)}[\phi+f_0]+\Rk\right)\vp\, +\cdots}\;.
\label{eq:SaddlePExpMain}
\end{equation}
In appendix \ref{app:OneLoop} we show by a careful analysis that \textbf{(i)} all higher order terms in
\eqref{eq:SaddlePExpMain} indicated by the dots do not contribute to the final result at one-loop level and vanish in
the large cutoff limit, \textbf{(ii)} the Jacobian $J_\UV \equiv \det_\UV\left(\frac{\delta f}{\delta\vp}\right)$ is
field independent and can be pulled out of the integral, \textbf{(iii)} the remaining Gaussian integral can be computed
exactly, giving rise to a determinant which can be written as a trace by using $\ln\det(\cdot)=\Tr\ln(\cdot)$, and
\textbf{(iv)} the stationary point $f_0$ is found to be of first order in $\hbar$, a result that can be exploited for a
subsequent $\hbar$-expansion. For further details we refer the reader to the appendix. Employing
\textbf{(i)}--\textbf{(iv)} we finally obtain
\begin{equation}
 \GkL[\phi] = \SB[\phi]+\frac{\hbar}{2}\,\Tr_\UV \ln\left[\textstyle\frac{1}{\hbar\mku M^2}\left(\SB^{(2)}[\phi]
 +\Rk\right)\right] +\mO(\hbar^{3/2}/\UV)+\mO(\hbar^2).
\label{eq:OneLoopScalar}
\end{equation}
Here and in the following, we use the definition $\Tr_\UV\big[(\mku\cdot\mku)\big] \equiv
\Tr\big[(\mku\cdot\mku)\mku \theta(\UV^2+\Box)\big]$ for the regularized trace. In eq.\ \eqref{eq:OneLoopScalar} the
terms of higher than linear order in $\hbar$ correspond to higher-loop contributions.

Moreover, we argue in appendix \ref{app:Measure} and \ref{app:OneLoop} that the above scalar field consideration can be
extended to the general case of arbitrary fields by taking into account the canonical mass dimensions of all fields
involved.\footnote{Note that raising and lowering indices leads to a change of mass dimension. This affects $\SB^{(2)}$
which must have as many upper indices as lower ones. Therefore, the power of $M$ in \eqref{eq:OneLoopScalar} needed
to make the argument of the logarithm dimensionless depends on both the canonical mass dimension of the fields
and the number of their upper and lower indices.} This amounts to replacing $M^{-2}$ in \eqref{eq:OneLoopScalar} with
$\mathcal{N}^{-1}$, where $\mathcal{N}$ denotes the block diagonal matrix whose dimension equals the number of
different fields and whose diagonal elements are given by the parameter $M$ raised to some power, determined by the
corresponding field type: We know already that the entry of $\mathcal{N}$ in the scalar field sector is given by $M^2$,
while it is, for instance, $M^d$ for gravitons and $M^2$ in the ghost sector.
Using this matrix $\mathcal{N}$ and setting $\hbar = 1$ again yields our final one-loop result,
\begin{equation}[b]
 \GkL = \SB + \frac{1}{2}\,\STr_\UV \ln\left[ \mathcal{N}^{-1}\left(\SB^{(2)}+\Rk\right) \right],
\label{eq:OneLoopFullMain}
\end{equation}
where the supertrace includes a summation over all field types and a minus sign for each Grassmann-valued field.

We emphasize that, due to the occurrence of the free parameter $M$ in eq.\ \eqref{eq:OneLoopFullMain}, bare couplings
will in general depend on $M$. Thus, the bare couplings may be adjusted (to an extent that is yet to be determined) by
tuning $M$. A particularly intriguing implementation of this possibility will be discussed in Sections
\ref{sec:BareMechanism} and \ref{sec:Bare2D} for the Einstein--Hilbert action.

%----------------------------------------------------------------------------------------------------------------------
\subsection{Exactness beyond one-loop in the large cutoff limit?}
\label{sec:BareOneLoopExact}
%----------------------------------------------------------------------------------------------------------------------

In this subsection we investigate the question whether the reconstruction formula
\eqref{eq:OneLoopFullMain}, which is inherently one-loop exact, actually becomes a fully exact relation once the limit
$\UV\rightarrow\infty$ is taken. As shown in Appendix \ref{app:BareOneLoopExact} this is not true in general.
Nevertheless, it turns out that for certain terms to be specified in a moment the relation becomes indeed exact in the
large cutoff limit.

For our argument we assume that any functional can be expanded in terms of linearly independent basis functionals of
theory space. With regard to a given functional equation this means that the equation holds true for each term of the
expansion separately. In this sense, the reconstruction formula can be analyzed term-wise. Then it is perfectly
possible that the one-loop relation is fully exact at large $\UV$ for one class of terms while there are nonvanishing
higher-loop contributions for another class of terms. As the full derivation is rather tedious, we work out the details
in the appendix in Section \ref{app:BareOneLoopExact}. Here we present only the final result including its meaning and
applications.

In the limit $k=\UV\to\infty$ the relation between bare and effective average action is given by
\begin{equation}[b]
 \PrDiv\Big\{\GLL -\SB\Big\} = \PrDiv\Big\{ {\textstyle\frac{\hbar}{2}}\mku
 \STr_\UV \ln\!\Big[\textstyle\frac{1}{\hbar}\,\mathcal{N}^{-1}\big(\SB^{(2)}+\RL\big)\Big]\Big\}.
\label{eq:RecExact}
\end{equation}
This is an \emph{exact identity} rather than a one-loop approximation. In \eqref{eq:RecExact} the projection $\PrDiv$
is to be understood as follows. In the intermediate steps leading to \eqref{eq:RecExact} (see Appendix
\ref{app:BareOneLoopExact}), particular terms are divergent in the limit $\UV\to\infty$ and would require higher-loop
corrections.
These terms must be excluded from our analysis in order to establish exactness of the reconstruction formula. We
achieve this by projecting onto a suitable subspace of theory space, namely the orthogonal complement to all divergent
terms. Specifically, which of the terms have to be ``projected away'' depends on the spacetime dimension:
\begin{itemize}
 \item $\bm{2<d\leq 4}$: In this case the projection operator amounts to $\PrDiv\equiv\PrFull$. Its application
 projects onto the orthogonal complement to all $\sg$- and $\sg\mku R$-terms. This means that all terms of the type
 $\int\!\sg$, $\int\!\sg\,\phi\mku\Box\mku\phi$, $\int\!\sg\,\phi^2$, $\int\!\sg\,\phi^4$, $\int\sg\, R$, $\int\sg\,
 R\mku \phi^2$, $\int\!\sg\,\Box\mku\phi\mku D^\mu\phi\mku D_\mu R$, etc.\ are projected away.
 \item $\bm{d=2}$: The projection is similar to the case $2<d\leq 4$ except that the $\sg\mku R$-terms do not have to
 be projected away this time: $\PrDiv\equiv\text{Pr}_{\perp(\sg)}$. Hence, only such terms that involve no curvature at
 all are affected by $\PrDiv$.
 \item $\bm{d>4}$: The higher the dimension the more terms have to be projected away. For $d>4$ all $\sg\mku R^2$-terms
 and possibly further higher dimensional operators become relevant as well, and we have
 $\PrDiv\equiv\text{Pr}_{\perp(\sg,\sg\mku R,\sg\mku R^2,\dotsc)}$.
\end{itemize}

Finally, let us briefly discuss how eq.\ \eqref{eq:RecExact} can be applied, when it is useful and when it is not.
In the case of scalar fields the additional information contained in \eqref{eq:RecExact} as compared
with \eqref{eq:OneLoopFullMain} is very little: Eq.\ \eqref{eq:RecExact} does not concern any of the terms
$\int\!\sg\,\phi\mku\Box\mku\phi$, $\int\!\sg\,\phi^2$, $\int\!\sg\,\phi^4$, $\int\sg\, R\mku \phi^2$ and so forth,
and thus the corresponding bare action terms cannot be determined on an exact level in this manner.
As these are the main terms a standard effective average action is composed of, identity \eqref{eq:RecExact} seems
inappropriate to find the most relevant part of the bare action. Therefore, we have to resort to the one-loop
approximation \eqref{eq:OneLoopFullMain} in that case. The same conclusion holds for other matter fields.

For pure metric gravity, however, eq.\ \eqref{eq:RecExact} contains a considerable amount of additional information,
at least as far as single-metric truncations are concerned. In this case, for $2<d\leq 4$ the projection $\PrDiv$
excludes only two terms from the equation: the cosmological constant term, $\int\!\sg$, and the first curvature term,
$\int\!\sg\,R$. Moreover, for $d=2$ the equation even holds true for all terms but the cosmological constant term. To
sum up, in the limit $\UV\to\infty$ we find that the identity $\GLL -\SB = \frac{\hbar}{2}\mku
 \STr_\UV \ln\!\big[\textstyle\frac{1}{\hbar}\,\mathcal{N}^{-1}\big(\SB^{(2)}+\RL\big)\big]$ is \emph{fully exact
except for the cosmological constant term} in $d=2$ (\emph{except for} $\int\!\sg$ \emph{and} $\int\!\sg\,R$ in
$2<d\leq 4$).

If we want to determine how the excluded terms enter the bare action, we can make use of the
one-loop approximation \eqref{eq:OneLoopFullMain} again which is valid for all terms.

As a last point we would like to mention a recently found simplification emerging for scalar fields in flat space
\cite{MS15}. It is based upon a different regularization scheme: Only the massless kinetic parts of the underlying
actions are regularized (leaving their interaction parts unmodified), and the various cutoffs involved have to satisfy
a certain sum rule as well as a compatibility condition. In this special case the trace in eq.\
\eqref{eq:OneLoopFullMain} amounts to a (divergent but irrelevant) field independent constant, and so do all
higher-loop terms. Thus, provided that the regulators satisfy all constraints, the reconstruction formula
\eqref{eq:OneLoopFullMain} at $k=\UV$ reduces to \cite{MS15} (cf.\ also \cite{VZ11})
\begin{equation}
 \GLL[\phi] = \SB[\phi]\qquad\text{for scalar fields.}
\label{eq:SEqualsGamma}
\end{equation}
It should be borne in mind, though, that the modified regulators imply a modification of the functional measure
as compared with our definition in Appendix \ref{app:Measure}. The authors of Ref.\ \cite{MS15} argue that their
discussion can be generalized to the case of other, for instance fermionic, matter fields.
Moreover, it can be verified that the results hold true in curved spacetime, too. In (the QFT approach to)
quantum gravity, however, where the integration variable of the functional integral is given by the dynamical metric,
the simple relation \eqref{eq:SEqualsGamma} is spoiled by additional correction terms. These further contributions
originate from Gaussian integrals one encounters in the proof of \eqref{eq:SEqualsGamma}. They can be treated as
irrelevant constants in the case of scalar fields \cite{MS15}, while they give rise to crucial field dependent terms
in gravity.\footnote{More precisely, in the background field approach these additional terms depend on the background
metric. This becomes particularly problematic for single-metric truncations.} Similar obstacles can occur in other
gauge theories as well.

In conclusion, the bare action may be determined by eq.\ \eqref{eq:SEqualsGamma} in the matter field
sector, and by eq.\ \eqref{eq:OneLoopFullMain} for gauge theories, in particular for gravity.

%----------------------------------------------------------------------------------------------------------------------
\section{Bare action for the Einstein--Hilbert truncation}
\label{sec:BareEH}
%----------------------------------------------------------------------------------------------------------------------

In this section we aim at applying the reconstruction formula discussed in the previous sections to metric gravity.
Our analysis will extend the results of Ref.\ \cite{MR09} where a map between bare and effective couplings was
considered for a twofold Einstein--Hilbert (EH) truncation. Using the same setting, we will prove the existence of a
fixed point in the bare sector for any choice of the measure parameter $M$ and any dimension $d$, we will investigate
the flow of the bare couplings in more detail, in particular near $2$ dimensions, and we try to simplify the map by
choosing a suitable value of $M$. This way we will demonstrate that $M$ can always be fixed such that the bare
cosmological constant vanishes. As we will show, this implies in $d=2+\ve$ that at first order the bare Newton
constant equals the effective one.

%----------------------------------------------------------------------------------------------------------------------
\subsection{Mapping between bare and effective couplings}
\label{sec:BareMap}
%----------------------------------------------------------------------------------------------------------------------

For pure (metric) gravity, both the EAA and the total bare action depend on four arguments in general, $\GkL\equiv
\GkL[g,\bg,\xi,\bx\mku]$ and $\SB\equiv\SB[g,\bg,\xi,\bx\mku]$, respectively, with the dynamical metric $g_\mn$, the
background metric $\bg_\mn$ and the ghost fields $\xi^\mu$, $\bx_\mu$. We employ optimized regulators $\Rk$ and set
$k=\UV$, implying the relation $\Gamma_{\UV,\UV} = \Gamma_{k=\UV}$, i.e.\ $\Gamma_{\UV,\UV}$ equals the UV cutoff-free
EAA \cite{MR09}. Our ansatz for $\Gamma_\UV$ reads
\begin{equation}
\begin{split}
\Gamma_\UV[g,\bg,\xi,\bx\mku]  = &-(16\pi G_\UV)^{-1}\int\dd x\sqrt{g}\, \big(R-2\Ldim_\UV \big) 
+S_{\text{gh}}[g,\bg,\xi,\bx\mku]
\\
& {} + (32\pi G_\UV)^{-1}\int\dd x \sqrt{\bg}\,\bg^{\mu\nu}(\mathcal{F}_{\mu}^{\alpha\beta}g_{\alpha\beta})
(\mathcal{F}_{\nu}^{\rho\sigma}g_{\rho\sigma}) ,
\end{split}
\label{eq:EHTrunc}
\end{equation}
where the last term on the RHS is the gauge fixing action corresponding to the harmonic coordinate condition with
$\mathcal{F}_{\mu}^{\alpha\beta}\equiv \delta^{\beta}_{\mu}\mku\bar{g}^{\alpha\gamma}\bD_{\gamma}
-\tfrac{1}{2}\mku\bar{g}^{\alpha\beta}\bar{D}_{\mu}$, and the second term is the associated ghost action.
Equation \eqref{eq:EHTrunc} involves the dimensionful running parameters $G_\UV$ and $\mathlarger{\Ldim}_\UV$, where
the symbol $\mathlarger{\Ldim}$ is used for the cosmological constant here in to order to avoid confusion with the
scale $\UV$.

We make an ansatz analogous to \eqref{eq:EHTrunc} also for the bare action:
\begin{equation}
\begin{split}
\SB[g,\bg,\xi,\bx\mku] = &-(16\pi \check{G}_\UV)^{-1}\int\dd x \sg\,
\big(R-2\check{\Ldim}_\UV\big) +S_{\textrm{gh}}[g,\bg,\xi,\bx\mku]
\\
& {} + (32\pi \check{G}_\UV)^{-1}\int\dd x \sbg\,\bg^{\mu\nu}
(\mathcal{F}_{\mu}^{\alpha\beta}g_{\alpha\beta})(\mathcal{F}_{\nu}^{\rho\sigma}g_{\rho\sigma}),
\end{split}
\label{eq:EHTruncBare}
\end{equation}
with the corresponding bare Newton and bare cosmological constant, $\check{G}_\UV$ and
$\check{\mathlarger{\Ldim}}_\UV$, respectively. Note that by virtue of the reconstruction formula the bare couplings
will exhibit a $\UV$-dependence, too.

In order to find the map relating bare to effective couplings, it is
sufficient to set $g_\mn=\bg_\mn$ and $\xi^\mu=0=\bx_\mu$ in \eqref{eq:OneLoopFullMain} after having computed the
second functional derivatives w.r.t.\ $g_\mn$, $\xi^\mu$ and $\bx_\mu$. Since there is only one metric left then, we
can omit the ``bar'' over background quantities for reasons of clarity from now on. Following Ref.\ \cite{MR09}, we
decompose the metric fluctuations into a traceless and a trace part, and without loss of generality we assume a
maximally symmetric background. Then \eqref{eq:OneLoopFullMain} leads to
\begin{equation}
\begin{split}
 \quad\,&\!\!\!\!\!\!\!\!\!\!\!\Gamma_{\UV}[g,g,0,0]-\SB[g,g,0,0] \\
 {}= &+\frac{1}{2}\,\Tr_\UV^\text{T}\ln \bigg\{ \frac{M^{-d}}{32\pi\mku\check{G}_{\UV}}
\Big[-\Box+\UV^2 \;R^{(0)}(-\Box/\UV^2)-2\check{\Ldim}_\UV+C_\text{T}R \Big]\bigg\}
\\
{} &+ \frac{1}{2}\,\Tr_\UV^\text{S}\ln \bigg\{\frac{M^{-d}}{32\pi\mku\check{G}_{\UV}} \Big(\frac{d-2}{2d}\Big)
\Big[-\Box+\UV^2\;R^{(0)}(-\Box/\UV^2)-2\check{\Ldim}_\UV+C_\text{S}R \Big]\bigg\}
\\
{} &- \Tr_\UV^\text{V}\ln \bigg\{ M^{-2}
\Big[-\Box+\UV^2\;R^{(0)}(-\Box/\UV^2)+C_\text{V}R \Big]\bigg\},
\end{split}
\label{eq:OneLoopEH}
\end{equation}
where the sub- and superscripts T, S and V refer to symmetric traceless tensors, scalars and vectors,
respectively. The constants in \eqref{eq:OneLoopEH} are defined by
\begin{equation}
C_\text{T}\equiv \frac{d(d-3)+4}{d(d-1)},\quad C_\text{S}\equiv\frac{d-4}{d},
\quad C_\text{V}\equiv-\frac{1}{d} \; ,
\end{equation}
like in Ref.\ \cite{Reuter1998}.
Using the heat kernel techniques introduced in appendix \ref{app:Heat}, we can expand the traces in terms of the
curvature $R$, collect all terms proportional to $\int\!\dd x\sg$ and $\int\!\dd x\sg\,R$, and compare the
corresponding coefficients. This yields the following map between bare and effective couplings, which was first
obtained in \cite{MR09}:
\begin{boxalign}
 \frac{1}{\cg_\UV}\left(\frac{6}{d}+\cl_\UV\right) - \frac{1}{g_\UV}\left(\frac{6}{d}+\lambda_\UV\right)
 &= 12\mku C_d\,\frac{d(d-1)+4(1-2\cl_\UV)}{d^2(1-2\cl_\UV)}\; ,
\label{eq:MapI}\\
 \frac{\cl_\UV}{\cg_\UV}-\frac{\lambda_\UV}{g_\UV} &= C_d \left[(d+1)\ln\left(\frac{\cg_\UV}{1-2\cl_\UV}\right)
 -Q_\UV\right].
\label{eq:MapII}
\end{boxalign}
Here, $\cg_\UV$ and $\cl_\UV$ ($g_\UV$ and $\lambda_\UV$) are the dimensionless bare (effective) Newton constant and
cosmological constant, respectively, and we have introduced the constant
\begin{equation}
 C_d \equiv \frac{1}{(4\pi)^{d/2-1}\mku\Gamma(d/2)} \;.
\end{equation}
The system $\{$\eqref{eq:MapI},\eqref{eq:MapII}$\}$ depends on a parameter $Q_\UV$ which is defined by
\begin{equation}
 Q_\UV \equiv \big[d(d+1)-8\big]\ln(\UV/M)-(d+1)\ln(32\pi)+{\textstyle\frac{2}{d}\ln\left(\frac{d-2}{2d}\right)}.
\label{eq:DefQ}
\end{equation}
As a consequence, the bare couplings are not completely determined in terms of the effective ones but rather depend on
this parameter. We observe that $Q_\UV$ --- besides its $\UV$-dependence --- depends on the measure parameter $M$.
Therefore, choosing different values of $M$ amounts to modifying $\cg_\UV$ and $\cl_\UV$, even if $\UV$, $g_\UV$ and
$\lambda_\UV$ are fixed. This confirms our general argument concerning the nonuniqueness of bare couplings. Unlike in
Ref.\ \cite{MR09} we will not confine ourselves to the case $M\propto\UV$ in the following but discuss arbitrary
choices as well.

Apart from the special dimension $d\approx 2.3723$ where the prefactor $[d(d+1)-8]$ of $\ln(\UV/M)$ in \eqref{eq:DefQ}
vanishes so that the $M$-dependence disappears, there is a one-to-one correspondence between $Q_\UV$ and $M$. Thus,
we may consider $Q_\UV$ a free parameter as well.

From a conceptual point of view, eqs.\ \eqref{eq:MapI} and \eqref{eq:MapII} continuously map any RG trajectory of the
effective side to an RG trajectory of the bare side, where the latter depends on the parameter $Q_\UV$. This way we
can obtain a $Q_\UV$-dependent family of flow diagrams for the bare couplings. The construction of each ``bare
trajectory'' involves five steps:
\textbf{(i)} We choose and fix some $Q_\UV$-value.
\textbf{(ii)} Then we pick an arbitrary point of the $(\cl,\cg)$-plane which serves as an initial condition for the
sought-after trajectory.
\textbf{(iii)} After inserting this point into eqs.\ \eqref{eq:MapI} and \eqref{eq:MapII}, the system is solved for the
effective couplings.
\textbf{(iv)} The resulting effective couplings serve, in turn, as an initial condition for the FRGE \eqref{eq:FRGE},
giving rise to an RG trajectory on the EAA side, $\UV\to(\lambda_\UV,g_\UV)$, where we employ the optimized cutoff
here.
\textbf{(v)} Using eqs.\ \eqref{eq:MapI} and \eqref{eq:MapII} again, each point of the effective trajectory is mapped
to a point in the bare sector, which finally leads to a trajectory $\UV\to(\cl_\UV,\cg_\UV)$.

By means of this construction we obtain a characteristic flow diagram corresponding to the chosen $Q_\UV$-value.

In Figure \ref{fig:BareFlow} we demonstrate to what extent the flow diagrams of the bare couplings in $d=4$ dimensions
depend on $Q_\UV$.
\begin{figure}[tp]
\small
 \begin{minipage}{0.47\columnwidth}
  \centering
  $Q_\UV=20$\\[0.4em]
  \includegraphics[width=\columnwidth]{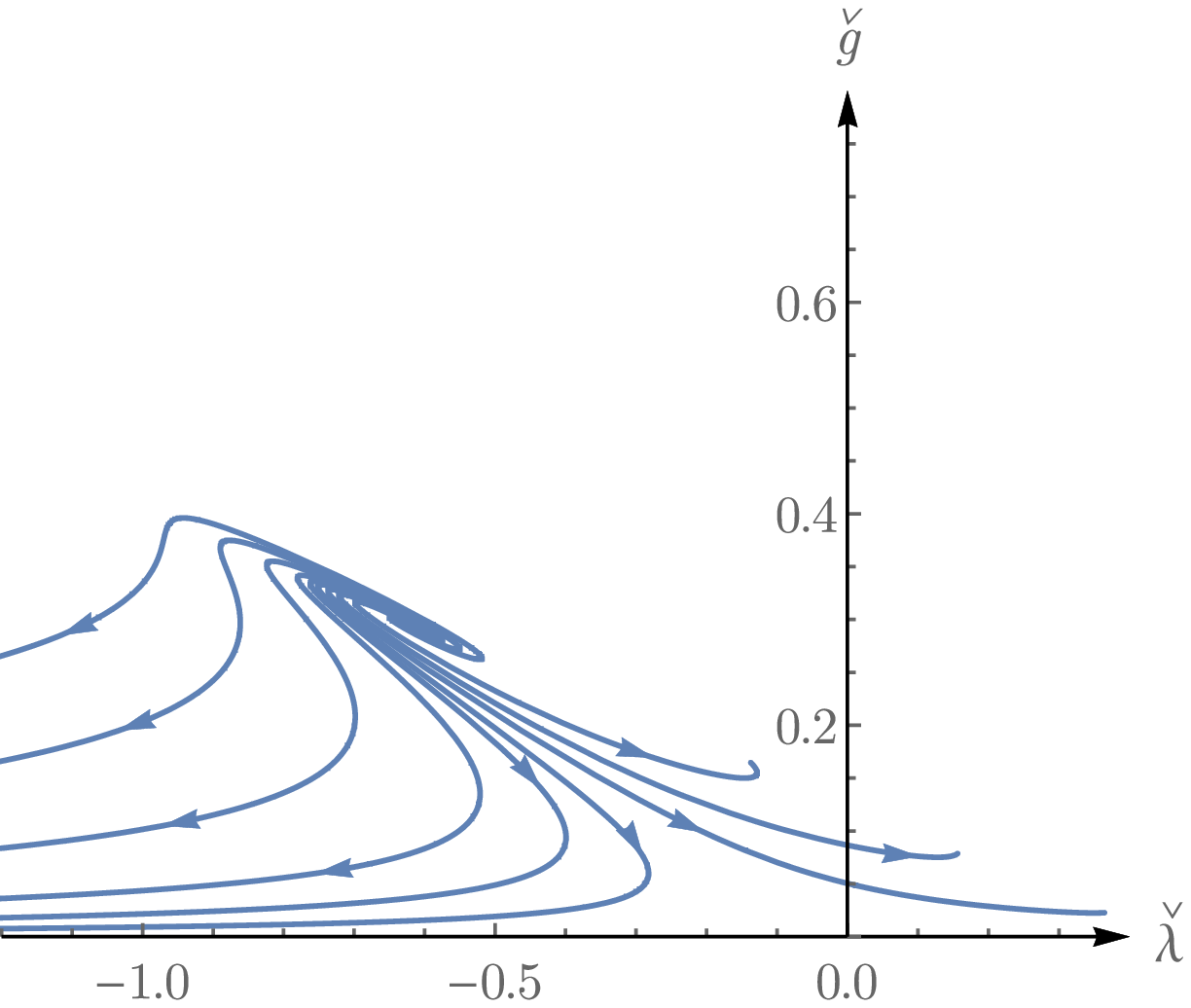}
 \end{minipage}
 \hfill
 \begin{minipage}{0.47\columnwidth}
  \centering
  $Q_\UV=10$\\[0.4em]
  \includegraphics[width=\columnwidth]{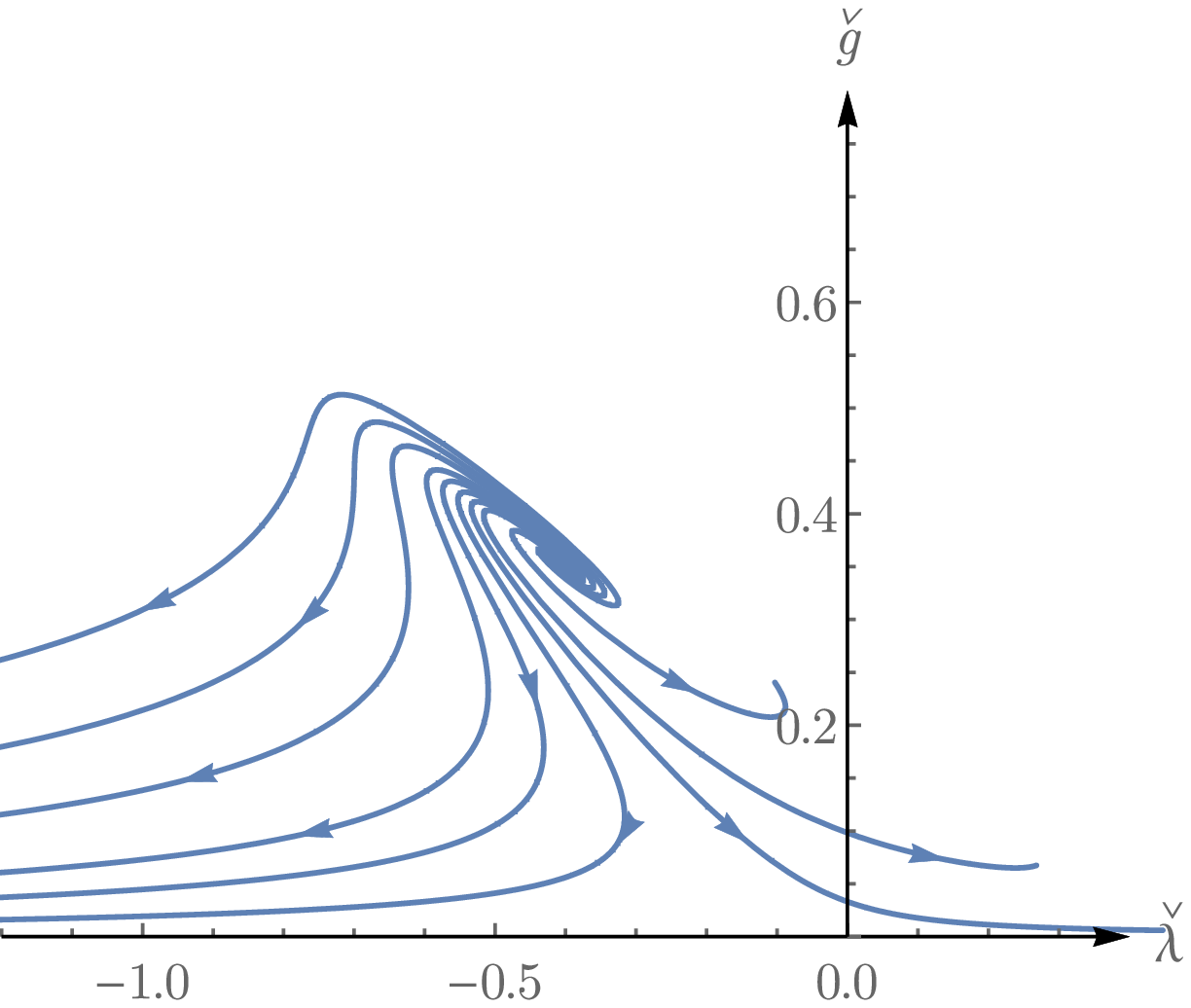}
 \end{minipage}\\[2.6em]
 \begin{minipage}{0.47\columnwidth}
  \centering
  $Q_\UV=2$\\[0.4em]
  \includegraphics[width=\columnwidth]{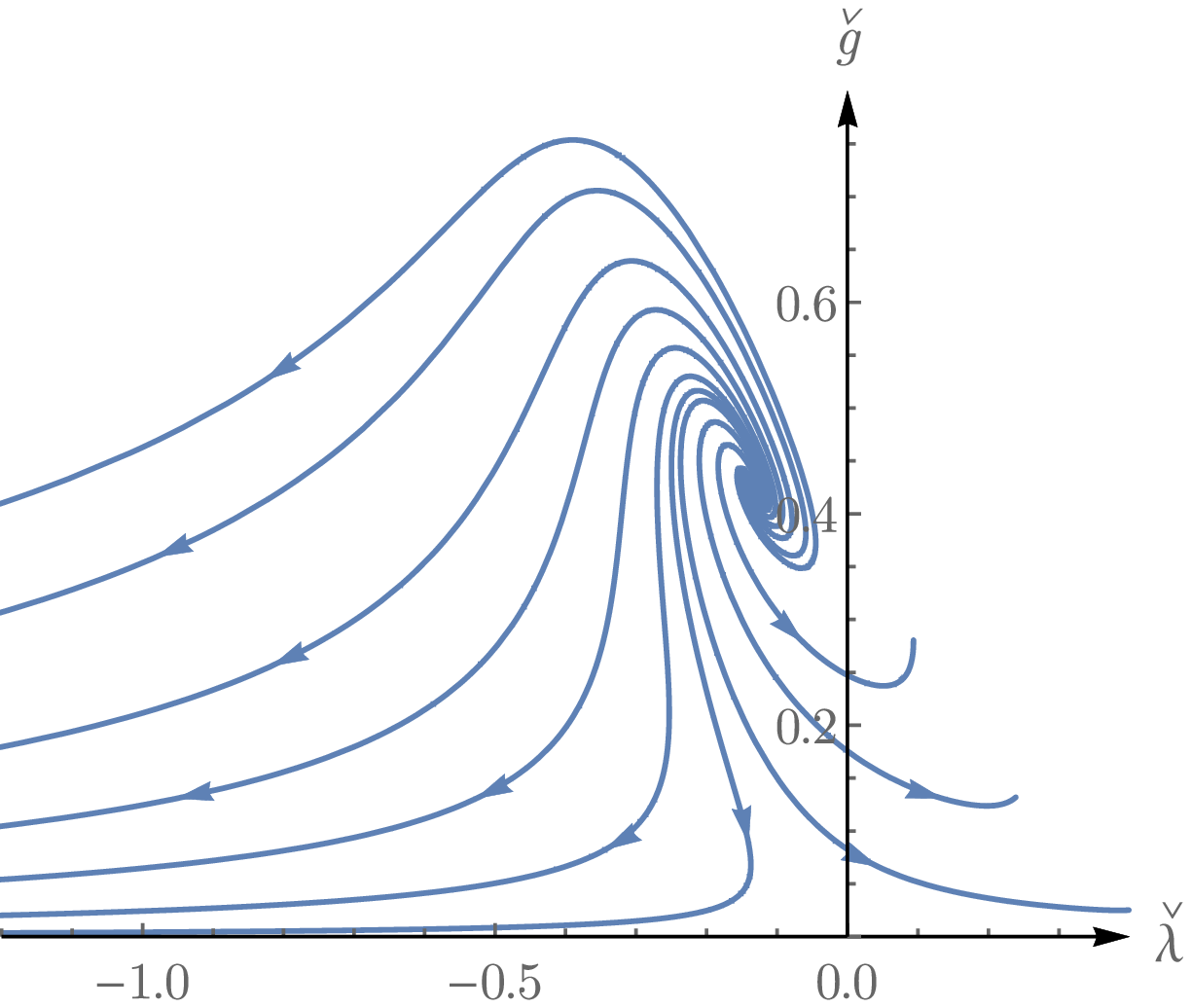}
 \end{minipage}
 \hfill
 \begin{minipage}{0.47\columnwidth}
  \centering
  $Q_\UV=-0.583183$\\[0.4em]
  \includegraphics[width=\columnwidth]{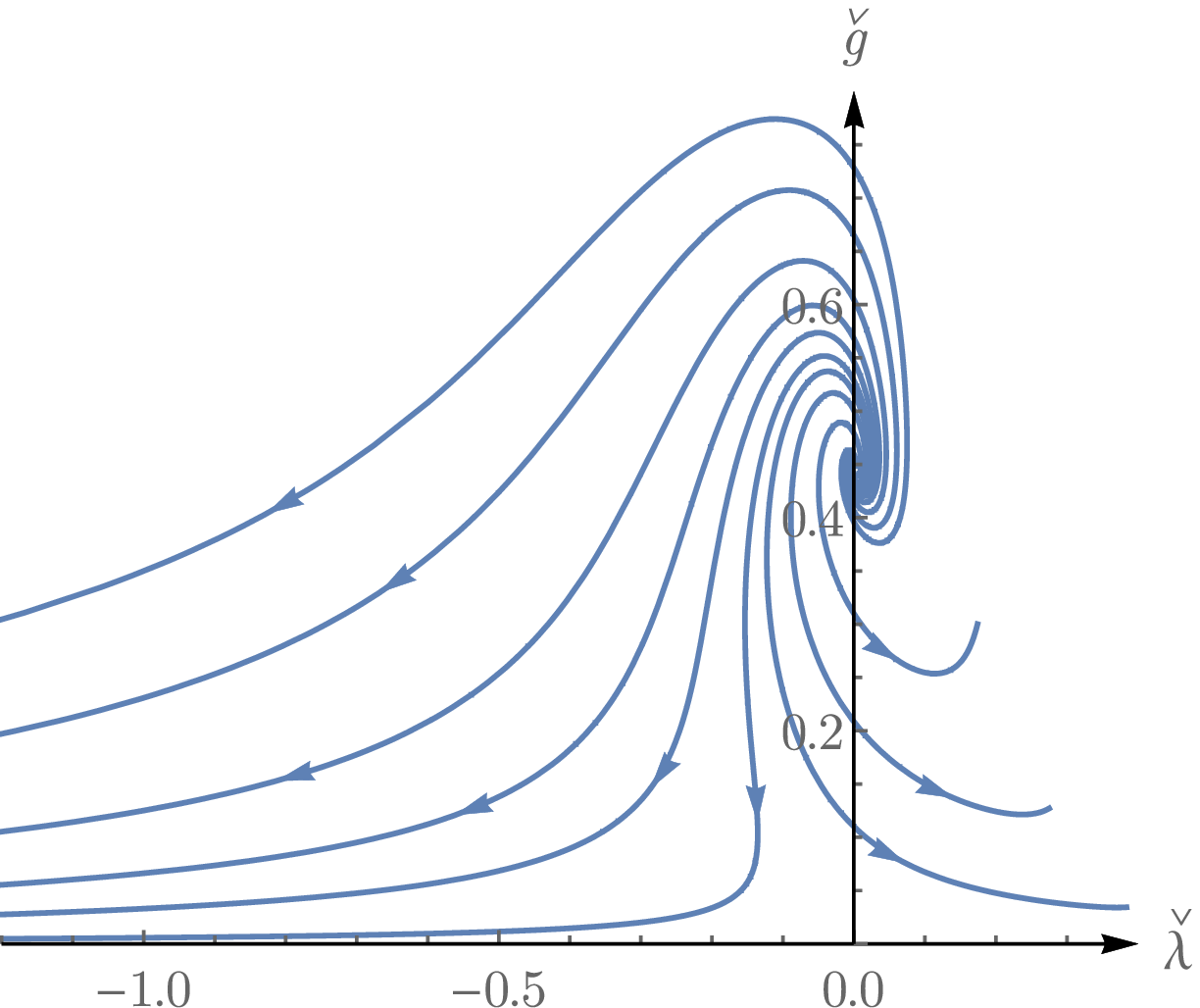}
 \end{minipage}\\[2.6em]
 \begin{minipage}{0.47\columnwidth}
  \centering
  $Q_\UV=-3$\\[0.4em]
  \includegraphics[width=\columnwidth]{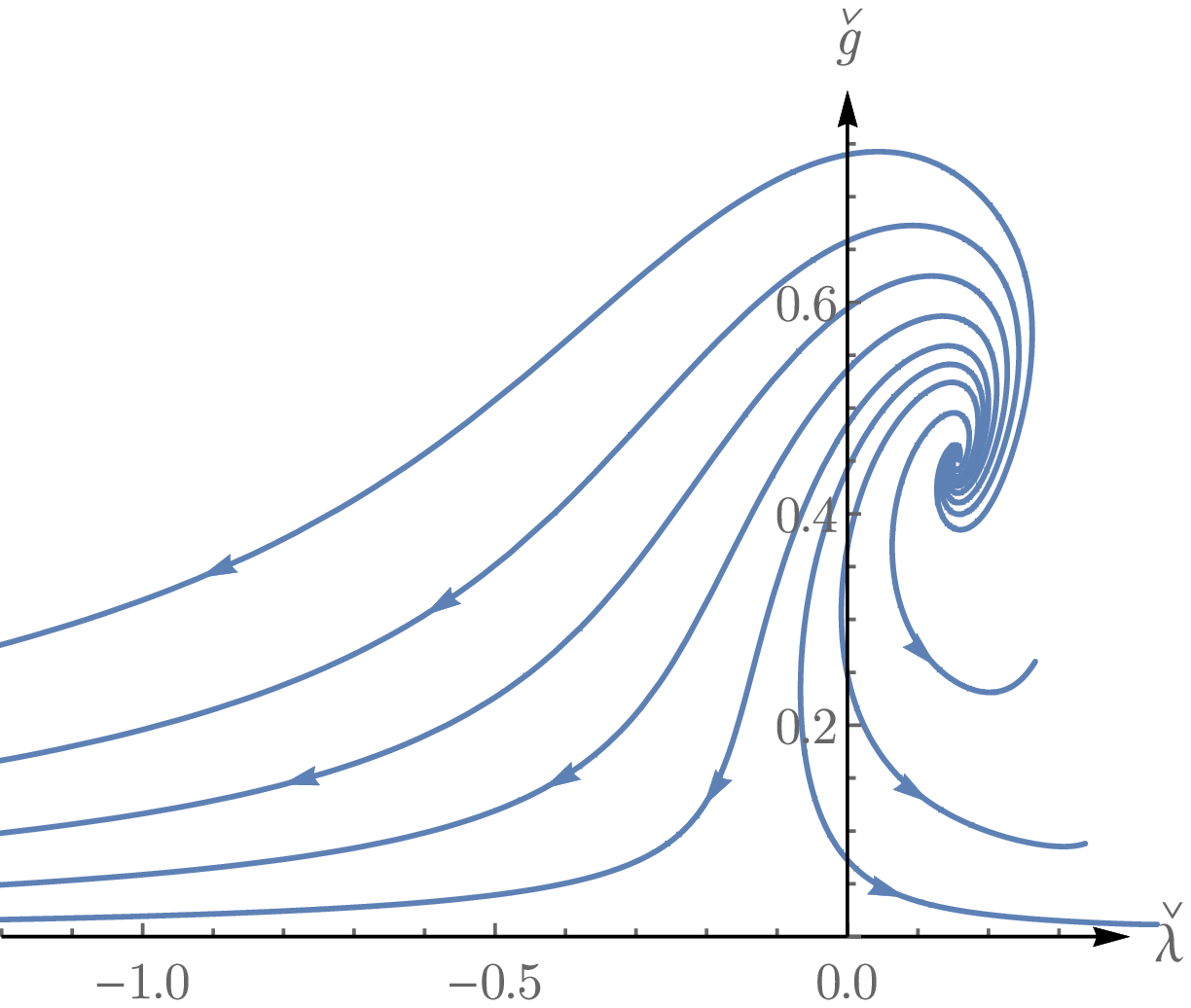}
 \end{minipage}
 \hfill
 \begin{minipage}{0.47\columnwidth}
  \centering
  $Q_\UV=-8$\\[0.4em]
  \includegraphics[width=\columnwidth]{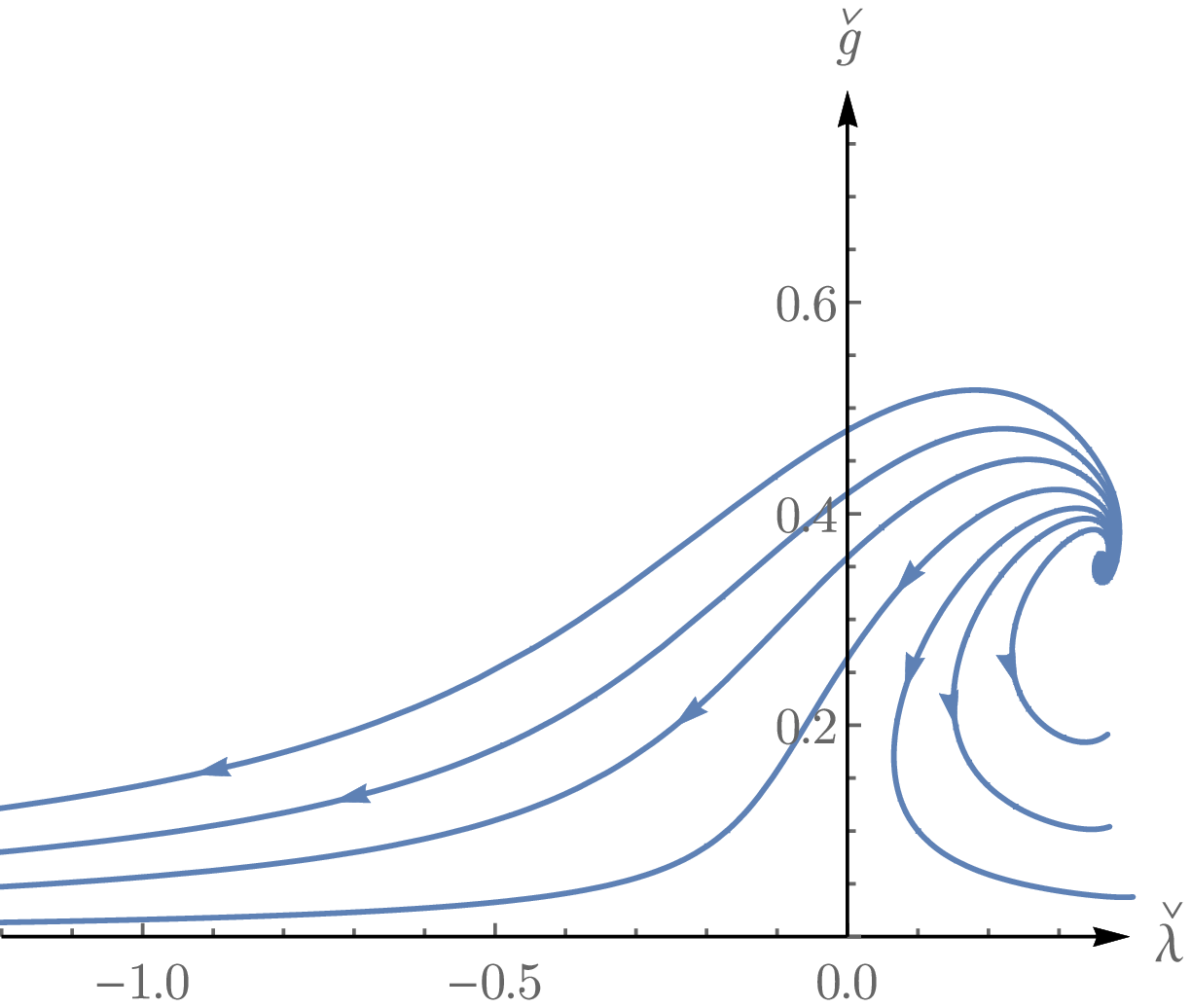}
 \end{minipage}
\vspace{2em}
\caption{Flow diagrams in the space of the \emph{bare} couplings $\cl$ and $\cg$ for several constant values of $Q_\UV$
 in $d=4$ dimensions.}
\label{fig:BareFlow}
\end{figure}
It seems that quantitative features like the position of the ``bare NGFP'' and the shape of the streamlines are
modified when $Q_\UV$ changes, while qualitative features like the mere existence of the fixed point and its critical
exponents are independent of $Q_\UV$. Whether this is indeed true, will be discussed in the next subsections, where
we investigate the existence of the NGFP for any choice of $Q_\UV$ and for all dimensions $d>2$. In particular,
the analysis will include the cases $Q_\UV\to\infty$ and $Q_\UV\to -\infty$.

%----------------------------------------------------------------------------------------------------------------------
\subsection{Existence of the bare NGFP}
\label{sec:BareExNGFP}
%----------------------------------------------------------------------------------------------------------------------

We restrict ourselves to the case $d>2$ as the EH action gives rise to a topological invariant in strictly  $d=2$
dimensions. From the RG studies of the EH truncation we know that the $\beta$-functions of $\lambda$ and $g$ possess a
nontrivial fixed point for any $d>2$ (see Ref.\ \cite{Nink2011} for instance). The corresponding coordinates
$\lambda_*$ and $g_*$ are to be inserted into the fixed point version of eqs.\ \eqref{eq:MapI} and \eqref{eq:MapII}.
The question about the existence of a fixed point for the bare couplings then boils down to the question if the system
can be solved for $\cl_*$ and $\cg_*$. Whether or not the answer depends on the underlying $Q_\UV$-value will be
investigated in this subsection.
% For the sake of brevity we use the shortcuts $\lambda\equiv\lambda_*$, $g\equiv g_*$, $\cl\equiv\cl_*$ and
% $\cg\equiv \cg_*$ in what follows.

Being the most natural assumption for the bare Newton constant we start with the relation $\cg_*>0$.\footnote{Only for
$\cg_*>0$ the kinetic term of the (traceless part of the) metric fluctuations in the bare action has the correct sign.
Furthermore, $\cg_*>0$ is in accordance with $g_*>0$, which is a necessary condition for the fixed point value of the
effective Newton constant since otherwise there would not exist any RG trajectory connecting the NGFP to the classical
regime.} In that case the logarithm in eq.\ \eqref{eq:MapII} requires that $1-2\cl_*>0$ for any finite $Q_\UV$. This
can be used in eq.\ \eqref{eq:MapI} in turn:
\begin{equation}
 \underbrace{\frac{1}{\cg_*}}_{{}>0}\left(\frac{6}{d}+\cl_*\right) =
 \underbrace{\vphantom{\frac{d(d-1)+4(1-2\cl_*)}{1-2\cl_*}}\frac{12\, C_d}{d^2}}_{{}>0}\,
 \underbrace{\frac{d(d-1)+4(1-2\cl_*)}{1-2\cl_*}}_{{}>0} + \frac{1}{g_*}\left(\frac{6}{d}+\lambda_*\right).
\label{eq:Estimate}
\end{equation}
For $2<d\lesssim 2.56$ the effective cosmological constant becomes negative at the fixed point \cite{CPR09,Nink2011},
but its absolute value remains sufficiently small such that $\frac{1}{g_*}\left(\frac{6}{d}+\lambda_*\right)>0$.
Clearly, this latter relation holds true also for larger dimensions where $\lambda_*>0$. Therefore, we can conclude
that the RHS of \eqref{eq:Estimate} is positive for all $d>2$, which implies on the LHS that $6/d+\cl_*>0$. To sum up,
we have found that the fixed point values of the bare couplings, if any, are confined to the
restricted domain
\begin{equation}[b]
  -\frac{6}{d}<\cl_*<\frac{1}{2} \quad\text{and}\quad \cg_*>0\,, \qquad\text{for $Q_\UV$ finite.}
\end{equation}
Moreover, from eq.\ \eqref{eq:Estimate}, i.e.\ from $\frac{1}{\cg_*}\left(\frac{6}{d}+\cl_*\right)=\text{finite}>0$,
follows that $\cg_*$ is finite as well. Thus, $\cg_*$ is bounded from above, too.

Whether the bare fixed point exits in fact can be clarified by reducing the system
$\{$\eqref{eq:MapI},\eqref{eq:MapII}$\}$ to a single equation. For that purpose we solve \eqref{eq:MapI} for $\cg_\UV$,
insert the result into \eqref{eq:MapII} and replace $g_\UV$ and $\lambda_\UV$ by their fixed point values. Then the
system boils down to the equation
\begin{equation}
 f(\cl_*)=0\,,
\label{eq:fZero}
\end{equation}
where the function $f(\cl)$ is given by
\begin{equation}
\begin{split}
 f(\cl) \equiv C_d\, Q_\UV -\frac{\lambda_*}{g_*} + \frac{\cl}{6/d+\cl}
 \left[ 12\mku C_d\,\frac{d(d-1)+4(1-2\cl)}{d^2(1-2\cl)} +\frac{1}{g_*}\left(\frac{6}{d}+\lambda_*\right)\right]\\
 {} + C_d\, (d+1) \ln\left\{  \frac{1-2\cl}{6/d+\cl} \left[ 12\mku C_d\,\frac{d(d-1)+4(1-2\cl)}{d^2(1-2\cl)}
 +\frac{1}{g_*}\left(\frac{6}{d}+\lambda_*\right)\right] \right\},
\end{split}
\label{eq:Definitionf}
\end{equation}
so it depends parametrically on $Q_\UV$. The existence of a bare NGFP is equivalent to the \emph{existence of a zero
of} $f$, and by eq.\ \eqref{eq:fZero} the zero is located at the yet unknown fixed point value $\cl_*$. Remarkably
enough, for the proof of existence we can proceed analytically by means of the following simple argument.

Let us first consider the case where $Q_\UV$ remains finite.
Recalling that $-6/d<\cl_*<1/2$, it turns out useful to study the asymptotic behavior of $f$ for $\cl\searrow -6/d$
and for $\cl\nearrow 1/2$. Both the third term in the definition \eqref{eq:Definitionf} of $f$,
$\frac{\cl}{6/d+\cl}\big[\cdots\big]$, and the logarithm are divergent in these limits. Since linear terms always
predominate over logarithmic ones when being divergent, it is the term $\frac{\cl}{6/d+\cl}\big[\cdots\big]$
that decides on the asymptotic running in either limit. The square bracket is always positive, while its prefactor
$\frac{\cl}{6/d+\cl}$ is negative for $\cl\searrow -6/d$. Taking all contributions together we find
\begin{equation}
 \lim_{\cl\searrow -6/d}\, f(\cl) = -\infty\,.
\end{equation}
On the other hand, $\frac{\cl}{6/d+\cl}$ is positive and remains finite for $\cl\nearrow 1/2$, while the square
bracket tends to infinity. This leads to
\begin{equation}
 \lim_{\cl\nearrow 1/2}\, f(\cl) = +\infty\,.
\end{equation}
Therefore, the function $f$ must change its sign between $-6/d$ and $1/2$. Furthermore, it is smooth in its domain of
definition. In conclusion, $f$ must have a zero. This proves the existence of a bare fixed point for any $d>2$ at any
finite $Q_\UV$.

Although the exact position of this zero of $f$ changes when $Q_\UV$ is varied, its mere existence is independent of
$Q_\UV$. Figure \ref{fig:fOfLambda} illustrates the situation. It shows the graph of $f$ in four dimensions for the
exemplary choice $Q_\UV=20$. By the definition of $f$, given in eq.\ \eqref{eq:Definitionf}, increasing $Q_\UV$ means
shifting the entire graph upwards, which, in turn, moves the zero $\cl_*$ towards the left boundary at $\cl=-6/d$.
Similarly, decreasing $Q_\UV$ amounts to shifting $\cl_*$ towards the right boundary at $\cl=1/2$. This suggests the
two relations $\lim_{Q_\UV \to\infty} \cl_* = -6/d$ and $\lim_{Q_\UV \to-\infty} \cl_* = 1/2$, which we would like to
prove now.
\begin{figure}[tp]%\begin{wrapfigure}{R}{0.58\textwidth}
\centering
%\normalcaption
%\captionwidth{0.53\textwidth}
\includegraphics[width=0.6\textwidth]{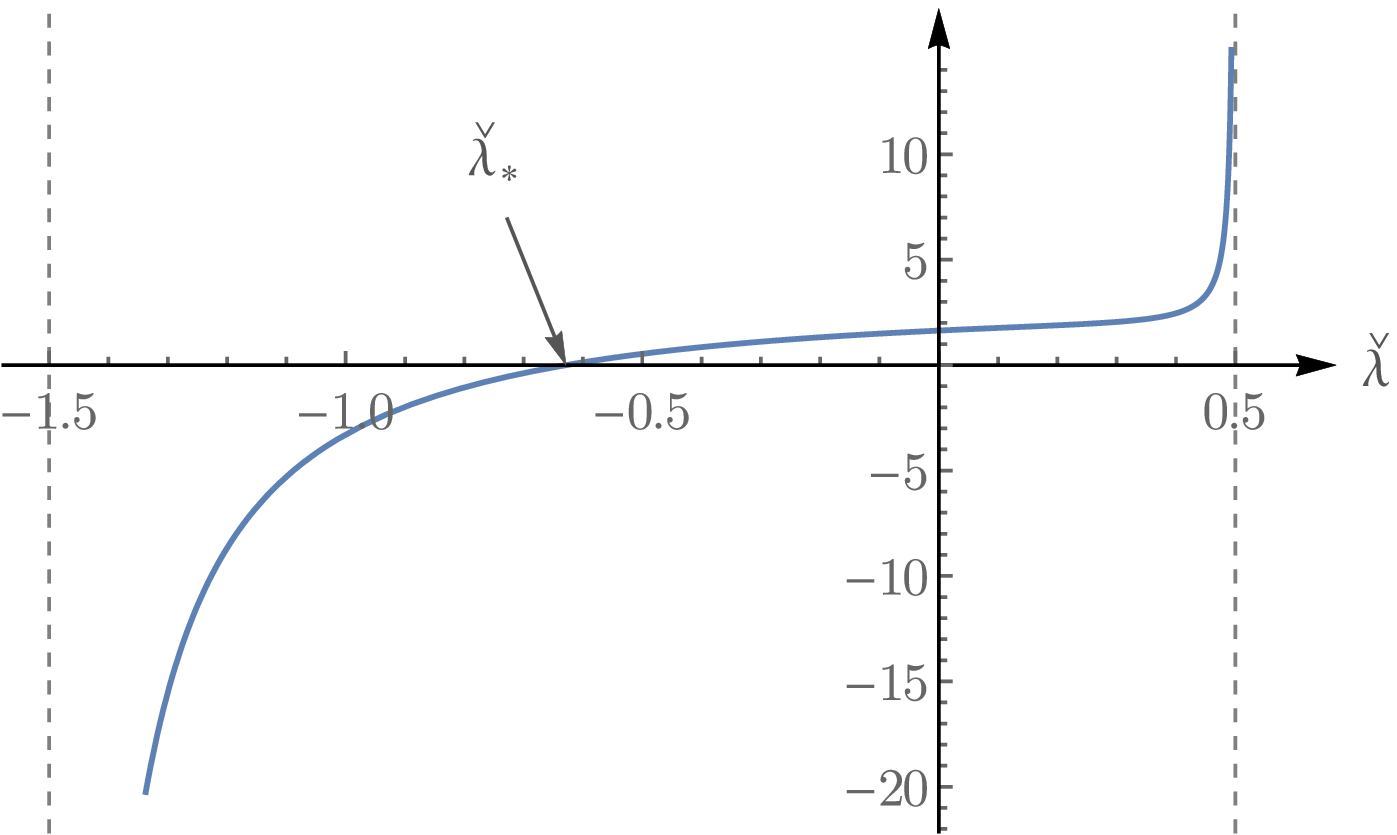}%\includegraphics[width=0.52\textwidth]{Graphics/fOfLambda}
\caption{The function $f(\cl)$ in $d=4$ dimensions for $Q_\UV=20$, having a zero at $\cl=\cl_*$.}
\label{fig:fOfLambda}
\end{figure}%\end{wrapfigure}

We begin with the limit $Q_\UV \to\infty$. By a careful analysis of eqs.\ \eqref{eq:MapI} and \eqref{eq:MapII} in this
limit we find that the bare fixed point couplings can be determined consistently only if $\cg_*\searrow 0$ and
$\cl_*\to\text{finite}<0$. Then we can deduce the precise $Q_\UV$-dependence of $\cg_*$ and $\cl_*$ as follows.

At leading order, the divergent behavior of $Q_\UV$ on the RHS of \eqref{eq:MapII} is compensated solely by the first
term on the LHS due to its denominator $\propto\cg_*$. Hence, we obtain
\begin{equation}
 \cg_* = -\frac{\cl_*}{C_d\, Q_\UV} + \mO\big(Q_\UV^{-2}\big)\,.
\label{eq:baregLargeQ}
\end{equation}
Inserting this into \eqref{eq:MapI} yields
\begin{equation}
 \cl_* = -\frac{6}{d} - \left[12\mku C_d\,\frac{d(d-1)+4(1-2\cl_*)}{d^2(1-2\cl_*)} +
 \frac{1}{g_*}\left(\frac{6}{d}+\lambda_*\right)\right]\frac{\cl_*}{C_d\, Q_\UV} + \mO\big(Q_\UV^{-2}\big)\,.
\label{eq:barelambdaLargeQ}
\end{equation}
At first order in $1/Q_\UV$, we have $\cl_*=-6/d$. This can be inserted back into the RHS of eq.\
\eqref{eq:barelambdaLargeQ} in order to determine the subleading order, and into \eqref{eq:baregLargeQ}. In this way,
we arrive at
\begin{equation}
\begin{split}
 \cg_* &= \frac{6/d}{C_d\, Q_\UV} + \mO\big(Q_\UV^{-2}\big)\,,\\
 \cl_* &= -\frac{6}{d} + \left[12\mku C_d\,\frac{d(d-1)+4+48/d}{d(d+12)} +
 \frac{6/d+\lambda_*}{g_*}\right]\frac{6/d}{C_d\, Q_\UV} + \mO\big(Q_\UV^{-2}\big)\,,
\end{split}
\end{equation}
in the limit $Q_\UV\to\infty$.

The limit $Q_\UV\to -\infty$ can be analyzed in a very similar way. Requiring that the divergent behavior of $Q_\UV$ be
compensated by $\cl_*$ and $\cg_*$ in order to satisfy eqs.\ \eqref{eq:MapI} and \eqref{eq:MapII} consistently we find
\begin{equation}
\begin{split}
 \cg_* &= \frac{1}{2\mku C_d\mku(-Q_\UV)} + \mO\big(Q_\UV^{-2}\big)\,,\\
 \cl_* &= \frac{1}{2} - \frac{6(d-1)}{d+12}\,\frac{1}{(-Q_\UV)} + \mO\big(Q_\UV^{-2}\big)\,,
\end{split}
\end{equation}
in the limit $Q_\UV\to -\infty$.

The preceding considerations prove our conjecture concerning the bare NGFP for divergent $Q_\UV$ which we have read
off from the graph of $f$ and which we can summarize as follows:
\begin{boxalign}
 &\cg_* \searrow 0\,, \qquad \cl_* \searrow -6/d\,,&&\text{for }Q_\UV\to\infty,
\label{eq:LimitQInfty}\\
 &\cg_* \searrow 0\,, \qquad \cl_* \nearrow 1/2\,,&&\text{for }Q_\UV\to -\infty.
\end{boxalign}

In order to illustrate how the position of the bare NGFP depends on $Q_\UV$ we can solve the system
$\{$\eqref{eq:MapI},\eqref{eq:MapII}$\}$ numerically for $\cl_*$ and $\cg_*$ at some $Q_\UV$ and repeat the procedure
for different $Q_\UV$'s. Then the result can be plotted as a parametric curve $\gamma:\mku Q_\UV\mapsto
\big(\cl_*(Q_\UV),\cg_*(Q_\UV)\big)$. The shape of such a curve as well as its endpoints depend on the spacetime
dimension. Figure \ref{fig:BareNGFPd4} depicts the situation in $d=4$. The curve starts at $(\cl_*,\cg_*)=(1/2,0)$
corresponding to $Q_\UV=-\infty$. For increasing $Q_\UV$ it moves to the left, where it increases first, before it
decreases again, until it finally approaches $(\cl_*,\cg_*)=(-3/2,0)$ for $Q_\UV\to\infty$.

\begin{figure}[tp]
  \begin{textblock}{14}(1.33,3.1)
   \small
   $Q_\UV\to\infty$
   \hspace{20em}
   $Q_\UV\to -\infty$
 \end{textblock}
 \centering
 \includegraphics[width=0.8\columnwidth]{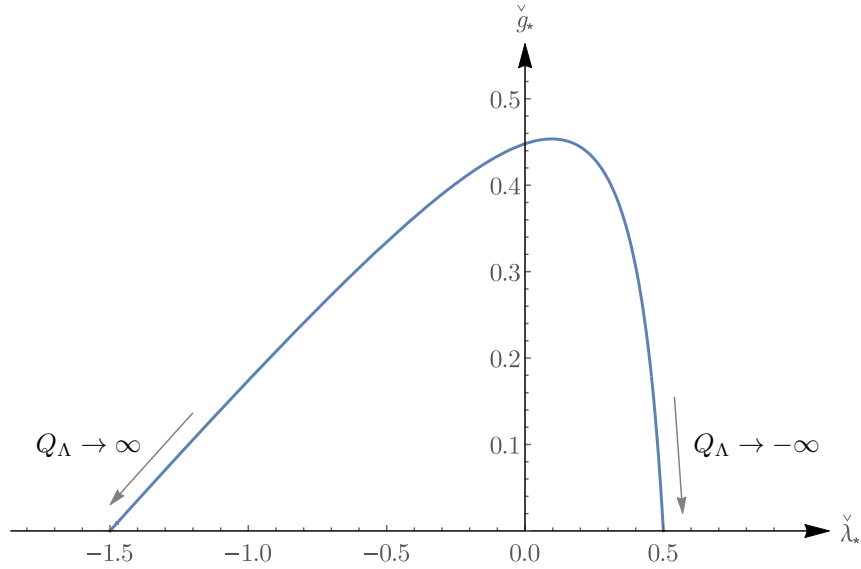}
\caption{Parametric plot showing the position of the bare NGFP dependent on $Q_\UV$ in $d=4$ dimensions, including the
 asymptotic fixed point positions in the limits $Q_\UV\to\infty$ and $Q_\UV\to -\infty$ at $(-3/2,0)$ and $(1/2,0)$,
 respectively.}
\label{fig:BareNGFPd4}
\end{figure}

\begin{figure}[tp]
\begin{minipage}{0.48\columnwidth}
  \begin{textblock}{8}(0,1.9)
  \scriptsize
  $Q_\UV\to\infty$
 \end{textblock}
 \begin{textblock}{2}(3.55,1.69)
  \scriptsize
  \centering
  $Q_\UV$\\ $\downarrow$\\ $-\infty$
 \end{textblock}
 \includegraphics[width=\columnwidth]{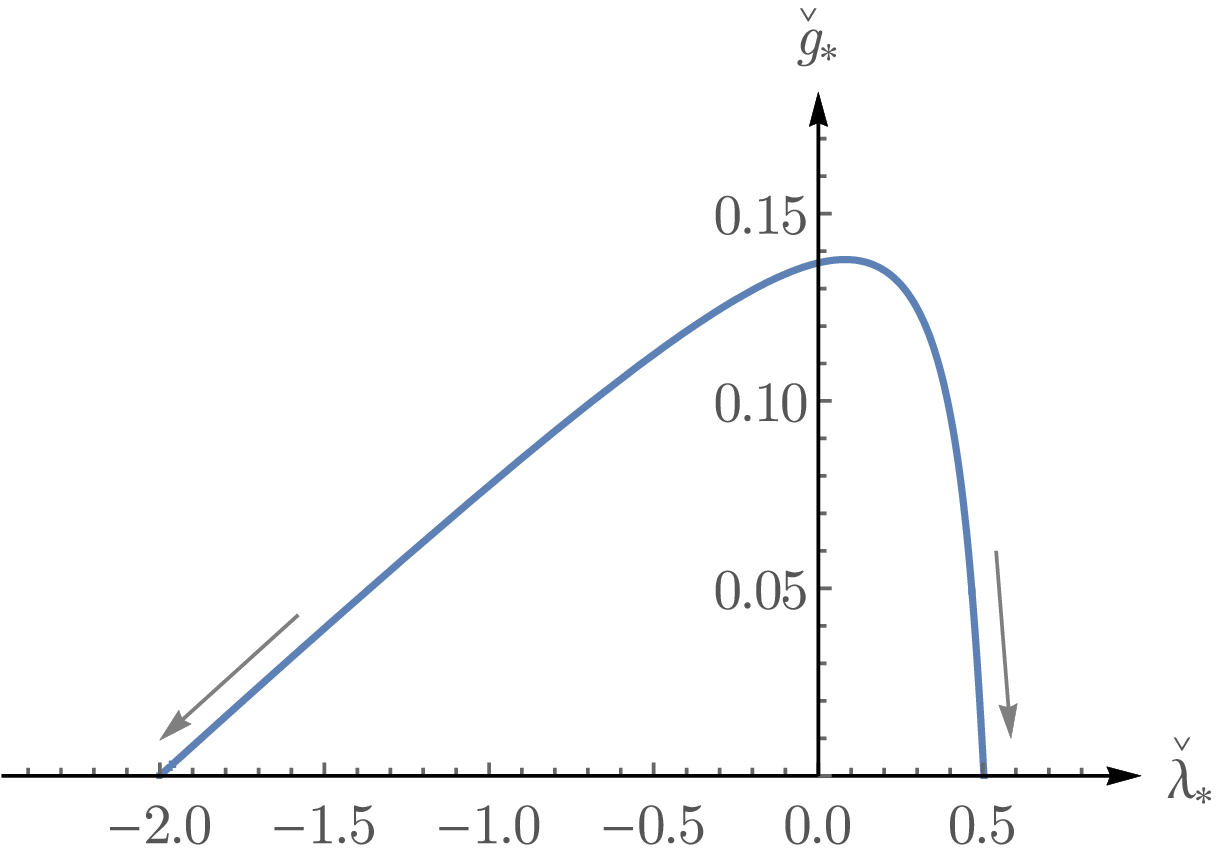}
\end{minipage}
\hfill
\begin{minipage}{0.48\columnwidth}
  \begin{textblock}{8}(0.1,1.9)
  \scriptsize
  $Q_\UV\to\infty$
 \end{textblock}
 \begin{textblock}{2}(3.55,1.7)
  \scriptsize
  \centering
  $Q_\UV$\\ $\downarrow$\\ $-\infty$
 \end{textblock}
 \includegraphics[width=\columnwidth]{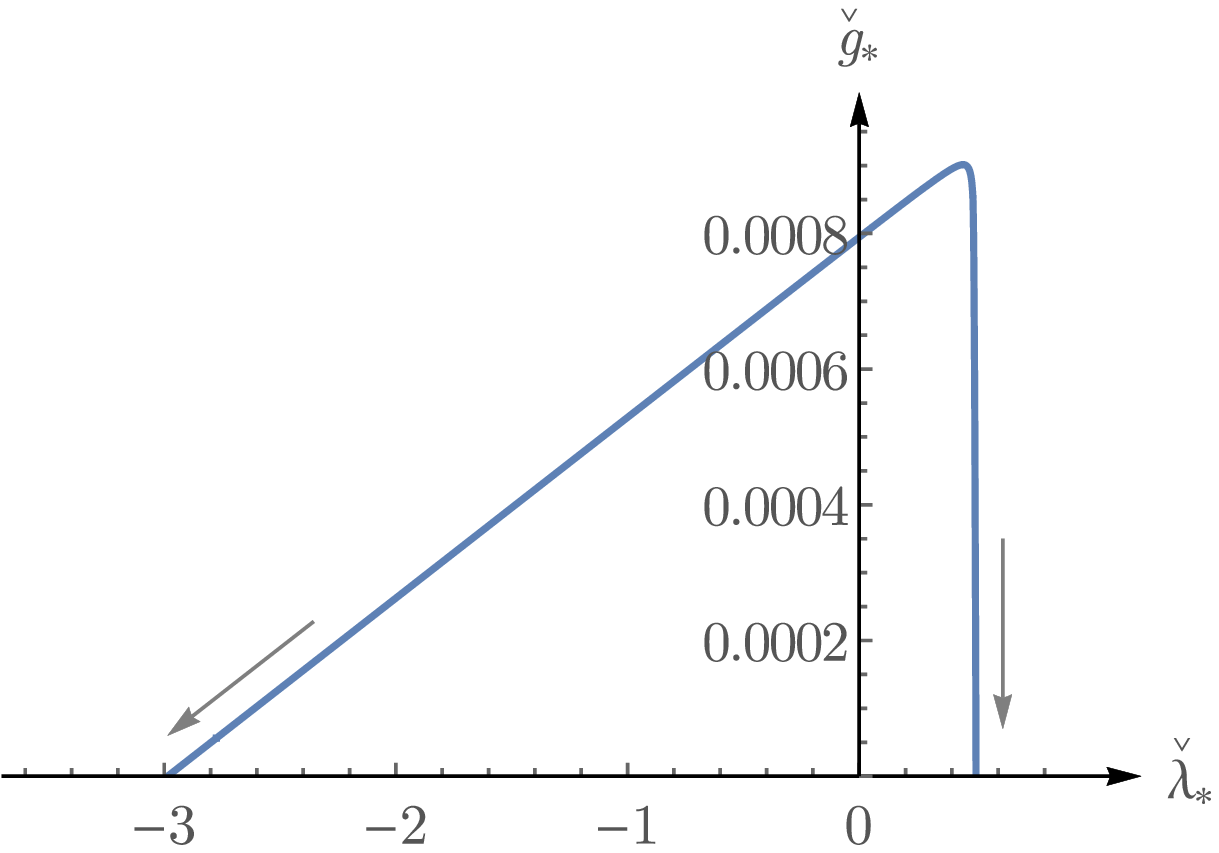}
\end{minipage}
\vspace{0.4em}
\caption{Parametric plots showing the position of the bare NGFP dependent on $Q_\UV$ in $d=3$ dimensions (left diagram)
 and $d=2+\ve$ dimensions with $\ve=0.01$ (right diagram).}
\label{fig:BareNGFPd3and2}
\end{figure}
For other dimensions we obtain qualitatively very similar pictures. The left diagram in Figure
\ref{fig:BareNGFPd3and2} shows the $3$-dimensional case while the right diagram is a representative of the
$2+\ve$-class, here for $\ve=0.01$. We make three important observations: When the dimension is lowered towards $2$,
(i) the left end point of the curve moves further to the left, in agreement with eq.\ \eqref{eq:LimitQInfty},
(ii) the height of the curve decreases, and (iii) the maximum point gets more and more peaked,
rendering the curve rather triangular. In the limit $d\to 2$ we ultimately obtain a perfect triangle with the right
side perpendicular to the baseline.

We would like to emphasize that, for any dimension $d>2$, these curves exhibit a smooth transition from $(\cl_*,\cg_*)
= (1/2,0)$ to $(\cl_*,\cg_*) = (-6/d,0)$, demonstrating once again the existence of the bare NGFP for any value of
$Q_\UV$.

%----------------------------------------------------------------------------------------------------------------------
\subsection{Critical exponents of the bare NGFP}
\label{sec:BareCritExp}
%----------------------------------------------------------------------------------------------------------------------

As usual, critical exponents are obtained by linearizing the flow in the vicinity of a fixed point. Let us start
with the effective couplings, here denoted by $\{u_\alpha\}$. Their linearized flow can be written as
\begin{equation}
 \p_t u_\alpha \equiv \UV\p_\UV\mku u_\alpha = \beta_\alpha(u_1,u_2,\ldots)
 \approx \sum_\sigma B_{\alpha\sigma} (u_\sigma-u_\sigma^*)\,,
\label{eq:LinFlow}
\end{equation}
with $B_{\alpha\sigma}\equiv \frac{\p\beta_\alpha}{\p u_\sigma}(u_1^*,u_2^*,\ldots)$, where the last relation
(``$\approx$'') in eq.\ \eqref{eq:LinFlow} means equality up to linear order. The critical exponents corresponding to
the fixed point $(u_1^*,u_2^*,\ldots)$ are defined to be minus one times the eigenvalues of the matrix $B$, i.e.\
they are solutions for $\theta$ to the equation
\begin{equation}
 \det(B+\theta\mku\Id)=0\,.
\label{eq:DefCritExp}
\end{equation}

In order to obtain the critical exponents for the \emph{bare} NGFP it is necessary to linearize the map
$(u_1,u_2,\ldots)\leftrightarrow(\cu_1,\cu_2,\ldots)$ as well because each bare coupling is considered to be a function
of the effective couplings, $\cu_\alpha \equiv \cu_\alpha(u_1,u_2,\ldots)$, and the flow originates from the effective
side:
\begin{equation}
 \p_t\cu_\alpha\equiv\UV\p_\UV\cu_\alpha = \sum_\rho \frac{\p\cu_\alpha}{\p u_\rho}\,\p_t u_\rho(u_1,u_2,\ldots).
\label{eq:BareFlowFromEff}
\end{equation}
Now, linearization must be applied to three parts in each term of the sum in \eqref{eq:BareFlowFromEff}: to
$\frac{\p\cu_\alpha}{\p u_\rho}$, to $\p_t u_\rho$ as in \eqref{eq:LinFlow}, and to the arguments $(u_1,u_2,\ldots)$
that have to be re-expressed in terms of the bare couplings again. For the first contribution we consider the following
linearization in the neighborhood of a fixed point:
\begin{equation}
 \cu_\alpha\equiv\cu_\alpha(u_1,u_2,\ldots) = \cu_\alpha(u_1^*,u_2^*,\ldots)
 + \sum_\rho \frac{\p\cu_\alpha}{\p u_\rho}(u_\rho-u_\rho^*) + \mO\big( (u-u^*)^2\big),
\end{equation}
so with $C_{\alpha\rho}\equiv \frac{\p\cu_\alpha}{\p u_\rho}(u_1^*,u_2^*,\ldots)$ we have, at linear order,
\begin{equation}
 \cu_\alpha-\cu_\alpha^* = \sum_\rho C_{\alpha\rho}(u_\rho-u_\rho^*),
\end{equation}
and similarly for the inverse,
\begin{equation}
 u_\sigma-u_\sigma^* = \sum_\kappa C_{\sigma\kappa}^{-1}(\cu_\kappa-\cu_\kappa^*).
\end{equation}
Thus, eq.\ \eqref{eq:BareFlowFromEff} in combination with \eqref{eq:LinFlow} yields
\begin{equation}
\begin{split}
 \p_t\cu_\alpha &= \sum_{\rho,\sigma} C_{\alpha\rho} B_{\rho\sigma}(u_\sigma-u_\sigma^*)+\mO\big((u-u^*)^2\big) \\
 &= \sum_{\rho,\sigma,\kappa} C_{\alpha\rho} B_{\rho\sigma} C_{\sigma\kappa}^{-1} (\cu_\kappa-\cu_\kappa^*)
 +\mO\big((\cu-\cu^*)^2\big).
\end{split}
\label{eq:LinFlowBare}
\end{equation}
From eq.\ \eqref{eq:LinFlowBare} we can finally read off the defining relation for the ``bare critical exponents'':
\begin{equation}
 \det\big(CBC^{-1}+\check{\theta}\mku\Id\big) = 0.
\end{equation}
Using $\det\!\big(CBC^{-1}+\check{\theta}\mku\Id\big)=\det\!\big[C(B+\check{\theta}\mku\Id)C^{-1}\big]=
\det(C)\det(B+\check{\theta}\mku\Id)\det^{-1}(C)$, we find that $\check{\theta}$ actually satisfies the same condition
as $\theta$, see \eqref{eq:DefCritExp}:
\begin{equation}
 \det(B+\check{\theta}\mku\Id)=0\,.
\end{equation}
This proves that \emph{bare fixed points have the same critical exponents as their counterparts on the EAA side}.

Regarding flow diagrams for bare couplings, for instance the ones in Figure \ref{fig:BareFlow}, this means that the
typical spiraling (or non-spiraling, for real critical exponents) form of the RG trajectories is preserved under the
map $(u_1,u_2,\ldots)\leftrightarrow(\cu_1,\cu_2,\ldots)$. The altered shapes of these spirals near the NGFP originate
from a change of the eigenvectors of the linearized flow which --- unlike the critical exponents --- are affected
by the map between effective and bare couplings. This phenomenon manifests itself as a squeezing of the spirals
in Figure \ref{fig:BareFlow} for large values of $Q_\UV$.

%----------------------------------------------------------------------------------------------------------------------
\subsection[A strategy to adjust bare couplings]{A strategy to adjust bare couplings:\linebreak critical
 \texorpdfstring{$\bm{Q_\UV}$}{Q}-value and vanishing cosmological constant}
\label{sec:BareMechanism}
%----------------------------------------------------------------------------------------------------------------------

In this section we would like to exploit the remaining freedom in setting up the functional integration measure,
associated with the free parameter $M$, in order to conveniently adjust the couplings in the bare action, in particular
the bare cosmological constant. Note that the $M$-dependence occurs in the measure and the bare action separately;
their combination in the path integral, however, gives rise to an $M$-independent effective action, so that no
physical quantity derived from it can depend on $M$. This holds true also for the FRGE \eqref{eq:FRGE} where any
potential $M$-dependence has dropped out. As already mentioned in Section \ref{sec:BareMap}, the free parameter $M$
translates into the parameter $Q_\UV$ which underlies the following discussion.

In Section \ref{sec:BareExNGFP} we showed that the flow of the bare couplings possesses an NGFP for any $d>2$ and for
any $Q_\UV$. Furthermore, we have seen that the position of this NGFP depends on $Q_\UV$: it starts at
$(\cl_*,\cg_*)=(1/2,0)$, corresponding to $Q_\UV=-\infty$, then it ``moves'' along an asymmetric arc, until it
ultimately approaches $(\cl_*,\cg_*)=(-6/d,0)$ as $Q_\UV\to\infty$. This implies a transition from positive to negative
bare cosmological constants. Hence, for reasons of continuity there must be \emph{a finite value of $Q_\UV$ at which
the bare cosmological constant vanishes}.

Before determining this critical $Q_\UV$-value, a comment regarding the significance of the bare \emph{fixed point} (as
compared with arbitrary points in the space of bare couplings) is in order: As we would like to remove the UV cutoff
ultimately by taking $\UV\to\infty$, it is in fact the bare NGFP that represents bare couplings in the common
sense.\footnote{Here the term ``bare NGFP'' refers to the NGFP of the effective couplings mapped into the space of bare
couplings. This notion includes two cases: The bare NGFP (i) is strictly a point, (ii) is divergent. Case (ii) means
that the effective couplings are mapped to such bare couplings which contain divergent contributions. (These divergent
parts exactly cancel out potential divergences in Feynman diagrams.) In both cases we can safely remove the cutoff in
the end.} Thus, although being unphysical it plays an important part at a computational level, which justifies an
investigation about how it can be adjusted conveniently. Nevertheless, in spite of the distinct role of the bare NGFP
we would like to keep our discussion as general as possible and consider also those bare couplings that do not
correspond to a fixed point.

In our Einstein--Hilbert setting a possible ``convenient adjustment'' entails fixing the bare cosmological constant to
zero. Let us denote the critical $Q_\UV$-value where this happens by $Q_\UV^{(0)}$. It can be obtained by setting
$\cl_\UV=0$ in eqs.\ \eqref{eq:MapI} and \eqref{eq:MapII}, and solving the system for $Q_\UV$. In this way we find that
\emph{the bare cosmological constant vanishes if} $Q_\UV=Q_\UV^{(0)}$, \emph{with}
\begin{equation}[b]
 Q_\UV^{(0)} \equiv \frac{1}{C_d}\,\frac{\lambda_\UV}{g_\UV}
 - (d+1)\ln\!\left[\frac{2\mku C_d}{d}\big(d(d-1)+4\big)+\frac{1}{g_\UV}+\frac{d}{6}\frac{\lambda_\UV}{g_\UV}\right].
\label{eq:Q0}
\end{equation}
Clearly, the statement remains valid at the NGFP, where the effective couplings $\lambda_\UV$ and $g_\UV$ have to be
replaced by their fixed point counterparts. In $d=4$, for instance, based on the NGFP values $\lambda_*$ and $g_*$ for
the Einstein--Hilbert truncation and the optimized cutoff, we obtain $Q_*^{(0)}\approx -0.583$. The $d$-dependence of
$Q_*^{(0)}$ is illustrated in Figure \ref{fig:QCrit}. We find that the critical value $Q_*^{(0)}$ exists in any
dimension $d>2$.
\begin{figure}[tp]
\centering
\includegraphics[width=0.6\columnwidth]{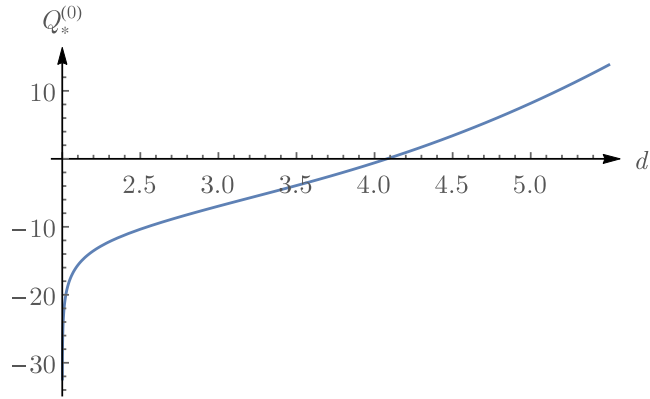}
\caption{Dependence of the critical value $Q_*^{(0)}$ on the dimension $d$ (taking the fixed point values based on the
 optimized cutoff for the effective couplings in \eqref{eq:Q0}).}
\label{fig:QCrit}
\end{figure}

As a remark we restate this result in terms of $M$. Using the definition of $Q_\UV$, given by eq.\ \eqref{eq:DefQ}, we
see that the bare cosmological constant vanishes if $M=M^{(0)}$, where $M^{(0)}$ satisfies\footnote{The critical value
$M^{(0)}$ exists for any $d>2$ with $d\neq 2.3723$. For $d = 2.3723$ the denominator of \eqref{eq:M0} becomes zero. In
this case the bare couplings are independent of $M$, i.e.\ they cannot be adjusted by tuning $M$. Most probably this
phenomenon is merely an artifact of the truncation and the approximate one-loop character of the reconstruction
formula.}
\begin{equation}
\begin{split}
 \ln\!\bigg(\frac{M^{(0)}}{\UV}\bigg) ={} &\frac{1}{8-d(d+1)}\bigg\{ (d+1)\ln(32\pi)
 -\frac{2}{d}\ln\!\left(\frac{d-2}{2d}\right)\\
 &+ \frac{1}{C_d}\,\frac{\lambda_\UV}{g_\UV} - (d+1)\ln\!\left[\frac{2\mku C_d}{d}\big(d(d-1)+4\big)
 + \frac{1}{g_\UV}+\frac{d}{6}\frac{\lambda_\UV}{g_\UV}\right] \bigg\}.
\end{split}
\label{eq:M0}
\end{equation}

As demonstrated in the next subsection, the consequences of a vanishing bare cosmological constant are particularly
interesting in $d=2+\ve$ dimensions.

%----------------------------------------------------------------------------------------------------------------------
\subsection[The bare couplings in \texorpdfstring{$2+\ve$}{2 + epsilon} dimensions]{The bare couplings in
 \texorpdfstring{\bm{$2+\ve$}}{2 + epsilon} dimensions}
\label{sec:Bare2D}
%----------------------------------------------------------------------------------------------------------------------

Let us review the above results and elaborate in more detail which simplifications emerge in $d=2+\ve$ dimensions.
By analogy with Figure \ref{fig:BareFlow} which showed several flow diagrams of the bare couplings in $d=4$ dimensions,
the $(2+\ve)$-dimensional case is depicted in Figure \ref{fig:BareFlow2D} where we choose $\ve=0.01$ as an example
here.
\begin{figure}[tp]
 \small
 \begin{minipage}{0.47\columnwidth}
  \centering
  $Q_\UV=4000$\\[0.4em]
  \includegraphics[width=\columnwidth]{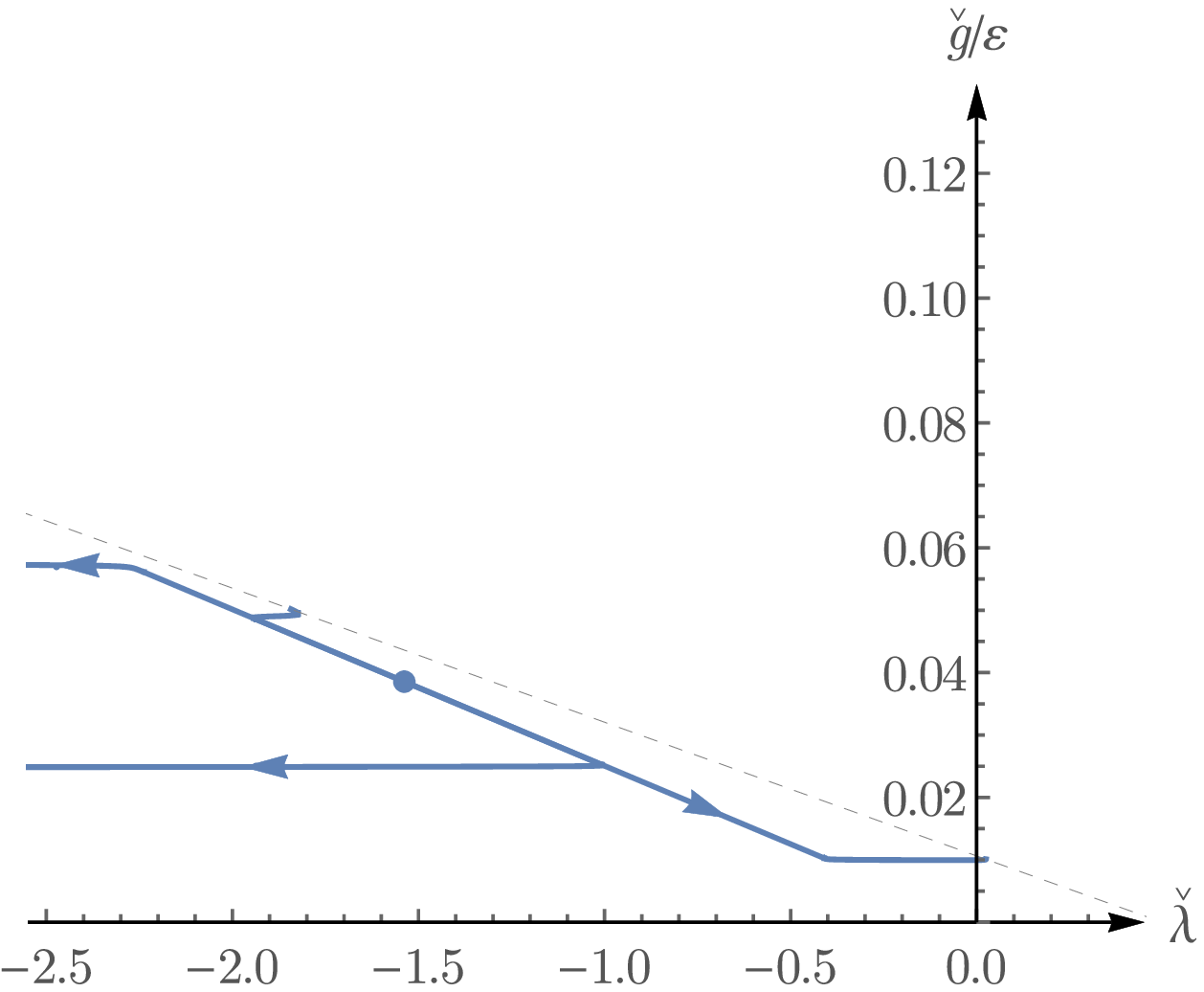}
 \end{minipage}
 \hfill
 \begin{minipage}{0.47\columnwidth}
  \centering
  $Q_\UV=1000$\\[0.4em]
  \includegraphics[width=\columnwidth]{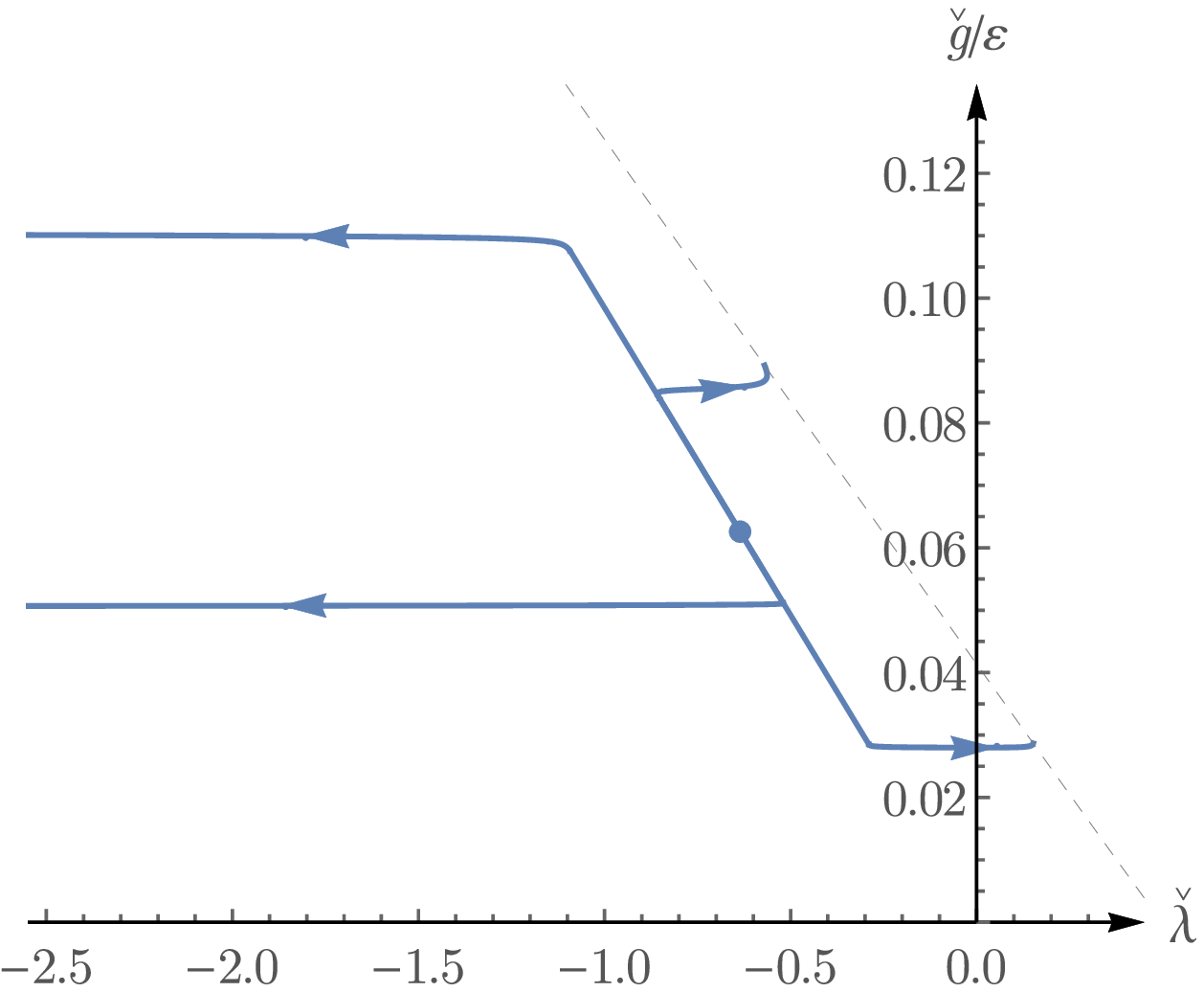}
 \end{minipage}\\[2.6em]
 \begin{minipage}{0.47\columnwidth}
  \centering
  $Q_\UV=-22.4671$\\[0.4em]
  \includegraphics[width=\columnwidth]{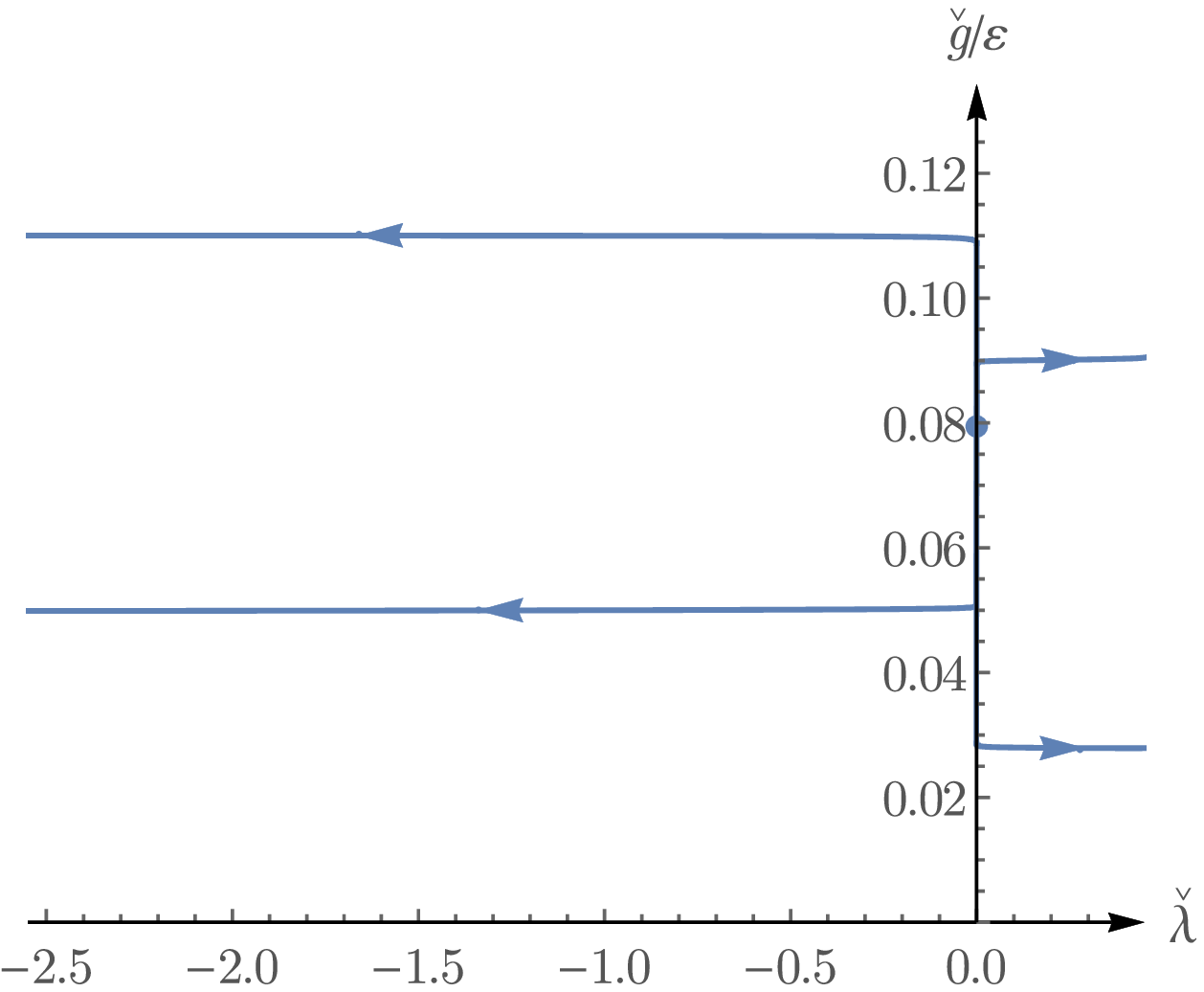}
 \end{minipage}
 \hfill
 \begin{minipage}{0.47\columnwidth}
  \centering
  $Q_\UV=-300$\\[0.4em]
  \includegraphics[width=\columnwidth]{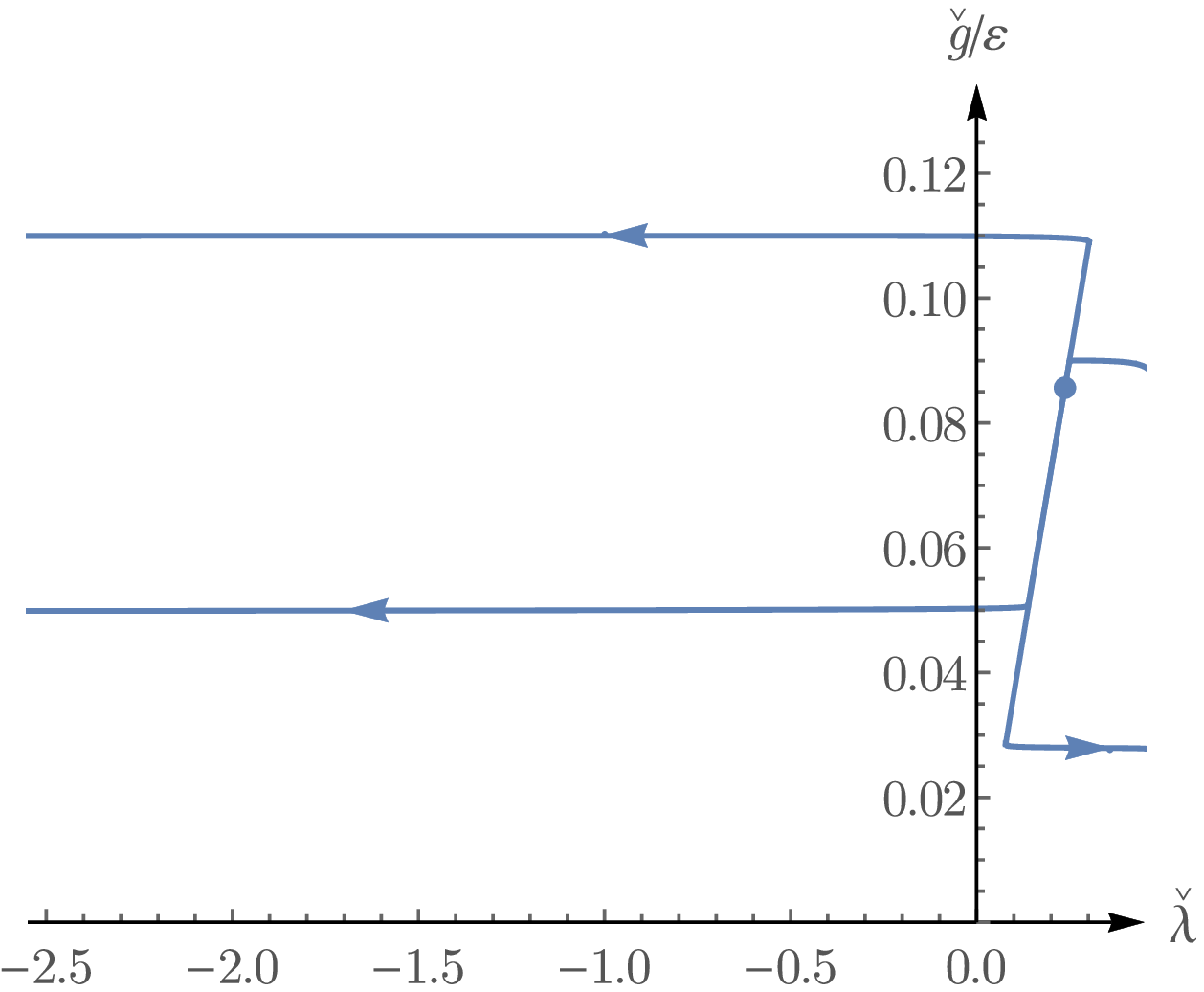}
 \end{minipage}
\vspace{2em}
\caption{Flow diagrams of the bare couplings $\cl$ and $\cg$ for several constant values of $Q_\UV$ in $d=2.01$
 dimensions. The bare NGFP is marked by a blue dot, and the gray dashed lines in the upper two figures represent the
 singularity lines known from the flow diagrams of the effective couplings (cf.\ Figure \ref{fig:StdSingle}, for
 instance), mapped into the space of bare couplings. For the sake of clarity we show only four representative
 trajectories for each diagram.}
\label{fig:BareFlow2D}
\end{figure}
We observe a $Q_\UV$-dependence of the flow similar to the one in $d=4$, including the ``moving'' bare fixed point.
Note that the qualitative structure of the trajectories is very similar to the one for the effective couplings, cf.\
Figure \ref{fig:Flow2D}: each trajectory consists of an almost horizontal part (in the IR), then a very sharp bend, and
finally a line that connects it to the bare NGFP (in the UV). Since the bare cosmological constant at the fixed point,
$\cl_*$, is not proportional to $\ve$, we do not normalize $\cl$ by the factor $1/\ve$. For that reason the singularity
line characterized by diverging $\beta$-functions is still present in Figure \ref{fig:BareFlow2D}, while it is shifted
to infinity for the effective couplings, see Figure \ref{fig:Flow2D}.
Apart from this numerical analysis we demonstrate in the following that it is possible to draw some important
conclusions at the analytical level, too.

We have already seen in the previous chapters that the effective couplings in an Einstein--Hilbert type EAA are of
the order $\ve$ at the fixed point:
\begin{equation}
 \lambda_*=\mO(\ve),\qquad g_*=\mO(\ve).
\end{equation}
In the vicinity of the NGFP the main $\ve$-order of the couplings does not change. Thus, we can assume
$\lambda_\UV=\mO(\ve)$ and $g_\UV=\mO(\ve)$ there, which can be exploited in an $\ve$-expansion in \eqref{eq:Q0}.
Moreover, we have $\frac{\lambda_\UV}{g_\UV}=\text{finite}+\mO(\ve)$ and $C_d=1+\mO(\ve)$, leading to the critical
value
\begin{equation}[b]
 Q_\UV^{(0)} = \frac{\lambda_\UV}{g_\UV} + 3\ln(g_\UV) + \mO(\ve\ln\ve),
\end{equation}
provided that both $\lambda_\UV$ and $g_\UV$ are of first order in $\ve$.

As above, we can express this result in terms of the parameter $M$. We find that the bare cosmological constant
vanishes if $M=M^{(0)}$, where $M^{(0)}$ satisfies
\begin{equation}[b]
 M^{(0)} = \alpha\mku\mku \ve\mku \UV\,.
\label{eq:M0Epsilon}
\end{equation}
In \eqref{eq:M0Epsilon}, $\alpha\equiv\alpha(\lambda_\UV,g_\UV)$ is a positive finite constant that depends only on the
effective couplings and whose leading order is given by
\begin{equation}
 \alpha = \exp\left[\textstyle\frac{1}{2}\frac{\lambda_\UV}{g_\UV}
 +\frac{3}{2}\ln\left(32\pi\mku\frac{g_\UV}{\ve}\right)+\ln(2)\right].
\end{equation}
Remarkably enough, we found $M^{(0)}\propto\UV$, which might be considered the expected behavior for a mass parameter,
but here it is not the result of any dimensional analysis. It has rather been derived by requiring a vanishing bare
cosmological constant. After all, $M\propto\UV$ seems to be the most natural choice.

There are two possible orders of taking limits in our setting: (i) $\UV\to\infty$ before $\ve\to 0$, and
(ii) $\UV\to\infty$ after $\ve\to 0$. The order must be considered part of the definition of the theory under
consideration. As we have seen in Chapter \ref{chap:EHLimit}, the limit $d\to 2$ of the Einstein--Hilbert action leads
to a new action with a reduced number of degrees of freedom. Therefore, taking the dimensional limit first before
reconstructing the bare action and taking $\UV\to\infty$ might give a different result (see Chapter
\ref{chap:BareLiouville}) than the one obtained by reconstructing $\SB$ first and taking the 2D limit afterwards. We
would like to point out that there is even a third possibility: a simultaneous limit, in particular with regard to eq.\
\eqref{eq:M0Epsilon}. For that purpose, we introduce a fixed reference scale, say $\UV^{(0)}$, and write the cutoff
scale as $\UV=\UV^{(0)}/\ve$. Then the limit $\UV\to\infty$ is equivalent to the limit $\ve\to 0$. By eq.\
\eqref{eq:M0Epsilon} we find that the bare cosmological vanishes at the critical value $M=M^{(0)}=\alpha\mku
\UV^{(0)}$. This establishes the possibility of a constant parameter $M$.

Finally, let us work out the most important consequence of a vanishing bare cosmological constant in
$d=2+\ve$ dimensions. It turns out that $\cl_\UV=0$ implies a particularly simple relation between bare and
effective Newton constant: Reconsidering equation \eqref{eq:MapI} with $\cl_\UV=0$, we obtain
\begin{equation}
\frac{1}{\cg_\UV}\big(3+\mO(\ve)\big) - \frac{1}{g_\UV}\big(3+\lambda_\UV+\mO(\ve)\big)= 18 + \mO(\ve).
\end{equation}
Choosing the effective couplings to lie in a neighborhood of the NGFP, i.e.\ assuming $\lambda_\UV=\mO(\ve)$ and
$g_\UV=\mO(\ve)$ again, multiplication by $g_\UV/3$ yields
\begin{equation}
 \frac{g_\UV}{\cg_\UV}-1=\mO(\ve),
\end{equation}
or $\cg_\UV = g_\UV +\mO(\varepsilon^2)$. Hence, for the special choice $M=M^{(0)}$, given by eq.\
\eqref{eq:M0Epsilon}, \emph{the bare Newton constant agrees with the effective Newton constant}.

To sum up, we have found a strategy to reconstruct the bare action in a specific way such that the bare coupling
constants are adjusted conveniently. The method relies on an appropriate choice of the measure parameter $M$:
If $M$ is chosen as in \eqref{eq:M0Epsilon} the bare couplings at the NGFP are given by
\begin{boxalign}
 \cl_* &= 0, \\
 \cg_* &= g_* + \mO(\ve^2).
\end{boxalign}
This powerful argument demonstrates that the freedom in defining a functional measure, i.e.\ the freedom in choosing
$M$, can be exploited to fix one of the bare couplings to a suitable value (here $\cl_*=0$), and possibly to obtain a
simpler map from the effective couplings to the remaining bare couplings. The result $\cg_* = g_* + \mO(\ve^2)$ is
crucial with regard to our discussion of the 2D limit of the Einstein--Hilbert action in Chapter \ref{chap:EHLimit},
and it lays the foundation for a reconstruction of the functional integral corresponding to a full gravity+matter
system, to be studied in more detail in Chapter \ref{chap:FullReconstruction}.

%----------------------------------------------------------------------------------------------------------------------
\addtocontents{toc}{\protect\newpage}
%----------------------------------------------------------------------------------------------------------------------
\chapter[The reconstructed path integral in 2D Asymptotic Safety]%
[The reconstructed path integral in 2D asymptotically safe gravity]%
{The reconstructed path integral in 2D asymptotically safe gravity}
\label{chap:FullReconstruction}
%----------------------------------------------------------------------------------------------------------------------

\begin{summary}
We combine the results of Chapters \ref{chap:NGFPCFT} and \ref{chap:Bare} by taking the asymptotically safe fixed point
theory pertaining to the EAA in $d=2$ dimensions and by reconstructing its corresponding functional integral. The
discussion is not restricted to the purely gravitational bare action but takes into account matter and ghosts
contributions as well, thus giving rise to the complete functional integral of all fields under consideration.
We find that it amounts to a CFT whose total central charge adds up to zero. In particular, we uncover a compensation
mechanism for the matter fields: They enter both the gravitational part and the matter part of the NGFP theory where
the two contributions exactly cancel each other. As a consequence, the gravitational dressing of matter field operators
is trivial, i.e.\ the matter system is not affected by its coupling to quantum gravity. This leads to a complete
quenching of the a priori expected Knizhnik--Polyakov--Zamolodchikov (KPZ) scaling. A possible connection of this
prediction to Monte Carlo results obtained in the discrete approach to 2D quantum gravity based upon causal dynamical
triangulations is mentioned. Furthermore, we describe similarities of the fixed point theory to, and differences from,
noncritical string theory.

\noindent
\textbf{What is new?} Showing the compensation of matter contributions, the vanishing of the total central charge
and the quenching of the KPZ scaling in 2D Asymptotic Safety.

\noindent
\textbf{Based on:} Ref.\ \cite{NR16a}.
\end{summary}

\pagebreak

%----------------------------------------------------------------------------------------------------------------------
\section{Remark on the reconstruction process}
%----------------------------------------------------------------------------------------------------------------------

Starting with an effective average action $\Gamma_k$ of the full system (including gravitational, ghost and matter
fields) we search for a functional integral representation that reproduces a given complete $\Gamma_k$-trajectory. In
our setting, this reconstruction can be considered for each sector (gravity, ghost, matter) separately.

Concerning the gravitational part we employ the results of the previous chapter, where we have seen that the map
between effective and bare couplings depends on the measure parameter $M$. As demonstrated in Section \ref{sec:Bare2D},
in $d=2+\ve$ dimensions there is one particular value of $M$ that leads to a vanishing bare cosmological constant,
$\cl_*=0$, and a bare Newton constant $\cg_\UV$ which equals precisely the effective one at the NGFP:
\begin{equation}
 \cg_*=g_*\,.
\label{eq:BareEqualsEffective}
\end{equation}
For the exponential parametrization of the metric this amounts to $\cg_*=\ve/b$ with $b=\frac{2}{3}(25-N)$.
After having reconstructed the gravitational functional integral in $d=2+\ve$, where the bare action is given by
$-\frac{1}{16\pi \check{G}_\UV}\int\td^{2+\ve} x \sg\, R$ with $\check{G}_\UV = \UV^{-\ve}\cg_*=\UV^{-\ve}g_*$, we take
its 2D limit by employing the methods of Section \ref{sec:IndGravityFromEH}. As a result we obtain a \emph{bare action
which is proportional to the induced gravity action},
\begin{equation}[b]
 \SB^\text{grav}[g] = \frac{(25-\ns)}{96\pi} \,I[g] +\cdots
\label{eq:SBareGrav}
\end{equation}
The dots indicate that there might appear additional terms originating from the zero modes, according to eq.\
\eqref{eq:LimitResultGen} in the appendix. For our present purposes they are irrelevant, though; all properties of the
functional integral that are considered here can be studied on the basis of the term $\propto I[g]$.

For the ghost system we avail ourselves of the argument presented in Section \ref{sec:UnitaryCFT}, point \textbf{(7)}:
In our setting, it is only the gauge invariant gravity+matter part of the EAA that is ``handed over'' from $d>2$ to
$d=2$, while we can fix the gauge directly in 2D. Being particularly convenient, we choose the conformal gauge and the
corresponding Faddeev-Popov determinant \cite{Polyakov1981}. The integration over the metric then boils down to an
integration over the Liouville field and the moduli parameters (cf.\ Sec.\ \ref{sec:ConfGauge}).

The bare action of the matter system can be reconstructed according to the results of Ref.\ \cite{MS15}: For
cutoffs satisfying certain constraints the bare action equals precisely the EAA when the respective cutoff scales are
identified. Thus, the bare matter action agrees with the RHS of eq.\ \eqref{eq:matter}, i.e.\ it is given by
\begin{equation}
 \SB^\text{m}[g,A] \equiv \frac{1}{2} \sum_{i=1}^N\int\!\dd x\sg\; g^\mn\,\p_\mu A^i\mku\p_\nu A^i\,,
\label{eq:SBareMatter}
\end{equation}
in agreement with eq.\ \eqref{eq:SEqualsGamma}.

We would like to point out that, by equations \eqref{eq:SBareGrav} and \eqref{eq:SBareMatter}, the number $\ns$ enters
\emph{both the gravitational and the matter part of the bare action}, respectively, the former being a consequence of
the $\ns$-dependence of the fixed point value $g_*\mku$.

%----------------------------------------------------------------------------------------------------------------------
\section{A functional integral for 2D asymptotically safe gravity}
%----------------------------------------------------------------------------------------------------------------------

\textbf{(1) The partition function}. Based on the above considerations we obtain the full reconstructed partition
function:
\begin{equation}
 Z = \int[\td\tau] \int\mD_{\gfull}\phi\;\, Z_\text{gh}\big[\gfull\big]\,\Zm\big[\gfull\big]\, \Yg\big[\gfull\big]\,.
\label{eq:PartFunc}
\end{equation}
The integrand of \eqref{eq:PartFunc} comprises the following factors:
the exponential of the gravitational part of the fixed point action,
\begin{equation}
 \Yg[g] \equiv \exp\!\left(-\frac{(25-\ns)}{96\pi}\,I[g]+\cdots\right),
\label{eq:Ygrav}
\end{equation}
the partition function of the matter system (cf.\ Appendix \ref{app:Weyl}),
\begin{equation}
\begin{split}
 \Zm[g] &\equiv \int\mD A\;\exp\!\Bigg(-\frac{1}{2}\sum\limits_{i=1}^{\ns} \int \td^2 x\sg\,
  g^\mn\mku\p_\mu A^i\,\p_\nu A^i\Bigg)\\
	&= {\det}^{-\ns/2}\big(-\Box_g\big)
	= \exp\left(-\frac{\ns}{96\pi}\,I[g]+\cdots\right),
\end{split}
\label{eq:Zmatter}
\end{equation}
the partition function of the $b$-$c$ ghost system, $Z_\text{gh}$, the split symmetry invariant measure for the
integration over the Liouville field, $\mD_{\gfull}\phi$, and finally the measure $[\td\tau]$ for the integration over
the moduli that are implicit in the reference metric pertaining to a given topological type of the spacetime manifold
(cf.\ Sec.\ \ref{sec:ConfGauge}). In eqs.\ \eqref{eq:Ygrav} and \eqref{eq:Zmatter} we suppressed possible contributions
to the bare cosmological constant. Here and in the following, we indicate them by the dots.

The behavior under Weyl transformations of the various factors is well known. Using in particular
eq.\ \eqref{eq:decomp1} with the (noncosmological constant part of the) renormalized Liouville action, $\Delta I$, as
defined in \eqref{eq:DeltaIDef}, we have
{\allowdisplaybreaks
\begin{subequations}
\begin{align}
 \Yg\big[\gfull\big] &= \Yg[\mku\hg]\;\exp\!\bigg(\!+\frac{(25-\ns)}{12\pi}\,\Delta I[\phi;\hg]\bigg), \\
 \Zm\big[\gfull\big] &= \Zm[\mku\hg]\;\exp\!\bigg(\!+\frac{\ns}{12\pi}\,\Delta I[\phi;\hg]\bigg), \\
 Z_\text{gh}\big[\gfull\big] &= Z_\text{gh}[\mku\hg]\;\exp\!\bigg(\!+\frac{(-26)}{12\pi}\,\Delta I[\phi;\hg]\bigg),
 \label{eq:WeylGhosts}\\
 \mD_{\gfull}\phi &= \mD_{\hg}\phi\;\exp\!\bigg(\!+\frac{1}{12\pi}\,\Delta I[\phi;\hg]\bigg).
 \label{eq:WeylDPhiMeasure}
\end{align}
\label{eq:WeylPartition}
\end{subequations}
}%
As before, possible (measure dependent) terms involving the bare cosmological constant are suppressed in eqs.\
\eqref{eq:WeylPartition}. On the RHS of \eqref{eq:WeylDPhiMeasure}, $\mD_{\hg}\phi$ is the translational invariant
measure now.

Up to this point, the discussion is almost the same as in noncritical string theory \cite{Polyakov1981}. The profound
difference lies in the purely gravitational part of the bare action, $\Yg$. Contrary to what happens in any
conventional field theory, whose bare action is a \emph{postulate} rather than the result of a \emph{calculation},
asymptotically safe gravity in $2$ dimensions is based upon a gravitational action which \emph{depends explicitly on
properties of the matter system}. In the example at hand, this dependence is via the number $\ns$ of $A^i$-fields that
makes its appearance in the fixed point action and hence in the ``Boltzmann factor'' \eqref{eq:Ygrav}.

\medskip
\noindent
\textbf{(1a) Matter refuses to matter: a compensation mechanism}. Remarkably enough, the integrand of
\eqref{eq:PartFunc} depends on $\ns$ only via the product $\Zm\cdot\Yg$ in which the $\ns$-dependence cancels between
the two factors. Multiplying \eqref{eq:Ygrav} and \eqref{eq:Zmatter} we obtain a result which, for any $\ns$, equals
that of pure gravity. It is always the same no matter how many scalar fields there are:
\begin{equation}
 \Zm[g]\,\Yg[g]= \exp\!\left(-\frac{25}{96\pi}\,I[g]+\cdots\right).
\label{eq:ZMatterYg}
\end{equation}
Under a Weyl rescaling this expression transforms as $\Zm\big[\gfull\big]\mku\Yg\big[\gfull\big]=\Zm[\hg]\mku\Yg[\hg]\,
\exp\left(+\frac{25}{12\pi}\Delta I[\phi;\hg]\right)$. As a consequence of eq.\ \eqref{eq:ZMatterYg}, the reconstructed
functional integral coincides always with that of \emph{pure gravity} (as long as we do not evaluate the expectation
value of observables depending on the $A$'s and as long as cosmological constant terms do not play a role):
\begin{equation}[b]
 \!Z = \int[\td\tau]\;\Zm[\hg]\,\Yg[\hg] \int\!\mD_{\gfull}\phi\;\, Z_\text{gh}\big[\gfull\big]\,
 \exp\left(\!+\frac{25}{12\pi}\,\Delta I[\phi;\hg]+\cdots\mkern-1mu\right).\!\!
\end{equation}

\medskip
\noindent
\textbf{(1b) Zero total central charge}. Over and above the specific form of its matter dependence, the fixed point
action displays a second miracle: Its central charge equals precisely the critical value $25$. Up to a cosmological
constant term possibly, this leads to a \emph{complete cancellation of the entire $\phi$-dependence} of the integrand
once the ghost contribution \eqref{eq:WeylGhosts} and the ``Jacobian'' factor in \eqref{eq:WeylDPhiMeasure} are taken
into account:
\begin{equation}[b]
 Z = \int[\td\tau]\;Z_\text{gh}[\hg]\,\Zm[\hg]\,\Yg[\hg] \int\mD_{\hg}\phi \; \exp(0+\cdots)\,.
\label{eq:ZNonCritical}
\end{equation}
Hence, for every choice of the matter sector, the total system described by the reconstructed functional integral of
asymptotically safe 2D gravity is a conformal field theory with central charge zero. The various sectors of this system
contribute to the total central charge as follows:
\begin{equation}
 c_\text{tot} = \underbrace{(25-\ns)}_{\text{NGFP, grav.\ part}}+\underbrace{\phantom{|}\ns\phantom{|}}_{\text{matter}}
 + \underbrace{\phantom{|}1\phantom{|}}_{\text{Jacobian}} + \underbrace{(-26)}_{\text{ghosts}} = 0 \,.
\label{eq:ctotZero}
\end{equation}

Actually, the result \eqref{eq:ctotZero} is even more general than we have indicated so far. In addition to the scalar
matter fields underlying our considerations up to this point, we can also bring massless free Dirac fermions into play
and couple them (minimally) to the dynamical metric by adding a corresponding term to the matter action
\eqref{eq:matter}. The contribution of each of such fermions to the $\beta$-function of Newton's constant in $d=2+\ve$
dimensions is the same as for a scalar field \cite{DP13,DEP14}, that is, fermions and scalars enter the central charge
in the same way. Hence, in all above equations for $\beta$-functions and central charges we may identify $\ns$ with
\begin{equation}
 \ns\equiv N_\text{S}+N_\text{F}\,,
\end{equation}
where $N_\text{S}$ and $N_\text{F}$ denote the number of real scalars and Dirac fermions, respectively.
In particular, we recover the same cancellation in the total central charge as in eq.\ \eqref{eq:ctotZero}: The central
charge of the matter system, $+\ns$, removes exactly a corresponding piece in the pure gravity contribution enforced by
the fixed point, $25-\ns$.

\medskip
\noindent
\textbf{(2) Observables}. By inserting appropriate functions $\bar{\ob}[\phi,A;\hg]$ into the path integral
\eqref{eq:PartFunc} we can in principle evaluate the expectation values of arbitrary observables $\ob[\mku g,A]=
\ob[\gfull,A]$. The insertion of $\bar{\ob}$ instead of $\ob$ is required due to the change of variables,
$g\mapsto \big(\phi,\{\tau\}\big)$, where in general $\bar{\ob}[\phi,A;\hg]\neq \ob[\gfull,A]$. In the case when the
observables do not involve the matter fields, their expectation values read
\begin{equation}
 \langle\ob\rangle=\frac{1}{Z}\int\mkern-1mu[\td\tau]\;\Zm[\hg]\,\Yg[\hg] \int\!\mD_{\gfull}\phi\;\, Z_\text{gh}
 \big[\gfull\big]\,\bob[\phi;\hg]\mkuu\exp\!\left(\frac{25}{12\pi}\mkuu\Delta I[\phi;\hg]\right)\mkern-2mu.
\label{eq:expObs}
\end{equation}
Without actually evaluating the $\phi$-integral we see that when the cosmological constant term is negligible \emph{the
expectation value of purely gravitational observables does not depend on the presence or absence of matter and its
properties}, provided the background factor $\Zm[\hg]$ in \eqref{eq:expObs} cancels against the corresponding piece in
the denominator of \eqref{eq:expObs}. At the very least, this happens if one considers expectation values at a fixed
point of the moduli space.

\medskip
\noindent
\textbf{(3) Gravitational dressing}. As it is well known \cite{David1988,DK89,Watabiki1993}, it is not completely
straightforward to find the functional $\bob[\phi;\hg]$ which one must use under a conformally gauge-fixed path
integral in order to represent a given diffeomorphism (and, trivially, Weyl) invariant observable $\ob[g]=\ob[\gfull]$.
The association $\ob\rightarrow\bob$ should respect the following conditions \cite{Watabiki1993}: $\bob[\phi;\hg]$ must
be invariant under diffeomorphisms, it must approach $\ob[\gfull]$ in the classical limit and $\ob[\hg]$ in the limit
$\phi\to 0$, and most importantly it must be such that the expectation value computed with its help is independent of
the reference metric chosen, $\hg_\mn$.

Let us briefly recall the David--Distler--Kawai (DDK) solution to this problem \cite{David1988,DK89}. For this purpose,
we consider 2D gravity coupled to an arbitrary matter system described by a CFT with central charge $c$ and partition
function $\Zmc[g]$. First we want to evaluate the partition function for a fixed volume (area) of spacetime, $V$:
\begin{equation}
 Z_V=\int\frac{\mD g}{\text{vol(Diff)}}\;\Zmc[g]\;\delta\!\left(V-\int\td^2 x\sg\right).
\end{equation}
This integral involves the observable $\ob[g]\equiv\int\td^2 x\sg\equiv\int\td^2 x\shg\,\exp(2\phi)$. The associated
$\bob$ satisfying the above conditions turns out to require only a ``deformation'' of the prefactor of $\phi$ in the
exponential:
\begin{equation}
 \bob[\phi;\hg] = \int\td^2 x\shg\,\exp(2\mku\alpha_1 \phi)\,.
\label{eq:AreaOp}
\end{equation}
The modified prefactor $\alpha_1$ depends on the central charge of the matter CFT according to
\begin{equation}
 \alpha_1 = \frac{2\mku\sqrt{25-c}}{\sqrt{25-c}+\sqrt{1-c}} = \frac{1}{12}\left[25-c-\sqrt{(25-c)(1-c)}\right]\,.
\label{eq:AlphaOne}
\end{equation}
Thus, in the conformal gauge, $Z_V$ reads as follows:
\begin{equation}
 Z_V = \int\mkern-1mu[\td\tau]\;Z_\text{gh}[\hg]\,\Zmc[\hg] \int\mD_{\hg}\phi\;\delta\!\left(\! V-\!\int\!\td^2 x\shg
 \,\e^{2\mku\alpha_1\phi}\right)\,\exp\!\left(\!-\frac{(25-c)}{12\pi}\mkuu\Delta I[\phi;\hg]\right)\mkern-2mu.
\label{eq:PartFuncZV}
\end{equation}
Similarly, the expectation value of an arbitrary observable $\ob[g]$ at fixed volume is given by $\langle\ob[g]\rangle
= Z_V^{-1}\langle\bob[\phi;\hg]\rangle'$. Here $\langle\cdots\rangle'$ is defined by analogy with \eqref{eq:PartFuncZV}
but with the additional factor $\bob[\phi;\hg]$ under the $\phi$-integral.

The DDK approach to the gravitational dressing of operators from the matter sector was developed as a conformal
gauge-analogue to the work of Knizhnik, Polyakov and Zamolodchikov (KPZ) \cite{Polyakov1987,KPZ88} based upon the light
cone gauge.

To study gravitational dressing, let us consider an arbitrary spinless primary field $\ob_n[g] \equiv \int\td^2 x\sg\;
\mP_{n+1}(g)$, where $\mP_n(g)$ is a generic scalar involving the matter fields with conformal weight $(n,n)$, that is,
it responds to a rescaling of the metric according to $\mP_n(\e^{-2\sigma}g) = \e^{2 n \sigma}\,\mP_n(g)$. Under the
functional integral, the observables $\ob_n$ are then represented by
\begin{equation}
 \bob_n[\phi;\hg] = \int\td^2 x\shg\;\exp(2\mku\alpha_{-n}\mku\phi)\,\mP_{n+1}(\hg)\,,
\label{eq:OpN}
\end{equation}
where the $c$-dependent constants in the dressing factors generalize eq.\ \eqref{eq:AlphaOne}:
\begin{equation}
 \alpha_n = \frac{2\mku n\mku\sqrt{25-c}}{\sqrt{25-c}+\sqrt{25-c-24n}} \;.
\label{eq:AlphaN}
\end{equation}
Using \eqref{eq:AlphaN} it is straightforward now to write down the modified conformal dimensions corrected by the
quantum gravity effects.

The results of the DDK approach reproduce those of KPZ (valid for spherical topology) and generalize them for
spacetimes of arbitrary topology. Within the framework of the EAA and its functional RG equations, the KPZ relations
were derived from Liouville theory in Ref.\ \cite{RW97}; for a review see \cite{CD15}.

\medskip
\noindent
\textbf{(4) Quenching of the KPZ scaling}. Let us apply the general DDK--KPZ formulae to the NGFP theory of
asymptotically safe gravity. We must then replace
\begin{equation}
 c\;\longrightarrow\; \cgr + N \equiv (25-N)+N = 25\,,
\end{equation}
since the relevant bare action now arises from both the integrated-out matter fluctuations and the pure-gravity NGFP
contribution, $\Yg$. Setting $c=25$ in eqs.\ \eqref{eq:AlphaOne} and \eqref{eq:AlphaN} we obtain
\begin{equation}
 \alpha_1=0\qquad \text{and}\qquad \alpha_n=0 \mku ,
\label{eq:AlphaZero}
\end{equation}
respectively. This implies that \emph{the Liouville field completely decouples from the area operator \eqref{eq:AreaOp}
and any of the observables \eqref{eq:OpN}}.

As a consequence, the dynamics of the matter system is unaffected by its coupling to quantum gravity. In particular,
its critical behavior is described by the properties (critical exponents, etc.) of the matter CFT defined on a
nondynamical, rigid background spacetime. Thus, the specific properties of the NGFP lead to a perfect ``quenching''
of the a priori expected KPZ scaling.

\medskip
\noindent
\textbf{(5) Relation to noncritical string theory}. The functional integral \eqref{eq:ZNonCritical} is identical to
the partition function of noncritical string theory in 25 Euclidean dimensions. This theory is equivalent to the usual
critical bosonic string living in a (25+1)-dimensional Minkowski space whereby the Liouville mode plays the role of
the time coordinate in the target space \cite{DG87,DG88,DNW89}. Whether we consider pure asymptotically safe gravity in
two dimensions, or couple any number of scalar and fermionic matter fields to it, the resulting partition function
equals always the one induced by the fluctuations of precisely 25 string positions $X^m(x^\mu)$.

There is, however, a certain difference between asymptotically safe gravity and noncritical string theory in the way
the special case of vanishing total central charge, i.e.\ of precisely 25 target space dimensions, is approached.
To see this, note that in the present work we related the Liouville field to the metric by the equation $g_\mn =
\e^{2\phi}\hg_\mn$, and at no point did we redefine $\phi$ by absorbing any constant factors in it. In this connection,
the Liouville action for a general central charge $c$ has the structure $\GL = -\frac{c}{24\pi} \int \big( \hD_\mu \phi
\mku \hD^\mu \phi + \hR\phi \big) + \cdots$.

\noindent
\textbf{(i)} In order to combine $\GL$ with the action of the string positions, $+\frac{1}{8\pi}\mkern-2mu\int\!
\hD_\mu X^m \mku \hD^\mu\mkern-2mu X^m\mkern-2mu$, it is natural to introduce the redefined field
\begin{equation}
 \phi' \equiv Q\mku\phi\qquad \text{with}\qquad Q\equiv \sqrt{\frac{c}{3}} \,,
\end{equation}
in terms of which $\GL = -\frac{1}{8\pi}\int \big( \hD_\mu \phi' \mku \hD^\mu \phi' + Q\hR\phi' \big) + \cdots$.
It is this new field $\phi'$ that plays the role of time in target space and combines with the $X^m$'s in the
conventionally normalized action $\frac{1}{8\pi}\int \big(-\hD_\mu \phi' \mku \hD^\mu \phi' + \hD_\mu X^m
\mku \hD^\mu X^m - Q\hR\phi' \big) + \cdots$ which enhances the original $\mathrm{O}(25)$ symmetry to the full Lorentz
group in target space, $\mathrm{O}(1,25)$ \cite{DNW89}.

In string theory, conformal invariance requires the total central charge to vanish, $c_\text{tot}=0$. Hence, arguing
that the combined $(X^0\equiv\phi',X^m)$-quantum system is equivalent to the usual bosonic string theory in the
critical dimension involves taking the limit $c\equiv c_\text{tot} \rightarrow 0$ in the above formulae. Obviously this
requires some care in calculating correlation functions as the relationship $\phi'\equiv \sqrt{c/3}\,\phi$ breaks down
in this limit. Considering vertex operators for the emission of a tachyon of 26-dimensional momentum $(P_0,P_m)$, say,
this involves combining the rescaling $\phi\rightarrow\sqrt{c/3}\,\phi$ with a corresponding rescaling of $P_0$ with
the inverse factor, $P_0\rightarrow\sqrt{3/c}\,P_0$, rendering their product $P_0 X^0\equiv P_0\phi'$ independent of
$c$. The vertex operator $\exp\mkern-1mu\big\{i(-P_0 X^0+P_m X^m)\big\}$ also displays the full $\mathrm{O}(1,25)$
invariance. (See Refs.\ \cite{DG87,DG88} for a detailed discussion.)

\noindent
\textbf{(ii)} In 2D asymptotically safe quantum gravity, too, the total central charge was found to vanish, albeit for
entirely different reasons than in string theory. However, here there is no obvious reason or motivation for any
rescaling before letting $c\rightarrow 0$. In all of the above equations, including \eqref{eq:AreaOp} and
\eqref{eq:OpN}, $\phi$ still denotes the Liouville field introduced originally. In quantum gravity we let
$c\rightarrow 0$ in the most straightforward way, setting in particular $c=0$ directly in \eqref{eq:AlphaOne} and
\eqref{eq:AlphaN}. This is what led us to \eqref{eq:AlphaZero}, that is, the disappearance of $\phi$ from the
exponentials $\exp(2\mku\alpha_{-n}\phi)$ multiplying the matter operators and the ``quenching'' of the KPZ-scaling.

%----------------------------------------------------------------------------------------------------------------------
\section{Comparison with Monte Carlo results}
%----------------------------------------------------------------------------------------------------------------------

In earlier work \cite{DR09,LR05,RS11} indications were found that suggest that Quantum
Einstein Gravity in the continuum formulation based upon the EAA might be related to the discrete approach employing
causal dynamical triangulation (CDT) \cite{AL98,AGJL12}. In particular, the respective predictions for the fractal
dimensions of spacetime were compared in detail and turned out similar \cite{LR05,RS11}. It is therefore natural to
ask whether the quenching of the KPZ-scaling due to the above compensation mechanism can be seen in 2D CDT simulations.
And in fact, the Monte Carlo studies indeed seem to suggest a picture that looks quite similar at first sight:
Coupling several copies of the Ising model \cite{AAL00} or the Potts model \cite{AALP09} to $2$-dimensional
Lorentzian quantum gravity in the CDT framework, there is strong numerical evidence that the critical behavior of the
combined system, in the matter sector, is described by \emph{the same} critical exponents as on a fixed, regular
lattice. Under the influence of the quantum fluctuations in the geometry the critical exponents do not get shifted
to their KPZ values.

While this seems a striking confirmation of our Asymptotic Safety-based prediction, one should be careful in
interpreting these results. In particular, it is unclear whether the underlying physics is the same in both cases.
In CDT, the presence (absence) of quantum gravity corrections of the matter exponents is attributed to the presence
(absence) of baby universes in Euclidean (causal Lorentzian) dynamical triangulations. In our approach instead, the
quantum gravity corrections that could in principle lead to the KPZ exponents are exactly compensated by the
\emph{explicit matter dependence} of the \emph{pure gravity}-part in the bare action. This matter dependence is an
immediate consequence of the very Asymptotic Safety requirement.

As yet, we considered conformal matter only which was exemplified by massless, minimally coupled scalar fields. In the
nonconformal case when those fields are given a mass for instance, the compensation between the matter contributions
to the bare NGFP action and those resulting from integrating them out will in general no longer be complete.
On the EAA side, this situation is described by a trajectory $k\mapsto\Gamma_k$ that runs away from the fixed point
as $k$ decreases, and typically the resulting ordinary effective action of the gravity+matter system, $\Gamma_{k=0}$,
will indeed be affected by the presence of matter.

This expected behavior seems to be matched by the results of very recent 2D Monte Carlo simulations of CDT coupled to
more than one massive scalar field \cite{AGJZ15a,AGJZ15b,AGJZ15c}. It was found that, above a certain value of their
mass, the dynamics of the CDT+matter system is significantly different from the massless case. In particular, a
characteristic ``blob + stalk'' behavior was observed, well known from 4D pure gravity CDT simulations, but absent in
2D with conformal matter.

%----------------------------------------------------------------------------------------------------------------------
\section{Summarizing remarks}
\label{sec:SumUpRecPI}
%----------------------------------------------------------------------------------------------------------------------

\noindent
\textbf{(1)}
We reconstructed the partition function for the complete 2D fixed point theory, whose gravitational part is governed by
the fixed point value of the Newton coupling. Interestingly enough, this value receives contributions from both gravity
and matter sector: $g_*= 3\mku\ve/\big(2\mku(25-\ns)\big)$, where the ``$+25$'' is of purely gravitational origin, and
``$-\ns$'' represents the matter portion. In this manner, the bare action of the pure gravity sector has a reminiscence
of matter by means of the number parameter $\ns$. On the other hand, $\ns$ clearly enters the bare action of the
matter sector, too. Considering gravity and matter in combination in the functional integral, there is a cancellation
of terms involving $\ns$.

\noindent
\textbf{(2)}
Due to this compensation of matter effects, and since the gravitational ``$+25$'' neutralizes the ``$-26$'' from the
ghosts and the ``$+1$'' from the measure of the Liouville field, the NGFP theory amounts to a CFT with vanishing total
central charge.

\noindent
\textbf{(3)}
Another consequence of the compensation mechanism can be observed for the gravitational dressing of operators from the
matter sector: There is a complete decoupling of the Liouville field from matter operators of the type
\eqref{eq:AreaOp} and \eqref{eq:OpN}. As a result, this leads to a full quenching of the KPZ-scaling, in distinction
from what one might have expected a priori. Remarkably enough, this quenching is precisely what is found in Monte Carlo
simulations of analogous systems in the framework of causal dynamical triangulation.

\noindent
\textbf{(4)}
Although these results are surprising and encouraging, they should be handled with care. Our arguments relied upon
numerous approximations at different stages of their derivation. (i) We employed the single-metric
Einstein--Hilbert truncation in $d>2$ for the gravitational EAA. (ii) For the bare action in $d>2$ we made an
Einstein--Hilbert ansatz, too, which is probably the most precarious approximation. (iii) The bare action was
reconstructed at one-loop level only. (iv) The matter sector is based on the simplest possible truncation ansatz. (v)
The running of the matter and the ghost action was neglected. (vi) In this chapter we neglected bare cosmological
constant terms, (vii) topological terms and (viii) zero mode contributions. (ix) The number $\ns$ enters some of the
neglected terms other than $I[g]$, which might spoil the perfect cancellation.

%----------------------------------------------------------------------------------------------------------------------
\chapter{The bare action in Liouville theory}
\label{chap:BareLiouville}
%----------------------------------------------------------------------------------------------------------------------

\begin{summary}
The results of Chapter \ref{chap:Bare}, in particular the reconstruction formula, are applied to Liouville theory.
That is, we aim at reconstructing the bare action for a theory whose effective average action is of the
Liouville type, $\int\td^2 x\sg\,(a\mku D_\mu\phi\mku D^\mu\phi\,+b\mku R\phi+c\,\e^{2\phi})$. This chapter
basically contains a collection of attempts, including setbacks, rather than a presentation of \emph{the} solution:
We test several ans\"{a}tze for the bare Liouville action all of which come with their characteristic advantages and
drawbacks, listed in Table \ref{tab:Liouville} in Section \ref{sec:BareLiouCon}. Our analysis
includes a numerical computation of bare couplings and an analytical argument to demonstrate their convergence in
one case. Finally, we specify the Ward identity corresponding to a Weyl transformation applied to the bare action
and evaluate its pure cutoff contributions for an optimized regulator.

\noindent
\textbf{What is new?} The application of the reconstruction formula to a bare action of
pure Liouville type (Sec.\ \ref{sec:BareLiouLiou}), to a bare potential consisting of a power series (Sec.\
\ref{sec:BareLiouPower}) and a series of exponentials (Sec.\ \ref{sec:BareLiouExpSeries}), and to an arbitrary
potential (Sec.\ \ref{sec:BareLiouGeneral}); the form of the Ward identity (Sec.\ \ref{sec:WeylWardIdentity}).
\end{summary}
\vspace{1em}
%----------------------------------------------------------------------------------------------------------------------

Its close connection to 2D quantum gravity and noncritical string theory as discussed in Chapters \ref{chap:Intro},
\ref{chap:EHLimit}, \ref{chap:NGFPCFT} and \ref{chap:FullReconstruction} renders Liouville field theory an interesting
topic to study. In what follows, we would like to shed some light on the relation between the effective average action
and the bare action in this theory. We have seen in Chapter \ref{chap:EHLimit} how an EAA of the Liouville type,
$\GkLiou$, emerges from an EAA in the Einstein--Hilbert (EH) truncation in $d>2$ dimensions, $\Gk^\text{EH}$, when the
limit $d\to 2$ is taken. This leaves us with the somewhat unusual situation of having a Liouville action on the
``already quantized'' \emph{EAA side}. By contrast, in the existing studies on Liouville theory (see for instance
Refs.\ \cite{RW97,Nakayama2004,Seiberg1990,GM93}) it is the \emph{bare action} that has the Liouville form and that
is yet to be quantized, while the corresponding effective (average) action is searched for.

The question we will focus on in this chapter is how the bare action must be chosen in order to be compatible (in the
sense of the reconstruction discussed in Chapter \ref{chap:Bare}, setting $k=\UV$) with an EAA of the Liouville type:
\begin{equation}
 \Gamma_\UV^\text{L} \xrightarrow{\text{reconstr.}} \SB ?
\end{equation}
In Ref.\ \cite{RW97} the inverse problem has been investigated, where the authors start with a Liouville action on the
bare side, $\SB^\text{L}$, make an ansatz for the EAA and determine its couplings at the UV scale $\UV$ by means of
Ward identities: $\SB^\text{L} \xrightarrow{\text{WI}} \Gamma_\UV$. An important result of this analysis is that the
EAA cannot have the standard Liouville form the bare action has, and thus $\Gamma_\UV\neq\SB^\text{L}$. Therefore, with
regard to our current setting that starts with a Liouville-type EAA, we expect that the bare theory cannot be given by
a pure Liouville action.

Before addressing the reconstruction procedure, we would like to point out a subtlety we encounter in our approach.
We know from Chapter \ref{chap:EHLimit} that Einstein--Hilbert actions in $d>2$ give rise to Liouville actions in the
2D limit. As a consequence, there are different possibilities for obtaining a bare action when starting out from an
Einstein--Hilbert-based effective average action. Figure \ref{fig:CommDiag} illustrates the two options we have.
\tikzstyle{int}=[rounded corners=3pt, draw, fill=gray!10, minimum width=4em, minimum height=3em]
\begin{figure}[bp]
\centering
\begin{tikzpicture}[node distance=10em,auto,>=triangle 45]
 \node [int] (GammaEH) at (0,3) {$\Gamma_{k=\UV}^\text{EH}$};
 \node [int] (GammaLiou) at (0,0) {$\Gamma_{k=\UV}^\text{L}$};
 \node [int] (SEH) at (4.5,3) {$\SB^\text{EH}$};
 \node [int] (SLiou) at (4.5,0) {$\SB^\text{L}$};
 \path[->] (GammaEH) edge node {reconstr.} (SEH);
 \path[->] (GammaEH) edge node [left] {$\lim\limits_{d\to 2}$} (GammaLiou);
 \path[->] (SEH) edge node {$\lim\limits_{d\to 2}$} (SLiou);
 \path[->] (GammaLiou) edge node [below] {reconstr.} (SLiou);
 \draw[->,dashed] (0.6,2.2) .. controls (0.6,0.2)  and (2.3,0.4) .. (3.5,0.4);
 \draw[->,dashed] (1,2.6) .. controls (2.2,2.6)  and (3.9,2.8) .. (3.9,0.8);
 \node (a) at (2.9,2) {(a)};
 \node (b) at (1.6,1) {(b)};
\end{tikzpicture}
\caption{Relation between Einstein--Hilbert and Liouville action, on both the effective and the
 bare side (left and right column, respectively), and the two ways to obtain the bare
 action when starting out from an Einstein--Hilbert-type effective average action.}
\label{fig:CommDiag}
\end{figure}
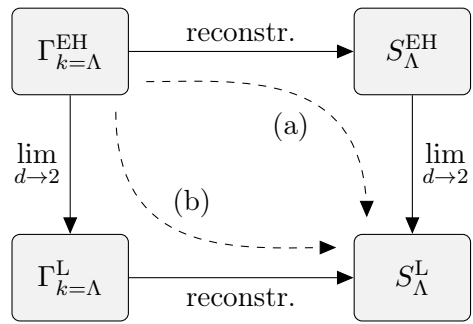
Given the EAA in the Einstein--Hilbert truncation, way (a) means reconstructing the bare action first, and then taking
the limit $d\to 2$ in order to obtain a Liouville-type bare action. Possibility (b) on the other hand, refers to the
way where the limit $d\to 2$ is taken first, yielding a Liouville EAA, and from this new action the bare action is
reconstructed.

A priori, it is not clear whether the diagram commutes, even if there were a way to perform the computations in a
full, i.e.\ untruncated, theory space for the bare action. This can be understood as follows. The reconstruction
in way (a) is based on the \emph{full metric} $g_\mn$ as arguments of the EAA and the bare action, and the underlying
functional integration variable is given by the metric fluctuations. By contrast, the \emph{conformal factor}
$\phi$ is the only argument of the actions at the bottom of way (b), and the corresponding functional integral is over
$\mD\phi$. Therefore, unless the functional measure satisfies additional requirements, say, some sort of generalized
version of uniform convergence in the limit $d\to 2$, the resulting bare action will probably depend on the order of
reconstruction and change of variables.

Once we have to resort to truncations, this effect will certainly become even more distinct. These general
arguments suggest that the bare action obtained in way (b) does not have the standard Liouville form (in agreement with
Ref.\ \cite{RW97}), while the one of way (a) does. Furthermore, way (b) violates the invariance under the Weyl
split-symmetry transformations \eqref{eq:SplitSymmetry} in general, while way (a) is Weyl split-symmetry preserving.
The one-loop results of this chapter will confirm these considerations.

For the sake of completeness, let us extent the picture shown in Figure \ref{fig:CommDiag} in order to clarify the
intermediate steps and relations as well, including the connection to the respective effective actions
$\Gamma\equiv\Gamma_{k=0}$. The result is contained in Figure \ref{fig:CommDiagFull}, where we show in detail which
relations have already been studied in the literature or in this thesis. As indicated by the dashed lines, a direct
evaluation of path integrals is a formidable task. Although it is possible to compute certain correlation functions
within a simple setting in Liouville theory \cite{GL91}, a general recipe for the calculations seems to be beyond
reach. In this chapter we take a first small step towards bridging one gap by investigating the reconstruction problem
at the bottom of Figure \ref{fig:CommDiagFull}.

\tikzstyle{int}=[rounded corners=3pt, draw, fill=gray!10, minimum width=4em, minimum height=3em]
\begin{figure}[tp]
\centering
\begin{tikzpicture}[auto,>=triangle 45]
  \node [int] (a) at (0,6) {$\Gamma_{k=\UV}^\text{EH}$};
  \node [int] (b) at (0,1.7) {$\Gamma_{k=\UV}^\text{ind}$};
  \node [int] (g) at (0,-2.6) {$\Gamma_{k=\UV}^\text{L}$};
  \node [int] (c) at (8,6) {$\SB^\text{EH}$};
  \node [int] (d) at (8,1.7) {$\SB^\text{ind}$};
  \node [int] (h) at (8,-2.6) {$\SB^\text{L}$};
  \node [int] (e) at (4,4.3) {$\Gamma_{k=0}^\text{EH}$};
  \node [int] (f) at (4,0) {$\Gamma_{k=0}^\text{ind}$};
  \node [int] (i) at (4,-4.3) {$\Gamma_{k=0}^\text{L}$};
  \path[->,very thick] (a) edge node [above left,xshift=-10pt] {\textbf{reconstr.}} (c);
  \path[->] (b) edge node [above left,xshift=-10pt] {reconstr.} (d);
  \path[->] (g) edge node [above left,xshift=-10pt] {reconstr.} (h);
  \path[->,very thick] (a) edge node [above left] {\bm{$\lim\limits_{d\to 2}$}} (b);
  \path[->,very thick] (c) edge node [above right] {\bm{$\lim\limits_{d\to 2}$}} (d);
  \path[->,very thick] (b) edge node [above left] {\bm{$\e^{2\phi}\hg$}} (g);
  \path[->,very thick] (d) edge node [above right] {\bm{$\e^{2\phi}\hg$}} (h);
  \path[->,very thick] (a) edge node [below left] {\textbf{FRGE}\!\!} (e);
  \path[->,very thick] (b) edge node [below left] {\textbf{FRGE}\!\!} (f);
  \path[->,very thick] (g) edge node [below left] {\textbf{FRGE}\!\!} (i);
  \path[->, dashed] (c) edge (e);
  \node [below right] at (5.5,5.05) {functional};
  \node [below right] at (5.5,4.55) {integral};
  \path[->, dashed] (d) edge (f);
  \node [below right] at (5.5,0.75) {functional};
  \node [below right] at (5.5,0.25) {integral};
  \path[->, dashed] (h) edge (i);
  \node [below right] at (5.5,-3.55) {functional};
  \node [below right] at (5.5,-4.05) {integral};
  \draw [very thick] (e) -- node [below right] {\bm{$\lim\limits_{d\to 2}$}} (4,1.9);
  \draw [very thick] (4,1.5) arc (-90:90:0.2);
  \draw [->,very thick] (4,1.5) -- (f);
  \draw [very thick] (f) -- node [below right] {\bm{$\e^{2\phi}\hg$}} (4,-2.4);
  \draw [very thick] (4,-2.8) arc (-90:90:0.2);
  \draw [->,very thick] (4,-2.8) -- (i);
  \draw [dotted] (-0.8,3) -- (10,3);
  \draw [dotted] (-0.8,-1.3) -- (10,-1.3);
  \draw [decorate,decoration={brace,amplitude=10pt,mirror,raise=4pt},yshift=0pt]
    (10,3.1) -- (10,6.6) node [black,right,midway,xshift=20pt] {\parbox[t]{4.5em}{Einstein--\\ Hilbert}};
  \draw [decorate,decoration={brace,amplitude=10pt,mirror,raise=4pt},yshift=0pt]
    (10,-1.2) -- (10,2.9) node [black,right,midway,xshift=20pt] {\parbox[t]{4.5em}{Induced\\ gravity}};
  \draw [decorate,decoration={brace,amplitude=10pt,mirror,raise=4pt},yshift=0pt]
    (10,-4.8) -- (10,-1.4) node [black,right,midway,xshift=20pt] {Liouville};
\end{tikzpicture}
\caption{Relation between Einstein--Hilbert, induced gravity and Liouville action, concerning the EAA for
 $k=\UV\to\infty$ (left vertical arrows), the effective action for $k=0$ (column in the middle) and the bare action
 (right vertical arrows). Thick arrows and bold-faced labels refer to relations that are either known in the literature
 or have been worked out in this thesis. (Reconstructing $\SB^\text{EH}$ from $\Gamma_{\UV}^\text{EH}$: Ref.\
 \cite{MR09} \& Chap.\ \ref{chap:Bare}; the 2D limit of EH-type actions: Chap.\ \ref{chap:EHLimit}; FRGE for
 $\Gk^\text{EH}$: Ref.\ \cite{Reuter1998} \& Chap.\ \ref{chap:ParamDep}; FRGE for $\Gk^\text{ind}$: Ref.\ \cite{CD15};
 FRGE for $\Gk^\text{L}$: Ref.\ \cite{RW97}; getting Liouville actions from induced gravity actions by inserting
 $g_\mn=\e^{2\phi}\hg_\mn\mku$: known transformation rules can be used, see e.g.\ App.\ \ref{app:Weyl}.) This
 chapter is dedicated to the horizontal arrow at the bottom, the reconstruction problem in Liouville theory.
}
\label{fig:CommDiagFull}
\end{figure}
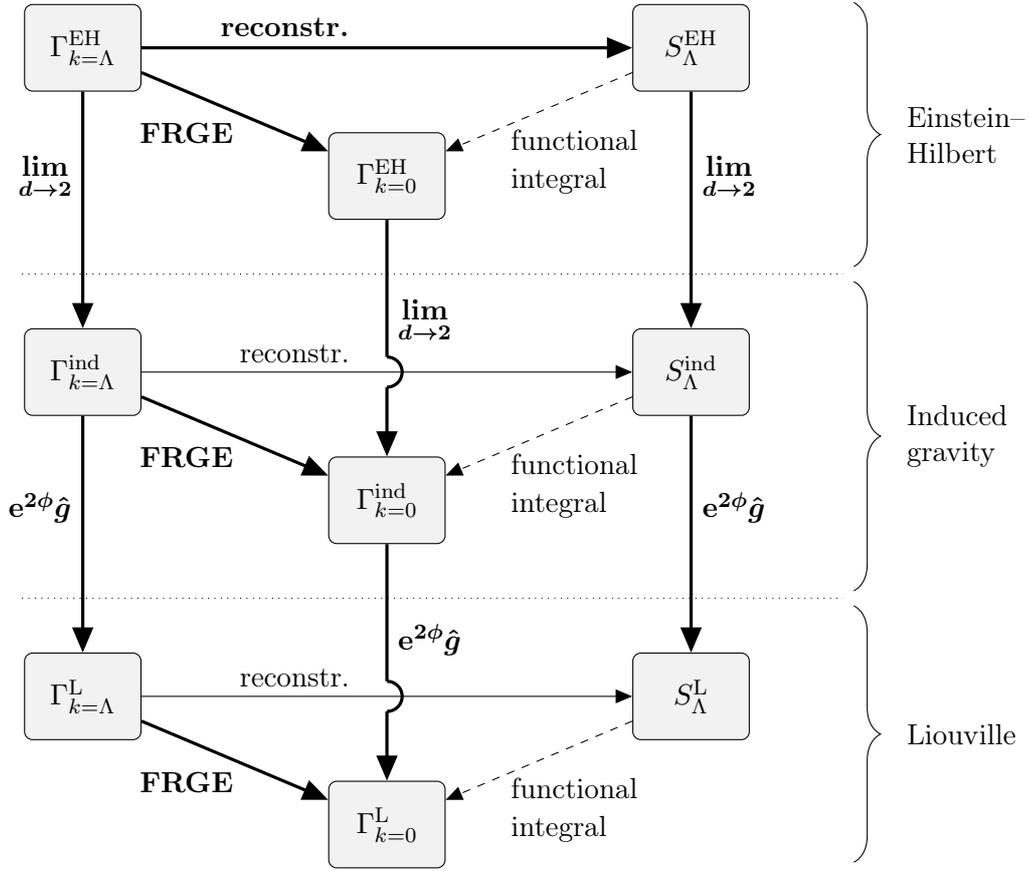

Our starting point is the Liouville EAA, $\Gamma_\UV^\text{L}\mku$, which is obtained by taking the 2D limit of the
Einstein--Hilbert EAA at the NGFP as described in Chapter \ref{chap:EHLimit}:
\begin{equation}
 \Gamma_\UV[\phi]\equiv\Gamma_\UV^\text{L}[\phi]=-\frac{b}{16\pi}\int\td^2 x\shg\,
 \left[ \phi\big(-\hB\big)\phi+\hR\mku\phi+\mu\mku\UV^2\,\e^{2\phi}\right],
\label{eq:GammaUVLiou}
\end{equation}
where $b$ and $\mu$ are determined by the fixed point values of the Newton constant and the cosmological constant in
$d=2+\ve$ dimensions:
\begin{equation}
 b=\lim_{\ve\to 0}\, \frac{\ve}{g_*}\qquad\text{and}\qquad \mu=-2\mku\lim_{\ve\to 0}\mku\frac{\lambda_*}{\ve}.
\label{eq:EAAbandmu}
\end{equation}
The numerical values of $b$ and $\mu$ depend on the underlying metric parametrization, see Chapter \ref{chap:ParamDep}.
For the linear parametrization we found the universal result $b=\frac{38}{3}$ and the cutoff dependent value
$\mu=\frac{3}{19}\mku \Phi_1^1(0)$, which amounts to $\mu=\frac{3}{19}$ for the optimized cutoff. For the exponential
parametrization, on the other hand, both $b$ and $\mu$ depend on the chosen regulator, where the optimized cutoff
leads to $b\approx\frac{50.45}{3}$ and $\mu\approx\frac{3}{20.58}$. Note that the common prefactor in
\eqref{eq:GammaUVLiou} is negative, that is, both the kinetic term and the potential involving $\mu>0$ have the
``wrong'' sign, irrespective of the parametrization. This means that the potential term must be taken into account in
addition to the kinetic term when discussing the conformal factor instability along the lines of Section
\ref{sec:UnitaryCFT}.

The analysis in the subsequent sections yields the same qualitative results for the two parametrizations; only in
Section \ref{sec:BareLiouExpSeries} a more precise distinction becomes necessary.
We will make several ans\"{a}tze for the bare action now and determine its bare couplings by inserting it together with
the EAA \eqref{eq:GammaUVLiou} into the reconstruction formula \eqref{eq:OneLoopFullMain}, i.e.\ into
$\Gamma_\UV = \SB + \frac{1}{2}\,\Tr_\UV \ln\big[ M^{-2}\big(\SB^{(2)}+\RL\big) \big]$.

%----------------------------------------------------------------------------------------------------------------------
\section{Liouville ansatz for the bare action}
\label{sec:BareLiouLiou}
%----------------------------------------------------------------------------------------------------------------------

To begin with, we consider an ansatz for the bare action which is purely of the Liouville type, but with modified
coefficients:
\begin{equation}
 \SB[\phi] = \frac{1}{2}\int\td^2 x\shg\left[\cZ\mku\phi\big(-\hB\big)\phi+ \cx\,\hR\mku\phi
 +\cgamma\mku\UV^2\,\e^{2\phi}\right],
\label{eq:BareLiouFirstAnsatz}
\end{equation}
where couplings with the inverse hat ($\,\check{{}}\,$) refer to bare couplings again, and, as above, we do not list
the reference metric $\hg_\mn$ as an argument explicitly. For the cutoff $\RL$ we chose an optimized regulator function
with the wave function renormalization included:
\begin{equation}
 \RL = \cZ\mku\big(\UV^2+\hB\big)\mku\theta\big(\UV^2+\hB\big).
\label{eq:BareOptCut}
\end{equation}
Since we have $\Tr_\UV\big[(\mku\cdot\mku)\big] \equiv\Tr\big[(\mku\cdot\mku)\mku \theta(\UV^2+\Box)\big]$, the
$\theta$-function in \eqref{eq:BareOptCut} evaluates to $1$ whenever $\RL$ appears inside a regularized trace.

The second derivative of the bare action \eqref{eq:BareLiouFirstAnsatz} is given by
\begin{equation}
 \SB^{(2)} = -\cZ\mku\hB + 2\mku\cgamma\mku\UV^2\,\e^{2\phi}.
\end{equation}
Thus, the trace term of the reconstruction formula can be written as
\begin{equation}
 \frac{1}{2}\,\Tr_\UV \ln\Big[ M^{-2}\big(\SB^{(2)}+\RL\big) \Big] 
 = \frac{1}{2}\,\Tr \Big[f_\UV(\phi)\mku\theta\big(\UV^2+\hB\big)\Big],
\label{eq:BareLiouTrace}
\end{equation}
with $f_\UV(\phi)\equiv \ln\big[\UV^2 M^{-2}\big(\cZ+2\cgamma\,\e^{2\phi}\big)\big]$. The trace in
\eqref{eq:BareLiouTrace} can be computed as usual by projecting it onto curvature invariants with the help of heat
kernel techniques, as introduced in Appendix \ref{app:Heat}, in particular eq.\ \eqref{eq:Heat3}. Employing the
generalized Mellin transforms \eqref{eq:Mellin} we obtain
\begin{align}
 &\frac{1}{2}\,\Tr_\UV \ln\Big[ M^{-2}\big(\SB^{(2)}+\RL\big) \Big] \nonumber\\
 &= \frac{1}{8\pi}\left\{ Q_1\big[\theta\big(\UV^2-(\cdot)\big)\big]\int\!\shg\,f_\UV(\phi) 
 + \frac{1}{6}\,Q_0\big[\theta\big(\UV^2-(\cdot)\big)\big]\int\!\shg\,\hR\,f_\UV(\phi)+\cdots \right\} \nonumber\\
 &= \frac{1}{8\pi}\left\{ \UV^2\!\int\!\shg\,f_\UV(\phi) 
 + \frac{1}{6}\int\!\shg\,\hR\,f_\UV(\phi)+\cdots \right\}.
\label{eq:BareLiouTrace2}
\end{align}
By the reconstruction formula \eqref{eq:OneLoopFullMain} this expression must agree with
\begin{equation}
\begin{split}
 \Gamma_\UV-\SB = {}&-\frac{b}{16\pi}\int\td^2 x\shg\,
 \left[ \phi\big(-\hB\big)\phi+\hR\mku\phi+\mu\mku\UV^2\,\e^{2\phi}\right]\\
 &- \frac{1}{2}\int\td^2 x\shg\left[\cZ\mku\phi\big(-\hB\big)\phi+ \cx\,\hR\mku\phi
 +\cgamma\mku\UV^2\,\e^{2\phi}\right].
\end{split}
\label{eq:GammaMinusSB}
\end{equation}
The couplings of the bare action can now be determined by equating \eqref{eq:GammaMinusSB} with
\eqref{eq:BareLiouTrace2} and comparing the coefficients of corresponding invariants.

First of all, the coefficients of the $\phi(-\hB)\phi$-terms dictate
\begin{equation}
 \cZ = -\frac{b}{8\pi}\,,
\end{equation}
for the truncation considered. The computation of $\cx$ and $\cgamma$ requires an expansion of the function $f_\UV$.
Interestingly enough, we are forced to consider two different expansions here: In order to determine $\cx$ we must
expand $f_\UV$ in terms of $\phi$, while for $\cgamma$ the expansion parameter is $\e^{2\phi}$ instead. The two cases
read
\begin{align}
 f_\UV(\phi) &= \ln\big[\UV^2 M^{-2}\big(\cZ+2\mku\cgamma\big)\big] + \frac{4\cgamma}{\cZ+2\cgamma}\,\phi
 + \mO\big(\phi^2\big),\\
 f_\UV(\phi) &= \ln\big(\cZ\UV^2 M^{-2}\big)+2\mku\cgamma\mku\cZ^{-1}\,\e^{2\phi}+\mO\big(\e^{4\phi}\big).
\label{eq:fExpansion1}
\end{align}
Then the coefficients of the $\hR\phi$-term give rise to the equation
\begin{equation}
 -b - 8\pi\mku\cx = \frac{4}{3}\,\frac{\cgamma}{\cZ+2\cgamma}\,.
\label{eq:CoeffRel1}
\end{equation}
In a similar manner, the coefficients of the $\e^{2\phi}$-terms have to satisfy
\begin{equation}
 -b\mku\mu- 8\pi\mku\cgamma = 4\mku\cgamma\mku\cZ^{-1}.
\label{eq:CoeffRel2}
\end{equation}
Note that the $M$-dependence has dropped out for these coefficients.
Equations \eqref{eq:CoeffRel1} and \eqref{eq:CoeffRel2} can easily be solved for $\cx$ and $\cgamma$. Let us express
the solutions in terms of the redefined bare couplings
\begin{equation}
 \cb \equiv - 8 \pi\,\cx,\quad\text{and}\quad \cm \equiv -\frac{8\pi\mku\cgamma}{\cb}\,,
\end{equation}
by analogy with $b$ and $\mu$ of the EAA. We obtain
\begin{equation}[b]
 \cb \approx \frac{38.63}{3}\,,\quad\text{and}\quad \cm \approx 0.227
\end{equation}
for the linear metric parametrization, and
\begin{equation}[b]
 \cb \approx \frac{51}{3}\,,\quad\text{and}\quad \cm \approx 0.189
\end{equation}
for the exponential parametrization. These values are strikingly close to their counterparts of the EAA,
$b=\frac{38}{3}$, $\mu\approx 0.158$, and $b\approx\frac{50.45}{3}$, $\mu\approx 0.146$ for the linear and the
exponential parametrization, respectively. Hence, \emph{the one-loop correction in the reconstruction formula has a
rather small effect} on the couplings considered in our setting.

There is a certain inconsistency inherent in the above equations, though. It traces back to eq.\
\eqref{eq:fExpansion1}, an expansion in terms of $\e^{2\phi}$ around $\e^{2\phi}=0$, i.e.\ around $\phi=-\infty$.
Only with that expansion we managed to project the trace onto a term proportional to $\e^{2\phi}$. Taken by itself,
this does not pose a problem. However, the computation should be consistent with an expansion in terms $\phi$ and a
subsequent resummation to get back the $\e^{2\phi}$-term. As we will argue now, this cannot be attained within the
underlying truncation.

From eq.\ \eqref{eq:GammaMinusSB} we read off the $\e^{2\phi}$-terms under the integral, adding up to
\begin{equation}
 -\left(\frac{b\mu}{16\pi}+\frac{\cgamma}{2}\right)\UV^2\,\Big\{ 1+2\phi
 +2\mku\phi^2+\cdots\Big\}.
\label{eq:ephiterms}
\end{equation}
This is to be compared with all terms in eq.\ \eqref{eq:BareLiouTrace2} of the type $\int\!\shg\,\phi^n$ without any
contribution from the curvature. For that purpose we expand $f_\UV$ in terms of $\phi$. We find that
\eqref{eq:ephiterms} must agree with
\begin{equation}
 \frac{\UV^2}{8\pi}\left\{\ln\Big[\UV^2 M^{-2}\big(\cZ+2\cgamma\big)\Big] + \frac{4\mku\cgamma}{\cZ+2\cgamma}\,\phi
 +\frac{4\mku\cgamma\mku\cZ}{(\cZ+2\cgamma)^2}\,\phi^2+\cdots \right\}.
\label{eq:ephiterms2}
\end{equation}
The crucial point is that there is no possibility to achieve \eqref{eq:ephiterms}${}={}$\eqref{eq:ephiterms2} for each
expansion term. In fact, the linear term in \eqref{eq:ephiterms2} enters $\e^{2\phi}$ only in part, while the remaining
part might be thought of to be distributed among $\e^{4\phi}$, $\e^{6\phi}$, etc. The same holds true for the quadratic
and all further terms. But since we have truncated the bare action theory space such that $\e^{2\phi}$ is the only
invariant of that type, we do not know which amount of each term in \eqref{eq:ephiterms2} must be split off as a
contribution to $\e^{2\phi}$. Thus, eqs.\ \eqref{eq:ephiterms} and \eqref{eq:ephiterms2} cannot be checked for
consistency this way. This consideration rather suggests taking into account \emph{a more complete set of basis
invariants}. Consequently, we study a series of invariants of the type $\phi^n$ in Section \ref{sec:BareLiouPower} and
invariants of the type $\e^{2n\phi}$ in Section \ref{sec:BareLiouExpSeries}.

As already mentioned in the introduction of this chapter, we expected some kind of inconsistency for the chosen
truncation in advance: Our ansatz was such that both EAA and bare action were of the Liouville type. This, however, is
ruled out by the Ward identities with respect to Weyl transformations \cite{RW97} that predict different forms of the
two actions. In combination with the above arguments this indicates that a different and more complete truncation for
the bare action has to be considered.

%----------------------------------------------------------------------------------------------------------------------
\section{Power series ansatz for the bare potential}
\label{sec:BareLiouPower}
%----------------------------------------------------------------------------------------------------------------------

Motivated by the previous arguments we start with a more general ansatz for the bare action now: We write the bare
potential as a power series,
\begin{equation}
 \SB[\phi] = \frac{1}{2}\int\td^2 x\shg\left[\cZ\mku\phi\big(-\hB\big)\phi+ \cx\,\hR\mku\phi
 + 2\mku\UV^2 \sum\limits_{n=0}^{\Nmax}\ca_n\,\phi^n\right],
\label{eq:BareLiouSecondAnsatz}
\end{equation}
where the number of terms in the series is given by $\Nmax+1$. We refer to $\Nmax$ as \emph{truncation parameter} as
it gives the highest power of $\phi$ in our truncation. The ultimate goal would be to consider the limit
$\Nmax\to\infty$. Due to restricted computational capacity and the lack of a suitable analytical mechanism, however, we
clearly cannot determine infinitely many bare couplings but have to resort to a finite truncation parameter $\Nmax$.
Nonetheless, we can study to what extent the results change when $\Nmax$ is increased.

The analysis is conducted as in the previous section. We insert the EAA \eqref{eq:GammaUVLiou}, the bare action
\eqref{eq:BareLiouSecondAnsatz} and its second derivative,
\begin{equation}
 \SB^{(2)} = -\cZ\mku\hB + \UV^2 \sum\limits_{n=2}^{\Nmax} n(n-1)\ca_n\,\phi^{n-2},
\end{equation}
into the reconstruction formula \eqref{eq:OneLoopFullMain}. The trace is expanded as above, the only difference
consisting in the choice of basis invariants where, as compared to Section \ref{sec:BareLiouLiou}, $\e^{2\phi}$ is
replaced by the set $\big\{\phi^0,\phi^1,\dotsc,\phi^{\Nmax}\big\}$:
\begin{equation}
\begin{split}
 \frac{1}{2}\,\Tr_\UV \ln\Big[ M^{-2}\big(\SB^{(2)}+\RL\big) \Big]
 = \frac{1}{8\pi}\int\td^2 x\shg\,\bigg\{\textstyle \hR\,\frac{\ca_3}{\cZ+2\mku\ca_2}\,\phi+\cdots\bigg\} \\
 +\frac{\UV^2}{8\pi}\int\td^2 x\shg\,\bigg\{ \ln\left[\UV^2 M^{-2}\big(\cZ+2\mku\ca_2\big)\right] \\
 +\textstyle\frac{6\mku\ca_3}{\cZ+2\mku\ca_2}\,\phi + \left[\frac{12\mku\ca_4}{\cZ+2\mku\ca_2}
 -18\left(\frac{\ca_3}{\cZ+2\mku\ca_2}\right)^{\!\!\mku 2}\,\right]\phi^2 + \cdots \bigg\}.
\end{split}
\label{eq:BareLiouTrace3}
\end{equation}
Reading off the coefficients in \eqref{eq:OneLoopFullMain} using \eqref{eq:BareLiouTrace3} yields a system of
equations, the first few of which are given by
\begin{equation}
\begin{split}
 -b &= 8\pi\mku\cZ,\qquad -b=8\pi\mku\cx + \frac{2\mku\ca_3}{\cZ+2\mku\ca_2}\,, \\
 -b\mu &= 16\pi\mku\ca_0 + 2\mku\ln\left[\UV^2 M^{-2}\big(\cZ+2\mku\ca_2\big)\right], \\
 -b\mu &= 8 \pi\mku\ca_1 + \frac{6\mku\ca_3}{\cZ+2\mku\ca_2} , \\
 -b\mu &= 8 \pi\mku\ca_2 - 18\left(\frac{\ca_3}{\cZ+2\mku\ca_2}\right)^{\!\!\mku 2} 
 + \frac{12\mku\ca_4}{\cZ+2\mku\ca_2}\,, \quad\text{etc.}
\end{split}
\label{eq:BarePowerSys}
\end{equation}
We find that the determining equation for a coupling $\ca_n$ is of the general form
$-b\mu = (\text{some number})\cdot\ca_n +\text{(some function of }\ca_2,\ca_3,\dotsc,\ca_{n+2}\text{)}$. In particular,
the calculation of $\ca_n$ requires the knowledge of $\ca_{n+1}$ and $\ca_{n+2}$. Note that due the finite truncation
parameter $\Nmax$ these latter couplings may be zero: $\ca_{\Nmax+1}=0$ and $\ca_{\Nmax+2}=0$. As a consequence, we do
not have to go to higher and higher orders to find a solution since the system of equations is actually closed.

Once we have chosen a truncation parameter we can perform a numerical analysis to solve \eqref{eq:BarePowerSys} for the
couplings. We refrain from presenting their precise numerical values as these are insignificant for the present
discussion. What is important, though, is how the couplings change when the truncation parameter $\Nmax$ is varied.

Let us illustrate the issue by means of a simple Taylor series of some analytic function. All coefficients are fixed by
the derivatives of the function at the expansion point. If we truncate the series after a finite amount of terms, there
will be a finite residual describing the deviation between the series and the function. The more terms are taken into
account, the smaller the residual gets. Furthermore, and this is the crucial point, the coefficients are independent of
the total number of terms in the truncated series.

With regard to this Taylor series example, we might hope that bare couplings in \eqref{eq:BareLiouSecondAnsatz} do not
depend on the truncation parameter $\Nmax$. This would allow us to justify our bare action ansatz with the finite
series a posteriori. Our second hope is that higher order couplings eventually tend to zero, $\ca_n\to 0$ for
$n\to\infty$ (which would require taking $\Nmax\to\infty$, too). As far as our numerical computation is concerned, both
points seem not to come true.

In Figure \ref{fig:BarePower} we demonstrate what happens. The plots show the dependence of $\ca_0,\dotsc,\ca_4$ and
$\cx$ on the truncation parameter $\Nmax$, where we use those values for $b$ and $\mu$ in the EAA that are based on the
linear metric parametrization --- similar results are obtained with the exponential parametrization. We observe heavy
fluctuations of all couplings when $\Nmax$ is varied. Remarkably enough, this holds true for $\cx$, too, even if that
one is not a coefficient of the power series. Moreover, it is surprising that the lower order couplings still depend
strongly on $\Nmax$ even if $\Nmax$ is already large.
\begin{figure}[tp]
 \begin{minipage}{0.3\columnwidth}
 \includegraphics[width=\columnwidth]{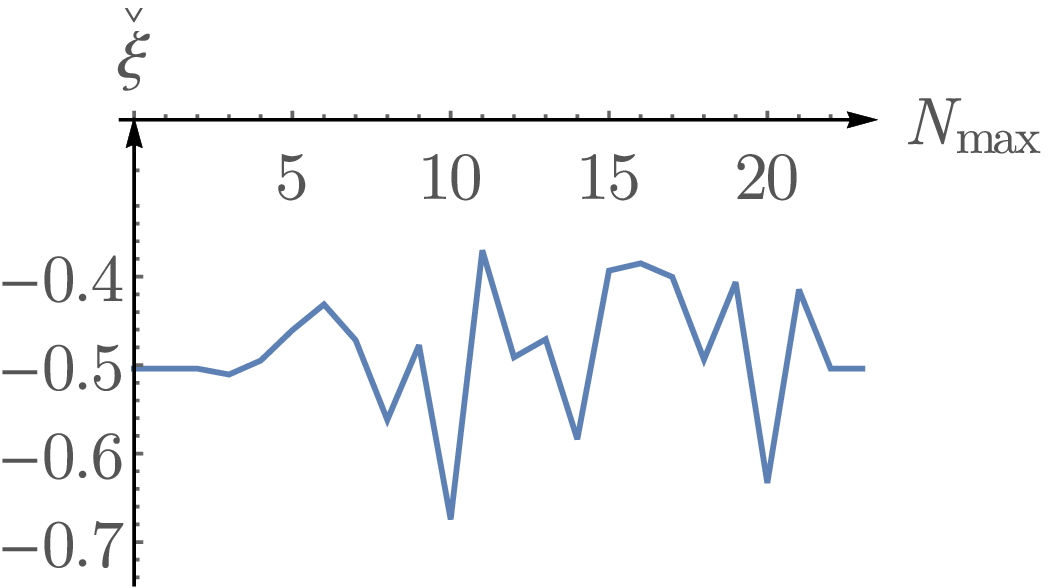}
 \end{minipage}
 \hfill
 \begin{minipage}{0.3\columnwidth}
 \includegraphics[width=\columnwidth]{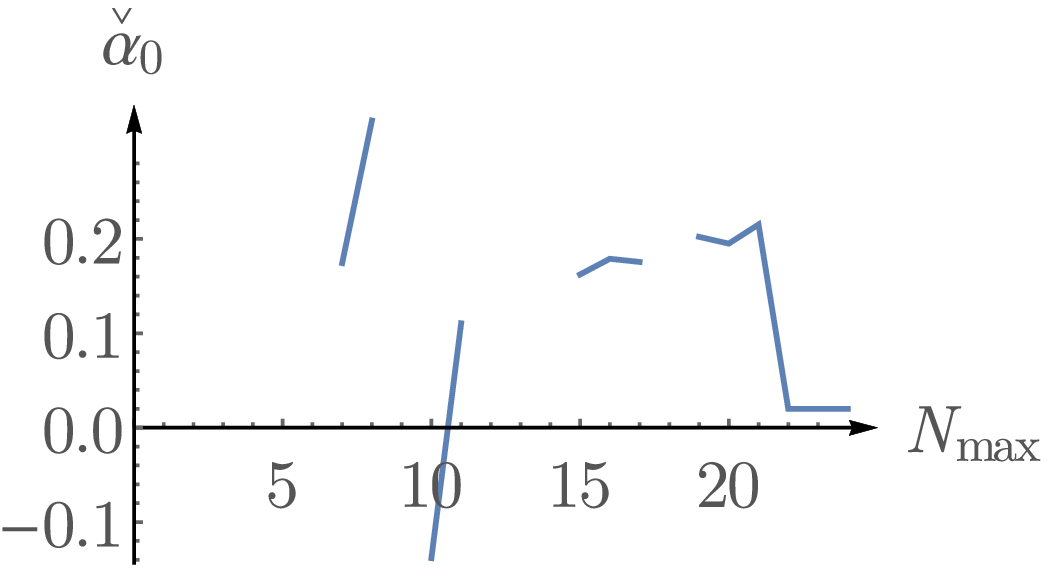}
 \end{minipage}
 \hfill
 \begin{minipage}{0.3\columnwidth}
 \includegraphics[width=\columnwidth]{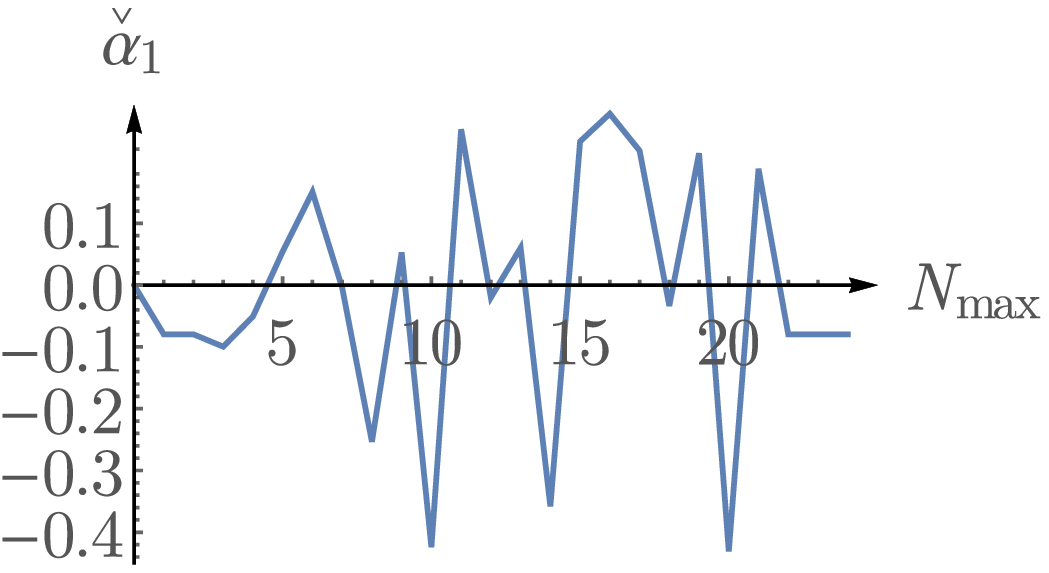}
 \end{minipage}
 
 \vspace{1em}
 \begin{minipage}{0.3\columnwidth}
 \includegraphics[width=\columnwidth]{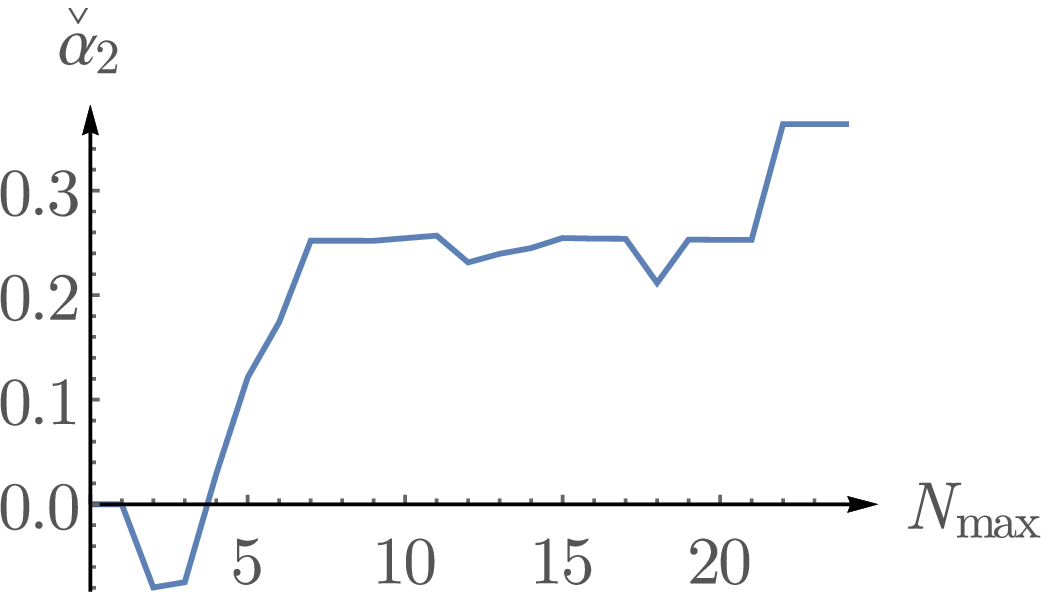}
 \end{minipage}
 \hfill
 \begin{minipage}{0.3\columnwidth}
 \includegraphics[width=\columnwidth]{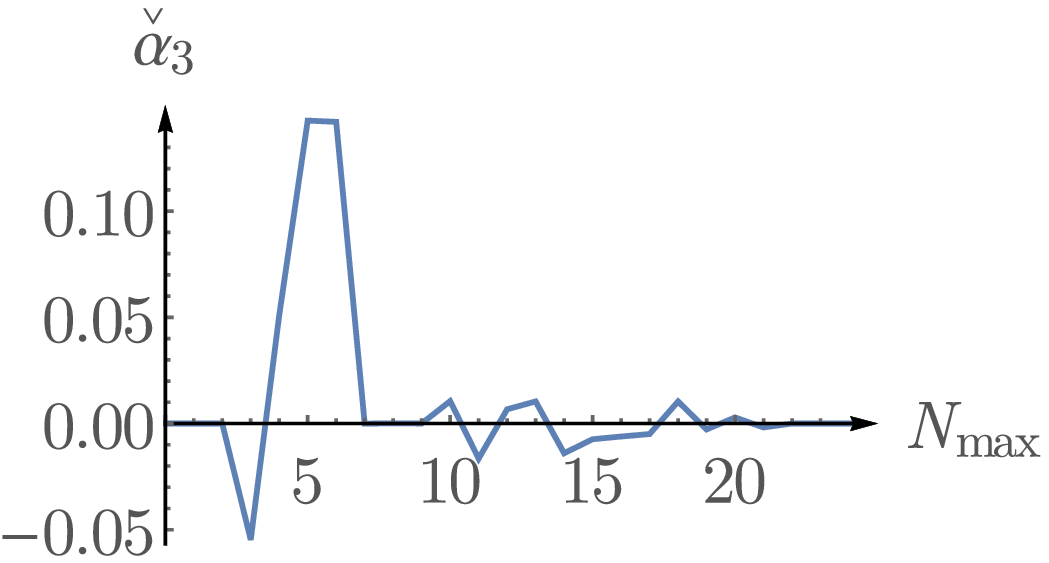}
 \end{minipage}
 \hfill
 \begin{minipage}{0.3\columnwidth}
 \includegraphics[width=\columnwidth]{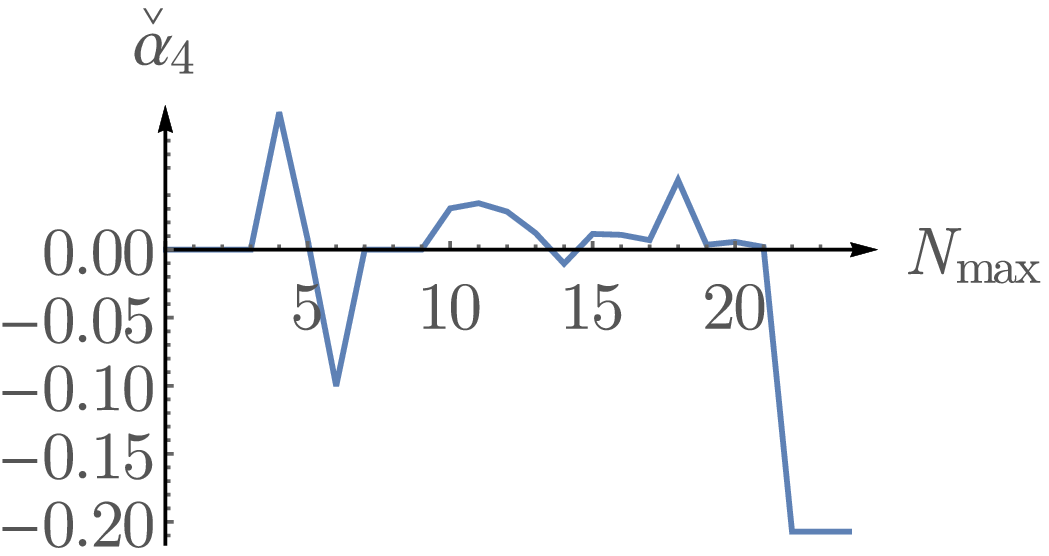}
 \end{minipage}
\vspace{0.5em}
\caption{The coupling $\cx$ and the first $5$ series coefficients of the bare potential, $\ca_0,\dotsc,\ca_4$,
 dependent on the truncation parameter $\Nmax$, i.e.\ dependent on the total number of terms in the power series
 minus one, cf.\ eq.\ \eqref{eq:BareLiouSecondAnsatz}. We observe that all couplings fluctuate heavily when $\Nmax$
 is varied. The coupling $\ca_0$ may even become complex for certain values of $\Nmax$, as indicated by the gaps in
 the corresponding plot. (Note that $\ca_0$ depends also on the measure parameter $M$, see \eqref{eq:BarePowerSys}.
 Here we chose $M=\UV$.) There is no indication of convergence of the couplings for increasing $\Nmax$.}
\label{fig:BarePower}
\end{figure}
The analysis goes up to the value $\Nmax=24$ beyond which the numerical results get unreliable. Clearly, the graphs
of all $\ca_n$ with $n\ge 1$ start at the origin (where $\Nmax=0$) since $\ca_n=0$ for $n>\Nmax$. For instance, in the
diagram for $\ca_4$ in Figure \ref{fig:BarePower} we see that $\ca_4$ can get nonzero only when $\Nmax\ge 4$.
Although Figure \ref{fig:BarePower} shows only six bare couplings, we have done the calculation for
$\ca_0,\dotsc,\ca_{24}$, and all resulting pictures show the same characteristic fluctuations. Here, we would like to
emphasize that higher order coefficients seem not to tend to zero eventually: Averaging over the absolute values of
the couplings $\ca_n$ we do not observe any significant decrease for increasing $n$. Due to their connection to the
power of $\phi$ in the series, these higher order couplings become more and more important. Therefore, both of our two
hopes vented above are not satisfied.

In summary, we have seen that a finite \emph{power series ansatz} for the bare potential appears to be
\emph{inappropriate for reconstructing the bare action} on the basis of \eqref{eq:OneLoopFullMain}. The resulting bare
couplings depend strongly on the number of terms in the series. We do not observe any convergence: neither do couplings
of some fixed index approach a stable value in the large $\Nmax$ limit, nor do higher order couplings $\ca_n$ become
small in the large $n$ limit. An equally heavy $\Nmax$-dependence is found for the form and the stability (boundedness)
of the total potential.

%----------------------------------------------------------------------------------------------------------------------
\section{The bare potential as a series of exponentials}
\label{sec:BareLiouExpSeries}
%----------------------------------------------------------------------------------------------------------------------

Motivated by our results of Section \ref{sec:BareLiouLiou} we would like to make an ansatz for the bare action which
consists of a Liouville action plus correction terms. The latter are organized as a \emph{series of exponentials} of
the type $\e^{2n\phi}$. Hence, the bare action within this truncation reads
\begin{equation}
 \SB[\phi] = \frac{1}{2}\int\td^2 x\shg\left[\cZ\mku\phi\big(-\hB\big)\phi+ \cx\,\hR\mku\phi
 + \UV^2 \sum\limits_{n=1}^{\Nmax}\cgamma_n\,\e^{2n\phi}\right].
\label{eq:BareLiouThirdAnsatz}
\end{equation}

This ansatz for the bare potential closely resembles a Fourier series. (For imaginary $\phi$ it \emph{is} a Fourier
series.) Just like $\big\{\e^{2 i\mku n x}\big\}$ is a basis for the space of square-integrable functions on
$[-\pi/2,\pi/2]$, we assume here that the terms $\int\td^2 x\shg\,\e^{2\mku n\mku \phi(x)}$ are linearly independent
and part of a basis of theory space. With regard to the inconsistencies found in Section \ref{sec:BareLiouLiou},
these terms certainly constitute a more complete set of invariants and we expect that some of the above issues
might get resolved.

Besides, we observe a certain similarity to the truncation ansatz for the sine-Gordon model considered in Refs.\
\cite{NNPS09,NNST09} where the potential term in the action is given by $V(\phi)=\sum_n u_n \cos(n\phi)$. This is a
further motivation to study such trun\-ca\-tions that comprise a series of exponentials, justifying our choice in
\eqref{eq:BareLiouThirdAnsatz}.

In order to determine the bare couplings in \eqref{eq:BareLiouThirdAnsatz} we proceed precisely as in the previous
sections. First, we compute the Hessian,
\begin{equation}
 \SB^{(2)} = -\cZ\mku\hB + 2\mku \UV^2 \sum\limits_{n=1}^{\Nmax} n^2\cgamma_n\,\e^{2n\phi}\,,
\end{equation}
which is inserted into the reconstruction formula \eqref{eq:OneLoopFullMain}. Second, we compute the trace analogously
to eq.\ \eqref{eq:BareLiouTrace2}. We obtain
\begin{equation}
 \frac{1}{2}\,\Tr_\UV \ln\Big[ M^{-2}\big(\SB^{(2)}+\RL\big) \Big] = \frac{1}{8\pi}\left\{ \UV^2\!\int\!\shg\,f_\UV(\phi) 
 + \frac{1}{6}\int\!\shg\,\hR\,f_\UV(\phi)+\cdots \right\},
\label{eq:BareLiouTracef}
\end{equation}
with
\begin{equation}
 f_\UV(\phi) = \ln\big(\UV^2 M^{-2}\cZ\big)
 + \ln\left(1 + 2\mku \cZ^{-1} \sum_{n=1}^{\Nmax} n^2\mku\cgamma_n\,\e^{2n\phi}\right).
\end{equation}
Third, we apply two different kinds of expansions to $f_\UV$: In the $\shg\,f_\UV(\phi)$-term in
\eqref{eq:BareLiouTracef} we must expand $f_\UV$ in terms of $\e^{2\phi},\e^{4\phi},$ etc.\ , while for the
$\shg\,\hR\,f_\UV(\phi)$-term it is sufficient to project $f_\UV$ onto its contribution linear in $\phi$.

\noindent
\textbf{(a) Expansion in terms of exponentials.}
Let us introduce the abbreviations
\begin{equation}
 a_n \equiv 2\mku\cZ^{-1}\mku n^2\,\cgamma_n,\quad x\equiv\e^{2\phi}\quad\text{and}\quad N\equiv\Nmax\,.
\end{equation}
Then $f_\UV$ assumes the form $f_\UV = \ln\big(\UV^2 M^{-2}\cZ\big) + \ln\big(1+\sum_{n=1}^N a_n x^n\big)$.
Employing the Taylor series of the logarithm leads to
\begin{equation}
 f_\UV = \ln\big(\UV^2 M^{-2}\cZ\big)
 - \sum_{k=1}^\infty \frac{(-1)^k}{k}\left(\textstyle\sum_{n=1}^N a_n x^n\right)^k .
\label{eq:BareLiouf}
\end{equation}
The $k$-th power of a sum can be calculated by means of the multinomial theorem:
\begin{equation}
 (y_1+\dots+y_N)^k = \sum_{|\alpha|=k} \frac{k!}{\alpha_1!\cdots\alpha_N!}\,y_1^{\alpha_1}\cdots y_N^{\alpha_N},
\end{equation}
where we use the multi-index notation, i.e.\ $\alpha\in\mathds{N}_0^N$. Applying this to \eqref{eq:BareLiouf} and
combining all powers of $x\equiv\e^{2\phi}$ we obtain
\begin{equation}
 f_\UV=\ln\big(\UV^2 M^{-2}\cZ\big)-\sum_{k=1}^\infty\sum_{|\alpha|=k}\frac{(-1)^k(k-1)!}{\alpha_1!\cdots\alpha_N!}\,
 a_1^{\alpha_1}\cdots a_N^{\alpha_N}\, x^{\sum_{n=1}^N n\mku \alpha_n}.
\label{eq:fExpExp1}
\end{equation}

\noindent
\textbf{(b) Expansion in terms of $\bm{\phi}$.} Up to linear order the expansion of $f_\UV$ in terms of $\phi$ reads
\begin{equation}
\begin{split}
 f_\UV=\ln\big(\UV^2 M^{-2}\cZ\big)
 + \ln\left(\textstyle 1 + 2\mku \cZ^{-1} \sum_{n=1}^{\Nmax} n^2\mku\cgamma_n\right)\\
 + \frac{4\mku\cZ^{-1}\sum\nolimits_{n=1}^{\Nmax}n^3\cgamma_n}{1+2\mku\cZ^{-1}
 \sum\nolimits_{n=1}^{\Nmax} n^2\mku\cgamma_n}\,\phi + \mO(\phi^2).
\end{split}
\label{eq:fExpExp2}
\end{equation}

Inserting \eqref{eq:fExpExp1} and \eqref{eq:fExpExp2} into eq.\ \eqref{eq:BareLiouTracef} yields
\begin{equation}
\begin{split}
 \frac{1}{2}\,&\Tr_\UV \ln\Big[ M^{-2}\big(\SB^{(2)}+\RL\big) \Big] \\
 &=\frac{\UV^2}{8\pi}\int\!\shg\, \sum_{k=1}^\infty\sum_{|\alpha|=k}\frac{(-1)^{k-1}(k-1)!}{\alpha_1!\cdots\alpha_N!}\,
 a_1^{\alpha_1}\cdots a_N^{\alpha_N}\, \e^{2\mku\phi\sum_{n=1}^N n\mku \alpha_n} \\
 &\phantom{={}}+\frac{1}{12\pi}\int\!\shg\,\hR\,\frac{\sum\nolimits_{n=1}^{\Nmax}n^3\cgamma_n}{\cZ+2
 \sum\nolimits_{n=1}^{\Nmax} n^2\mku\cgamma_n}\,\phi +\cdots,
\end{split}
\end{equation}
According to eq.\ \eqref{eq:OneLoopFullMain}, this expression must agree with $\Gamma_\UV[\phi]-\SB[\phi]$. As usual,
we can read off the coefficients belonging to the same invariant and set up a system of equations defining the bare
couplings. By suitably rearranging these equations, each coupling $\cgamma_n$ can be expressed in terms of
$\cZ,\cgamma_1,\dotsc,\cgamma_{n-1}$, whereas $\cx$ depends on all other couplings involved in our truncation:
\begin{boxalign}
 \cZ &= -\frac{b}{8\pi}\,,
\label{eq:BareZ}\\
 \cx &= -\frac{b}{8\pi} - \frac{1}{6\pi}\,\frac{\sum\nolimits_{n=1}^{N}n^3\cgamma_n}{\cZ+2
 \sum\nolimits_{n=1}^{N} n^2\mku\cgamma_n}\,,
\label{eq:cxi}\\
 \cgamma_1 &= -\frac{b\mu\cZ}{4+8\pi\cZ}\,,
\label{eq:cgamma1}\\
 \cgamma_n &= \frac{\cZ}{2n^2+4\pi\cZ}\, \sum_{k=2}^n \!\!
 \sum_{{\substack{\,\alpha\in\mathds{N}_0^N\\ |\alpha|=k\\ \sum_i i\alpha_i=n}}} \!\!\!\!
 \frac{(-1)^k(k-1)!}{\alpha_1!\cdots\alpha_N!}\, a_1^{\alpha_1}\cdots a_{n-1}^{\alpha_{n-1}}\quad\text{for }
 2\le n \le N \mku ,
\label{eq:cgamman}\\
 \cgamma_n &= 0 \quad \text{for } n>N,\qquad
 \text{with } N\equiv\Nmax \text{ and } a_n\equiv 2\mku\cZ^{-1}n^2\,\cgamma_n \, .
\end{boxalign}

Before calculating the bare couplings numerically a couple of remarks are in order.
\smallskip

\noindent
\textbf{(1)} The second sum in eq.\ \eqref{eq:cgamman} is over all vectors $\alpha\in\mathds{N}_0^N$ that satisfy the
two constraints $|\alpha|\equiv\sum_i \alpha_i = k$ and $\sum_i i\alpha_i = n$. These constraints reduce the number of
contributing terms considerably. They dictate that $\alpha_i=0$ for $i\ge n$, so instead of $\alpha\in\mathds{N}_0^N$
we could write $\alpha\in\mathds{N}_0^{n-1}$ as well.

As an example for how the constraints restrict the sum, let us
consider the case $n=2=k$: There is only one possible vector $\alpha$ left, namely $\alpha_1=2,\,\alpha_2=0$,
$\alpha_3,\dotsc,\alpha_N=0$. Since the first sum in \eqref{eq:cgamman} requires $k\le n$, the defining equation for
$\cgamma_2$ involves only one term, and we finally obtain $\cgamma_2 = \frac{1}{4+2\pi\cZ}\,\cZ^{-1}\,\cgamma_1^2$.
\medskip

\noindent
\textbf{(2)} As long as $n\le\Nmax$, \emph{the bare couplings} $\cgamma_n$ \emph{are independent of the number}
$\Nmax$. This is a tremendous advantage as compared with the situation in Section \ref{sec:BareLiouPower} where the
resulting bare couplings depended strongly on $\Nmax$, which led to significant fluctuations and an instable behavior.
Here, on the other hand, we find that a coupling $\cgamma_n$ is determined once the lower order couplings
$\cZ,\cgamma_1,\dotsc,\cgamma_{n-1}$ are known, and increasing $\Nmax$ does not have any effect on $\cgamma_n$.
Having calculated a coupling at one point fixes it ``for all times'' (that is, for all $\Nmax$, in particular for
$\Nmax\to\infty$).
\medskip

\noindent
\textbf{(3)} Related to our second remark, we observe that \emph{the bare couplings can be computed iteratively}:
Inserting $Z=-b/(8\pi)$ into eq.\ \eqref{eq:cgamma1} determines $\cgamma_1$, which can be used, in turn, to calculate
$\cgamma_2$, and so forth. Only $\cx$ depends on all other couplings. We might hope, however, that the $\cgamma_n$'s
decrease sufficiently fast such that $\cx$ actually converges. As we will see, this seems indeed to be the case.
\medskip

Clearly, the numerical values of the bare couplings are sensitive to the effective couplings $b$ and $\mu$. According
to the discussion below eq.\ \eqref{eq:EAAbandmu} the latter depend on the underlying metric parametrization. As the
linear and the exponential parametrization lead to different results for the bare potential, we study the two cases
separately.

%----------------------------------------------------------------------------------------------------------------------
\subsection{Results for the linear parametrization}
\label{sec:BareLiouExpSeriesLin}
%----------------------------------------------------------------------------------------------------------------------

In the case of the linear parametrization we insert $b=\frac{38}{3}$ and $\mu=\frac{3}{19}$ into the system
\eqref{eq:BareZ} - \eqref{eq:cgamman} and solve numerically for the bare couplings. The result for the first
48 couplings $\cgamma_n$ is shown in Figure \ref{fig:BareGammaLinParam}.\footnote{The computation time
grows exponentially. It took approximately 10 hours in \Mma\ to calculate $\cgamma_{48}$. During the calculation
of $\cgamma_{49}$, \Mma\ ran into a memory overflow error after about 15 hours. Surely it is possible to find
faster and more reliable algorithms and programming languages, but for our purposes knowing the first 48 couplings is
more than enough.}
\begin{figure}[tp]
 \centering
 \includegraphics[width=0.75\columnwidth]{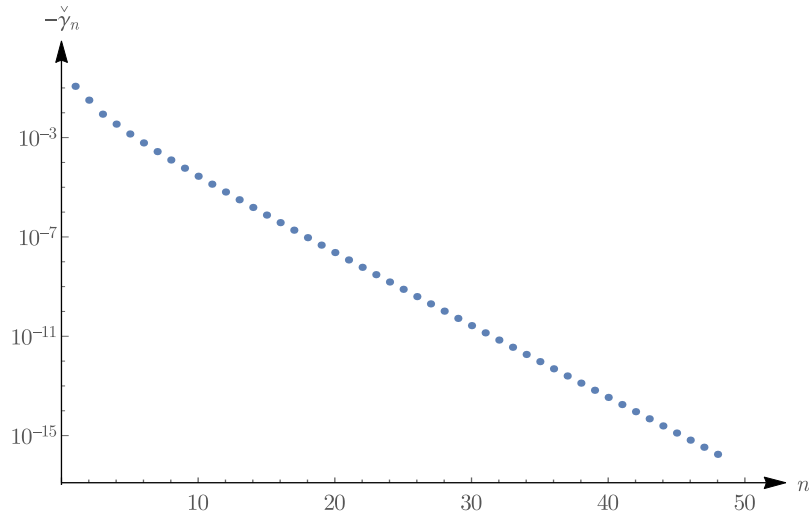}
\caption{Logarithmic plot showing the absolute values of the bare couplings $\cgamma_n$ dependent on their
 index $n$, in the range $n=1,\ldots,48$, based on the linear parametrization. We observe an approximately exponential
 decrease towards larger $n$. All couplings have the same sign.}
\label{fig:BareGammaLinParam}
\end{figure}
It reveals a surprising and very important feature of the couplings: \emph{for increasing $n$ we observe a fast and
monotonic decrease of the $\cgamma_n$'s}. This decrease seems to exhibit an \emph{exponential behavior} at large $n$,
as suggested by the approximately linear decrease in the logarithmic plot in Figure \ref{fig:BareGammaLinParam}.

This observation is another advantage of the truncation \eqref{eq:BareLiouThirdAnsatz} as compared with the power
series ansatz in Section \ref{sec:BareLiouPower} where all couplings were of the same order of magnitude. Here the
situation is different as higher order couplings decrease sufficiently fast. We would like to point out, however, that
our numerical analysis does not prove the convergence in a mathematically rigorous sense. This raises the question to
what extent the discussion can be brought to a rigorous analytical level.

The significance of such a consideration resides in the fact that truncations of the type
\eqref{eq:BareLiouThirdAnsatz} are justified only if higher order couplings get less and less important, such that the
finite series already encapsulates the most essential information. Otherwise, computing $\cx$ according to
\eqref{eq:cxi} would be pointless as long as $\Nmax$ remains finite. Therefore, a more thorough analysis serves as a
consistency check for the truncation.
In Appendix \ref{app:ProofOfConvergence} we present an argument that provides strong evidence for the convergence of
the couplings $\cgamma_i$ as $i\to\infty$. In terms of $a_i\equiv 2\mku\cZ^{-1}i^2\,\cgamma_i$ the statement reads:
Provided that the first $n$ couplings $a_i$, $i=1,\ldots,n$, decrease exponentially, say $a_i=A\,\e^{-\lambda i}$, then
the value of $a_{n+1}$ is less than or equal to $A\,\e^{-\lambda (n+1)}$. This result supports the convergence
conjecture. However, since the decrease of the first $n$ couplings deviates slightly from an exact exponential
fall-off, in particular at small $n$, see Figure \ref{fig:BareGammaLinParam}, the assumption of the proof is not
strictly satisfied.\footnote{The proof in Appendix \ref{app:ProofOfConvergence} is carried out in terms of
$a_n\equiv 2\mku\cZ^{-1}n^2\,\cgamma_n$ instead of $\cgamma_n$. The additional factor $n^2$ is irrelevant for the
discussion of the fall-off behavior: Once we know that $a_n$ decreases exponentially with $n$, the $\cgamma_n$'s are
dominated by an exponential decrease as well (and vice versa). The diagrams for both $\cgamma_n$ (Figure
\ref{fig:BareGammaLinParam}) and $a_n$ (Figure \ref{fig:UpperBoundForA}) show the characteristic exponential behavior
for increasing $n$ while there are deviations from the exponential for small $n$.} Hence, we must rely on a numerical
computation of the first couplings. This constitutes a gap in the proof. Nonetheless, all indications coming from
Appendix \ref{app:ProofOfConvergence} and Figure \ref{fig:BareGammaLinParam} point towards converging couplings.

By virtue of Figure \ref{fig:BareXiLinParam}, our conjecture receives additional support. It shows the coupling $\cx$
dependent on $\Nmax$. Once $\Nmax$ is greater than about 15, $\cx$ is approximately constant. In this region,
increasing $\Nmax$ further, i.e.\ including more terms in the bare potential and in eq.\ \eqref{eq:cxi}, has no
observable effect on $\cx$. The last ten entries in the diagram differ only by the number
$\big(\cx|_{\Nmax=39}-\cx|_{\Nmax=48}\big)\approx 1.7\cdot 10^{-10}$. We emphasize that such a \emph{fast and stable
convergence} behavior is a striking result which might not have been expected in advance.
\begin{figure}[tp]
 \centering
 \includegraphics[width=0.75\columnwidth]{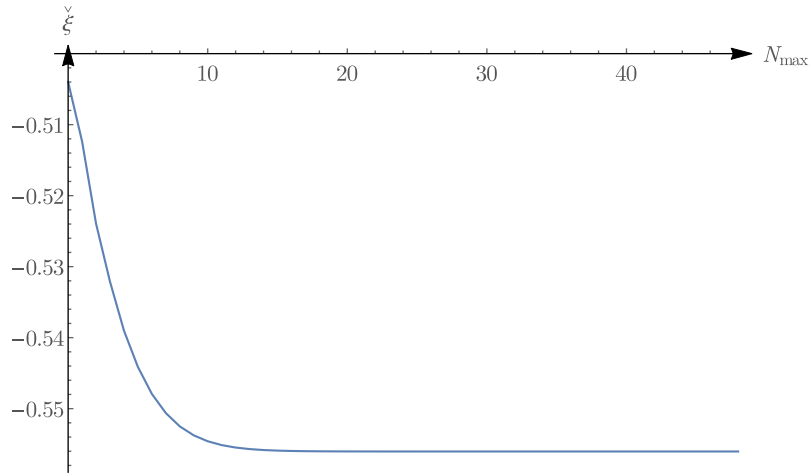}
\caption{Dependence of $\cx$ on $\Nmax$, i.e.\ on the number of exponential terms in the bare potential. (Note that the
 discrete set of points is joined by line segments for illustrative purposes.) For increasing $\Nmax$ the curve
 converges to the value $\cx\to -0.55604$.}
\label{fig:BareXiLinParam}
\end{figure}
After determining a fit function based on an exponential decrease of the couplings and a subsequent extrapolation we
find $\cx\to -0.55604$ in the large $\Nmax$-limit. For comparison with the EAA coupling $b=\frac{38}{3}$ we compute its
bare counterpart by their relation to the $\hR\mku\phi$-term in the actions. We obtain
$\cb \equiv -8\pi\mku\cx\approx\frac{41.92}{3}$, so the bare and the effective coupling are reasonably close together.

At this point a remark concerning the bare potential is in order. As can be seen in Figure \ref{fig:BareGammaLinParam},
all couplings $\cgamma_i$ come with a negative sign. For that reason, the bare potential, $\cV(\phi)=\frac{1}{2} \UV^2
\sum_{n=1}^{\Nmax}\cgamma_n\,\e^{2n\phi}$, is negative for all $\phi$. Moreover, it is \emph{not bounded from below}.
This observation is independent of the number of terms included in the bare potential. Figure
\ref{fig:BarePotExpLinParam} shows the dimensionless version of $\cV$ for $\Nmax=1$, $\Nmax=2$ and $\Nmax=48$. We see
that $\cV$ does not possess any minimum but it tends to $-\infty$ in the large field limit.

Whether or not this apparent instability of the conformal factor poses a physical problem is a different question,
though. In fact, we see from the action \eqref{eq:BareLiouThirdAnsatz} and from \eqref{eq:BareZ} that the kinetic term
is negative, too, since $\cZ<0$. Therefore, the kinetic term and the bare potential $\cV$ have the same sign. This is
precisely what was observed for the effective average action \eqref{eq:GammaUVLiou}, where we mentioned that both
sources of negativity should be taken into account in our discussion. Again, as argued in Section \ref{sec:UnitaryCFT},
the conformal factor instability is not an unmistakable sign for a physical deficiency but it can be cured by imposing
appropriate constraints to cut out negative norm states.\footnote{As mentioned previously, a consideration at the
technical level might require special attention (regularization, analytic continuation, or similar) at intermediate
steps of the calculation such that the functional integral can be made sense of (cf.\ Ref.\ \cite{MM90}, for instance).
We leave this point for future investigations.} In this regard, an unbounded potential might be unproblematic after
all.
\begin{figure}[tp]
 \centering
 \includegraphics[width=0.65\columnwidth]{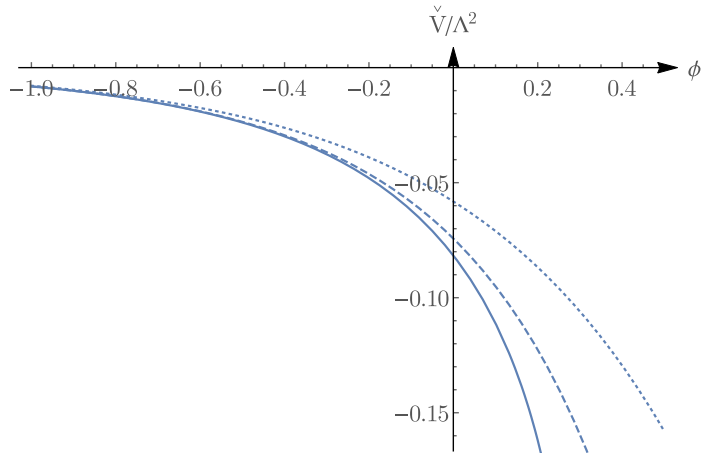}
\caption{Bare potential for $\Nmax=1$ (dotted), $\Nmax=2$ (dashed), $\Nmax=48$ (plain), based on the linear
 parametrization.}
\label{fig:BarePotExpLinParam}
\end{figure}

%----------------------------------------------------------------------------------------------------------------------
\subsection{Results for the exponential parametrization}
\label{sec:BareLiouExpSeriesExp}
%----------------------------------------------------------------------------------------------------------------------

In order to study the differences that arise from using the (fixed point version of the) EAA based on the exponential
parametrization, we simply replace the effective couplings $b$ and $\mu$ by their modified values,
$b\approx\frac{50.45}{3}$ and $\mu\approx 0.145772$, while, apart from this, we proceed as in the previous subsection,
i.e.\ we solve eqs.\ \eqref{eq:BareZ} - \eqref{eq:cgamman} numerically for the bare couplings. The result for
$\cgamma_n$, $n=1,\dotsc,48$, is depicted in Figure \ref{fig:BareGammaExpParam}.
It shows a fall-off behavior of the couplings very similar to the one corresponding to the linear parametrization:
The absolute values of the $\cgamma_n$'s seem again to decrease exponentially on average as $n$ increases. As compared
with Figure \ref{fig:BareGammaLinParam} there are two differences, though. First, the deviations from a perfect
exponential fall-off are more distinct, and second, the sign of the couplings fluctuates. The latter is indicated by
the two different colors of the points in Figure \ref{fig:BareGammaExpParam}. It appears that there are as many
positive as negative signs which alternate without following any obvious regular pattern. This phenomenon renders a
rigorous discussion about the couplings' convergence more involved, cf.\ Appendix \ref{app:CheckInit}.
\begin{figure}[tp]
 \centering
 \includegraphics[width=0.75\columnwidth]{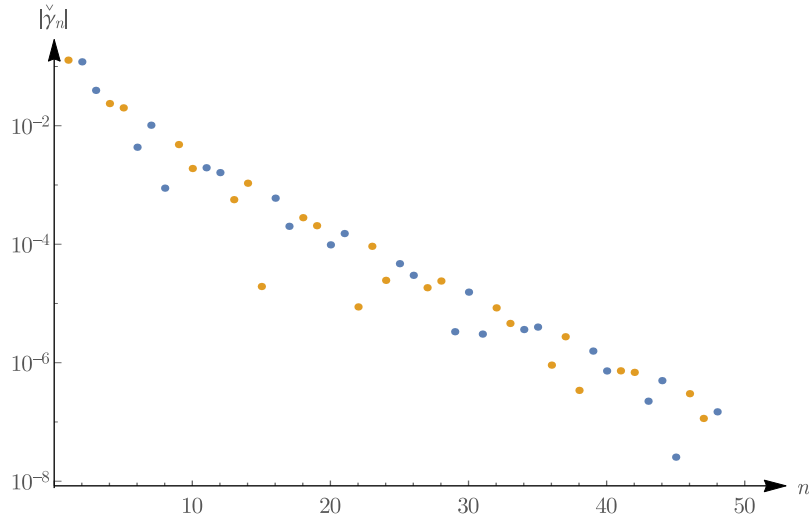}
\caption{Logarithmic plot showing the absolute values of the bare couplings $\cgamma_n$ dependent on their
 index $n$, in the range $n=1,\ldots,48$, based on the exponential parametrization. The average decrease behavior
 towards larger $n$ is still approximately exponential, although there are larger fluctuations as compared with
 Figure \ref{fig:BareGammaLinParam}. Couplings represented by a blue dot have a positive sign, while dark yellow
 dots refer to negative signs.}
\label{fig:BareGammaExpParam}
\end{figure}

The dependence of $\cx$ on the number $\Nmax$ is shown in Figure \ref{fig:BareXiExpParam}. We observe an oscillation
whose amplitude decreases towards larger $\Nmax$. Ultimately, $\cx$ seems to converge in the large $\Nmax$ limit.
In comparison with Figure \ref{fig:BareXiLinParam} (which did not show any oscillation) this convergence is slower.
The limit that $\cx$ approaches can be obtained by fitting a damped oscillation to the points in Figure
\ref{fig:BareXiExpParam} and applying an extrapolation at large $\Nmax$ subsequently.\footnote{More precisely, it
turned out that the data points in Figure \ref{fig:BareXiExpParam} are most efficiently approximated by a function of
the type $f(x)=c_2\,\e^{-\lambda_2\mku x}\mku \sin(\omega x+x_0)+c_1\,\e^{-\lambda_1\mku x}+c_0$ with $x\equiv\Nmax$.}
This way we find that $\cx\to -0.6019$ for $\Nmax\to\infty$. In order to compare this value with the effective
coupling $b\approx\frac{50.45}{3}$ we consider $\cb \equiv -8\pi\mku\cx$ again, yielding $\cb \approx \frac{45.38}{3}$.
\begin{figure}[tp]
 \centering
 \includegraphics[width=0.75\columnwidth]{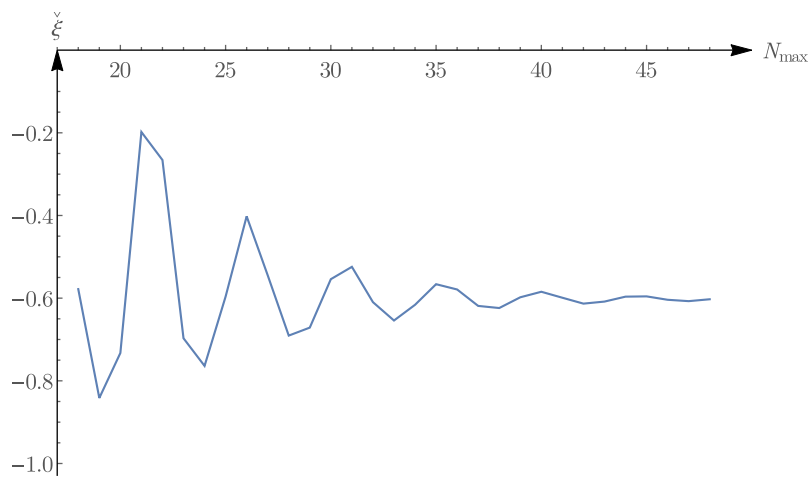}
\caption{Dependence of $\cx$ on $\Nmax$. (Again, the discrete set of points has been joined by line segments for
 illustrative purposes.) The diagram starts at $\Nmax=18$ as this captures the significant region concerning
 convergence; for smaller $\Nmax$ the fluctuations are stronger and more irregular. Fitting a curve to the depicted
 points shows that $\cx$ converges to $-0.6019$ for $\Nmax\to\infty$.}
\label{fig:BareXiExpParam}
\end{figure}

\begin{figure}[tp]
 \centering
 \includegraphics[width=0.65\columnwidth]{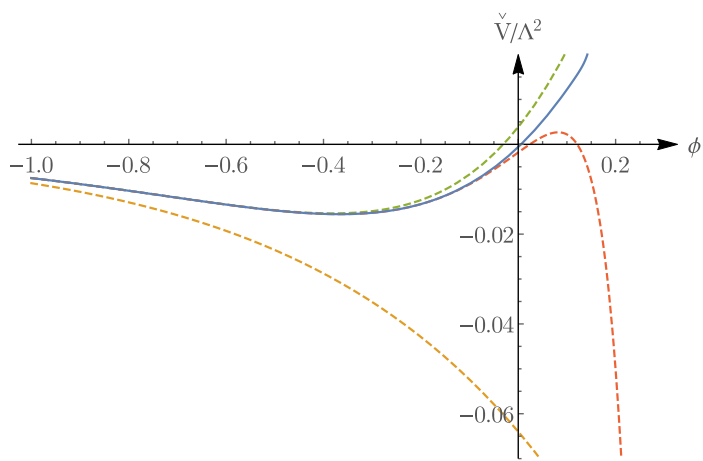}
\caption{Bare potential for $\Nmax=1$ (dark yellow, dashed), $\Nmax=4$ (green, dashed), $\Nmax=10$ (orange, dashed),
 and $\Nmax=48$ (blue), using the exponential parametrization.}
\label{fig:BarePotExpExpParam}
\end{figure}
At last, let us investigate how the bare potential changes as $\Nmax$ is increased. In Figure
\ref{fig:BarePotExpExpParam} we show its dimensionless version, $\cV/\UV^2$, for $\Nmax=1$, $\Nmax=4$, $\Nmax=10$ and
$\Nmax=48$. We observe that the bare potential possesses a minimum for all $\Nmax\ge 2$, which is located at
$\phi\approx -0.37$ at large $\Nmax$. For increasing numbers $\Nmax$ the potential seems to converge pointwise to a
limit function which is given approximately by the blue curve (in the depicted region $\cV|_{\Nmax=48}$ is supposed to
be close to $\cV|_{\Nmax\to\infty}$) and whose minimum becomes a \emph{global minimum}.\footnote{If $\Nmax$ corresponds
to a coupling with negative sign, see Figure \ref{fig:BareGammaExpParam}, then $\cV(\phi)\to -\infty$ for
$\phi\to\infty$, so the minimum is only a local one. If, on the other hand, the last coupling of the series in the
potential is positive, then the minimum is a global one. The limit potential $\cV|_{\Nmax\to\infty}$ seems to have
a unique global minimum, too.} Hence, \emph{the bare potential becomes bounded from below}, i.e., unlike the one for
the linear parametrization, cf.\ Figure \ref{fig:BarePotExpLinParam}, it has a stabilizing character now. The minimum
breaks scale invariance, in accordance with the Ward identities w.r.t.\ combined Weyl transformations (cf.\ Ref.\
\cite{RW97} and Sections \ref{sec:BareLiouCon} and \ref{sec:WeylWardIdentity}). Note that, with regard to the conformal
factor instability, the kinetic term ``counteracts'' the potential this time since the former is negative and the
latter is bounded from below.

%----------------------------------------------------------------------------------------------------------------------
\section{Bare action with a general potential}
\label{sec:BareLiouGeneral}
%----------------------------------------------------------------------------------------------------------------------

As mentioned in the introduction of this chapter, Ref.\ \cite{RW97} is focused on the computation of the \emph{EAA}
provided that the \emph{bare action} is given, i.e.\ it concerns the opposite direction as compared with our preceding
discussion. There the authors find that, if the bare potential has a pure Liouville form, $\cm\,\e^{2\phi}$, then a
calculation of the effective potential based on the truncation ansatz $\mu\,\e^{\alpha\mku\phi}$ shows that $\alpha$
cannot equal $2$, so the bare and the effective potential are different.

This consideration applied to our present case suggests studying a truncation ansatz for the bare potential which is
of the type $\cm\,\e^{\ca\mku\phi}$ if the effective potential is given by $\mu\,\e^{2\phi}$. However, it is not
possible to obtain such a bare potential by means of the reconstruction formula \eqref{eq:OneLoopFullMain}: We have to
know which terms the trace must be projected onto, e.g.\ $\int\shg\,\e^{2\phi}$, $\int\shg\,\e^{4\phi}$, etc.
Only then we can determine their coefficients consistently. Thus, we do not investigate such truncations with modified
exponents like $\ca\phi$.
Nonetheless, we can study a truncation for the bare action whose \emph{potential is left completely arbitrary}.
The idea is to leave the logarithm appearing in the reconstruction formula unexpanded rather than to extract any terms
($\propto \e^{2\phi}$, $\propto\phi$, or similar). This leads to a second order differential equation for the bare
potential $\cV(\phi)$ which can be solved numerically and whose asymptotic behavior can be determined analytically.

We start out with the general ansatz
\begin{equation}
 \SB[\phi] = \frac{1}{2}\int\td^2 x\shg\left[\cZ\mku\phi\big(-\hB\big)\phi+ \cx\,\hR\mku\phi+\cV(\phi)\right].
\label{eq:BareLiouGenAnsatz}
\end{equation}
The corresponding Hessian reads
\begin{equation}
 \SB^{(2)} = -\cZ\,\hB +\frac{1}{2}\mku\cV''(\phi).
\end{equation}
This is to be inserted into \eqref{eq:OneLoopFullMain} where the trace is treated as in the previous sections.
As a result, the trace term is the same as in eq.\ \eqref{eq:BareLiouTracef}, the only difference being a modification
of the function $f_\UV\mku$ according to
\begin{equation}
 f_\UV(\phi) = \ln\big[\UV^2 M^{-2}\cZ + \textstyle\frac{1}{2} M^{-2}\mku\cV''(\phi)\big].
\label{eq:fUVArbPot}
\end{equation}
Then the reconstruction formula $\Gamma_\UV=\SB+\frac{1}{2}\mku\Tr\ln\big[M^{-2}\big(\SB^{(2)}+\RL\big)\big]$ at lowest
order in the curvature, $\mO(R^0)$, amounts to
$-\frac{b\mku\mu}{16\pi}\int\shg\,\e^{2\phi} = \frac{1}{2}\int\shg\,\cV(\phi)+\frac{\UV^2}{8\pi}\int\shg\,f_\UV(\phi)$.
Comparing coefficients yields
\begin{equation}
 -\frac{b\mku\UV^2\mu}{16\pi}\,\e^{2\phi}= \frac{1}{2}\,\cV(\phi)+\frac{\UV^2}{8\pi}\ln\big[\UV^2 M^{-2}\cZ +
 \textstyle\frac{1}{2} M^{-2}\mku\cV''(\phi)\big],
\label{eq:BarePotDiffEqUnsolved}
\end{equation}
and by solving for $\cV''(\phi)$ we obtain
\begin{equation}[b]
 \cV''(\phi) = 2\mku M^2\mku\exp\left[-\textstyle\frac{1}{2}b\mku\mu\,\e^{2\phi} - 4\pi\mku\UV^{-2}\cV(\phi)\right]
 -2\mku\UV^2\mku\cZ\,.
\label{eq:BarePotDiffEq}
\end{equation}
This equation fixes $\cV(\phi)$ up to two unknown initial conditions, say $\cV(0)$ and $\cV'(0)$.

Before solving the differential equation \eqref{eq:BarePotDiffEq} numerically, we try to assess the asymptotic behavior
of the potential for $\phi\to -\infty$ and $\phi\to\infty$ at an analytical level. As we search for bounded potentials,
it turns out convenient to distinguish between the case where $\cV$ is bounded from below and the case where $\cV$ is
bounded from above. Although these properties concerning boundedness are used as assumptions, we test a posteriori
whether they are satisfied by the resulting solution for $\cV$.
\medskip

\noindent
\textbf{(a) Assumption: \bm{$\cV$} is bounded from below.} Let us consider the limit of very small fields and very
large fields separately in our analysis.
\begin{itemize}
 \item \textbf{The case \bm{$\phi\ll -1$}}: In this limit we may assume $\e^{2\phi}\approx 0$ such that eq.\
 \eqref{eq:BarePotDiffEq} reduces to $\cV''(\phi) = 2\left(M^2\,\e^{- 4\pi\mku\UV^{-2}\cV(\phi)} -\UV^2\mku\cZ\right)$.
 Furthermore, boundedness of $\cV$ requires $\cV(\phi)\to\infty$ or $\cV(\phi)\to\text{const}$ for $\phi\to -\infty$.
 Thus, for $\phi\ll -1$, the differential equation simplifies to $\cV''(\phi)\approx\text{const}$, leading to
 $\cV(\phi)\sim\phi^2$ asymptotically. Here, the afore-mentioned requirement dictates a positive sign in front of the
 $\phi^2$-term. As a consequence, $\e^{- 4\pi\mku\UV^{-2}\cV(\phi)}\to 0$ for $\phi\to -\infty$. In this limit we have
 $\cV''(\phi)=-2\UV^2\cZ$. Integration yields
 \begin{equation}
  \cV(\phi) = -2\UV^2\cZ\mku\phi^2 + \cV'(0)\phi + \cV(0).
  \label{eq:BarePotAsSol}
 \end{equation}
 This asymptotic solution meets the above requirement only if $\cZ<0$. Since $\cZ$ is not modified as compared to the
 previous subsections, this is indeed the case: Both for the linear and for the exponential parametrization we have
 $\cZ<0$, so the solution \eqref{eq:BarePotAsSol} is consistent.
 \item \textbf{The case \bm{$\phi\gg 1$}}: Let us assume for a moment that the term $\e^{2\phi}$ in eq.\
 \eqref{eq:BarePotDiffEq} dominates over $4\pi\mku\UV^{-2}\cV(\phi)$, an assumption that is to be check for consistency
 once we have found an asymptotic solution. In this case we find $\e^{-\frac{1}{2}b\mku\mu\,\e^{2\phi}
 - 4\pi\mku\UV^{-2}\cV(\phi)}\to 0$ for $\phi\to\infty$. Therefore, the large $\phi$ limit amounts to $\cV''(\phi)=
 - 2\UV^2\cZ$ again, so we find precisely the same solution as in eq.\ \eqref{eq:BarePotAsSol}. Again, this result
 is consistent with our above assumption.
\end{itemize}

\noindent
\textbf{(b) Assumption: \bm{$\cV$} is bounded from above.} Actually, there is no solution to eq.\
\eqref{eq:BarePotDiffEq} which satisfies the assumption consistently. To see this, it is sufficient to consider the
case $\phi\ll -1$, that is, $\e^{2\phi}\approx 0$. Then the differential equation becomes $\cV''(\phi) = 2\mku M^2\,
\e^{- 4\pi\mku\UV^{-2} \cV(\phi)} -2\mku\UV^2\mku\cZ$ again. Now, boundedness of $\cV$ dictates $\cV(\phi)\to-\infty$
or $\cV(\phi)\to \text{const}$ for $\phi\to -\infty$.

If $\lim_{\phi\to-\infty}\cV(\phi)=\text{const}$, the differential equation boils down to $\cV''(\phi)=\text{const}$
in the limit of small $\phi$. This is in contradiction with $\cV(\phi)= \text{const}$, though.

On the other hand, if $\lim_{\phi\to-\infty}\cV(\phi)=-\infty$, the differential equation reduces to $\cV''(\phi) =
2\mku M^2\,\e^{- 4\pi\mku\UV^{-2}\cV(\phi)}$. This case would require $\cV(\phi)\to-\infty$ and $\cV''(\phi)\to+\infty$
at the same time. However, there is no smooth function satisfying both conditions simultaneously.
Hence, $\cV$ cannot be bounded from above.
\medskip

Taking all cases together, we have demonstrated that \emph{the bare potential approaches the parabola given by eq.\
\eqref{eq:BarePotAsSol} asymptotically}, for both $\phi\to -\infty$ and $\phi\to \infty$. We emphasize in particular
that this asymptotic behavior is independent of the measure parameter $M$. 

For small values of $|\phi|$ we expect deviations of $\cV$ from a perfect parabola form. The magnitude of these
deviations is revealed by a numerical analysis in the following.

All numerical computations are performed with \Mma. We use the initial conditions $\cV(0)=0$ and $\cV'(0)=0$. Different
choices would merely amount to shifted graphs for the resulting potentials. The values $b$ and $\mu$ are chosen to
correspond to the linear parametrization; the ones for the exponential parametrization would qualitatively lead to the
same picture. For the measure parameter we choose $M=\UV$. The result is shown in Figure \ref{fig:BareLiouArbPot}.
It confirms our expectations remarkably well. We observe that the bare potential noticeably deviates from a parabola
form for small values of $|\phi|$. For large $|\phi|$, on the other hand, it converges to the parabola given by
$\cV(\phi) \sim -2\UV^2\cZ\mku\phi^2$. Note that the degree of deviation depends on the measure parameter $M$:
For increasing $M$, the deviations become more distinct, in particular in the small $|\phi|$ regime, while they are
completely absent for $M\to 0$, as can be seen from eq.\ \eqref{eq:BarePotDiffEq}. The asymptotic behavior is the same
for all values of $M$, though.
\begin{figure}[tp]
\begin{minipage}{0.49\textwidth}
 \includegraphics[width=\columnwidth]{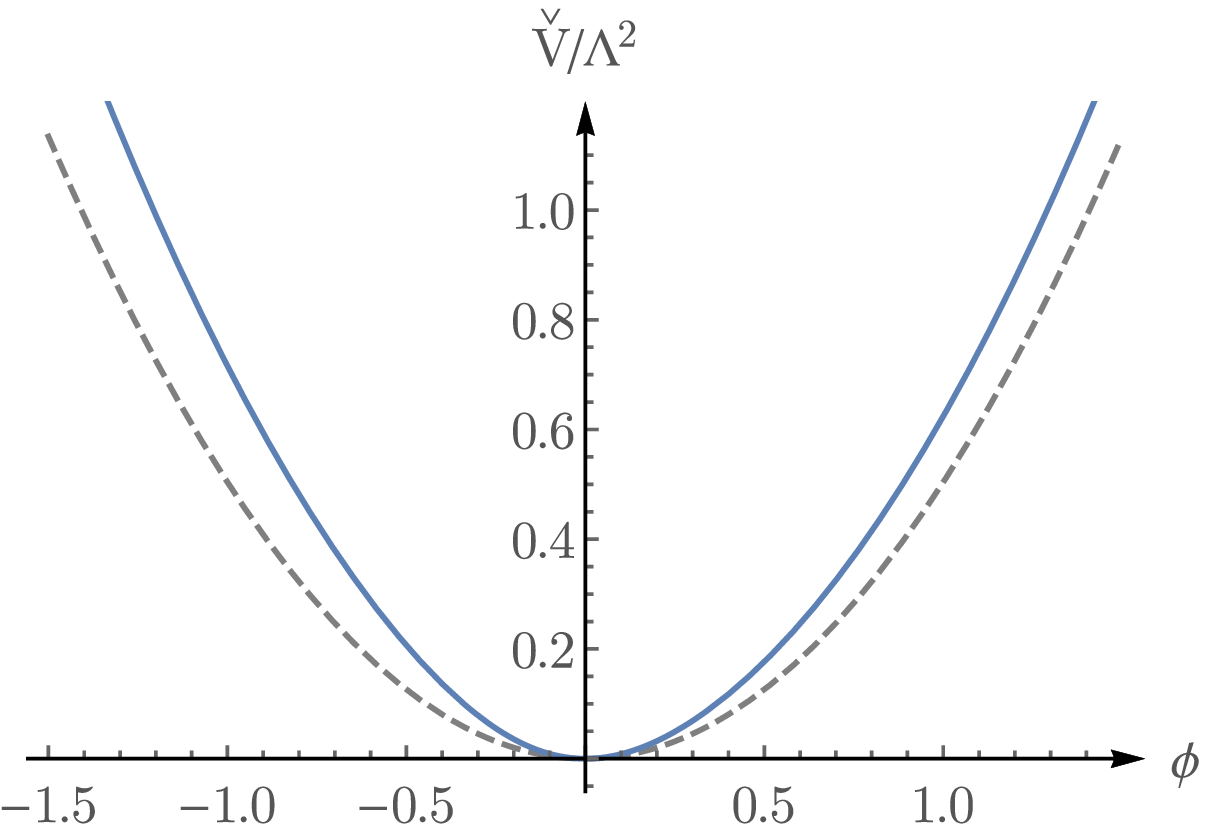}
\end{minipage}
\hfill
\begin{minipage}{0.49\textwidth}
 \includegraphics[width=\columnwidth]{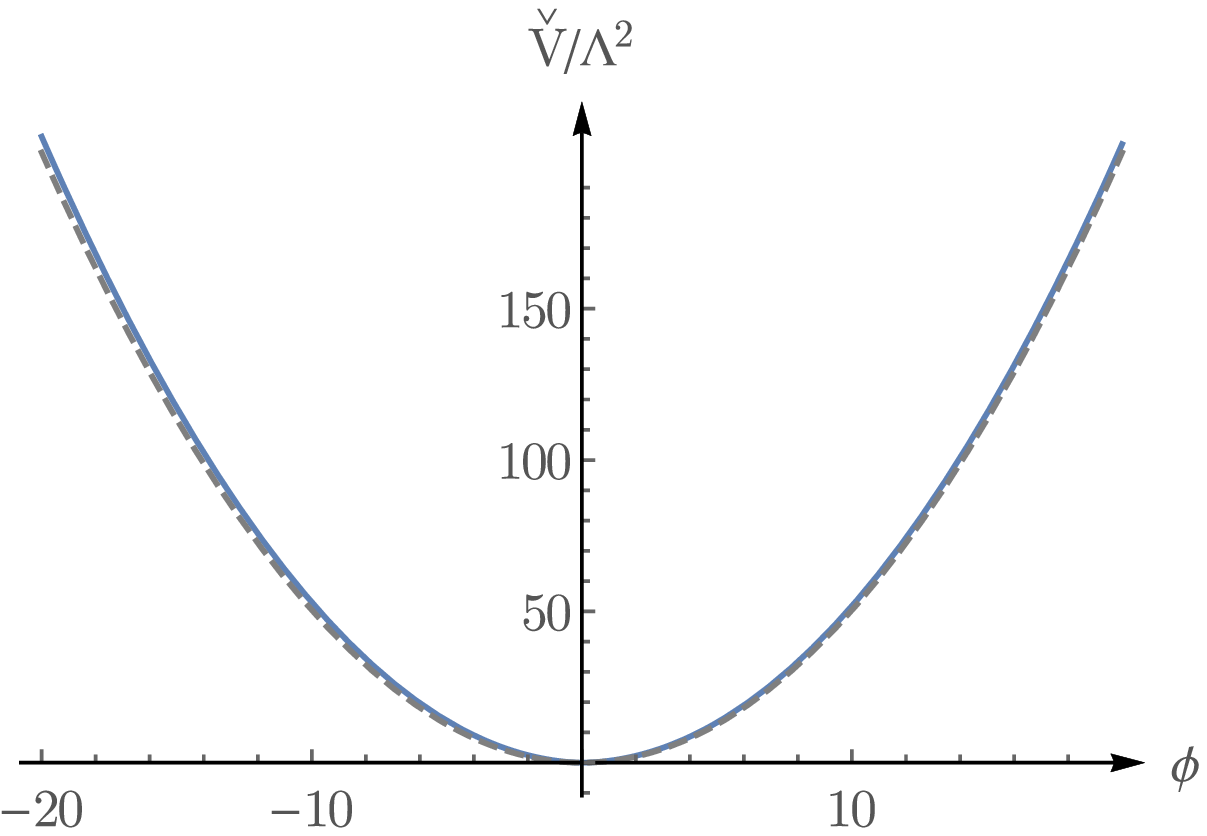}
\end{minipage} 
\caption{Bare potential (blue) in comparison with a perfect parabola (gray, dashed). In the regime of small absolute
 field values (left diagram) there are observable deviations, while the effect weakens towards larger values of
 $|\phi|$ (right diagram).}
\label{fig:BareLiouArbPot}
\end{figure}

Once we know the function $f_\UV$ in eq.\ \eqref{eq:fUVArbPot} it is straightforward to extract an equation for the
coefficients of the $\hR\mku\phi$-terms, too, by using the same strategy as in the previous sections. This determines
the bare coupling $\cx\mku$:
\begin{equation}
 \cx = \frac{b}{8\pi}\left(-1+\frac{\mu}{3}\right)+\frac{1}{6}\UV^{-2}\cV'(0).
\end{equation}
For the values of $b$ and $\mu$ based on the linear parametrization, and the initial condition $\cV'(0)=0$, we obtain
$\cx\approx -0.477$. In terms of $\cb \equiv -8\pi\mku\cx$ this amounts to $\cb=\frac{36}{3}$.

Up to this point, the above results seem to be quite promising. However, a note of caution is in order. The issue can
be understood by reviewing eq.\ \eqref{eq:BarePotDiffEqUnsolved}. Our investigation has revealed the asymptotically
quadratic form of the bare potential, which implies the relation $\cV''(\phi)\approx -2\UV^2\cZ$. Inserting this into
\eqref{eq:BarePotDiffEqUnsolved} shows that the argument of the logarithm is close to zero, $\UV^2 M^{-2}\cZ +
\frac{1}{2} M^{-2}\mku\cV''(\phi)\approx 0$. This indicates a high degree of \emph{fine-tuning}: Eq.\
\eqref{eq:BarePotDiffEqUnsolved} can be solved only if the argument of the logarithm is extremely small compared with
$\cV(\phi)$ and $\e^{2\phi}$. At the same time, it must not become exactly zero. Such a solution appears to be rather
unnatural: All large terms are induced by a small fine-tuned term.

Moreover, this means that \emph{all contributions to the effective potential stem from the one-loop term}, in
disagreement with the conventional picture which assumes that the bare action represents an essential part of the
EAA, according to $\Gamma_\UV=\SB+\text{correction}$. The major significance of the one-loop term suggests that
higher-loop orders might become even more important. Therefore, we do not consider the above results reliable. In a
sense, the one-loop reconstruction formula predicts its own breakdown when applied to the setting discussed in this
subsection.

%----------------------------------------------------------------------------------------------------------------------
\section{Summarizing remarks}
\label{sec:BareLiouCon}
%----------------------------------------------------------------------------------------------------------------------

The preceding sections concerned the reconstruction problem in Liouville theory. We tried to determine the bare action
by applying eq.\ \eqref{eq:OneLoopFullMain} to a Liouville-type effective average action. Recall that there are
different ways to obtain a bare action when starting from an Einstein--Hilbert-type EAA, as shown in Figures
\ref{fig:CommDiag} and \ref{fig:CommDiagFull}. In this chapter we studied the last step in the chain
\begin{equation}
 \Gamma_\UV^\text{EH}[g] \to \Gamma_\UV^\text{ind}[g] \to \Gamma_\UV^\text{L}[\phi;\hg]+\Gamma_\UV^\text{ind}[\hg]
 \to \SB[\phi;\hg] + \Gamma_\UV^\text{ind}[\hg].
\label{eq:BareChain}
\end{equation}
In \eqref{eq:BareChain} we explicitly state the remaining part $\Gamma_\UV^\text{ind}[\hg]$ that does not
contain any contributions from the conformal factor and that is not involved in the reconstruction process. It is
mentioned here since the combination $\Gamma_\UV^\text{L}[\phi;\hg]+\Gamma_\UV^\text{ind}[\hg]$ can be interpreted
as a \emph{conformal field theory} whose central charge $c$ can be read off from $\Gamma_\UV^\text{ind}[\hg]$ or,
equivalently, from the $\hR\mku\phi$-term in $\Gamma_\UV^\text{L}[\phi;\hg]$. In terms of the effective coupling
$b$ we have $c=\frac{3}{2}b$. Now, the crucial point is that \emph{after} the reconstruction process, i.e.\ after the
last step in \eqref{eq:BareChain}, the sum $\SB[\phi;\hg] + \Gamma_\UV^\text{ind}[\hg]$ is \emph{no} conformal field
theory because $\SB[\phi;\hg]$ is not a pure Liouville action. Hence, although we can compute $\cb$ as the coefficient
of the $\hR\mku\phi$-term in the bare action, the quantity $\frac{3}{2}\cb$ does \emph{not} represent a central charge.

Having said this, let us briefly sum up the results of this chapter obtained so far. We considered several truncation
ans\"{a}tze for $\SB[\phi;\hg]$ with different bare potentials, viz., a pure Liouville potential, a power series, a
series of exponentials, and an arbitrary function. Apart from some interesting results, we uncovered also a couple of
drawbacks. It turned out that the most promising among the studied candidates for the bare potential is a series of
exponentials, $\cV(\phi)= \UV^2 \sum_{n=1}^{\Nmax}\cgamma_n\,\e^{2n\phi}$. We were able to compute the bare couplings
$\cgamma_n$ iteratively. They do not depend on $\Nmax$ and they tend to zero as $n\to \infty$. Including an increasing
number of terms in the potential affects the bare coupling $\cx$, but we observed a fast convergence. Depending on
the underlying metric parametrization and on $\Nmax$ the total bare potential can be bounded from below or bounded from
above, affecting the instability of the conformal factor.
It has been discussed in Section \ref{sec:UnitaryCFT}, however, that the conformal factor instability may be cured
by imposing appropriate constraints in order to project onto physical states only.

{%
\renewcommand{\arraystretch}{1.6}
\begin{table}[tp]
\centering
\begin{tabular}{ccc}
 \hline
 Ansatz for $\cV$  &  \bm{$+$}  &  \bm{$-$} \\
 \hline
 $\cgamma_{\UV}\,\e^{2\phi}$  &  \begin{minipage}{0.3358\textwidth}\vspace{0.4em}\begin{itemize}[nolistsep,leftmargin=1em]
   \item Simple, natural ansatz
   \item Same form as $\Gamma_\UV$
 \end{itemize} \end{minipage} & \begin{minipage}{0.315\textwidth}\vspace{0.9em}\begin{itemize}[nolistsep,leftmargin=1em]
   \item No closure: $\Tr\ln$-terms\\ do not combine to $\e^{2\phi}$
   \item Disagrees with Ward\\ identities \cite{RW97}
 \end{itemize} \end{minipage}\vspace{0.7em} \\
 \hline
 Power series &  \begin{minipage}{0.3358\textwidth}\vspace{0.4em}\begin{itemize}[nolistsep,leftmargin=1em]
   \item Simple extension
   \item High-dim.\ theory space
 \end{itemize}\end{minipage} & \begin{minipage}{0.315\textwidth}\vspace{0.9em}\begin{itemize}[nolistsep,leftmargin=1em]
   \item No convergence: coefficients depend heavily on \# of terms in series
   \item $R\phi$-term not convergent
   \item Higher order terms more and more important
 \end{itemize} \end{minipage}\vspace{0.7em} \\
 \hline
 $\sum_n \cgamma_{n,\UV}\, \e^{2n\mku\phi}$ &
 \begin{minipage}{0.3358\textwidth}\vspace{0.9em}\begin{itemize}[nolistsep,leftmargin=1em]
   \item Similar to Fourier series
   \item Similar to sine-Gordon
   \item High-dim.\ theory space
   \item ``Liouville action plus\\ higher order terms''
   \item Series coeffs.\ converge
   \item $R\phi$-term converges
 \end{itemize}
 \vspace{0.5em}
 \hspace*{0.5\textwidth}
 \begin{minipage}{0.4\textwidth}\begin{itemize}[nolistsep,leftmargin=1em]
   \item For lin.\ parametrization:$\mkern-150mu$\\ \mbox{$\cV$ bounded from above}$\mkern-150mu$
   \item For exp.\ parametrization:$\mkern-150mu$\\ \mbox{$\cV$ bounded from below}$\mkern-150mu$
 \end{itemize}
 \end{minipage}
 \vspace{0.7em}
 \end{minipage}
 & \\
 \hline
 \vspace{0.4em}\begin{minipage}{0.26\textwidth}\centering General potential\\[0em] (numerical analysis)\end{minipage} &
 \begin{minipage}{0.3358\textwidth}\vspace{0.4em}\vspace{0.7em}\begin{itemize}[nolistsep,leftmargin=1em]
   \item Most general form
   \item $\infty$-dim.\ theory space
   \item Simple asymptotic\\ behavior: $\cV \sim\phi^2$
 \end{itemize}
 \vspace{2.1em}
 \hspace*{0.5\textwidth}
 \begin{minipage}{0.4\textwidth}\begin{itemize}[nolistsep,leftmargin=1em]
   \item \mbox{$\cV$ bounded from below}$\mkern-150mu$
 \end{itemize}
 \end{minipage}
\end{minipage}\vspace{0.5em} & 
 \begin{minipage}{0.315\textwidth}\vspace{0.7em}\begin{itemize}[nolistsep,leftmargin=1em]
   \item Fine tuning required:\\ argument of $\Tr\ln$-term is close to zero
   \item Importance of one-loop term suggests considering higher-loop orders
 \end{itemize} \vspace{1.2em}\end{minipage}\vspace{0em} \\
 \hline
\end{tabular}
\vspace{0.3em}
\caption{Assets and drawbacks of four ans\"{a}tze for the potential $\cV$ of the bare Liouville action, based on the
 one-loop reconstruction performed in this chapter.}
\label{tab:Liouville}
\end{table}%
}%

In Table \ref{tab:Liouville} we summarize advantages and disadvantages of the four different ans\"{a}tze. In either
case it remained unclear to what extent we can actually rely on the calculations performed in this chapter. We
emphasize that all results were obtained on the basis of the reconstruction formula \eqref{eq:OneLoopFullMain}. Thus,
our findings suggest that the approximate character inherent in the one-loop formula \eqref{eq:OneLoopFullMain} might
prevent us from determining the essential part of the bare action in Liouville theory: The one-loop term might possibly
not contain enough information, while higher-loop orders might be more important in this case. In this regard,
different methods like the use of Ward identities may be expected to lead to more reliable results. For that reason,
we derive the Ward identity corresponding to Weyl split-symmetry transformations in the next section.

%----------------------------------------------------------------------------------------------------------------------
\section{Ward identity with respect to Weyl split-symmetry}
\label{sec:WeylWardIdentity}
%----------------------------------------------------------------------------------------------------------------------

Being a quantum version of Noether's theorem, Ward identities\footnote{Some authors differentiate between the terms
``Ward identity'' and ``Ward--Takahashi identity'', where the former is considered a special case of the latter. Here,
on the other hand, we always think of the general version when speaking about ``Ward identities''.} (WIs) describe the
relation between correlation functions arising from the symmetries of (the bare action of) a quantum field theory.
Their derivation is based on the invariance of the functional measure under a symmetry transformation. If the measure
is noninvariant, it contributes an additional term to the WIs, which are then called ``anomalous Ward identities''.
In both cases, the transformation behavior of the \emph{bare action} and the \emph{measure} is known, while relations
for correlation functions, encoded in the \emph{effective (average) action}, are searched for.

In the reconstruction process considered in this chapter, the situation is different: We now start out from the
\emph{effective average action} and its symmetries, and we specify the functional \emph{measure} for the
reconstruction, but we do not know how the \emph{bare action} changes under the corresponding symmetry transformations.
This raises the question whether it is possible to deduce certain identities that the bare action has to satisfy upon
transformation. In a sense, such relations may be considered as \emph{reverse Ward identities}.

Here, we consider the \emph{Weyl split-symmetry transformation}, or combined Weyl transformation,
\begin{equation}
 \hg_\mn \to \e^{2\sigma}\hg_\mn\,,\qquad \phi \to \phi-\sigma\,,
\label{eq:SplitSym}
\end{equation}
which leaves the full metric $g_\mn=\e^{2\phi}\hg_\mn$ unaltered. Any functional $F[\phi;\hg]$ which is invariant under
the Weyl split-symmetry transformation \eqref{eq:SplitSym} can be written as a functional $\tilde{F}[g]$ of the full
metric, and any functional which can be expressed completely in terms of the full metric is Weyl split-symmetry
invariant.

As recalled in the previous section in eq.\ \eqref{eq:BareChain}, the reconstruction started with the sum
$\Gamma_\UV^\text{L}[\phi;\hg]+\Gamma_\UV^\text{ind}[\hg]$ which can be written in the form $\Gamma_\UV^\text{full}[g]
=\Gamma_\UV^\text{ind}[g]+c\int\!\sg$, a strictly Weyl split-symmetry invariant functional. In what follows, we will
show that, after having reconstructed the bare action with respect to the Liouville field, the sum $\SB[\phi;\hg]+
\Gamma_\UV^\text{ind}[\hg]$ is \emph{Weyl split-symmetry violating}. For that purpose, we will derive a WI in the
reverse sense that governs the transformation behavior of $\SB[\phi;\hg]$.

For our discussion we make use of the results of Appendix \ref{app:Weyl} (in particular the transformation rules) and
Chapter \ref{chap:EHLimit}. The full functional we start with is given by the induced gravity action plus a
cosmological constant term,
\begin{equation}
 \Gamma_\UV^\text{full}[g] = \Gamma_\UV^\text{ind}[g] - \frac{b\mu}{16\pi}\,\UV^2\int\td^2 x\sg \;,
\label{eq:GammaFullGammaInd}
\end{equation}
with $\Gamma_\UV^\text{ind}[g]=\frac{b}{64\pi}\mku I[g]$ plus zero mode contributions. As shown in Chapter
\ref{chap:EHLimit}, $\Gamma_\UV^\text{full}$ can be interpreted as the 2D limit of the Einstein--Hilbert action.
Inserting the metric $g_\mn=\e^{2\phi}\hg_\mn$ yields
\begin{equation}
 \Gamma_\UV^\text{full}[\e^{2\phi}\hg]=\Gamma_\UV^\text{ind}[\hg] + \Gamma_\UV^\text{L}[\phi;\hg],
\label{eq:GammaUVfull}
\end{equation}
with $\Gamma_\UV^\text{L}[\phi;\hg]=-\frac{b}{16\pi}\int\td^2 x\shg\,\big[ \phi\big(-\hB\big)\phi
+\hR\mku\phi+\mu\mku\UV^2\,\e^{2\phi}\big]$, as given in eq.\ \eqref{eq:GammaUVLiou}. The behavior of the first term
on the RHS of \eqref{eq:GammaUVfull} under Weyl transformations reads $\Gamma_\UV^\text{ind}[\e^{2\sigma}\hg]=
\Gamma_\UV^\text{ind}[\hg] - \frac{b}{8\pi}\,\Delta I[\sigma;\hg]$,
see eq.\ \eqref{eq:ItoDeltaI} in the appendix,\footnote{Although $\Gamma_\UV^\text{ind}$ contains --- apart from the
functional $I$ --- additional terms due to topological and zero mode contributions in general, see Appendix
\ref{app:Zero}, its above-stated transformation behavior is exact: $\Gamma_\UV^\text{ind}[\e^{2\sigma}\hg]=
\Gamma_\UV^\text{ind}[\hg] - \frac{b}{8\pi}\,\Delta I[\sigma;\hg]$.
The reason why there are correction terms to be added to $I$ but no ones to $\Delta I$ is that
the construction of $\Gamma_\UV^\text{ind}$ was actually \emph{based} on the exact transformation rule, see Chapter
\ref{chap:EHLimit}, so the rule must hold irrespective of the precise form of $\Gamma_\UV^\text{ind}$.}
with
\begin{equation}
 \Delta I[\sigma;\hg] \equiv \frac{1}{2}\int\td^2x\shg\left[\hD_\mu\sigma\hD^\mu\sigma+\hR\sigma\right].
\end{equation}
Besides, the Liouville action transforms as
\begin{equation}
 \Gamma_\UV^\text{L}[\phi-\sigma;\e^{2\sigma}\hg] =\Gamma_\UV^\text{L}[\phi;\hg]+\frac{b}{8\pi}\,\Delta I[\sigma;\hg],
\label{eq:TransLawLiou}
\end{equation}
under \eqref{eq:SplitSym}. Note that in the sum of these transformation laws the terms involving $\Delta I$ cancel each
other. Hence, the sum $\Gamma_\UV^\text{ind}[\hg] + \Gamma_\UV^\text{L}[\phi;\hg]$ is indeed Weyl split-symmetry
invariant, as it should be.

%----------------------------------------------------------------------------------------------------------------------
\subsection{Derivation of the Ward identity}
\label{sec:WIDerivation}
%----------------------------------------------------------------------------------------------------------------------

Let $\SB[\phi;\hg]$ denote the bare action that corresponds to the Liouville EAA, $\Gamma_\UV^\text{L}[\phi;\hg]$.
In order to derive a WI describing the transformation behavior of $\SB[\phi;\hg]$ we consider the functional integral
representation of the Liouville part of $\Gamma_\UV^\text{full}$:
\begin{equation}
\begin{split}
 \e^{-\Gamma_\UV^\text{full}[\e^{2\phi}\hg]} &= \e^{-\Gamma_\UV^\text{ind}[\hg]}\,\e^{-\Gamma_\UV^\text{L}[\phi;\hg]}\\
 &= \e^{-\Gamma_\UV^\text{ind}[\hg]}\int\mD_\UV^{[\hg]}\chi\;\e^{-\SB[\chi;\hg]+(\chi-\phi)
\cdot(\Gamma_\UV^\text{L})^{(1)}[\phi;\hg] - \frac{1}{2}(\chi-\phi)\cdot\RL(\chi-\phi)}\,.
\end{split}
\label{eq:FuncIntLiou}
\end{equation}
In \eqref{eq:FuncIntLiou} we explicitly indicate the metric dependence of the (translation invariant) measure by
writing $\mD_\UV^{[\hg]}\chi$ (cf.\ definition in App.\ \ref{app:Measure}),
and we bear in mind that the cutoff $\RL\equiv\RL(-\hB)$ depends on $\hg_\mn$, too. Furthermore, $(\Gamma_\UV^\text{L}
)^{(1)}[\phi;\hg]\equiv\frac{1}{\shg}\frac{\delta\Gamma_\UV^\text{L}[\phi;\hg]}{\delta\phi}$ is the first functional
derivative w.r.t.\ the Liouville field, and the dot refers to a spacetime integration, $f\cdot g\equiv\int\td^2 x\shg
\,f(x)g(x)$. Note that in \eqref{eq:FuncIntLiou} the induced gravity action part decouples from the functional
integral. Applying the transformation \eqref{eq:SplitSym} to the remaining (pure Liouville) part, we observe that the
shift of the Liouville field, $\phi\to\phi-\sigma$, is most conveniently taken into account by simultaneously changing
the integration variable,
\begin{equation}
 \chi\to\chi-\sigma,
\label{eq:ShiftMeasure}
\end{equation}
since $\phi$ makes its appearance in \eqref{eq:FuncIntLiou} as $(\chi-\phi)$ several times. Then this difference is
invariant under the combined transformations \eqref{eq:SplitSym} and \eqref{eq:ShiftMeasure}:
$(\chi-\phi)\to(\chi-\phi)$. Note that --- due to its translational invariance --- the measure is not modified by the
shift \eqref{eq:ShiftMeasure}: $\mD_\UV^{[\hg]}\chi'=\mD_\UV^{[\hg]}\chi$.

The transformation behavior of $\SB[\phi;\hg]$ is governed by the transformation laws of all those terms in
\eqref{eq:FuncIntLiou} that are changed by \eqref{eq:SplitSym} and \eqref{eq:ShiftMeasure}, viz:
\begin{equation}
 \bullet\; \Gamma_\UV^\text{L}[\phi;\hg] \qquad
 \bullet\; \frac{\delta\Gamma_\UV^\text{L}[\phi;\hg]}{\delta\phi} \qquad
 \bullet\; \mD_\UV^{[\hg]}\chi \qquad
 \bullet\; \shg\,\RL
\end{equation}
Since the behavior of $\Gamma_\UV^\text{L}[\phi;\hg]$ under \eqref{eq:SplitSym} has already been stated in eq.\
\eqref{eq:TransLawLiou}, it is only the last three terms that are to be investigated.
\medskip

\noindent
\textbf{(1) Transformation of \bm{$\delta\Gamma_\UV^\text{L}/\delta\phi\,$}:}\\
The first functional derivative of the Liouville action w.r.t.\ $\phi$ is given by
\begin{equation}
 \frac{\delta\Gamma_\UV^\text{L}}{\delta\phi}[\phi;\hg] = 
 -\frac{b}{16\pi}\shg\left[-2\mku\mku\hB\mku\phi+\hR+2\mku\mu\mku\UV^2\,\e^{2\phi}\right].
\label{eq:FirstDerOfLiou}
\end{equation}
Using the Weyl transformation rules of Appendix \ref{app:Weyl} we find that \eqref{eq:FirstDerOfLiou} is actually
invariant under \eqref{eq:SplitSym}:
\begin{equation}
 \frac{\delta\Gamma_\UV^\text{L}}{\delta\phi}[\phi-\sigma;\mku\mku \e^{2\sigma}\hg]
 = \frac{\delta\Gamma_\UV^\text{L}}{\delta\phi}[\phi;\hg].
\end{equation}
\smallskip

\noindent
\textbf{(2) Transformation of the measure \bm{$\mD_\UV^{[\hg]}\chi\,$}:}\\
In appendix \ref{app:WeylTransMeasure} we derive the transformation of the measure under the change $\hg_\mn \to
\hg'_\mn \equiv \e^{2\sigma}\hg_\mn\mku$. It is given by
\begin{equation}
 \mD_\UV^{[\hg']}\chi = \e^{-\Delta\Gi[\hg',\hg]}\; \mD_\UV^{[\hg]}\chi\;.
\label{eq:TransMeasure}
\end{equation}
In \eqref{eq:TransMeasure} the exponent of the crucial transformation factor, $\Delta\Gi[\hg',\hg]$, reads
\begin{equation}
 \Delta\Gi[\hg',\hg]\equiv -\frac{1}{12\pi}\Delta I[\sigma;\hg]+ \frac{1}{2}\ln\bigg(\frac{\hV'}{\hV}\bigg)-
 \frac{\UV^2}{8\pi}\big(\hV'-\hV\big),
\label{eq:DeltaGammaInd}
\end{equation}
with $\Delta I[\sigma;\hg] \equiv \frac{1}{2}\int\td^2x\shg\big[\hD_\mu\sigma\hD^\mu\sigma+\hR\sigma\big]$, and the
volume terms are defined by $\hV\equiv\int\td^2 x\shg$ and $\hV'\equiv\int\td^2 x\sqrt{\hg'}$. The term
$\frac{1}{2}\ln\big(\hV'/\hV\big)$ is present in \eqref{eq:DeltaGammaInd} only if the Laplacians $\hB$ and $\hB'$
have zero modes. The divergent contributions $\frac{\UV^2}{8\pi}\hV$ and $\frac{\UV^2}{8\pi}\hV'$ may be absorbed in
the cosmological constant term of the bare action later on.
\medskip

\noindent
\textbf{(3) Transformation of \bm{$\shg\,\RL\,$}:}\\
It turns out that for the derivation of the searched-for Ward identity it is sufficient to consider the
transformations only up to linear order in $\sigma$, since knowing the behavior under an infinitesimal transformation,
$\hg_\mn\to\hg_\mn+2\hg_\mn\mku\delta\sigma$, already fixes the full transformation law. To find the corresponding
relation for $\shg\,\RL$ we exploit a functional identity which is valid for any functional of the metric:
\begin{equation}
\begin{split}
 F\big[\hg'\big] &= F\big[\e^{2\sigma}\hg\big]=F\big[\hg+2\sigma\mku\hg+\mO(\sigma^2)\big] \\
 &= F[\hg] +2\int\td^2 x\,\sigma(x)\mku\hg_\mn(x)\frac{\delta}{\delta \hg_\mn(x)}F[\mku\hg] + \mO(\sigma^2).
\end{split}
\end{equation}
Thus, the cutoff operator transforms as
\begin{equation}
 \big(\shg\,\RL\big)' = \big(\shg\,\RL\big) +2\int\td^2 x\,\sigma\,\hg_\mn\,\frac{\delta}{\delta \hg_\mn}
 \big(\shg\,\RL\big) + \mO(\sigma^2).
\end{equation}
\smallskip

In a very similar way we can express the transformation of the bare action as
\begin{equation}
 \SB[\chi';\hg'] = \SB[\chi-\sigma\mku;\,\e^{2\sigma}\hg] = \SB[\chi;\hg]+ \int\!\td^2 x\left(2\,\hg_\mn
 \frac{\delta\SB}{\delta\hg_\mn}-\frac{\delta \SB}{\delta\chi}\right)\! \sigma + \mO(\sigma^2).
\label{eq:SBTrans}
\end{equation}
\smallskip

\noindent
\textbf{Resulting transformation of the functional integral:}\\
Now that we have collected all pieces that contribute to the Ward identity, we can divide \eqref{eq:FuncIntLiou} by
$\e^{-\Gamma_\UV^\text{ind}[\hg]}$ and apply the transformations \eqref{eq:SplitSym} and \eqref{eq:ShiftMeasure} to the
remainder:
\begin{equation}
 \e^{-\Gamma_\UV^\text{L}[\phi';\hg']} = \int\mD_\UV^{[\hg']}\chi'\;\e^{-\SB[\chi';\hg']+(\chi'-\phi')
 \cdot(\Gamma_\UV^\text{L})^{(1)}[\phi';\hg'] - \frac{1}{2}(\chi'-\phi')\cdot\RL'(\chi'-\phi')}\,.
\label{eq:FuncIntLiouPrime}
\end{equation}
By eq.\ \eqref{eq:TransLawLiou} the LHS of \eqref{eq:FuncIntLiouPrime} amounts to
\begin{equation}
 \e^{-\Gamma_\UV^\text{L}[\phi';\hg']} =\e^{-\Gamma_\UV^\text{L}[\phi;\hg]}\;\e^{-\frac{b}{8\pi}\,\Delta I[\sigma;\hg]}
 =\e^{-\Gamma_\UV^\text{L}[\phi;\hg]}\Big[\textstyle 1-\frac{b}{16\pi}\,\hR\cdot\sigma+\mO(\sigma^2)\Big].
\label{eq:GammaUVLPrime}
\end{equation}
Using the above list of transformation laws, the RHS of \eqref{eq:FuncIntLiouPrime} becomes
\begin{equation}
\begin{split}
 \int\mD_\UV^{[\hg]}\chi\;\exp\bigg\{&-\Delta\Gi[\hg',\hg] -\SB[\chi;\hg]- \left( \textstyle 2\,\frac{\hg_\mn}{\shg}
 \frac{\delta\SB}{\delta\hg_\mn}-\frac{1}{\shg}\frac{\delta \SB}{\delta\chi}\right)\cdot \sigma \\
 &+(\chi-\phi) \cdot(\Gamma_\UV^\text{L})^{(1)}[\phi;\hg]
 - \frac{1}{2}(\chi-\phi)\cdot\RL(\chi-\phi) \\
 &-(\chi-\phi)\cdot \left(\textstyle \sigma\cdot\frac{\hg_\mn}{\shg}\,\frac{\delta}{\delta \hg_\mn}
 \big(\shg\,\RL\big)\right)(\chi-\phi) + \mO(\sigma^2)
 \bigg\}.
\end{split}
\end{equation}
With $\Delta\Gi[\hg',\hg]= -\frac{1}{24\pi}\,\hR\cdot\sigma +\big(\frac{1}{\hV}-\frac{\UV^2}{4\pi}\big)\int\shg\,\sigma
+\mO(\sigma^2)$ we can expand the exponential in terms of $\sigma$, yielding
\begin{equation}
\begin{split}
 \int\mD_\UV^{[\hg]}&\chi\;\exp\Big\{-\SB[\chi;\hg]+(\chi-\phi) \cdot(\Gamma_\UV^\text{L})^{(1)}[\phi;\hg]
 - \textstyle\frac{1}{2}(\chi-\phi)\cdot\RL(\chi-\phi)\Big\}\\
 &\times\bigg[1+\textstyle\frac{1}{24\pi}\,\hR\cdot\sigma -\Big(\frac{1}{\hV}-\frac{\UV^2}{4\pi}\Big)\int\shg\,\sigma
 - \left( \textstyle 2\,\frac{\hg_\mn}{\shg}
 \frac{\delta\SB}{\delta\hg_\mn}-\frac{1}{\shg}\frac{\delta \SB}{\delta\chi}\right)\cdot \sigma\\
 &\qquad-(\chi-\phi)\cdot \left(\textstyle \sigma\cdot\frac{\hg_\mn}{\shg}\,\frac{\delta}{\delta \hg_\mn}
 \big(\shg\,\RL\big)\right)(\chi-\phi)
 \bigg] + \mO(\sigma^2).
\end{split}
\label{eq:PIPrime}
\end{equation}
Since we know from eq.\ \eqref{eq:FuncIntLiouPrime} that \eqref{eq:GammaUVLPrime} agrees with \eqref{eq:PIPrime}, the
difference of these latter two expressions must vanish: $\eqref{eq:PIPrime} - \eqref{eq:GammaUVLPrime}=0$. This leads
to
\begin{equation}
\begin{split}
 0=&\int\mD_\UV^{[\hg]}\chi\;\exp\Big\{-\SB[\chi;\hg]+(\chi-\phi) \cdot(\Gamma_\UV^\text{L})^{(1)}[\phi;\hg]
 - \textstyle\frac{1}{2}(\chi-\phi)\cdot\RL(\chi-\phi)\Big\}\\
 &\times\int\td^2 x\shgx\,\bigg[\textstyle\left(\frac{b}{16\pi}+\frac{1}{24\pi}\right)\hR(x)
 -\Big(\frac{1}{\hV}-\frac{\UV^2}{4\pi}\Big) - \left( \textstyle 2\,\frac{\hg_\mn}{\shg}
 \frac{\delta\SB}{\delta\hg_\mn}-\frac{1}{\shg}\frac{\delta \SB}{\delta\chi}\right) \\
 &\quad-\int\td^2 y \;(\chi-\phi)(y) \left(\textstyle \frac{\hg_\mn(x)}{\shgx}\,\frac{\delta}{\delta \hg_\mn(x)}
 \big(\shgy\,\RL\big)\right)(\chi-\phi)(y)
 \bigg]\sigma(x) + \mO(\sigma^2).
\end{split}
\label{eq:PITransDiff}
\end{equation}
Upon dividing eq.\ \eqref{eq:PITransDiff} by the normalization factor $\e^{-\Gamma_\UV^\text{L}[\phi;\hg]}$ we observe
that it becomes in fact an identity for the expectation value of $\int\td^2 x\shgx\,\big[\cdots\big]\sigma(x)$.
Furthermore, as we kept $\sigma$ completely arbitrary, we conclude that the expectation value of the square bracket in
\eqref{eq:PITransDiff} must be equal to zero. We thus obtain
\begin{equation}
\begin{split}
 \bigg\langle&\textstyle\left(\frac{b}{16\pi}+\frac{1}{24\pi}\right)\hR(x)
 -\Big(\frac{1}{\hV}-\frac{\UV^2}{4\pi}\Big) - \left( \textstyle 2\,\frac{\hg_\mn}{\shg}
 \frac{\delta\SB}{\delta\hg_\mn}-\frac{1}{\shg}\frac{\delta \SB}{\delta\chi}\right) \\
 &\quad-\int\td^2 y \;(\chi-\phi)(y) \left(\textstyle \frac{\hg_\mn(x)}{\shgx}\,\frac{\delta}{\delta \hg_\mn(x)}
 \big(\shgy\,\RL\big)\right)(\chi-\phi)(y)
 \bigg\rangle=0\,.
\end{split}
\label{eq:PreWI}
\end{equation}

In Appendix \ref{app:WeylTransCutoff} we show that the cutoff contribution to \eqref{eq:PreWI} can be rephrased by two
simple terms involving the propagator $\big(\Gamma_\UV^\text{L}{}^{(2)}+\RL\big)^{-1}$. Moreover, we express the number
$b$, i.e.\ the EAA coupling $\propto \frac{1}{g_*}$ at the NGFP, in terms of the gravitational central charge (cf.\
Chapter \ref{chap:NGFPCFT} in the pure gravity case): We have $b=\frac{2}{3}\mku c$, with $c\equiv\cgr=25$ for the
exponential metric parametrization\footnote{As shown in Section \ref{sec:singleExp2}, for the exponential
parametrization the fixed point value of Newton's constant is cutoff scheme dependent if the cosmological constant is
taken into account, and so is $c$. Based on the optimized cutoff, for instance, we found $c=25.226$. However, when the
cosmological constant is set to zero, we obtain the cutoff independent result $c=25$.} and $c=19$ for the linear
parametrization. With these modifications, we arrive at the main result of this section, the Ward identity for the bare
action $\SB[\chi;\hg]$ concerning Weyl split-symmetry transformations:
\begin{equation}[b]
\begin{aligned}
 \frac{1}{\shgx}\left\langle\frac{\delta \SB}{\delta\chi(x)}\right\rangle
 - 2\,\frac{\hg_\mn(x)}{\shgx}\left\langle\frac{\delta\SB}{\delta\hg_\mn(x)}\right\rangle
 +\frac{c+1}{24\pi}\,\hR(x) +\bigg(\frac{\UV^2}{4\pi}-\frac{1}{\hV}\bigg) & \\[0.6em]
 -\big\langle x\big|\,\RL\,\big(\Gamma_\UV^\text{L}{}^{(2)}+\RL\big)^{-1}\big|x\big\rangle
 -\Tr_\UV\!\Big[\hRL(x)\,\big(\Gamma_\UV^\text{L}{}^{(2)}+\RL\big)^{-1}\Big] &=0\,.
\end{aligned}
\label{eq:WISplitSym}
\end{equation}
The abbreviation $\hRL(x)$ which we introduced in \eqref{eq:WISplitSym} is defined by
\begin{equation}
 \hRL(x) \equiv  \frac{\hg_\mn(x)}{\shgx}\,\frac{\delta}{\delta \hg_\mn(x)}\,\RL\,,
\end{equation}
with $\RL \equiv \RL[\hg_\mn(y)] \equiv \RL(-\hB_y)$, where the argument $y$ corresponds to the variable of spacetime
integration which is implicit in the trace. Note that we kept the regulator function arbitrary up to this point.

Before trying to simplify the Ward identity further by specifying the regulator shape, we would like to mention some
important general aspects.
\medskip

\noindent
\textbf{Remarks}\\[0.3em]
\textbf{(1)} Eq.\ \eqref{eq:WISplitSym} describes the \emph{change of the bare action under infinitesimal Weyl
split-symmetry transformations}, $\chi\to\chi-\sigma$, $\hg_\mn\to\e^{2\sigma}\hg_\mn\,$: According to
\eqref{eq:SBTrans} we have
\begin{equation}
 \Delta\SB[\chi;\hg] \equiv \SB[\chi-\sigma\mku;\,\e^{2\sigma}\hg]-\SB[\chi;\hg] = \int\!\mkern-1mu\td^2 x\mkern-1mu
 \left(\!2\,\hg_\mn \frac{\delta\SB}{\delta\hg_\mn}-\frac{\delta \SB}{\delta\chi}\right)\! \sigma + \mO(\sigma^2).
\end{equation}
Hence, it is the expectation value of this variation that is fixed by the WI. Note that the expectation value is with
respect to the field $\chi$ only.
\smallskip

\noindent
\textbf{(2)} The bare action must strictly satisfy the WI. Therefore, any candidate for $\SB$ we can think of can be
\emph{checked for validity} by inserting it into \eqref{eq:WISplitSym}. In this regard, the WI may be used in addition
to the reconstruction formula \eqref{eq:OneLoopFullMain} in order to determine $\SB$. While this might be a powerful
tool in certain simple cases, the WI seems to be too complex to fully compute the bare action in general since it
involves expectation values which, in turn, depend on the bare action itself.
\smallskip

\noindent
\textbf{(3)}
\emph{The bare action $\SB[\chi;\hg]$ is not Weyl split-symmetry invariant}. This follows immediately from the Ward
identity \eqref{eq:WISplitSym} and the first remark. If $\SB$ were Weyl split-symmetry invariant, it would satisfy
\begin{equation}
 \textstyle \Big\langle\frac{1}{\shgx}\frac{\delta \SB}{\delta\chi(x)}
 - 2\,\frac{\hg_\mn(x)}{\shgx}\frac{\delta\SB}{\delta\hg_\mn(x)}\Big\rangle = 0\,.
\label{eq:IfSplitSym}
\end{equation}
However, the Ward identity dictates that the right-hand side of \eqref{eq:IfSplitSym} must be nonzero: there are terms
proportional to the curvature, a pure number contribution and cutoff terms. The sum of these additional terms is
cutoff dependent and does not equal zero in general. This can already be seen in the vanishing cutoff limit.
\smallskip

\noindent
\textbf{(4)}
\emph{The sum $\Gamma_\UV^\text{ind}[\hg]+\SB[\chi;\hg]$ is not Weyl split-symmetry invariant}: In Section
\ref{sec:BareLiouCon} and in the beginning of the current section we have discussed that the combination
$\Gamma_\UV^\text{ind}[\hg] + \Gamma_\UV^\text{L}[\chi;\hg]$ is invariant under Weyl split-symmetry transformations.
This invariance is a manifestation of the interplay of $\Gamma_\UV^\text{ind}$ and $\Gamma_\UV^\text{L}$, whose changes
under the transformations exactly cancel each other. At linear order in $\sigma$, this requires the transformation
law $\Gamma_\UV^\text{L}[\phi-\sigma;\mkuu\e^{2\phi}\hg] = \Gamma_\UV^\text{L}[\phi;\hg]+\frac{b}{16\pi}\int\td^2 x
\shg\,\hR\mku\sigma$, or, in terms of derivatives w.r.t.\ $\hg_\mn$ and the Liouville field,
\begin{equation}
 \frac{1}{\shg}\frac{\delta \Gamma_\UV^\text{L}}{\delta\phi}- 2\,\frac{\hg_\mn}{\shg}
 \frac{\delta\Gamma_\UV^\text{L}}{\delta\hg_\mn} = -\frac{b}{16\pi}\,\hR \equiv -\frac{c}{24\pi}\,\hR\,.
\end{equation}
Now, if the sum $\Gamma_\UV^\text{ind}[\hg]+\SB[\chi;\hg]$ were Weyl split-symmetry invariant, then $\SB$ would have to
satisfy an equivalent relation: $\frac{1}{\shg}\frac{\delta\SB}{\delta\chi}- 2\,\frac{\hg_\mn}{\shg}
\frac{\delta\SB}{\delta\hg_\mn} \stackrel{!}{=} -\frac{c}{24\pi}\,\hR$. Taking the expectation value of both sides
yields the requirement
\begin{equation}
 \bigg\langle\frac{1}{\shg}\frac{\delta\SB}{\delta\chi}- 2\,\frac{\hg_\mn}{\shg}
 \frac{\delta\SB}{\delta\hg_\mn}\bigg\rangle \stackrel{!}{=} -\frac{c}{24\pi}\,\hR\,.
\label{eq:IfSumSplitSym}
\end{equation}
Clearly, this possibility is ruled out by the Ward identity \eqref{eq:WISplitSym}: There must be
additional terms on the RHS of \eqref{eq:IfSumSplitSym}, in particular additional curvature contributions.
Thus, $\Gamma_\UV^\text{ind}[\hg]+\SB[\chi;\hg]$ is Weyl split-symmetry violating.
\smallskip

\noindent
\textbf{(5)} The pure number terms in \eqref{eq:WISplitSym}, $\frac{\UV^2}{4\pi}$ and $\frac{1}{\hV}$, which stem from
the divergent part of the functional measure and from the zero modes, respectively, can be \emph{absorbed by a
redefinition of the cosmological constant term} in the bare action: Suppose that the bare action can be written as
$\SB[\chi;\hg]=\cl\int\td^2 x\shg+X[\chi;\hg]$, where $X[\chi;\hg]$ comprises all remaining terms. Then
$\Big\langle\frac{1}{\shg}\frac{\delta\SB}{\delta\chi}- 2\,\frac{\hg_\mn}{\shg}\frac{\delta\SB}{\delta\hg_\mn}
\Big\rangle=-2\mku\cl+X\text{-terms}$. Now, let us consider the redefinition
\begin{equation}
 \widetilde{S}_\UV[\chi;\hg] \equiv \left(\cl-\frac{\UV^2}{4\pi}\right)\int\td^2 x\shg + \frac{1}{2}
 \ln\big(\hV/V_0\big) + X[\chi;\hg]\,,
\end{equation}
where $V_0$ is an arbitrary reference volume. This leads to
\begin{equation}
 \bigg\langle\frac{1}{\shg}\frac{\delta\widetilde{S}_\UV}{\delta\chi}- 2\,\frac{\hg_\mn}{\shg}
 \frac{\delta\widetilde{S}_\UV}{\delta\hg_\mn}\bigg\rangle = -2\mku\cl -\bigg(\frac{\UV^2}{4\pi}-\frac{1}{\hV}\bigg)
 + X\text{-terms}.
\label{eq:TransSTilde}
\end{equation}
We conclude that the additional term in \eqref{eq:TransSTilde}, $\Big(\frac{\UV^2}{4\pi}-\frac{1}{\hV}\big)$,
precisely annihilates the corresponding contribution in \eqref{eq:WISplitSym}. Thus, the redefined bare action
$\widetilde{S}_\UV$ satisfies eq.\ \eqref{eq:WISplitSym} with the term $\Big(\frac{\UV^2}{4\pi}-\frac{1}{\hV}\big)$
missing and with $\SB$ replaced by $\widetilde{S}_\UV$. 
\smallskip

\noindent
\textbf{(6)} In Chapter \ref{chap:Bare} we have demonstrated that the EAA actually depends on two scales, as indicated
by the notation $\GkL\mku$. However, since we were interested in the EAA with its couplings at the UV fixed point
throughout the current chapter, we have identified $k$ with $\UV$ here (having in mind the large-$\UV$ limit). This
scale identification thus underlies also our derivation of \eqref{eq:WISplitSym}. The \emph{generalization to the case
of two independent scales $k$ and $\UV$} is straightforward, though. We merely have to repeat all steps that led to
\eqref{eq:WISplitSym}, the only modifications being the replacements $\RL\to\Rk$ and $\Gamma_\UV^\text{L}\to
\Gamma_{k,\UV}^\text{L}$. The Ward identity then reads
\begin{equation}
\begin{aligned}
 \frac{1}{\shgx}\left\langle\frac{\delta \SB}{\delta\chi(x)}\right\rangle
 - 2\,\frac{\hg_\mn(x)}{\shgx}\left\langle\frac{\delta\SB}{\delta\hg_\mn(x)}\right\rangle
 +\frac{c+1}{24\pi}\,\hR(x) +\bigg(\frac{\UV^2}{4\pi}-\frac{1}{\hV}\bigg) & \\[0.3em]
 -\big\langle x\big|\,\Rk\,\big(\GkL^\text{L}{}^{(2)}+\Rk\big)^{-1}\big|x\big\rangle
 -\Tr_\UV\!\Big[\hRk(x)\,\big(\GkL^\text{L}{}^{(2)}+\Rk\big)^{-1}\Big] &=0\,.
\end{aligned}
\end{equation}
\smallskip

In the last two subsections of this chapter we will compute the cutoff terms appearing in \eqref{eq:WISplitSym} for the
optimized regulator and try to make a general statement about the form of the bare action.

%----------------------------------------------------------------------------------------------------------------------
\subsection{The Ward identity for the optimized cutoff}
\label{sec:WIOptCut}
%----------------------------------------------------------------------------------------------------------------------

Upon employing the optimized cutoff, eq.\ \eqref{eq:WISplitSym} reduces to a much simpler identity. Here we briefly
outline the main reason for the special status of the optimized cutoff, while further details and all underlying
calculations can be found in Appendix \ref{app:WeylTransCutoffOpt}.

The second functional derivative of the EAA reads $\Gamma_\UV^\text{L}{}^{(2)} = \ZL\big(-\hB
+2\mku\mu\mku\UV^2\,\e^{2\phi}\big)$, with $\ZL\equiv -\frac{b}{8\pi}$. According to the standard convention, the
cutoff is chosen to have the same prefactor as $-\hB$ in $\Gamma_\UV^\text{L}{}^{(2)}$. Then the optimized
cutoff is given by
\begin{equation}
 \RL\equiv\RL(-\hB)=\ZL\mku\big(\UV^2+\hB\big)\,\theta\big(\UV^2+\hB\big),
\end{equation}
leading to the inverse propagator 
\begin{equation}
 \Gamma_\UV^\text{L}{}^{(2)} + \RL = \ZL\Big[-\hB+2\mku\mu\mku\UV^2\,\e^{2\phi}+\big(\UV^2+\hB\big)\,\theta
 \big(\UV^2+\hB\big)\Big].
\label{eq:GammaLiou2}
\end{equation}
Suppose that this operator acts on an eigenmode of $-\hB$ with the eigenvalue $\omega^2\le\UV^2$. In this case the
$\theta$-function in \eqref{eq:GammaLiou2} evaluates to $1$, and we have, symbolically,
\begin{equation}
 \big(\Gamma_\UV^\text{L}{}^{(2)}+\RL\big)\Big|_{\omega^2\le\UV^2} = \ZL\big(\UV^2+2\mku\mu\mku\UV^2\,\e^{2\phi}\big).
\label{eq:InvPropOpt}
\end{equation}
Now the crucial point is that $\big(\Gamma_\UV^\text{L}{}^{(2)}+\RL\big)$ appears in the WI \eqref{eq:WISplitSym} only
in combination with another cutoff term, either with $\RL$ or with $\hRL(x)$. When using the optimized cutoff, these
terms strictly suppress all those eigenmodes whose squared ``momenta'', i.e.\ eigenvalues of $-\hB$, are larger than
$\UV^2$. Therefore, we can replace $\big(\Gamma_\UV^\text{L}{}^{(2)}+\RL\big)$ in \eqref{eq:WISplitSym} \emph{for all
modes} by the RHS of eq.\ \eqref{eq:InvPropOpt}, not only for the low momentum modes. As a consequence, $\big(
\Gamma_\UV^\text{L}{}^{(2)}+\RL\big)^{-1}$ does no longer contain any differential operators, so, broadly speaking, it
can be pulled out of the trace and out of $\langle x|\cdot|x\rangle$ in \eqref{eq:WISplitSym}. This circumstance is a
tremendous simplification. It allows us to calculate the cutoff terms in the WI at an exact level.
We emphasize that such a simplification occurs only if the optimized cutoff is used.

As worked out in Appendix \ref{app:WeylTransCutoffOpt}, we find that the Ward identity \eqref{eq:WISplitSym} in
case of an optimized cutoff reduces to
\begin{equation}
\begin{aligned}
 \frac{1}{\shgx}&\left\langle\frac{\delta \SB}{\delta\chi(x)}\right\rangle
 - 2\,\frac{\hg_\mn(x)}{\shgx}\left\langle\frac{\delta\SB}{\delta\hg_\mn(x)}\right\rangle
 +\frac{c+1}{24\pi}\,\hR(x) +\bigg(\frac{\UV^2}{4\pi}-\frac{1}{\hV}\bigg) \\
 - \frac{1}{4\pi}\Bigg\{&\frac{\UV^2}{1+2\mku\mu\,\e^{2\mku\phi(x)}} +\frac{1}{6}\;\frac{\hR(x)}{1+2\mku\mu\,
 \e^{2\mku\phi(x)}} - \frac{1}{6}\,\hB\left[\frac{1}{1+2\mku\mu\,\e^{2\mku\phi(x)}}\right] \\
 &+ \frac{1}{30}\,\UV^{-2}\,\frac{1}{1+2\mku\mu\,\e^{2\mku\phi(x)}}\,\hB\mkuu\hR(x)
 -\frac{1}{30}\mku\UV^{-2}\mkuu\hR(x)\,\hB \left[\frac{1}{1+2\mku\mu\,\e^{2\mku\phi(x)}}\right] \\
 &- \frac{1}{30}\mku\UV^{-2}\,\hB\bigg[\frac{\hR(x)}{1+2\mku\mu\,\e^{2\mku\phi(x)}}\bigg]
 - \frac{1}{30}\mku\UV^{-2}\,\hB^2\left[\frac{1}{1+2\mku\mu\,\e^{2\mku\phi(x)}}\right] \Bigg\} = 0 \,.
\end{aligned}
\label{eq:WISplitOpt}
\end{equation}
Note that eq.\ \eqref{eq:WISplitOpt} is an exact result; there are no higher order curvature or derivative terms.
Moreover, we observe that the last two lines of \eqref{eq:WISplitOpt} are suppressed in the limit $\UV\to\infty$.
Therefore, the contribution from the cutoff operator $\RL$ to the WI reduces to only three terms, given by the second
line of \eqref{eq:WISplitOpt}: a pure potential term, a term of first order in the curvature, and a term involving
derivatives of the Liouville field.

In spite of the simplifications entailed by the optimized cutoff, there is still no easy way to solve eq.\
\eqref{eq:WISplitOpt} for $\SB$ since the occurring expectation values depend implicitly on the bare action again.
That means, the WI is a functional integro-differential equation whose solutions cannot be found by our methods in
general. Nonetheless, we will demonstrate in the next subsection that we can draw some important conclusions about
the term in $\SB$ linear in $\hR$ and about the bare potential.

%----------------------------------------------------------------------------------------------------------------------
\subsection{A note on central charges and the bare potential}
\label{sec:WIImp}
%----------------------------------------------------------------------------------------------------------------------

As we have mentioned in the beginning of this section, the starting point of our analysis was given by the induced
gravity action plus a cosmological constant term, $\Gamma_\UV^\text{ind}[g] - \frac{b\mu}{16\pi}\,\UV^2\int\td^2 x\sg$,
see eq.\ \eqref{eq:GammaFullGammaInd} for instance. We have seen in Chapter \ref{chap:NGFPCFT} that
$\Gamma_\UV^\text{ind}[g]$ is linked to a CFT since it can be written as a functional integral over a
conformally invariant action, $\e^{-\Gamma_\UV^\text{ind}[g]}=\int\mD_\UV\chi\;\e^{-S[\chi]}$. Furthermore, it can be
expressed in terms of the functional $I[g]$ (defined in Appendix \ref{app:Weyl}): $\Gamma_\UV^\text{ind}[g]=
\frac{c}{96\pi}\mku I[g]$ (modulo corrections due to topological terms and zero modes), with the corresponding central
charge $c=\cgr$ as defined in Chapter \ref{chap:NGFPCFT}. By decomposing the metric into conformal factor and reference
metric, $g_\mn=\e^{2\phi}\hg_\mn$, the full EAA assumes the form $\Gamma_\UV^\text{ind}[g] - \frac{b\mu}{16\pi}\,\UV^2
\int\td^2 x\sg = \Gamma_\UV^\text{ind}[\hg] + \Gamma_\UV^\text{L}[\phi;\hg]$. 

The point we want to make here is that the central charge
can be read off from three different terms: from the prefactor of $I[g]$ in $\Gamma_\UV^\text{ind}[g]$, from the
prefactor of $I[\hg]$ in $\Gamma_\UV^\text{ind}[\hg]$, as well as from the prefactor of $\int\td^2 x\shg\,\hR\mku\phi$
and of $\int\td^2 x\shg\,\phi\mku(-\hB)\mku\phi$ in $\Gamma_\UV^\text{L}[\phi;\hg]$. As we are focusing on Liouville
theory in this chapter, we would like to extract $c$ from $\Gamma_\UV^\text{L}[\phi;\hg]$, where $c=\frac{3}{2}b$. For
this purpose, the relation
\begin{equation}
 \frac{1}{\shg}\,\frac{\delta \Gamma_\UV^\text{L}}{\delta\phi} - 2\,\frac{\hg_\mn}{\shg}\,
 \frac{\delta\Gamma_\UV^\text{L}}{\delta\hg_\mn}= - \frac{b}{16\pi}\,\hR \equiv -\frac{c}{24\pi}\,\hR
\label{eq:GammaDevToCC}
\end{equation}
seems to be most appropriate to indicate the central charge in our case.

When reconstructing the bare action that belongs to the Liouville EAA, the full action changes according to
$\Gamma_\UV^\text{ind}[\hg] + \Gamma_\UV^\text{L}[\phi;\hg] \to \Gamma_\UV^\text{ind}[\hg] + \SB[\phi;\hg]$.
It is crucial to recognize that \emph{the reconstructed side does not correspond to a CFT} because of the Weyl
split-symmetry violating behavior of the sum $\Gamma_\UV^\text{ind}[\hg] + \SB[\phi;\hg]$, a direct
consequence of the WI \eqref{eq:WISplitSym}, cf.\ remark \textbf{(4)} at the end of subsection \ref{sec:WIDerivation}.
This sum cannot be written as a functional of the full metric alone, and there is no way to express it as a functional
integral over a conformally invariant action. Thus, not being a CFT, \emph{there is no central charge associated to the
bare action}.

Nevertheless, we may analyze to what extent eq.\ \eqref{eq:GammaDevToCC} gets changed during the transition
from the effective to the bare side. By analogy with \eqref{eq:GammaDevToCC} we define $\check{c}$ by
\begin{equation}
 \frac{1}{\shg}\left\langle\frac{\delta \SB}{\delta\chi}\right\rangle
 - 2\,\frac{\hg_\mn}{\shg}\left\langle\frac{\delta\SB}{\delta\hg_\mn}\right\rangle
 \equiv -\frac{\check{c}}{24\pi}\,\hR + \text{remainder}\,,
\label{eq:DefCheckC}
\end{equation}
where ``remainder'' refers to all contributions that do not contain the curvature $\hR$ alone, i.e.\ $\text{remainder}
=\text{const}+\mO(\hR^2)+\mO(\hD_\mu\hR)+\mO(\phi)$, with $\phi=\langle\chi\rangle$. Bearing in mind that $\check{c}$
has no interpretation of a central charge we can, loosely speaking, use the difference $(\check{c}-c)$ as a
measure for the ``deviation of $\SB$ from a CFT''. This difference can be inferred from the WI.

Collecting all terms in eq.\ \eqref{eq:WISplitOpt} proportional to $\hR$ we obtain
\begin{equation}
 -\frac{\check{c}}{24\pi}\,\hR +\frac{c+1}{24\pi}\,\hR - \frac{1}{24\pi}\,\frac{1}{1+2\mku\mu} \,\hR  
 + \text{const}+\mO\big(\hR^2,\hD_\mu\hR,\phi\big) = 0\,.
\end{equation}
Therefore, we conclude
\begin{equation}[b]
 \check{c} = c + 1 - \frac{1}{1+2\mku\mu}\,.
\end{equation}

For the exponential metric parametrization and a nonzero cosmological constant we observe the transition
\begin{align}
 c\approx25.226\;&\longrightarrow\;\check{c}\approx 25.452\,,
\label{eq:CCBTransExp}
\intertext{while setting the cosmological constant to zero by hand ($\lambda_*=0$, $\mu=0$) leads to}
c=25\;&\longrightarrow\;\check{c}= 25\,.
\intertext{For the linear parametrization, on the other hand, we find}
 c=19\;&\longrightarrow\;\check{c}=19.24\,,
\label{eq:CCBTransLin}
\end{align}
in the general case, and $c=19\longrightarrow\check{c}=19$ if the cosmological constant is set to zero.
The numbers in \eqref{eq:CCBTransExp} and \eqref{eq:CCBTransLin} are based on the optimized cutoff again (thus
$c\neq 25$ in \eqref{eq:CCBTransExp}, cf.\ Section \ref{sec:singleExp2}). They can be
used as reference values since the bare action $\SB$ must strictly satisfy the Ward identity, and they should be
reproduced when reconstructing $\SB$ by whatever method. In particular, we can test in principle the validity of the
one-loop approximation \eqref{eq:OneLoopFullMain} in combination with the ans\"{a}tze we made for the bare action in
Sections \ref{sec:BareLiouLiou}--\ref{sec:BareLiouGeneral}.

Evaluating the expectation values on the LHS of \eqref{eq:DefCheckC} is a formidable task in general, even if we
knew the bare action. For the truncations studied in Sections \ref{sec:BareLiouLiou}--\ref{sec:BareLiouGeneral}
the methods we have at hand are in fact not sufficient to compute $\check{c}$. Therefore, we resort to the following
assumption.

We have mentioned that the central charge associated to the EAA $\Gamma_\UV^\text{L}$ can be read off
from the $\hR\mku\phi$-term as well: $c=\frac{3}{2}\mku b$ where $\Gamma_\UV^\text{L}[\phi;\hg]=-\frac{b}{16\pi}\int
\td^2 x\shg\,\hR\mku\phi+\cdots$. In this respect let us define the number $\check{c}'\equiv\frac{3}{2}
\mku\cb$ if the bare action is of the form $\SB[\chi;\hg]=-\frac{\cb}{16\pi}\int\td^2 x\shg\,\hR\mku\chi+\cdots$,
resulting in $\frac{1}{\shg}\,\frac{\delta \SB}{\delta\chi}- 2\,\frac{\hg_\mn}{\shg}\,\frac{\delta\SB}{\delta\hg_\mn}
=-\frac{\check{c}'}{24\pi}\,\hR + \cdots$. Upon taking the expectation value of the latter equation, it might happen
that the dots give rise to yet another contribution to $\hR$. Hence, according to definition \eqref{eq:DefCheckC} we
expect $\check{c}'\neq\check{c}$ in general. Now the assumption we make is that the additional contribution to $\hR$ is
comparatively small, implying $\check{c}'\approx\check{c}$. The validity of this approximation can be checked
within different truncations for the bare action.

In Table \ref{tab:CompBareCC} we list the numbers $\check{c}'$ entailed by the truncation ans\"{a}tze considered in
Sections \ref{sec:BareLiouLiou}, \ref{sec:BareLiouExpSeries} and \ref{sec:BareLiouGeneral} (excluding the truncation
studied in Section \ref{sec:BareLiouPower} which was already ruled out) and compare it to the exact
result $\check{c}$ from the WI.
{%
\renewcommand{\arraystretch}{1.2}
\begin{table}[tp]
\centering
\begin{tabular}{lcccc}
 \hline
   & (WI) & (a)  &  (b) & (c) \\
 \hline
 Exponential parametrization$\;$ & $\;25.45\;$ & $\;25.50\;$ & $\;22.69\;$ & $\;24\;$\\
 Linear parametrization & $19.24$ & $19.32$ & $20.96$ & $18$\\
 \hline
\end{tabular}
\vspace{0.3em}
\caption{Comparison of the numbers $\check{c}$ and $\check{c}'$ obtained in four different approaches, for both the
 exponential and the linear parametrization. The columns refer to: (WI) the number $\check{c}$ from the Ward
 identity, (a) the number $\check{c}'$ from the reconstruction formula in combination with a pure Liouville ansatz for
 the bare action, cf.\ Section \ref{sec:BareLiouLiou}, (b) the number $\check{c}'$ from the reconstruction formula with
 an ansatz for the bare potential that consists of a series of exponentials, cf.\ Section \ref{sec:BareLiouExpSeries},
 (c) the number $\check{c}'$ from the reconstruction formula with a general bare potential, cf.\ Section
 \ref{sec:BareLiouGeneral}. In (a)--(c) we used $\check{c}'\equiv\frac{3}{2}\mku\cb\equiv-12\mku\pi\mku\cx\mku$.}
\label{tab:CompBareCC}
\end{table}%
}%
It is surprising that the deviations among the different approaches are rather small within each parametrization.
Remarkably enough, the numbers $\check{c}'$ resulting from the truncation based on a pure Liouville ansatz lie closest
to their counterparts $\check{c}$. Although this appears to be an argument in favor of the Liouville ansatz for the
bare action, it remains unclear how conclusive it is. It might very well be possible that the other truncations are
more appropriate after all, while only the approximation $\check{c}'\approx\check{c}$ is less good. The main conclusion
we want to draw here is that for all three truncations (Secs.\ \ref{sec:BareLiouLiou}, \ref{sec:BareLiouExpSeries} and
\ref{sec:BareLiouGeneral}) the numbers $\check{c}'$ are ``not too inconsistent'' with the WI.

Finally, we would like to briefly comment on the form of the bare potential favored by the Ward identity. Let us assume
that the bare action is of the form $\SB[\chi;\hg]=\int\td^2 x\shg\,\big[\mku\frac{1}{2}\mku\cZ\,\chi(-\hB)\chi 
-\frac{\check{c}'}{24\pi}\,\hR\mku\chi+\cV(\chi)\big]$. Then we have
\begin{equation}
 \frac{1}{\shg}\,\frac{\delta \SB}{\delta\chi}- 2\,\frac{\hg_\mn}{\shg}\,\frac{\delta\SB}{\delta\hg_\mn}
 = -\cZ\mku\hB\mku\chi- \frac{\check{c}'}{24\pi}\,\hR+ \cV'(\chi)- \frac{\check{c}'}{12\pi}\,\hB\mku\chi- 2\cV(\chi)\,.
\end{equation}
By collecting all those terms in the WI for the optimized cutoff, eq.\ \eqref{eq:WISplitOpt}, that do not contain any
contribution from the curvature or from the derivatives of the field, we obtain\footnote{The reader should not confuse
the bare potential $\cV$ with the volume $\hV\equiv\int\td^2 x\shg\mku$.}
\begin{equation}
\begin{split}
 \big\langle\cV'(\chi)\big\rangle-2\mkuu\big\langle\cV(\chi)\big\rangle &=-\bigg(\frac{\UV^2}{4\pi}-\frac{1}{\hV}\bigg)
 +\frac{1}{4\pi}\,\frac{\UV^2}{1+2\mku\mu\,\e^{2\phi}} + \mO(\hD_\mu\phi) + \mO(\hR) \\
 &= \frac{1}{\hV} - \frac{\mu\mku\UV^2}{2\pi}\,\e^{2\phi} + \frac{\mu^2\mku\UV^2}{\pi}\,\e^{4\phi}
 - \frac{2\mku\mu^3\mku\UV^2}{\pi}\,\e^{6\phi} + \cdots\,.
\end{split}
\label{eq:EqForBarePot}
\end{equation}
As already mentioned previously, the expectation values $\big\langle\cV'(\chi)\big\rangle-2\mkuu\big\langle\cV(\chi)
\big\rangle$ cannot be computed in general by our methods, so we cannot solve \eqref{eq:EqForBarePot} for $\cV(\chi)$.
However, two important statements can be made here. First, the bare action cannot have the pure Liouville form. If it
were so, $\cV(\chi)$ would be proportional to $\e^{2\chi}$, which would lead to $\big\langle\cV'(\chi)\big\rangle
-2\mkuu\big\langle\cV(\chi)\big\rangle=0$, in contradiction to \eqref{eq:EqForBarePot}. Second, the RHS of
\eqref{eq:EqForBarePot} suggests that the bare potential might involve a series of exponentials, providing yet another
justification of the ansatz chosen in Section \ref{sec:BareLiouExpSeries}.

%----------------------------------------------------------------------------------------------------------------------
\chapter{Summary, conclusions and outlook}
\label{chap:Conclusion}
%----------------------------------------------------------------------------------------------------------------------

In this thesis we elaborated several fundamental aspects of Quantum Einstein Gravity. We started by discussing
a number of basic level questions concerning the structure of the space of metrics. In this context we provided a fresh
look at the role played by different metric parametrizations. With regard to the Asymptotic Safety program it was
explained that RG flows and corresponding fixed points can depend on the way the metric is parametrized. For two
parametrizations the compatibility of Asymptotic Safety and background independence was demonstrated within a bimetric
setting.
Furthermore, we constructed a manifestly two-dimensional theory of asymptotically safe gravity which was shown to
correspond to a unitary conformal field theory. This result is a major achievement of this work since it allows for
studying unitarity in combination with Asymptotic Safety for the first time.
Finally, we argued that there is a one-loop relation between the effective average action and the bare action, and we
proposed a strategy for adjusting bare couplings conveniently by means of an appropriate choice of the functional
measure.

Let us summarize our most important results and class their extensibility.
\medskip

\noindent
\textbf{(1) Field parametrizations and RG studies.}
What is the structure of the field space under consideration? How should the field variables be parametrized? Does it
make any physical difference if we change the parametrization? To what extent do RG flows and fixed points depend on
parametrizations? These questions were studied and answered in Chapters \ref{chap:SpaceOfMetrics} and
\ref{chap:ParamDep}. While Chapter \ref{chap:SpaceOfMetrics} concerned the mathematical foundations, Chapter
\ref{chap:ParamDep} focused on the physical implications.

\noindent
\textbf{(1a)} We contrasted the space of metrics, $\mF$, with the space of symmetric rank-$2$ tensor fields, $\SymT$.
While $\SymT$ is a vector space, $\mF$ is a nonlinear, open, path-connected subset of $\SymT$.
Here, the most important advancement consisted in the introduction of a novel connection on the space of metrics: In
local coordinates, $\mF$ at a given spacetime point is isomorphic to $\GLd/\Opq(p,q)$. The canonical connection on this
latter bundle, providing the most natural definition of a horizontal direction, can be lifted to a spacetime dependent
connection on $\mF$.

Geodesics with respect to this proposed connection are parametrized by a simple exponential,
$g_\mn=\bg_{\mu\rho}\mku\big(\e^h\big)^\rho{}_\nu\mku$, where $h_\mn$ is a symmetric tensor field. Every $g_\mn$
described in this way defines a proper metric on $\mF$ with the same signature as $\bg_\mn\mku$.
On the other hand, geodesics with respect to the trivial (flat) connection are parametrized linearly by
$g_\mn = \bg_\mn + h_\mn\mku$. If $h_\mn$ is not further constrained, then $g_\mn$ can ``leave'' the space of metrics:
In this case, the linear split does not parametrize a proper metric on $\mF$ but rather a general symmetric tensor in
$\SymT$.

Hence, the exponential and the linear parametrization describe different objects. They cannot be
converted into each other by field redefinitions, and their use may very well lead to physically inequivalent theories.

\noindent
\textbf{(1b)} In fact, RG flows are parametrization dependent. Within the Einstein--Hilbert truncation we found that
the coordinates as well as further properties of the non-Gaussian fixed point depend on the choice of parametrization.
This study comprises the first nonperturbative RG analysis based on the exponential parametrization.

Numerical results can most conclusively be discussed in $d=2+\ve>2$ dimensions since the fixed point value of the
dimensionless Newton constant becomes universal (scheme independent) in the limit of small $\ve$. Leaving
the cosmological constant aside for a moment, we derived the universal results $g_* = \frac{3}{38}\mku\ve$ for the
linear parametrization, and $g_* = \frac{3}{50}\mku\ve$ for the exponential parametrization.
We uncovered a close relation between these fixed point values and the critical central charge
$c^\text{crit}=25$ known from conformal field theory and bosonic string theory. For the exponential parametrization
we reproduced $c^\text{crit}=25$, whereas the linear split gives rise to $c^\text{crit}=19$, indicating that the
exponential parametrization might be more appropriate in the 2D limit.

\noindent
\textbf{(1c)} Within a bimetric setting we demonstrated that Asymptotic Safety can be reconciled with the requirement
for background independence. To this end, we singled out a specific RG trajectory, characterized by (i) an
asymptotically safe behavior in the UV limit and (ii) the property that background couplings are located at a fixed
point in the IR limit. Then the non-gauge part of the effective average action at vanishing RG scale becomes
independent of the background metric. We showed that such trajectories exist for both parametrizations considered.

\noindent
\textbf{(1d) Outlook.}
Although having presented arguments in favor of the use of the exponential parametrization in and near $d=2$ dimensions,
particularly in view of comparisons with 2D conformal field theory, the linear parametrization might be suited
equally well for the application to other cases. Thus, we do not promote any general preference. Our message is merely
that the choice of parametrization does indeed matter. As long as it is unclear what the fundamental variables of
quantum gravity are, one should be open towards either kind of parametrization.

By now it is an active research area to find modified parametrizations that are specifically designed for particular
applications, their motivation ranging from a reduction of technical complexity, to a simplification of Ward
identities, to a simpler treatment of gauge degrees of freedom. For instance, constructing an explicit parametrization
on the basis of the Vilkovisky--DeWitt formalism in combination with RG methods might turn out an extremely useful tool
for studying quantum gravity in a gauge independent way.

Furthermore, it would be interesting to work out in a future project whether different parametrizations actually refer
to different universality classes. In the present context this would mean that there is a second pure gravity fixed
point suitable for the Asymptotic Safety program, but with different properties such as critical exponents.
Investigating this possibility would require considering enlarged truncation spaces as compared with the ones covered
in this thesis.

Finally, advanced studies on background independence should take into account the full geometric split-Ward identities.
We have argued that the (untruncated) gravitational effective average action depends only seemingly on two metrics
independently since a change of the dynamical metric can in principle be compensated for by a variation of the
background metric and vice versa. This link opens up the potential possibility to formulate the complete theory in
terms of one single metric and a redefined effective average action which would then be background independent by
construction but whose evolution equation might not have the familiar form of the FRGE.
\medskip

\noindent
\textbf{(2) The unitary conformal field theory behind 2D Asymptotic Safety.}
In Chapters \ref{chap:EHLimit} and \ref{chap:NGFPCFT} we investigated whether the theory defined directly at the fixed
point belonging to an asymptotically safe RG trajectory in $d=2$ dimensions represents a conformal field theory, and if
so, whether it admits unitary representations of the corresponding Virasoro algebra. Chapter \ref{chap:EHLimit} focused
on establishing the form of the action functional at the fixed point, whereas Chapter \ref{chap:NGFPCFT} addressed its
conformal properties and unitarity.

\noindent
\textbf{(2a)} We argued that, within the Einstein--Hilbert truncation in $d=2+\ve>2$ dimensions, the decisive part of
both the effective average action and the bare action is of the form $\frac{1}{\ve}\int\td^{2+\ve}\sg\, R$. In the
limit $\ve\to 0$ we observed a kind of compensation between the integral and the prefactor: While the integral tends
to a trivial, metric independent term, the prefactor $1/\ve$ tends to infinity. We demonstrated that the essential part
of the common limit actually remains finite. Our key result is that the local Einstein--Hilbert action in $d>2$
dimensions approaches Polyakov's nonlocal induced gravity action in the 2D limit.

\noindent
\textbf{(2b)} With the analysis described in \textbf{(2a)} we paved the way for a detailed study of the 2D fixed point
theory. The most important contribution to the corresponding effective average action functional was shown to be given
by $\frac{c}{96\pi}\int\td^2 x \sg\,R\,\Box^{-1} R\mku$, with $c=25-N$ ($c=19-N$) for the exponential (linear)
parametrization. Here, $N$ denotes the number of additionally included scalar or fermionic matter fields. From
conformal field theory considerations we know that such an induced gravity action can be interpreted as the effective
action of a conformally invariant theory with central charge $c$.

\noindent
\textbf{(2c)} Provided that the number of matter fields is not too large, $N\le 24$, this conformal field theory at the
fixed point is indeed unitary as the associated Virasoro algebra with $c\ge 1$ possesses representations with a
positive state space. This result constitutes the first proof of unitarity in an asymptotically safe theory of quantum
gravity.

Finally, we showed that unitarity is closely connected to the conformal factor instability. The theory can be unitary
only if the kinetic term of the conformal factor has the ``wrong'' sign. We argued, however, that this observation is
not only physically acceptable but even expected since that sign is crucial for the universal attractivity of gravity.

\noindent
\textbf{(2d) Outlook.}
In the introduction (Chapter \ref{chap:Intro}) we raised the question if there is a theory of the gravitational field
which is asymptotically safe and background independent and unitary at the same time. For the bimetric truncation
considered in Chapter \ref{chap:ParamDep}, Asymptotic Safety was shown to be reconcilable with background independence,
and our 2D fixed point theory example demonstrated the compatibility of Asymptotic Safety and unitary. It remains
an open problem, however, whether all three properties can be combined in a single theory. We conjecture that sticking
with the 2D setting is the most promising way to deal with this problem.
In any case, such an investigation would call for a bimetric treatment and the inclusion of Ward identities, though.
As yet, we do not know if a fully bimetric fixed point theory can be interpreted as a conformal field theory.

The next step would consist in generalizing the arguments to $d=4$ dimensions. Many open questions could be studied in
this context, about the possibility to unmask a 4D conformal field theory at a nontrivial RG fixed point or about the
form of the corresponding action, for example. Anyhow, one should bear in mind that a theory may very well be unitary
without featuring the conformal symmetry. Thus, proving unitarity might require employing additional techniques after
all.
\medskip

\noindent
\textbf{(3) Reconstructing the functional integral.}
In the FRG approach to asymptotically safe gravity, calculations are usually based upon the effective average action
rather than a bare action. Chapters \ref{chap:Bare}, \ref{chap:FullReconstruction} and \ref{chap:BareLiouville} were
devoted to the question how the corresponding functional integral, comprising the functional measure and the bare
action, can be reconstructed from the effective average action.

\noindent
\textbf{(3a)} We started in Chapter \ref{chap:Bare} by specifying the measure and deriving a general one-loop relation
between the bare action and the effective average action. It was demonstrated that, after having expanded the relation
in terms of basis functionals, the one-loop approximation actually becomes an exact equation in the large cutoff limit
for certain expansion terms.

As an example, we considered the Einstein--Hilbert truncation of the effective average action and reconstructed the
associated bare action by making an Einstein--Hilbert ansatz as well. We proved the existence of a nontrivial fixed
point in the bare sector, irrespective of the dimension and the underlying functional measure. Over and above, we
revealed the intriguing opportunity to adjust bare couplings conveniently by means of a suitable choice of the measure.
For instance, the bare cosmological constant at the fixed point can be made vanish in any dimension, and in $2+\ve$
dimensions one can achieve that the fixed point values of the effective and the bare Newton constant agree.

\noindent
\textbf{(3b)} In Chapter \ref{chap:FullReconstruction} we applied these result to the 2D conformal fixed point theory
discussed in points \textbf{(2b)} and \textbf{(2c)} and reconstructed the corresponding functional integral. The
induced gravity action part of the partition function was shown to be independent of the number of included matter
fields. This has the surprising consequence that the total central charge of the gravity+matter system vanishes.
Besides, it leads to a decoupling of the conformal factor from observables under the functional integral and a
quenching of the KPZ relations. Finally, we compared and contrasted 2D asymptotically safe quantum gravity with
noncritical string theory and the causal dynamical triangulation approach.

\noindent
\textbf{(3c)} Chapter \ref{chap:BareLiouville} was dedicated to the reconstruction of the bare action in Liouville
theory. We found that, if the effective average action is of the Liouville type, the most auspicious ansatz made for
the bare action includes a series of exponentials of the form $\e^{2n\phi}\mku$. Our results were supported by
specifically derived Ward identities.

\noindent
\textbf{(3d) Outlook.} In particular cases the approximative character of the one-loop reconstruction relation may
prevent access to the correct form of the bare action or set us on the wrong track when trying to find suitable
truncation ans\"{a}tze. This may happen if higher loop orders become too significant. In this regard, it would be
interesting to assess the range of validity of the reconstruction formula in more detail. Furthermore, we do not
exclude the possibility that the measure and the regularization prescription can be modified in such a way that one can
derive an exact relation. As discussed in Chapter \ref{chap:Bare}, this can be done for scalar fields under certain
conditions, whereas the understanding of the general case is still vague, in particular for the gravitational field.

Nevertheless, in future works the bare actions reconstructed by means of the one-loop relation can be used to compare
the FRG results to other approaches and to gain further insight into the underlying microscopic systems. In Liouville
theory, for instance, this may guide lattice simulations into the right way to guessing a qualified discretized bare
theory and taking the continuum limit in a suitable manner. Moreover, for theories involving 2D asymptotically safe
gravity coupled to matter we laid the foundations for further studies concerning the quenching of the KPZ relations
and its possible implications for related physical models.

\appendix

%----------------------------------------------------------------------------------------------------------------------
\chapter{Variations of geometric quantities}
\label{app:Variations}
%----------------------------------------------------------------------------------------------------------------------

In this appendix we list variation formulae for all geometric quantities relevant to this work, i.e.\ for the metric
determinant and the various curvature tensors. Here we consider general variations of the metric,
$g_\mn\mapsto g_\mn+\delta g_\mn\mku$. (The special case of Weyl variations implies a couple of simplifications,
see Appendix \ref{app:Weyl}.) Throughout this thesis we employ the following definitions:
\begin{align}
 R_{\phantom{e}\rho\mu\nu}^\sigma &= \p_\mu\Gamma_{\nu\rho}^\sigma-\p_\nu\Gamma_{\mu\rho}^\sigma
 + \Gamma_{\mu\tau}^\sigma \Gamma_{\nu\rho}^\tau - \Gamma_{\nu\tau}^\sigma \Gamma_{\mu\rho}^\tau\,,\\
 R_{\mu\nu} &= R_{\mu\sigma\nu}^\sigma\,, \\
 R &= g^{\mu\nu}R_{\mu\nu}\,.
\end{align}
The Riemann tensor satisfies the identities
\begin{alignat}{2}
 [D_\mu,D_\nu]V^\sigma &= R_{\phantom{e}\rho\mu\nu}^\sigma V^\rho &&\qquad\mbox{for vectors,}\\
 [D_\mu,D_\nu]A_\rho &= -R_{\phantom{e}\rho\mu\nu}^\sigma A_\sigma &&\qquad\mbox{for 1-forms,}\\
 [D_\mu,D_\nu]H_{\alpha\beta} &= -R_{\phantom{e}\alpha\mu\nu}^\tau H_{\tau\beta}
 -R_{\phantom{e}\beta\mu\nu}^\tau H_{\alpha\tau} &&\qquad\mbox{for $(0,2)$-tensors,}
\end{alignat}
which can be used to derive its variation in a straightforward way. Here, we merely present the result, though.
We have:
\begin{align}
 \delta g^{\mu\nu} = {}& -g^{\mu\alpha}g^{\nu\beta}\delta g_{\alpha\beta} \,,
\\[0.7em]
 \delta g	= {}& g\, g^{\mu\nu}\delta g_\mn \,,
\\[0.7em]
 \delta \sqrt{g} = {}& \textstyle\frac{1}{2}\,\sqrt{g}\, g^\mn\delta g_\mn \,,
\\[0.7em]
 \delta^2 \sqrt{g} = {}& \textstyle\frac{1}{2}\sqrt{g}\left(\,\frac{1}{2}\,g^\mn g^{\alpha\beta}\,\delta g_\mn
 \delta g_{\alpha\beta} - g^{\mu\alpha}g^{\nu\beta}\delta g_{\alpha\beta} \delta g_\mn\right)\,,
\\[0.7em]
 \delta\Gamma_{\mu\nu}^\sigma = {}& \textstyle \frac{1}{2}\, g^{\sigma\beta}\left(D_\mu \delta g_{\nu\beta}
 + D_\nu \delta g_{\mu\beta}-D_\beta \delta g_{\mu\nu}\right)\,,
\\[0.7em]
 \delta R_{\phantom{e}\rho\mu\nu}^\lambda = {}& \textstyle \frac{1}{2}\,\big(-R_{\phantom{e}\rho\mu\nu}^\sigma
 g^{\lambda\alpha}\delta g_{\alpha\sigma}+R_{\phantom{e}\sigma\mu\nu}^\lambda g^{\sigma\alpha}\delta g_{\alpha\rho}
 + g^{\lambda\alpha} D_\mu D_\rho\delta g_{\alpha\nu}\nonumber\\
 & \phantom{\textstyle \frac{1}{2}\,\big(} - g^{\lambda\alpha} D_\nu D_\rho\delta g_{\alpha\mu}+D_\nu D^\lambda
 \delta g_{\mu\rho} - D_\mu D^\lambda\delta g_{\nu\rho}\big)\,,
\displaybreak\\
 \delta R_\mn = {}& \textstyle \frac{1}{2}\,\big(-g^{\sigma\beta} R_{\phantom{e}\mu\sigma\nu}^\alpha \delta
 g_{\alpha\beta} + R^\alpha{}_\nu\mku \delta g_{\mu\alpha} + D^\sigma D_\mu\delta g_{\nu\sigma}\nonumber\\
 & \phantom{\textstyle \frac{1}{2}\,\big(} - g^{\sigma\alpha} D_\nu D_\mu\delta g_{\sigma\alpha}+D_\nu D^\sigma
 \delta g_{\sigma\mu} -D_\sigma D^\sigma\delta g_{\nu\mu}\big)\,,
\\[0.7em]
 \delta R = {}& -R^\mn\delta g_\mn+D^\mu\big(D^\nu\delta g_{\nu\mu}-g^{\nu\alpha} D_\mu\delta g_{\nu\alpha}\big)\,,
\\[0.7em]
 \delta^2 R = {}& g^{\sigma\alpha}R^\mn\delta g_{\mu\alpha}\delta g_{\sigma\nu}-R^{\mu\nu\rho\sigma}
 \delta g_{\nu\rho}\delta g_{\mu\sigma} + 2\mku g^{\sigma\alpha}\delta g_\mn D^\mu D^\nu\delta g_{\sigma\alpha}
 \nonumber\\
 & + 2\mku g^{\mu\alpha}g^{\nu\beta}\delta g_{\alpha\beta} D_\sigma D^\sigma \delta g_\mn
 -3\mku g^{\mu\alpha}\delta g_{\alpha\nu} D^\nu D^\sigma \delta g_{\sigma\mu} \nonumber\\
 & - g^{\nu\alpha}\delta g_{\mu\alpha} D^\sigma D^\mu \delta g_{\sigma\nu}
 - 2\mku g^{\mu\alpha}(D^\nu\delta g_{\alpha\nu})(D^\sigma\delta g_{\sigma\mu}) \nonumber\\
 & - g^{\nu\alpha}(D^\sigma\delta g_{\mu\alpha})(D^\mu\delta g_{\sigma\nu})
 + 2\mku g^{\sigma\alpha}(D^\mu\delta g_\mn)(D^\nu\delta g_{\sigma\alpha}) \nonumber\\
 & + \textstyle\frac{3}{2}\mku g^{\mu\alpha}g^{\nu\beta}(D_\sigma\delta g_\mn)(D^\sigma\delta g_{\alpha\beta})
 - \textstyle\frac{1}{2}\mku g^\mn g^{\alpha\beta}(D_\sigma\delta g_\mn)(D^\sigma\delta g_{\alpha\beta}) \,.
\end{align}
Note that indices are lowered and raised by $g_{\mu\nu}$ and $g^{\mu\nu}$, respectively, and $g$ denotes the
determinant of the metric.
The above variations are used in Appendix \ref{app:TransBetaExp} in order to derive the Hessians belonging to two
different truncations of the effective average action, encountered in the RG analysis of Chapter \ref{chap:ParamDep}.

%----------------------------------------------------------------------------------------------------------------------
\chapter{Matrix representation of operators in curved spacetime}
\label{app:OperatorRep}
%----------------------------------------------------------------------------------------------------------------------

In this appendix we briefly summarize some important conventions for the representation of operators and functional
derivatives in curved spacetime.
\medskip

\noindent
\textbf{(1) Orthogonality and completeness in curved spacetime}.
In curved space, $\frac{1}{\sbg}\,\delta(x-y)$ replaces the $\delta$-function of flat space. Orthogonality and
completeness relations thus involve the background metric $\bg_\mn\mku$, too:
\begin{align}
 \langle x|y\rangle &= \frac{1}{\sbgy}\,\delta(x-y) \,, \\
 \mathds{1} &= \int\dd x \sbgx\; |x\rangle\langle x| \,.
\end{align}
%\smallskip

\noindent
\textbf{(2) Matrix representation of operators}.
Let $\mO$ be a local operator. Then its matrix representation $\mO_{xy}$ in position space (differential operator
representation) reads
\begin{equation}
 \mO_{xy}\equiv \langle x|\mO|y\rangle \equiv\mO\,\frac{1}{\sbgy}\,\delta(x-y)\equiv\frac{1}{\sbgy}\,\mO\,\delta(x-y).
\label{eqn:opxy}
\end{equation}
In the middle and the RHS we assumed that $\mO\equiv\mO_{(x)}^\text{diff-op}$ is a differential operator acting on $x$
so that it commutes with $\sbgy$. In this setting the identity operator is given by
\begin{equation}
 I_{xy}\equiv\mathds{1}_{xy}\equiv\langle x|\mathds{1}|y\rangle= \langle x|y\rangle = \frac{1}{\sbgy}\,\delta(x-y) \,.
\end{equation}
We abbreviate $\int_y \equiv \int\dd y\sbgy$ and $\psi_x=\psi(x)$ in the following.
Using $\psi(x)=\langle x|\psi\rangle$, equation \eqref{eqn:opxy} is consistent with
$\int_y \mO_{xy}\psi_y=\int_y\langle x|\mO|y\rangle\langle y|\psi\rangle = \langle x|\mO|\psi\rangle =(\mO\psi)_x
= \mO\psi(x)$. As an example for equation \eqref{eqn:opxy}, let us consider the operator $\mO=\bar{\Box}$ acting on a
field inside an integral. In this case we have
\begin{equation}
 \int_y \bar{\Box}_{xy}\phi_y = \int\dd y\sbgy \frac{1}{\sbgy}\,\bar{\Box}\,\delta(x-y)\phi(y) = \bB\mku \phi(x)\,.
\end{equation}
%\medskip

\noindent
\textbf{(3) Relation to functional derivatives of action functionals}.
We define
\begin{equation}
 \Gamma^{(2)}\equiv\Gamma^{(2)}(x,y) \equiv(\Gamma^{(2)})_{xy} \equiv \frac{1}{\sqrt{\bg(x)\bg(y)}}\;
 \frac{\delta^2\Gamma}{\delta\phi(x)\delta\phi(y)}\;.
\end{equation}
Considering the EAA $\Gamma_k\equiv\Gamma_k[\phi]=\frac{1}{2}\int\dd x\sbgx\, \phi(x)(-\bar{\Box})\phi(x)$ for
instance, we have
\begin{equation}
\Gamma_k^{(2)}(x,y) = -\bar{\Box}_{xy}=-\frac{1}{\sbgy}\,\bar{\Box}\,\delta(x-y) \,,
\end{equation}
and according to the above convention we write $\Gamma_k^{(2)}=-\bar{\Box}$.
\medskip

\noindent
\textbf{(4) Functional traces}.
We define the functional trace by
\begin{equation}
 \Tr(\mO)\equiv \int\dd x\sbgx\,\langle x|\mO|x\rangle \equiv \int_x \mO_{xx} \,.
\label{eq:DefOpTrace}
\end{equation}
Note that if there is a nontrivial internal index space, eq.\ \eqref{eq:DefOpTrace} must be replaced by $\Tr(\mO)\equiv
\int_x \tr\mO_{xx}$, where `$\tr$' denotes the trace over internal indices.
\medskip

\noindent
\textbf{(5) Notation for inverse operators}.
Using the relations $\phi(x)=\frac{1}{\sbgx}\frac{\delta W_k}{\delta J(x)}$ and $J(x)=\frac{1}{\sbgx}\frac{\delta
\tilde\Gamma_k}{\delta \phi(x)}\mku$, with $\tilde\Gamma_k = \Gamma_k+\frac{1}{2}\int\dd x \sbg\,\phi\, \Rk\, \phi$,
and thus $\tilde\Gamma^{(2)}=\Gamma^{(2)}+\Rk$ (cf.~Section \ref{sec:FRG}), yields the relation
\begin{align}
\int_y \big(W^{(2)}_k\big)_{xy}&\big(\Gamma^{(2)}_k+\Rk\big)_{yz} = \int_y \big(W^{(2)}_k\big)_{xy}
 \big(\tilde\Gamma^{(2)}_k\big)_{yz} \nonumber \\
 &= \int\dd y\sbgy\,\frac{1}{\sqrt{\bg(x)\bg(y)}}\;\frac{\delta^2 W_k}{\delta J(x)\delta J(y)}\,
 \frac{1}{\sqrt{\bg(y)\bg(z)}}\;\frac{\delta^2\tilde\Gamma_k}{\delta\phi(y)\delta\phi(z)} \nonumber\\
 &= \int\dd y\sbgy\,\frac{1}{\sbgy}\,\frac{\delta\phi(x)}{\delta J(y)}\,\frac{1}{\sbgz}\,
 \frac{\delta J(y)}{\delta\phi(z)} 
 = \frac{1}{\sbgz}\;\frac{\delta\phi(x)}{\delta \phi(z)} \nonumber\\ 
 &= \frac{1}{\sbgz}\;\delta(x-z)\,.
\end{align}
Since $\frac{1}{\sbg}\,\delta(x-y)$ is the $\delta$-function of curved space (i.e.\ the identity), we can
write
\begin{equation}
 \big(\Gamma^{(2)}_k+\Rk\big)^{-1} (x,y) \,=\, W_k^{(2)}(x,y) \,,
\label{eq:RelBetWkAndGammakInv}
\end{equation}
where $\big(\Gamma^{(2)}_k+\Rk\big)^{-1} (x,y) \equiv \big\langle x\big | \big(\Gamma^{(2)}_k+\Rk\big)^{-1} \big|
y\big\rangle\mku$ (which is possibly nonlocal).

With $\langle\chi(x)\rangle=\phi(x)$, the connection between eq.\ \eqref{eq:RelBetWkAndGammakInv} and the expectation
value $\langle\chi(x)\chi(y)\rangle$ is given by
\begin{equation}
 \langle\chi(x)\chi(y)\rangle -\phi(x)\phi(y) = W_k^{(2)}(x,y) \equiv \frac{1}{\sqrt{\bg(x)\bg(y)}}\;
 \frac{\delta^2 W_k}{\delta J(x)\delta J(y)}\;,
\end{equation}
or, equivalently
\begin{equation}
 \langle\chi(x)\chi(y)\rangle -\phi(x)\phi(y) = \big(\Gamma^{(2)}_k+\Rk\big)^{-1} (x,y) \;.
\end{equation}

%----------------------------------------------------------------------------------------------------------------------
\chapter{Heat kernel expansion}
\label{app:Heat}
%----------------------------------------------------------------------------------------------------------------------

In this appendix we introduce the heat kernel and present an expansion formula for its trace.
For derivations and further details we refer the reader to the pertinent literature, for instance
\cite{Vassilevich2003,DeWitt1965,DeWitt2003,BV87,BV90a,BV90b,Vilkovisky1992,Avramidi2000,GSZ11}.

Let $M$ be a manifold of dimension $d$ and $H$ a second order partial differential operator on $M$ of the Laplace type,
that is, covariant derivatives in $H$ are contracted with the metric, and the internal index structure of the second
derivative term is trivial. Then $H$ can be written in the form
\begin{equation}
 H =  \Id\mku\Box + E\,,
\label{eq:LaplaceWithE}
\end{equation}
where the identity in $\Id\mku\Box\equiv \Id\mku g^\mn D_\mu D_\nu$ corresponds to the internal index space, and $E$ is
an endomorphism, i.e.\ a (generally matrix-valued) function on $M$ acting on internal indices.

We define the heat kernel $K\equiv K(s;x,y)$ as a solution to the heat equation
\begin{equation}
 \frac{\p K}{\p s} = H\mku K\,,\qquad
 \text{with initial condition}\quad K(s=0;x,y)={\textstyle\frac{1}{\sg}}\,\delta(x-y).
\label{eq:HeatEquation}
\end{equation}
The formal solution to \eqref{eq:HeatEquation} reads
\begin{equation}
 K(s;x,y) = \e^{s H}\Big[{\textstyle\frac{1}{\sg}}\,\delta(x-y)\Big]
 \equiv \big\langle x\big|\e^{s H}\big|y\big\rangle,
\end{equation}
or short, $K=\e^{s H}$. It possesses a so-called early time expansion, a power series in terms of $s$
around $s=0$. While this expansion is nonlocal (as it involves geodesic distances and their derivatives), there
exists a local early time expansion once the coincidence limit $y\rightarrow x$ is taken:
\begin{equation}
 K(s;x,x) = \left(\frac{1}{4\pi s}\right)^{\!d/2}\,\sum\limits_{n=0}^\infty s^n\mku \tr a_n(x).
\label{eq:HeatExpCoincidence}
\end{equation}
The first three of the so-called Seeley--DeWitt coefficients in eq.\ \eqref{eq:HeatExpCoincidence} are given by
\begin{align}
 a_0(x) &= \Id\,,
\label{eq:coeff0}\\
 a_1(x) &= P\,,
\label{eq:coeff1}\\
 a_2(x) &= \textstyle \frac{1}{180}\left ( R_{\mu\nu\rho\sigma}R^{\mu\nu\rho\sigma}-R_{\mu\nu}R^{\mu\nu}
  +\Box R\right)\Id +\frac{1}{2}P^2+\frac{1}{12}\mathcal{R}_{\mu\nu}\mathcal{R}^{\mu\nu}+ \frac{1}{6}\Box P\,,
\label{eq:coeff2}
\end{align}
where $P\equiv E+\frac{1}{6}\mku R\,\Id$, and the commutator curvature $\mathcal{R}_{\mu\nu}\equiv[D_\mu,D_\nu]$
is associated with the full (spacetime plus gauge etc.) connection. Note that ``tr'' in eq.\
\eqref{eq:HeatExpCoincidence} denotes the trace over internal indices only.

As we will see in a moment, the trace of the heat kernel is of particular importance since it can be used to compute
very general operator traces. Let $f$ be a square integrable function on $M$. Then from \eqref{eq:HeatExpCoincidence}
follows that
\begin{equation}
 \Tr\Big[f\,\e^{s H}\Big] = \left(\frac{1}{4\pi s}\right)^{\!d/2}\,
 \sum\limits_{n=0}^\infty s^n\int\dd x\sg\, \tr a_n(x)f(x).
\end{equation}
This result can be employed to calculate traces of functions of $H$, or more general, to calculate
$\Tr\big[f\,W(-H)\big]$, where $W$ is a function that decreases sufficiently fast regarding convergence of the trace.
For this purpose, we write $W(-H)$ as a Laplace transform, $W(-H)=\int_0^\infty \td s\;\e^{s H}\, \widetilde{W}(s)$,
insert the early time expansion for $\Tr\left[f\,\e^{s H}\right]$, and perform the $s$-integration for each term in
the series separately. This yields
\begin{equation}[b]
 \Tr\big[f\,W(-H)\big] = \left(\frac{1}{4\pi}\right)^{\!d/2}\,
 \sum\limits_{n=0}^\infty  Q_{d/2-n}[W] \int\dd x\sg\, \tr a_n(x)f(x).
\label{eq:Heat1}
\end{equation}
Here we introduced the ``$Q$-functionals'' \cite{Reuter1998} (generalized Mellin transforms) $Q_m[W]$, defined by
\begin{equation}
 Q_m[W] \equiv
\begin{cases}\displaystyle
  \frac{1}{\Gamma(m)}\,\int_0^\infty dz\; z^{m-1}\, W(z)\quad\quad &\text{for }m>0, \\[1em]
  (-1)^{-m}\,W^{(-m)}(0) &\text{for }m\leq 0.
\end{cases}
\label{eq:Mellin}
\end{equation}
If there is an additional uncontracted covariant derivative, the first terms of the heat kernel expansion are given by
\cite{GSZ11}
\begin{equation}
 \Tr\big[f\,D_\mu\mku W(-H)\big] = \left(\textstyle\frac{1}{4\pi}\right)^{\!d/2}\,Q_{d/2-1}[W]\int\!\sg\,
 f\mku \tr\big[\textstyle \frac{1}{12}\mku D_\mu R + \frac{1}{2}\mku D_\mu E
 -\frac{1}{2}\mku D^\nu\mathcal{R}_{\nu\mu} \big] + \cdots
\label{eq:Heat2}
\end{equation}

For the special case of a vanishing endomorphism in \eqref{eq:LaplaceWithE} we obtain
\begin{equation}
 \Tr\big[f\,W(-\Box)\big] = \left(\textstyle\frac{1}{4\pi}\right)^{\!d/2}\,\tr(\Id) \left\{
  Q_{d/2}[W]\int\!\sg\,f + {\textstyle\frac{1}{6}}\,Q_{d/2-1}[W]\int\!\sg\,R\,f \right\},
\label{eq:Heat3}
\end{equation}
up to terms of higher order in the curvature.

%----------------------------------------------------------------------------------------------------------------------
\chapter{Cutoff shape functions and threshold functions}
\label{app:Cutoffs}
%----------------------------------------------------------------------------------------------------------------------

In this appendix we list three possible cutoff shape functions which are used throughout this thesis: the optimized
cutoff \cite{Litim2001}, an exponential cutoff \cite{LR02a,LR02c}, and the sharp cutoff\cite{RS02}. We define threshold
functions as in Ref.\ \cite{Reuter1998} and evaluate them for the cutoffs considered. (See Ref.\ \cite{Nink2011} for a
more detailed discussion.)

The cutoff operator $\Rk$ can be written in terms of a dimensionless function $\Rz$:
\begin{equation}
 \Rk(-\Box) = \mZ_k\mkuu k^2\mku\Rz\big(-\Box/k^2\big),
\label{eq:RkInRz}
\end{equation}
where the (possibly matrix-valued) function $\mZ_k$ is usually chosen to agree with the wave function renormalization,
and $\Rz$ is referred to as the \emph{cutoff shape function}. Since $\Rk$ is meant to be an IR cutoff, we impose the
conditions
\begin{alignat}{2}
 &\text{(i)} \qquad\quad &&\Rz(0)=1\,, \label{eq:CutoffCond1}\\
 &\text{(ii)}\qquad\quad &&\lim_{\mathclap{z\to\infty}}\;\Rz(z)=0 \,, \label{eq:CutoffCond2}
\end{alignat}
where the latter is often combined with the requirement that the decrease be sufficiently fast in order that mainly IR
modes are suppressed. Specifically, we consider:
\begin{itemize}
 \item \textbf{The optimized cutoff}
 \begin{equation}
  \Rz(z)\equiv(1-z)\mkuu\theta(1-z).
 \end{equation}
 \item \textbf{The ``$\bm{s}$-class exponential cutoff''}
 \begin{equation}
  \Rz(z;s)\equiv\frac{sz}{e^{sz}-1}\,,\qquad s>0.
 \end{equation}
 \item \textbf{The sharp cutoff}
 \begin{equation}
  \Rk(-\Box)\equiv \tilde{R}\,\theta\big(1+\Box/k^2\big),
 \end{equation}
 where $\tilde{R}$ has mass dimension $2$, and the limit $\tilde{R}\to\infty$ is to be taken in the end (i.e.\ after
 evaluating traces / performing momentum integrals that involve the cutoff). Note that \emph{the sharp cutoff is
 not a standard regulator} since it cannot be written in the form \eqref{eq:RkInRz} and it is not finite at vanishing
 argument.
\end{itemize}

\section{Threshold functions and their properties}

Throughout this thesis we use the threshold functions $\Phi_n^p(w)$ and $\widetilde{\Phi}_n^p(w)$ defined by
\begin{align}
 \Phi_n^p(w) &\equiv \frac{1}{\Gamma(n)}\int_0^\infty \td z\; z^{n-1}\frac{\Rz(z)-z\mku \Rz{}'(z)}{\left[z+\Rz(z)
 +w\right]^p}\;, \label{eq:Phinp}\\[0.5em]
 \widetilde{\Phi}_n^p(w) &\equiv \frac{1}{\Gamma(n)}\int_0^\infty \td z\; z^{n-1}\frac{\Rz(z)}{\left[z+\Rz(z)
 +w\right]^p}\;,\label{eq:tPhinp}
\end{align}
for $n>0$, as well as $\Phi_0^p(w)\equiv \lim_{n\to 0}\Phi_n^p(w)$ and $\tPhi_0^p(w)\equiv \lim_{n\to 0}\tPhi_n^p(w)$.
(For the sharp cutoff these definitions have to be expressed in terms of $\Rk$, cf.\ \cite{RS02}.) Based on the
conditions \eqref{eq:CutoffCond1} and \eqref{eq:CutoffCond2} it is possible to deduce the following \emph{general,
universal (i.e.\ cutoff shape independent) properties} (see e.g.\ Ref.\ \cite{Nink2011} for proofs):
\begin{align}
 &\bullet\qquad \lim_{w\to\infty}\Phi_n^p(w)=0,\quad \lim_{w\to\infty}\tPhi_n^p(w)=0 ,\\
 &\bullet\qquad \vphantom{\lim_{w\to\infty}\tPhi_n^p}\textstyle \frac{\td}{\td w}\Phi_n^p(w)
 = -p\mkuu\Phi_n^{p+1}(w),\quad \frac{\td}{\td w}\tPhi_n^p(w) = -p\mkuu\tPhi_n^{p+1}(w) ,\qquad\\
 &\bullet\qquad \Phi_0^p(w) = (1+w)^{-p}\,,\quad \tPhi_0^p(w) = (1+w)^{-p}\,,
 \vphantom{\lim_{w\to\infty}\tPhi_n^p}\\
 &\bullet\qquad \vphantom{\lim_{w\to\infty}\tPhi_n^p}\Phi_n^{n+1}(0) = \textstyle \frac{1}{\Gamma(n+1)} \,.
\end{align}

For the \emph{optimized cutoff} all threshold functions can be evaluated analytically:
\begin{align}
 \Phi_n^p(w)^\text{opt} &= \frac{1}{\Gamma(n+1)}(1+w)^{-p}\,,\\
 \tPhi_n^p(w)^\text{opt} &= \frac{1}{\Gamma(n+2)}(1+w)^{-p}\,.
\end{align}

When the \emph{exponential cutoff} is employed, the threshold functions can be expressed in terms of
\emph{polylogarithms}. We refrain from listing the lengthy results here, but refer to Ref.\ \cite{LR02c} instead.

For the \emph{sharp cutoff} the threshold functions have to be redefined in terms of $\Rk$ before they can be computed
analytically \cite{RS02}. This results in
\begin{alignat}{3}
 \Phi_n^p(w)^\text{sh}&= \frac{1}{\Gamma(n)}\,\frac{1}{p-1}\,\frac{1}{(1+w)^{p-1}}\,,\quad
 \tPhi_n^p(w)^\text{sh}&&=0, &&\text{for }p>1,\\
 \Phi_n^1(w)^\text{sh}&= -\frac{1}{\Gamma(n)}\,\ln(1+w)+\vp_n\,,\quad
 \tPhi_n^1(w)^\text{sh}&&=\frac{1}{\Gamma(n+1)}\,,\quad &&\text{for }p=1,
\end{alignat}
where the $\vp_n$'s are constants of integration that can be chosen conveniently.

%----------------------------------------------------------------------------------------------------------------------
\chapter{The exponential parametrization and the space of metrics}
\label{app:ExpParam}
%----------------------------------------------------------------------------------------------------------------------

In this appendix we want to establish the connection between the exponential metric parametrization and the space of
metrics. As we will see, this requires a distinction between Euclidean and Lorentzian metrics. Therefore, we specify
metric signatures explicitly in the following. Recall that the space of metrics is defined by
\begin{equation}
 \mF_{(p,q)} \equiv \Big\{g\in\SymT\,\Big |\; g \text{ has signature } (p,q)\Big\},
\end{equation}
where $\SymT$ is the space of symmetric rank-$2$ tensor fields. In what follows, we compare $\mFpq$ to the
space that is generated by the exponential parametrization, henceforth denoted by $\tFpq(\bg)$, i.e.\ the set of all
those tensors having the representation $\bg\,\e^{\mku\bg^{-1} h}$ for a given background metric $\bg\mku$:
\begin{equation}
 \tFpq(\bg) \equiv \Big\{ g = \bg\,\e^{\mku\bg^{-1} h}\;\Big|\; h\in\SymT\Big\}\quad
 \text{with }\bg\in\mFpq \,.
\end{equation}
Here and in the following, we use the (matrix form of the) local coordinate representation of metrics, and we do not
write the spacetime dependence explicitly. This is admissible due to the pointwise character of the exponential
parametrization, cf.\ Chapter \ref{chap:SpaceOfMetrics}, in particular Section \ref{sec:DetConnections}.

Ultimately, we would like to find out whether $\tFpq(\bg)\subset\mFpq$ and $\mFpq\subset\tFpq(\bg)$. That is, we
investigate (a) if the exponential parametrization gives rise to a metric with signature $(p,q)$ again, and (b) if
every signature-$(p,q)$ metric can be parametrized by $\bg\,\e^{\mku\bg^{-1} h}$. We will show that $\tFpq(\bg)=\mFpq$
holds only for positive definite (Euclidean) and negative definite metrics. For indefinite (Lorentzian) metrics, on the
other hand, we will see that $\tFpq(\bg)\subset\mFpq$, but $\mFpq\not\subset\tFpq(\bg)$.

Let us start with a remark. Proving that $\bg\,\e^{\mku\bg^{-1} h}$ represents a proper metric requires proving
symmetry and positive definiteness. We emphasize that these statements are not obvious: The product of two symmetric
positive definite matrices is in general neither positive definite nor symmetric. In addition, a hypothetical proof of
$\mFpq\subset\tFpq(\bg)$ would require determining $h$ such that $g=\bg\,\e^{\mku\bg^{-1} h}$ for $g$ and $\bg$ given,
but in general only little is known about existence and uniqueness of real logarithms of products of matrices,
and $\bg^{-1}h = \ln(\bg^{-1}g)$ might not exist.

The following four lemmas turn out to be useful, though. They finally lead to the main results of this appendix,
Theorems \ref{theo:ExpParamIsMetric}--\ref{theo:NotSurNotIn}.

\begin{lemma}
\label{lem:PosDefMatSol}
 Let $C$ be a real symmetric positive definite matrix. Then there exists a unique real symmetric solution $H$ to the
 equation $C=e^H$.
\end{lemma}
\noindent
\textbf{Proof.}\\
\emph{Existence:} With $C\in\text{Sym}_{n\times n\,}$, there exists an orthogonal matrix $S\in\On$ and a diagonal
matrix $\Lambda=\diag(\lambda_1,\ldots,\lambda_n)$, with $\{\lambda_i\}$ the eigenvalues of $C$, such that
$C=S^T\Lambda S$. Positive definiteness of $C$ implies that all $\lambda_i$ are positive. Now, let us set $H\equiv S^T
\diag(\ln\lambda_1,\ldots,\ln\lambda_n)S$. Then $H$ is real and symmetric. Exponentiating $H$ yields
\begin{align*}
 e^H = S^T e^{\diag(\ln\lambda_1,...,\ln\lambda_n)}S
 = S^T\diag(\lambda_1,...,\lambda_n)S =C,
\end{align*}
proving the existence of a real symmetric solution.\\
\emph{Uniqueness:} Assume that $H$ is a real symmetric matrix satisfying $C=e^H$. Assume that $H'$ is another real
symmetric matrix with the same exponential, $C=e^{H'}$. Due to their symmetry, there are matrices $O\in\On$ and $O'\in
\On$ together with the diagonal matrices $D=\diag(d_1,\ldots,d_n)$ and $D'=\diag(d_1',\ldots,d_n')$, where
$d_i$ are the eigenvalues of $H$ and $d_i'$ are the eigenvalues of $H'$, such that $H=O^TDO$ and $H'={O'}^TD'O'$. Then
we have $C=e^H=e^{O^TDO}=O^Te^DO$, and, similarly, $C={O'}^Te^{D'}O'$. Equating these expression leads to $e^D\big
(O{O'}^T\big)=\big(O{O'}^T\big)e^{D'}$, or, rewritten,
\begin{equation}
 e^D U=U e^{D'} ,
\label{eq:UD}
\end{equation}
with $U=O{O'}^T\in\On$. The matrix entries on the LHS of \eqref{eq:UD} read
\begin{equation}
 \big( e^{D} U \big)_{ij} = \sum\limits_{k=1}^{n} e^{d_{i}} \delta_{i k} u_{k j}
 = e^{d_{i}} u_{i j} \, ,
\end{equation}
and, analogously for the RHS, $\big( U e^{D'} \big)_{ij} = e^{d'_{j}} u_{i j}$. For any pair $(i,j)$ this gives the
relation $(e^{d_{i}} - e^{d'_{j}} ) u_{i j} = 0$. Since all $d_{i}$ are real, we conclude that $( d_{i} - d'_{j} )
u_{i j} = 0$. Back to matrix form again, this yields $D U -  U D' = 0$. Re\-instating $U=O{O'}^T$ and rearranging
finally results in
\begin{equation}
H=O^TDO={O'}^TD'O'=H' \, ,
\end{equation}
which proves the uniqueness of $H$. \hfill $\Box$

\begin{lemma}
\label{lem:RootsPol}
 The $n$ roots of a polynomial $p(z)=\sum_{k=0}^n a_k\mku z^k$ of degree $n$ depend continuously on the coefficients
 $\{a_k\}$.
\end{lemma}
\noindent
For a proof, see for instance Refs.\ \cite{Marden1966,Zedek1965}.

\begin{lemma}
\label{lem:MatContEigen}
 The eigenvalues of a matrix depend continuously on the matrix entries.
\end{lemma}
\noindent
\textbf{Proof.}\\
Follows immediately from Lemma \ref{lem:RootsPol} and the fact that the coefficients of the characteristic polynomial
of a matrix depend continuously on the matrix entries. \hfill $\Box$

\begin{lemma}
\label{lem:Culver}
 Let $C$ be a real square matrix. Then there exists a real solution $X$ to the equation $C = e^X$ if and only if $C$
 is nonsingular and each elementary divisor (Jordan block) of $C$ belonging to a negative eigenvalue occurs an even
 number of times.
\end{lemma}
\noindent
For a proof, see Ref.\ \cite{Culver1966}.
\bigskip

Now, let us come back to the space of metrics and the exponential parametrization. We will exploit the above lemmas to
reveal a number of important properties. Let us begin with a theorem which is valid for all signatures.

\begin{theorem}
\label{theo:ExpParamIsMetric}
 Let $h\in\SymT$ and $\bg\in\mFpq$. Then $g$ defined by $g\equiv \bg\, e^{\bg^{-1}h}$ belongs to $\mFpq$, too.
 Equivalently, if $\bg\in\mFpq$, then
 \begin{equation}[b]
  \tFpq(\bg)\subset\mFpq\quad\forall\,p,q\,.
 \end{equation}
 This means that the exponential parametrization gives rise to a proper metric.
\end{theorem}
\noindent
\textbf{Proof.}\\
We have to show that $g=\bg\, e^{\bg^{-1}h}$ is symmetric and has signature $(p,q)$.\\
\emph{Symmetry:}
\begin{equation}
 g^T=\big(e^{\bg^{-1}h}\big)^T\bg^T=e^{h^T(\bg^{-1})^T}\bg=e^{\bg\,\bg^{-1}h\,\bg^{-1}}\,\bg
 =\bg\, e^{\bg^{-1}h}\,\bg^{-1}\bg=\bg\, e^{\bg^{-1}h}=g \, .
\label{eq:AppProofSym}
\end{equation}
\emph{Signature:} Let us define the $s$-dependent matrix
\begin{equation}
g(s)=\bg \, e^{s\,\bg^{-1}h} \,,
\label{eq:gs}
\end{equation}
with $s\in\mathds{R}$. We notice that $g(s)$ depends continuously on $s$. Thus, $g(s)$ interpolates continuously
between $\bg$ and $g$:
\begin{equation}
g(0)=\bg \,,\qquad g(1)=g \,.
\label{eq:g0and1}
\end{equation}
By analogy with eq.\ \eqref{eq:AppProofSym} we conclude that $g(s)$ is symmetric, too. Hence, all its eigenvalues are
real for all $s$. Obviously, $g(s)$ has the same eigenvalues as $\bg$ at $s=0$, while it has the same eigenvalues as
$g$ at $s=1$. Now, let us consider the determinant of $g(s)$. Using the matrix relation $\det\exp(M)=\exp\Tr(M)$ we
find
\begin{equation}
\det\big(g(s)\big)=\det\left(\bg \, e^{s\,\bg^{-1}h}\right)
=\det(\bg)\det\left( e^{s\,\bg^{-1}h}\right)=\det(\bg)\, e^{s\Tr(\bg^{-1}h)}.
\end{equation}
Since $s\Tr(\bg^{-1}h)\in\mathds{R}$, we have $e^{s\Tr(\bg^{-1}h)}>0$. Therefore, the determinants of $g(s)$ and $\bg$
have the same sign, for all $s$. In particular, $\det(g(s))\neq 0$ for all $s$. That is, according to $\det(g(s))=
\lambda_1^s\mku\lambda_2^s\cdots\lambda_n^s$ (where $\lambda_i^s$ denotes the $i$-th eigenvalues of $g(s)$), no
eigenvalue $\lambda_i^s$ can get zero, regardless of which value of $s$ is taken:
\begin{equation}
 \lambda_i^s \neq 0 \quad \forall\,s\,.
\end{equation}
From Lemma \ref{lem:MatContEigen} we know that the $\lambda_i^s$ depend continuously on $g(s)$, so they depend
continuously on $s$. As a consequence, the $\lambda_i^s$ cannot change their signs when varying $s$ from $0$ to $1$.
That means that the total number of positive (negative) eigenvalues $\lambda_i^s$ at $s=0$ agrees with the total number
of positive (negative) eigenvalues $\lambda_i^s$ at $s=1$. With \eqref{eq:g0and1} we conclude that $g$ and
$\bg$ have the same number of positive (and negative) eigenvalues, so they have the same signature. \hfill $\Box$

\bigskip

For the final part of this appendix a distinction between Euclidean and Lorentzian signatures becomes necessary. More
precisely, positive definite and negative definite metrics fall into one class, $(p,q)=(d,0)$ and $(p,q)=(0,d)$,
respectively, while indefinite metrics with signature $(p,q)$, $p\ge 1$, $q\ge 1$, fall into the second class. We would
like to answer the question whether a symmetric tensor $h\in\SymT$ exists for all $g\in\mFpq$ and all $\bg\in\mFpq$
such that $g=\bg\, e^{\mku\bg^{-1}h}$.

\begin{theorem}
 Let $g\in\mFpq$ and $\bg\in\mFpq$ with $(p,q)=(d,0)$ or $(p,q)=(0,d)$, corresponding to positive or negative definite
 metrics, respectively. Then there exists a unique $h\in\SymT$ satisfying $g=\bg\, e^{\mku\bg^{-1}h}$. Therefore,
 \begin{equation}[b]
  \mF_{(d,0)}=\widetilde{\mF}_{(d,0)}(\bg) \quad\text{and}\quad \mF_{(0,d)}=\widetilde{\mF}_{(0,d)}(\bg)\,,
 \end{equation}
 where the correspondence is one-to-one. This means that every positive definite (Euclidean) metric and every negative
 definite metric can be represented uniquely by the exponential parametrization, and that the exponential
 parametrization uniquely defines a proper metric.
\end{theorem}
\noindent
\textbf{Proof.}\\
We know already from Theorem \ref{theo:ExpParamIsMetric} that $\tFpq(\bg)\subset\mFpq$. Moreover, for each $h\in\SymT$
and $\bg\in\mFpq$ there is one and only one $g\in\mFpq$ such that the defining equation given by the exponential
parametrization is satisfied (since it is already solved for $g$). Hence, it remains to be shown that for each
$g\in\mFpq$ and $\bg\in\mFpq$ there exists a unique $h\in\SymT$ satisfying $g=\bg\, e^{\mku\bg^{-1}h}$.

\noindent
\textbf{The case }\bm{$(p,q)=(d,0).$}\\
\emph{Existence:} Since $\bg$ is symmetric and positive definite, we can define $\bg^{1/2}$ to be the (unique)
principal square root. Note that $\bg^{1/2}$ is real and symmetric again. The key idea is to rewrite the exponential
parametrization as follows:
\begin{equation}
 g = \bg\, e^{\bg^{-1}h} = \bg\, e^{\bg^{-1/2}\bg^{-1/2}h\,\bg^{-1/2}\bg^{1/2}}
 = \bg^{1/2} e^{\bg^{-1/2}h\,\bg^{-1/2}}\bg^{1/2},
\end{equation}
leading to
\begin{equation}
\bg^{-1/2} g\, \bg^{-1/2} = e^{\bg^{-1/2}h\,\bg^{-1/2}}.
\label{eq:paramMatrixModified}
\end{equation}
We observe that the LHS of equation \eqref{eq:paramMatrixModified} is real and symmetric. Furthermore, it is positive
definite, as follows from
\begin{equation}
 z^T\big(\bg^{-1/2} g\, \bg^{-1/2}\big)z=(\bg^{-1/2}z)^T g\, (\bg^{-1/2}z)=y^Tg\mku y>0\,,
\end{equation}
for $y=\bg^{-1/2}z$ and $z\in\mathbb{R}^d$ arbitrary. Thus, Lemma \ref{lem:PosDefMatSol} is applicable to eq.\
\eqref{eq:paramMatrixModified}: There exists a unique real symmetric matrix $H$ satisfying $\bg^{-1/2} g\, \bg^{-1/2}
=e^H$. Setting $h\equiv\bg^{1/2}H\,\bg^{1/2}$ and noting that $h$ is real and symmetric proves the existence.

\noindent
\emph{Uniqueness:} Since there is more than one square root of $\bg$ in general, it remains to be shown that the $h$
constructed above does not depend on the choice
of the root. Let us assume that there exists another symmetric solution $h'$ corresponding to
another square root $(\bg^{1/2})'$, i.e.\ $g = \bar{g}\, e^{\bar{g}^{-1} h'}$.
In the manner of equation \eqref{eq:paramMatrixModified} we rewrite again
\begin{equation}
 \bg^{-1/2} g\, \bg^{-1/2} = e^{\bg^{-1/2} h'\, \bg^{-1/2}}
\stackrel{!}{=} e^{\bg^{-1/2} h\, \bg^{-1/2}},
\end{equation}
where we use the principal root $\bg^{1/2}$ on all sides. We already know from Lemma \ref{lem:PosDefMatSol} that the symmetric
logarithm of the LHS is unique. Therefore, the exponents on the RHS have to agree,
$\bg^{-1/2} h'\, \bg^{-1/2} = \bg^{-1/2} h\, \bg^{-1/2}$, and
finally $h'=h$, completing the proof of uniqueness.

\noindent
\textbf{The case }\bm{$(p,q)=(0,d).$}\\
Let us define $\tilde{g}\equiv -g$ and $\tilde{\bg}\equiv -\bg$. Then both $\tilde{g}$ and $\tilde{\bg}$ are positive
definite. Thus, we can apply the above results concerning the case $(p,q)=(d,0)$: There exists a unique $\tilde{h}\in
\SymT$ satisfying
\begin{equation}
 \tilde{g}=\tilde{\bg}\, e^{\mku\tilde{\bg}^{-1}\tilde{h}}\,.
\end{equation}
After setting $h\equiv -\tilde{h}$ we conclude that $g=\bg\, e^{\mku\bg^{-1}h}$ and that this $h$ is unique.
\hfill $\Box$

\begin{theorem}
\label{theo:NotSurNotIn}
 Let $g\in\mFpq$ and $\bg\in\mFpq$ with $p\ge 1$, $q\ge 1$, corresponding to indefinite (i.e.\ Lorentzian) metrics.
 Then, in general there exists no $h\in\SymT$ such that $g=\bg\, e^{\mku\bg^{-1}h}$ is satisfied. Equivalently,
 \begin{equation}[b]
  \mFpq\not\subset\tFpq\quad \text{for } p\ge 1,\, q\ge 1\,.
 \end{equation}
 This means that the map
 \begin{equation}
  \SymT \to \mFpq,\quad h \mapsto g=\bg\, e^{\mku\bg^{-1}h}\,,
 \label{eq:Maphg}
 \end{equation}
is not surjective for $p\ge 1$, $q\ge 1$. Moreover, it is also not injective for $p\ge 1$, $q\ge 1$.
\end{theorem}
\noindent
\textbf{Proof.}\\
Non-surjectivity of \eqref{eq:Maphg} immediately implies $\mFpq\not\subset\tFpq$. Thus, in order to prove Theorem
\ref{theo:NotSurNotIn} we only have to find counterexamples against surjectivity and injectivity. As argued above it
is sufficient to specify these examples as matrices, i.e.\ as the local representation of rank-$2$ tensors at a fixed
spacetime point.

\noindent
\emph{Surjectivity:}
We rewrite the exponential parametrization as
\begin{equation}
 \bg^{-1} g = \e^{\mku\bg^{-1} h}\;.
\label{eq:RewrittenExpParam}
\end{equation}
The idea is to find $\bg$ and $g$ such that the LHS of \eqref{eq:RewrittenExpParam} cannot be expressed as an
exponential. For this purpose let us consider the following matrices:
% To use \hdashline load arydshln (\RequirePackage{arydshln})
\begin{align}
 \bg &=\begin{array}{c@{\!\!\!}l}
 \begin{pmatrix}
  1\\
  &-1\\
  &&1\\
  &&&\ddots\\
  &&&&1\\
  &&&&&-1\\
  &&&&&&\ddots\\
  &&&&&&&-1
 \end{pmatrix}
 &
 \begin{array}[c]{@{}l@{\,}l}
  \vphantom{1}& \\
  \vphantom{-1}& \\
  \left. \begin{array}{c} \vphantom{1} \\ \vphantom{\ddots} \\ \vphantom{1}\end{array}\right\} & \text{$p-1$ times} \\
  \left. \begin{array}{c} \vphantom{-1} \\ \vphantom{\ddots} \\ \vphantom{-1}\end{array}\right\} & \text{$q-1$ times}
 \end{array}
 \end{array}
\label{eq:bgCounterex}
 \\[1em]
 g &=\begin{array}{c@{\!\!\!}l}
 \begin{pmatrix}
  -2\\
  &1\\
  &&1\\
  &&&\ddots\\
  &&&&1\\
  &&&&&-1\\
  &&&&&&\ddots\\
  &&&&&&&-1
 \end{pmatrix}
 &
 \begin{array}[c]{@{}l@{\,}l}
  \vphantom{1}& \\
  \vphantom{-1}& \\
  \left. \begin{array}{c} \vphantom{1} \\ \vphantom{\ddots} \\ \vphantom{1}\end{array}\right\} & \text{$p-1$ times} \\
  \left. \begin{array}{c} \vphantom{-1} \\ \vphantom{\ddots} \\ \vphantom{-1}\end{array}\right\} & \text{$q-1$ times}
 \end{array}
 \end{array}
\end{align}
Then the product $\bg^{-1} g$ is given by
\begin{equation}
 \bg^{-1} g = \begin{array}{c@{\!\!\!}l}
 \begin{pmatrix}
  -2\\
  &-1\\
  &&1\\
  &&&\ddots\\
  &&&&1
 \end{pmatrix}
 &
 \begin{array}[c]{@{}l@{\,}l}
  \vphantom{-2}& \\
  \vphantom{-1}& \\
  \left. \begin{array}{c} \vphantom{1} \\ \vphantom{\ddots} \\ \vphantom{1}\end{array}\right\} & \text{$p+q-2$ times}
 \end{array}
 \end{array}
\end{equation}
Since this matrix is diagonal, it is already in Jordan normal form, so we can read off its Jordan blocks. There is
one block belonging to the eigenvalue $-2$, one block belonging to the eigenvalue $-1$ and one block belonging to the
eigenvalue $1$. Thus, according to Lemma \ref{lem:Culver} there is no \emph{real} solution to the equation
$\bg^{-1} g = \e^X$ because both of the two negative eigenvalues of $\bg^{-1} g$ occur an odd number of times. As a
consequence, there is no $h\in\SymT$ satisfying $\bg^{-1} g = \e^{\mku\bg^{-1} h}$. This proves the non-surjectivity of
the map \eqref{eq:Maphg} for $p\ge 1$, $q\ge 1$.

\noindent
\emph{Injectivity:}
Let us consider the same $\bg$ as given in eq.\ \eqref{eq:bgCounterex}, together with the following family of symmetric
matrices parametrized by $\alpha\in\mathbb{R}\mku$:
\begin{equation}
 h_\alpha = \begin{array}{c@{\!\!\!}l}
 \begin{pmatrix}
  0&\alpha&0&\cdots&0\\
  \alpha&0&&&\\
  0&&\ddots&&\vdots\\
  \vdots&&&\ddots&\\
  0&&\cdots&&0
 \end{pmatrix}
 &
 \begin{array}[c]{@{}l@{\,}l}
  \vphantom{-2}& \\
  \vphantom{-1}& \\
  \left. \begin{array}{c} \vphantom{1} \\ \vphantom{\ddots} \\ \vphantom{1}\end{array}\right\} & \text{$p+q-2$ times}
 \end{array}
 \end{array}
\end{equation}
Then we find $\bg^{-1} h_\alpha = \alpha\mku J_{12}$, where $J_{12}$ is amongst the generators of the rotation group
$\Od$, with $1,2$ denoting the variant coordinates. The matrix exponential of $\bg^{-1} h_\alpha$ amounts to
\begin{equation}
 \e^{\mku\bg^{-1} h_\alpha} = \begin{pmatrix}
  \cos\alpha&-\sin\alpha\\
  \sin\alpha&\cos\alpha\\
  &&1\\
  &&&\ddots\\
  &&&&1
 \end{pmatrix}
\end{equation}
This gives rise to an $\alpha$-dependent metric $g_\alpha\mkuu$:
\begin{equation}
 g_\alpha = \bg\,\e^{\mku\bg^{-1} h_\alpha} =
 \begin{array}{c@{\!\!\!}l}
 \begin{pmatrix}
  \cos\alpha&-\sin\alpha\\
  -\sin\alpha&-\cos\alpha\\
  &&1\\
  &&&\ddots\\
  &&&&1\\
  &&&&&-1\\
  &&&&&&\ddots\\
  &&&&&&&-1
 \end{pmatrix}
 &
 \begin{array}[c]{@{}l@{\,}l}
  \vphantom{1}& \\
  \vphantom{-1}& \\
  \left. \begin{array}{c} \vphantom{1} \\ \vphantom{\ddots} \\ \vphantom{1}\end{array}\right\} & \text{$p-1$ times} \\
  \left. \begin{array}{c} \vphantom{-1} \\ \vphantom{\ddots} \\ \vphantom{-1}\end{array}\right\} & \text{$q-1$ times}
 \end{array}
 \end{array}
\label{eq:InjCounterex}
\end{equation}
Obviously, eq.\ \eqref{eq:InjCounterex} defines a \emph{periodic} solution $g_\alpha\in\mFpq$. There are infinitely
many $\alpha$ that lead to the same $g_\alpha$. In particular, we have $g_\alpha=\bg$ for all $\alpha \in \{2\pi\mku k
\,|\, k \in\mathbb{Z}\}$.
This completes the proof of non-injectivity of \eqref{eq:Maphg} for $p\ge 1$, $q\ge 1$. \hfill $\Box$

\medskip

More illustrative counterexamples against surjectivity and injectivity on the basis of eqs.\
\eqref{eq:bgCounterex}--\eqref{eq:InjCounterex} can be found in the body of this thesis in Section \ref{sec:EuLor}.
\medskip

While all proofs in this appendix made use of purely algebraic arguments, they are reviewed in a differential-geometric
language in Section \ref{sec:GroupTheory}, revealing the basic origin of the corresponding statements.

%----------------------------------------------------------------------------------------------------------------------
\chapter{Split-Ward identities for the geometric effective average action}
\label{app:SplitWard}
%----------------------------------------------------------------------------------------------------------------------

In this appendix we derive the split-Ward identities for the geometric effective average action $\Gamma_k\mku$,
introduced in Section \ref{sec:Applications}. These identities imply that the dependence of $\Gamma_k$ on its arguments
is intertwined: A variation of $\Gamma_k$ with respect to the background field, say, $\bp$, can be compensated for by
a variation with respect to the dynamical field, say, $\vp$. The subsequent derivation is independent of the
underlying field space connection. In this sense it generalizes References \cite{MR10} (flat field space connection in
a conformally reduced setting) and \cite{Pawlowski2003} (Vilkovisky--DeWitt connection).
\smallskip

\noindent
\textbf{(1) The defining functional integral.}
Our starting point is given by the functional integro-differential equation determining $\Gamma_k\mku$, where we employ
a modified version according to point \textbf{(4)} of Section \ref{sec:Applications} in order to define $\Gamma_k$ in a
covariant manner. Here, ``covariance'' means ``covariance with respect to field space $\mF\mkuu$''. Since we would like
to keep the discussion as general as possible, we allow for an extra $\bp$-dependence in $\Gamma_k\mku$.
Our arguments are phrased in terms of the ``tilde-version'' of $\Gamma_k$ (cf.\ Section \ref{sec:Applications}),
$\tilde{\Gamma}_k[h;\bp]\equiv \Gamma_k\big[\vp[h;\bp],\bp\big]$, but we omit the tilde in the
following since the semicolon notation, $\Gamma_k[h;\bp]$, is already sufficient to distinguish it from
$\Gamma_k\big[\vp,\bp\big]$. At the level of $\Gamma_k\big[\vp,\bp\big]$ the extra
$\bp$-dependence is explicitly visible, while for $\Gamma_k[h;\bp]$ it is encoded in the split-Ward identities.

Note that all tangent vectors are elements of $T_{\bp}\mF$ now. Generalizing point \textbf{(4)} of Section
\ref{sec:Applications}, the source couples no longer to the tangent vector to the geodesic connecting the dynamical
field $\vp$ to the integration variable $\hvp$, but rather to $\big(\hat{h}-h\big)$, where $\hat{h}\equiv
\hat{h}[\bp,\hvp]$ denotes the tangent vector to the geodesic connecting $\bp$ to $\hvp$, and $h$ is the independent
argument of $\Gamma_k$ which is interpreted as a tangent vector to the geodesic connecting $\bp$ to $\vp$. That is,
we can write the source term (in DeWitt index notation) as $S^\mathrm{source} = J_a\mku \big(\hat{h}^a-h^a\big)
\equiv J_a\mku \big(\hat{h}^a[\bp,\hvp]-h^a\big)$, where $\hat{h}$ and $h$ are elements of $T_{\bp}\mF$, and the source
$J\in T^*_{\bp}\mF$ can be expressed in terms of $\delta\Gamma_k/\delta h$. These considerations lead to the following
functional integro-differential equation defining $\Gamma_k\mku$:
\begin{equation}
\begin{split}
 \e^{-\Gamma_k[h;\bp]} = \int\dmu\;\exp\bigg\{-S[\hvp]-\Sgf[\hvp,\bp]-\Sgh[\hvp,\bp,C,\bar{C}\mku]&\\
 -\Delta S_k\big[\hat{h}[\bp,\hvp]-h;\bp\big]
 + \frac{\delta\Gamma_k}{\delta h^a}\big(\hat{h}^a[\bp,\hvp]-h^a\big) &\, \bigg\}.
 %- \Delta S_k^\text{gh}[C-\xi,\bar{C}-\bx;\bp] + \frac{\delta\Gamma_k}{\delta \xi^a}\big(C^a-\xi^a\big)
 %+ \big(\bar{C}_a-\bx_a\big)\mku\frac{\delta\Gamma_k}{\delta \bx_a}
\end{split}
\label{eq:GeoFuncInt}
\end{equation}
Here, $\dmu\equiv\mD\hvp\sqrt{\det G_{ij}[\hvp]}\,\mD C\mku\mD\bar{C}\sqrt{\det (G^\text{gh})^a{}_b}$ is the
covariantly defined and background field independent measure for the quantum field $\hvp$ and the ghosts $C$ and
$\bar{C}$ (where $G_{ij}[\hvp]$ is the usual field space metric, and
$\sqrt{\det (G^\text{gh})^a{}_b}$ is merely a constant factor since the ghost field space metric $(G^\text{gh})^a{}_b$
is assumed to be field independent). The cutoff action is given by \mbox{$\Delta S_k\big[\hat{h}-h;\bp\big]\equiv
\frac{1}{2}\big(\hat{h}^a-h^a\big)(\Rk)_{ab} \big(\hat{h}^b-h^b\big)$}.
% for standard fields and $\Delta S_k^\text{gh}[C-\xi,\bar{C}-\bx;\bp]\equiv \big(\bar{C}_a
%-\bx_a\big)\big(\Rk^\text{gh}\big)^a{}_b\mkuu \big(C^b-\xi^b\big)$ for the ghosts.
In this version of the effective average action, the relation between $\hh$ and $h$ is given by $h=\langle\hh\rangle$.
We would like to point out that this entails $\vp\neq\langle\hvp\rangle$ in general; the dynamical field $\vp$ is
rather defined through a geodesic, $\vp\equiv\vp[h;\bp]=\vp\big[\big\langle\hh\big\rangle; \bp\big]$.

Equation \eqref{eq:GeoFuncInt} is obtained by constructing $\Gamma_k$ as the Legendre transform of $W_k\equiv\ln Z_k$
plus a cutoff contribution, as discussed in Section \ref{sec:EAAFRGE}, and by replacing the source according to $J_a=
\frac{\delta\Gamma_k}{\delta h^a}+(\Rk)_{ab}h^b$. Note that the Legendre transform concerns only the
fields $J\leftrightarrow h$. It does not involve the ghosts, though. (Also, we did not include any source terms for the
ghost fields and ghost cutoff terms in the functional integral.) We chose this version of $\Gamma_k$ here for a better
comparison with the existing works on split-Ward identities \cite{BK87,Kunstatter1992,BMV03,Pawlowski2003}.
The alternative version of $\Gamma_k$, which includes a Legendre transform with respect to the ghosts and is thus a
functional of $h$, $\bp$, $\xi$ and $\bx$, with $\xi\equiv\langle C\rangle$ and $\bx\equiv\langle\bar{C}\rangle$,
leads to very similar split-Ward identities to the ones derived below (the main difference being a sum over all
field types considered and a replacement of traces by supertraces).
\smallskip

\noindent
\textbf{(2) Expectation values.}
In this setting, expectation values can be determined by using the relation
\begin{equation}
 \langle F\rangle = \frac{1}{A_k} \int\dmu\; F\; \e^{-S-\Sgf-\Sgh-\Delta S_k %-\Delta S_k^\text{gh}
 +\frac{\delta\Gamma_k}{\delta h^a}\hat{h}^a} \,,
 %+\frac{\delta\Gamma_k}{\delta\xi^a}C^a + \bar{C}_a\frac{\delta\Gamma_k}{\delta\bx_a}
\label{eq:DefExpValue}
\end{equation}
with
\begin{equation}
 A_k \equiv \int\dmu\; \e^{-S-\Sgf-\Sgh-\Delta S_k%-\Delta S_k^\text{gh}
 +\frac{\delta\Gamma_k}{\delta h^a}\hat{h}^a} \,.
 %+\frac{\delta\Gamma_k}{\delta\xi^a}C^a +\bar{C}_a\frac{\delta\Gamma_k}{\delta\bx_a}
\end{equation}
Up to a factor, $A_k$ agrees with the partition function $Z_k\mku$. Note that $S$, $\Sgf$, $\Sgh$ and $\Delta S_k$ are
the same as in eq.\ \eqref{eq:GeoFuncInt}, whereas the source terms are different.
\smallskip

\noindent
\textbf{(3) Reexpressing the auxiliary term $\bm{\big\langle\big(\hat{h}^a-h^a\big)\hat{h}^i{}_{;\mku l}\big\rangle}$.}
For later use, let us consider the expression $\frac{\delta}{\delta h^j}\big\langle\hat{h}^i{}_{;\mku l}\big\rangle$,
which we would like to relate to $\big\langle\big(\hat{h}^a-h^a\big)\hat{h}^i{}_{;\mku l}\big\rangle$. Here, we use a
semicolon to denote a covariant derivative with respect to the background field $\bp$, for instance
$\hat{h}^i{}_{;\mku l}\equiv \bar{\mD}_l\hat{h}^i \equiv \frac{\delta}{\delta\bp^l}\hat{h}^i + \Gamma^i_{lj}[\bp]
\hat{h}^j$ with a general field space connection $\Gamma^i_{lj}[\bp]$. Employing eq.\ \eqref{eq:DefExpValue} we obtain
\begin{equation}
\begin{split}
 \frac{\delta}{\delta h^j}\big\langle\hat{h}^i{}_{;\mku l}\big\rangle = {} &\Big\langle (\Rk)_{ja}\big(\hat{h}^a
 -h^a\big)\hat{h}^i{}_{;\mku l}\Big\rangle + \bigg\langle\frac{\delta^2\Gamma_k}{\delta h^j\mku\delta h^a}\,\hh^a\mku
 \hh^i{}_{;\mku l}\bigg\rangle \\
 &- \frac{1}{A_k^2}\,\frac{\delta A_k}{\delta h^j} \int\dmu\; \hat{h}^i{}_{;\mku l}\; \e^{-S-\Sgf-\Sgh-\Delta S_k
 + \frac{\delta\Gamma_k}{\delta h^a}\hat{h}^a}\;.
\end{split}
\end{equation}
The second term on the RHS can be written as $\Gamma_{k,ja}\big\langle\hh^a\mku\hh^i{}_{;\mku l}\big\rangle$, with the
comma in $\Gamma_{k,ja}$ denoting derivatives with respect to $h$, while the third term amounts to
\begin{align}
 &-\big\langle\hh^i{}_{;\mku l}\big\rangle\,\frac{1}{A_k}\int\dmu\, \Big((\Rk)_{ja}\big(\hh^a-h^a\big)
 +\Gamma_{k,ja}\mku\hh^a\Big)\, \e^{-S-\Sgf-\Sgh-\Delta S_k + \frac{\delta\Gamma_k}{\delta h^a}\hat{h}^a} \nonumber\\
 &= -\big\langle\hh^i{}_{;\mku l}\big\rangle(\Rk)_{ja}\big\langle\hh^a-h^a\big\rangle
 - \big\langle\hh^i{}_{;\mku l}\big\rangle\Gamma_{k,ja} \big\langle\hh^a\big\rangle
 = - \big\langle\hh^i{}_{;\mku l}\big\rangle\mku\Gamma_{k,ja}\mkuu h^a \nonumber\\
 &= - \Gamma_{k,ja}\mku\big\langle h^a\mku\hh^i{}_{;\mku l}\big\rangle \,,
\end{align}
where we have exploited that $\big\langle\hh^a-h^a\big\rangle=0$. Taking all pieces together we have
\begin{equation}
\begin{split}
 \frac{\delta}{\delta h^j}\big\langle\hat{h}^i{}_{;\mku l}\big\rangle &= (\Rk)_{ja}\big\langle \big(\hat{h}^a-h^a\big)
 \hat{h}^i{}_{;\mku l}\big\rangle + \Gamma_{k,\mku ja}\mku\big\langle \big(\hat{h}^a-h^a\big)
 \hat{h}^i{}_{;\mku l}\big\rangle \\
 &= \big(\Gamma_k^{(2)}+\Rk\big)_{ja}\mku\big\langle \big(\hat{h}^a-h^a\big) \hat{h}^i{}_{;\mku l}\big\rangle \,,
\end{split}
\end{equation}
where $\Gamma_k^{(2)}$ is the Hessian of $\Gamma_k$ with respect to $h$. This can be rewritten by introducing the
propagator
\begin{equation}
 \mG_k \equiv \big(\Gamma_k^{(2)}+\Rk\big)^{-1}\,.
\end{equation}
Here (and only here) we denote the propagator by $\mG_k$ in order to avoid confusion with the field space metric $G$.
(Usually the propagator is labeled by $G_k\mku$.) We finally obtain
\begin{equation}
 \big\langle \big(\hh^a-h^a\big) \hat{h}^i{}_{;\mku l}\big\rangle = \mG_k^{aj}\,\frac{\delta}{\delta h^j}
 \big\langle\hat{h}^i{}_{;\mku l}\big\rangle \,.
\label{eq:AuxExpVal}
\end{equation}
This auxiliary equation is needed for the following point.
\smallskip

\noindent
\textbf{(4) Deriving the split-Ward identities.}
We proceed by computing the covariant derivative $\bar{\mD}_j\equiv (\cdot)_{;\mku j}$ of $\Gamma_k$ with respect to
the background field, where $\Gamma_k$ is determined by taking the logarithm of eq.\ \eqref{eq:GeoFuncInt}. Since
$\Gamma_k$ is a scalar, the covariant derivative amounts to an ordinary functional derivative: $\Gamma_{k\mku ;\mku j}
= \frac{\delta\Gamma_k}{\delta\bp^j}\mku$, but the vector-valued expressions inside the functional integral will be
affected by the field space connection, so there the covariant derivative does not reduce to a usual one. We find
\begin{equation}
\begin{split}
 -\frac{\delta\Gamma_k}{\delta\bp^j} = {}&-\bigg\langle\frac{\delta\Sgf}{\delta\bp^j}\bigg\rangle
 - \bigg\langle\frac{\delta\Sgh}{\delta\bp^j}\bigg\rangle - \frac{1}{2}(\Rk)_{il\mku ;\mku j}\big\langle
 \big(\hh^i-h^i\big)\big(\hh^l-h^l\big)\big\rangle \\
 &- (\Rk)_{il}\mku\big\langle\big(\hh^i-h^i\big)\mku \hh^l{}_{;\mku j}\mku\big\rangle
 + \bigg(\frac{\delta\Gamma_k}{\delta h^a}\bigg)_{\! ;\mku j}\mkuu\big\langle\hh^a-h^a\big\rangle
 + \frac{\delta\Gamma_k}{\delta h^a} \mkuu\big\langle\hh^a{}_{;\mku j}\big\rangle\,.
\end{split}
\end{equation}
Using $\big\langle\hh^a-h^a\big\rangle=0$ and $\big\langle\big(\hh^i-h^i\big)\big(\hh^l-h^l\big)\big\rangle =
\big(W_k^{(2)}\big)^{il} = \mG_k^{il}$ (cf.\ point \textbf{(5)} of Appendix \ref{app:OperatorRep}) as well as
eq.\ \eqref{eq:AuxExpVal} yields
\begin{equation}
\begin{split}
 \frac{\delta\Gamma_k}{\delta\bp^j} + \frac{\delta\Gamma_k}{\delta h^a} \mkuu\big\langle\hh^a{}_{;\mku j}\big\rangle
 = {}& \frac{1}{2}(\Rk)_{il\mku ;\mku j}\mku \mG_k^{il}
 + (\Rk)_{il}\mku\mG_k^{im}\,\frac{\delta}{\delta h^m}\big\langle\hh^l{}_{;\mku j}\big\rangle
 + \bigg\langle\frac{\delta\Sgf}{\delta\bp^j}\bigg\rangle
 + \bigg\langle\frac{\delta\Sgh}{\delta\bp^j}\bigg\rangle \,.
\end{split}
\label{eq:PreSplitWard}
\end{equation}
We observe that the first two terms on the RHS of \eqref{eq:PreSplitWard} can be represented as operator traces since
the summation ``closes''. This leads to our final result:
\begin{equation}[b]
\begin{aligned}
 \frac{\delta\Gamma_k}{\delta\bp^j} + \frac{\delta\Gamma_k}{\delta h^a} \mkuu\big\langle\bar{\mD}_j\mku\hh^a\big\rangle
 = {}& \frac{1}{2}\Tr\Big[(\bar{\mD}_j \mku\Rk)\mku \mG_k\Big]
 + \Tr\bigg[\Rk \mku\mG_k\,\frac{\delta\big\langle\bar{\mD}_j\mku\hh\big\rangle}{\delta h}\bigg] \\
 & + \bigg\langle\frac{\delta\Sgf}{\delta\bp^j}\bigg\rangle
 + \bigg\langle\frac{\delta\Sgh}{\delta\bp^j}\bigg\rangle \,.
\end{aligned}
\label{eq:SplitWard}
\end{equation}
Here, the matrix representation of the term $\frac{\delta\langle\bar{\mD}_j\mku\hh\rangle}{\delta h}$ is given by its
components $\frac{\delta\langle\bar{\mD}_j\mku\hh^l\rangle}{\delta h^m}\mku$.
\smallskip

\noindent
\textbf{(5) Special cases of field space connections.}\\
\emph{Metric connection:} By noticing that the index structure of the cutoff operator is provided by the field space
metric alone, $(\Rk)_{il} \equiv G_{il}[\bp] \Rk[\bp]$, we see that its covariant derivative in \eqref{eq:SplitWard}
reduces to an ordinary derivative,
\begin{equation}
 (\Rk)_{il\mku ;\mku j} \equiv \big(G_{il}[\bp] \Rk[\bp]\big)_{;\mku j} = G_{il}[\bp] \frac{\delta\Rk}{\delta\bp^j}\,.
\end{equation}
\emph{Flat/trivial connection:} For a flat field space we have $\hh^a\equiv\hh^a[\bp,\hvp]=\hvp^a-\bp^a$ and thus
$\bar{\mD}_j\mku\hh^a=-\delta^a_j\mku$. Then the second trace term in \eqref{eq:SplitWard} vanishes:
\begin{equation}
 \frac{\delta\Gamma_k}{\delta\bp^j} - \frac{\delta\Gamma_k}{\delta h^j}
 = \frac{1}{2}\Tr\bigg[\frac{\delta\Rk}{\delta\bp^j}\, \mG_k\bigg]
 + \bigg\langle\frac{\delta\Sgf}{\delta\bp^j}\bigg\rangle
 + \bigg\langle\frac{\delta\Sgh}{\delta\bp^j}\bigg\rangle \,.
\end{equation}
\emph{Vilkovisky-DeWitt connection:} As shown in Reference \cite{Pawlowski2003}, the explicit gauge fixing and ghost
terms in \eqref{eq:SplitWard} vanish if the Vilkovisky-DeWitt connection is used.
\smallskip

\noindent
\textbf{(6) The split-Ward identities for $\bm{\Gamma}\mku$.} Since the effective average action $\Gamma_k$ at the
scale $k=0$ agrees with the conventional effective action $\Gamma$, it is straightforward to extract the split-Ward
identities for $\Gamma=\Gamma_{k=0}$ from eq.\ \eqref{eq:SplitWard}: Exploiting the fact that the cutoff operator
$\Rk$ vanishes for $k=0$ we obtain
\begin{equation}
 \frac{\delta\Gamma}{\delta\bp^j} + \frac{\delta\Gamma}{\delta h^a}\mkuu\big\langle\bar{\mD}_j\mku\hh^a\big\rangle
 = \bigg\langle\frac{\delta\Sgf}{\delta\bp^j}\bigg\rangle + \bigg\langle\frac{\delta\Sgh}{\delta\bp^j}\bigg\rangle \,.
\end{equation}

%----------------------------------------------------------------------------------------------------------------------
\chapter[The \texorpdfstring{\bm{$\beta$}}{beta}-functions for the exponential parametrization]%
[Transformation laws \& \texorpdfstring{$\beta$}{beta}-functions for the exponential parametrization]%
{Transformation laws and \texorpdfstring{\bm{$\beta$}}{beta}-functions for the exponential parametrization}
\label{app:TransBetaExp}
%----------------------------------------------------------------------------------------------------------------------

In this appendix we derive $\beta$-functions both for the single-metric truncation considered in Section
\ref{sec:single} and for the bimetric truncation covered in Section \ref{sec:bi}. We begin with a discussion on the
transformation behavior of $h$ under diffeomorphisms assuming that $g$ and $\bg$ transform as tensor fields.

%----------------------------------------------------------------------------------------------------------------------
\section[Transformation behavior of \texorpdfstring{$h$}{h}]{Transformation behavior of \texorpdfstring{\bm{$h$}}{h}}
\label{app:Trans}
%----------------------------------------------------------------------------------------------------------------------

Let $g_\mn$ and $\bg_\mn$ transform as proper tensor fields under diffeomorphisms, i.e.\ they satisfy $\delta g_\mn
= \mL_\xi g_\mn$ and $\delta \bg_\mn =\mL_\xi \bg_\mn$. Here $\mL_\xi$ denote the Lie derivative along the vector field
$\xi$ which generates the underlying diffeomorphism. Using the linear parametrization, $g_\mn=\bg_\mn+h_\mn$, implies
directly that $h_\mn$ transforms as a tensor field, too: $\delta h_\mn = \mL_\xi h_\mn$. For the exponential
parametrization, on the other hand, it requires more effort to come to that conclusion.
We will need the following two lemmas.
\begin{lemma}
\label{lem:VarMatrixExp}
 The variation of the matrix exponential of a square matrix $A$ is given by
 \begin{equation}
  \delta\left(\e^A\right) = \int_0^1 \e^{tA}\;\delta A\;\e^{(1-t)A}\,\mathrm{d} t \, .
 \end{equation}
\end{lemma}
\noindent
\textbf{Proof:}
We exploit two mathematical identities.\\[0.5em]
(i) We employ the summation formula
\begin{equation}
 \sum\limits_{n=1}^\infty\; \sum\limits_{m=0}^{n-1} = \sum\limits_{m=0}^\infty\; \sum\limits_{n=m+1}^\infty \;,
\end{equation}
which follows from simple reordering arguments as illustrated in Figure \ref{fig:ordSum}.
\begin{figure}[htp]
\begin{minipage}[t][][t]{0.6\columnwidth}
 \captionwidth{\columnwidth}
 \caption{There are two possibilities to sum over all discrete points in the shaded area (where the origin in the
 diagram is located at $n=1$, $m=0$): First, from $n=1$ to
 $n=\infty$ and  from $m=0$ to $m=n-1$, and second, from $m=0$ to $m=\infty$ and from $n=m+1$ to $n=\infty$.}
 \label{fig:ordSum}
\end{minipage}
\hfill
\begin{minipage}[t][][t]{0.22\columnwidth}
 \vspace{0em}
 \includegraphics[width=\columnwidth]{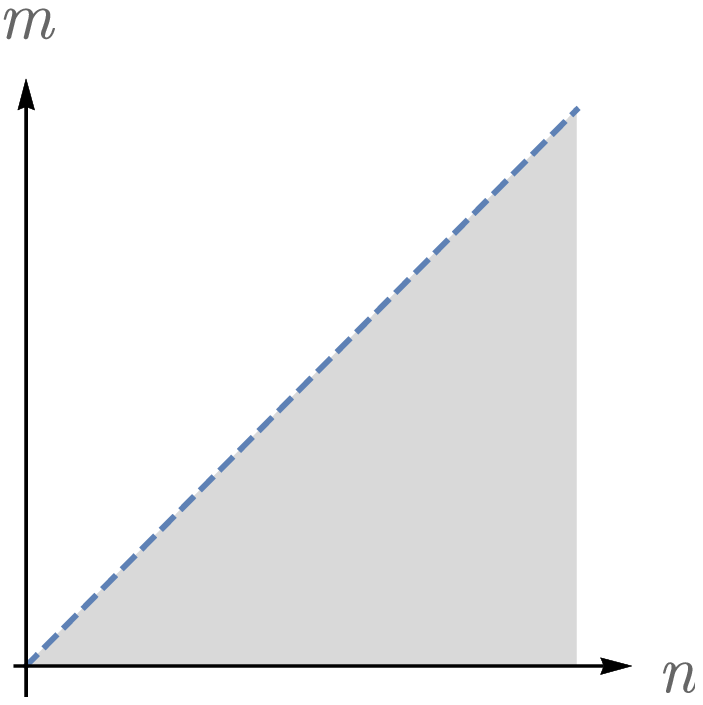}
\end{minipage}
\hspace{1em}
\end{figure}
\smallskip

\noindent
(ii) We make use of the integral representation of the Euler beta function (Euler integral of the first kind) and its
value in terms of factorials for integer numbers:
\begin{equation}
 B(m+1,p+1) = \int_0^1 t^m (1-t)^p\, \td t = \frac{m!\,p!}{(m+p+1)!}
\end{equation}

With these two formulae we find
\begin{align}
 \delta\left(\e^A\right) &= \delta\left(\sum\limits_{n=0}^\infty \frac{1}{n!}\, A^n\right)
    = \sum\limits_{n=1}^\infty \frac{1}{n!}\; \sum\limits_{m=0}^{n-1} A^m\,\delta A\; A^{n-m-1} \nonumber\\
    &\stackrel{\mathclap{(\text{i})}}{=} \sum\limits_{m=0}^\infty \; \sum\limits_{n=m+1}^\infty\,\frac{1}{n!}\,
    A^m\,\delta A\; A^{n-m-1}
    = \sum\limits_{m=0}^\infty \; \sum\limits_{p=0}^\infty\,\frac{1}{(m+p+1)!}\, A^m\,\delta A\; A^p \nonumber\\
    &= \sum\limits_{m=0}^\infty \; \sum\limits_{p=0}^\infty\,\frac{m!\,p!}{(m+p+1)!}\, \frac{A^m}{m!}\,\delta A\;
    \frac{A^p}{p!} \nonumber\\
    &\stackrel{\mathclap{(\text{ii})}}{=} \sum\limits_{m=0}^\infty \; \sum\limits_{p=0}^\infty\,\int_0^1 t^m (1-t)^p\,
    \td t\,\; \frac{A^m}{m!}\,\delta A\; \frac{A^p}{p!} \nonumber\\
    &= \sum\limits_{m=0}^\infty \; \sum\limits_{p=0}^\infty\,\int_0^1 \frac{(tA)^m}{m!} \,\delta A\;
    \frac{\big[(1-t)A\big]^p}{p!}\, \td t \nonumber\\
    &= \int_0^1 \e^{tA}\;\delta A\;\e^{(1-t)A}\,\td t \;,
\end{align}
where summation and integration commute due to the convergence properties of the exponential function.\hfill $\Box$

\begin{lemma}
\label{lem:MatrixLog}
 If existent, the real matrix logarithm of a real square matrix $A$ can be represented by the expression
 \begin{equation}
  \ln(A) = -\int_\epsilon^\infty \frac{\e^{-sA}}{s}\,\mathrm{d} s -\ln(\epsilon)\Id-\gamma\,\Id+\mO(\epsilon),
 \label{eq:lnAFormula}
 \end{equation}
 where $\gamma$ denotes the Euler--Mascheroni constant.
\end{lemma}
\noindent
\textbf{Proof:} Let us begin with the special case of a positive real number $A$. Then we can rewrite the logarithm as
\begin{align}
 \ln(A) &=\int_1^A\frac{1}{t}\,\td t = \int_1^A\!\td t\left[-\frac{1}{t}\,\e^{-st}\right]_{s=0}^{s=\infty}
 = \int_1^A\!\td t\int_0^\infty\!\td s\,\e^{-st} \nonumber\\
 &= \int_0^\infty\!\td s\int_1^A\!\td t\,\e^{-st} = \int_0^\infty\!\td s\left(\frac{1}{s}\,\e^{-s}-\frac{1}{s}
 \,\e^{-sA}\right) \nonumber\\
 &= \int_0^\epsilon\!\td s\,\frac{1}{s}\left(\e^{-s}-1\right) + \int_0^\epsilon\!\td s\,\frac{1}{s}\left(1-\e^{-sA}
 \right)  +\int_\epsilon^\infty\!\td s\,\frac{\e^{-s}}{s} - \int_\epsilon^\infty\!\td s\,\frac{\e^{-sA}}{s} \nonumber\\
 &= - \int_\epsilon^\infty\!\td s\,\frac{\e^{-sA}}{s} + \int_\epsilon^\infty\!\td s\,\frac{\e^{-s}}{s} +\mO(\epsilon)\,,
\label{eq:lnOfA}
\end{align}
where the mean value theorem for integration, employed in the last equality, is applicable since both $\frac{1}{s}
\left(\e^{-s}-1\right)$ and $\frac{1}{s}\left(1-\e^{-sA}\right)$ are continuous functions.

The term $\int_\epsilon^\infty\td s\,\frac{\e^{-s}}{s}$ can be evaluated as follows. Substituting $s\to s\epsilon$
we observe
\begin{equation}
 \int_\epsilon^\infty\!\td s\,\frac{\e^{-s}}{s} = \int_1^\infty\!\td s\,\frac{\e^{-s\epsilon}}{s}\,.
\label{eq:integralSubstituted}
\end{equation}
Furthermore, defining $f(s)=\ln(s)\e^{-s\epsilon}\mku$, we can exploit that $f'(s)=\frac{\e^{-s\epsilon}}{s}-\epsilon
\ln(s)\e^{-s\epsilon}$ and that $\int_1^\infty f'(s)\td s=f(\infty)-f(1)=0$, so we have
\begin{align}
 \int_1^\infty\frac{\e^{-s\epsilon}}{s}\,\td s &= \epsilon\int_1^\infty\ln(s)\e^{-s\epsilon}\,\td s
 = \int_\epsilon^\infty\ln\!\left(\frac{t}{\epsilon}\right)\e^{-t}\,\td t \nonumber\\
 &= \int_0^\infty\ln(t)\e^{-t}\,\td t-\underbrace{\int_0^\epsilon
 \underbrace{\ln(t)\e^{-t}}_\text{integrable}\,\td t}_{=\,\mO(\epsilon)}  - \ln(\epsilon)
 \underbrace{\int_\epsilon^\infty \e^{-t}\,\td t}_{\mathclap{=\,\e^{-\epsilon}\,=\,1+\mathcal{O}(\epsilon)}} \nonumber\\
  &= \int_0^\infty\ln(t)\e^{-t}\,\td t -\ln(\epsilon)+\mathcal{O}(\epsilon)\,.
\end{align}
Finally, with
\begin{equation}
\begin{split}
  -\gamma &= \Gamma'(1)=\frac{\td}{\td z}\int_0^\infty \e^{(z-1)\ln(t)}\,\e^{-t}\,\td t\;\bigg|_{z=1} = \int_0^\infty
  \ln(t)t^{z-1}\,\e^{-t}\,\td t\;\bigg|_{z=1}\\
  &=\int_0^\infty\ln(t)\e^{-t}\,\td t\;,
\end{split}
\end{equation}
we obtain
\begin{equation}
 \int_1^\infty\frac{\e^{-s\epsilon}}{s}\,\td s = -\ln(\epsilon)-\gamma+\mathcal{O}(\epsilon)\,,
\end{equation}
and thus, using \eqref{eq:lnOfA} and \eqref{eq:integralSubstituted},
\begin{equation}
 \ln(A) = -\int_\epsilon^\infty \frac{\e^{-sA}}{s}\,\td s -\ln(\epsilon)-\gamma+\mathcal{O}(\epsilon)\,.
\label{eq:lnOfA2}
\end{equation}
Note that the divergence at the lower limit of integration for $\epsilon\to 0$ is canceled by the term $\ln(\epsilon)$.

Now let $A$ be a square matrix (or an operator). Since the exponential is defined both for matrices and operators,
relation \eqref{eq:lnOfA2} remains valid in this generalized case. For the argument it is sufficient to know that the
logarithm is the inverse function of the exponential and that the calculation rules for the usual exponential
hold true for the matrix exponential as well, provided that commuting matrices are considered. (The latter requirement
is satisfied as $A$ and $\Id$ are the only matrices that can occur here.) Existence of a real logarithm on
the LHS of \eqref{eq:lnAFormula} is equivalent to convergence of the RHS. This completes the proof.
\hfill $\Box$

\bigskip

Lemmas \ref{lem:VarMatrixExp} and \ref{lem:MatrixLog} now allow us to prove the following theorem.
\begin{theorem}
 Let $\bg$ be a metric tensor and let $g$ be related to $\bg$ and $h$ by the exponential parametrization,
 $g=\bg\,\e^{\mku\bg^{-1}h}$. Then $h$ transforms as a tensor field if and only if $g$ transforms as a tensor field.
\end{theorem}
\noindent
\textbf{Proof:}\\[0.5em]
``$\Rightarrow$'': We begin with the case where $h$ transforms as a tensor field, $\delta h=\mL_\xi h$. Then
\begin{equation}
\begin{split}
 \delta\big(\e^{\mku\bg^{-1}h}\big) &= \int_0^1\td t\; \e^{t\mku\bg^{-1}h}\;\delta \left(\bg^{-1}h\right)\;
 \e^{(1-t)\bg^{-1}h} \\
 &= \int_0^1\td t\; \e^{t\mku\bg^{-1}h}\;\mL_\xi \left(\bg^{-1}h\right)\;
 \e^{(1-t)\bg^{-1}h} = \mL_\xi \big(\e^{\mku\bg^{-1}h}\big).
\end{split}
\end{equation}
since both $\bg^{-1}$ and $h$ transform as tensor fields. Hence, $\e^{\mku\bg^{-1}h}$ transforms as a tensor field,
too, and so does $g=\bg\,\e^{\mku\bg^{-1}h}$.\\[0.5em]
``$\Leftarrow$'': Now let us consider the case where $g$ transform as a tensor field, while the transformation behavior
of the symmetric field $h$ is a priori unknown. Clearly, the exponential $\e^{\mku\bg^{-1}h}=\bg^{-1}\mku g$ transforms
as a tensor field since both $g$ and $\bg$ are tensor fields. Therefore, $X$ defined by
\begin{equation}
 X \equiv \e^{\mku\bg^{-1}h} - \Id
\end{equation}
transforms as a tensor field, too, as $\delta\Id=0=\mL_\xi\Id$. As proven in Appendix \ref{app:ExpParam}, there exists
a unique real logarithm of $\e^{\mku\bg^{-1}h}\mku$, namely $\bg^{-1}h = \ln(\Id+X)$.

Let us assume for a moment that the matrix norm of $X$ is sufficiently small. Then we can expand $\ln(\Id+X)$ according
to
\begin{equation}
 \bg^{-1}h = \ln(\Id+X) = -\sum_{n=1}^\infty\frac{(-1)^n}{n} X^n\,.
\label{eq:gmoneh}
\end{equation}
Applying a transformation to \eqref{eq:gmoneh} leads to
\begin{equation}
\begin{split}
 \delta(\bg^{-1}h) &= -\delta\sum_{n=1}^\infty\frac{(-1)^n}{n} X^n
 = -\sum_{n=1}^\infty\frac{(-1)^n}{n} \delta(X^n) \\
 &= -\sum_{n=1}^\infty\frac{(-1)^n}{n} \mL_\xi(X^n) = -\mL_\xi\sum_{n=1}^\infty\frac{(-1)^n}{n} X^n
 = \mL_\xi (\bg^{-1}h),
\end{split}
\end{equation}
where we assumed in the second equality that $||\delta X||$ is sufficiently small, guaranteeing uniform
convergence of the last term in the first row, so that the variation can be commuted with the sum. This proves that
$\bg^{-1}h$ transforms as a tensor field, and so does $h\mku$: $\delta h=\mL_\xi h$.

In the general case, if the matrix norm of $X$ can become arbitrarily large, we can make use of the representation
formula for matrix logarithms, as given in Lemma \ref{lem:MatrixLog}: If a real square matrix $A$ possesses a real
logarithm, it satisfies the relation $\ln(A) = -\int_\epsilon^\infty\frac{\e^{-sA}}{s}\td s -\ln(\epsilon)\Id
-\gamma\Id + \mO(\epsilon)$. Now, if $A$ transforms as a tensor field, then we know from the case
``$\Rightarrow$'' that the matrix exponential $\e^{-sA}$ is a proper tensor field, too. Hence, also $\ln(A)$ must
transforms as a tensor field. Identifying $A$ with $\Id+X$ proves the statement, i.e.\ $\ln(\Id+X)=\bg^{-1}h$
transforms as a tensor field, and therefore $\delta h=\mL_\xi h$. \hfill $\Box$

\medskip

For the trace part of $h$, defined by $\phi\equiv\Tr(\bg^{-1}h)$, this result can be checked in a different way.
Applying a transformation to the RHS of $g=\bg\,\e^{\mku\bg^{-1}h}$ yields
\begin{align}
 \delta g &= (\delta\bg)\,\e^{\mku\bg^{-1}h} + \bg\,\delta\big(\e^{\mku\bg^{-1}h}\big) \nonumber\\
 &= (\mL_\xi \bg)\,\e^{\mku\bg^{-1}h} + \bg \int_0^1\td t\; \e^{t\mku\bg^{-1}h}\;\delta \left(\bg^{-1}h\right)\;
 \e^{(1-t)\bg^{-1}h}\,.
\label{eq:Deltag1}
\end{align}
On the other hand, we also know that $\delta g= \mL_\xi g$, so
\begin{align}
 \delta g &= \mL_\xi\big(\bg\,\e^{\mku\bg^{-1}h}\big) =
 (\mL_\xi \bg)\,\e^{\mku\bg^{-1}h} + \bg\,\mL_\xi\big(\e^{\mku\bg^{-1}h}\big) \nonumber\\
 &= (\mL_\xi \bg)\,\e^{\mku\bg^{-1}h} + \bg \int_0^1\td t\; \e^{t\mku\bg^{-1}h}\;\mL_\xi \left(\bg^{-1}h\right)\;
 \e^{(1-t)\bg^{-1}h}\,.
\label{eq:Deltag2}
\end{align}
Comparing \eqref{eq:Deltag1} with \eqref{eq:Deltag2} leads to
\begin{equation}
 \int_0^1\td t\; \e^{t\mku\bg^{-1}h}\,\Big[\delta\left(\bg^{-1}h\right)-\mL_\xi \left(\bg^{-1}h\right)\Big]\,
 \e^{(1-t)\bg^{-1}h} = 0\,.
\label{eq:RelForVarh}
\end{equation}
Since the exponents in eq.\ \eqref{eq:RelForVarh} do in general not commute with the variations, it is not obvious that
$\delta\left(\bg^{-1}h\right)$ must agree with $\mL_\xi \left(\bg^{-1}h\right)$. However, upon taking the trace of
\eqref{eq:RelForVarh} we obtain
\begin{equation}
\begin{split}
 0 &= \int_0^1\td t\,\Tr\bigg\{\e^{t\mku\bg^{-1}h}\,\Big[\delta\left(\bg^{-1}h\right)
 -\mL_\xi \left(\bg^{-1}h\right)\Big]\,\e^{(1-t)\bg^{-1}h}\bigg\} \\
 &= \int_0^1\td t\,\Tr\bigg\{\Big[\delta\left(\bg^{-1}h\right)-\mL_\xi \left(\bg^{-1}h\right)\Big]\,\mathds{1}\bigg\}
 = \Tr\Big[\delta\left(\bg^{-1}h\right)-\mL_\xi \left(\bg^{-1}h\right)\Big],
\end{split}
\end{equation}
and with $\phi=\Tr(\bg^{-1}h)$ finally $\delta\phi=\mL_\xi \phi$.

%----------------------------------------------------------------------------------------------------------------------
\section[Hessians and \texorpdfstring{$\beta$}{beta}-functions in the single-metric case]%
{Hessians and \texorpdfstring{\bm{$\beta$}}{beta}-functions in the single-metric case}
\label{app:single}
%----------------------------------------------------------------------------------------------------------------------

In order to derive $\beta$-functions we follow the steps outlined in Section \ref{sec:Recipe}, adopting the notation of
Reference \cite{Reuter1998}. We consider the gravitational EAA
\begin{equation}
 \Gamma_k^\text{grav}\big[g,\bg\big] \equiv \frac{1}{16\pi G_k} \int \! \dd x \sg \,\big( -R + 2\Lambda_k \big),
\end{equation}
along with the gauge fixing action
\begin{equation}
 \Gamma_k^\text{gf}[g,\bg] = \frac{\alpha^{-1}}{32\pi\mku G_k}\int\dd x\sbg\,\bg^\mn
 \big(\mF_\mu^{\alpha\beta}[\bg] g_{\alpha\beta}\big)\big(\mF_\nu^\rs[\bg] g_\rs\big),
\label{eq:SingleGfAction}
\end{equation}
with $\alpha=1$ and $\mF_\mu^{\alpha\beta}[\bg]\equiv\delta_\mu^\beta \bg^{\alpha\tau}\bD_\tau - \frac{1}{2}
\bg^{\alpha\beta}\bD_\mu\mku$. Note that equation \eqref{eq:SingleGfAction} represents a ``$g_\mn$-type'' gauge
fixing action, cf.\ Section \ref{sec:ParamDepFramework}.

Now the exponential metric parametrization, $g_\mn= \bg_{\mu\rho}(\e^h)^\rho{}_\nu\mku$, is inserted into
$\Gamma_k^\text{grav}$ and into $\Gamma_k^\text{gf}$. Their sum, $\Gamma_k= \Gamma_k^\text{grav} +
\Gamma_k^\text{gf}$, is to be expanded in terms of $h_\mn$ then. The quadratic term of $\Gamma_k$ can be obtained by
employing the variation relations specified in Appendix \ref{app:Variations} and by some lengthy algebraic reshaping.
The result reads
\begin{equation}
 \Gamma_k^\text{quad} = \frac{1}{32 \pi G_k} \int \dd x \sbg \, h_\mn \! \left( 
 -K^\mn{}_\rs \bar{D}^2 + U^\mn{}_\rs \right) h^\rs,
\end{equation}
with $K^\mn{}_\rs \equiv\frac{1}{2}\big(\delta^\mu_{(\rho}\delta^\nu_{\sigma)}-\frac{1}{2}\mku\bg^\mn
\bg_{\rho\sigma}\big)$ and
\begin{equation}
 U^\mn{}_\rs \equiv -\frac{1}{4} \, \bg^\mn \bg_\rs \bar{R} + \frac{1}{2}\big(\bg^\mn \bar{R}_\rs + \bg_\rs
 \bar{R}^\mn\big) - \bar{R}^\mu{}_{(\rho}{}^\nu{}_{\sigma)} + \frac{1}{2} \, \bg^\mn \bg_\rs \Lambda_k \,,
\end{equation}
where round brackets enclosing index pairs denote symmetrization. We observe that the additional terms resulting from
the use of the exponential parametrization cancel some of those which are already present in the standard
calculation (cf.\ Ref.\ \cite{Reuter1998}).\footnote{For the linear parametrization one finds the same
$K^\mn{}_\rs$ as above, while $U^\mn{}_\rs$ is given by the tensor
$U^\mn{}_\rs\equiv \frac{1}{2}\big(\delta^\mu_{(\rho}\delta^\nu_{\sigma)}-\frac{1}{2}\mku\bg^\mn \bg_{\rho\sigma}\big)
\left(\bar{R}- 2\mku\Lambda_k\right) + \frac{1}{2}\left(\bg^\mn\bar{R}_\rs +\bg_\rs\bar{R}^\mn\right)
- \delta^{(\mu}_{(\rho} \bar{R}^{\mku\nu)}{}_{\sigma)} - \bar{R}^{\mku\mu}{}_{(\rho}{}^\nu{}_{\sigma)}$.}

After splitting the field $h_\mn$ into trace and traceless part, $h_\mn=\hat{h}_\mn+\frac{1}{d}\mkuu\bg_\mn \phi$,
where $\phi=\bg^\mn h_\mn$ and $\bg^\mn\hat{h}_\mn=0$, and inserting a maximally symmetric background for
$\bg_\mn$,\footnote{A maximally symmetric background $\bg_\mn$ implies $\bR_{\mu\nu\rho\sigma}= \frac{1}{d(d-1)}
\big(\bg_{\mu\rho}\bg_{\nu\sigma} - \bg_{\mu\sigma}\bg_{\nu\rho}\big)\bR\mkuu$ for the Riemann tensor and
$\bR_\mn = \frac{1}{d}\,\bg_\mn\bR\mkuu$ for the Ricci tensor.} we obtain
\begin{equation}
\begin{split}
\Gamma_k^\text{quad} = \frac{1}{64\pi G_k} \int\dd x\sbg \, \bigg\{ \hat{h}_\mn \Big( -\bar{D}^2
+ C_\text{T} \bar{R} \Big) \hat{h}^\mn \\
- \bigg(\frac{d-2}{2d}\bigg) \phi \Big(-\bar{D}^2 + C_\text{S} \bar{R} - \mu \Lambda_k \Big)\phi
\bigg\},
\end{split}
\label{eq:HessianTraceTraceless}
\end{equation}
with the constants $C_\text{T}\equiv\frac{2}{d(d-1)}$ and $C_\text{S}\equiv\frac{d-2}{d}$ (which are modified in
comparison with Ref.\ \cite{Reuter1998}), as well as
\begin{equation}
 \mu\equiv\frac{2d}{d-2} \,.
\end{equation}
As argued on general grounds in Section \ref{sec:singleExpd} on the basis of eq.\
\eqref{eq:sgExpParam}, the cosmological constant does indeed drop out of the traceless sector.

By the methods of Section \ref{sec:Recipe} (choosing the same cutoff as in Ref.\ \cite{Reuter1998}) we find that the
resulting anomalous dimension of Newton's constant, $\eta_N \equiv G_k^{-1}\mkuu k\p_k G_k$, is given by
\begin{equation}
 \eta_N = \frac{g\mku B_1(\lambda)}{1-g\mku B_2(\lambda)}\, ,
\end{equation}
where $g$ and $\lambda$ denote the dimensionless versions of the Newton constant and the cosmological constant,
respectively,\footnote{Here, $g$ and $\lambda$ play the role of independent arguments, so they carry no index $k$.}
and $B_1$, $B_2$ are functions of $\lambda$:
\begin{align}
 &\begin{aligned}
 B_1(\lambda) = \frac{1}{3}(4\pi)^{1-d/2} \bigg\{ \big(d^2-3d-2\big) \Phi_{d/2-1}^1(0) 
 - 12\,\frac{3d+2}{d}\, \Phi_{d/2}^2(0)\quad\, \\
 + 2\,\Phi_{d/2-1}^1(-\mu\lambda) - 12\,\frac{d-2}{d}\,\Phi_{d/2}^2(-\mu\lambda) \bigg\} , 
 \end{aligned} 
 \label{eq:B1SingleExpParam}\\
 &\begin{aligned}
 B_2(\lambda) = - \frac{1}{6}(4\pi)^{1-d/2} \bigg\{ (d-1)(d+2) \tPhi_{d/2-1}^1(0)
 - 12\,\frac{d+2}{d}\, \tPhi_{d/2}^2(0)\quad\, \\
 + 2\,\tPhi_{d/2-1}^1(-\mu\lambda) -12\,\frac{d-2}{d}\,\tPhi_{d/2}^2(-\mu\lambda) \bigg\}.
 \end{aligned}
 \label{eq:B2SingleExpParam}
\end{align}
The threshold functions $\Phi^p_n$ and $\tPhi^p_n$ are defined in Appendix \ref{app:Cutoffs}.
Finally, we find the following result for the $\beta$-functions of $g_k=k^{d-2}G_k$ and $\lambda_k=k^{-2}\Lambda_k$:
\begin{align}
 \beta_g = {}&(d-2+\eta_N)g,
 \label{eq:beta_g_FRG}\\
 \beta_\lambda = {}&-(2-\eta_N)\lambda + {\textstyle \frac{1}{2}}(4\pi)^{1-d/2} g \, \Big \{
 2\big(d^2-3d-2\big) \Phi_{d/2}^1(0) \nonumber\\ &- (d-1)(d+2)\eta_N \tPhi_{d/2}^1(0)
 + 4\Phi_{d/2}^1(-\mu\lambda) - 2\eta_N \tPhi_{d/2}^1(-\mu\lambda) \Big\}.
 \label{eq:beta_lambda_FRG}
\end{align}
The special cases $d=4$ and $d=2+\ve$ and their main consequences are treated in detail in Sections
\ref{sec:singleExp4} and \ref{sec:singleExp2}, respectively.

If the matter action \eqref{eq:matter} is included in the truncation ansatz for the EAA, we obtain the modified
quadratic term
\begin{equation}
 \Gamma_k^\text{quad,full} = \Gamma_k^\text{quad} + \frac{1}{2}\int\dd x\sbg\; A^i\big(-\delta_{ij}\bB\big)A^j\,,
\end{equation}
where $\Gamma_k^\text{quad}$ denotes the pure gravity result \eqref{eq:HessianTraceTraceless}, and we have already
identified $g_\mn$ with $\bg_\mn\mku$. The sum both over $i$ and over $j$ is from $1$ to $N$. This changes the
functions $B_1(\lambda)$ and $B_2(\lambda)$ given by eqs.\ \eqref{eq:B1SingleExpParam} and
\eqref{eq:B2SingleExpParam}, respectively, into
\begin{align}
 B_1^\text{full}(\lambda) &= B_1(\lambda)+\frac{1}{3}(4\pi)^{1-d/2} \bigg\{2\mku N\mku \Phi_{d/2-1}^1(0)\bigg\} ,\\
 B_2^\text{full}(\lambda) &= B_2(\lambda)\,,
\end{align}
leading to the modified anomalous dimension
\begin{equation}
 \eta_N^\text{full} = \frac{g\mku B_1^\text{full}(\lambda)}{1-g\mku B_2^\text{full}(\lambda)}\, .
\end{equation}
Finally, the corresponding $\beta$-functions read
\begin{align}
 \beta_g^\text{full} &= \big(d-2+\eta_N^\text{full}\big)g \,,
\label{eq:beta_g_FRG_N}\\
 \beta_\lambda^\text{full} &= -\big(2-\eta_N^\text{full}\big)\lambda + {\textstyle \frac{1}{2}}(4\pi)^{1-d/2} g \,
 \Big \{ 2\big(d^2-3d-2\big) \Phi_{d/2}^1(0) + 4 \mku N \Phi_{d/2}^1(0) \nonumber\\
 &\qquad\quad\mku - (d-1)(d+2)\eta_N^\text{full}\mku \tPhi_{d/2}^1(0) + 4\mku\Phi_{d/2}^1(-\mu\lambda)
 - 2\mku\eta_N^\text{full}\mku \tPhi_{d/2}^1(-\mu\lambda) \Big\}.
\label{eq:beta_lambda_FRG_N}
\end{align}

%----------------------------------------------------------------------------------------------------------------------
\section[Hessians and \texorpdfstring{$\beta$}{beta}-functions in the bimetric case]%
{Hessians and \texorpdfstring{\bm{$\beta$}}{beta}-functions in the bimetric case}
\label{app:bi}
%----------------------------------------------------------------------------------------------------------------------

We consider the truncation ansatz
\begin{align}
 \Gamma_k\big[g,\bg,\xi,\bx\, \big] =\; &\frac{1}{16\pi G_k^\text{Dyn}} \int\! \dd x \sg
 \big(\! -R + 2\Lambda_k^\text{Dyn} \big) + \Gamma_k^\text{gf}\big[g,\bg \big]
 + \Gamma_k^\text{gh}\big[g,\bg,\xi,\bx\, \big] \nonumber\\
 &+\frac{1}{16\pi G_k^\text{B}} \int\! \dd x \sbg \big(\! -\bar{R} + 2\Lambda_k^\text{B}\big)\,,
\end{align}
consisting of one Einstein--Hilbert-type action for the dynamical ('Dyn') sector and one for the background ('B')
sector. For reasons explained in Section \ref{sec:bi}, we employ the conformal projection technique \cite{BR14}. It
consists in setting the dynamical metric to $g_\mn=\e^{2\Omega}\bg_\mn$ (after having taken functional derivatives).
In the following, we denote this projection by $(\cdots)|_\text{pr}\,$.
For the exponential parametrization, $g_{\mu\nu}=\bg_{\mu\rho}(\e^h)^\rho{}_\nu$, it is equivalent to setting
$h^\rho{}_\nu=2\Omega\,\delta^\rho_\nu\,$. This affects the derivatives of $g_\mn$ w.r.t.\
$h_{\rho\sigma}$ appearing in equation \eqref{eq:2ndVar} as follows:
\begin{align}
 \frac{\delta g_\mn(x)}{\delta h_{\rho\sigma}(y)}\, \bigg|_\text{pr}
 &= \e^{2\Omega} \, \delta^\rho_{(\mu} \, \delta^\sigma_{\nu)} \, \delta(x-y), \displaybreak[0]\\[0.2em]
 \frac{\delta^2 g_\mn(u)}{\delta h_{\rho\sigma}(x) \, \delta h_{\lambda\gamma}(y)}
\, \bigg|_\text{pr} &= {\textstyle \frac{1}{2}}\, \e^{2\Omega}
\left(\bg^{\lambda(\sigma} \delta^{\rho)}_{(\mu} \, \delta^\gamma_{\nu)} + \bg^{\rho(\gamma} \delta^{\lambda)}_{(\mu}
\, \delta^\sigma_{\nu)} \right) \delta(u-x) \delta(u-y) .
\end{align}
Now, the Hessian $(\Gamma_k)^{(2)}_{hh}$ (where derivatives are w.r.t.\ $h_\mn\mku$, and ghost fields are set to
zero) is obtained by inserting these relations into eq.\ \eqref{eq:2ndVar} and by computing the remaining derivatives
of $\Gamma_k$ w.r.t.\ $g_\mn$ by means of the formulae given in Appendix \ref{app:Variations}. The result can be
simplified by applying the conformal projection again and by choosing the ``$\Omega$ deformed $\alpha=1$ gauge'' as in
Ref.\ \cite{BR14}. For the ``$\Omega$ deformed $\alpha=1$ gauge'' and the harmonic coordinate condition the gauge
fixing action reads
\begin{equation}
 \Gamma_k^\text{gf}[g,\bg] = \frac{\alpha^{-1}}{32\pi\mku G_k^\text{Dyn}}\int\dd x\sbg\,\bg^\mn
 \big(\mF_\mu^{\alpha\beta}[\bg] g_{\alpha\beta}\big)\big(\mF_\nu^\rs[\bg] g_\rs\big),
\label{eq:BiGfAction}
\end{equation}
with $\alpha^{-1} \equiv \e^{(d-6)\Omega}$ and $\mF_\mu^{\alpha\beta}[\bg]\equiv\delta_\mu^\beta \bg^{\alpha\tau}
\bD_\tau - \frac{1}{2}\bg^{\alpha\beta}\bD_\mu\mku$. Like in the single-metric case, eq.\ \eqref{eq:BiGfAction}
 represents a ``$g_\mn$-type'' gauge fixing action (see Section \ref{sec:ParamDepFramework}). Putting all contributions
together yields the Hessian
\begin{equation}
\begin{split}
 \big((\Gamma_k)_{hh}^{(2)}\big)^{\mn\rho\sigma}&\Big|_\text{pr} =
 \frac{\e^{(d-2)\Omega}}{32\pi G_k^\text{Dyn}} \Big\{ \big(-\bg^{\mu(\rho}\bg^{\sigma)\nu}+
 {\textstyle\frac{1}{2}}\bg^\mn\bg^\rs \big)\bar{D}^2 \\
 & -{\textstyle\frac{1}{2}}\big(\bar{R}-2\, \e^{2\Omega}\Lambda_k^\text{Dyn}\big) \bg^\mn \bg^\rs
 + 2\mkuu\bar{R}^{\rho(\mu\nu)\sigma} + \bg^\rs \bar{R}^\mn +\bg^\mn \bar{R}^\rs \Big\}
\end{split}
\label{eq:biHessian}
\end{equation}
in the graviton sector, as well as
\begin{equation}
\big(\big(\Gamma_k^\text{gh}\big)_{\xi\bx}^{(2)}\big)^\mu_{~\nu}\Big|_\text{pr} = \sqrt{2}\, \e^{2\Omega}
\big(\bar{R}^\mu{}_\nu + \delta^\mu_\nu \bar{D}^2\big)
\end{equation}
and $\big(\Gamma_k^\text{gh}\big)_{\bx\xi}^{(2)}=-\big(\Gamma_k^\text{gh}\big)_{\xi\bx}^{(2)}$
in the ghost sector.

Compared with Ref.\ \cite{BR14}, the Hessians for the ghosts are not modified, but the one for the graviton sector is
different:
(a) The terms in the curly brackets in \eqref{eq:biHessian} have changed, in particular, the cosmological constant term
is proportional to $\bg^\mn\bg^\rs$ now, so it drops out of the traceless sector as it did in the single-metric
computation of Section \ref{app:single}. (b) The numerator of the prefactor has changed from $\e^{(d-6)\Omega}$ into
$\e^{(d-2)\Omega}$, signaling the special role of $d=2$ dimensions.

Upon decomposing $h_\mn$ into trace and traceless parts, $h_\mn\equiv\hat{h}_\mn+\frac{1}{d}\mku\bg_\mn \phi$, with
$\phi=\bg^\mn h_\mn$ and $\bg^\mn\hat{h}_\mn=0$, and choosing a maximally symmetric background, eq.\
\eqref{eq:biHessian} boils down to
\begin{align}
 \big((\Gamma_k)_{\hat{h}\hat{h}}^{(2)}\big)^{\mn\rho\sigma}\Big|_\text{pr}
 &= \frac{\e^{(d-2)\Omega}}{32\pi G_k^\text{Dyn}}\; \bg^{\mu(\rho}\bg^{\sigma)\nu}\left[-\bD^2 
 + \frac{2}{d(d-1)}\,\bR\right], \\
 (\Gamma_k)_{\phi\phi}^{(2)}\mku\Big|_\text{pr} &= -\left(\!\frac{d-2}{2d}\!\mku\right)
 \frac{\e^{(d-2)\Omega}}{32\pi G_k^\text{Dyn}}\left[-\bD^2 - \frac{2\mku d}{d-2}\,\e^{2\Omega}\Lambda_k^\text{Dyn}
 + \frac{d-2}{d}\,\bR\right]\!,
\end{align}
where the off-diagonal parts of the Hessian, $(\Gamma_k)_{\hat{h}\phi}^{(2)}$ and $(\Gamma_k)_{\phi\hat{h}}^{(2)}$,
vanish identically. Similarly, we find for the ghost sector:
\begin{equation}
 \big(\big(\Gamma_k^\text{gh}\big)_{\xi\bx}^{(2)}\big)^\mu_{~\nu}\Big|_\text{pr}
 = -\big(\big(\Gamma_k^\text{gh}\big)_{\bx\xi}^{(2)}\big)^\mu_{~\nu}\Big|_\text{pr}
 = -\sqrt{2}\, \e^{2\Omega}\,\delta^\mu_\nu \big(-\bar{D}^2-\textstyle\frac{1}{d}\mku\bar{R}\mku\big)\,.
\end{equation}

Unlike in Ref.\ \cite{BR14}, we include the factor $\e^{(d-2)\Omega}$ ($\e^{2\Omega}$) in the cutoff operator $\Rk$ for
the gravitons (ghosts). Projected onto the various sectors we have
\begin{align}
 (\Rk)_{\hat{h}\hat{h}} &= \frac{\e^{(d-2)\Omega}}{32\pi G_k^\text{Dyn}}\; k^2 R^{(0)}\big(-\bD^2/k^2\big)\,, \\
 (\Rk)_{\phi\phi} &= -\left(\!\frac{d-2}{2d}\!\mku\right)\frac{\e^{(d-2)\Omega}}{32\pi G_k^\text{Dyn}}\; k^2 R^{(0)}
 \big(-\bD^2/k^2\big)\,, \\
 (\Rk^\text{gh})_{\xi\bx} &= -(\Rk^\text{gh})_{\bx\xi} = -\sqrt{2}\,\e^{2\Omega}\mku k^2 R^{(0)}\big(-\bD^2/k^2\big)\,.
\end{align}
The reason for the inclusion of $\e^{(d-2)\Omega}$ ($\e^{2\Omega}$) in $\Rk$ is given by the requirement that cutoff
operators be compatible with the standard replacement rule \cite{CPR09} of Laplacians occurring in inverse propagators
when the regularization is switched on, which, in our case, reads: $-\bD^2 \mapsto -\bD^2
+ k^2 R^{(0)}\big(-\bD^2/k^2\big)$.

Based on the above foundations we can finally apply the steps specified in Section \ref{sec:Recipe} in order to derive
the $\beta$-functions. The separation between dynamical and background quantities is realized by means of an expansion
in terms of $\Omega$ and a subsequent comparison of coefficients \cite{BR14}.

For the 'Dyn' couplings we find the following results: The anomalous dimension of $G_k^\text{Dyn}$, defined by
$\eta^\text{Dyn} \equiv k\p_k G_k^\text{Dyn}/G_k^\text{Dyn}$, is given by
\begin{equation}
 \eta^\text{Dyn} = \frac{\gDyn\mku B_1(\lDyn)}{1+\gDyn\mku B_2(\lDyn)}\,,
\end{equation}
with
\begin{align}
&\begin{aligned}
 B_1(\lDyn) = 8\mkuu (4\pi)^{1-d/2} \lDyn \Big\{& \textstyle\frac{d}{3 (d-2)^2}\, \Phi_{d/2-1}^2\left(-\mu \lDyn
 \right) \\
 &- \textstyle\frac{4}{d-2}\, \Phi_{d/2}^3 \left(-\mu\lDyn\right) \Big\},
\end{aligned}\\
&\begin{aligned}
 B_2(\lDyn) = 4\mkuu (4\pi)^{1-d/2} \lDyn \Big\{ &\textstyle\frac{d}{3 (d-2)^2}\,
 \tPhi_{d/2-1}^2\left(-\mu \lDyn\right) \\
 & - \textstyle \frac{4}{d-2}\,
 \tPhi_{d/2}^3 \left(-\mu\lDyn\right) \Big\},
\end{aligned}
\end{align}
where the constant $\mu$ is defined by $\mu\equiv\frac{2d}{d-2}$ again.
The $\beta$-function of the dimensionless dynamical Newton constant, $\gDyn_k =k^{d-2}G_k^\text{Dyn}$, then reads
\begin{equation}
 \beta_g^\text{Dyn} = \big(d-2+\eta^\text{Dyn}\big)\gDyn\,,
\label{eq:betagDynBi}
\end{equation}
and for the dimensionless dynamical cosmological constant, $\lDyn_k=k^{-2}\Lambda_k^\text{Dyn}$, we find
\begin{equation}
\begin{split}
 \beta_\lambda^\text{Dyn} = {}&\big(-2+\eta^\text{Dyn}\big)\lDyn \\
 &\!+ {\textstyle\frac{4}{d-2}}\mku (4\pi)^{1-d/2} \lDyn\mku\gDyn\Big\{
 2\mkuu\Phi_{d/2}^2\left(-\mu\lDyn\right) - \eta^\text{Dyn}\mkuu\tPhi_{d/2}^2\left(-\mu\lDyn\right) \Big\}.
\end{split}
\end{equation}

In the background sector, on the other hand, the anomalous dimension of $G_k^\text{B}$ is given by
\begin{equation}
\begin{split}
 \eta^\text{B} = - \textstyle\frac{1}{6}\,(&4\pi)^{1-d/2}\mku \gB\mku
 \bigg\{ 8d\,\Phi_{d/2-1}^1(0) - 4 \Phi_{d/2-1}^1\left(-\mu\lDyn\right) + 48\,\Phi_{d/2}^2(0) \\
 &- (d-1)(d+2)\left[ 2\,\Phi_{d/2-1}^1(0) - \eta^\text{Dyn}\,\tPhi_{d/2-1}^1(0)\right] \\
 &+ 2\mku\eta^\text{Dyn}\,\tPhi_{d/2-1}^1\left(-\mu\lDyn\right)
 + \textstyle\frac{12(d+2)}{d} \left[ 2\,\Phi_{d/2}^2(0) - \eta^\text{Dyn}\,\tPhi_{d/2}^2(0)\right] \\
 &+ \textstyle\frac{12(d-2)}{d} \left[ 2\,\Phi_{d/2}^2\left(-\mu\lDyn\right) - \eta^\text{Dyn}\,\tPhi_{d/2}^2
 \left(-\mu\lDyn\right)\right] \\
 &+ \textstyle\frac{8}{(d-2)^2}\,\lDyn\mku \Big[ 
 2d\,\Phi_{d/2-1}^2\left(-\mu\lDyn\right) - 24(d-2)\mku\Phi_{d/2}^3\left(-\mu\lDyn\right) \\
 &\qquad\qquad\qquad\; + 12(d-2)\eta^\text{Dyn}\mkuu\tPhi_{d/2}^3\left(-\mu\lDyn\right)\\
 &\qquad\qquad\qquad\; -\eta^\text{Dyn}\mku d\,\tPhi_{d/2-1}^2\left(-\mu\lDyn\right)
 \;\Big] \bigg\},
\end{split}
\end{equation}
and the $\beta$-functions of $\gB_k =k^{d-2}G_k^\text{B}$ and $\lB_k=k^{-2}\Lambda_k^\text{B}$ read, respectively,
\begin{equation}
 \beta_g^\text{B} = \big(d-2+\eta^\text{B}\big)\gB \,,
\end{equation}
\begin{equation}
\begin{split}
 \beta_\lambda^\text{B} ={} &\big(-2+ \eta^\text{B}\big)\lB + (4\pi)^{1-d/2}\mkuu\gB \Big \{
 -4d\,\Phi_{d/2}^1(0)+2\,\Phi_{d/2}^1\left(-\mu\lDyn\right) \\
 &+ (d-1)(d+2)\left[\Phi_{d/2}^1(0) - \textstyle\frac{1}{2}\mkuu\eta^\text{Dyn}\tPhi_{d/2}^1(0) \right]
 -\eta^\text{Dyn}\,\tPhi_{d/2}^1\left(-\mu\lDyn\right) \\
 &+ {\textstyle\frac{4}{d-2}}\,\lDyn\mku\left[ -2\,\Phi_{d/2}^2\left(-\mu\lDyn\right)
 + \eta^\text{Dyn}\mkuu \tPhi_{d/2}^2\left(-\mu\lDyn\right) \right]
 \Big\}.
\end{split}
\label{eq:betalBBi}
\end{equation}

Note the characteristic hierarchy of the above system of $\beta$-functions:
\begin{equation}
\begin{split}
 \beta_g^\text{Dyn} &\equiv \beta_g^\text{Dyn}\big( \gDyn, \lDyn \big)\,, \\
 \beta_\lambda^\text{Dyn} &\equiv \beta_\lambda^\text{Dyn}\big( \gDyn, \lDyn \big)\,, \\
 \beta_g^\text{B} &\equiv \beta_g^\text{B}\big( \gDyn, \lDyn, \gB \big)\,, \\
 \beta_\lambda^\text{B} &\equiv \beta_\lambda^\text{B}\big( \gDyn, \lDyn, \gB, \lB \big)\,, \\
\end{split}
\end{equation}
in agreement with the general consideration that led to \eqref{eq:BiHierarchy}.
In particular, the dynamical couplings form a closed subsystem which can be solved separately.
We show the resulting flow diagrams and analyze their properties in Section \ref{sec:bi}.

%----------------------------------------------------------------------------------------------------------------------
\chapter[Weyl transformations, zero modes and induced gravity action]%
{Weyl transformations, zero modes and the induced gravity action}
\label{app:Weyl}
%----------------------------------------------------------------------------------------------------------------------

In this appendix we list the behavior of various geometric objects under Weyl transformations, including the induced
gravity functional, which is needed in the main part of this thesis. Weyl transformations are given by
$\hg_\mn \to g_\mn$ with
\begin{equation}
 g_\mn = \e^{2\sigma} \hg_\mn \,,
\label{eq:DefWeylTransf}
\end{equation}
where $\sigma$ is a scalar function on the spacetime manifold.

\medskip
\noindent
\textbf{(1)}
From the definition of the Christoffel connection we immediately obtain
\begin{equation}
 \Gamma^\alpha_\mn=\hat{\Gamma}^\alpha_\mn + \delta^\alpha_\mu \hD_\nu\sigma + \delta^\alpha_\nu \hD_\mu\sigma
  - \hg_\mn\hD^\alpha\sigma\, .
\label{eq:WeylChrist}
\end{equation}
Note that indices (on the right hand side) are raised and lowered by means of $\hg^\mn$ and $\hg_\mn$, respectively.
From \eqref{eq:WeylChrist} we easily deduce the Riemann tensor and its contractions,
\begin{align}
\begin{split} R^\alpha_{\mu\nu\rho} &= \hR^\alpha_{\mu\nu\rho} + 2\,\hg_{\mu[\nu}\hD_{\rho]}\hD^\alpha \sigma
  - 2\,\delta^\alpha_{[\nu}\hD_{\rho]}\hD_\mu \sigma - 2\,\hg_{\mu[\nu}\hD_{\rho]}\sigma \hD^\alpha \sigma\\
  &\hspace{3.34em}+ 2\,\delta^\alpha_{[\nu}\hD_{\rho]}\sigma \hD_\mu \sigma
  + 2\,\hg_{\mu[\nu}\delta^\alpha_{\rho]} \hD_\beta \sigma \hD^\beta \sigma \, ,
\end{split}\\
R_\mn &=  \hR_\mn - (d-2)\Big(\hD_\mu \hD_\nu \sigma - \hD_\mu \sigma \hD_\nu \sigma\Big)
   - \hg_\mn \Big[\hB\sigma +(d-2)\hD_\alpha\sigma\hD^\alpha \sigma \Big] , \\
R &= \e^{-2\sigma}\left[\hR-(d-1)(d-2)\hD_\mu\sigma \hD^\mu\sigma -2(d-1)\hB\sigma\right] ,
\end{align}
where $\hB\equiv\hD_\alpha\hD^\alpha$ and the square brackets enclosing indices denote antisymmetrization,
$A_{[\mn]} = \frac{1}{2}(A_\mn-A_{\nu\mu})$. Note that since the underlying connection is given by the Christoffel
symbols, i.e.\ it is torsion free, we have $\hD_\mu\hD_\nu \sigma = \hD_\nu\hD_\mu \sigma$. For the Einstein tensor we
find
\begin{equation}
 G_\mn = \hat{G}_\mn + (d-2)\left[ -\hD_\mu \hD_\nu \sigma + \hg_\mn\hB\sigma + \hD_\mu\sigma \hD_\nu\sigma
  + \frac{d-3}{2}\hg_\mn \hD_\alpha\sigma \hD^\alpha\sigma \right].
\label{eq:WeylEinstein}
\end{equation}
Furthermore, the metric determinant transforms as
\begin{equation}
 \sg = \shg\, \e^{d\sigma} \, .
\label{eq:TransMetricDet}
\end{equation}
Hence, we arrive at the useful relations
\begin{align}
 \sg\,R &= \e^{(d-2)\sigma}\shg \left[\hR-(d-1)(d-2)\hD_\mu\sigma \hD^\mu\sigma-2(d-1)\hB\sigma\right], \\
 \int\dd x\sg\,R &= \int\dd x\shg\,\e^{(d-2)\sigma} \left[\hR+(d-1)(d-2)\hD_\mu\sigma \hD^\mu\sigma
 \right].\label{eq:EHexpanded}
\end{align}
The transformation behavior of the Laplacian is given by
\begin{equation}
\Box f = \e^{-2\sigma}\hB f +(d-2)\e^{-2\sigma} \hD_\mu \sigma\mku \hD^\mu f \, ,
\label{eq:WeylLaplace}
\end{equation}
where $f$ is an arbitrary scalar function.

\medskip
\noindent
\textbf{(2)}
In the special case of two dimensions, $d=2$, we obtain
\begin{align}
 R &= \e^{-2\sigma}\left[\hR-2\hB\sigma\right],
 \label{eq:WeylR}\\
 \sg\,R &= \shg \left[\hR-2\hB\sigma\right],
 \label{eq:WeylgR}\\
 \Box f &= \e^{-2\sigma}\hB f\,,
 \label{eq:TransLaplace2D}\\
 \sg\,\Box f &= \shg\,\hB f \,.
 \label{eq:TransSgLaplace2D}
\end{align}

\medskip
\noindent
\textbf{(3)}
Due to its relevance to the induced gravity action we are particularly interested
in the transformation behavior of $\Box^{-1}R\mku$, with the inverse Laplacian (Green's function)
$\Box^{-1}\equiv\Box^{-1}(x,y)$, where $(\Box^{-1}R)(x)$ refers to
\begin{equation}
 (\Box^{-1}R)(x) \equiv \int\dd y\sg\;\Box^{-1}(x,y)R(y).
\label{eq:BoxMOne}
\end{equation}
If $\Box$ has no zero modes, its inverse is defined by $\Box\big[\Box^{-1}(x,y)\big]=\frac{1}{\sg}\delta(x-y)$,
cf.\ App.~\ref{app:OperatorRep}. On the other hand, if $\Box$ has normalizable
zero modes, then $\Box^{-1}$ is defined as the inverse of $\Box$ on the orthogonal complement to its kernel, where the
delta function has to be modified appropriately, that is, $\Box\,\Box^{-1}(x,y)=\frac{1}{\sg}\delta(x-y)-
\text{Pr}_0(x,y)$, and $\text{Pr}_0$ denotes the projection onto zero modes. Whenever we write $\Box^{-1}$ in this
thesis, this definition is meant implicitly.
  
\medskip
\noindent
\textbf{(4)}
Since the consideration of zero modes requires a more careful treatment, we first consider the situation where zero
modes are absent in the following subsection, before investigating the general case in Subsection \ref{app:Zero}.

%----------------------------------------------------------------------------------------------------------------------
\section{The induced gravity action in the absence of zero modes}
\label{app:NoZero}
%----------------------------------------------------------------------------------------------------------------------

If the Laplacian has no zero modes, then the equation $\Box f=h$ can be solved for $f$ by direct inversion of $\Box$,
that is, $f=\Box^{-1}h$. In this case the transformation behavior of the Green's function $\Box^{-1}$ is given by
\begin{equation}
 \Box^{-1}\big(\e^{-2\sigma}\,h\big) = \hB^{-1}h\,.
\end{equation}
This gives rise to
\begin{equation}
 \Box^{-1}R = \hB^{-1} \hR -2\sigma.
 \label{eq:WeylInvBoxR}
\end{equation}

For our arguments in Section \ref{sec:EpsilonLimit} we need to determine the transformation behavior of the induced
gravity functional $I[g]$ which can be defined as the normalized finite part of Polyakov's induced effective action
\cite{Polyakov1981}:
\begin{equation}
\textstyle
 \Gi[g] = \frac{1}{2}\Tr\ln(-\Box)\,.
\label{eq:GiWithoutZero}
\end{equation}
In the absence of zero modes, the trace in \eqref{eq:GiWithoutZero} can be computed explicitly. The result, $\Gi[g]$,
consists of a universal finite part and a regularization scheme dependent divergent part. Regularizing by means of a
proper time cutoff \cite{BV87,BV90a,BV90b,Vilkovisky1992}, for instance, one obtains from eq.\ \eqref{eq:GiWithoutZero}:
\begin{equation}
 \Gi[g] = \frac{1}{96\pi}\int\td^2 x\sg\, R\,\Box^{-1}R - \frac{1}{8\pi s}\int\td^2 x\sg\,.
\label{eq:GammaInd}
\end{equation}
The second term on the RHS of eq.\ \eqref{eq:GammaInd} is scheme dependent and divergent in the limit
$s\rightarrow 0$. It might be absorbed by a redefinition of the cosmological constant. The first term, on the other
hand, contains all relevant information, so we focus on it for our further investigations. We define the induced
gravity functional $I[g]$ to be proportional to \emph{the finite part of} $\Gi[g]$,
\begin{equation}
 I[g] \equiv 96\pi\; \Gi[g]\big|_\text{finite} = \int\td^2 x\sg\, R\,\Box^{-1}R \,.
\label{eq:IDefinition}
\end{equation}
Using \eqref{eq:WeylgR} and \eqref{eq:WeylInvBoxR} we now obtain, after integrating by parts,
\begin{equation}
  I[g] = \int\td^2x\shg\left[\hR\,\hB^{-1}\hR-4\hR\sigma+4\sigma\hB\sigma\right].
\end{equation}
This can be written as
\begin{equation}[b]
 I[g] - I[\hg] = - 8\,\Delta I[\sigma;\hg],
\label{eq:ItoDeltaI}
\end{equation}
with the functional $\Delta I$ defined by
\begin{equation}
 \Delta I[\sigma;\hg] \equiv \frac{1}{2}\int\td^2x\shg\left[\hD_\mu\sigma\hD^\mu\sigma+\hR\sigma\right].
\label{eq:DeltaI}
\end{equation}

These results prove useful for calculating the 2D limit of the Einstein--Hilbert action, as applied
in Sections \ref{sec:genRem} and \ref{sec:EpsilonLimit}.

%----------------------------------------------------------------------------------------------------------------------
\section{The treatment of zero modes}
\label{app:Zero}
%----------------------------------------------------------------------------------------------------------------------
What is different and which results of Section \ref{app:NoZero} remain valid when the scalar Laplacian has one or more
zero modes? To illustrate the issue let us start from scratch and consider a functional integral over a simple scalar
field $X$ minimally coupled to the metric. Integrating out $X$ will ``induce'' a gravity action for the metric then.
The corresponding partition function is given by
\begin{equation}
 \tilde{Z}[g] \equiv \int\mD X\;\e^{-\frac{1}{2}\int\td^2 x\sg\,g^\mn\mku\p_\mu X\,\p_\nu X}
 = \int\mD X\;\e^{-\frac{1}{2}\int\td^2 x\sg\,X(-\Box) X}\,.
\label{eq:ZDiv}
\end{equation}
(The notation with the tilde is chosen since definition \eqref{eq:ZDiv} is pathological and has to be modified as shown
in the following.) Let us expand the field $X$ in terms of normalized eigenmodes $\vp^{(n)}$ of the Laplacian
$-\Box$, that is, $X = \sum_n c_n\, \vp^{(n)}$, where $-\Box\mku \vp^{(n)}=\lambda_n\vp^{(n)}$, with the normalization
$\int\td^2 x\sg\;\vp^{(n)}(x)\,\vp^{(m)}(x)=\delta_{mn}$. Then the integral in \eqref{eq:ZDiv} can be written as
\begin{equation}
 \tilde{Z}[g] = \int\prod\limits_n \frac{\td c_n}{\sqrt{2\pi}}\;\e^{-\frac{1}{2}\sum_n \lambda_n\,c_n^2}\,.
\end{equation}
Now let us suppose that the Laplacian has a zero mode, $-\Box\mku\vp^{(0)}=0$, i.e.\ $\lambda_0=0$. In this case the
integration over its Fourier coefficient, $\int\td c_0 \;\e^{-\frac{1}{2}\lambda_0\,c_0^2} =\int\td c_0\;1$, is
\emph{divergent}, and so is $\tilde{Z}[g]$. Thus, the zero mode(s) has to be \emph{excluded from the path integral}
in the first place. The correct definition reads
\begin{equation}
 Z[g] \equiv \int\mD' X\;\e^{-\frac{1}{2}\int\td^2 x\sg\,g^\mn\mku\p_\mu X\,\p_\nu X}\,.
\label{eq:ZCorrect}
\end{equation}
Here and in the following, the prime denotes the exclusion of zero modes.

We will consider only connected manifolds with vanishing boundary. In that case the Laplacian has (at most) one single
normalized zero mode. It is given by
\begin{equation}
 \vp^{(0)} = 1/\sqrt{V}\,,
\end{equation}
with the volume, or area, $V=\int\td^2 x\sg\,$.

Performing the Gaussian integral in eq.\ \eqref{eq:ZCorrect} one obtains\footnote{As we will see in App.\
\ref{app:Recon}, eq.\ \eqref{eq:ZTildeComp} actually receives a contribution from the functional measure, too, which
may be indicated by $Z[g]=\big[\det_\UV'(-\Box/M^2)\big]^{-1/2}$. In the present case, this modification merely gives
rise to additional, inessential constants which we do not write explicitly henceforth.}
\begin{equation}
 Z[g]=\big[\detp(-\Box)\big]^{-\frac{1}{2}}.
\label{eq:ZTildeComp}
\end{equation}
The corresponding effective action $\Gi$ is determined by $Z\equiv\e^{-\Gi}$, leading to
\begin{equation}
\textstyle
 \Gi[g] = \frac{1}{2}\ln\detp(-\Box) = \frac{1}{2}\Trp\ln(-\Box)\,,
\label{eq:GiDef}
\end{equation}
which is Polyakov's induced gravity action, adapted to taking account of zero modes. In order to find an integral
representation for $\Gi$ similar to eq.\ \eqref{eq:GammaInd} it turns out convenient to consider the variation of $\Gi$
under a finite Weyl transformation, giving rise to a strictly local term and a term involving the logarithm of the
volume (see e.g.\ \cite{FV11}): The finite part of the variation reads
\begin{equation}
 \Gi[g] - \Gi[\hg] = - \frac{1}{12\pi}\,\Delta I[\sigma;\hg] + \frac{1}{2}\,\ln\left(V/\hat{V}\right) ,
\label{eq:GitoDeltaIZero}
\end{equation}
with the volume terms $V\equiv\int\td^2 x\sg$ and $\hat{V}\equiv\int\td^2 x\shg$, and with $\Delta I[\sigma;\hg]$ as
defined in eq.\ \eqref{eq:DeltaI}. The second term on the RHS of \eqref{eq:GitoDeltaIZero} originates from the zero
mode contribution contained in the conformal factor.

To extract an explicit expression for $\Gi$ from \eqref{eq:GitoDeltaIZero} that depends only on one metric, we aim at
eliminating the conformal factor and rewrite also the RHS of \eqref{eq:GitoDeltaIZero} as the difference between
some functional evaluated at $g$ and the same functional evaluated at $\hg$.
Although the existence of such a representation can be proven \cite{Dowker1994},
the explicit form of $\Gi[g]$ with only one argument is (to the best of our knowledge) not known in general.
As already pointed out in Ref.\ \cite{Duff1994}, the problem occurs for uniform rescalings when the conformal factor is
a constant, i.e.\ proportional to the zero mode: In this case even the formula $\int g^\mn\,\frac{\delta S[g]}{\delta
g_\mn}= \frac{1}{2}\frac{\p S[\e^{2\sigma}g]}{\p \sigma}\big|_{\sigma=0}\,$, where $\sigma$ is a constant, does not
apply, a counterexample being the induced gravity functional \eqref{eq:IDefinition} which is invariant under uniform
rescalings but whose metric variation gives rise to the anomaly proportional to $R$.

To eliminate the conformal factor in \eqref{eq:GitoDeltaIZero} we would like to solve the equation
\begin{equation}
 \Box\mku\sigma = \textstyle\frac{1}{2\,\sg}\left(\shg\hR-\sg R\right)
\label{eq:sigmaInTermsOfgandhg}
\end{equation}
for $\sigma$, where \eqref{eq:sigmaInTermsOfgandhg} follows from \eqref{eq:WeylgR} and the identity
$\shg\,\hB = \sg\,\Box$, valid in 2D. The existence of a solution is guaranteed by the fact that the RHS of
\eqref{eq:sigmaInTermsOfgandhg} is orthogonal to the zero mode, thanks to topological invariance. 
However, the conformal factor itself could have a contribution from the zero mode. As a consequence, the solution
for $\sigma$ is not unique. Employing the Green's function $\Box^{-1}$ as defined below eq.\ \eqref{eq:BoxMOne}
we obtain
\begin{equation}
 \sigma = \textstyle\frac{1}{2}\,\Box^{-1}\frac{1}{\sg}\left(\shg\hR-\sg R\right) + \frac{1}{V}\int\sg\,\sigma,
\label{eq:SolSigma}
\end{equation}
where the second term is the constant zero mode part. (Recall that $\Box^{-1}$ is the inverse of $\Box$ on the
orthogonal complement to the kernel of $\Box$, and it satisfies $\Box\,\Box^{-1}(x,y)=\frac{1}{\sg}\delta(x-y)-
\frac{1}{V}$.)
Making use of the relation $\sigma = \frac{1}{2}\ln(\sg/\shg)$ the last term in \eqref{eq:SolSigma} can be expressed
in terms of the metrics $g_\mn$ and $\hg_\mn$, too. Then eq.\ \eqref{eq:GitoDeltaIZero} becomes
\begin{equation}
 \Gi[g]-\Gi[\hg] = \Gi[g,\hg],
\end{equation}
with the both $g_\mn$- and $\hg_\mn$-dependent functional \cite{Dowker1994}
\begin{equation}
\begin{split}
 \Gi[g,\hg] \equiv\; &\frac{1}{96\pi}\int\textstyle\left(\sg R+\shg\hR\right)\Box^{-1}\frac{1}{\sg}
 \left(\sg R-\shg\hR\right)\\
 &-\textstyle\frac{\chi}{12\mku V}\int\sg\,\ln\left(\frac{\sg}{\shg}\right)
 +\frac{1}{2}\,\ln\left(\frac{V}{\hat{V}}\right),
\end{split}
\label{eq:IWeylTransGeneral}
\end{equation}
where we have used $\int\td^2 x\sg\,R=4\pi\mku\chi\mku$ again. In this expression it does not seem possible to
disentangle $g$ from $\hg$.

Nevertheless, by introducing a fiducial metric $g_0$ in \eqref{eq:IWeylTransGeneral} we could define $\Gi[g]$ formally
up to an additive constant by
\begin{equation}
 \Gi[g]\equiv \Gi[g,g_0].
\end{equation}
Employing this definition, $\Gi[g]$ indeed satisfies eq.\ \eqref{eq:GitoDeltaIZero}. The corresponding functional
$I^\text{full}[g]$ (where $I^\text{full}$ refers to the general case, with zero mode and arbitrary rescalings) can be
obtained by applying rule \eqref{eq:IDefinition}, $I^\text{full}[g] \equiv 96\pi\; \Gi[g]|_\text{finite}\,$,
resulting in
\begin{equation}
 I^\text{full}[g] \equiv I[g] + R[g,g_0],
\label{eq:DefIFull}
\end{equation}
with $I[g]=\int\!\sg\,R\,\Box^{-1}R$ as above, and with the residue
\begin{equation}
 R[g,g_0] \equiv -\int\sqrt{g_0}\,R(g_0)\Box^{-1}{\textstyle\frac{\sqrt{g_0}}{\sg}}\,R(g_0)
 - \frac{8\pi\chi}{V}\int\sg\,\ln\left(\textstyle\frac{\sg}{\sqrt{g_0}}\right)
 +48\pi\,\ln\left(\textstyle\frac{V}{V_0}\right).
\label{eq:Residue}
\end{equation}
This residue is due to the zero mode contribution to the conformal factor relating $g$ with $g_0$.
Using eq.\ \eqref{eq:GitoDeltaIZero} leads to a transformation behavior of $I^\text{full}[g]$ similar to the one found
in Section \ref{app:NoZero}. We obtain
\begin{equation}[b]
 I^\text{full}[g] - I^\text{full}[\hg] = - 8\,\Delta I[\sigma;\hg] + 48\pi\,\ln\left(V/\hat{V}\right) .
\label{eq:ItoDeltaIZero}
\end{equation}
Thus, apart from the pure volume terms we recover the same result as in eq.\ \eqref{eq:ItoDeltaI}, the modification
being due to the zero modes of $\Box$ and $\hB$, $\vp^{(0)}=1/\sqrt{V}$ and $\hat{\vp}^{(0)}=1/\sqrt{\hat{V}}$,
respectively.

Concerning our results of Section \ref{sec:IndGravityFromEH}, we observe that $I[g]$ is to be replaced according to
\begin{equation}
 I[g] \;\rightarrow\; I^\text{full}[g]-48\pi\,\ln(V/V_0),
\label{eq:IReplace}
\end{equation}
where the corresponding behavior under Weyl transformations is given by eq.\ \eqref{eq:ItoDeltaIZero}. Thus, in the
general case there are additional correction terms in consequence of the zero modes. In particular,
eq.\ \eqref{eq:LimitResult} generalizes to
\begin{equation}
 \frac{1}{\ve}\int\td^{2+\ve}x\sg\, R
 = -\frac{1}{4}I[g] + Q[g,g_0] +\frac{4\pi\chi}{\ve}+C\big(\{\tau\}\big)+\mO(\ve),
\label{eq:LimitResultGen}
\end{equation}
with the correction terms $Q[g,g_0]\equiv\frac{1}{4}\int\!\sqrt{g_0}\,R(g_0)\Box^{-1}\frac{\sqrt{g_0}}{\sg}\,R(g_0)
 + \frac{2\pi\chi}{V}\int\!\sg\,\ln\left(\frac{\sg}{\sqrt{g_0}}\right)$. We point out that the crucial
result in eq.\ \eqref{eq:LimitResult}, the appearance of the nonlocal action $I[g]$, is contained in its extension
\eqref{eq:LimitResultGen}, too. All conclusions in the main part of this thesis that relied on the emergence of
$I[g]$ in the 2D limit of the Einstein--Hilbert action remain valid in the presence of zero modes. The correction terms
in \eqref{eq:LimitResultGen} do not change our main results; in particular the central charge, which is read off from
the prefactor of $I[g]$, remains unaltered.

Finally, two comments are in order.

\medskip
\noindent
\textbf{(1) Nonvanishing Euler characteristics}. We would like to point out the following subtlety concerning the
induced gravity functional $I[g]$. As argued above, $\Box^{-1}$ is defined such that it affects only nonzero modes
while it ``projects away'' the zero modes of the objects it acts on. In particular, the function $(\Box^{-1}R)(x)$
satisfies $\Box\,\Box^{-1}R = R - \frac{1}{V}\,4\pi\chi$. Hence, for manifolds with vanishing Euler characteristic,
$\chi=0$, we recover the usual feature of an inverse operator, $\Box\,\Box^{-1}R=R$, as long as $\Box^{-1}$ acts on
$R$. The reason behind this property is that the Fourier expansion of $R$ cannot contain any contribution $\propto
c_0\vp^{(0)}$ from the zero mode if $\chi=0$. As a consequence $\Box^{-1}R$ is nonzero provided that $R$ does not
vanish, and, in turn, $I[g]$ is a nonzero functional.

On the other
hand, if $\chi\neq 0$, then it might happen that $I[g]$ vanishes. As an example, let us consider a sphere with constant
curvature $R>0$. Since $R$ is proportional to the constant zero mode in this case, we have $\Box^{-1}R=0$, and thus
$I[g]=0$. With regard to eq.\ \eqref{eq:ItoDeltaIZero} this means that all nontrivial contributions to the LHS must
come from $I^\text{full}[\hg]$ and from the residue contained in $I^\text{full}[g]$.

\medskip
\noindent
\textbf{(2) A modified induced gravity functional}.
The occurrence of the volume term in eq.\ \eqref{eq:ItoDeltaIZero} can be understood as follows. We removed the zero
modes from the path integral \eqref{eq:ZCorrect}, and this exclusion affects the transformation behavior,
replacing \eqref{eq:ItoDeltaI} with \eqref{eq:ItoDeltaIZero}. However, there is the possibility to redefine the
partition function in order to absorb the volume terms. Let us briefly sketch the idea.

As above, we expand the scalar field $X$ in the partition function in terms of normalized eigenmodes $\vp^{(n)}$ of
the Laplacian, $X=\sum_n c_n\,\vp^{(n)}$, and insert this into eq.\ \eqref{eq:ZCorrect}. Then it is easy to show (see
e.g.\ \cite{ID89}) that the transformation behavior of $\ln Z$ under an infinitesimal Weyl variation according to
eq.\ \eqref{eq:DefWeylTransf}, $\delta g_\mn=2\sigma\, g_\mn$, is given by
\begin{equation}
 \delta\ln Z =\int\td^2 x\sg\,\left(\frac{1}{4}\frac{\delta g}{g}\right)\sum_{n=0}^\infty\big[\vp^{(n)}\big]^2
 -\frac{1}{2}\frac{\delta V}{V}\,.
\end{equation}
Rearranging terms yields
\begin{equation}
 \delta\ln\left(\sqrt{V/V_0}\,Z\right)
 = \int\td^2 x\sg\,\left(\frac{1}{4}\frac{\delta g}{g}\right)\sum_{n=0}^\infty\big[\vp^{(n)}\big]^2\,,
\label{eq:Rearranged}
\end{equation}
where $V_0$ is an arbitrary reference volume introduced merely to render the argument of the logarithm dimensionless.
The advantage of eq.\ \eqref{eq:Rearranged} is that its RHS does no longer contain any distinction between zero and
nonzero modes, hence the combination $\sqrt{V/V_0}\,Z$ is more appropriate for a treatment of all modes on an equal
footing.

These observations suggest introducing the modified definition
\begin{equation}
 Z^\text{mod}[g] \equiv \sqrt{V/V_0}\int\mD' X\;\e^{-\frac{1}{2}\int\td^2 x\sg\,g^\mn\mku\p_\mu X\,\p_\nu X}\,.
\label{eq:ZDef}
\end{equation}
The corresponding effective action reads
\begin{equation}
\textstyle
 \Gamma^\text{ind,mod}[g] = \frac{1}{2}\ln\detp(-\Box)-\frac{1}{2}\ln\frac{V}{V_0} \,.
\label{eq:GiMod}
\end{equation}
This modified effective action is often used in the literature \cite{CKHT87}. Applying the rule \eqref{eq:IDefinition}
to \eqref{eq:GiMod} and using \eqref{eq:DefIFull} yields the \emph{modified induced gravity functional}
\begin{equation}
 I^{\mku\text{mod}}[g] \equiv I^\text{full}[g]-48\pi\,\ln{\textstyle\frac{V}{V_0}} \,,
\end{equation}
consistent with \eqref{eq:IReplace}.
Employing eq.\ \eqref{eq:ItoDeltaIZero} we find that it transforms according to
\begin{equation}[b]
 I^{\mku\text{mod}}[g] - I^{\mku\text{mod}}[\hg] = - 8\,\Delta I[\sigma;\hg],
\end{equation}
with $\Delta I$ as defined in eq.\ \eqref{eq:DeltaI}. Thus, for $I^{\mku\text{mod}}[g]$ we recover the same behavior
under Weyl transformations as for $I[g]$ in eq.\ \eqref{eq:ItoDeltaI}, which was the transformation law for the case
without zero modes.

In conclusion, zero modes can be taken into account by employing a modified definition of the path integral, where the
behavior of the (generalized) induced gravity functional under Weyl rescalings remains essentially the same.

%----------------------------------------------------------------------------------------------------------------------
\chapter{Reconstructing the bare action from the effective average action}
\label{app:Recon}
%----------------------------------------------------------------------------------------------------------------------

We have seen that solutions to the FRGE do not depend on any underlying path integral description. Nonetheless, in
particular cases the bare action appearing in the exponent of a suitably defined functional integral may be of
interest, too. This raises the following question: Given an effective average action $\Gamma_k$ which solves the FRGE,
can we find a bare action and a functional measure such that the functional integration reproduces $\Gamma_k\mku$?
In this appendix we give a detailed derivation of a one-loop ``reconstruction formula'' which can be used to determine
the bare action approximately provided that $\Gamma_k$ is known.

Before we can reconstruct the bare action, however, we have to specify the measure of the corresponding functional
integral. It turns out that the definition is usually not unique but depends on a tunable free parameter instead. This
will be worked out in Section \ref{app:Measure}. Thereafter we derive the reconstruction formula in Section
\ref{app:ReconGeneral}, and we prove that it becomes an exact relation for certain terms when the large cutoff limit is
taken. The results are applied to a gravitational EAA of Einstein--Hilbert type and to
Liouville theory in Chapters \ref{chap:Bare} and \ref{chap:BareLiouville}, respectively, in the body of this thesis.

%----------------------------------------------------------------------------------------------------------------------
\section{Definition of the functional measure}
\label{app:Measure}
%----------------------------------------------------------------------------------------------------------------------

Let $\vp$ denote a generic field. We have argued in Chapter \ref{chap:Bare} that the bare action $\SB[\vp]$ alone has
no significance at all. It is rather a combination of measure and bare action, $\td\mu[\vp]\,\exp(-\SB[\vp])$, which
defines a meaningful quantity. In other words, stating $\SB$ would be pointless without knowing the measure.

There is an elegant but not unambiguous way to define the measure by employing Gaussian integrals \cite{Mottola1995}.
This method relies on a given inner product on field space,\footnote{More precisely, in Ref.\ \cite{Mottola1995} the
construction is based on an inner product on the cotangent space of infinitesimal deformations of the underlying field
space. For the sake of our argument and for simplicity, however, we regard the field space as a vector space with a
scalar product here, the generalization being straightforward.} denoted by $\langle \vp,\vp \rangle$. Then the measure
$\td\mu$ is fixed by requiring $\int\td\mu[\vp]\,\e^{-\frac{1}{2}\langle \vp,\vp \rangle}=1$. However, there is a
subtlety in this argument that demands further investigations.

The crucial point is that the exponent in this definition as well as the overall result of the path integral should
be pure numbers without any mass dimension. This has to be reconciled with the fact that a generic field usually comes
with a canonical mass dimension which may be determined by dimensional analysis of the kinetic term in an associated
action.\footnote{We point out that the mass dimensions of fields should be considered as inputs,
depending on allowed field space monomials and on the dimensions of coupling constants.}
Therefore, it is necessary in general to include a mass scale in the inner product. For scalar fields with their
inherent mass dimension $[\vp]=(d-2)/2$, for instance, a suitable definition would be $\langle \vp_1,\vp_2 \rangle
\equiv \int\dd x\sg\,M^2\mku\vp_1(x)\mku\vp_2(x)$, involving some external mass scale $M$. That means, the inner
product can be used to measure distances in field space in units of $M$. A priori, $M$ is not related to any cutoff
scale but serves as a free parameter. Given $M$, the functional measure can now be fixed by the modified requirement
$\int\td\mu_M[\vp]\,\e^{-\frac{1}{2}\int\dd x\sg\,M^2\vp^2}=1$, where we allow an explicit $M$-dependence in
$\td\mu_M[\vp]$.

Note that this defining expression is invariant under rescalings of $M$ if $\vp$ and the metric $g_\mn$ are
rescaled as well. However, when including a second scale, say $k$, for the renormalization procedure, such a metric
rescaling is not desired as it would also change the eigenvalues of modes which are suppressed. Thus, in general there
is no invariance under rescalings of $M$, and the measure remains $M$-dependent. Only in terms of dimensionless fields
and couplings this dependence drops out. Our main observation here is that $M$ may be considered a free parameter which
can be tuned to adjust the measure, giving rise to a change of the bare action in turn. We emphasize that this freedom
signals the ``unphysicalness'' of the bare action.

In order to make the construction of the measure more explicit, we avail ourselves of an argument used previously in
Refs.\ \cite{Fujikawa1979,Fujikawa1980,Fujikawa1981,FS2004}. We aim at computing a functional integral of the type
\begin{equation}
 \int\td\mu_M[\vp]\;\e^{-\frac{1}{2}\int\dd x\sg\,\vp\,\mO\,\vp},
\label{eq:GaussianIntegral}
\end{equation}
where $\mO$ is an arbitrary positive operator which appears in the integral in its differential operator
representation, the case of the scalar Laplacian, $\mO=-\Box$, being of primary importance for our studies. It is
assumed that there is a complete set of orthonormal eigenfunctions, $\{\vp_n\}$, satisfying
\begin{equation}
 \mO\mku\vp_n=\lambda_n\vp_n\,,
\end{equation}
where the orthonormality condition is with respect to the above inner product, i.e.\ we have 
$\langle \vp_i,\vp_j \rangle = \int\dd x\sg\,M^2\mku\vp_i(x)\mku\vp_j(x)=\delta_{ij}$. As pointed out in Ref.\
\cite{Fujikawa1981}, the requirement for manifest covariance under general coordinate transformations dictates choosing
a measure which is constructed from the modified field $\tilde{\vp}\equiv g^{1/4}\mku\vp$ with weight $\frac{1}{2}$:
\begin{equation}
 \td\mu_M[\vp] \equiv \mC\prod_x \frac{\td\tilde{\vp}(x)}{M^\kappa} \,,
\label{eq:PreMeasure}
\end{equation}
with a normalization constant $\mC$ to be determined in a moment and with the mass dimension 
$\kappa = \big[\tilde{\vp}\big]$, which amounts to $\kappa = -1$ if $\vp$ is a standard scalar field.

The reason for this choice of the measure can be understood as
follows. Let us expand the field $\vp$ in terms of eigenmodes of the operator $\mO$,
\begin{equation}
 \vp(x)=\sum\limits_{i=1}^\infty a_i\mku \vp_i(x).
\end{equation}
Then the measure \eqref{eq:PreMeasure} receives contributions from the Jacobians, formally leading to
\cite{FS2004,Fujikawa1981}
\begin{align}
 \td\mu_M[\vp] &= \mC\prod_x \frac{\td\tilde{\vp}(x)}{M^\kappa}=\mC\,\det\!\bigg[\frac{g^{1/4}\vp_i(x)}{M^\kappa}\bigg]
 \prod_n\td a_n = \mC \, \det\!\bigg[\frac{g^{1/4}\langle x|\vp_i\rangle}{M^\kappa}\bigg] \prod_n\td a_n\qquad
 \nonumber\\
 &= \mC\,\bigg\{\det\!\bigg[\frac{g^{1/4}\langle\vp_i|x\rangle}{M^\kappa}\bigg]
  \det\!\bigg[\frac{g^{1/4}\langle x|\vp_j\rangle}{M^\kappa}\bigg] \bigg\}^{1/2} \, \prod_n\td a_n \nonumber\\
 &= \mC \, \det{}^{\!1/2}\Big[\sum\nolimits_x \sg\,M^2\mku\langle\vp_i|x\rangle\langle x|\vp_j\rangle\Big]
 \prod_n\td a_n \\
 &= \mC \, \det{}^{\!1/2}\Big[\int\dd x\sg\,M^2\mku\vp_i(x)\vp_j(x)\Big] \prod_n\td a_n
 = \mC \, \det{}^{\!1/2}(\delta_{ij})\mku \prod_n\td a_n \nonumber\\ &= \mC \prod_n\td a_n \,.\nonumber
\end{align}
Thus $\td\mu_M[\vp]$ can be written in terms of the standard translation invariant measures $\td a_n$ alone, i.e.\ it
does no longer involve any $x$-dependent terms, satisfying the general covariance condition in this way. Furthermore,
in this representation the $M$-dependence in $\td\mu_M$ has dropped out completely. (We keep the index $M$, though,
since $M$ enters another term which can be seen as part of the measure. This is shown in a moment.)

A generic QFT usually has to cope with UV divergences and needs to be regularized. The most straightforward way to
regularize the functional integral is to restrict the contributing modes by cutting off the high momentum parts at some
UV scale, say, $\UV$. In our setting this translates into restricting the modes with respect to a ``cutoff index'' $N$,
and the measure becomes
\begin{equation}
 \td\mu_M^N[\vp] = \mC \prod_{n=1}^N \td a_n \,.
\end{equation}
Consequently, all appearances of $\vp$ in the path integral must be projected onto low momentum modes, too
\cite{ABG1982}: $\vp(x) = \sum_{n=1}^N a_n \vp_n(x)$. The Gaussian integral \eqref{eq:GaussianIntegral} can now be
evaluated, and we find
\begin{equation}
\begin{split}
 \int\td\mu_M^N[\vp]\;\e^{-\frac{1}{2}\int\dd x\sg\,\vp\,\mO\,\vp}
 &= \mC\int \prod_{n=1}^N \td a_n\;
  \e^{-\frac{1}{2}\int\dd x\sg\sum_{i=1}^N a_i\vp_i(x)\,\mO\mku\sum_{j=1}^N a_j\vp_j(x)} \\
 &= \mC\int \prod_{n=1}^N \td a_n\; \e^{-\frac{1}{2}\sum_{i,j}^Na_i a_j \lambda_j M^{-2}\delta_{ij}} \\
 &= \mC\, \sqrt{\frac{(2\pi)^N M^{2N}}{\lambda_1 \cdots \lambda_N}}
 = \mC\, (2\pi)^{\frac{N}{2}}\,\det{}_N^{-\frac{1}{2}}\left(\mO/M^2\right),
\end{split}
\end{equation} 
where the index $N$ in the determinant indicates the exclusion of high momentum modes. Choosing the normalization
$\mC \equiv (2\pi)^{-N/2}$, we finally obtain
\begin{equation}
 \int\td\mu_M^N[\vp]\;\e^{-\frac{1}{2}\int\dd x\sg\,\vp\,\mO\,\vp} = \det{}_{\!N}^{-\frac{1}{2}}\left(\mO/M^2\right).
\label{eq:GaussianIntegralRes}
\end{equation}

With this result we understand the above remark on the $M$-dependence of the measure: First, it is possible to absorb
all $M$-factors appearing inside the determinant on the RHS of \eqref{eq:GaussianIntegralRes} into the measure by an
appropriate redefinition. Since we would like to have a dimensionless argument in the determinant, however, we keep
our current definition of the measure. But second, the index $N$ may be regarded as a function of the cutoff scale
$\UV$, a convenient choice being $N=\UV/M$. In any case, whenever the regularization is based on the scale $\UV$,
the measure inevitably receives a contribution from the parameter $M$. For convenience, we use the notation
$\mD_\UV\vp$ in the subsequent sections, defined by
\begin{equation}
 \mD_\UV\vp \equiv \td\mu_M^{N=\UV/M}[\vp],
\end{equation}
without writing the present $M$-dependence explicitly. By analogy with eq.\ \eqref{eq:GaussianIntegralRes}, we denote
the determinant restricted to modes with momenta below $\UV$ by $\det_\UV$, and similarly we write $\Tr_\UV$ for the
corresponding trace.

As a consistency check we can choose $\mO$ in eq.\ \eqref{eq:GaussianIntegralRes} to be $M^2$ times the identity.
Then the exponent amounts to $-\frac{1}{2}\langle\vp,\vp\rangle$ with the inner product $\langle\cdot,\cdot\rangle$
defined above, so the functional integral becomes $\int\td\mu_M^N[\vp]\,\exp\big(-\frac{1}{2}\langle\vp,\vp\rangle\big)
= \det{}_{\!N}^{-1/2} (\Id) = 1$, as it should be \cite{Mottola1995}.

Finally, let us comment on the case where the exponent in the functional integral contains terms of higher than
quadratic order in $\vp$. In anticipation of our calculation in the subsequent section, we consider integrals of the
type\\
$\int\mD_\UV\vp\,\exp\left\{-\frac{1}{2}\int\!\!\sg\, \vp A\mku\vp+\int\!\!\sg\, B\mku \UV^{-1}\vp^3+\int\!\!\sg\,
C\mku \UV^{-2}\vp^4+\mO(\UV^{-3})\right\}$, where the operators $A$, $B$ and $C$ are of the order $\UV^0$ at large
cutoff scales. Without further restrictions, this has no well-behaved UV limit. The issue
can be illustrated by means of the usual integral $\int_{-\infty}^\infty\td x \,\exp\left\{-\frac{1}{2}a x^2+b\mku
\UV^{-1}x^3+c\mku\UV^{-2}x^4+\mO(\UV^{-3})\right\}$, which is divergent for all values of $\Lambda$ if $c>0$.
However, there is the possibility of restricting the domain of integration according to $\int_{-\infty}^\infty
\rightarrow\int_{-L}^L$ and take the limit $L\rightarrow\infty$ only after taking the UV limit $\UV\rightarrow\infty$.
A particularly convenient choice is a simultaneous limit because this method involves taking only one limit
effectively, namely $\UV\rightarrow\infty$. The idea is to set $L=\sqrt[4]{\UV/M}$, where the $4^\text{th}$ root is
essentially chosen in order to achieve convergence of the integral under consideration. Then we find that
$\int_{-\sqrt[4]{\UV/M}}^{\sqrt[4]{\UV/M}}\td x \,\exp\left\{-\frac{1}{2}a x^2+b\mku \UV^{-1}x^3+c\mku\UV^{-2}x^4+
\mO(\UV^{-3})\right\}$ remains finite as $\UV\rightarrow\infty$ for all $b$ and $c$ if $a>0$, and the result is
independent of the $x^3$-, $x^4$- and higher order terms in the exponent.
The same can be done for the functional integral. This justifies the modified definition
\begin{equation}
 \int\mD_\UV \vp\equiv (2\pi)^{-N(\UV)/2} \prod\limits_{n=1}^{N(\UV)}
 \int_{-\sqrt[4]{\UV/M}}^{\sqrt[4]{\UV/M}}\td a_n\,,\quad \text{with } N(\UV)\equiv \UV/M \,.
\end{equation}
With this definition, all higher (than quadratic) order terms in the exponent in the functional integral can be
dropped provided that these terms are accompanied by an appropriate power of $\UV$. We obtain the result
\begin{equation}
\begin{split}
 \int\mD_\UV\vp\,\e^{-\frac{1}{2}\int\!\!\sg\, \vp A\mku\vp+\int\!\!\sg\, B\mku \UV^{-1}\vp^3+\int\!\!\sg\,
C\mku \UV^{-2}\vp^4+\mO(\UV^{-3})}\\ = \int\mD_\UV\vp\,\e^{-\frac{1}{2}\int\!\!\sg\, \vp A\mku\vp}
= \det{}_{\!\UV}^{-\frac{1}{2}}\left(A/M^2\right) ,
\end{split}
\label{eq:PIResult}
\end{equation}
when the limit $\UV\rightarrow\infty$ is taken. Again, for large $\UV$ all scale dependence of the terms in the
exponent on the LHS is stated explicitly, i.e.\ we assume that $A$, $B$ and $C$ are of the order $\UV^0$ in the limit.

In conclusion, we have seen that both the functional measure and exponents in the integral, in particular any bare
action, depend on a free parameter $M$. Therefore, we expect this parameter to enter the reconstruction formula for
the bare action as well.\footnote{Note that in our approach to gravity the details of the regularization depend on the
background metric $\bg_\mn$ since high momentum modes are cut off with respect to the background Laplacian $\bB$. As a
consequence, functional integrals and determinants exhibit a background dependence, too, before the UV limit
$\UV\rightarrow\infty$ is taken. This can be made explicit by writing $\det_\UV(\cdot)\equiv \det\big[(\cdot)
\Theta(\UV+\bB)\big]$. In the limit $\UV\rightarrow\infty$ this additional source of background dependence is absent.}
As a final remark we would like to point out that the arguments presented above are valid for scalar fields, but they
can easily be extended to arbitrary fields such as the metric fluctuations by defining a suitable inner product in the
corresponding field space and by correctly taking into account all mass dimensions. Clearly, since we can have
different field types with different mass dimensions in general, we can think of $\vp$ in eq.\ \eqref{eq:PIResult} as a
vector with one component for each field type, and the real number $M^{-2}$ on the RHS of \eqref{eq:PIResult} must be
replaced by a block diagonal matrix, say $\mathcal{N}^{-1}$, whose diagonal entries read $M^{-\alpha}$. Here, $\alpha$
is adapted to the associated field type, e.g.\ $\alpha=2$ for scalars and $\alpha=d$ for gravitons.

%----------------------------------------------------------------------------------------------------------------------
\section{The reconstruction formula}
\label{app:ReconGeneral}
%----------------------------------------------------------------------------------------------------------------------

%----------------------------------------------------------------------------------------------------------------------
\subsection{Derivation of the one-loop reconstruction formula}
\label{app:OneLoop}
%----------------------------------------------------------------------------------------------------------------------

The following derivation is based on and extends the one of Ref.\ \cite{MR09}. According to the arguments of Chapter
\ref{chap:Bare}, the effective average action $\GkL$ is determined by the defining functional integral
\begin{align}
 \exp\left\{-\textstyle\frac{1}{\hbar}\GkL[\phi]\right\} &= \int\mD_\UV f\,\exp\left\{\textstyle\frac{1}{\hbar}
 \left(-\SB[\phi+f] +\int\frac{\delta\GkL[\phi]}{\delta\phi}\mku f -\frac{1}{2}\int\sg\, f\mku\Rk\mku f \right)\right\}
 \nonumber\\
 &\equiv \int\mD_\UV f\,\exp\left\{-\textstyle\frac{1}{\hbar}\St[f;\phi]\right\}\,,
\label{eq:RelGammaSB}
\end{align}
with the bare action $\SB$, the functional measure $\mD_\UV f$ as defined in Section \ref{app:Measure}, and the total
action
\begin{equation}
 \St[f;\phi]\equiv \SB[\phi+f]-\int\frac{\delta\GkL[\phi]}{\delta\phi}\mku f +\frac{1}{2}\int\sg\, f\mku\Rk\mku f\,.
\end{equation}
The bare action $\SB$ depends on $\UV$ and $M$, while the total action depends on all three scales, $\UV$, $M$ and $k$.
In the present section we state $\hbar$ explicitly as it will serve as a bookkeeping parameter.

In order to ``solve'' eq.\ \eqref{eq:RelGammaSB} for the bare action (up to one-loop level), we perform a saddle
point expansion in the integral. For that purpose, we need an extreme value of the total action: We define $f_0$ as a
stationary point: $\frac{\delta\St}{\delta f}[f_0;\phi]=0$, or equivalently,
\begin{equation}
 \frac{\delta \SB}{\delta\phi}[\phi+f_0]- \frac{\delta\GkL}{\delta\phi}[\phi]+\sg\,\Rk\mku f_0 = 0\,.
\label{eq:DefStatPoint}
\end{equation}
The existence of such a stationary point is guaranteed by the properties of $\SB$ and $\Rk$ which are bounded from
below provided that $\SB$ behaves like a generic action, an assumption to be checked a posteriori. Now we can expand
$f$ around $f_0$ using the parametrization
\begin{equation}
 f \equiv f_0 + \sqrt{\hbar}\,\frac{M}{\UV}\,\vp \,.
\label{eq:ChangeOfVariables}
\end{equation}
This choice is particularly convenient for our subsequent expansion since it allows using $\hbar$ to count loop orders
and suppressing fluctuations by letting $\UV/M\rightarrow\infty$. As the first variation of $\St$ vanishes at $f_0$,
we obtain the series
\begin{equation}
 \St[f;\phi] = \St[f_0;\phi] + \hbar\,\frac{M^2}{\UV^2}\,\frac{1}{2}\int\vp\,\frac{\delta^2\St}{\delta f^2}[f_0]\,\vp
 + \mO\left(\hbar^{3/2}\St^{(3)}/\UV^3\right),
\label{eq:SaddlePExp}
\end{equation}
with the second order derivative given by
\begin{equation}
 \frac{\delta^2\St}{\delta f(x)\delta f(y)}[f_0] = \frac{\delta^2\SB}{\delta \phi(x)\delta\phi(y)}[\phi+f_0]
 +\sg\,\Rk\delta(x-y) \;.
\end{equation}
We can make the natural assumption that\footnote{Note that $\SB^{(2)}[\phi](x,y)\equiv g^{-1/2}(x)\,g^{-1/2}(y)
\frac{\delta^2\SB[\phi]}{\delta\phi(x)\delta\phi(y)}$, while in its representation as a differential operator,
$\SB^{(2)}[\phi](x,y)\equiv g^{-1/2}(x)\big(\SB^{(2)}[\phi]\big)^\text{diff-op}\,\delta(x-y)$, one of the two factors
$\sg$ drops out (cf.\ Appendix \ref{app:OperatorRep}). Thus, $\SB^{(2)}[\phi]$ and $\Rk$ always occur with the same
power of $\sg$.}
\begin{equation}
 \SB^{(2)}+\Rk = \mO(\UV^2)\quad\text{at fixed fields for } k^2\le\UV^2 \, .
\end{equation}
This assumption is reasonable since $\Rk\propto k^2\le\UV^2$ for all standard regulators, and $\SB^{(2)}[\phi]=
\mO(\UV^2)$ is usually satisfied by any standard action as can be seen by dimensional analysis. Thus, we find that
$\delta^2\St/\delta f^2$ is at most of order $\mO(\UV^2)$. In turn, this holds true for higher order derivatives as
well, i.e.\ $\delta^3\St/\delta f^3,\,\delta^4\St/\delta f^4,\,\cdots=\mO(\UV^2)$. In the expansion
\eqref{eq:SaddlePExp} any higher order term involving $\delta^n\St/\delta f^n[f_0]$ goes along with the factor
$\hbar^{n/2}\frac{M^n}{\UV^n}\mku\vp^n$, so their combination is of the order $\mO(\hbar^{n/2}/\UV^{n-2})$. Therefore,
the remainder in \eqref{eq:SaddlePExp} can be replaced according to
\begin{equation}
 \mO\left(\hbar^{3/2}\St^{(3)}/\UV^3\right) = \mO\left(\hbar^{3/2}/\UV\right).
\end{equation}
By our argument at the end of Section \ref{app:Measure}, these higher order terms which contribute to the exponent in
the path integral by $\UV^{-1}\vp^3$, $\UV^{-2}\vp^4$, etc.\ will ultimately vanish as $\UV$ is sent to $\infty$.
Hence, for large cutoff scales $\UV$ all nontrivial contribution comes indeed from the quadratic term in eq.\
\eqref{eq:SaddlePExp}.

The Jacobian induced by the change of variables \eqref{eq:ChangeOfVariables} can be written as
\begin{equation}
  \mD_\UV f = \left|\det{}_{\!\UV}\! \left(\textstyle\frac{\delta f}{\delta\vp}\right)\right|\mD_\UV\vp
  = \det{}_{\!\UV}\!\left(\textstyle\sqrt{\hbar}\,\frac{M}{\UV}\Id\right)\mD_\UV\vp 
  = \e^{-\frac{1}{2}\ln\det_\UV\left(\hbar^{-1}\frac{\UV^2}{M^2}\Id\right)}\mD_\UV\vp\,.
\end{equation}
By the identity $\ln\det(\cdot)=\Tr\ln(\cdot)$ we can express this as
\begin{equation}
 \mD_\UV f = J_\UV\,\mD_\UV\vp\,,
\label{eq:ChangeOfMeasure}
\end{equation}
with the Jacobian $J_\UV$ defined by
\begin{equation}
 J_\UV \equiv \e^{-\frac{1}{2}\Tr_\UV\ln\left(\hbar^{-1}\frac{\UV^2}{M^2}\Id\right)}.
\label{eq:JacobianJ}
\end{equation}
Note that $J_\UV$ is independent of $\vp$ (or $f$) and can be pulled out of the path integral, giving rise to an
additional factor. Furthermore, since $\Tr_\UV\ln\left(\hbar^{-1}\frac{\UV^2}{M^2}\Id\right)$ is strictly monotonically
increasing for increasing ratio $\UV/M$, we find that $J_\UV$ is bounded in the UV regime, and thus the large cutoff
limit exists.

Combining \eqref{eq:RelGammaSB} with \eqref{eq:SaddlePExp} and \eqref{eq:ChangeOfMeasure} yields
\begin{equation}
\begin{split}
 \e^{-\frac{1}{\hbar}\GkL[\phi]} &= J_\UV\,\e^{-\frac{1}{\hbar}\St[f_0;\phi]}
  \int\mD_\UV \vp\,\e^{-\frac{1}{2}\frac{M^2}{\UV^2}\int\!\sg\,\vp\left(\SB^{(2)}[\phi+f_0]+\Rk\right)\vp\, 
   +\,\mO(\hbar^{1/2}/\Lambda)}\\
 &= J_\UV\,\e^{-\frac{1}{\hbar}\St[f_0;\phi]}\mku \det{}_{\!\UV}^{-\frac{1}{2}}\!\left[\textstyle\frac{1}{M^2}
  \frac{M^2}{\UV^2}\left(\SB^{(2)}[\phi+f_0]+\Rk\right)\right]\! \cdot\e^{\mO(\hbar^{1/2}/\Lambda)}.
\end{split}
\end{equation}
At this point we can reinsert $\St[f_0;\phi]$ and take the logarithm:
\begin{equation}
\begin{split}
 \GkL[\phi] = &\SB[\phi+f_0]-\int{\textstyle\frac{\delta\GkL}{\delta\phi}}\mku f_0 +\frac{1}{2}\int\!\sg\,
  f_0\mku\Rk\mku f_0 -\hbar\mku\ln J_\UV \\
 &+\frac{\hbar}{2}\,\Tr_\UV \ln\left[\textstyle\frac{1}{\UV^2}\left(\SB^{(2)}[\phi+f_0]+\Rk\right)\right]
 + \mO(\hbar^{3/2}/\Lambda).
\end{split}
\end{equation}
Expanding $\SB[\phi+f_0]$ in terms of $f_0$ we obtain the intermediate result
\begin{equation}
\begin{split}
 \GkL[\phi] -\SB[\phi] = &\int\left({\textstyle\frac{\delta\SB[\phi]}{\delta\phi}-\frac{\delta\GkL[\phi]}{\delta\phi}}
 \right) f_0 +\frac{1}{2}\int\!\sg\, f_0\left(\SB^{(2)}[\phi]+\Rk\right) f_0\\
 &+\frac{\hbar}{2}\,\Tr_\UV \ln\left[\textstyle\frac{1}{\UV^2}\left(\SB^{(2)}[\phi]
 +\int\!\sg\, \SB^{(3)}[\phi]f_0+\cdots+\Rk\right)\right]\\
 &-\hbar\mku\ln J_\UV + \mO(f_0^3) + \mO(\hbar^{3/2}/\Lambda).
\end{split}
\label{eq:IntRes1}
\end{equation}

Moreover, from the definition of $f_0$, eq.\ \eqref{eq:DefStatPoint} we derive a second important relation,
based upon an expansion in terms of $f_0$ again:
\begin{equation}
 \sg\left(\SB^{(2)}[\phi]+\Rk\right)f_0 = \frac{\delta\GkL}{\delta\phi}[\phi]
 -\frac{\delta\SB}{\delta\phi}[\phi]+\mO(f_0^2).
\label{eq:IntRes2}
\end{equation}
Now we can combine \eqref{eq:IntRes1} and \eqref{eq:IntRes2}, leading to
\begin{equation}
\begin{split}
 &\sg\left(\SB^{(2)}[\phi]+\Rk\right)f_0+\mO(f_0^2) =  \frac{\delta\GkL}{\delta\phi}[\phi]
 -\frac{\delta\SB}{\delta\phi}[\phi]\\
 &= \int\sg\left(\SB^{(2)}[\phi]-\GkL^{(2)}[\phi]\right)f_0
 +\int{\textstyle\left(\frac{\delta\SB}{\delta\phi}-\frac{\delta\GkL}{\delta\phi}\right)\frac{\delta f_0}{\delta\phi}}
 \\
 &\quad +\int\sg\,{\textstyle\frac{\delta f_0}{\delta\phi}\left(\SB^{(2)}[\phi]+\Rk\right)f_0}
 -\frac{\delta}{\delta\phi}\big(\hbar\mku\ln J_\UV\big)\\
 &\quad +\frac{\hbar}{2}\,\frac{\delta}{\delta\phi}\Tr_\UV\ln\left[\textstyle\frac{1}{\UV^2}\left(\SB^{(2)}[\phi]
 +\int\sg\,\SB^{(3)}[\phi]\mku f_0+\mO(f_0^2)+\Rk\right)\right]\\
 &\quad +\mO(f_0^2)+\hbar\,\mO(\delta f_0/\delta\phi)+\mO(\hbar^{3/2}/\UV).
\end{split}
\label{eq:ExpF0Hbar}
\end{equation}
From this expression we can draw an important conclusion: We observe that each term in eq.\ \eqref{eq:ExpF0Hbar} is
proportional to $f_0$ and/or $\hbar$ and/or $\delta f_0/\delta\phi$. Furthermore, there are terms that involve $f_0$
but no factor $\hbar$ and vice versa. Hence, $f_0$ must be of the order $\hbar$, and $\hbar$ must be of the order
$f_0$,
\begin{equation}
 f_0 = 0+\mO(\hbar)\quad\text{and}\quad \hbar = 0+ \mO(f_0).
\label{eq:OFIsOH}
\end{equation}
Consequently, we have $\mO(f_0^2)=\mO(\hbar^2)$, $\hbar\,\mO(f_0)=\mO(\hbar^2)$ and $\hbar\,\mO(\delta f_0/\delta\phi)
=\mO(\hbar^2)$ in eq.\ \eqref{eq:ExpF0Hbar}. Inserting relation \eqref{eq:OFIsOH} into \eqref{eq:IntRes1} we find
\begin{equation}
 \GkL[\phi] -\SB[\phi]=\mO(\hbar).
\end{equation}
With this result, we conclude that the first term on the RHS of \eqref{eq:IntRes1} is in fact of order $\mO(\hbar^2)$.
Collecting all terms up to linear order in $\hbar$ and using \eqref{eq:JacobianJ}, we arrive at our final result:
\begin{equation}
\begin{split}
 \GkL[\phi]-\SB[\phi] &= \frac{\hbar}{2}\,\Tr_\UV \ln\left[\textstyle\frac{1}{\UV^2}\left(\SB^{(2)}[\phi]
 +\Rk\right)\right]-\hbar\mku\ln J_\UV+\mO(\hbar^{3/2}/\UV)+\mO(\hbar^2) \\
 &= \frac{\hbar}{2}\,\Tr_\UV \ln\left[\textstyle\frac{1}{\hbar\mku M^2}\left(\SB^{(2)}[\phi]
 +\Rk\right)\right] +\mO(\hbar^{3/2}/\UV)+\mO(\hbar^2).
\end{split}
\end{equation}
In the large cutoff limit all terms of order $\mO(\hbar^{3/2}/\UV)$ vanish, and the order $\mO(\hbar^2)$ represents
second and higher loop contributions. At one-loop level, setting $\hbar=1$, we obtain the reconstruction formula
\begin{equation}[b]
 \GkL[\phi] = \SB[\phi] + \frac{1}{2}\,\Tr_\UV \ln\left[\textstyle\frac{1}{M^2}\left(\SB^{(2)}[\phi]
 +\Rk\right)\right] .
\label{eq:OneLoopReconstruction}
\end{equation}

As we have already pointed out at the end of Section \ref{app:Measure}, our consideration can be generalized to
arbitrary fields in a straightforward way by taking into account the canonical mass dimensions of all fields involved.
Let $\mathcal{N}$ be the block diagonal matrix which contains for each field the parameter $M$ raised to the
corresponding power. For instance, its entry in the graviton sector equals $M^d$, while it is $M^2$ in the ghost
sector as well as for scalar fields. With this matrix, \eqref{eq:OneLoopReconstruction} extends to
\begin{equation}[b]
 \Gamma_{k,\UV} = \SB + \frac{1}{2}\,\STr_\UV \ln\left[ \mathcal{N}^{-1}\left(\SB^{(2)}+\Rk\right) \right].
\label{eq:OneLoopFull}
\end{equation}

For completeness we simplify eq.\ \eqref{eq:ExpF0Hbar} by observing that the $\phi$-derivative of the field independent
Jacobian $J_\UV$ vanishes and by combining all irrelevant orders. This yields
\begin{equation}
 \sg\left(\SB^{(2)}[\phi]+\Rk\right)f_0+\mO(f_0^2) =
 \frac{\hbar}{2}\,\frac{\delta}{\delta\phi}\Tr_\UV\ln\left[\textstyle\frac{1}{\UV^2}\left(\SB^{(2)}[\phi]
 +\Rk\right)\right] +\mO(\hbar^{3/2}/\UV),
\label{eq:IntermediateBare}
\end{equation}
a relation that is used in the next subsection to study the limit $\UV\rightarrow\infty$.

%----------------------------------------------------------------------------------------------------------------------
\subsection{Exactness beyond one-loop in the large cutoff limit}
\label{app:BareOneLoopExact}
%----------------------------------------------------------------------------------------------------------------------

The identity \eqref{eq:OneLoopFull} derived in the previous subsection is inherently one-loop exact.
In what follows we would like to investigate whether or not this one-loop relation actually becomes fully exact once
the limit $\UV\rightarrow\infty$ is taken. In order to answer this question we will decompose \eqref{eq:OneLoopFull}
into different types of terms. We will then see that the reconstruction formula is indeed fully exact in the large
cutoff limit for certain terms, while we must settle for one-loop exactness for the remaining terms.

As usual, we assume that there is a set of basis functionals $\{P_\alpha[\mku\cdot\mku]\}$ which can be used to expand
elements of theory space. In particular, the effective average action can be written as
$\GkL[\phi] = \sum_\alpha c_\alpha(k,\UV) P_\alpha[\phi]$ where $c_\alpha(k,\UV)$ are the running couplings. In this
regard we can expand the RHS of eq.\ \eqref{eq:OneLoopFull}, too, in terms of basis functionals. The question
concerning exactness beyond one-loop level can then be approached for each term separately.

The starting point is provided by eq.\ \eqref{eq:IntermediateBare}, an intermediate result of the previous subsection
which ultimately led to \eqref{eq:OneLoopFull}, and which can be written as
\begin{equation}
 \left(\SB^{(2)}[\phi]+\Rk\right)f_0+\mO(f_0^2) =
 \frac{\hbar}{2}\,\frac{1}{\sg}\,\frac{\delta}{\delta\phi}\Tr_\UV\ln\left[\textstyle\frac{1}{\UV^2}
 \left(\SB^{(2)}[\phi] +\Rk\right)\right] +\mO(\hbar^{3/2}/\UV).
\label{eq:IntermRes}
\end{equation}
In this equation the variation $\frac{1}{\sg}\mku\frac{\delta}{\delta\phi}$ can be pulled into the trace now. Note that
the relation $\delta\mku\ln(A) = A^{-1}\delta A$, valid for pure numbers, does not hold true for a general operator $A$
and an arbitrary variation $\delta A$ since $A$ and $\delta A$ do not commute in general. Due to the cyclicity of the
trace, however, the traced version of this identify remains valid also for operators: $\Tr\big[\delta\mku\ln(A)\big]
= \Tr\big[A^{-1} \delta A\big]$. Applying this to \eqref{eq:IntermRes} yields
\begin{equation}
 \left(\SB^{(2)}[\phi]+\Rk\right)f_0+\mO(f_0^2) =
 \frac{\hbar}{2}\,\Tr_\UV \!\left[\frac{\SB^{(3)}[\phi]}{\SB^{(2)}[\phi]+\Rk}\right] +\mO(\hbar^{3/2}/\UV).
\label{eq:IntermRes2}
\end{equation}
The asymptotic behavior of $\SB^{(3)}[\phi]$ at large $\UV$ is at most of the same order as the one of
$\SB^{(2)}[\phi]$. Thus, the argument of the trace on the RHS of \eqref{eq:IntermRes2} remains finite in
the limit $\UV\rightarrow\infty$ at fixed $\phi$.

In general, $\SB^{(2)}[\phi]+\Rk$ is a function of $-\Box$ plus $\phi$-dependent terms. Hence, when expanding
$\big(\SB^{(2)}[\phi]+\Rk\big)^{-1}$ in terms of $\phi$ we must take into account that the Laplacian commuted to
the rightmost position in each term gives rise to additional derivative terms proportional to $D_\mu\phi$,
$\Box\phi$, etc. Taking all terms together, we can write symbolically:
\begin{equation}
 \frac{\SB^{(3)}[\phi]}{\SB^{(2)}[\phi]+\Rk} = \sum_i V_i\big(\phi,D_\mu\phi,\cdots;\UV)\,W_i(-\Box,D_\mu,\cdots;\UV),
\label{eq:ExpS3S2Rk}
\end{equation}
with some functions $V_i$ and $W_i$ that do not have to be specified in more detail here; for our argument it suffices
to know that their combination as in \eqref{eq:ExpS3S2Rk} remains finite in the limit $\UV\to\infty$. We insert this
expression into the trace in eq.\ \eqref{eq:IntermRes2} now. Recalling that $\Tr_\UV\big[(\mku\cdot\mku)\big] \equiv
\Tr\big[(\mku\cdot\mku)\mku \theta(\UV^2+\Box)\big]$ we obtain
\begin{equation}
 \left(\SB^{(2)}[\phi]+\Rk\right)f_0+\mO(f_0^2) =
 \frac{\hbar}{2}\,\Tr \!\left[(\text{finite})\,\theta(\UV^2+\Box)\right] +\mO(\hbar^{3/2}/\UV).
\label{eq:IntermRes3}
\end{equation}

If ``(finite)'' in \eqref{eq:IntermRes3} were a pure number, say $c$, the trace could be determined by making use of
eq.\ \eqref{eq:Heat3} of Appendix \ref{app:Heat}, with the generalized Mellin transforms \eqref{eq:Mellin}, giving rise
to
\begin{equation}
\begin{split}
 \Tr \!&\left[ c\,\theta(\UV^2+\Box)\right]\\ &= c\left(\textstyle\frac{1}{4\pi}\right)^{d/2}\tr(\Id)\left\{
 {\textstyle\frac{1}{\Gamma(d/2+1)}}\UV^d\int\!\sg + {\textstyle\frac{1}{6}\frac{1}{\Gamma(d/2)}}\UV^{d-2}\int\!\sg\,R
 + \mO(R^2)\right\},
\end{split}
\label{eq:IfNumber}
\end{equation}
where the terms of the order $R^2$, $R^4$, etc.\ are accompanied with factors $\UV^{d-4}$, $\UV^{d-6}$, and so forth,
respectively, so provided that $d\leq 4$ these terms remain finite in the limit $\UV\to\infty$.

However, the term ``(finite)'' in \eqref{eq:IntermRes3} contains functions of $\Box$ and $\phi$ in general. This
modifies the result \eqref{eq:IfNumber} in that the coefficients of $\int\!\sg$, $\int\!\sg\, R$, etc.\ are no longer
constant but rather functions of $\phi(x)$, $\Box\mku\phi(x)$ and further derivative terms. The important point is that
the asymptotic behavior for large $\UV$ remains unaltered for the various terms in the heat kernel series. As a result,
we find
\begin{equation}
 \Tr_\UV \!\left[\frac{\SB^{(3)}[\phi]}{\SB^{(2)}[\phi]+\Rk}\right] = \text{finite}
 + \UV^d\int\!\sg\,F_0(\phi,D_\mu\phi,...) + \UV^{d-2}\int\!\sg\,F_1(\phi,D_\mu\phi,...) R\mkuu .
\label{eq:Trfinite}
\end{equation}
Here $F_0$ and $F_1$ are finite scalar densities that do not have to be determined in detail to advance our
argument.\footnote{More precisely, $F_0$ and $F_1$ are scalar densities of weight $-1$ w.r.t.\ the point $x$ and
scalar densities of weight $0$ w.r.t.\ the integration variable, say $y$. The additional appearance of the metric
determinant, $1/\!\sqrt{g(x)}$, stems from the LHS of eq.\ \eqref{eq:Trfinite} since $\SB^{(3)}[\phi]$ is defined as
$1/\!\sqrt{g(x)}\,\frac{\delta}{\delta\phi(x)}\SB^{(2)}[\phi]$.}
The only information we need at this point is that they do not contain any curvature terms.

It is known that $\int\!\sg$, $\int\!\sg\, R$, $\int\!\sg\, R^2$, etc., are linearly independent basis functionals
in a pure metric gravity theory space \cite{FKWC92}. Thus, we can make the plausible assumption that
$\int\!\sg\,F_0(\phi,D_\mu\phi,...)$, $\int\!\sg\, R\,F_1(\phi,D_\mu\phi,...)$, $\int\!\sg\,R^2\,F_2(\phi,D_\mu\phi,
...)$, etc., are linearly independent, too. In this regard it is possible to project any functional onto the orthogonal
complement to all functionals of the type $\int\!\sg\,(\cdot)$ and $\int\!\sg\, R\,(\cdot)$, i.e.\ we ``project away''
the divergent terms according to eq.\ \eqref{eq:Trfinite}. Henceforth we denote such a projection by $\PrFull$. Its
application to eq.\ \eqref{eq:Trfinite} yields
\begin{equation}
 \PrFull\left\{ \Tr_\UV \!\left[\frac{\SB^{(3)}[\phi]}{\SB^{(2)}[\phi]+\Rk}\right]\right\} = \text{finite}.
\label{eq:Trfinite2}
\end{equation}
Thus, by means of eq.\ \eqref{eq:IntermRes2} we obtain
\begin{equation}
 \PrFull\left\{\left(\SB^{(2)}[\phi]+\Rk\right)f_0+\mO(f_0^2)\right\} = \text{finite}.
\label{eq:LHSfnite}
\end{equation}

At this point it is convenient to identify the scales $k$ and $\UV$ such that a simultaneous limit
$k=\UV\to\infty$ can be considered. We now assume that $\big(\SB^{(2)}[\phi]+\RL\big)$ is of the order $\UV^2$.
Therefore, apart from those terms that are ``projected away'' in eq.\ \eqref{eq:LHSfnite}, we can conclude that
$f_0$ is of the order $\UV^{-2}$ or lower. Using in addition that $f_0\propto\hbar$ we may reexpress it as
\begin{equation}
 f_0 = \hbar\,\frac{M^2}{\UV^2}\,\tilde{f}_0\,,
\end{equation}
where $\tilde{f}_0=\mO(\hbar^0)$ and $\lim_{\UV\to\infty}\tilde{f}_0=\text{finite}$, bearing in mind that this result
holds true only for the ``projected version'' of $f_0\mku$.

This crucial result can be used to simplify eq.\ \eqref{eq:IntRes1} of the previous subsection: Since
$\big(\SB^{(2)}[\phi]+\RL\big)f_0$ is finite upon projection, the term $f_0\left(\SB^{(2)}[\phi]+\RL\right)f_0$
approaches $0$ in the limit $\UV\to\infty$. Furthermore, all higher order terms in the trace on the RHS of
\eqref{eq:IntRes1}, $\int f_0\,\SB^{(3)}[\phi]$, etc., remain finite for large $\UV$, and with the prefactor
$1/\UV^2$ these terms vanish as $\UV\to\infty$. Thus, for large $\UV$ eq.\ \eqref{eq:IntRes1} reduces to
\begin{equation}
\begin{split}
 \GLL[\phi] -\SB[\phi] = \; &\hbar\, M^2\int \tilde{f}_0\,\frac{1}{\UV^2}\,\frac{\delta}{\delta\phi}
 \Big(\SB[\phi]-\GLL[\phi]\Big)\\
 &+\frac{\hbar}{2}\,\Tr_\UV \ln\left[\textstyle\frac{1}{\hbar M^2}\left(\SB^{(2)}[\phi]+\RL\right)\right],
\end{split}
\label{eq:IntermRes4}
\end{equation}
up to the terms that have been projected away. To proceed with this expression, let us denote the asymptotic behavior
of $\GLL[\phi] -\SB[\phi]$ at high cutoff scales $\UV$ by $A(\UV)$, i.e.\ for the quotient we have $\lim_{\UV\to\infty}
\big(\GLL[\phi] - \SB[\phi]\big)/A(\UV) = \text{finite}$. Dividing \eqref{eq:IntermRes4} by $A(\UV)$ we observe that
the first term on the RHS vanishes in the limit $\UV\to\infty$ since $\int\tilde{f}_0\,\frac{1}{\UV^2}\,
\frac{\delta}{\delta\phi} \frac{\SB[\phi]-\GLL[\phi]}{A(\UV)}\to\big(\frac{1}{\UV^2}\cdot\text{finite}\big)$ after having applied
the projection as above. Hence, all nonvanishing contributions to the RHS of \eqref{eq:IntermRes4} must stem from the
trace part:
\begin{equation}
 \frac{1}{A(\UV)}\,\frac{\hbar}{2}\,\Tr_\UV \ln\left[\textstyle\frac{1}{\hbar M^2}\left(\SB^{(2)}[\phi]
 +\RL\right)\right] = \text{finite},
\end{equation}
so this trace term must have the same asymptotic behavior as $\GLL[\phi] -\SB[\phi]$. In conclusion, the first term on
the RHS of \eqref{eq:IntermRes4} can be dropped at large $\UV$ since it becomes small compared with the other ones.
Writing the projection explicitly again, we arrive at our final result:
\begin{equation}[b]
 \PrFull\Big\{\GLL -\SB\Big\} = \PrFull\Big\{ {\textstyle\frac{\hbar}{2}}\mku
 \Tr_\UV \ln\!\Big[\textstyle\frac{1}{\hbar M^2}\big(\SB^{(2)}+\RL\big)\Big]\Big\}.
\label{eq:OneLoopExactIdentity}
\end{equation}
Remarkably enough, this identity is \emph{exact} in the limit $\UV\to\infty$, that is, it is \emph{not} a one-loop
approximation. The meaning of \eqref{eq:OneLoopExactIdentity} is the following: Once we project onto the orthogonal
complement to all $\sg$- and $\sg\mku R$-terms, the one-loop equation $\GLL -\SB =
\frac{\hbar}{2}\mku \Tr_\UV \ln\!\big[\frac{1}{\hbar M^2}(\SB^{(2)}+\RL)\big]$ turns into an exact equation in the
limit of large cutoff scales.

As in the previous subsection, the result can be extended beyond scalar field level. For general fields the factor
$M^{-2}$ in eq.\ \eqref{eq:OneLoopExactIdentity} must be replaced by $\mathcal{N}^{-1}$ as in \eqref{eq:OneLoopFull},
and the trace becomes a supertrace.

We would like to point out another interesting result: Among the divergent terms in eq.\ \eqref{eq:Trfinite} the ones
involving $R$ assume a special role in that they become actually finite in $d=2$ dimensions. Therefore, in $2$
dimensions we have to ``project away'' only the $\sg$-terms in order to achieve exactness of the reconstruction formula
in the limit $\UV\to\infty$.

%----------------------------------------------------------------------------------------------------------------------
\chapter[On the convergence of higher order Liouville couplings]%
{On the convergence of higher order couplings when the bare potential is a series of exponentials}
\label{app:ProofOfConvergence}
%----------------------------------------------------------------------------------------------------------------------

This appendix supplements the discussion in Section \ref{sec:BareLiouExpSeries} which concerned reconstructing the bare
action for a Liouville-type effective average action. The truncation ansatz for the bare action included a potential
term consisting of a series of exponentials, $\cV(\phi)=\frac{1}{2}\UV^2 \sum_{n=1}^{\Nmax}\cgamma_n\,\e^{2n\phi}$.
A numerical reconstruction of the bare couplings indicated that the $\cgamma_n$ decrease approximately exponentially
for increasing $n$. In what follows, we present an argument that supports the convergence conjecture. Although most
steps will be proven rigorously, the application to the actual couplings $\cgamma_n$ relies on a certain assumption and
a numerical computation of initial values, rendering our observations less conclusive. Nonetheless, our statements
reveal the reason behind the fast decrease of higher order couplings.

All numerical estimates are based on the EAA couplings $b$ and $\mu$ for the linear metric parametrization (using the
optimized cutoff); at the end of this appendix we briefly mention the differences the use of the exponential
parametrization entails.

For convenience we perform our analysis in terms of
\begin{equation}
 a_n\equiv 2\mku\cZ^{-1}n^2\,\cgamma_n \,,
\label{eq:anFromGamman}
\end{equation}
with $\cZ = -b/(8\pi)$. Then eqs.\ \eqref{eq:cgamma1} and \eqref{eq:cgamman} can be written as
\begin{align}
 a_1 &= -\frac{b\mu}{2+4\pi\cZ}\,,
\label{eq:Defa1}\\
 a_n &= \frac{n^2}{n^2+2\pi\cZ}\, \sum_{k=2}^n\!
 \sum_{{\substack{\,\alpha\in\mathds{N}_0^{n}\\ |\alpha|=k\\ \sum_i i\alpha_i=n}}} \!\!\!
 \frac{(-1)^k(k-1)!}{\alpha_1!\cdots\alpha_{n}!}\, a_1^{\alpha_1}\cdots a_{n-1}^{\alpha_{n-1}} \,.
\label{eq:Defan}
\end{align}
Let us consider the case where the couplings $a_1,\dotsc,a_n$ are already known, and where an estimate for the coupling
$a_{n+1}$ is sought after. In order to proceed we make an important \emph{assumption}:
Motivated by the fall-off behavior of the couplings, see Figure \ref{fig:BareGammaLinParam}, we assume
\begin{equation}
 a_i = A\,\e^{-\lambda i}\quad\text{for } 1\le i\le n.
\label{eq:BareLiouAssumpFallOff}
\end{equation}
Furthermore, we assume that the constants $A$ and $\lambda$ satisfy
\begin{equation}
 A > 0, \quad \lambda > 0, \quad \text{and}\quad |A-1|<1.
\label{eq:BareLiouAssumpConsts}
\end{equation}
We have already noticed in Sec.\ \ref{sec:BareLiouExpSeriesLin} that the first assumption, eq.\
\eqref{eq:BareLiouAssumpFallOff}, is valid only approximately since there are slight deviations from an exact
exponential decrease. It can be thought of as an upper bound, though. In this regard, it will be checked numerically
later on whether \eqref{eq:BareLiouAssumpConsts} is satisfied. We will indeed determine $A$ and $\lambda$
respecting \eqref{eq:BareLiouAssumpConsts} such that $a_i \le A\,\e^{-\lambda i}$ for the first couplings, see
Sec.\ \ref{app:CheckInit}.

Based on assumption \eqref{eq:BareLiouAssumpFallOff} we aim at proving $a_{n+1}\le A\,\e^{-\lambda (n+1)}$.

Our argument makes use of (a) an important \emph{combinatorial identity}, and (b) an \emph{inequality} involving
$A$ and $\cZ$. The combinatorial identity is given by
\begin{equation}
\sum_{{\substack{\,\alpha\in\mathds{N}_0^n\\ |\alpha|=k\\[0.1em] \sum_i i\alpha_i=n}}} \!\!\!\!
 \frac{(k-1)!}{\alpha_1!\cdots\alpha_n!} = \frac{1}{n}\binom{n}{k}\,,
\label{eq:CombId}
\end{equation}
for $k\le n$. We will prove eq.\ \eqref{eq:CombId} in Sec.\ \ref{app:ProofCombId}. (To the best of our knowledge,
neither the identity itself nor its proof can be found in the literature.) The inequality reads
\begin{equation}
 \frac{n^2}{n^2+2\pi\mku\cZ}\left[A+\frac{1}{n}(1-A)^n-\frac{1}{n}\right] \le A,
\label{eq:Inequality}
\end{equation}
where $n\in\mathds{N}$, $A>0$ and $|A-1|<1$.
We show in Sec.\ \ref{app:ProofInequality} that it is satisfied for all $n$ greater than some threshold value, in
particular it holds true in the limit $n\to\infty$. For our setting we will determine an estimate for $A$ numerically
in Sec.\ \ref{app:CheckInit}, on the basis of which the inequality \eqref{eq:Inequality} is satisfied for all $n\ge 5$.
\medskip

\noindent
\textbf{Proof of }$\bm{a_{n+1}\le A\,\e^{-\lambda (n+1)}}$ \textbf{assuming that (\ref{eq:BareLiouAssumpFallOff})
holds true.}

\noindent
By eq.\ \eqref{eq:Defan} we have
\begin{equation}
 a_{n+1} = \frac{(n+1)^2}{(n+1)^2+2\pi\cZ}\, \sum_{k=2}^{n+1}\!
 \sum_{{\substack{\,\alpha\in\mathds{N}_0^{n+1}\\ |\alpha|=k\\[0.15em] \sum_i i\alpha_i=n+1}}} \!\!
 \frac{(-1)^k(k-1)!}{\alpha_1!\cdots\alpha_{n+1}!}\, a_1^{\alpha_1}\cdots a_{n}^{\alpha_{n}}\,.
\label{eq:Defanp1}
\end{equation}
Now assumption \eqref{eq:BareLiouAssumpFallOff} can be used to simplify the product
$a_1^{\alpha_1}\cdots a_{n}^{\alpha_{n}}\mku$ in the sum:
\begin{equation}
\begin{split}
 a_1^{\alpha_1}\cdots a_{n}^{\alpha_{n}} &= A^{\alpha_1}\mku\e^{-\lambda\mku\alpha_1}\,
 A^{\alpha_2}\mku\e^{-2\mku\lambda\mku\alpha_2}\,\cdots A^{\alpha_n}\mku\e^{-n\mku\lambda\mku\alpha_n}\\
 &= A^{|\alpha|}\mku\e^{-\lambda\sum_i i\mku\alpha_i}
 = A^k\,\e^{-\lambda(n+1)}\,.
\end{split}
\end{equation}
Thus, eq.\ \eqref{eq:Defanp1} reduces to
\begin{equation}
 a_{n+1} = \frac{(n+1)^2}{(n+1)^2+2\pi\cZ}\, \sum_{k=2}^{n+1} (-A)^k\,\e^{-\lambda(n+1)}\!\!\!\!\!
 \sum_{{\substack{\,\alpha\in\mathds{N}_0^{n+1}\\ |\alpha|=k\\[0.15em] \sum_i i\alpha_i=n+1}}} \!\!\!
 \frac{(k-1)!}{\alpha_1!\cdots\alpha_{n+1}!} \, .
\end{equation}
At this point the inner sum on the RHS can be replaced by means of the combinatorial identity \eqref{eq:CombId}:
\begin{equation}
 a_{n+1} = \frac{(n+1)^2}{(n+1)^2+2\pi\cZ}\;\e^{-\lambda(n+1)}\,\frac{1}{n+1}\; \sum_{k=2}^{n+1} 
 \binom{n+1}{k} 1^{(n+1)-k} (-A)^k \,,
\end{equation}
where we have inserted a factor $1\equiv 1^{(n+1)-k}$. Applying the binomial theorem to the remaining sum,
$\sum_{k=2}^{n+1}\binom{n+1}{k} 1^{(n+1)-k} (-A)^k = (1-A)^{n+1}-(n+1)(-A)-1$, yields
\begin{equation}
 a_{n+1} = \e^{-\lambda(n+1)}\;\frac{(n+1)^2}{(n+1)^2+2\pi\cZ}
 \left[A+\frac{1}{n+1}(1-A)^{n+1}-\frac{1}{n+1}\right].
\label{eq:anp1int}
\end{equation}
As mentioned above and proven in Sec.\ \ref{app:ProofInequality}, inequality \eqref{eq:Inequality} is valid for all
$n$ greater than a yet to be determined threshold value. We assume here that $n$ is already large enough, so that the
inequality holds true for $n+1$, too. Hence, the last two factors on the RHS of \eqref{eq:anp1int} taken together are
bounded from above by $A$, and we obtain
\begin{equation}[b]
 a_{n+1}\le A\,\e^{-\lambda (n+1)}\,.
\label{eq:anUpperBound}
\end{equation}
This completes our proof.\hfill$\Box$

\medskip
Since we assumed $|A-1|<1$, cf.\ eq.\ \eqref{eq:BareLiouAssumpConsts}, the term $(1-A)^{n+1}$ in \eqref{eq:anp1int}
tends to zero in the large $n$ limit, and we have $-1<(1-A)^{n+1}<1$ for all $n$. Thus, the square bracket in
\eqref{eq:anp1int} satisfies $[\cdots]>A-\frac{2}{n+1}$. This leads to $[\cdots]>0$ for all $n>\frac{2}{A}-1$.
Furthermore, the factor $\frac{(n+1)^2}{(n+1)^2+2\pi\cZ}$ is always positive. Combining these results, eq.\
\eqref{eq:anp1int} yields a second estimate:
\begin{equation}
 a_{n+1}>0.
\end{equation}
Moreover, considered the fact that the fraction and the square bracket in \eqref{eq:anp1int} in the limit $n\to\infty$
satisfy $\frac{(n+1)^2}{(n+1)^2+2\pi\cZ}\to 1$ and $\left[A+\frac{1}{n+1}(1-A)^{n+1}-\frac{1}{n+1}\right]\to A$,
respectively, we conclude that $a_{n+1}$ lies close to the upper bound given by eq.\ \eqref{eq:anUpperBound}, i.e.\
$a_{n+1}\approx A\,\e^{-\lambda (n+1)}$, provided that $n$ is sufficiently large and that
\eqref{eq:BareLiouAssumpFallOff} is given.
\medskip

\noindent
\textbf{Remarks:} The above argument mimics a proof by induction. If we had obtained $a_{n+1}= A\,\e^{-\lambda (n+1)}$
instead of \eqref{eq:anUpperBound}, we could have concluded immediately that all couplings are given by the same
exponential law, so that $a_n\to 0$ exponentially for $n\to 0$. However, we have only obtained an inequality for
$a_{n+1}$. Therefore, the inductive chain is interrupted when going to $n+2$ since \eqref{eq:BareLiouAssumpFallOff}
might no longer be satisfied for $i=1,\dotsc,n+1$, and convergence of the couplings cannot be proven this
way.\footnote{Relaxing the assumption in \eqref{eq:BareLiouAssumpFallOff} by requiring $a_i \le A\,\e^{-\lambda i}$
for $1\le i\le n$ is not an option. The conclusion \eqref{eq:anUpperBound} would no longer be admissible. This is
due to the fact that there is an alternating sign, $(-1)^k$, in the sum in eq.\ \eqref{eq:Defanp1}, which prevents us
from estimating the sum of all terms by means of an inequality.

Moreover, trying to find a similar statement as \eqref{eq:anUpperBound} with $a_i$ and $a_{n+1}$ replaced by their
absolute values in \eqref{eq:BareLiouAssumpFallOff} and \eqref{eq:anUpperBound}, respectively, does not work either:
In this case, $(1-A)^{n+1}$ in \eqref{eq:anp1int} is substituted by $(1+|A|)^{n+1}$ which is divergent in the large $n$
limit.} Nonetheless, \eqref{eq:anUpperBound} means that an exponential decrease of the first $n$ couplings leads to the
same or an even larger fall-off for $a_{n+1}$, which strongly suggests that the couplings do in fact converge.

%----------------------------------------------------------------------------------------------------------------------
\section{Proof of the combinatorial identity}
\label{app:ProofCombId}
%----------------------------------------------------------------------------------------------------------------------

In this section we would like to prove the combinatorial identity \eqref{eq:CombId}. It involves a sum over a
multi-index $\alpha\in\mathds{N}_0^n$ whose absolute value is fixed by $|\alpha|\equiv\sum_i\alpha_i=k$ and which
satisfies the additional constraint $\sum_i\mku i\alpha_i=n$. These two constraints reduce the number of possible terms
considerably and turn the sum into a combinatorial problem. To the best of our knowledge, the identity has not yet been
mentioned in the literature, so we present a detailed proof here.

Prior to this, let us consider an example of the sum in order to understand how it is computed: Let $n=4$ and $k=2$.
Then the only possible multi-indices $\alpha\in\mathds{N}_0^4$ whose absolute value equals $2$ are given by
$(1,1,0,0)$, $(1,0,1,0)$, $(1,0,0,1)$, $(0,1,1,0)$, $(0,1,0,1)$, $(0,0,1,1)$, $(2,0,0,0)$, $(0,2,0,0)$, $(0,0,2,0)$ and
$(0,0,0,2)$. Among these vectors there are only two that satisfy $\sum_i\mku i\alpha_i=4$, namely $(1,0,1,0)$ and
$(0,2,0,0)$. Hence, in this case the LHS of eq.\ \eqref{eq:CombId} is given by
\begin{equation}
 \frac{1!}{1!\,0!\,1!\,0!}+\frac{1!}{0!\,2!\,0!\,0!}=1+\frac{1}{2}=\frac{3}{2}\,.
\end{equation}
The RHS of \eqref{eq:CombId} gives $\frac{1}{4}\binom{4}{2} = \frac{1}{4}\frac{4!}{2!\,2!}
= \frac{3}{2}\mku$, too, so the identity is satisfied.
\medskip

\noindent
\textbf{Proof of (\ref{eq:CombId}).}

\noindent
It is shown that the RHS and the LHS of \eqref{eq:CombId} satisfy the same recurrence relation and the same initial
conditions.

We define
\begin{equation}
 \Omega_{n,k} \equiv \sum_{{\substack{\,\alpha\in\mathds{N}_0^n\\ |\alpha|=k\\[0.1em] \sum_i i\mku\alpha_i=n}}}
 \!\! \frac{1}{\alpha_1!\cdots\alpha_n!}\;.
\label{eq:DefOmegank}
\end{equation}
Since the multi-index is restricted by $|\alpha|=k$, its components are less than or at most equal to $k$, and we can
think of the multi-sum as $n$ sums, $\sum_{\alpha_1=0}^k\cdots\sum_{\alpha_n=0}^k\,$, where the $\alpha_i$'s are still
subjected to the two constraints. Now we split off the first sum and shift the remaining indices. We obtain
\begin{equation}
 \Omega_{n,k} = \sum_{\alpha_1=0}^k\frac{1}{\alpha_1!}\;\;\mathop{\sum_{\alpha_2}\,\cdots\,\sum_{\alpha_n}}_{%
 \substack{\sum_{i=1}^n \alpha_i = k \\[0.1em] \sum_{i=1}^n\mku i\mku\alpha_i = n}}\;\frac{1}{\alpha_2!\cdots\alpha_n!}
 = \sum_{j=0}^k\frac{1}{j!}\;\mathop{\sum_{\alpha_2}\,\cdots\,\sum_{\alpha_n}}_{%
 \substack{\sum_{i=2}^n\alpha_i=k-j\\[0.1em]\sum_{i=2}^n\mku i\mku\alpha_i=n-j}}\;\frac{1}{\alpha_2!\cdots\alpha_n!}\;,
\label{eq:SumProofIntStep}
\end{equation}
where we have relabeled $\alpha_1$ by $j$. Defining $\tilde{\alpha}_i \equiv \alpha_{i+1}$ we can write
\eqref{eq:SumProofIntStep} as
\begin{equation}
 \Omega_{n,k} = \sum_{j=0}^k\frac{1}{j!}\;\mathop{\sum_{\tilde{\alpha}_1}\,\cdots\,\sum_{\tilde{\alpha}_{n-1}}}_{%
 \substack{\sum_{i=1}^{n-1}\tilde{\alpha}_i=k-j\\[0.1em]\sum_{i=1}^{n-1}\mku i\mku\tilde{\alpha}_i=n-k}}\;
 \frac{1}{\tilde{\alpha}_1!\cdots\tilde{\alpha}_{n-1}!}\;.
\label{eq:SumProofIntStep2}
\end{equation}
The second constraint under the sums in \eqref{eq:SumProofIntStep2} has been obtained by rearranging its counterpart on
the RHS of eq.\ \eqref{eq:SumProofIntStep}, $\sum_{i=2}^n\mku i\mku\alpha_i=n-j$, as follows:
\begin{equation}
 n-j = \sum_{i=2}^n i\mku\alpha_i = \sum_{i=2}^n i\mku\tilde{\alpha}_{i-1} = 
 \sum_{i=1}^{n-1} (i+1)\tilde{\alpha}_{i} = \sum_{i=1}^{n-1} i\mku\tilde{\alpha}_{i}
 + \sum_{i=1}^{n-1} \tilde{\alpha}_{i} = \sum_{i=1}^{n-1} i\mku\tilde{\alpha}_{i} + k-j,
\end{equation}
leading to $\sum_{i=1}^{n-1}\mku i\mku\tilde{\alpha}_i=n-k$. In fact, this constraint dictates that all
$\tilde{\alpha}_i$ with $i>n-k$ must vanish. Therefore, we can consider the multi-index $\tilde{\alpha}$ as an element
of $\mathds{N}_0^{n-k}$ effectively rather than $\mathds{N}_0^{n-1}$, and the two constraints in
\eqref{eq:SumProofIntStep2} amount to $\sum_{i=1}^{n-k}\tilde{\alpha}_i=k-j$ and
$\sum_{i=1}^{n-k}\mku i\mku\tilde{\alpha}_i=n-k$. This enables us to identify the $\tilde{\alpha}$-sums in
\eqref{eq:SumProofIntStep2} with $\Omega_{n-k,k-j}$. As a result we find the \emph{recurrence relation}
\begin{equation}
 \Omega_{n,k} = \sum_{j=0}^k \frac{1}{j!}\,\Omega_{n-k,k-j}\,.
\label{eq:RecOmega}
\end{equation}

Furthermore, we have the \emph{initial values}
\begin{equation}
 \text{(i)}\;\,\Omega_{n,n}=\frac{1}{n!}\,,\qquad\text{(ii)}\;\,\Omega_{n,k}=0\;\text{ for }k>n,
 \qquad\text{(iii)}\;\,\Omega_{n,0}=0.
\end{equation}
These equations can be shown as follows.\\
(i) Setting $k=n$ in \eqref{eq:DefOmegank} we notice that the only possible
multi-index $\alpha$ satisfying both constraints is the one with $\alpha_1=n$ and $\alpha_2=\cdots=\alpha_n=0$.
Thus, the main sum over $\alpha$ consists of one term only: $\Omega_{n,n}=\frac{1}{n!\,0!\cdots 0!}=\frac{1}{n!}$.\\
(ii) The constraints imply $n=\sum_i i\mku\alpha_i\ge \sum_i\mku\alpha_i = k$, so for $k>n$ the main sum over $\alpha$
contains no term at all and amounts to zero.\\
(iii) For $k=0$ the constraint $\sum_i\mku\alpha_i = k$ forces all $\alpha_i$ to vanish. In that case, the constraint
$\sum_i i\mku\alpha_i=n$ can not be satisfied since $n\ge 1$, and so the main sum over $\alpha$ contains no term
either.
\smallskip

Next, we define
\begin{equation}
 \Psi_{n,k} \equiv \frac{1}{(k-1)!}\,\frac{1}{n}\binom{n}{k} = \frac{1}{k!}\binom{n-1}{k-1},
\label{eq:DefPsink}
\end{equation}
for $n\ge k\ge 1$, as well as
\begin{equation}
 \Psi_{n,k} \equiv 0\;\text{ for }k>n\quad\text{and}\quad \Psi_{n,0}\equiv 0\,.
\label{eq:DefPsink2}
\end{equation}
With regard to eq.\ \eqref{eq:CombId}, we have to prove $\Omega_{n,k}=\Psi_{n,k}$. For that purpose it suffices to show
that $\Omega_{n,k}$ and $\Psi_{n,k}$ satisfy the same recurrence relation and the same initial conditions. (By means of
eq.\ \eqref{eq:RecOmega}, all $\Omega_{n,k}$'s can be expressed in terms of the initial values. This statement would
hold true for $\Psi_{n,k}$, too, if we found the same recurrence relation and initial conditions.) Using
\eqref{eq:DefPsink} we have
\begin{equation}
\begin{split}
 \sum_{j=0}^k \frac{1}{j!}\,&\Psi_{n-k,k-j} = \sum_{j=0}^{k-1}\;\frac{(n-k)!}{j!\,(k-j-1)!\,(n-k)\,(k-j)!\,(n-2k+j)!}\\
 &= \sum_{j=0}^{k-1}\; \frac{1}{k!}\,\frac{k!}{j!\,(k-j)!}\,\frac{(n-k-1)!}{(k-j-1)!\,[(n-k-1)-(k-j-1)]!}\\
 &= \frac{1}{k!} \sum_{j=0}^{k-1} \binom{k}{j}\binom{n-k-1}{k-1-j} \stackrel{(*)}{=} \frac{1}{k!}\binom{k+n-k-1}{k-1}
 = \frac{1}{k!}\binom{n-1}{k-1}\\
 &= \Psi_{n,k}\;.
\end{split}
\label{eq:RecPsi}
\end{equation}
In \eqref{eq:RecPsi} the equality labeled by $(*)$ makes use of Vandermonde's identity which is given by
$\binom{m+n}{r}=\sum_{i=0}^r \binom{m}{i}\binom{n}{r-i}$ for $m,n,r\in\mathds{N}_0$. Thus, $\Psi_{n,k}$ indeed
satisfies the same recurrence relation as $\Omega_{n,k}$.

Finally, we convince ourselves of the validity of the initial conditions: With
$\Psi_{n,n}=\frac{1}{n!}\binom{n-1}{n-1}=\frac{1}{n!}$ and with the definitions in \eqref{eq:DefPsink2} we have
in fact the same initial values for $\Psi_{n,k}$ as the ones for $\Omega_{n,k}\mku$.

In conclusion, $\Psi_{n,k}$ \emph{and} $\Omega_{n,k}$ \emph{satisfy the same recurrence relation and the same
initial conditions, so} $\Omega_{n,k} = \Psi_{n,k}$. This proves the combinatorial identity \eqref{eq:CombId}.
\hfill$\Box$

%----------------------------------------------------------------------------------------------------------------------
\section{Proof of the inequality}
\label{app:ProofInequality}
%----------------------------------------------------------------------------------------------------------------------

In this section we will prove that inequality \eqref{eq:Inequality} is satisfied for all $n$ greater than a certain
threshold value which is to be determined. As $|1-A|<1$ by assumption, we can make use of
\begin{equation}
 (1-A)^n<1\quad\forall n\in\mathds{N}.
\label{eq:EstFor1mA}
\end{equation}
\begin{itemize}
\item \textbf{The case} \bm{$\cZ\ge 0$}. In this case the statement is obvious since, first,
$\frac{n^2}{n^2+2\pi\mku\cZ}\le 1$, and second, $\frac{1}{n}(1-A)^n-\frac{1}{n}<0$, by eq.\ \eqref{eq:EstFor1mA}.
Thus, \eqref{eq:Inequality} is satisfied for all $n\in\mathds{N}$ without further ado.
\item \textbf{The case} \bm{$\cZ< 0$}. This is the interesting case since in our analysis in Section
\ref{sec:BareLiouExpSeries} we have $\cZ< 0$ for either parametrization. We want to determine a threshold value
$\ntr$ such that \eqref{eq:Inequality} is satisfied for all $n>\ntr\,$. Since the factor $\frac{n^2}{n^2+2\pi\mku\cZ}$
has a pole at $n=\sqrt{-2\pi\mku\cZ}$, our first requirement for the threshold value is $n>\ntr\ge\sqrt{-2\pi\mku\cZ}$.
Unlike the case $\cZ\ge 0$, here the problem which hampers a straightforward estimate in \eqref{eq:Inequality} arises
from the different behavior of the two factors,
\begin{equation}
 \underbrace{\frac{n^2}{n^2+2\pi\mku\cZ}\vphantom{\bigg|}}_{\ge 1}\,
 \underbrace{\left[A+\frac{1}{n}(1-A)^n-\frac{1}{n}\right]}_{\le A},
\label{eq:EstForA}
\end{equation}
so the product is not less than or equal to $A$ for \emph{all} $n$. Hence, a more careful argument is required.
Subtracting \eqref{eq:EstForA} from $A$ yields
\begin{equation}
\textstyle
 A - \frac{n^2}{n^2+2\pi\mku\cZ}\left[A+\frac{1}{n}(1-A)^n-\frac{1}{n}\right]
 = \frac{n}{n^2+2\pi\mku\cZ}\left[\frac{2\pi\mku\cZ\mku A}{n}-(1-A)^n+1\right],
\end{equation}
and showing that this expression is greater than zero is equivalent to proving \eqref{eq:Inequality}. As we required
$n>\sqrt{-2\pi\mku\cZ}$, $n\in\mathds{N}$, we have $\frac{n}{n^2+2\pi\mku\cZ}>0$, so it remains to be shown that
\begin{equation}
 \frac{2\pi\mku\cZ\mku A}{n}-(1-A)^n+1\stackrel{!}{>}0.
\label{eq:IneqToProve}
\end{equation}
The idea is to determine threshold values with respect to $n$ for the first two terms separately, such that both
$2\pi\mku\cZ\mku A\mku\frac{1}{n}>-\frac{1}{2}$ and $-(1-A)^n>-\frac{1}{2}$.

For the first term in \eqref{eq:IneqToProve} we require $n>-4\pi\mku\cZ\mku A$. Then rearranging yields indeed
$2\pi\mku\cZ\mku A\mku\frac{1}{n}>-\frac{1}{2}$.

Regarding the second term, we differentiate between $A=1$ and $A\neq 1$. For $A=1$, obviously $-(1-A)^n=0>-\frac{1}{2}$
without further conditions on $n$. For $A\neq 1$ we require $n>-\frac{\ln(2)}{\ln|1-A|}$. This is equivalent to
$|1-A|^n<\frac{1}{2}$, which implies $(1-A)^n<\frac{1}{2}$.

Taking all requirements together we can define the threshold value now:
\begin{equation}
 \ntr \equiv \max\left(\sqrt{-2\pi\mku\cZ},\,-4\pi\mku\cZ\mku A,\,-\frac{\ln(2)}{\ln|1-A|}\right),
\label{eq:ntr}
\end{equation}
for $A\neq 1$, and $\ntr \equiv \max\left(\sqrt{-2\pi\mku\cZ},\,-4\pi\mku\cZ\mku A\right)$ for $A=1$. Then we find
\begin{equation}
 \frac{2\pi\mku\cZ\mku A}{n}-(1-A)^n+1 > -\frac{1}{2} -\frac{1}{2} +1 = 0\qquad\forall\, n>\ntr\,.
\end{equation}
As a consequence, we obtain the desired inequality,
\begin{equation}
 \frac{n^2}{n^2+2\pi\mku\cZ}\left[A+\frac{1}{n}(1-A)^n-\frac{1}{n}\right] \le A\qquad\forall\, n>\ntr\,,
\label{eq:InequalityNtr}
\end{equation}
where the ``$=$''-case included in ``$\le$'' applies to $n\to\infty$ only.\hfill$\Box$
\end{itemize}

%----------------------------------------------------------------------------------------------------------------------
\section{Numerical check of initial conditions}
\label{app:CheckInit}
%----------------------------------------------------------------------------------------------------------------------

Finally, we would like to check if and to what extent the first couplings obtained by numerical computation satisfy
assumption \eqref{eq:BareLiouAssumpFallOff}. If they do, at least approximately, we have to make sure that the
corresponding values of $A$ and $\lambda$ meet the conditions \eqref{eq:BareLiouAssumpConsts}. Furthermore, we want to
determine the threshold value $\ntr$ beyond which \eqref{eq:InequalityNtr} holds true. It should be a value that is
easily accessible by our numerical analysis; otherwise the above proofs would be pointless.

We use the results of Section \ref{sec:BareLiouExpSeriesLin}, more precisely, the bare couplings $\cgamma_n$ calculated
on the basis of the EAA couplings $b$ and $\mu$ for the linear metric parametrization ($b=38/3$, $\mu=0.1579$), see
Figure \ref{fig:BareGammaLinParam}. By eq.\ \eqref{eq:anFromGamman} we express those couplings in terms of $a_n$,
i.e.\ we determine $a_n$ for $n=1,\dotsc,48$.

Figure \ref{fig:UpperBoundForA} shows the first 10 couplings $a_n$, $n=1,\dotsc,10$. We find that their fall-off
behavior for increasing $n$ is not exactly given by a straight line in the logarithmic diagram, so the assumed
exponential decrease is observed only at an approximate level, $a_n\approx A\,\e^{-\lambda n}$. Although lacking an
exact relation, we might determine an upper bound for $a_n$ in terms of $A$ and $\lambda$ such that
\begin{equation}
 a_n \le A\,\e^{-\lambda n}\,.
\end{equation}
For this purpose we proceed as follows. We fit a linear function of the type $f(n)=c_1\mku n+c_0$ to the set of points
$(n,\ln a_n)$ for $n=2,\dotsc,10$.\footnote{We excluded $a_1$ here because it is the only coupling among the $a_n$'s
which is determined by a different formula, eq.\ \eqref{eq:Defa1}, and because its corresponding point in the diagram
deviates stronger from the line.} The result reads
\begin{equation}
 f(n) = -0.477\, n + 0.350 \,.
\label{eq:Fitan}
\end{equation}
Then we shift this function slightly upwards, $f(n)\to\tilde{f}(n)=f(n)+\tilde{c}\mku$, such that $\ln a_n \le
\tilde{f}(n)$ for all $n=1,\dotsc,10$, yielding an upper bound for $\ln a_n$. Here we find that $\tilde{c}=0.1$ is a
sufficiently large shift. Exponentiating $\tilde{f}(n)$ finally leads to the desired bound for $a_n$. Based on the
fitting data \eqref{eq:Fitan} we obtain
\begin{equation}
 A = 1.568\qquad\text{and}\qquad \lambda=0.477\,,
\label{eq:ResForAAndLambda}
\end{equation}
such that $a_n \le A\,\e^{-\lambda n}$ is indeed satisfied for the first 10 couplings. This upper bound is shown in
Figure \ref{fig:UpperBoundForA} as well.
\begin{figure}[tp]
 \centering
 \includegraphics[width=0.75\columnwidth]{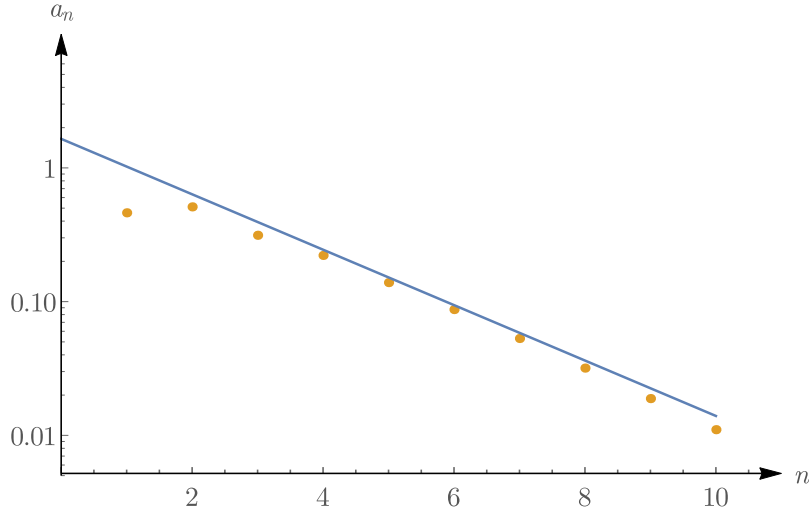}
\caption{Logarithmic plot of the first 10 couplings $a_n$ (dark yellow points) and a line serving as an upper bound
 (blue). The bound was obtained by fitting a linear function, $c_1\mku n+c_0$, to $\ln(a_n)$ for $n=2,\dotsc,10$ and
 shifting it slightly upwards ($c_0\to c_0+0.1$).}
\label{fig:UpperBoundForA}
\end{figure}

Remarkably enough, the values in \eqref{eq:ResForAAndLambda} meet the conditions \eqref{eq:BareLiouAssumpConsts}:
$A > 0$, $\lambda > 0$ and $|A-1|<1$.

In summary, we have not been able to show that the required assumption $a_n = A\,\e^{-\lambda n}$ is strictly satisfied
for the first couplings, nor did we find a more general proof with relaxed and less restrictive assumptions. However,
we have found an upper bound, which actually serves as a good approximation for the couplings at the same time:
$a_n \lesssim A\,\e^{-\lambda n}\mku$. Taking all of the above arguments together, we collected strong evidence for the
convergence of the couplings as $n\to\infty$.

It remains to be checked if the threshold value corresponding to inequality \eqref{eq:InequalityNtr} is accessible by
our numerical analysis, i.e.\ if we can compute all $a_n$ with $n\le\ntr$. (Note that we have calculated the $a_n$'s up
to $n=48$.) Previously, we have tested the compatibility of the first 10 couplings with the requirements
for the proof of \eqref{eq:anUpperBound}. 
In this respect, it would be desirable if \eqref{eq:InequalityNtr} were satisfied for all $n>10$.

Using the result for the threshold value, eq.\ \eqref{eq:ntr}, and inserting the numerically determined parameter $A$,
given by \eqref{eq:ResForAAndLambda}, we obtain
\begin{equation}
 \ntr=9.93\,.
\end{equation}
This remarkable result proves that \eqref{eq:InequalityNtr} is satisfied for all $n\ge 10$, in perfect
agreement with our wish expressed in the previous paragraph.

We can even find a lower threshold value. (The one in eq.\ \eqref{eq:ntr} is sufficient for \eqref{eq:InequalityNtr},
but it has been derived by very careful estimates that might be undercut.) This possibility is illustrated in Figure
\ref{fig:CheckInequality}. It shows the values resulting from the LHS of \eqref{eq:InequalityNtr},
$\frac{n^2}{n^2+2\pi\mku\cZ}\left[A+\frac{1}{n}(1-A)^n-\frac{1}{n}\right]$, dependent on $n$. For comparison, the
reference value $A=1.568$ is represented by the dashed horizontal line. All points in Figure \ref{fig:CheckInequality}
below this line, i.e.\ all $n\ge 5$, satisfy the inequality \eqref{eq:InequalityNtr}.
\begin{figure}[tp]
 \centering
 \includegraphics[width=0.6\textwidth]{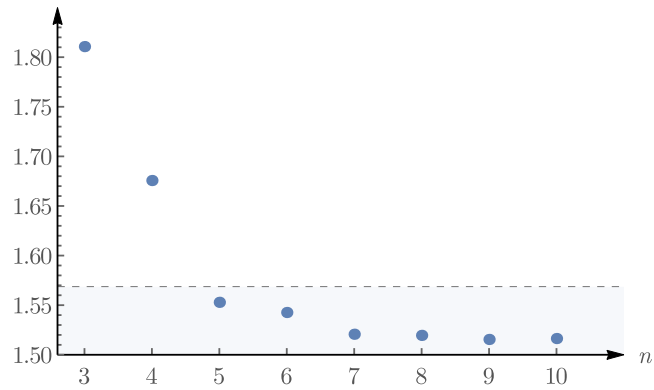}
\caption{Check of inequality \eqref{eq:InequalityNtr}: The blue points show
 $\frac{n^2}{n^2+2\pi\mku\cZ}\left[A+\frac{1}{n}(1-A)^n-\frac{1}{n}\right]$ plotted against $n$. The dashed horizontal
 line is located at the height $A$. Thus, we observe that the inequality holds true for $n\ge 5$.}
\label{fig:CheckInequality}
\end{figure}

At last, we would like to briefly point out the differences arising from the use of the exponential parametrization as
compared with the above results. For the linear parametrization all bare couplings $\cgamma_n$ are negative (all
$a_n$ are positive). This fact rendered the above considerations possible. As we have seen in Section
\ref{sec:BareLiouExpSeriesExp}, on the other hand, the exponential parametrization results in a set of $\cgamma_n$
characterized by changing signs. Although being evenly distributed on average, these sign fluctuations seem to be
irregular, see Figure \ref{fig:BareGammaExpParam}. Therefore, the requirement $a_i = A\,\e^{-\lambda i}$ for all
$i\le n$ with some $n\in\mathds{N}$, cf.\ eq.\ \eqref{eq:BareLiouAssumpFallOff}, cannot be satisfied in this case.
Figure \ref{fig:BareGammaExpParam} rather suggests that it is the absolute values of the couplings that decrease
exponentially. In the beginning of this appendix we have already mentioned, however, that our proofs do not
appropriately generalize to a formulation in terms of absolute values. Hence, we must rely on the numerical analysis
at this point. Having said this, it is surprising that the couplings $\cgamma_n$ and $\cx$ seem to converge almost
equally well as observed for the linear parametrization.

%----------------------------------------------------------------------------------------------------------------------
\chapter{Weyl transformation of the functional measure and the cutoff}
\label{app:WeylMeasureCutoff}
%----------------------------------------------------------------------------------------------------------------------

This appendix addresses the transformation law of the functional measure, $\mD_\UV^{[\hg]}\chi$, and the cutoff
contribution to the Ward identity w.r.t.\ Weyl split-symmetry. The latter requires a computation of the term
\begin{equation}
 \Big\langle \int\td^2 y\;(\chi-\phi)(y)\,\frac{\hg_\mn(x)}{\shgx}\mku\frac{\delta}{\delta\hg_\mn(x)}\left[\left(\shg
 \,\RL\right)(y)\right](\chi-\phi)(y)\Big\rangle.
\end{equation}
We will simplify this expression for general regulators in Section
\ref{app:WeylTransCutoff} and evaluate it explicitly for the optimized cutoff in Section \ref{app:WeylTransCutoffOpt}.
These considerations supplement the discussion of Weyl split-symmetry transformations and Ward identities contained in
Section \ref{sec:WeylWardIdentity}.

%----------------------------------------------------------------------------------------------------------------------
\section{Weyl transformation of the functional measure}
\label{app:WeylTransMeasure}
%----------------------------------------------------------------------------------------------------------------------

Since the measure defined in Appendix \ref{app:Measure} is translational invariant, the change $\chi\to\chi\mku{}'=
\chi-\sigma$ leaves it unaltered, $\mD_\UV^{[\hg]}\chi\mku{}'=\mD_\UV^{[\hg]}\chi$. Thus, it remains to be investigated
how the measure transforms under Weyl transformations, $\hg_\mn\to\hg'_\mn$, with
\begin{equation}
 \hg'_\mn=\e^{2\sigma}\hg_\mn\,.
\end{equation}
For that purpose, we define the two functionals
\begin{equation}
 \hS[\chi] = \frac{1}{2}\int\td^2 x\shg\;\chi\big(-\hB\big)\chi\,,\qquad
 \hS'[\chi] = \frac{1}{2}\int\td^2 x\sqrt{\hg\mku{}'}\;\chi\big(-\hB\mku{}'\big)\chi \,.
\end{equation}
By eq.\ \eqref{eq:TransSgLaplace2D} we observe that $\hS'=\hS$. For our discussion we are going to exploit known
identities for functional integrals, the connection of $\hS$ and $\hS'$ to the induced gravity action $\Gi$, and the
transformation laws of $\Gi$ considered in Appendix \ref{app:Weyl}. We proceed in four steps.
\smallskip

\noindent
\textbf{(1)} We recall that the induced gravity action is defined by
\begin{equation}
 \e^{-\Gi[\hg]} \equiv \int \mD_\UV^{[\hg]}\chi\;\e^{-\hS[\chi]}\,,
\label{eq:GammaIndDefPI}
\end{equation}
where we use the shorthand notation $\Gi[\hg]\equiv\Gi_{k=0,\UV}[\hg]$.
\smallskip

\noindent
\textbf{(2)} We know from Appendix \ref{app:Weyl}, in particular eq.\ \eqref{eq:GitoDeltaIZero}, that the
transformation behavior of the finite part of $\Gi[\hg]$ is given by $\Gi[\hg']\big|_\text{finite} = \Gi[\hg]
\big|_\text{finite} - \frac{1}{12\pi}\Delta I[\sigma;\hg] +\frac{1}{2}\ln\big(\hV'/\hV\big)$, where the functional
$\Delta I[\sigma;\hg]$ has been defined in eq.\ \eqref{eq:DeltaI}, and $\hV\equiv\int\td^2 x\shg$ and $\hV'\equiv
\int\td^2 x\sqrt{\hg'}$ denote the respective volume terms. These volume terms are purely due to possible zero mode
contributions; if the Laplacians do not have any zero modes, they cancel each other. As discussed in Section
\ref{app:Zero}, the divergent part of $\Gi[\hg]$ depends on the underlying regularization scheme. Regularizing the
measure as in Appendices \ref{app:Measure} and \ref{app:NoZero}, the transformation law of the full induced gravity
action reads
\begin{equation}
 \Gi[\hg']=\Gi[\hg]- \frac{1}{12\pi}\Delta I[\sigma;\hg] + \frac{1}{2}\ln\bigg(\frac{\hV'}{\hV}\bigg) 
 - \frac{\UV^2}{8\pi}\big(\hV'-\hV\big).
\label{eq:GiWeylTrans}
\end{equation}
Applying \eqref{eq:GiWeylTrans} to \eqref{eq:GammaIndDefPI} and using $\hS'=\hS$ yields
\begin{equation}
 \int \mD_\UV^{[\hg']}\chi\;\e^{-\hS'[\chi]} = \e^{-\Delta\Gi[\hg',\hg]}\int \mD_\UV^{[\hg]}\chi\;\e^{-\hS'[\chi]}\,,
\label{eq:TransWeylPI}
\end{equation}
with $\Delta\Gi[\hg',\hg]\equiv -\frac{1}{12\pi}\Delta I[\sigma;\hg]+ \frac{1}{2}\ln\left(\frac{\hV'}{\hV}\right)-
\frac{\UV^2}{8\pi}\big(\hV'-\hV\big)$. From eq.\ \eqref{eq:TransWeylPI} we can read off that the measure must transform
as $\mD_\UV^{[\hg']}\chi = \e^{-\Delta\Gi[\hg',\hg]}\, \mD_\UV^{[\hg]}\chi$ provided that the integrand is given by
$\e^{-\hS'[\chi]}$. We would like to prove next that this relation is actually independent of the integrand.
\smallskip

\noindent
\textbf{(3)} We repeat the above integration, but we include an arbitrary functional this time, i.e.\ we aim at
calculating $\int \mD_\UV^{[\hg']}\chi\;\e^{-\hS'[\chi]}\,F[\chi;\hg']$. For that purpose, we are going to need two
functional identities. First, observe that the argument $\chi$ in $F[\chi;\hg']$ can be replaced according to
\begin{equation}
 F[\chi;\hg'] = F\left[{\textstyle\frac{1}{\sqrt{\hg'}}\frac{\delta}{\delta J}}\,;\hg'\right]\,
 \e^{\int\td^2 x\sqrt{\hg'}\, J\,\chi}\,\Big|_{J=0}\;.
\label{eq:FReplArg}
\end{equation}
For the second identity, let $\tg_\mn$ denote an arbitrary metric which is merely used to specify the measure. Then,
by completing the square in the ensuing functional integral, we find
\begin{equation}
\begin{split}
 \int\mD_\UV^{[\tg]}\chi\; \e^{-\hS+ J\cdot\chi}
 &= \int\mD_\UV^{[\tg]}\chi\;\e^{-\frac{1}{2}\big[\chi\cdot(-\hB)\chi-2\, J\cdot\chi\big]} \\
 &= \int\mD_\UV^{[\tg]}\chi\;\e^{-\frac{1}{2}(\chi+\hB^{-1}J)\cdot(-\hB)(\chi+\hB^{-1}J)}\,
 \e^{-\frac{1}{2}J\cdot\hB^{-1}J} \\
 &= \int\mD_\UV^{[\tg]}\chi\;\e^{-\frac{1}{2}\chi\cdot(-\hB)\chi}\,\e^{-\frac{1}{2}J\cdot\hB^{-1}J}
 = \e^{-\frac{1}{2}J\cdot\hB^{-1}J} \int\mD_\UV^{[\tg]}\chi\; \e^{-\hS}\,,
\end{split}
\label{eq:QuadPI}
\end{equation}
and an equivalent relation in terms of $\hS'$ is obtained by replacing $\hB$ with $\hB{\mku}'$.
From the second to the third line we shifted the integration variable according to $\chi\to\chi-\hB^{-1}J$ and
exploited the translational invariance of the measure. Note that \eqref{eq:QuadPI} holds for any metric $\tg_\mn$
in the measure.
\smallskip

\noindent
\textbf{(4)} Combining the above results we obtain
\begin{align}
 \int \mD_\UV^{[\hg']}&\chi\;\e^{-\hS'[\chi]}\,F[\chi;\hg'] \stackrel{\eqref{eq:FReplArg}}{=}
 \int \mD_\UV^{[\hg']}\chi\;\e^{-\hS'[\chi]}\,F\left[{\textstyle\frac{1}{\sqrt{\hg'}}\frac{\delta}{\delta J}}\,;\hg'
 \right]\, \e^{\int\td^2 x\sqrt{\hg'}\, J\,\chi}\,\Big|_{J=0}\nonumber\\
 &\stackrel{\mathclap{\eqref{eq:QuadPI}}}{=}\;\; F\left[{\textstyle\frac{1}{\sqrt{\hg'}}\frac{\delta}{\delta J}}\,;\hg'
 \right] \e^{-\frac{1}{2}\int\td^2 x\sqrt{\hg'}\,J\mku\hB'^{-1}J} \int\mD_\UV^{[\hg']}\chi\; \e^{-\hS'[\chi]}\,
 \Big|_{J=0}\nonumber\\
 &\stackrel{\mathclap{\eqref{eq:TransWeylPI}}}{=}\;\; \e^{-\Delta\Gi[\hg',\hg]}\,
 F\left[{\textstyle\frac{1}{\sqrt{\hg'}}\frac{\delta}{\delta J}}\,;\hg' \right] \e^{-\frac{1}{2}\int\td^2 x\sqrt{\hg'}
 \,J\mku\hB'^{-1}J} \int \mD_\UV^{[\hg]}\chi\;\e^{-\hS'[\chi]}\, \Big|_{J=0}\nonumber\\
 &\stackrel{\mathclap{\eqref{eq:QuadPI}}}{=}\;\;\e^{-\Delta\Gi[\hg',\hg]}\, \int\mD_\UV^{[\hg]}\chi\;\e^{-\hS'[\chi]}\;
 F\left[{\textstyle\frac{1}{\sqrt{\hg'}}\frac{\delta}{\delta J}}\,;\hg' \right]\, \e^{\int\td^2 x\sqrt{\hg'}\, J\,\chi}
 \,\Big|_{J=0}\nonumber\\
 &\stackrel{\mathclap{\eqref{eq:FReplArg}}}{=}\;\;\e^{-\Delta\Gi[\hg',\hg]}\,
 \int \mD_\UV^{[\hg]}\chi\;\e^{-\hS'[\chi]}\,F[\chi;\hg'],
\end{align}
for an arbitrary functional $F[\chi;\hg']$. Therefore, we conclude that the measure transforms as
\vspace{0.5em}
\begin{equation}[b]
 \mD_\UV^{[\hg']}\chi = \e^{-\Delta\Gi[\hg',\hg]}\; \mD_\UV^{[\hg]}\chi\;.
\end{equation}
\vskip0.5em
\noindent
The exponent of the crucial transformation factor, $\Delta\Gi[\hg',\hg]$, is given by
\begin{equation}
 \Delta\Gi[\hg',\hg]\equiv -\frac{1}{12\pi}\Delta I[\sigma;\hg]+ \frac{1}{2}\ln\left(\frac{\hV'}{\hV}\right)-
 \frac{\UV^2}{8\pi}\big(\hV'-\hV\big),
\label{eq:DeltaGammaIndDef}
\end{equation}
with $\Delta I[\sigma;\hg] \equiv \frac{1}{2}\int\td^2x\shg\big[\hD_\mu\sigma\hD^\mu\sigma+\hR\sigma\big]$.
Again, the term $\frac{1}{2}\ln\big(\hV'/\hV\big)$ occurs only in the presence of zero modes.
%\footnote{As an aside, let us prove the
% statement concerning the zero modes. For that purpose we define the metric $\hg_\mn^{(\alpha)}\equiv\e^{2\alpha\sigma}
% \hg_\mn\mku$, which interpolates between $\hg_\mn^{(0)}=\hg_\mn$ and $\hg_\mn^{(1)}=\hg'_\mn$. By eq.\
% \eqref{eq:TransLaplace2D} the corresponding Laplacian is given by $\hB^{(\alpha)} = \e^{-2\alpha\sigma}\mku\hB\mku$.
% Hence, if $\hB$ has no zero modes, then $\hB^{(\alpha)}$ has no zero modes either. This means in particular that the
% projection of the field $\sigma$ onto its constant part must vanish: $\int\td^2 x\sqrt{\hg^{(\alpha)}}\,\sigma=0$.
% Rewriting this equation yields $0=\int\td^2 x\shg\,\sigma\,\e^{2\alpha\sigma}=\frac{1}{2}\frac{\td}{\td\alpha}\int\shg
% \,\e^{2\alpha\sigma}\equiv \frac{1}{2}\frac{\td}{\td\alpha}\hV^{(\alpha)}$, and integrating from $\alpha=0$ to
% $\alpha=1$ we obtain
% \begin{equation}
%  0 = \int_0^1\td\alpha\,\frac{1}{2}\frac{\td}{\td\alpha}\hV^{(\alpha)}=\frac{1}{2}\left[\hV^{(\alpha)}\right]_0^1
%  = \frac{1}{2}\big(\hV^{(1)}-\hV^{(0)}\big)= \frac{1}{2}\big(\hV'-\hV\big).
% \end{equation}
% Thus, $\hV'=\hV$ in the absence of zero modes, leading to a cancellation of the volume terms in
% \eqref{eq:DeltaGammaIndDef}.}

%----------------------------------------------------------------------------------------------------------------------
\section{Simplification of the cutoff contribution}
\label{app:WeylTransCutoff}
%----------------------------------------------------------------------------------------------------------------------

In this section we reexpress the cutoff contribution to the Ward identity as it occurs in eq.\ \eqref{eq:PreWI},
$\Big\langle\int\td^2 y\;(\chi-\phi)(y)\,\frac{\hg_\mn(x)}{\shgx}\mku\frac{\delta}{\delta\hg_\mn(x)}\left[\left(\shg\,
\RL\right)(y)\right](\chi-\phi)(y)\Big\rangle$, in terms of the propagator $\big(\Gamma_\UV^\text{L}{}^{(2)}+\RL
\big)^{-1}$. For this purpose, we exploit two well known identities. First,
\begin{equation}
 \big\langle(\chi-\phi)\,A\,(\chi-\phi)\big\rangle = \big\langle\chi\,A\,\chi\big\rangle
 - \big\langle\chi\big\rangle A\,\phi - \phi\,A\big\langle\chi\big\rangle + \phi\,A\,\phi
 = \big\langle\chi\,A\,\chi\big\rangle - \phi\,A\,\phi,
\end{equation}
and second, we observe that a contracted metric derivative, $\hg_\mn\mku\frac{\delta}{\delta\hg_\mn}$, can be
represented as a derivative with respect to $\sigma\mku$:
\begin{equation}
\begin{split}
 \frac{\delta}{\delta\sigma(x)}\mku F[\e^{2\sigma}\hg]\Big|_{\sigma=0}  &= \int\td y\;\frac{\delta F[\hg]}{\delta
 \hg_\mn(y)}\,\frac{\delta\big[\e^{2\sigma(y)}\hg_\mn(y)\big]}{\delta\sigma(x)}\,\bigg|_{\sigma=0} \\
 &= \int\td y\;\frac{\delta F[\hg]}{\delta\hg_\mn(y)}\;2\,\e^{2\sigma(x)}\,\hg_\mn(x)\,\delta(x-y)\,\bigg|_{\sigma=0}\\
 &= 2\,\hg_\mn(x)\,\frac{\delta F[\hg]}{\delta\hg_\mn(x)}\;.
\end{split}
\label{eq:TracedDerivative}
\end{equation}
The latter relation can be used, for instance, to compute the variation of the square root of the metric determinant
in an easy way, yielding
\begin{equation}
 \frac{\hg_\mn(x)}{\shgx}\mku\frac{\delta}{\delta\hg_\mn(x)}\,\shgy = \delta(x-y).
\end{equation}
In addition to that, we introduce the abbreviation
\begin{equation}
 \hRL(x) \equiv  \frac{\hg_\mn(x)}{\shgx}\,\frac{\delta}{\delta \hg_\mn(x)}\,\RL\,,
\label{eq:DefhRL}
\end{equation}
with $\RL \equiv \RL[\hg_\mn(y)] \equiv \RL(-\hB_y)$ where the argument $y$ agrees with the variable of integration
in the expression under consideration.

Based on this groundwork we obtain
\begin{align}
 &\bigg\langle\int\td^2 y\;(\chi-\phi)(y)\,\frac{\hg_\mn(x)}{\shgx}\mku\frac{\delta}{\delta\hg_\mn(x)}\left[\left(\shg\,
 \RL\right)(y)\right](\chi-\phi)(y)\bigg\rangle \nonumber\\
 &= \int\td^2 y\Big[\big\langle\chi(y)\RL\,\chi(y)\big\rangle- \phi(y)\RL\,\phi(y) \Big]
 \delta(x-y) \nonumber\\
 &\qquad{} + \int\td^2 y\shgy\Big[\big\langle\chi(y)\hRL(x)\mku\chi(y)\big\rangle- \phi(y)\hRL(x)\mku\phi(y) \Big]
 \nonumber\\
 &= \int\td^2 y\;\delta(x-y)\int\td^2 z\shgz\,{\textstyle\frac{1}{\shgz}}\,\RL\,\delta(y-z) \Big[\big\langle\chi(y)
 \chi(z)\big\rangle - \phi(y)\phi(z) \Big] \nonumber\\
 &\qquad + \int\td^2 y\shgy\int\td^2 z\shgz\,{\textstyle\frac{1}{\shgz}}\,\hRL(x)\,\delta(y-z) \Big[\big\langle\chi(y)
 \chi(z)\big\rangle - \phi(y)\phi(z) \Big] \nonumber\\
 &= \int\td^2 y\;\delta(x-y)\int\td^2 z\shgz\,\big(\RL\big)_{yz} \big(\Gamma_\UV^\text{L}{}^{(2)}+\RL\big)^{-1}_{zy}
 \nonumber\\
 &\qquad+\int\td^2 y\shgy\int\td^2 z\shgz\,\big[\hRL(x)\big]_{yz} \big(\Gamma_\UV^\text{L}{}^{(2)}+\RL\big)^{-1}_{zy}
 \nonumber\\
 &= \int\td^2 y\;\delta(x-y) \Big[\RL\big(\Gamma_\UV^\text{L}{}^{(2)}+\RL\big)^{-1}\Big]_{yy} \nonumber\\
 &\qquad + \int\td^2 y\shgy \Big[\hRL(x)\big(\Gamma_\UV^\text{L}{}^{(2)}+\RL\big)^{-1}\Big]_{yy} \nonumber\\
 &= \big\langle x\big|\,\RL\,\big(\Gamma_\UV^\text{L}{}^{(2)}+\RL\big)^{-1}\big|x\big\rangle
 +\Tr_\UV\!\Big[\hRL(x)\,\big(\Gamma_\UV^\text{L}{}^{(2)}+\RL\big)^{-1}\Big]
\label{eq:SimplCutoffTerm}
\end{align}
Here we have employed the operator conventions discussed in Appendix \ref{app:OperatorRep}. In particular, for the
third equality we have exploited that the propagator can be expressed as $\big(\Gamma_\UV^\text{L}{}^{(2)}+\RL
\big)^{-1}_{xy} = \big\langle\chi(x)\chi(y)\big\rangle - \phi(x)\phi(y)$.

The advantage of our result \eqref{eq:SimplCutoffTerm} lies in the fact that we do no longer have to compute any
involved expectation values. The latter are replaced by the propagator, an object which is obtained straightforwardly
in our case with the EAA given.

%----------------------------------------------------------------------------------------------------------------------
\section{The Ward identity for the optimized cutoff}
\label{app:WeylTransCutoffOpt}
%----------------------------------------------------------------------------------------------------------------------

Finally, we would like to evaluate the cutoff terms obtained in the previous section, $\big\langle x\big|\,\RL\,
\big(\Gamma_\UV^\text{L}{}^{(2)}+\RL\big)^{-1}\big|x\big\rangle$ and $\Tr_\UV\big[\hRL(x)\,\big(\Gamma_\UV^\text{L}
{}^{(2)}+\RL\big)^{-1}\big]$, when using the optimized cutoff, $\RL\equiv\RL(-\hB)=\ZL\mku\big(\UV^2+\hB\big)\,\theta
\big(\UV^2+\hB\big)$ with $\ZL\equiv -\frac{b}{8\pi}$. It is crucial for the argument that $\big(\Gamma_\UV^\text{L}
{}^{(2)}+\RL\big)^{-1}$ becomes diagonal in its spacetime representation when combined with $\RL$ or $\hRL(x)$.
Diagonality of an operator $\mO$ means $\langle x|\mO|y\rangle\propto \delta(x-y)$. The reason why the propagator
becomes diagonal is that it does no longer contain any differential operators provided that it is multiplied by a
cutoff term. We will clarify the details in a moment. We emphasize that this diagonality is a special feature of the
optimized cutoff; the general treatment is more involved.

The second functional derivative of $\Gamma_\UV^\text{L}$ is given by $\Gamma_\UV^\text{L}{}^{(2)} = \ZL\big(-\hB
+2\mku\mu\mku\UV^2\,\e^{2\phi}\big)$, with $\ZL=-\frac{b}{8\pi}$, so we have
\begin{equation}
 \Gamma_\UV^\text{L}{}^{(2)} + \RL = \ZL\Big[-\hB+2\mku\mu\mku\UV^2\,\e^{2\phi}+\big(\UV^2+\hB\big)\,\theta
\big(\UV^2+\hB\big)\Big].
\label{eq:WTCOProp}
\end{equation}
Upon multiplying this expression by either $\RL$ or $\hRL(x)$ we observe that the step-function $\theta\big(\UV^2
+\hB\big)$ contained in both of these cutoff terms effectively suppresses all modes with $\omega^2/\UV^2\ge 1$, where
$\omega^2$ is an eigenvalue of $-\hB$. For all
remaining modes the $\theta$-function in \eqref{eq:WTCOProp} equals $1$. From this we infer that
\begin{equation}
 \text{cutoff}{}\times\big(\Gamma_\UV^\text{L}{}^{(2)}+\RL\big)^{-1}
 = {}\text{cutoff}{}\times\Big[\ZL\big(\UV^2+2\mku\mu\mku\UV^2\,\e^{2\phi}\big)\Big]^{-1}\,,
\label{eq:WTCOPropDiag}
\end{equation}
where ``cutoff'' is a placeholder for $\RL$ or $\hRL(x)$. Hence, $\big(\Gamma_\UV^\text{L}{}^{(2)}+\RL\big)^{-1}$
is a pure number whenever it occurs in combination with a cutoff term, so it is indeed diagonal in $x$-space. Note
that, as usual, we employ the conventions for operator representations specified in Appendix \ref{app:OperatorRep}.
\medskip

\noindent
\textbf{(1) Evaluation of Tr\bm{${}_\UV\big[\hRL(x)\,\big(\Gamma_\UV^\text{L}{}^{(2)}+\RL\big)^{-1}\big]$}}.\\
Here the trace $\Tr_\UV$ reduces to a standard trace, $\Tr$, since $\hRL$ already suppresses all modes with momenta
larger than $\UV$. Using \eqref{eq:WTCOPropDiag} in addition, we obtain
\begin{equation}
\begin{split}
 &\Tr_\UV\Big[\hRL(x)\,\big(\Gamma_\UV^\text{L}{}^{(2)}+\RL\big)^{-1}\Big]
 =\Tr\Big[\hRL(x)\,\big(\Gamma_\UV^\text{L}{}^{(2)}+\RL\big)^{-1}\Big]\\
 &= \int\td^2 y\,\td^2 z\shgy\,\shgz\,\big\langle y\big|\hRL(x)\big|z\big\rangle\big\langle z\big|
 \big(\Gamma_\UV^\text{L}{}^{(2)}+\RL\big)^{-1}\big| y\big\rangle\\
 &= \int\td^2 y\,\td^2 z\left[\textstyle \frac{\hg_\mn(x)}{\shgx}\frac{\delta}{\delta\hg_\mn(x)}
 \Big(\RL\big(-\hB_y\big)\delta(y-z)\Big)\right]\Big[\ZL\mku\UV^2\big(1+2\mku\mu\,\e^{2\phi}\big)\Big]^{-1}\delta(z-y),
\end{split}
\label{eq:TrhRL}
\end{equation}
where we have inserted the definition \eqref{eq:DefhRL}.
Now we can separately analyze the remaining cutoff contribution, $\RL\big(-\hB_y\big)\delta(y-z)$. For that purpose
we express it in terms of a Laplace transform:
\begin{equation}
 \RL\big(-\hB_y\big)\delta(y-z) = \int_0^\infty\td s\;\widetilde{\mathcal{R}}_\UV(s)\,\e^{s\mku\hB}\mku\delta(y-z)
\label{eq:LaplTrhRL}
\end{equation}
At this point we can exploit the known results concerning heat kernel expansions, see Appendix \ref{app:Heat}.
Here the expansion has the form $\e^{s\mku\hB}\mku\delta(y-z)=\sum_n s^n\mku A_n(y,z)$. Since there is a second
delta function on the very right of eq.\ \eqref{eq:TrhRL}, we can take the coincidence limit $z\to y$ in the heat
kernel expansion.\footnote{Note that taking the coincidence limit, that is, letting $z\to y$, commutes with taking the
derivative $\delta/\delta \hg_\mn$ in \eqref{eq:TrhRL}. There are terms proportional to the squared geodesic distance,
$\sigma(y,z)$, appearing in the off-diagonal heat kernel expansion (i.e.\ the expansion before taking the
coincidence limit), which might potentially lead to noncommuting terms at first sight since $\lim_{z\to y}\sigma(y,z)
=0$. However, it is possible to show that $\frac{\delta}{\delta\hg_\mn}\sigma(y,z)\propto\sigma(y,z)$, and similarly
for all spacetime derivatives of $\sigma(y,z)$. Hence, applying $\frac{\delta}{\delta\hg_\mn}$ to the expansion does
not affect whether or not certain terms of the expansion vanish in the coincidence limit, and so, taking
$\frac{\delta}{\delta\hg_\mn}$ commutes with taking $z\to y$.} This leads to significant simplifications, and we
obtain \cite{Vassilevich2003,DeWitt1965,DeWitt2003,BV87,BV90a,BV90b,Vilkovisky1992,Avramidi2000,GSZ11}
\begin{equation}
\begin{split}
 \e^{s\mku\hB}\mku\delta(y-z)\big|_{z\to y} &= \frac{1}{4\pi\mku s}\shgy\;\sum_{n=0}^\infty s^n\mku a_n(y,y)\\
 &= \frac{1}{4\pi\mku s}\shgy\,\left[\textstyle 1+\frac{1}{6}s\mku \hR + \frac{1}{60}s^2\mku\hR^2
 + \frac{1}{30}s^2\mku\hB\mku\hR + \mO(s^3)\right].
\end{split}
\label{eq:HeatCoincidence}
\end{equation}
Furthermore, we can make use of the fact that the generalized Mellin transform $Q_n[W]$ --- defined by eq.\
\eqref{eq:Mellin} in Appendix \ref{app:Heat} --- of some function $W$ has an equivalent representation in terms of the
inverse Laplace transform $\widetilde{W}$:
\begin{equation}
 Q_n[W] = \int_0^\infty\td s\; \widetilde{W}(s) s^{-n} \qquad (\text{for all }n).
\label{eq:MellinAlt}
\end{equation}
Combining \eqref{eq:LaplTrhRL}, \eqref{eq:HeatCoincidence} and \eqref{eq:MellinAlt} we find
\begin{equation}
\begin{split}
 &\Big(\RL\big(-\hB_y\big)\delta(y-z)\Big)\Big|_{z\to y} \\
 &=\frac{1}{4\pi}\shgy\left(Q_1[\RL]+\frac{1}{6}\,\hR\, Q_0[\RL]
 +\frac{1}{60}\left(\hR^2+2\,\hB\mku\hR\right)Q_{-1}[\RL]+\dots\right),
\end{split}
\label{eq:HeatExpGen}
\end{equation}
where the dots refer to all terms proportional to $Q_n[\RL]$ with $n\le -2$. For the optimized cutoff, $\RL\equiv
\RL(-\hB)=\ZL\mku\big(\UV^2+\hB\big)\,\theta\big(\UV^2+\hB\big)$, the generalized Mellin transforms are computed most
easily by using eq.\ \eqref{eq:Mellin}. They read
\begin{equation}
 Q_n[\RL] = \begin{cases} \frac{1}{\Gamma(n+2)}\,\ZL\mkuu\UV^{2n+2} \; &\text{for } n>-2\mku ,\\[0.3em]
 0 & \text{for } n\le -2\mku ,\end{cases}
\label{eq:WIMellin}
\end{equation}
in particular $Q_1[\RL] = \frac{1}{2}\ZL\mkuu\UV^4$, $Q_0[\RL]= \ZL\mkuu\UV^2$ and $Q_{-1}[\RL] = \ZL$. Note that
the dots in eq.\ \eqref{eq:HeatExpGen} vanish identically for the optimized cutoff since $Q_n[\RL]= 0$ for all
$n\le -2$. Hence, the following equation is an exact identity in that case:
\begin{equation}
\begin{split}
 \Big(\RL\big(-\hB_y\big)\delta(y-z)\Big)\Big|_{z\to y}
 =\frac{1}{4\pi}\shgy\,\ZL\left(\textstyle \frac{1}{2}\UV^4 +\frac{1}{6}\UV^2\,\hR
 +\frac{1}{60}\hR^2 +\frac{1}{30}\mku\hB\mku\hR\right).
\end{split}
\label{eq:HeatExpOpt}
\end{equation}

This expression can be inserted into eq.\ \eqref{eq:TrhRL} now. Then the metric derivative
$\frac{\hg_\mn(x)}{\shgx}\frac{\delta}{\delta\hg_\mn(x)}$ acts on all terms on the RHS of eq.\ \eqref{eq:HeatExpOpt}.
From eq.\ \eqref{eq:TracedDerivative} we already know that $\frac{\hg_\mn(x)}{\shgx}\mku\frac{\delta}{\delta\hg_\mn(x)}
\,\shgy = \delta(x-y)$, and using \eqref{eq:TracedDerivative} yields
\begin{align}
 \frac{\hg_\mn(x)}{\shgx}\frac{\delta}{\delta\hg_\mn(x)}\left(\shgy\,\hR(y)\right) &= -\hB\mkuu\delta(x-y)\,, \\
 \frac{\hg_\mn(x)}{\shgx}\frac{\delta}{\delta\hg_\mn(x)}\left(\shgy\,\hR^2(y)\right) &= -2\,\hR\,\hB\mkuu\delta(x-y)
 -\hR^2\mku\delta(x-y)\,, \\
 \frac{\hg_\mn(x)}{\shgx}\frac{\delta}{\delta\hg_\mn(x)}\left(\shgy\,\hB_y\mku\hR(y)\right) &= -\hB\mku\big[\hR\,
 \delta(x-y)\big]- \hB^2\mku\delta(x-y)\,.
\end{align}
By means of these relations we can finally compute the integrals in \eqref{eq:TrhRL}. After integrating by parts all
those terms with a Laplace operator $\hB$ acting on a delta-function we obtain the result
\begin{equation}[b]
\begin{aligned}
 \Tr_\UV\!&\Big[\hRL(x)\,\big(\Gamma_\UV^\text{L}{}^{(2)}+\RL\big)^{-1}\Big]\\
 ={} &\frac{1}{8\pi}\Bigg\{ \frac{\UV^2}{1+2\mku\mu\,\e^{2\mku\phi(x)}} -\frac{1}{3}\,\hB\left[\frac{1}{1+2\mku\mu\,
 \e^{2\mku\phi(x)}}\right]\\
 &\phantom{\frac{1}{8\pi}\Bigg\{}- \frac{1}{15}\,\hB\bigg[\frac{\UV^{-2}\hR(x)}{1+2\mku\mu\,\e^{2\mku\phi(x)}}\bigg]
 - \frac{1}{30}\,\hR^2(x)\mku\frac{\UV^{-2}}{1+2\mku\mu\,\e^{2\mku\phi(x)}} \\
 &\phantom{\frac{1}{8\pi}\Bigg\{}-\frac{1}{15}\,\hR(x)\,\hB \left[\frac{\UV^{-2}}{1+2\mku\mu\,\e^{2\mku\phi(x)}}\right]
 - \frac{1}{15}\,\hB^2\left[\frac{\UV^{-2}}{1+2\mku\mu\,\e^{2\mku\phi(x)}}\right] \Bigg\}.
\end{aligned}
\label{eq:WITraceTermOpt}
\end{equation}
This is an \emph{exact relation for the optimized cutoff}; there are no further higher order terms. Note that the last
two lines in \eqref{eq:WITraceTermOpt} are suppressed in the limit $\UV\to\infty$. Moreover, we point out
that \emph{there is no contribution proportional to $\hR$ only}. This is crucial for a discussion concerning central
charges, see Section \ref{sec:WIImp}.
\medskip

\noindent
\textbf{(2) Evaluation of \bm{$\big\langle x\big|\,\RL\,\big(\Gamma_\UV^\text{L}{}^{(2)}+\RL\big)^{-1}\big|x\big
\rangle$}}.\\
Making use of eq.\ \eqref{eq:WTCOPropDiag} we find that the propagator can be pulled out of $\langle x|\cdot|x\rangle$:
\begin{equation}
\begin{split}
 \big\langle x\big|\,\RL\,\big(\Gamma_\UV^\text{L}{}^{(2)}+\RL\big)^{-1}\big|x\big\rangle
 &= \Big\langle x\Big|\,\RL\,\Big[\ZL\big(\UV^2+2\mku\mu\mku\UV^2\,\e^{2\phi}\big)\Big]^{-1}\Big|x\Big\rangle \\
 &= \frac{1}{\ZL\big(\UV^2+2\mku\mu\mku\UV^2\,\e^{2\mku\phi(x)}\big)}\,\big\langle x\big|\RL\big|x\big\rangle\,.
\end{split}
\end{equation}
The term $\big\langle x\big|\RL\big|x\big\rangle$ can be obtained by means of the heat kernel formalism. It is given by
\begin{equation}
 \big\langle x\big|\RL\big|x\big\rangle = \frac{1}{4\pi}\sum_{n=0}^\infty Q_{1-n}[\RL]\, a_n(x,x),
\end{equation}
where the Seeley--DeWitt coefficients $a_n(x,x)$ are defined in App.\ \ref{app:Heat}. The generalized Mellin
transforms have already been computed above, see eq.\ \eqref{eq:WIMellin}: $Q_{1-0}[\RL] = \frac{1}{2}\ZL\mkuu\UV^4$,
$Q_{1-1}[\RL] = \ZL\mkuu\UV^2$, $Q_{1-2}[\RL] = \ZL$ and $Q_{1-n}[\RL] = 0$ for all $n\ge 3$.

Putting all pieces together, we arrive at the final result
\begin{equation}[b]
\begin{aligned}
 &\big\langle x\big|\,\RL\,\big(\Gamma_\UV^\text{L}{}^{(2)}+\RL\big)^{-1}\big|x\big\rangle \\
 &= \frac{1}{8\pi}\;\frac{1}{1+2\mku\mu\,\e^{2\mku\phi(x)}}\left\{ \UV^2+\frac{1}{3}\mku\hR(x)
 +\frac{1}{30}\,\UV^{-2}\mku\hR^2(x)+\frac{1}{15}\,\UV^{-2}\mku\hB\mkuu\hR(x)\right\}
\end{aligned}
\label{eq:WIEVTermOpt}
\end{equation}
Again, this equation is \emph{exact for the optimized cutoff}. The last two terms on the RHS of eq.\
\eqref{eq:WIEVTermOpt} are suppressed in the limit $\UV\to\infty$.

Unlike \eqref{eq:WITraceTermOpt}, eq.\ \eqref{eq:WIEVTermOpt} contains a small contribution purely proportional to
$\hR$ alone: By expanding $\frac{1}{1+2\mku\mu\,\e^{2\mku\phi(x)}}=\frac{1}{1+2\mku\mu}+\mO(\phi)$ we find
\begin{equation}
 \big\langle x\big|\,\RL\,\big(\Gamma_\UV^\text{L}{}^{(2)}+\RL\big)^{-1}\big|x\big\rangle
 = \frac{1}{24\pi}\,\frac{1}{1+2\mku\mu}\,\hR + \text{const}+\mO\big(\phi,\hR^2,\hB\mku\hR\big).
\end{equation}
For the exponential metric parametrization we have $\frac{1}{1+2\mku\mu}\approx 0.774$, while the linear
parametrization amounts to $\frac{1}{1+2\mku\mu}=0.76$. These numbers are indeed ``small'' since they appear in the
Ward identity \eqref{eq:WISplitSym} as prefactors of $\frac{1}{24\pi}\,\hR$, so they are to be compared with $c+1=26$
for the exponential parametrization ($c+1=20$ for the linear parametrization).

%----------------------------------------------------------------------------------------------------------------------
\chapter*{Acknowledgments}
\addcontentsline{toc}{chapter}{Acknowledgments}
\markboth{Acknowledgments}{}
%----------------------------------------------------------------------------------------------------------------------

This thesis could not have been completed without the guidance, encouragement and support of many people to whom I am
indebted. I would like to thank my supervisor, Prof.\ Martin Reuter, for many invaluable suggestions and for his sheer
inexhaustible patience during our discussions. Also, I appreciated the constant interest shown by my second supervisor,
Prof.\ Stefan Scherer, in the current status and the topic of my work. Special thanks go to my colleagues, Daniel
Becker, Adriano Contillo, Jan-Eric Daum, Maximilian Demmel, Kai Groh, Ulrich Harst, Carlo Pagani, Stefan Rechenberger,
Gregor Schollmeyer and Omar Zanusso, for countless stimulating and helpful debates about physics and beyond.
I am very lucky to have good friends, in particular Christopher Busch, Jan Fischer, Christopher Heep and Konstantin
Ott, with whom I have spent so many pleasant hours. Furthermore, I am thankful to Ana Juri\v{c}i\'{c} for supporting
me and cheering me up brilliantly during the final stage of this work.
Finally, I would like to express my deepest gratitude to my family for always standing by my side: Thank you, Mom,
Dad, David, Simon and Lena!

%\include{Chapters/CurriculumVitae}

%%% Style of the Bilbliography
\bibliographystyle{phdbib}

\begin{raggedright}

%%% File with Bib-entries (*.bib)
\bibliography{Chapters/Literature}

\end{raggedright}

%----------------------------------------------------------------------------------------------------------------------

\end{Spacing}

\end{document}